\documentclass[iop]{emulateapj}
\usepackage[breaklinks,colorlinks,citecolor=blue,linkcolor=magenta,urlcolor=blue]{hyperref}
\usepackage{subfigure}

\newcommand*{\mysub}[2]{\ensuremath{#1_{\mathrm{#2}}}}

\newcommand*{\Omegam}{\mysub{\Omega}{m}}

\newcommand*{\Omegal}{\mysub{\Omega}{\Lambda}}

\newcommand*{\LCDM}{$\Lambda$CDM }

\newcommand*{\ltsim}{\ {\raise-.75ex\hbox{$\buildrel<\over\sim$}}\ }
\newcommand*{\gtsim}{\ {\raise-.75ex\hbox{$\buildrel>\over\sim$}}\ }
\newcommand*{\proptosim}{\ {\raise-.75ex\hbox{$\buildrel\propto\over\sim$}}\ }

\newcommand*{\Chandra}{{\it Chandra }}

\newcommand{\Herschel}{{\it Herschel }}

\newcommand{\Spitzer}{{\it Spitzer }}

\shorttitle{The Candidate Cluster and Protocluster Catalog (CCPC) II }
\shortauthors{Franck \& McGaugh}

\begin{document}
\title{The Candidate Cluster and Protocluster Catalog (CCPC) II: Spectroscopically identified structures spanning $2<\lowercase{z}<6.6$}


\author{J.R. Franck\altaffilmark{1},
	 S.S. McGaugh\altaffilmark{1}}

 \altaffiltext{1}{Case Western Reserve University, 10900 Euclid Ave., Cleveland, OH 44106}

\begin{abstract}
The Candidate Cluster and Protocluster Catalog (CCPC) is a list of objects at redshifts $z > 2$ composed of galaxies with spectroscopically confirmed redshifts that are coincident on the sky and in redshift. These protoclusters are identified by searching for groups in volumes corresponding to the expected size of the most massive protoclusters at these redshifts. In CCPC1 we identified 43 candidate protoclusters among 14,000 galaxies between $2.74 < z < 3.71$. Here we expand our search to more than 40,000 galaxies with spectroscopic redshifts $z > 2.00$, resulting in an additional 173 candidate structures. The most significant of these are 36 protoclusters with overdensities $\delta_{gal}>7$. We also identify three large proto-supercluster candidates containing multiple protoclusters at $z = 2.3,3.5$ and $z=6.56$. Eight candidates with $N\ge10$ galaxies are found at redshifts $z > 4.0$. The last system in the catalog is the most distant spectroscopic protocluster candidate known to date at $z=6.56$. 
\end{abstract}

\keywords{galaxies: clusters: general - galaxies: clusters: catalog}

\section{Introduction} \label{sec:intro}

Protoclusters are the diffuse, extended building blocks that will collapse into a galaxy cluster
at the current epoch. Structures at high redshift have power as tests of cosmology, as the maximum mass that can collapse and virialize
at a given epoch depends on the mass density (\Omegam), the power spectrum ($\sigma_{8}$), and the dark energy equation of state ($w$) \citep{1998ApJ...504....1B,vik09,mor11}. These primeval
 systems are also unique laboratories in which we can observe galaxies in the Universe assemble into
larger collections and evolve in dense environments. Galaxy formation and hierarchical accretion scenarios
 can be examined by compiling large numbers of high redshift galaxies at various times \citep{coo14,wyl14}.
Protocluster galaxies have been found to have enhanced mass assembly \citep{cas15} and evolution \citep{hat11}. Star formation rates of $10^{4}$ $M_{\odot}$ $yr^{-1}$ for a protocluster \citep{2014MNRAS.439.1193C} are in 
significant excess of hydrodynamic simulation expectations \citep{2015MNRAS.450.1320G,2015arXiv151201561C}. 
As a result, the search for high redshift clusters and protoclusters within the astrophysical community has 
been rapidly expanded in the last few years.


Less than two decades ago, protoclusters were 
 relatively unknown until rich, Lyman-Break Galaxy (LBG) overdensities were discovered by \citet{1998ApJ...492..428S} at $z\sim3$ and
$Ly\alpha$ emitters (LAEs) at $z\sim4$ \citep{ven02}. Until recently, few systematic surveys for $z\gtrsim2$ systems existed. 
With the advent of Clusters Around Radio Loud-AGN (CARLA) \Spitzer survey \citep{wyl13}, the Search for
Protoclusters with \Herschel \citep{rig14}, data mining the GOODS-N/GOODS-S fields \citep{2009ApJ...691L..33K,2009AnA...501..865S,2015JKAS...48...21K},
 and finding structures within the COSMOS/UltraVISTA field
\citep{die13,chi14}, the number of photometrically/spectroscopically identified candidate protoclusters is now in the hundreds.
Catalogs of pure spectroscopically identified protoclusters in the high redshift universe, as in \citet{2007AnA...461..823V},
typically have fewer than 10 such candidates, however.  Most of these previously identified structures are 
found within $z\sim3\pm1$ \citep{2007AnA...461..823V,wyl14,ccpc1}, with a few at $z\sim6$ \citep{utsumi10,toshikawa14}.
 These objects were identified with varying instruments,
selection techniques, and fields-of-view (FOV) that rarely encompass the entire system.

It was this understanding that prompted a systematic search of archival data in a simple manner to
identify high redshift ($z>2$) clusters and protoclusters. 
This work builds on the original harvest of 43 members in the Candidate Cluster and Protocluster Catalog (CCPC) 
between $2.74<z<3.71$ \cite[hereafter known as CCPC1][]{ccpc1}. These structures were identified as galaxy 
overdensities in fixed comoving volumes. In this second paper (hereafter CCPC2), we have extended the redshift range of our
search to $2.00<z<7.00$ using a similar search method. We eschew the common search technique of using High Redshift Radio Galaxies
 (HzRGs) as biased tracers of structure \citep{ven02,wyl13,rig14}, but we have recovered overdensities identified in this way.
  Presented in this work are 173 additional protocluster candidates, 23 of which have been found in the literature.

We have organized CCPC2 as follows: Section~\ref{sec:ccpc} discusses the archival data from which
these structures are gleaned, the algorithm that identifies structures, calculation of the overdensity
of each system, and the explanation of the two mass estimators. In Section~\ref{sec:dis}, we explore the nature
of these overdensities, their significance as structures, compare their properties to simulations, and discuss three 
supercluster candidates.

We assume a standard cosmology, adopting $H_0 = 70$ km s$^{-1}$  Mpc$^{-1}$, a matter density of $\Omegam = 0.3$, and $\Omegal = 0.7$. The Universe is 3.2 Gyr old
at the redshift of $z=2.0$, and has a comoving angular scale of 1.51 Mpc arcmin$^{-1}$. At $z=7$, the corresponding age is 0.75 Gyr, while the angular scale is 2.51 comoving Mpc arcmin$^{-1}$  \citep{wri06}. 

\section{The Candidate Cluster and Protocluster Catalog (CCPC) II} \label{sec:ccpc}

\begin{deluxetable*}{c c c c c c c c c c c}
\tablewidth{0pt}
\tablecolumns{11}
\tablecaption{Candidate Cluster and Protocluster Catalog (CCPC) II - Best Candidates}
\tablehead{
\colhead{Candidate}	&\colhead{RA}	&\colhead{DEC}&\colhead{Redshift}	&\colhead{$\sigma_z$}&\colhead{$N$}& \colhead{$N_{R\le10}$} & \colhead{Overdensity}	&\colhead{Cluster}	&\colhead{Q}	&	\colhead{Recovered}	\\
\colhead{Name}&\colhead{($deg$)}&\colhead{($deg$)}&\colhead{($z_{avg}$)}&	\colhead{}&	\colhead{}	&\colhead{cMpc}&\colhead{($\delta_{gal}$)}	&\colhead{Probability ($\%$)}	&	\colhead{}&	\colhead{Reference}
}	
\startdata																								
CCPC-z20-002	&	222.20	&	8.92	&	2.002	&	0.008	&	11	&	7	&	9.38	$\pm$	5.34		&	100.0	&	1	&	8\\
CCPC-z20-003	&	29.62	&	-25.05	&	2.018	&	0.004	&	10	&	10	&	19.43	$\pm$	13.06		&	100.0	&	1	&	1,2\\
CCPC-z20-009	&	150.04	&	2.21	&	2.098	&	0.005	&	10	&	4	&	13.15	$\pm$	6.54		&	100.0	&	1	&	4,5\\
CCPC-z21-004	&	175.15	&	-26.47	&	2.155	&	0.007	&	24	&	24	&	6.34	$\pm$	3.36		&	100.0	&	1	&	9,1\\
CCPC-z21-005	&	214.31	&	52.40	&	2.160	&	0.007	&	5	&	3	&	9.07	$\pm$	6.93		&	100.0	&	1	&	\\
CCPC-z21-006	&	334.35	&	0.32	&	2.172	&	0.005	&	4	&	3	&	18.85	$\pm$	13.55		&	100.0	&	1	&	\\
CCPC-z21-007	&	356.58	&	12.80	&	2.174	&	0.002	&	7	&	7	&	17.27	$\pm$	10.57		&	100.0	&	1	&	\\
CCPC-z21-008	&	149.98	&	2.11	&	2.179	&	0.002	&	5	&	1	&	9.41	$\pm$	5.38		&	100.0	&	2	&	4\\
CCPC-z22-007	&	255.20	&	64.26	&	2.296	&	0.008	&	32	&	30	&	7.77	$\pm$	2.90		&	100.0	&	1	&	6\\
CCPC-z23-002	&	334.46	&	0.14	&	2.309	&	0.009	&	4	&	3	&	11.45	$\pm$	9.05		&	100.0	&	1	&	\\
CCPC-z23-003	&	214.39	&	52.49	&	2.333	&	0.008	&	4	&	4	&	11.73	$\pm$	10.66		&	100.0	&	1	&	\\
CCPC-z23-007	&	258.53	&	50.27	&	2.390	&	0.005	&	7	&	6	&	16.63	$\pm$	14.52		&	100.0	&	1	&	1,20\\
CCPC-z24-003	&	164.20	&	-3.64	&	2.426	&	0.005	&	7	&	7	&	15.06	$\pm$	8.92		&	100.0	&	1	&	\\
CCPC-z24-005	&	150.00	&	2.26	&	2.442	&	0.009	&	14	&	8	&	9.27	$\pm$	4.93		&	100.0	&	1	&	4,10\\
CCPC-z25-002	&	255.18	&	64.17	&	2.537	&	0.002	&	4	&	4	&	19.86	$\pm$	13.41		&	100.0	&	1	&	\\
CCPC-z25-003	&	143.36	&	28.77	&	2.548	&	0.003	&	5	&	5	&	10.89	$\pm$	7.70		&	100.0	&	1	&	\\
CCPC-z25-007	&	216.14	&	22.84	&	2.581	&	0.007	&	5	&	2	&	10.90	$\pm$	6.72		&	100.0	&	1	&	\\
CCPC-z27-012	&	16.48	&	-25.81	&	2.758	&	0.008	&	4	&	3	&	12.83	$\pm$	10.91		&	100.0	&	1	&	\\
CCPC-z28-011	&	36.36	&	-4.32	&	2.820	&	0.006	&	4	&	2	&	9.10	$\pm$	7.53		&	100.0	&	1	&	\\
CCPC-z28-016	&	36.27	&	-4.28	&	2.866	&	0.006	&	5	&	1	&	15.46	$\pm$	11.42		&	100.0	&	1	&	\\
CCPC-z29-009	&	136.34	&	34.14	&	2.905	&	0.010	&	5	&	5	&	13.33	$\pm$	9.13		&	100.0	&	1	&	\\
CCPC-z29-011	&	339.95	&	11.87	&	2.925	&	0.008	&	13	&	3	&	9.62	$\pm$	5.14		&	100.0	&	1	&	\\
CCPC-z29-013	&	13.31	&	12.63	&	2.934	&	0.002	&	5	&	4	&	14.67	$\pm$	10.21		&	100.0	&	1	&	\\
CCPC-z31-015	&	339.87	&	11.88	&	3.148	&	0.008	&	9	&	2	&	8.15	$\pm$	4.56		&	100.0	&	1	&	\\
CCPC-z32-007	&	46.15	&	-0.19	&	3.233	&	0.007	&	5	&	4	&	11.47	$\pm$	9.32		&	100.0	&	1	&	\\
CCPC-z33-006	&	150.07	&	2.28	&	3.303	&	0.008	&	4	&	4	&	17.49	$\pm$	13.89		&	100.0	&	2	&	\\
CCPC-z33-007	&	334.35	&	0.07	&	3.310	&	0.012	&	7	&	5	&	11.52	$\pm$	9.15		&	100.0	&	1	&	\\
CCPC-z33-010	&	216.12	&	22.83	&	3.379	&	0.009	&	7	&	3	&	10.44	$\pm$	7.10		&	100.0	&	1	&	\\
CCPC-z36-007	&	34.54	&	-5.30	&	3.688	&	0.010	&	5	&	2	&	15.22	$\pm$	16.25		&	90.0	&	2	&	\\
CCPC-z44-003	&	189.15	&	62.23	&	4.424	&	0.010	&	5	&	3	&	13.79	$\pm$	12.03		&	90.0	&	1	&	\\
CCPC-z26-001	&	339.84	&	11.81	&	2.617	&	0.003	&	4	&	4	&	7.33	$\pm$	6.16		&	85.6	&	1	&	\\
CCPC-z26-006	&	255.14	&	64.22	&	2.688	&	0.005	&	5	&	5	&	7.24	$\pm$	5.18		&	85.6	&	1	&	\\
CCPC-z27-008	&	36.39	&	-4.51	&	2.729	&	0.006	&	6	&	2	&	7.27	$\pm$	4.83		&	85.6	&	1	&	\\
CCPC-z31-008	&	339.89	&	11.88	&	3.104	&	0.007	&	8	&	5	&	7.70	$\pm$	4.58		&	85.6	&	1	&	\\
CCPC-z31-017	&	36.76	&	-4.56	&	3.187	&	0.006	&	11	&	3	&	7.26	$\pm$	3.67		&	85.6	&	1	&	\\
CCPC-z34-006	&	36.54	&	-4.63	&	3.472	&	0.013	&	6	&	1	&	7.82	$\pm$	5.40		&	85.6	&	2	&	
\enddata
\tablecomments{The names and positions (1$^{st}$ through 3$^{rd}$ columns) of the most overdense candidate protoclusters, with redshifts corresponding to the average value for the system, with their dispersion $\sigma_z$. The number of galaxies within $R=10$ and 20 cMpc from the search center are included as $N_{R\le10}$ (7$^{th}$ column) and $N$ (6$^{th}$), respectively. The 8$^{th}$ column contains a measure of the galaxy overdensity ($\delta_{gal}$), which is explained in Section 2.3. A probability is assigned to each system that it will collapse into a cluster by $z=0.$ These probability distribution functions at the relevant redshifts can be found in Figure 8 of \citet{chi13} from analysis of protocluster $\delta_{gal}$ values in the Millennium simulation. Entries in the table are arranged by their collapse probabilities in descending order. The `Q' column provides a rating of the source $Q$uality of redshifts as identified within NED. The best quality redshift systems are listed as a `1'. If we recover an overdensity that was previously identified, the discovery references are listed in the last column. References: (1) \citet{2012ApJ...749..169G}, (2) \citet{2013AnA...559A...2G}, (4) \citet{die13}, (5) \citet{2014ApJ...795L..20Y}, (6) \citet{2005ApJ...626...44S}, (8) \citet{gob13}, (9) \citet{1997AnA...326..580P}, (10) \citet{chi14}, (20) \citet{1999AJ....118.2547K} } \label{tab:best_cat}
\end{deluxetable*}

\begin{deluxetable*}{c c c c c c c}
\tablewidth{0pt}
\tablecolumns{7}
\tablecaption{CCPC II: Mass Estimates-Best Candidates}
\tablehead{
\colhead{Candidate} & \colhead{$R_e$}	&\colhead{$\sigma$}	&\colhead{Virial Mass} &	\colhead{$\delta_{m}$}	& \colhead{Overdensity}	&	\colhead{Overdensity Mass}	\\
\colhead{Name} & \colhead{(Mpc)}	&\colhead{(km s$^{-1}$)}	&\colhead{Estimate ($10^{14}$ $M_{\odot}$)}	&\colhead{($b=3$)}	&\colhead{Volume (cMpc$^3$)}	&\colhead{Estimate ($10^{14}$ $M_{\odot}$)}
} 
\startdata
CCPC-z20-002	&	4.5	&	749	&	14.7	&	3.13	&	5361	&	9.8	\\
CCPC-z20-003	&	2.3	&	367	&	1.8	&	6.48	&	125	&	0.4	\\
CCPC-z20-009	&	4.3	&	446	&	4.9	&	4.38	&	589	&	1.4	\\
CCPC-z21-004	&	3.4	&	701	&	9.6	&	2.11	&	519	&	0.7	\\
CCPC-z21-005	&	5.0	&	689	&	13.8	&	3.02	&	799	&	1.5	\\
CCPC-z21-006	&	2.9	&	479	&	3.9	&	6.28	&	58	&	0.2	\\
CCPC-z21-007	&	1.2	&	156	&	0.2	&	5.76	&	13	&	$<0.1$	\\
CCPC-z21-008	&	3.7	&	211	&	1.0	&	3.14	&	176	&	0.3	\\
CCPC-z22-007	&	3.1	&	709	&	8.9	&	2.59	&	3032	&	4.7	\\
CCPC-z23-002	&	2.6	&	783	&	9.3	&	3.82	&	60	&	0.1	\\
CCPC-z23-003	&	2.3	&	711	&	6.8	&	3.91	&	171	&	0.4	\\
CCPC-z23-007	&	1.5	&	442	&	1.7	&	5.54	&	82	&	0.2	\\
CCPC-z24-003	&	2.4	&	399	&	2.2	&	5.02	&	327	&	0.8	\\
CCPC-z24-005	&	4.1	&	770	&	14.1	&	3.09	&	9110	&	15.5	\\
CCPC-z25-002	&	1.3	&	208	&	0.3	&	6.62	&	28	&	0.1	\\
CCPC-z25-003	&	2.2	&	295	&	1.1	&	3.63	&	116	&	0.2	\\
CCPC-z25-007	&	3.1	&	620	&	6.9	&	3.63	&	215	&	0.4	\\
CCPC-z27-012	&	1.3	&	632	&	2.9	&	4.28	&	217	&	0.5	\\
CCPC-z28-011	&	2.6	&	504	&	3.8	&	3.03	&	280	&	0.5	\\
CCPC-z28-016	&	4.0	&	429	&	4.2	&	5.15	&	458	&	1.2	\\
CCPC-z29-009	&	2.2	&	757	&	7.2	&	4.44	&	27	&	0.1	\\
CCPC-z29-011	&	4.1	&	628	&	9.3	&	3.21	&	4579	&	8.0	\\
CCPC-z29-013	&	1.7	&	189	&	0.3	&	4.89	&	56	&	0.1	\\
CCPC-z31-015	&	3.7	&	558	&	6.6	&	2.72	&	2011	&	3.0	\\
CCPC-z32-007	&	2.6	&	504	&	3.8	&	3.82	&	280	&	0.6	\\
CCPC-z33-006	&	1.8	&	578	&	3.5	&	5.83	&	115	&	0.3	\\
CCPC-z33-007	&	3.2	&	818	&	12.3	&	3.84	&	101	&	0.2	\\
CCPC-z33-010	&	3.4	&	611	&	7.3	&	3.48	&	730	&	1.4	\\
CCPC-z36-007	&	3.3	&	634	&	7.7	&	5.07	&	165	&	0.4	\\
CCPC-z44-003	&	1.4	&	575	&	2.6	&	4.60	&	456	&	1.0	\\
CCPC-z26-001	&	0.7	&	245	&	0.2	&	2.44	&	3	&	$<0.1$	\\
CCPC-z26-006	&	2.2	&	366	&	1.7	&	2.41	&	139	&	0.2	\\
CCPC-z27-008	&	3.5	&	463	&	4.3	&	2.42	&	5570	&	7.9	\\
CCPC-z31-008	&	2.6	&	541	&	4.3	&	2.57	&	1038	&	1.6	\\
CCPC-z31-017	&	3.7	&	462	&	4.5	&	2.42	&	10025	&	14.6	\\
CCPC-z34-006	&	4.0	&	878	&	17.7	&	2.61	&	270	&	0.4	
\enddata
\tablecomments{The mass estimates for the largest overdensities in CCPC2 (same order as Table~\ref{tab:best_cat}). The effective radii ($R_e$, in physical units) and velocity dispersions ($\sigma$) are used to compute a virial mass estimate, with the caveat that these systems are not expected to be in equilibrium at the relevant redshifts. We obtain the mass overdensity $\delta_{m}$ by assuming galaxies are linearly biased tracers of mass with a slope of $b=3$. From this and the volume contained within the galaxy distribution we can compute an expected mass of the cluster \citep{1998ApJ...492..428S}. This assumes that the total volume will collapse into a single halo at $z=0$.
} \label{tab:best_mass}
\end{deluxetable*}

\subsection{Data}

In CCPC1, a NASA Extragalactic Database (NED\footnote{\url{http://ned.ipac.caltech.edu/}}) search was used to compile a list of $\sim$14,000 galaxies with spectroscopic redshifts between $2.74<z<3.71$. Many of these galaxies were found in Hubble Deep Fields, and spectroscopic follow-up of Lyman Break Galaxies (LBGs) \citet{2003ApJ...592..728S}. In this work, we have expanded the redshift range to $2.00<z<10.23$, which constitutes a galaxy list of $\>$47,000 objects. The upper limit of $z=10.23$ is an extreme example; only 9 galaxies have $z> 7.5$ in our list. The NED database holds published redshift uncertainties for 813 of the $\sim1400$ candidate protocluster galaxies, with a mean value of $\sigma=0.001$. At a redshift of $z=3$, this represents a comoving distance uncertainty of $1.0$ Mpc \citep{wri06}.

Nearly 200 sources of spectroscopic measurements were used to identify candidate clusters (listed in Table~\ref{tab:tot_ref}), although many are concentrated in a few catalogs.
The single largest source of spectroscopic redshifts that were identified as protocluster galaxies is from \citet{2003ApJ...592..728S}. These galaxies were identified as Lyman Break Galaxies initially (with additional color criteria), and then followed up spectroscopically to a limiting magnitude of $R_{AB}=25.5$ using the Keck telescopes over a field of view of 0.38 square degrees. Many other spectroscopic surveys utilized very deep, multi-wavelength fields (e.g. CANDELS GOODS, Hubble Deep/Ultra Deep) to identify candidate high redshift galaxies with various color cuts, dropouts, or \Chandra emission \citep{2004AnA...418..885N,2005AnA...439..845L,2006ApJ...653.1004R,2010AnA...512A..12B}. These galaxies would then be targeted by VIMOS/FORS/MOIRCS (or a similar instrument) to confirm their redshift. 

Another common source of redshifts is from surveys that capitalize on strong line emission from faint star forming galaxies, either with  $Ly\alpha$ or $H\alpha$. By using a narrowband (NB) filter which selects out one of these lines at a specific redshift, one can effectively identify a large number of sources efficiently. These candidates can then be targeted using a large telescope to confirm their redshift. This technique has been used to identify a number of high redshift protoclusters \citep{2003AnA...407..147F,2007AnA...461..823V,coo14}. An interesting observation that was noticed by \citet{2013MNRAS.428.1551K} is that $Ly\alpha$ and $H\alpha$ are rarely simultaneously observed in star-forming galaxies within a protocluster at $z=2.16$, suggesting that they may be entirely different populations. They could also have varying amounts and distribution of dust. 
 Simulations suggest that roughly $90\%$ of protocluster galaxies are actively forming stars at these redshifts \citep{2015arXiv151201561C}.

\subsection{Candidate Protocluster Criteria}

While the most massive, $z=0$ clusters have radii on the order of a few Mpc, the components that form these systems (i.e. protoclusters) are much more extended at higher redshift. Previously identified structures at high redshift have observed sizes larger than 50 comoving Mpc (cMpc) \citep{2003ApJ...586L.111S,2005ApJ...634L.125M,lee14}. These seemingly large sizes have a theoretical basis within \LCDM simulations as well. \citet{chi13} and \citet{2015MNRAS.452.2528M} mined the Millennium simulation \citep{2005Natur.435..629S} for clusters (collapsed halos with $\gtrsim 10^{14}$ $M_{\odot}$) at $z=0$. Using the semi-analytic model of \citet{2011MNRAS.413..101G} to trace the galaxy distributions of the primordial clusters backwards, they analyzed the evolution of these systems. \citet{2015MNRAS.452.2528M} found that at $z=2$, 90$\%$ of the stellar mass of a protocluster can be found spread across 35 $h^{-1}$ cMpc, which grows to more than 40 $h^{-1}$ cMpc at $z=5$. This example is typical for the largest mass systems (clusters with $M_{z=0}>10^{15}$ $M_{\odot}$), while a cluster with $M_{z=0}=10^{14}$ $M_{\odot}$ might have a modest size of $\sim20$ $h^{-1}$ cMpc. On the sky, the most massive protoclusters can span more than 0.5 degrees, while the smallest mass clusters are roughly half of that \citep{2015MNRAS.452.2528M}.

This can be problematic when searching for such systems, in that many deep, spectroscopic surveys (with the exception of the CANDELS GOODS-S field) would not encompass the full breadth of the most massive protoclusters. In \citet{ccpc1}, we plotted the positions of field and CCPC1 protocluster galaxies within search radii of $20$ cMpc to illustrate the distribution of members with respect to the survey widths in the Appendix.  Surveys rarely extended beyond the probed search radius. Similarly, \citet{2015MNRAS.452.2528M} make an example of two candidate systems found within the literature \citep{2013MNRAS.428.1551K,coo14} that have particularly small fields of view which could only target the innermost halo of a protocluster. This imposes a selection effect, in that most ($\sim90\%$) simulated protoclusters do not have a single, `main' halo at these redshifts \citep{2015MNRAS.452.2528M}.


We took an initial list of $\sim$14,000 galaxies and identified galaxy groups of 3 or more within a radius of 2 arcminutes and a $\Delta z<0.03$ in CCPC1. Individual galaxies are required to be more than 3'' from one another to be considered unique. From these groups, we expanded our search to a radius of 20 cMpc on the sky and a redshift depth of $\pm 20$ cMpc. Any cylindrical volume that contained 4 or more galaxies and had a galaxy overdensity of $\delta_{gal}>0.25$ (calculation found in the following subsection) was considered a candidate protocluster and included in CCPC1.

These requirements are based on the need for a simple, adaptable, and effective means of identifying galaxy structures from a variety of surveys. The large search radius enables wide surveys to be adequately probed corresponding to the size of the largest protoclusters. The modest richness requirement of 4 or more galaxies is sensitive to surveys of small volumes with few expected galaxies. These two extremes of survey depths are moderated by the density requirement of $\delta_{gal}>0.25$. The $N\ge4$ galaxy requirement serves primarily as a signpost from which to calculate the overdensity. A more detailed discussion of the algorithm will follow.

In this paper, we have removed the intermediate step of finding groups of 3 or more galaxies within a 2 arcminute radius. The overdensity of these candidate protoclusters, which may not be accompanied by a dense, central knot of galaxies, is of greater importance than the possible chance alignment of galaxies on the sky. Therefore, any group of 4 or more galaxies that also exhibits a galaxy overdensity of $\delta_{gal}>0.25$ is considered to be part of the CCPC. The following subsection will describe the overdensity calculation in detail. In Section~\ref{sec:test} we justify the removal of the intermediate step statistically.

Identifying structure in large, high redshift data sets is not a novel exercise. \citet{die13} found infalling groups of spectroscopic galaxies separated by small physical separations in space in the zCOSMOS field. The GOODS-N and GOODS-S fields have been the subject of many searches for overdensities found with photometric redshifts and supplementary spectroscopic catalogs at $z>2$ \citep{2009ApJ...691L..33K,2009AnA...501..865S,2015JKAS...48...21K}. Indeed, we may have recovered a number of these overdensities, which have $0\le N\le4$ spectroscopic member galaxies in \citet{2015JKAS...48...21K}. In CCPC1 \citep{ccpc1}, our candidate protocluster CCPC-z37-001 was identified near a photometric redshift overdensity by \citet{2009ApJ...691L..33K}, while CCPC-z27-004,CCPC-29-004, and CCPC-z34-001 most likely coincide with overdensities found in \citet{2015JKAS...48...21K}. CCPC-z22-001,CCPC-z23-001 and CCPC-z24-001 (all featured in CCPC2) have similar positions and redshifts to the overdensities reported by \citet{2009AnA...501..865S}, while CCPC-z25-004, CCPC-z25-005, CCPC-z40-001, and CCPC-z42-001 may coincide with candidate structures in \citet{2015JKAS...48...21K}. CCPC-z28-002, CCPC-z34-002, and CCPC-z37-001 in \citet{ccpc1} are all overdensities originally found in the VIMOS GOODS-S spectroscopic survey by \citet{2010AnA...512A..12B}.

The CCPC is unique in that it applied a single and simple algorithm to all archival spectroscopic data available.
The number of galaxy overdensities identified in this paper (hereafter referred to as CCPC2) is 173, the 36 strongest of which are listed in Table~\ref{tab:best_cat}. The division of these `Best' candidates is explained in Section 2.3 (Overdensities). The total list of candidates can be found in Table~\ref{tab:tot_cat}. The RA/DEC coordinates in the Table are centered on the galaxy from which the number of members is maximized for a given overdensity. The redshift listed is the mean value of the system members, and may be minimally offset from the search galaxy's redshift in some cases ($\Delta z\sim0.0005$ on average). We have included the number of galaxies within $R=10,20$ cMpc from the central galaxy, which are equivalent in many instances. This is primarily dependent on the limited FOV of the surveys. In combination with the 43 protocluster candidates in CCPC1, the Candidate Cluster and Protocluster Catalog contains 216 systems between the redshifts of $2<z<7$. To the best of our knowledge, this represents the largest collection of spectroscopic, $z\ge2$ galaxy overdensities in the literature at the time of writing. Each CCPC candidate has an individual list of redshift measurement references in Table~\ref{tab:tot_ref}.

As in CCPC1, many of the protoclusters have more than the minimum number of members, with a median value of 6 galaxies per candidate. There are 40 CCPC2 systems with the minimum 4 galaxies, and 9 candidates with 23 or more members (Table~\ref{tab:tot_cat}). There does not appear to be a strong trend in numbers of galaxy members as a function of redshift for the bulk of the CCPC. However, all candidates with more than 20 galaxies are found at $z<4.5$. The median number of galaxies in protoclusters in the redshift bins of $2<z<3$, $3<z<4$, $4<z<5$, $5<z<7$ are 5, 6, 8, and 7, respectively within CCPC2. Although these are only slight variations, one would naively expect that the median numbers would decrease as a function of distance. Perhaps at high redshift we are only identifying the richest overdensities, and thus their median members are larger.

Each candidate structure in CCPC2 has a spectroscopic rating (`Q') associated with it in Table~\ref{tab:best_cat} and  Table~\ref{tab:tot_cat}. In many cases, redshift values cataloged by NED have an accompanying Qualifier flag that distinguishes the quality of a given redshift measurement. If a protocluster has 4 or more member galaxies with no quality flags raised, thus satisfying our criteria as a candidate structure, it is assigned the greatest rating of  `1'. If there are 4 or more galaxies that have either no flags or were identified based on a single line, the system is rated slightly lower with a `2'. A rating of `3' is assigned to collections of galaxies with redshift flags of a somewhat uncertain measurement or tentative result. Galaxies with redshifts flagged as photometric redshift, modeled from SEDs, or highly uncertain/questionable values, are not used in this work. In CCPC2, there are 135 (out of 173) systems identified with the highest quality spectroscopic data (RATING=1), and 32 that have a rank `2' rating. 

\subsection{Candidate Overdensities}

\begin{figure*}
\centering
\includegraphics[scale=0.35]{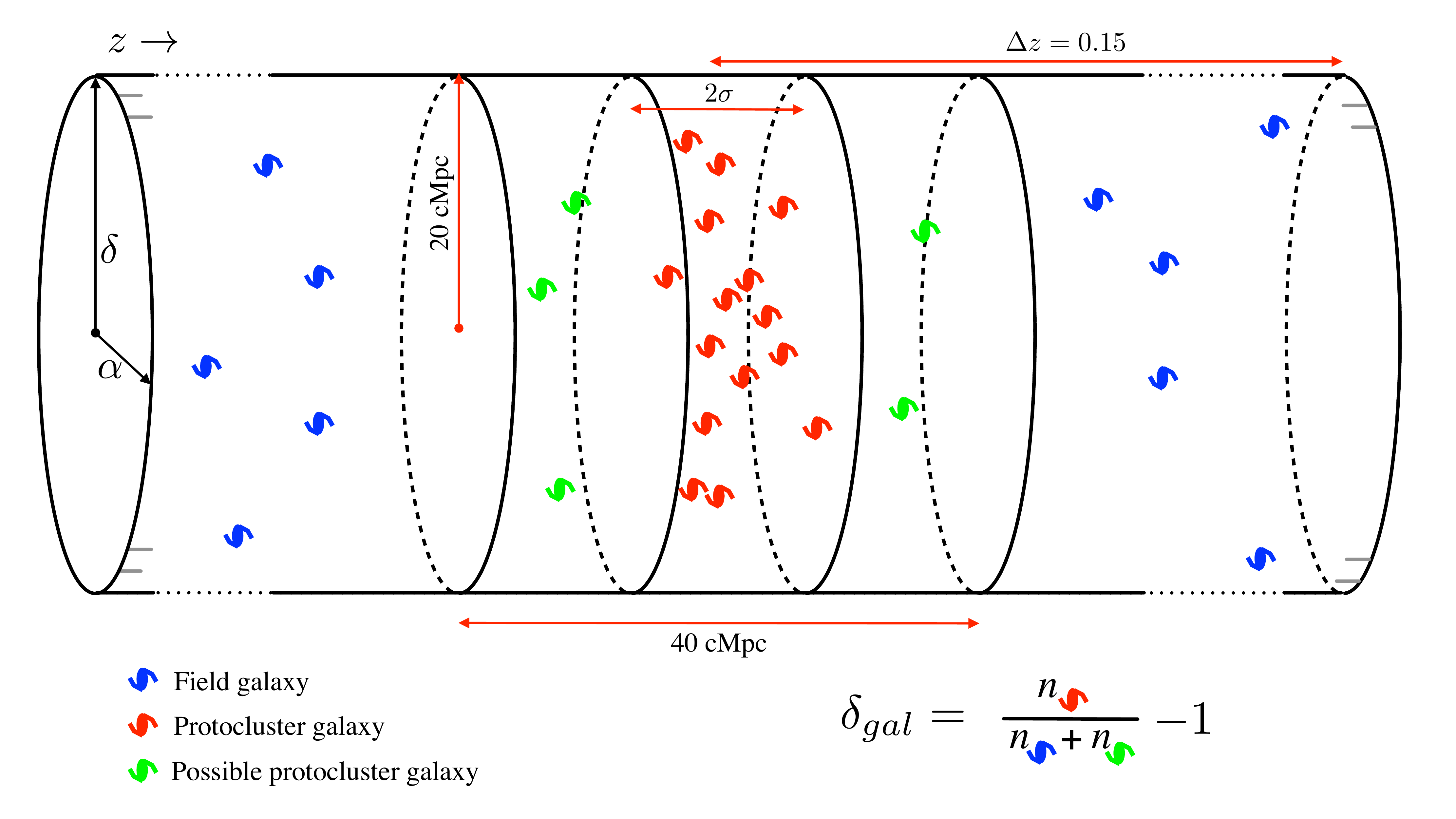}
\hfill
\caption{A visual representation of the overdensity ($\delta_{gal}$) measurement. Galaxies within $\pm20$ cMpc but outside of $z\pm\sigma$ are considered possible protocluster members, but are treated as field galaxies for the computation of the overdensity so that it is a conservative estimate.}
\label{fig:over}
\end{figure*}

To measure the strength of these candidate systems quantitatively, we estimate their galaxy overdensity using the simple formula $\delta_{gal} = (n_{proto}/n_{field})-1$. The number density of protocluster galaxies ($n_{proto}$) is based on the density of galaxies along the line of sight (LOS) within the dispersion ($\sigma_z$) from the center of the redshift distribution. Limiting the galaxies in the overdensity to those only within the redshift dispersion likely excludes some objects that are on the outskirts of the candidate structure, and so the $\delta_{gal}$ values are conservative estimates. These overdensities and their uncertainties can be found in Table~\ref{tab:best_cat} and Table~\ref{tab:tot_cat}. Fig~\ref{fig:over} illustrates the calculation of the galaxy overdensity.

To calculate the number density of the field ($n_{field}$), we use the same aperture on the sky as our search criteria ($R=20$ cMpc) and identify all galaxies along the line of sight (using the same galaxy surveys) within a maximum length of $\Delta z=0.15$ from the protocluster center. The overdense region is excised from the field counts. The choice of the field length is typically an order of magnitude longer than the overdense region to ensure a fair sample. A visual representation of the calculation is shown in Fig~\ref{fig:over}. It is important to note that in every step of calculating the overdensity ($\delta_{gal}$), we are adopting the most conservative values. The purpose of the CCPC was to investigate protoclusters using methods which did not `cherry-pick' the largest values of overdensities.

The uncertainties in field counts are estimated using the cosmic variance calculator from \citet{tre08}, while the uncertainty within the overdense region is $\sqrt{N_{proto}}$. For each structure, we input the volume probed by field systems and an assumption of the completeness of the spectroscopic survey. The output is not particularly sensitive to the completeness assumption ($ \sigma \pm 0.1$ galaxies between 10-90$\%$), and so we have adopted a $50\%$ completeness. The field and protocluster uncertainties are then added in quadrature. On average, the inclusion of cosmic variance and completeness adds 0.3 galaxies to the uncertainty of the CCPC when compared to the Poissonian treatment of uncertainties. 

When measuring $n_{field}$ or $n_{proto}$ for the calculation of $\delta_{gal}$, the length ($\Delta z$) that the density is computed over is limited by the extent of the data, and not the volume queried (e.g. $\Delta z\le0.15$ for the field). For example, NB filters do not extend the full possible width of the field distribution. In  $<15\%$ of cases in the CCPC, low numbers of galaxies clustered in redshift space increased $n_{proto}$  or $n_{field}$ to large levels (e.g. $\delta_{gal} > 50$). We set $\Delta z$ equal to the protocluster redshift dispersion ($\sigma_z$) for these low richness cases.
In some overlapping instances, the field galaxy counts were very low ($N_{field}<3$) compared to other candidate structures. For these systems in CCPC1 we injected seven more galaxies into the counts for the field number density $n_{field}$ to reflect the median field counts in the low-richness sample. This effectively decreases the overdensities to more reasonable values. In this work we have neglected this rather un-physical method, but instead put brackets around the overdensity values to reflect their low-richness status in Table~\ref{tab:tot_cat}. The resulting overdensities are highly uncertain as a result, and should not be relied on without further observations. Although these overdensity vales are questionable, the average number density of these systems ($n\sim 7\times10^{-2}$ $cMpc^{-3}$) exceeds that of the mean value for CCPC2. This suggests that these systems are likely overdensities, and need not be removed simply because they lack field counts.

The median value of $\delta_{gal}=2.9$ for CCPC2, only slightly larger than for CCPC1 ($\delta_{gal}\sim2$). Only $15\%$ of CCPC2 systems have $\delta_{gal}<1$.
Generally, $\delta_{gal}\sim2$ is typical of many protoclusters in the literature \cite[see Table 5 for a summary in][]{chi13}. These values have considerable breadth, from $0.7_{-0.6}^{+0.8}$ \citep{2007AnA...461..823V} to 16$\pm7$ \citep{toshikawa14}, which are consistent with the range in the CCPC. The mean number density of the $\delta_{gal}$ volume is $n=5.9\times10^{-2}$ $cMpc^{-3}$ in CCPC2. For a general comparison, the density of LAEs at $z\sim3$ is $n=1.5\times10^{-3}$ $cMpc^{-3}$ \citep{2007ApJ...667...79G}, but this value is dependent on the galaxy type. In simulated protoclusters within Millennium, analysis shows that overdensities of $\delta_{gal}=1\pm1$ from redshifts $z=2-5$ are consistent with low mass, $M_{z=0}\approx 10^{14}$ $M_{\odot}$ clusters \citep{chi13}. This assumes that galaxies in the overdensity have $SFR>1$ $M_{\odot}$ $yr^{-1}$ and are within boxes of [25 cMpc]$^3$. The most massive systems $M_{z=0}> 10^{15}$ $M_{\odot}$ have overdensities roughly a factor of 3 larger, as one would expect \citep{chi13}. 

\emph{These simulations also show that $\delta_{gal}$ expectations for protoclusters are inversely proportional to the volume probed. \citet{2003ApJ...592..728S} identified LBGs in windows of $R\sim5$ cMpc on the sky at these redshifts, which is only 25\% the size of the maximum CCPC search radius. The line-of-sight distance probed is not similarly hampered. Figure 3 in \citet{chi13} suggest that the stacked overdensity profiles of Millennium protoclusters can be factors of 6 or more larger when evaluated in such small boxes ($R_e\sim5$, $z=3$). Indeed, the CCPC candidates in the regions  of \citet{2003ApJ...592..728S} do have a larger typical overdensity with a median value of $\delta_{gal}>5$, 20 times larger than our minimum required overdensity and 70\% larger than that of the CCPC2 median value. This illustrates the inherent uncertainty when evaluating overdense regions.}   

A relatively compact overdensity region can be dwarfed by the diffuse nature of field galaxies along the line of sight in some instances (like CCPC-z21-003), resulting in $\delta_{gal}\sim0$. For four these special cases, we limit the sky aperture of our field counts to the surface area of the overdense region. The overdensities estimated in this way are noted in Table~\ref{tab:tot_cat}.  Occasionally, the maximum field length of $\Delta z=0.15$ will intersect other structures in the same field. As the Universe does not contain a simple, smooth distribution of galaxies, it can be quite difficult to measure the `field' surrounding an overdensity. Therefore, the value of $\delta_{gal}$ should be taken as an estimate of the strength of the structure, and should not be treated as an absolute metric. As an example of this, we have CCPC-z23-001 and CCPC-z23-005, which are two structures identified by our algorithm whose volumes are fused together in an unbroken $\sim 80$ cMpc long galaxy distribution. They have 43 and 23 galaxies, respectively, which is significantly larger than the median number of member galaxies in CCPC2 (6). Despite their richness, they have middling $\delta_{gal}$ values of 2.0 and 0.7, as their field counts intersect one another.  We will discuss the implications of this `superstructure' in Section~\ref{sec:dis} with more detail.

\subsection{Probable Reality of Protoclusters}

\begin{figure*}
\centering
\includegraphics[scale=0.33]{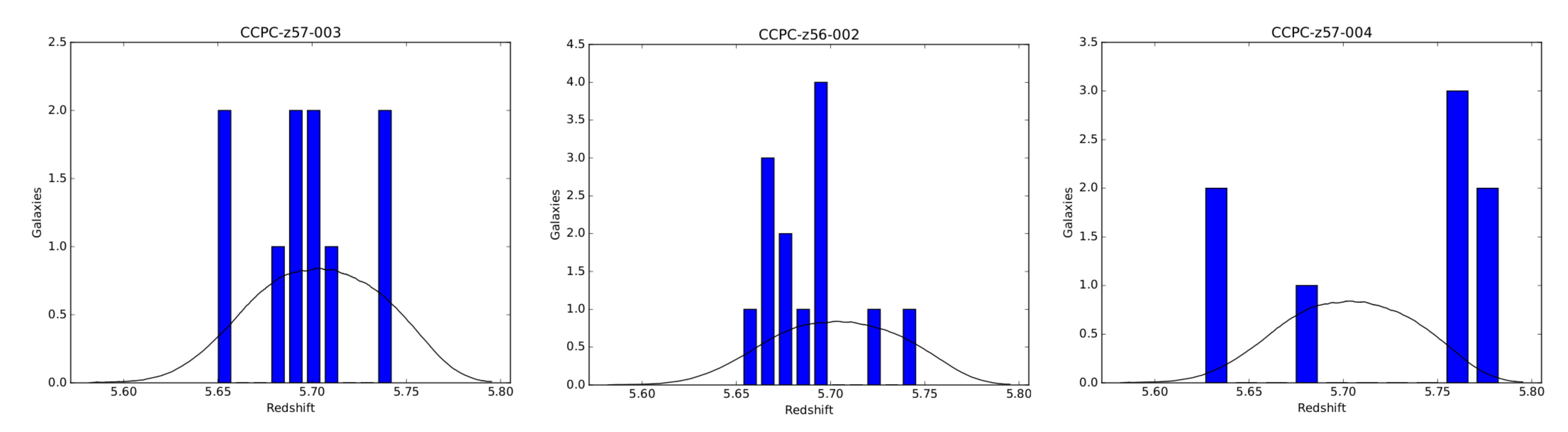}
\hfill
\caption{To test the possibility that galaxies observed through a narrowband filter could produce a false positive protocluster detection from its transmission function, we constructed a Monte Carlo simulation. Shown here are the distribution of LAE galaxies within the NB filter NB816 for three protocluster candidates. Plotted in black is the transmission curve of NB816 at this redshift, arbitrarily normalized to illustrate its shape. The simulated galaxies were distributed following the transmission function of the filter. The simulated distributions were then compared to the mean redshift ($<z_{obs}> - <z_{MC}>$) and dispersion ($\sigma_{z,obs} - \sigma_{z,MC}$) of the actual data. The left panel shows one extreme case in which the distribution of real galaxies show only a 0.5$\sigma$ offset from the Monte Carlo simulation. The center panel is a galaxy distribution in the filter that shares the median value difference (2$\sigma$) between the simulated LAEs and the observations, while the right plot shows the most extreme difference example ($\sim5\sigma$). With the possible exception of the leftmost case, the observed distributions do not follow simply from the shape of the filter transmission.}
\label{fig:NB}
\end{figure*}

It is possible that some of the structures we identify are real, and others are chance coincidences. To estimate the probable reality of each overdensity, we refer to simulations of structure formation.
\citet{chi13} computed the probability distribution function for overdensities of galaxies becoming a cluster at $z=0$  (their Fig 8) using the semi-analytic models of \citet{2011MNRAS.413..101G} in the Millennium simulation. For different values of $\delta_{gal}$, they calculated the fraction of volumes that would become halos of $M>10^{14}$ $M_{\odot}$ (i.e. clusters). They chose to identify galaxy overdensities in the simulation in [15 cMpc]$^3$ boxes at redshifts $z=2-5$ for different galaxy populations. Using their results for galaxies with $M_*>10^{10}$ $M_{\odot}$ (the most biased tracers of mass), we attribute a conservative probability that each CCPC candidate will collapse into a cluster at $z=0$ based on its overdensity. If the galaxies within these structures are less biased tracers (e.g. galaxies with $M_*<10^{10}$ $M_{\odot}$), these probabilities can increase by as much as 40$\%$. These percentages can be found in Tables~\ref{tab:best_cat},~\ref{tab:tot_cat} next to each CCPC candidate. Table~\ref{tab:best_cat} contain systems that have $\ge85\%$ probability of becoming a cluster at $z=0.$

There are indications that the overdensities and probabilities we calculate for the CCPC are too conservative. In Section~\ref{sec:dis}, we estimate the number of structures expected within the volume of the CANDELS GOODS-S field. Compared to the candidate protoclusters we identify (and the probability of collapse we assign), the discrepancy is at a minimum of a factor of two smaller, and may be an order of magnitude too low.

It has been noted that a spurious protocluster signal could be the result of the transmission curve of a NB filter \citep{2007AnA...461..823V}. A false-positive overdensity is possible if more galaxies are detected at the central wavelength of the filter because of its greater transmittance, while the less-responsive tails yield fewer detections. For instance, the FWHM of the Subaru Telescope's NB filters used to select LAEs at $z=3.1,5.7,6.5$ are only slightly larger than the expected diameter of the most massive protoclusters at these epochs. A smooth distribution of galaxies could appear as an overdensity at the central wavelength with respect to the edges, where `field' galaxies might reside. We find that this is not the case, and the galaxy distributions differ significantly ($>2\sigma$) from the transmission functions. This serves as a further confirmation that we are detecting actual structure. 

To model the possibility of false detections, we adopted a similar approach to \citet{2007AnA...461..823V}. We ran a Monte Carlo simulation based on the transmission curves of the Suprime-Cam's NB503, NB816, and NB921 filters with $10^4$ iterations for all of the 15 protoclusters detected from these observations. For each iteration, the number of galaxies observed within the filter were distributed according to its transmission probability. The difference between the real vs. simulated mean redshift ($<z_{obs}> - <z_{MC}>$) and observed vs. Monte Carlo redshift dispersions ($\sigma_{z,obs} - \sigma_{z,MC}$) were calculated in units of the standard deviation in the simulation. When compared to the actual data of galaxies identified through these filters, the average deviation from the Monte Carlo was 2.4$\sigma$ (combined $<z>$, dispersion deviations). This suggests that the galaxy distribution (and therefore the overdensities)  are not merely the result of the NB filter. In Fig~\ref{fig:NB}, the galaxy distributions of three protoclusters (CCPC-z57-004, CCPC-z56-002, CCPC-z57-004) are plotted with respect to the transmittance of NB816. These examples show the full range of the deviations from the Monte Carlo ($0.5-4.9\sigma$). 

\subsection{Mass Estimates}

The mass of clusters and protoclusters at various redshifts is a stringent probe of cosmological parameters (\Omegam, \Omegal, $\sigma_{8}$ and $w$), as the cluster mass function is tied to these values \citep{pre74}.  We attempt to provide two mass estimates in this work. The first is based on the volume and mass overdensity ($\delta_{m}$) that may collapse by $z=0$, and the second is a crude virial mass estimate based on the velocity dispersion $\sigma$ and effective radius of the system. However, these two values do not appear correlated and are highly uncertain.

\citet{1998ApJ...492..428S} and \citet{2007AnA...461..823V} have estimated the expected collapsed masses of protoclusters at high redshift by estimating a galaxy overdensity ($\delta_{gal}$), a volume encompassing the galaxies within the overdensity $V$, and the critical density of the Universe at that redshift $\rho_{crit}$. Galaxies are assumed to be mere tracers of the dark matter distribution, so it is required to assume a bias parameter $b$ to convert the observed galaxies into a mass overdensity $\delta_{m}$. Historically, linear bias parameters of $3\le b \le6$ have been used \citep{1998ApJ...492..428S,2007AnA...461..823V}, and the matter overdensity can be found via $\delta_{m} = \delta_{gal}/b$. In this work, we adopt a bias of $b=3$. At a redshift of $z=3$, $\rho_{crit}=4.2\times10^{10}$ $M_{\odot}$ cMpc$^{-3}$ in our assumed cosmology. If the entire volume is assumed to collapse into a single halo by $z=0$, that cluster will have an estimated mass of 
\begin{equation}
M = \rho_{crit,z}V(1+\delta_{m}).
\end{equation} Table~\ref{tab:best_mass} contains the mass estimates for the protoclusters with the most significant overdensities from Table~\ref{tab:best_cat}. All system masses are located in Table~\ref{tab:tot_mass}.

There are a number of assumptions that go into this calculation, the most critical of which is the volume estimate.  Galaxies on the outskirts of the extended protocluster distribution may not collapse into a single halo by $z=0$, or be bound to the structure at all, as is seen in simulations \citep{2015MNRAS.452.2528M}. In addition, the SAM used by \citet{2015arXiv151201561C} to investigate protocluster galaxies suggests that these may be indistinguishable observationally from their field counterparts. Therefore, including these as tracers of the volume that will collapse into a cluster can greatly increase the mass estimate of the system, possibly by orders of magnitude. Some previous works that have utilized this mass estimator include a corrective factor for redshift space distortions \citep{1998ApJ...492..428S,2005ApJ...626...44S}. This can result in a difference factor of $\sim2$ in some instances. As the volume assumptions can change the mass estimate by orders of magnitude, this space distortion calculation is neglected.

The bias parameter $b$ is an assumed value, and depends on the galaxies used as tracers, in that higher mass galaxies are more biased tracers of mass. If the galaxies that trace the protoclusters do not have $b=3$, the overdensity mass estimate will also be systematically affected. Bias parameters can also evolve over time, growing larger at higher redshifts. For LAEs at $z=3.1$, \citet{2007ApJ...671..278G} estimate a value of $b=1.7$, while LAEs at $z=4.86$ have $b\approx3$ \citep{2003ApJ...582...60O}. Biases of larger mass galaxies (LBGs, for instance) can have $b\gtrsim4$ at $z>4$ \citep{2004ApJ...611..685O}. Based on the no-merger model of \citet{1996ApJ...461L..65F}, strong evolution of the bias ($\Delta b\gtrsim3$ from $z=0$) is predicted for high redshift, large bias systems (e.g. $b=6$ at $z=5$). For systems of modest bias ($1<b<2$), the evolutionary difference is less than unity. If our algorithm is selecting only the most biased sources, the implied mass overdensities may not be sufficient to become clusters. However, if the bias is similar to the roughly constant, unevolving value of \citet{2007ApJ...671..278G}, our candidates might be more significant than we claim.

The mean estimated collapsed mass of candidates in CCPC2 is $M_{z=0}=1.8\times10^{14}$ $M_{\odot}$, consistent with low mass clusters found in the local universe. Systems of this mass comprise roughly $70\%$ of the population of clusters in the Millennium simulation \citep{chi13}, while clusters of $M_{z=0}>10^{15}$ $M_{\odot}$ should represent only $2\%$ of cases. However, our search methodology is not expected to be mass-blind, and preferentially selects the highest overdensity systems. 


For clusters at low redshift ($z\sim0$), a traditional method of estimating a mass was to assume the system was virialized, measure the velocity dispersion ($\sigma$) and effective radius $R_e$, and compute the system mass via 
\begin{equation}
M = \frac{2\sigma^2}{G} R_{hms},
\end{equation} where $R_{hms}\sim 1.25 R_e$.
However, this assumption is not expected to hold at higher redshifts in \LCDM. At $3$ Gyrs after the Big Bang ($z\sim2$), analysis of the Millennium simulation shows that the the progenitors of the most massive clusters at $z=0$ ($M_{z=0}>10^{15}$ $M_{\odot}$) have a dark matter halo of $\ge10^{14}$ $M_{\odot}$, while the lowest mass clusters at the present day may not have assembled this mass until $z\approx0.6$  \citep{chi13}. Prior to this epoch, it is likely that only subhalos have virialized.

For each object in the CCPC, we have calculated the effective radius ($R_e$) in which 50$\%$ of the total protocluster members reside, as well as the velocity dispersion of the system $\sigma$, and computed a `virial' mass estimate. Mass estimates for the entire CCPC2 list are in Table~\ref{tab:tot_mass}, while the most overdense protoclusters can be found in Table~\ref{tab:best_mass}.  We again urge caution in interpreting these results, as it is unlikely that such extended, diffuse systems are in virial equilibrium. We use the `virial' term only because we utilizie the virial mass equation. As there are few protocluster members in some CCPC systems, their diffuse nature can imply large values of both $R_e$ and $\sigma$, as is the case for CCPC-z24-007. It has only 4 galaxy members (the minimum number) and an effective radius of 4.8 (physical) Mpc, a dispersion of $760$ $km \, s^{-1}$. These yield a virial mass estimate of $1.6\times 10^{15}$ $M_{\odot}$, a factor of two larger than the average estimate in CCPC2 ($\sim8\times 10^{14}$ $M_{\odot}$), despite its minimal richness.

In contrast, there are a few systems (CCPC-z20-005, CCPC-z21-011, CCPC-z22-002) that have a large number of candidate galaxy members ($N\ge18$) with large implied virial masses ($\ge 10^{15}$ $M_{\odot}$). These systems are at an epoch ($z\le2.3$) that can theoretically host virialized clusters \citep{chi13}. With the significant increase in richness, these mass estimates may be at least more physically meaningful than the previous example (CCPC-z24-007), although these would still be an order of magnitude larger than predicted. Further examination and discussion of the mass estimates (via mass overdensities and velocity dispersions) can be found in Section~\ref{sec:mass_dis}.

\subsection{Objects of Interest}

We have compiled a list of targets that are of potential significance, but violate our strict spectroscopic redshift criteria. These Objects Of Interest (OOI) are generally the result of narrow-band or photometric redshift observations, but otherwise fulfill the requirements of a CCPC target.The last two entries (OOI-z65-001 and -002) are spectroscopic galaxies, but had no field sources with which to calculate a $\delta_{gal}$ value. Some of these will likely prove to be fictional if targeted spectroscopically. These are listed in Table~\ref{tab:ooi}.

\section{Discussion} \label{sec:dis}

\subsection{Tests of Structure}\label{sec:test}

The method in which we identify structures is relatively simple. In CCPC1 \citep{ccpc1}, we performed a
number of tests on the algorithm in which candidate protoclusters were identified. In addition, the significance
of the candidates were also evaluated. It is trivial to list positions of galaxy associations, but much
more difficult to find physical systems. We have three overarching diagnostics of structures, each with sub-tests
for significance: (1) the CCPC systems exist as overdensities, (2) they are statistically distinct spatially when
compared to the `field', and (3) their number densities in deep surveys are comparable to expectations 
from large simulations. That we recover a number of previously identified structures with our algorithm
is an additional confirmation of its fidelity.


One of the criteria for CCPC protoclusters is that candidate systems must show an overdensity of galaxies ($\delta_{gal}\ge0.25$)
when compared to the local field along the line-of-sight ($\Delta z\pm 0.15$). This criteria is simply a lower limit, and the 
median overdensity is $\delta_{gal}=2.9$. Within the literature, previously identified protoclusters (and simulations) at these redshifts
show similar overdensities \cite[Table 5 and Section 3.6, respectively, in][]{chi13}.  

Furthermore, these overdensities can be observed visually as spikes of galaxies along the line of sight in $N(z)$ plots (Fig~\ref{fig:CCPC-z20-001}).
These plots were constructed using the full aperture of the sky search radius ($R=20$ cMpc). Many systems, such as CCPC-z20-003 (Fig~\ref{fig:CCPC-z20-003}), have 
clear overdensities along the line of sight. Other overdensities, particularly those with centrally condensed galaxy distributions on the sky (e.g. CCPC-z20-008),
may appear as little more than noise. This underlines the importance of having more than a single protocluster identifier (e.g. only $N(z)$ spikes, $\delta_{gal}$,
or a minimum number of galaxies).

It is of interest that the $\delta_{gal}$ values and CCPC galaxy members show little correlation. It is logical
that spectroscopic surveys which have deeper limiting magnitudes would find more galaxies in both the field
and in structures, all galaxies being equal. However, if protoclusters are indeed regions of enhanced mass assembly \citep{cas15},
rapid star formation or galaxy evolution \citep{hat11,2014MNRAS.439.1193C}, it would be reasonable to assume that
some manifestation of this behavior would present itself.

In fact, some of the largest overdensities ($\delta_{gal}\ge10$) have fewer than 10 galaxy members within the search volume.
For many of these cases (e.g. CCPC-z21-007 and CCPC-z23-007), the Poissonian uncertainties are 
large ($\ge \pm 10$) because of the small number of galaxies in the overdensity calculation (especially field galaxies). Some
cases of these low $N$, high $\delta_{gal}$ cases are, interestingly, previously identified structures.  

CCPC1 and CCPC2 have nearly identical median overdensities of $\delta_{gal}\sim2.0,2.9$ over the
entirety of the redshift range. This is not a perfectly direct comparison, as the
CCPC1 criteria required 3 or more galaxies to be found within 2 arcminutes of 
the center of the search radius. Potentially, CCPC1 may have been selecting the minority of protoclusters
that have dominant main halos \citep{2015MNRAS.452.2528M}, which would represent 
stronger overdensities. Also, there is an expectation of some $\delta_{gal}$ evolution with 
redshift \citep{chi13}, but with larger variations depending on the galaxies observed as tracers. 

Despite these differences in redshift and methodology, the protoclusters at $2.74<z<3.71$ in the 
CCPC2 have a median $\delta_{gal}\sim1.7$ and are consistent within the uncertainties of $\delta_{gal}\sim2$ 
for CCPC1 and CCPC2. This is of two-fold importance: (1) there were 66 overdensities left undiscovered in the 
redshift range of CCPC1 (which contained 43 structures) by requiring an initial core group of galaxies to
be a criterion, and (2) that the slightly greater median $\delta_{gal}$ between CCPC1 and CCPC2 suggests we are not admitting poorer candidates by removing this step.


\begin{figure*}
\centering
\includegraphics[scale=0.5]{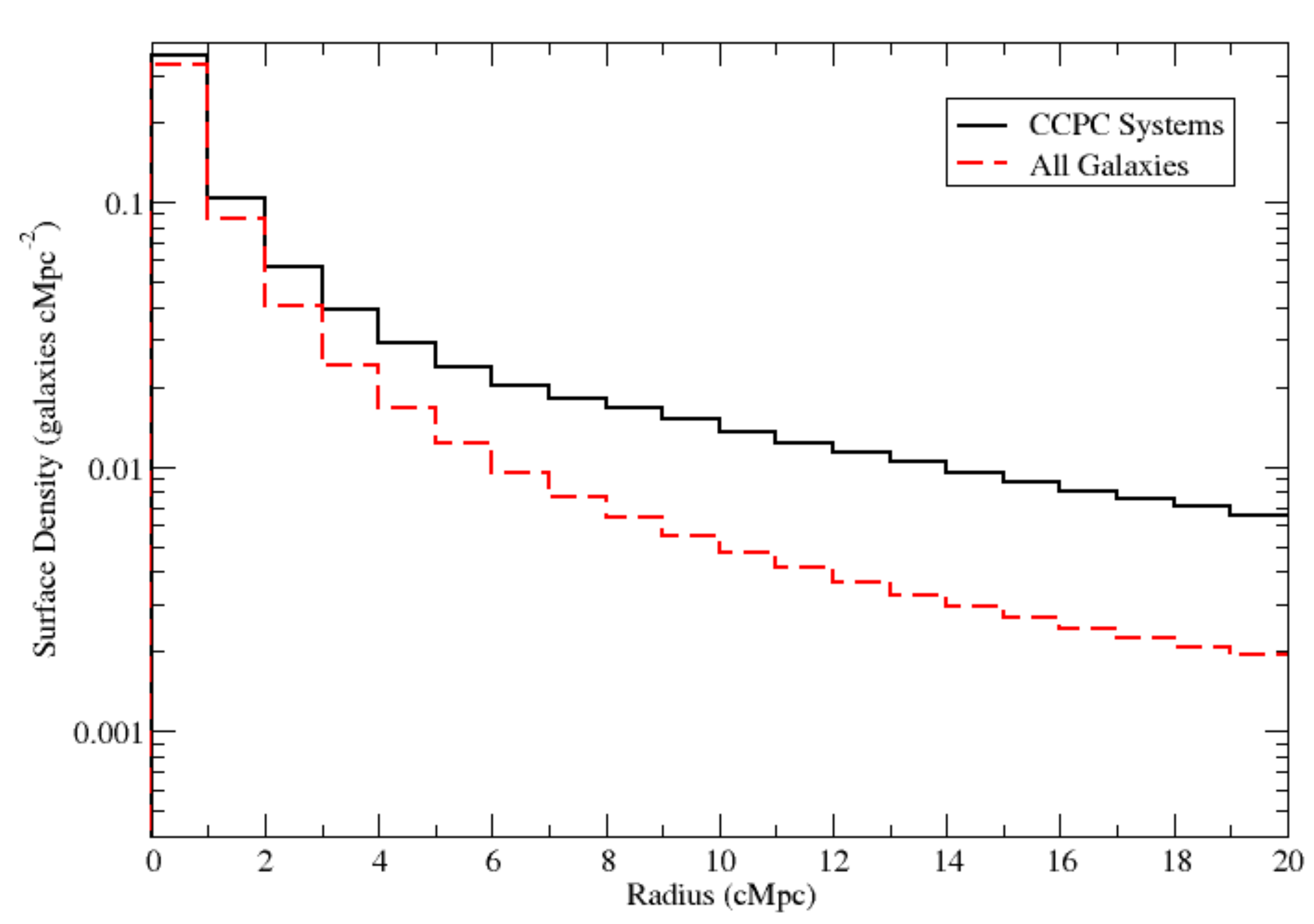}
\hfill
\caption{The mean galaxy density on the sky of CCPC2 galaxies is shown in black. In contrast is the mean surface density profile of All Galaxies in our initial list ($>$40,000 spectroscopic sources with $z>2.00$), which is shown as the red dashed profile. The `All Galaxies' distribution contains CCPC2 galaxies. This `field' proxy is consistent with there being a single galaxy in the center of a $R=20$ cMpc search radius with few companions. A two-sided KS test (0.6) suggests that these are distinct distributions.}
\label{fig:density}
\end{figure*}

\begin{figure*}
\centering
\begin{subfigure}
\centering
\includegraphics[scale=0.5]{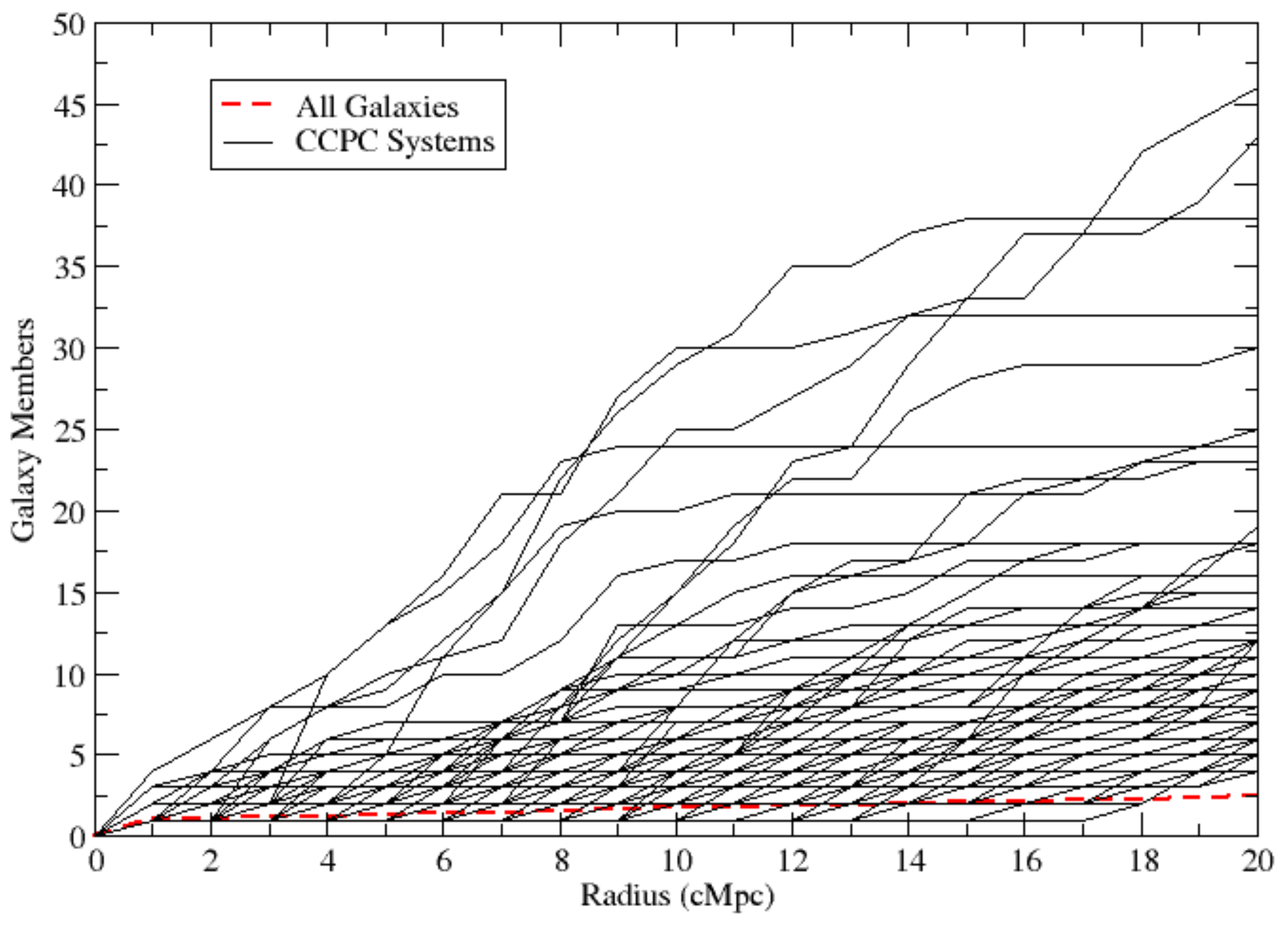}
\end{subfigure}
\hfill
\begin{subfigure}
\centering
\includegraphics[scale=0.5]{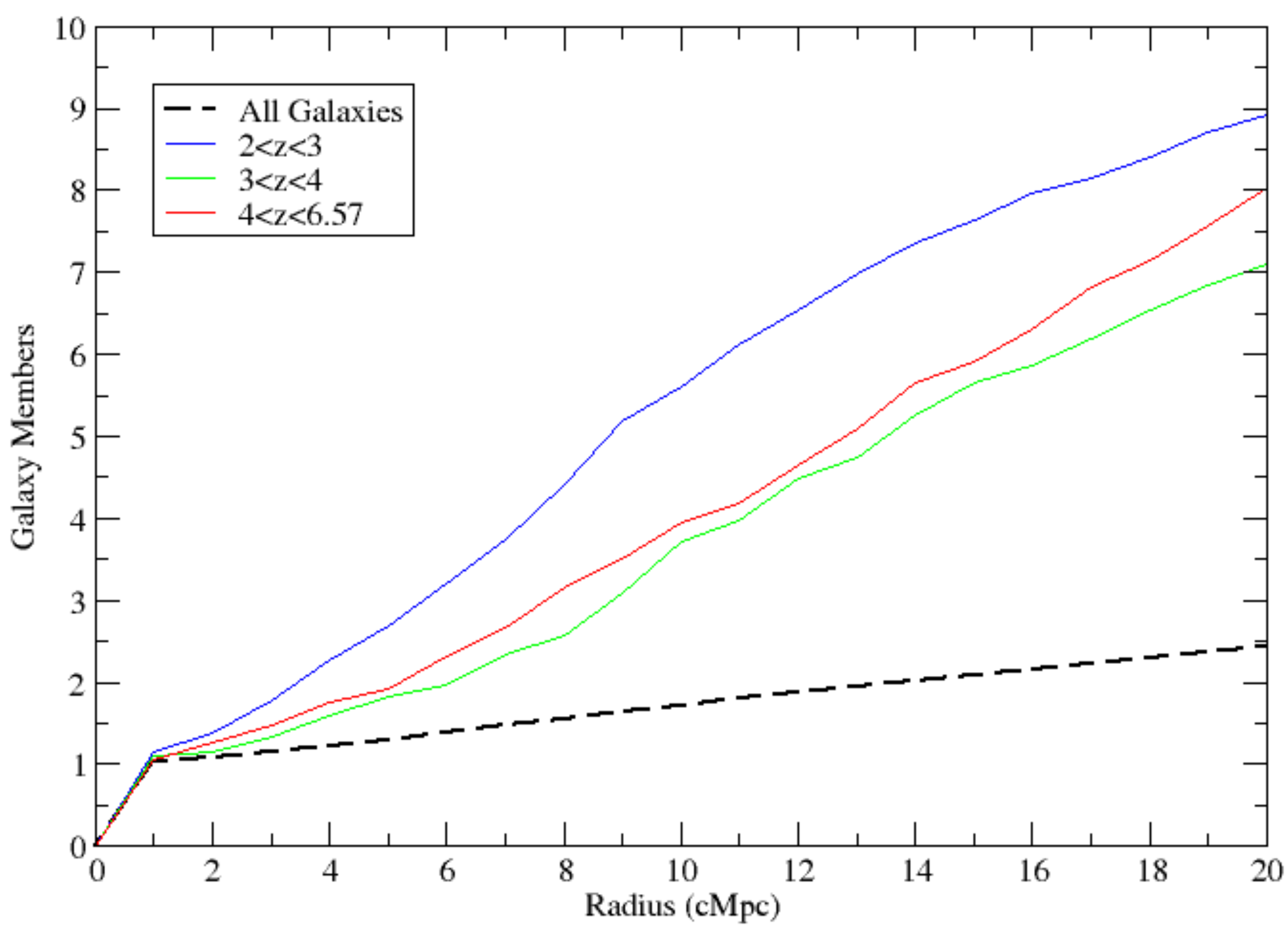}
\end{subfigure}
\hfill
\caption{$Top:$ The distribution of galaxies for each of the 173 CCPC2 systems are shown in black as a function of distance from the center of the search radius in comoving Mpc. The mean distribution of 44,000 spectroscopic galaxies from which we identified structures is shown as a dashed red line. For the CCPC systems, there is a clear difference in both the number and distribution of galaxies. In many cases, the survey edge can be seen as a flat line at $R<20$ cMpc. $Bottom:$ At different bins in redshift, we have plotted the mean number of galaxies as a function of radius, compared to the black dashed line representing All Galaxies at $z\ge2$. As shown by the top panel, there is considerable variation in the distribution of galaxies. The difference in survey widths also prevents meaningful analysis of the comoving distribution of the galaxies, especially at radii $>10$ cMpc. Primarily, this plot is effective at illustrating the difference between field galaxy distributions and those within protoclusters at all redshifts. A KS test of the two mean distributions (`All Galaxies' vs. CCPC galaxies) gives a value of 0.75, indicative that these two populations are not from the same group. }
\label{fig:dist}
\end{figure*}

Galaxy overdensities should not only exist along the line-of-sight (as observed in $N(z)$ plots and $\delta_{gal}>0$), but their spatial distributions should also be distinct from the field. As protoclusters exist as very extended, diffuse systems on the order of tens of cMpc \cite[Figs 1,2 in][]{2015MNRAS.452.2528M}, their profiles will naturally look more similar to isolated galaxies than $z=0$ clusters. However, in CCPC1 we showed that there were significant differences between the surface density and cumulative distribution of field and protocluster galaxies.

In this work, we present similar findings based on the mean surface density of all CCPC2 systems (Fig~\ref{fig:density}). In order to capture the essence of `field' galaxies, we used the initial list of $>47,000$ galaxies from which we identified the candidate protoclusters at $z>2$ to search for any companion galaxies within a radius of 20 cMpc, and $\pm20$ cMpc along the line of sight. This mirrors the search volume of the CCPC algorithm. Included in this galaxy list are all of the CCPC2 member objects. The mean sky surface density (galaxies $cMpc^{-2}$) is shown as a red dashed line in Fig~\ref{fig:density}. This is consistent with a single galaxy found within $R\le1$ cMpc  ($\Sigma =0.33 $  cMpc$^{-2})$. This is labeled `All Galaxies' instead of `Field Galaxies', as we did not separate out isolated galaxy systems from overdense regions. By comparison, there are more mean galaxies in the central regions of CCPC systems (black line), and a shallower slope (-1.3 in $log-log$ space) compared to the field galaxy slope (-1.7). A two-sample Kolmogrov-Smirnov test (KS) value of 0.6 shows that these are distinct distributions at the $99.93\%$ level.

Fig~\ref{fig:dist} illustrates the number of galaxies as a function of search radius $N(r)$ for each individual CCPC2 overdensity in the top panel. As in Fig~\ref{fig:density}, the red dashed line is representative of field galaxies, with clear differences in the distribution and mean number of sources at different comoving radii. The distribution of protocluster galaxies is highly variable, with some overdensities being very concentrated while others have lower central concentrations, but continually add galaxies to large radii. Plotting the number distributions instead of cumulative distributions (as in CCPC1) highlights the differences in survey widths more effectively, as some $N(r)$ profiles flatten out well before the maximum search radius of $R=20$ cMpc. It is therefore important to not infer a `characteristic' distribution of galaxies in protoclusters from such a plot, as some spectroscopic galaxy surveys will only target the innermost regions of structure, while others will trace out to larger radii. In addition, simulations suggest that there exists a menagerie of galaxy distributions within these high redshift systems, even among protoclusters that will have the same mass at $z=0$ \citep{chi13,2015MNRAS.452.2528M}.

The bottom panel of Fig~\ref{fig:dist} shows the mean `All Galaxies' distribution as a \emph{black} dashed line (instead of a red dashed line in the top panel), while CCPC2 galaxies in various redshift bins are shown in blue ($2<z<3$), green ($3<z<4$), and red ($4<z<6.57$). On average, `All Galaxies' will have roughly 1.5 companion galaxies surrounding it within the fixed CCPC volume, while at a minimum a CCPC system will have 3 other companions, and on average more than 7.

A KS test of the radial distributions of galaxies between `All Galaxies' and the mean CCPC2 galaxies gives a value of 0.75, more significantly different than the surface density comparison (KS$=0.60$). The various redshift bins, within the considerable scatter illustrated by the top panel, show little evolution. Simulations suggest there are large distribution variations of individual systems even at the same mass and redshift \citep{chi13,2015MNRAS.452.2528M}. When coupled with the heterogeneous spectroscopic data (survey width/depth differences) from which we draw our candidates, the lack of meaningful $N(r)$ variations is not particularly surprising. 




The most significant difference in selecting protocluster candidates between CCPC1 and CCPC2 was that the former contained a `group finding' intermediate step. Initial groups were selected by requiring at least 3 galaxies within 2' of the search center, and within a $\Delta z = 0.03$. Requiring this intermediate step does not appear to affect the quality of sources, as the median overdensities between CCPC1 and CCPC2 are equivalent $(\delta_{gal}\sim2)$. Interestingly, requiring a centrally concentrated group of galaxies does not seem to have a significant effect on the surface density profiles (KS$=0.35$ between CCPC1 and 2). Only the inner regions ($R\sim1$ cMpc) show significant differences, with CCPC1 having a larger surface density (as a criterion) of $\Delta \Sigma=0.12 $ cMpc$^{-2}$. We conclude that the structures identified using the selection methods of CCPC1 and CCPC2 are not significantly different from one another, and both are selecting plausible protocluster candidates.

The CANDELS GOODS-S field is a deep, multi-wavelength field in which 27 CCPC structures have been identified. Of these, 9 were identified in \citet{ccpc1} (in the redshift range $2.74<z<3.71$). It is the deepest, widest field from which we draw candidate spectroscopic galaxies to identify protoclusters. The majority of the spectroscopic footprint falls on an area of the sky of roughly 0.4 degrees on a side, which corresponds to a box with sides approximately 35 cMpc wide at $z=2$. At $z=5.7$, the box side length increases to $\sim55$ cMpc in our assumed cosmology, with a length along the line of sight ($z=2\rightarrow5.7$) of almost 3 cGpc. This length is the expanse of the field in which we identify protoclusters.

From this pencil beam survey, we estimate the volume that we have probed to be $\sim5.89\times 10^{6}$ cMpc$^3$. The Millennium Simulation has a volume of $[500\, h^{-1} \, cMpc]^3$ \citep{2005Natur.435..629S}, which is almost 2 orders of magnitude larger than the GOODS-S data. \citet{chi13} identified 2832 clusters at $z=0$ with masses $M > 10^{14}$ $h^{-1}$ $M_{\odot}$ within their analysis of the simulation. This corresponds to a number density of $7.8\times 10^{-6}$ cMpc$^{-3}$. Therefore, one would expect to find 46 clusters at $z=0$ in the volume of the GOODS-S field. As we have only identified 27 systems in this region, our algorithm is probably not over-identifying structure. Using the number density of low mass clusters ($M_{z=0} <3\times 10^{14}$ $M_{\odot}$) in Millennium, there should be an estimated 32 such systems in this deep field \citep{chi13}, while 14 Virgo-mass or larger ($M_{z=0} \ge3\times 10^{14}$ $M_{\odot}$) protoclusters are expected within this volume. It is probable that our algorithm is only identifying some of the richest overdensities that will collapse into Virgo or larger mass systems at $z=0$, while missing many of the smaller protoclusters in this field.


In all cases of protocluster candidates within the GOODS-S field, the probability that these objects will collapse into clusters by $z=0$, based on the overdensities of \citet{chi13}, is considerably less than 100$\%$ (see Table~\ref{tab:tot_cat}). If we sum their fractional collapse estimates (e.g. a candidate's $10\%$ chance of collapse can be approximated as $\frac{1}{10}$ of a cluster), for the entire CCPC we have identified only $\sim3.7$ clusters from $2<z<5.7$ in the CANDELS GOODS-S survey. As we expect more than 40 structures in this probed volume, there exists a serious discrepancy. One explanation could be that our conservative estimates for both $\delta_{gal}$ and its application in determining a collapse probability from the work of \citet{chi13} is much too stringent. This is a probable scenario, as explained in the outline of overdensities and probabilities within Section~\ref{sec:ccpc}. Another option could be that the number density estimates are for all clusters, while we are mainly identifying higher mass systems. However, 14 Virgo-like protoclusters are expected in this volume, and our probabilities are an order of magnitude smaller than that. It is also plausible that CANDELS GOODS-S field is not accurately represented within the Millennium simulation. Indeed, a combination of these elements are probable to span the gap of the excess number of simulated systems.

As mentioned earlier, many of these deep, pencil-beam surveys from which we identify protoclusters have sky widths smaller than the expected size ($\sim20$ cMpc) of the most massive protoclusters \citep{2015MNRAS.452.2528M}. Therefore, if the overdensity does not significantly fall within the footprint of the spectroscopic data, it may be missed by our search method. Intriguingly, \citet{2015JKAS...48...21K} delved into both the GOODS-S and GOODS-N combined spectroscopic and photometric data and found an excess of structures in the redshift range $0.6\le z\le4.5$ when compared with the Millennium Simulation protoclusters. While in the CCPC we have limited our comparison to the number density, they included the mass of their structures, finding a factor of 5 more systems with $M>7\times10^{13}$ $M_{\odot}$. \citet{2015JKAS...48...21K} provide some plausible explanations for the overabundance (elements of the input physics in the models could be simply incorrect, \LCDM may not be an accurate representation of the universe, overestimation of masses), but no definitive diagnosis for this complex problem.

That we derive the opposite conclusion from the same data set, similar search methodology, and identical simulation is puzzling. It could be possible that the precision of photometric redshift measurements produces an increase in false-positive structure detections. This seems unlikely, as we found more overdensities (24 versus 9) in the shared redshift space with \cite{2015JKAS...48...21K} in the GOODS-S field. We find it more likely that the authors' method is sufficiently distinct from our own so as to be difficult to compare results.

\subsection{Poissonian Expectation Model}

A further confirmation that these associations are more than chance groups of galaxies can be found by estimating the number of false positives that should be expected in a smooth density field (i.e. lacking structure) along the line of sight that could arise from Poissonian fluctuations. We can utilize the analytical formula 
\begin{equation}
N_M =N_g e^{-2nl} [1-e^{-nl}]^{M-1} 
\end{equation}
from they toy model of \citet{she01}, which approximates the number ($N$) of Poissonian fluctuations of $M$ galaxies that would manifest as protoclusters. This model uses the number density along the line of sight, $n$, for $N_{gal}$ total galaxies, separated by a linking length $l$. The CANDELS GOODS-S field is approximately a 1-D pencil beam survey in the context of this work ($\sim40\times40\times3000$ cMpc from $2<z<5.7$). The number density of spectroscopic sources varies from $2<z<5.7$, which makes it a linearly decreasing density field as a function of redshift, unlike the toy model. At $2<z<3$, $n=0.76$ galaxies $cMpc^{-1}$, while over the full range ($2<z<5.7$), this decreases to $n=0.40$ galaxies $cMpc^{-1}$. We adopt the latter value ($n=0.40$ galaxies $cMpc^{-1}$) as the estimate of the number density.

We require a suitable estimate of the linking length $l$ between galaxies in this survey. We can measure the strength of clustering at different length scales by using the auto-correlation function \citep{dav83}. 
\begin{equation}
\xi (l) = \frac{N_{DD}}{N_{DR}} \frac{n_R}{n_D} -1.
\end{equation}
This compares the number of pairs of actual galaxies ($N_{DD}$) found within shells separated by $l\pm \Delta l$ in the GOODS-S field, to the number of $D$ata-$R$andom pairs of galaxies ($N_{DR}$). When the value of $\xi(l)$ crosses $\xi=1$, the clustering strength at length $l$ is said to transition from strong to weak. At $l=1.1-1.75$ this occurs within the spectroscopic data set. This is similar to the linking length measured between $2.74<z<3.71$ found in CCPC1 \citep{ccpc1}.

The CCPC algorithm does not utilize linking lengths to identify structure, as $l$ would vary from survey to survey. To similarly match the fluctuations in the toy model of \citet{she01}, we identified structures using a one-dimensional Friends-of-Friends (FoF) algorithm \citep{huc82}. We also ran 500 Monte Carlo simulations of the field and computed the mean number of groups found using the FoF algorithm as a check on the toy model. For $l=1.2-1.75$ cMpc, an excess number of groups are identified above the Poissonian expectation for any choice of the minimum member galaxies ($M\ge2$). There were 15 FoF structures found within the CANDELS data set using separation $l=1.2$ cMpc  with $M\ge6$ galaxies and 6 associations of $M\ge8$ members. The toy model predicts 4 and  0.6 such groups of galaxies should exist (respectively) from Poissonian fluctuations, while the Monte Carlo finds a mean of 1.4 and 0.08 systems of such richness. Choosing a larger separation of $l=1.75$ cMpc reveals an excess of 21 FoF groups over the expectation of \citet{she01} for associations of 6 or more members, and an excess of 25 compared to the Monte Carlo value. For groups of 10 or more galaxies with separation of  $l=1.75$ cMpc, there should be fewer than one chance fluctuation in the Poissonian and Monte Carlo models, while 5 systems are identified in the GOODS-S volume. 

The clear excess of FoF groups using a range of values for $l$ and $M$ compared to what would be expected stochastically illustrates that there are physical associations within these data.
Values of $N_m$ and $l$ are unique to this survey and do not apply to CCPC protoclusters outside of the CANDELS GOODS-S sample. This example is only used to illustrate that an excess of clustering is found over Poissonian fluctuations in a simple toy model.

\subsection{Mass Estimates} \label{sec:mass_dis}

In Sec~\ref{sec:ccpc}, we outlined two distinct methods to estimate the mass of protoclusters at these high redshifts. Unlike the methods utilized for `nearby' clusters at $z\le1$ which use signatures from a massive $M > 10^{14}$  $M_{\odot}$, virialized halo (e.g. SZ effect, X-ray emission), protoclusters at $z\ge2$ are not expected to have main halos of this magnitude \citep{chi13,2015MNRAS.452.2528M}. More uncertain means must be employed to provide some metric of the mass of these systems. 

One method used by \citet{1998ApJ...492..428S} and \citet{ven02} is to calculate the volume of the overdensity that will collapse into a cluster at $z=0$. The mass is simply the density of the Universe ($\rho_{crit}$) multiplied by the mass overdensity ($\delta_{m}=\delta_{gal}/b$) and volume of the system (Equation 1). The mean overdensity mass is $M_{z=0}\sim 1.8\times10^{14}$ $M_{\odot}$ for the CCPC2 catalog. As the majority of clusters ($\sim70\%$) in the Universe are expected to be of this mass \citep{chi13}, this appears reasonable. We have also made the most conservative estimates of the volume and overdensities ($\delta_{m}$) which would enhance this expectation. 

There does not appear to be a significant trend in decreasing mass with increasing redshift. There is considerable scatter in the $\delta_{gal}-z$ distribution, which subsequently persists into the mass estimates. Most of the sources with the highest probability of collapsing into a structure (Tab~\ref{tab:best_mass}) are at low redshift and can have masses in excess of  $M_{z=0}> 5\times10^{14}$ $M_{\odot}$. A large mass estimate for CCPC-z65-005, the highest redshift system, is an example of the lack of mass evolution. Located at a redshift of $z=6.56$, its large volume ($\sim 7000$ cMpc$^3$) and overdensity imply a collapsed mass of nearly $M_{z=0}\approx 6\times10^{14}$ $M_{\odot}$. The CCPC1 has a mean mass estimate of $2.5\times10^{14}$ $M_{\odot}$, only slightly larger than CCPC2. 

It should be noted that the $\delta_{gal}$ estimator relies on a number of assumptions. Primarily, the tracer of both the volume and matter overdensity are galaxies, which are not a significant contributor to the density of the Universe in \LCDM. We must assume that all galaxies, especially those at the outskirts that define the volume of the overdensity, will collapse by $z=0$. Numerous simulations have shown that many galaxies within the comoving volume of the structure may not be bound to the cluster by $z=0$ \citep{2015MNRAS.452.2528M}. They also lack physical properties (e.g. enhanced SFRs, stellar colors, etc.) that could distinguish them as outliers observationally in the SAMs of \citet{2015arXiv151201561C}. We also must adopt a linear bias parameter $b$ that translates a galaxy overdensity into a mass overdensity ($\delta_{m} = \delta_{gal}/b$). This bias value is assumed to be in the range of 3-6 at these redshifts \citep{1998ApJ...492..428S}, and is dependent on the types of galaxies observed. Furthermore, the values of $\delta_{gal}$ are highly dependent on how the field is defined and the scales at which the overdensity exists, as we have pointed out here and in CCPC1.


The `virial' mass estimate also has a number of systematic uncertainties. These systems are the diffuse, primordial manifestations of clusters and are not expected to be in equilibrium in \LCDM. In the Millennium simulation, \citet{chi13} found that the first $M\ge10^{14}$ $M_{\odot}$ halos are not present until $z\le2.3$, and may not be virialized until a dynamical time later. However, there are some indications that subhalos in protoclusters can be virialized, as observed by \citet{2007AnA...461..823V,shi14} and \citet{2016arXiv160502079T}. These subhalos can be seen as bimodal distributions of protocluster galaxies along the LOS. Intriguingly, \citep{2016arXiv160407404W} recently identified a cluster core emitting X-rays at $z=2.5$. The cosmological implications of this discovery are uncertain.

Virial equilibrium is unlikely to hold, but we can query the data to see what mass is implied.
Equation 2 is the familiar virial mass estimate that only requires an effective radius ($R_{hms}=1.25R_e$) and velocity dispersion ($\sigma$) for a system in equilibrium. Based on the position of galaxies in the plane of the sky, we have calculated the radius in which $50\%$ of the member galaxies can be found. The mean $R_e$ is $2.8\pm1.0$ (physical) Mpc. The relatively small number of galaxies (minimum of 4) in our criteria is a further source of significant uncertainty, in that values of $R_e$ can be based on only two systems. The dispersion velocity is simply the standard deviation from the mean redshift of the system, and is on average 653 km s$^{-1}$. This is typical of dispersions found for other protoclusters, as compiled in Table 5 by \citet{chi13}.  We note that the few galaxy members involved in the calculation of the dispersion of these systems could introduce a large bias. Some candidates with $N=4$ members have $\sigma_z$ values based on only two galaxies. Richness alone does not appear to be a significant driver, however. The average velocity dispersion for systems with $N\ge23$ members is 661 km s$^{-1}$, the same as the mean value of the entire group.

Interestingly, these large velocity dispersions, like those at $z\ge3$ as found in \citet{2007AnA...461..823V}, CCPC1, and here, appear to be larger than those predicted by simulations. \citet{chi13} report that dispersions along the line of sight for overdensities at $z=3$ in the Millennium Simulation are $400\pm60$ km s$^{-1}$ for the progenitors of $M_{z=0}\ge10^{15}$ $M_{\odot}$ clusters. Our typical redshift uncertainty ($\sigma\sim 0.001$) represents a velocity uncertainty of 75 km s$^{-1}$ (at $z=3$), which cannot account for the $+100$ km s$^{-1}$ dispersion excess with respect to the simulated systems. Similarly, \citet{2005PhDT........14V} compared the protocluster dispersions in their sample to simulated dark matter halos of clusters within a similar window size. For $\sigma_{z=0}=1000$ km s$^{-1}$ systems (e.g. the Coma cluster), the dark matter velocity dispersion was found to be systematically lower than what was observed in the protoclusters at nearly all redshifts sampled.

It could be possible that galaxies not bound to the protocluster, or those at the outskirts, are boosting the dispersion significantly above expectations. \citet{2015arXiv151201561C} plot the velocity field for simulated protocluster galaxies in their Fig 2 for a system with $M_{z=0}\ge10^{15}$ $M_{\odot}$. At $z=3$, the members near the central galaxy ($R<5$ cMpc) have velocities of 200-400 km s$^{-1}$. Galaxies more than 20 cMpc away can have velocities in excess of 1000 km s$^{-1}$ \citep{2015arXiv151201561C}. Survey sizes within the CCPC1 and 2 vary in both width and depth, yet this phenomenon persists. In our smallest survey volumes, the inner regions should contain the brightest, most massive galaxies and the velocities are expected to be the smallest \citep{2015arXiv151201561C}. Greater spectroscopic completeness within some of the richest overdensities might shed light on this mystery. This would allow a more apt comparison to simulated protoclusters, which `observe' all galaxies and not just the brightest sources.

As an attempt to mitigate the weight of outliers for small numbers of galaxies when calculating a velocity dispersion for protoclusters, \citet{2007AnA...461..823V} utilized the biweight location estimator \citep{1990AJ....100...32B}. The dispersions discussed previously, both here and in CCPC1, were simple standard deviations. We employed this biweight estimator in the AstroPy package \citep{astropy} and iteratively solved for the velocity dispersion for each CCPC system. The median value of these biweight dispersions across the CCPC2 catalog is 717 km s$^{-1}$, which is $<1\sigma$ larger than a simple median value of all dispersions (677 km s$^{-1}$). Outlier galaxies do not appear to be biasing our results significantly.

We note that unlike CCPC1, this work did not require a knot of $\ge3$ galaxies to exist near the central search regions. This had the occasional effect of `centering' the galaxy distribution. For CCPC2, we sought only to maximize the number of galaxy members within the search radius of $R\le20$ cMpc. Therefore, some $R_e$ values could possibly be reduced if a different protocluster center were chosen (as in CCPC-z28-017), whereas others would be unaffected by this method change. However, for the bulk of the CCPC2 population, this is unimportant. The mean $R_e=2.2$ Mpc for CCPC1 is only marginally smaller than CCPC2 and within its standard deviation. The velocity dispersions are essentially equivalent for CCPC1 and 2 as well (668 and 653 km s$^{-1}$, respectively).

By limiting the distance we probe for structure along the line of sight, the calculated velocity dispersions are decreased, as one would expect. For example, restricting the search redshift range to half its original length 	($\Delta z \pm 10$ cMpc), only 32 protocluster candidates exceed the 400$\pm60$ value for the expected dispersion of a $M_{z=0}\ge10^{15}$ $M_{\odot}$ cluster at $z=3$ \citep{chi13}. These anomalous systems also have only 		$\sim6	$ galaxies, on average. To establish an expectation of the role window size affects the observed dispersion, we designed a Monte Carlo simulation of a protocluster at $z=3$. $N=6$, 10, and 200 galaxies were randomly distributed $10^4$ times with varying LOS distances ($1<d<20$ cMpc). We measured the mean dispersion and standard deviation of the simulation for each LOS distance. The Monte Carlo produces a linear relationship between LOS window size and velocity dispersion. When compared to the CCPC2 dispersion values in different window sizes, as well as the expectation value from the Millennium simulation \citep{chi13}, the Monte Carlo results were not distinct at a statistically significant level. These results suggest that the velocity dispersion excess could merely be the result of window size rather than a physical characteristic of the system.

The Monte Carlo simulation is only an effective modelling tool if we assume the correct, underlying distribution of galaxies and their velocities. For instance, we can make a simple assumption that all galaxies in the simulation have a true distance corresponding to $z=3$, and then imprint a Gaussian distribution of velocities of $\sigma=400$ km s$^{-1}$. These results can fit the data point of CCPC2 with a window size of $\pm10$ cMpc, as well as the expectation value of Chiang et al. (2013). We can modify the simulation again by allowing galaxies to be normally distributed within a specified radius, and then applying a random velocity (a pseudo-virial distribution) to each one individually. This can provide a fit to the CCPC2 for the full window size of the redshift distribution, but no other data points. Applying only infalling velocities to spatially dispersed galaxies can also match the observations. These examples are meant to illustrate the underlying degeneracy of the simulation and the complexity of the problem. Fundamentally, the menagerie of simulated protocluster sizes, velocity fields, and evolutionary states \citep{chi13,2015MNRAS.452.2528M,2015arXiv151201561C} make the prospect of a direct comparison of dispersions to real data a daunting task, even at a specific redshift and structure mass.

The average virial mass estimate is $M=8\times 10^{14}$ $M_{\odot}$ in CCPC2. Included are 61 structures (plus 13 in CCPC1) that have masses $M>\times 10^{15}$ $M_{\odot}$ at $z>2$. The Millennium simulation analysis by \citet{chi13} suggest that for the single most massive halo in a protocluster, none should have masses in excess of $\sim10^{14}$ prior to $z=2.3$. Likewise, \citet{mor11} predict that at $z\ge2$, there should not be a single collapsed structure in a \LCDM universe with $M > 6 \times 10^{14}$ $h^{-1}$ $M_{\odot}$. If any of these systems are in virial equilibrium, it could pose a serious challenge to the concordance cosmology.

\citet{chi13} also computed the radii (in comoving units) for the main halos in protoclusters at redshifts $z=2-5$ (their Fig 2). As noted earlier, our search algorithm was not designed to minimize the $R_e$ of candidate systems, and so the radii listed in Table~\ref{tab:best_mass} may be an overestimate. Subsequently, these large radii could bias our anomalously large virial mass estimates. To test this, we limited our entire sample to those with $R_e$ values less than the expected range for the most massive protoclusters \citep{chi13}. There were 66 such sources, with an average radius of 6.5 cMpc and redshift of $z\sim3.4$ (but spread between $2.00<z<6.5$). The average mass of these systems is $M_{vir}=2.7\times 10^{14}$ $M_{\odot}$. At $z=2.0$, this is approximately the mass expected for the largest main halos of the universe at this epoch. Observational mass estimates for structures at $z\sim2$ from SZ/X-ray emission are consistent with this result \citep{gob11,man14}. However, at larger redshifts, especially those at $z\ge6$, the discrepancy is on the order of $10^2$ from the predictions of \citet{chi13}. For at least this subsample, the radii do not appear to be the aberrant property in determining the mass.

It would seem that these discrepancies can be directly attributed to the inherent uncertainties in the velocity dispersions. \citet{gob11} identified a (proto)cluster at $z\sim2$ (identified here as CCPC-z20-002) and calculated its mass via X-Ray emission to be $M_{X-ray}=5-8\times 10^{13}$. In a follow-up spectroscopic study of the galaxies in the system, they obtained a velocity dispersion of 1300 km s$^{-1}$ \citep{gob13}, which would suggest a much larger mass than inferred by the emission from the intracluster medium. Considerable work needs to be done to establish reliable methods for estimating protocluster masses at high redshift.

The overdensity and virial mass estimates (Equations 1 and 2, respectively) do not correlate with one another in the CCPC. The two estimators are ostensibly measuring two different properties, so taken at face value this lack of a link is not too surprising. The overdensity method is attempting to quantify the amount of mass that will collapse at $z=0$, while the virial estimate is a representation of the current mass of the system in dynamical equilibrium. Despite the fact that even cluster progenitors of the same $z=0$ mass can exist in a variety of evolutionary stages at $z\ge2$ \citep{chi13,2015MNRAS.452.2528M}, one would expect some relationship to exist. The conclusion that can be drawn from this comparison, in addition to the previous paragraphs discussing both estimators, is that these mass values are highly uncertain; no reliable mass indicator is available at present. While each may represent some aspect of the physical nature for these systems, we caution the reader not to rely on these values.





\subsection{Protocluster Groups}

\begin{figure*}
\centering
\includegraphics[height=7.5cm,width=18.0cm]{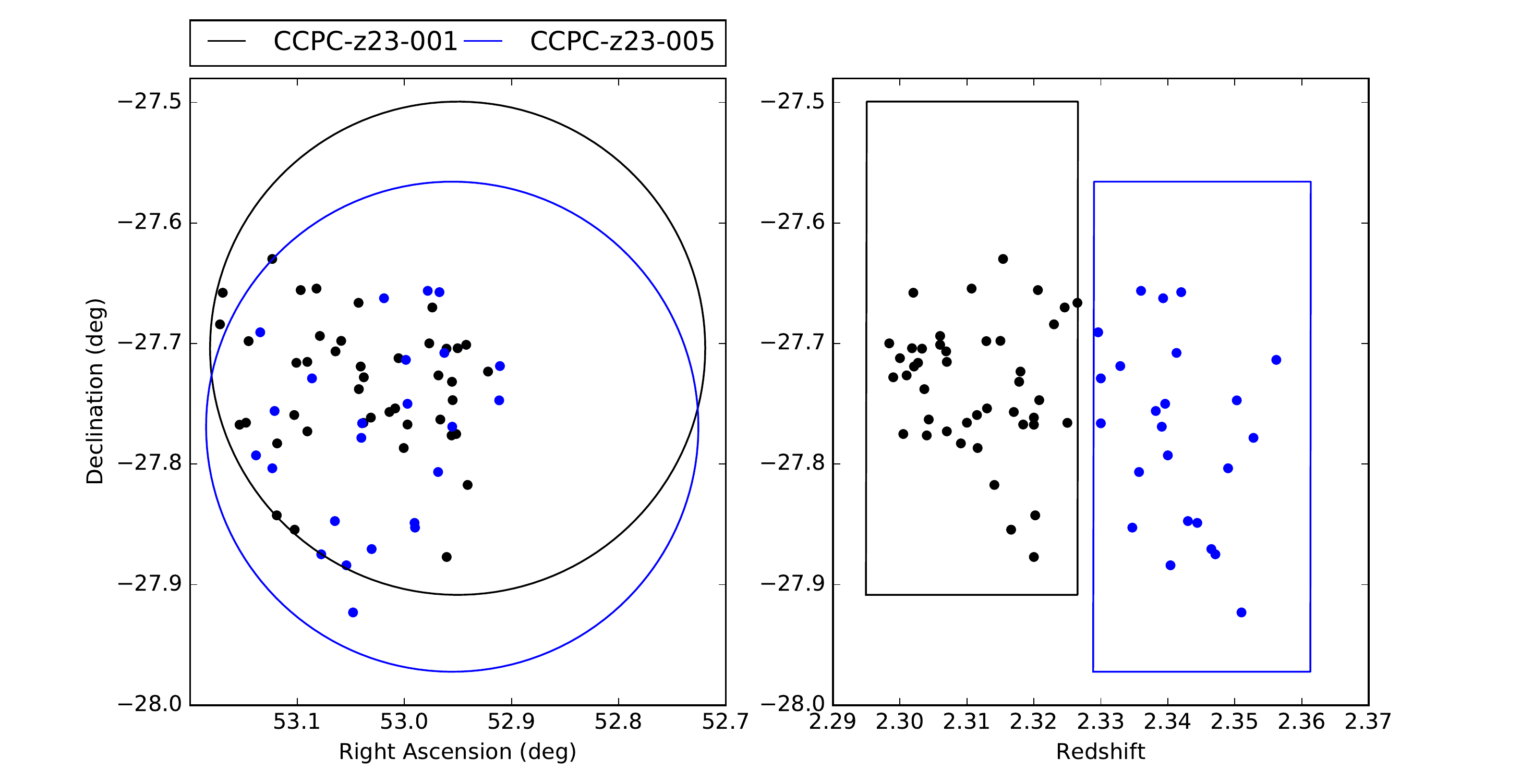}
\hfill
\caption{At $z\sim2.33$, our search algorithm selected the two galaxy associations (CCPC-z23-001 and CCPC-z23-005) as being strong protocluster candidates. $Left:$ Sky plot showing the overlap of the two structures on the sky, with the ellipses illustrating the boundary of our search volume ($R=20$ cMpc). $Right:$ Line-of-sight distribution of galaxies in the two candidates, which appear as a single, unbroken distribution. The boxes are the boundaries of $\Delta z$ corresponding to $\pm20$ cMpc. Further analysis shows that these objects appear as one continuous structure of volume $4.6\times10^{4}$ cMpc$^3$. The comoving volume is calculated by assuming a box with a length corresponding to the minimum and maximum redshifts, and width bounded by the galaxy positions in the sky plot ($Left$ panel).} There is also no break in the $N(z)$ distribution that would indicate two separate structures. A mass estimate based on its overdensity is $3\times10^{15}$ $M_{\odot}$, which would be one of the largest structures known.
\label{fig:SUPER1}
\end{figure*}


\emph{CCPC-z23-001 and CCPC-z23-005}: These overdensities were identified as two distinct systems by our search algorithm. After inspection, it was realized that these objects appear to be a single, extended system at $z=2.33$. Fig~\ref{fig:SUPER1} shows that the galaxy footprints overlap one another on the sky. However, the line-of-sight distribution of the galaxies appear to show a continuous distribution of sources of roughly 70 cMpc in length. The $N(z)$ plot exhibits a single, wide spike in galaxy counts (i.e. not a bimodal distribution). This spike was also identified using a combined catalog of primarily photometric with some spectroscopic galaxies by \citet{2009AnA...501..865S} as well as \citet{2015JKAS...48...21K}. On the sky, the system is approximately 22$\times$31 cMpc in RA/DEC, respectively. This is a large structure, but it is not unique. For example, \citet{lee14} found three overdensities at $z=3.78$ within a $75\times75\times25$ cMpc$^3$ volume. \citet{dey16} provided follow-up spectroscopic coverage to that extended system, which may be part of a filament stretching $\sim170$ cMpc. More recently, \citet{2016arXiv160607073Z} identifed four protoclusters of LAEs with volumes in excess of $[15 cMpc]^3$ around an overdensity discovered by \citet{2010AnA...512A..12B} and listed in the CCPC1 as CCPC-z28-002.

The galaxy overdensities of CCPC-z23-001 and CCPC-z23-005 are relatively modest ($\delta_{gal}=2$, $0.7$ respectively). However, this can be partially attributed to the strong `field' counts from the neighboring system. CCPC-z23-001 is one of the richest systems in the combined CCPC ($N=43$), while together they boast 66 member galaxies. These galaxies fill a volume of $4.6\times10^{4}$ cMpc$^3$ and have a combined overdensity of $\delta_m=0.52$. To compute the mass overdensity, we found the number density of galaxies within $z=2.329\pm0.03$ and a field length $\Delta z=0.15$. The volume is simply the rectangular region that encapsulates all galaxy members multiplied by the length of the box in comoving units. This volume and overdensity imply a mass of $3\times10^{15}$ $M_{\odot}$ if the superstructure were to collapse. However, this large of a volume is not expected to become one system by $z=0$ in a \LCDM Universe. Within this volume, assuming a number density of clusters to be $7.8\times 10^{-6}$ cMpc$^{-3}$  \citep{chi13}, we would expect 0.3, $M\ge 10^{14}$ $M_{\odot}$ clusters to be found. If they are each $\ge 10^{15}$ $M_{\odot}$ systems, as their overdensity and volume implies, the cluster number density decreases to $1.6\times 10^{-7}$ cMpc$^{-3}$. Within the estimated $5.893\times10^6$ cMpc$^3$ volume in which we identify structure in the GOODS-S field ($2<z<5.7$), we would expect to find fewer than one $\ge 10^{15}$ $M_{\odot}$ protocluster. That we find two systems in a volume $8\times10^{-3}$ smaller is intriguing.  The nature and fate of this system(s) is not yet understood.

\begin{figure*}
\centering
\includegraphics[height=7.5cm,width=18.0cm]{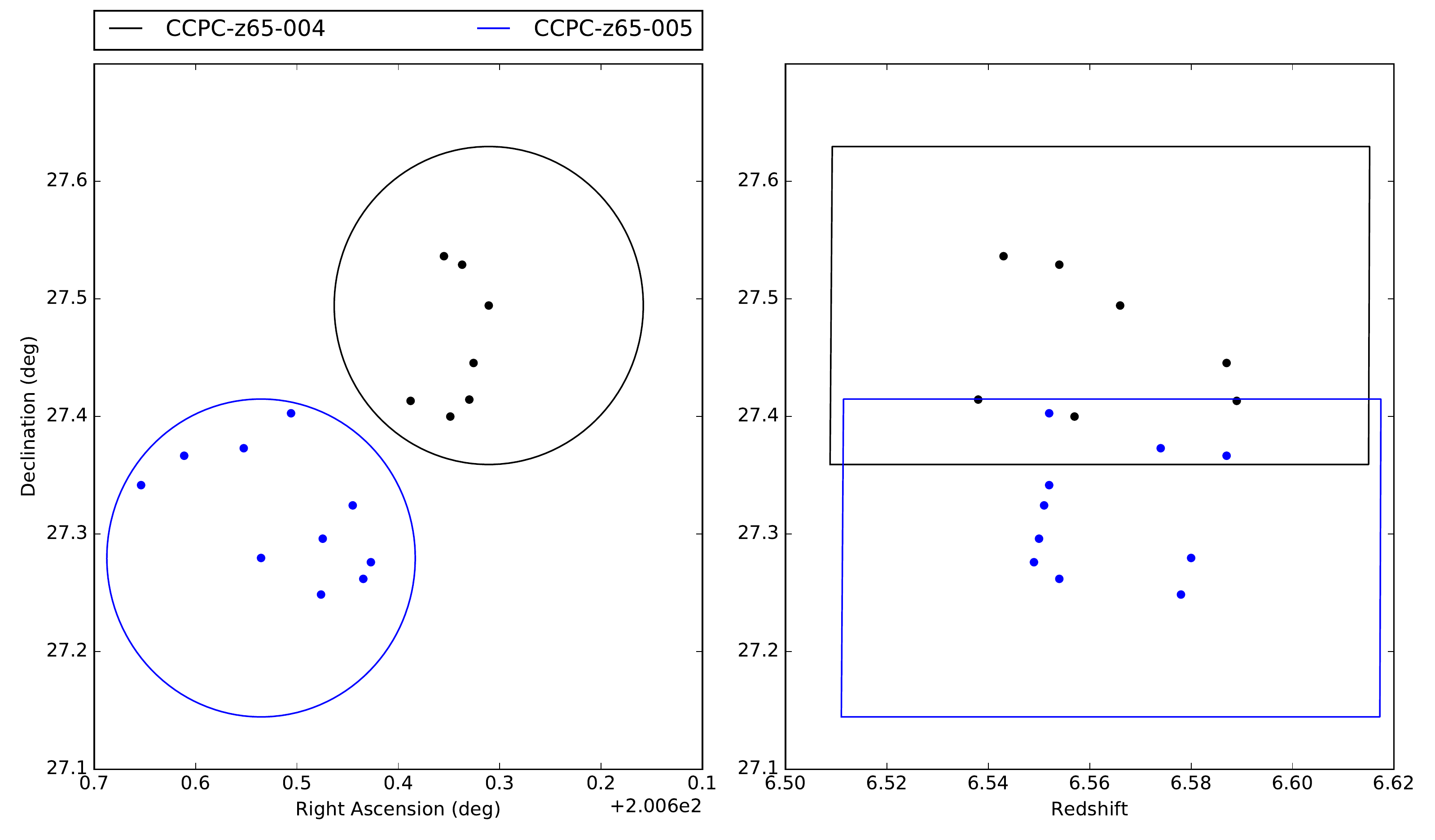}
\hfill
\caption{CCPC-z65-004 and CCPC-z65-005 are two systems separated by $\sim$40 cMpc on the sky ($Left$ panel), with their search volumes (ellipses) nearly touching. Indeed, a search center located at the midpoint (RA: 201 $deg$, DEC: +27.4 $deg$) would capture the majority of galaxies in the two distributions. In the $Right$ panel, the $\Delta z$ plots corresponding to a length of 40 cMpc, almost perfectly overlap, showing the galaxies along the line of sight. These could be two distinct associations, or one system with a geometric center in between the two groups. The mass of this system is estimated to be $M>2\times10^{15}$ $M_{\odot}$. These two (sub)protoclusters are the highest spectroscopically identified protoclusters at the time of writing.}
\label{fig:SUPER2}
\end{figure*}

\emph{CCPC-z65-004 and CCPC-z65-005}: This object is similar to the previous example in that the algorithm detected these two objects as separate sources. However, the redshift at which this object is found makes it even more interesting. Along the line of sight, these two systems essentially overlap (right panel of Fig~\ref{fig:SUPER2}), are much shorter than the $\Delta z=40$ cMpc search length ($<20 $ cMpc), and are thin. On the sky, their respective search centers are merely 40 cMpc offset from one another. This could mean that this is a single, very massive protocluster ($\delta_{gal}\sim4$, $M>2\times10^{15}$ $M_{\odot}$) with a geometric center in between the two coordinates listed. Of the two other $z\sim6$ protoclusters known \citep{utsumi10,toshikawa14}, this combined system has an overdensity mass estimate at least 5 times larger.

Regardless of whether CCPC-z65-004 and CCPC-z65-005 are a single system or two separate, high mass protoclusters, this detection represents the highest redshift association of galaxies that has been spectroscopically confirmed to the best of our knowledge. \citet{tre12} and \citet{2015arXiv150901751I} have both identified protocluster candidates at $z\sim8$ based on strong overdensities of $Y$-dropout galaxies, but these have yet to be spectroscopically confirmed.


\begin{figure*}
\centering
\includegraphics[height=7.5cm,width=18.0cm]{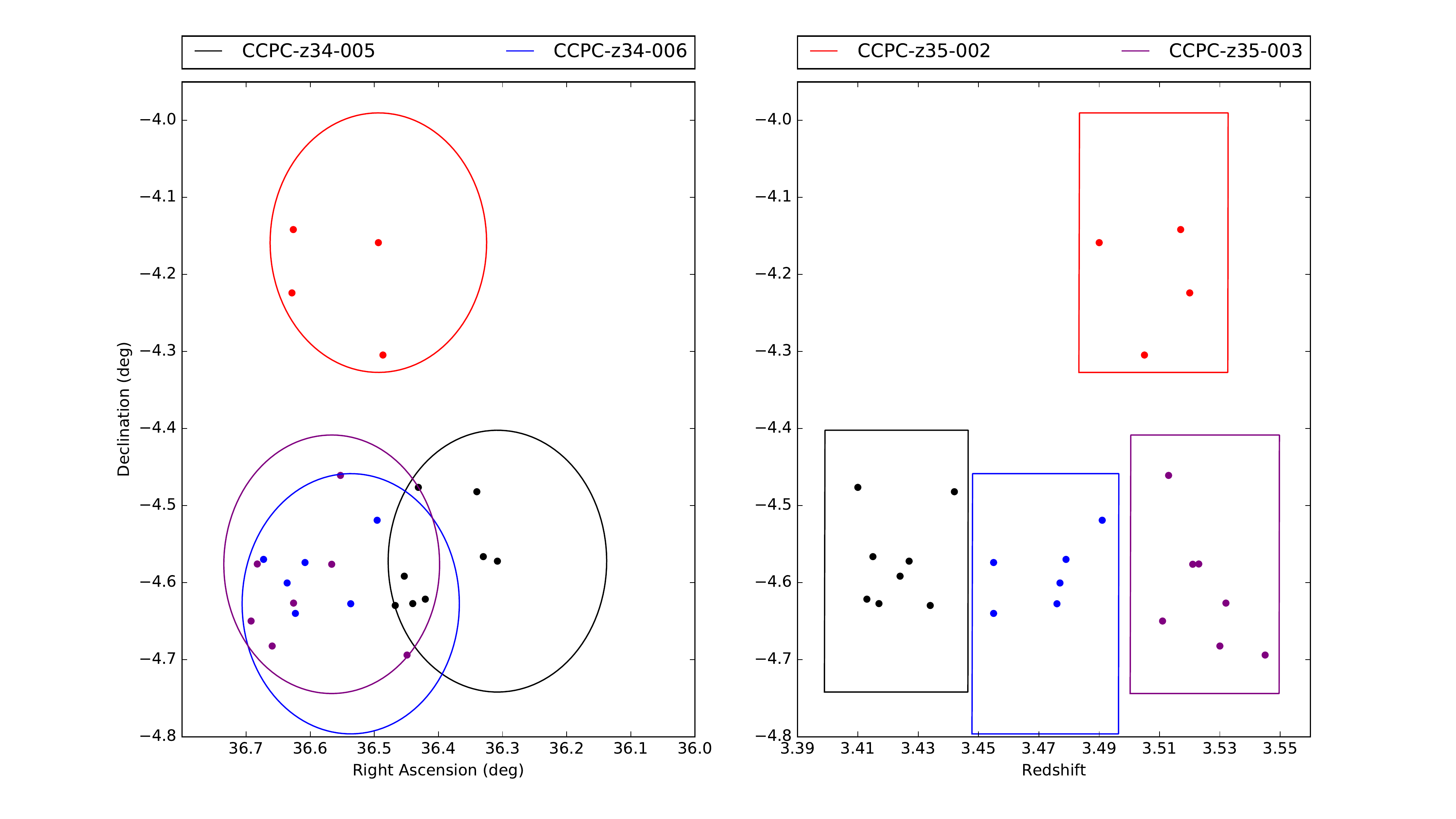}
\hfill
\caption{A group of 4 protocluster candidates at $z=3.5$ that are proximate in space. The $Left$ panel is the sky association of the candidates, while the $Right$ panel shows their galaxy distribution along the line of sight. The ellipses/boxes show the search volume boundaries ($R_{search}=20$ cMpc, $\Delta z \pm 20$ cMpc). Three of these (CCPC-z34-005, CCPC-z34-006, CCPC-z35-003) exist in a chain along the LOS stretching $\le 120$ cMpc. This may become a supercluster-sized structure at $z=0$.}
\label{fig:SUPER3}
\end{figure*}

\emph{CCPC-z34-005, CCPC-z34-006, CCPC-z35-002, and CCPC-z35-003 Complex}: These overdensities, although individually unremarkable, are part of a linked superstructure. These four separate systems almost touch on the sky in Fig~\ref{fig:SUPER3}.  CCPC-z34-005, CCPC-z34-006, and CCPC-z35-003 all exist along the same line of sight in a chain of $\le 120$ cMpc in length. The center of CCPC-z35-002 is separated from the chain by approximately 40 cMpc.  This complex may be a proto-supercluster in the process of assembly.

There are a number of other systems in CCPC2 that may be associated with one another. CCPC-z27-006 and CCPC-z27-010 appear to be nearly touching along the LOS and overlap one another on the sky. They appear similar to CCPC-z23-001 and CCPC-z23-005 (as seen in Fig~\ref{fig:SUPER1}), but are not nearly as rich. CCPC-z28-015 and CCPC-z28-016, and CCPC-z45-001-CCPC-z45-002 are more similar in nature to the distribution of galaxies in Fig~\ref{fig:SUPER2} with similar redshifts but separated on the sky by roughly the search radius of the algorithm. However, the associations are not as strong as that of CCPC-z65-004-CCPC-z65-005. Nevertheless, it appears that it is not uncommon for protocluster candidates to reside in very large scale associations. 

We emphasize again that the volumes estimated in this manner are highly uncertain. Candidate member galaxies on the outskirts of a distribution, whether projected on the sky or along the line of sight, can significantly enhance the volume implied. These volumetric tracers may not be bound to the structure at all, or may be sufficiently separated that they will not reside within the cluster's halo at $z=0$.

\section{Summary} \label{sec:sum}

We have extended the Candidate Cluster and Protocluster Catalog to redshifts $2<z<6.6$ in CCPC2, adding 173 protocluster candidates to the 43 in CCPC1. In the CCPC2, we identified galaxy overdensities ($\delta_{gal}>0.25$) of 4 or more galaxies within a search radius of $R=20$ cMpc and within a $\Delta z$ of $\pm 20$ cMpc. In Table~\ref{tab:tot_cat}, all candidate protoclusters are listed. The 36 systems that have the largest collapse probabilites ($\delta_{gal}>6$) are found in Table~\ref{tab:best_cat}. The median number of galaxy members is 6 in CCPC2, and 9 candidates have $N>23$ galaxies. Above a redshift of $z>4$, we have identified 40 structures. Prior to this work, fewer than 10 had been identified.  At the time of writing, this list includes the most distant spectroscopically-confirmed protocluster known (CCPC-z65-005). The combined CCPC is the largest known list of high redshift, spectroscopic protoclusters to date.

Following the examples of \citet{1998ApJ...492..428S} and \citet{ven02}, we estimate the strength of these protoclusters by computing their overdensities with respect to field counts of galaxies at similar redshifts.
These structures contain a median overdensity of $\delta_{gal}\sim2.9$, slightly larger than the median value found in CCPC1 and typical of the overdensities found in the literature \cite[summarized in Table 5 of][]{chi13}, as well as expectations from simulations. These overdensities and their visual counterparts ($N(z)$ plots for each CCPC member in the Appendix) are two pieces of evidence suggesting that these galaxy associations are indeed structures and not merely coincident on the sky. We emphasize that these overdensities are conservative estimates (Fig~\ref{fig:over}).

In Section~\ref{sec:dis}, we compare the spatial distribution and surface densities of CCPC systems to field galaxies. Using the list of $\ge40,000$ galaxies from which we identify overdensities, we built the mean surface density of `All galaxies' as a proxy for the field in Fig~\ref{fig:density}. Compared to the mean distribution of CCPC galaxies out to $R=20$ cMpc, there are significantly fewer galaxies at all radii in the field. A KS test of these two distributions shows a statistically significant difference (KS$=0.60$). Fig~\ref{fig:dist} illustrate a similar trend in the total number of galaxies as a function of the distance from search center. There are fewer field galaxies, at all redshifts and radii, than in CCPC systems.
We estimate the expected number of protoclusters that could be found within the CANDELS GOODS-S pencil beam survey volume based on the number density of clusters in Millennium \citep{chi13}. In CCPC1 and CCPC2, we found 27 candidate structures, whereas 46 clusters of $M\ge 10^{14}$ $M_{\odot}$ were expected in a similar volume of the \LCDM N-body simulation. This suggests that we are not over-identifying structure using our method, and may be recovering only the most significant overdensities. For systems of Virgo-mass and larger ($M\ge 3\times10^{14}$ $M_{\odot}$), 14 are expected in the GOODS-S volume.

We apply an analytic toy model by \citet{she01} to estimate the expected number of galaxy groups stochastically produced in a smooth density field. We then compare this expectation value to the number of FoF groups identified in CANDELS GOODS-S survey and found an excess of FoF groups compared to what would be expected stochastically from Poissonian fluctuations and Monte Carlo simulations. This test furthers the notion that these are physical structures and not mere chance overdensities.  

These tests were to ensure that the structures identified with the heterogeneous data available were legitimate structures. We have shown that: (1) these objects exist as galaxy overdensities (median $\delta_{gal}\sim 2.9$), (2) have $4\times$ larger number densities than LAEs (Section~\ref{sec:ccpc}), (3) differ spatially from field spectroscopic catalogs in Section~\ref{sec:dis}, (4) are not the result of Poissonian fluctuations in a smooth density field \citep{she01}, and (5) we do not over-identify structures in the deepest survey used (CANDELS GOODS-S). It is statistically unlikely that these tests, when taken together, would fail to distinguish the field from 
the candidates if a significant number of these systems are not actual structures.

For each CCPC member, we estimated the mass of the structure using two distinct methods. The first technique uses a linear bias parameter $b$ to transform the galaxy overdensity $\delta_{gal}$ into a mass overdensity $\delta_{m}$. If one assumes that the volume ($V$) traced by galaxies will collapse to a cluster, the mass of the system at $z=0$ can be approximated by the product of the volume and mass density \citep{1998ApJ...492..428S}. The average overdensity mass estimate for the CCPC2 is $1.8\times10^{14}$ $M_{\odot}$. We also computed the virial mass for each system based on its effective radius $R_e$ and velocity dispersion $\sigma$, making the crude assumption that these systems are in virial equilibrium. The mean `virial' mass is $M=8.4\times 10^{14}$ $M_{\odot}$ calculated for all CCPC2 members. With the combined CCPC1 and CCPC2 catalog, a total of 74 structures have mass estimates of $M\ge 10^{15}$ $M_{\odot}$. Such large masses are not expected at redshifts $z>2.3$ in N-body simulations \citep{chi13}. There is little agreement between the volumetric and virial mass estimators which emphasizes their uncertainty and the dubiousness of the necessary assumptions (e.g. virial equilibrium). 

There does appear to be a discrepancy between the observed and predicted velocity dispersions \citep{chi13,2015arXiv151201561C}. The median value is $\sigma\ge650$ km s$^{-1}$ in both CCPC1 and CCPC2, with some systems having dispersions as large as $900$ km s$^{-1}$, here and in \citet{2007AnA...461..823V}. This is more than double the expected value at the relevant redshifts for the most massive protoclusters \cite[$400\pm60$ km s$^{-1}$ in][]{chi13}. This is important because no assumption of equilibrium is made. However, it may not be possible to make an apples-to-apples comparison, as objects near the edge of the protocluster have large ($\sim1000$ km s$^{-1}$) simulated velocities \citep{2015arXiv151201561C}. By reducing the line-of-sight window size to $\pm10$ cMpc generally removes this observed excess to expected values. We note that it is difficult to map `observations' of simulations to real data. The observed excess of velocity dispersion in the CCPC is most likely not a physically-relevant result.

The CCPC2 also has three groups of protoclusters (Fig~\ref{fig:SUPER1}, Fig~\ref{fig:SUPER2}, and Fig~\ref{fig:SUPER3}) that may be primordial superclusters based on their small ($\le100$ cMpc) separations. CCPC-z23-001 and CCPC-z23-005 appear to be physically connected in a chain of length $\sim 70$ cMpc. Its nature is not understood currently, as it appears to be a single, extended structure of volume $\ge4\times10^4$ cMpc$^3$ and mass $\ge 10^{15}$ $M_{\odot}$. CCPC-z65-004 and CCPC-z65-005 are the two highest redshift spectroscopically confirmed protoclusters known to date, and are separated by a mere 40 cMpc. They may be a single, massive structure with a geometric center offset from their galaxy distributions, or two disparate protoclusters very near to one another. CCPC-z34-005, CCPC-z34-006, CCPC-z35-002, and CCPC-z35-003 form a protocluster complex, with three existing in a chain along the line of sight. 

In total, we have identified 173 protocluster candidates that appear to be genuine, spectroscopically confirmed, physical associations of galaxies at high redshift. Some of these systems reside in close proximity, as if part of (proto) super-clusters. Candidate protoclusters of high confidence are found up to $z\approx6.5$. There appears to be a rich amount of structure still to be revealed in the high redshift universe.

\acknowledgements
We would like to thank the anonymous referee for comments that improved this work.
JRF is grateful to Idit Zehavi for helpful discussion on Poissonian expectations for a smooth density field, and Ji Hoon Kim for providing Subaru NB filter curves. This research made use of Astropy, a community-developed core Python package for Astronomy (Astropy Collaboration, 2013). This research has made use of the NASA/IPAC Extragalactic Database (NED) which is operated by the Jet Propulsion Laboratory, California Institute of Technology, under contract with the National Aeronautics and Space Administration.

\bibliography{protos_omega}
\bibliographystyle{apj}


\clearpage
\appendix \label{sec:app}

\LongTables
\begin{deluxetable*}{c c c c c c c c c c c}
\tablewidth{0pt}
\tablecolumns{11}
\tablecaption{Candidate Cluster and Protocluster Catalog (CCPC) - All Candidates}
\tablehead{
\colhead{Candidate}	&\colhead{RA}	&\colhead{DEC}&\colhead{Redshift}	&\colhead{$\sigma_z$}&\colhead{$N$}& \colhead{$N_{R\le10}$} & \colhead{Overdensity}	&\colhead{Cluster}	&\colhead{Q}	&	\colhead{Recovered}	\\
\colhead{Name}&\colhead{($deg$)}&\colhead{($deg$)}&\colhead{($z_{avg}$)}&	\colhead{}&	\colhead{}	&\colhead{cMpc}&\colhead{($\delta_{gal}$)}	&\colhead{Probability ($\%$)}	&	\colhead{}&	\colhead{Reference}
}
\startdata	
CCPC-z20-001	&	189.00	&	62.18	&	1.997	&	0.006	&	30	&	15	&	5.07	$\pm$	1.52		&	79.4	&	1	&	\\
CCPC-z20-002	&	222.20	&	8.92	&	2.002	&	0.008	&	11	&	7	&	9.38	$\pm$	5.34		&	100.0	&	1	&	8\\
CCPC-z20-003	&	29.62	&	-25.05	&	2.018	&	0.004	&	10	&	10	&	19.43	$\pm$	13.06		&	100.0	&	1	&	1,2\\
CCPC-z20-004	&	36.62	&	-4.52	&	2.023	&	0.006	&	6	&	1	&	2.23	$\pm$	1.67		&	31.3	&	1	&	\\
CCPC-z20-005	&	52.97	&	-27.80	&	2.033	&	0.008	&	25	&	8	&	0.99	$\pm$	0.25		&	2.5	&	1	&	\\
CCPC-z20-006	&	189.19	&	62.29	&	2.031	&	0.008	&	8	&	6	&	0.48	$\pm$	0.22		&	2.5	&	1	&	\\
CCPC-z20-007	&	149.24	&	69.65	&	2.049	&	0.006	&	4	&	4	&	[2]	$\pm$	4.16		&	11.3	&	1	&	3\\
CCPC-z20-008	&	188.97	&	62.23	&	2.086	&	0.007	&	15	&	9	&	1.30	$\pm$	0.52		&	11.3	&	1	&	\\
CCPC-z20-009	&	150.04	&	2.21	&	2.098	&	0.005	&	10	&	4	&	13.15	$\pm$	6.54		&	100.0	&	1	&	4,5\\
CCPC-z21-001	&	246.49	&	26.75	&	2.107	&	0.007	&	5	&	4	&	4.17	$\pm$	3.02		&	75.6	&	1	&	\\
CCPC-z21-002	&	356.52	&	12.76	&	2.114	&	0.004	&	5	&	4	&	1.90	$\pm$	1.52		&	11.3	&	1	&	\\
CCPC-z21-003	&	189.18	&	62.21	&	2.129	&	0.008	&	9	&	7	&	1.91	$\pm$	0.87	\tablenotemark{a}  	&	11.3	&	1	&	\\
CCPC-z21-004	&	175.15	&	-26.47	&	2.155	&	0.007	&	24	&	24	&	6.34	$\pm$	3.36		&	100.0	&	1	&	9,1\\
CCPC-z21-005	&	214.31	&	52.40	&	2.160	&	0.007	&	5	&	3	&	9.07	$\pm$	6.93		&	100.0	&	1	&	\\
CCPC-z21-006	&	334.35	&	0.32	&	2.172	&	0.005	&	4	&	3	&	18.85	$\pm$	13.55		&	100.0	&	1	&	\\
CCPC-z21-007	&	356.58	&	12.80	&	2.174	&	0.002	&	7	&	7	&	17.27	$\pm$	10.57		&	100.0	&	1	&	\\
CCPC-z21-008	&	149.98	&	2.11	&	2.179	&	0.002	&	5	&	1	&	9.41	$\pm$	5.38		&	100.0	&	2	&	4\\
CCPC-z21-009	&	246.39	&	26.90	&	2.183	&	0.005	&	5	&	2	&	2.95	$\pm$	2.02		&	31.3	&	1	&	13\\
CCPC-z21-010	&	338.17	&	-60.52	&	2.187	&	0.006	&	6	&	5	&	3.59	$\pm$	2.22		&	52.5	&	1	&	\\
CCPC-z21-011	&	189.22	&	62.25	&	2.199	&	0.010	&	18	&	17	&	0.70	$\pm$	0.25		&	2.5	&	1	&	\\
CCPC-z22-001	&	53.09	&	-27.94	&	2.205	&	0.008	&	10	&	4	&	0.63	$\pm$	0.22	\tablenotemark{a}  	&	2.5	&	1	&	\\
CCPC-z22-002	&	189.05	&	62.18	&	2.234	&	0.009	&	18	&	8	&	0.32	$\pm$	0.13		&	2.5	&	1	&	\\
CCPC-z22-003	&	198.03	&	42.66	&	2.239	&	0.006	&	5	&	5	&	5.83	$\pm$	4.96		&	79.4	&	1	&	\\
CCPC-z22-004	&	255.20	&	64.22	&	2.243	&	0.007	&	6	&	6	&	1.06	$\pm$	0.61		&	11.3	&	1	&	\\
CCPC-z22-005	&	16.48	&	-25.78	&	2.251	&	0.004	&	5	&	5	&	4.53	$\pm$	2.80		&	75.6	&	1	&	\\
CCPC-z22-006	&	149.93	&	2.20	&	2.283	&	0.006	&	5	&	1	&	1.94	$\pm$	1.31		&	11.3	&	2	&	4\\
CCPC-z22-007	&	255.20	&	64.26	&	2.296	&	0.008	&	32	&	30	&	7.77	$\pm$	2.90		&	100.0	&	1	&	6\\
CCPC-z23-001	&	52.95	&	-27.70	&	2.311	&	0.008	&	43	&	25	&	2.06	$\pm$	0.57		&	31.3	&	1	&	22\\
CCPC-z23-002	&	334.46	&	0.14	&	2.309	&	0.009	&	4	&	3	&	11.45	$\pm$	9.05		&	100.0	&	1	&	\\
CCPC-z23-003	&	214.39	&	52.49	&	2.333	&	0.008	&	4	&	4	&	11.73	$\pm$	10.66		&	100.0	&	1	&	\\
CCPC-z23-004	&	255.17	&	64.17	&	2.337	&	0.008	&	6	&	5	&	0.45	$\pm$	0.28		&	2.5	&	1	&	7\\
CCPC-z23-005	&	52.96	&	-27.77	&	2.341	&	0.007	&	23	&	11	&	0.68	$\pm$	0.22		&	2.5	&	1	&	\\
CCPC-z23-006	&	16.47	&	-25.76	&	2.346	&	0.007	&	5	&	5	&	4.65	$\pm$	3.13		&	75.6	&	1	&	\\
CCPC-z23-007	&	258.53	&	50.27	&	2.390	&	0.005	&	7	&	6	&	16.63	$\pm$	14.52		&	100.0	&	1	&	1,20\\
CCPC-z24-001	&	53.00	&	-27.85	&	2.400	&	0.007	&	12	&	1	&	0.57	$\pm$	0.20		\tablenotemark{a}  	&	2.5	&	1	&	\\
CCPC-z24-002	&	189.08	&	62.19	&	2.416	&	0.009	&	14	&	11	&	0.72	$\pm$	0.29		&	2.5	&	1	&	\\
CCPC-z24-003	&	164.20	&	-3.64	&	2.426	&	0.005	&	7	&	7	&	15.06	$\pm$	8.92		&	100.0	&	1	&	\\
CCPC-z24-004	&	255.24	&	64.21	&	2.430	&	0.007	&	6	&	6	&	0.61	$\pm$	0.39		&	2.5	&	1	&	\\
CCPC-z24-005	&	150.00	&	2.26	&	2.442	&	0.009	&	14	&	8	&	9.27	$\pm$	4.93		&	100.0	&	1	&	4,10\\
CCPC-z24-006	&	16.49	&	-25.73	&	2.443	&	0.006	&	5	&	5	&	3.69	$\pm$	2.32		&	52.5	&	1	&	\\
CCPC-z24-007	&	46.11	&	-0.28	&	2.457	&	0.009	&	4	&	1	&	2.52	$\pm$	1.92		&	31.3	&	1	&	\\
CCPC-z24-008	&	246.39	&	26.90	&	2.475	&	0.010	&	4	&	2	&	2.89	$\pm$	2.33		&	31.3	&	1	&	\\
CCPC-z24-009	&	189.13	&	62.27	&	2.487	&	0.007	&	16	&	13	&	5.61	$\pm$	2.06		&	79.4	&	1	&	\\
CCPC-z24-010	&	316.81	&	23.53	&	2.486	&	0.002	&	4	&	4	&	[2]	$\pm$	4.16		&	11.3	&	1	&	1\\
CCPC-z25-001	&	246.37	&	26.80	&	2.529	&	0.005	&	6	&	2	&	2.01	$\pm$	1.40		&	18.1	&	1	&	\\
CCPC-z25-002	&	255.18	&	64.17	&	2.537	&	0.002	&	4	&	4	&	19.86	$\pm$	13.41		&	100.0	&	1	&	\\
CCPC-z25-003	&	143.36	&	28.77	&	2.548	&	0.003	&	5	&	5	&	10.89	$\pm$	7.70		&	100.0	&	1	&	\\
CCPC-z25-004	&	189.34	&	62.15	&	2.550	&	0.006	&	4	&	2	&	2.49	$\pm$	1.59		&	18.1	&	1	&	\\
CCPC-z25-005	&	53.13	&	-27.83	&	2.567	&	0.009	&	46	&	15	&	2.10	$\pm$	0.53		&	18.1	&	1	&	 23\\
CCPC-z25-006	&	255.17	&	64.17	&	2.574	&	0.005	&	4	&	4	&	6.29	$\pm$	5.02		&	73.1	&	1	&	\\
CCPC-z25-007	&	216.14	&	22.84	&	2.581	&	0.007	&	5	&	2	&	10.90	$\pm$	6.72		&	100.0	&	1	&	\\
CCPC-z25-008	&	189.07	&	62.25	&	2.589	&	0.009	&	11	&	9	&	0.47	$\pm$	0.23		&	1.3	&	1	&	\\
CCPC-z26-001	&	339.84	&	11.81	&	2.617	&	0.003	&	4	&	4	&	7.33	$\pm$	6.16		&	85.6	&	1	&	\\
CCPC-z26-002	&	216.12	&	22.83	&	2.636	&	0.004	&	4	&	4	&	2.63	$\pm$	2.01		&	18.1	&	1	&	\\
CCPC-z26-003	&	189.26	&	62.36	&	2.645	&	0.009	&	4	&	1	&	0.58	$\pm$	0.45		&	1.3	&	1	&	\\
CCPC-z26-004	&	53.03	&	-27.88	&	2.666	&	0.010	&	18	&	10	&	1.12	$\pm$	0.36		&	10.6	&	1	&	\\
CCPC-z26-005	&	13.30	&	12.59	&	2.669	&	0.010	&	7	&	3	&	4.63	$\pm$	3.27		&	66.9	&	1	&	\\
CCPC-z26-006	&	255.14	&	64.22	&	2.688	&	0.005	&	5	&	5	&	7.24	$\pm$	5.18		&	85.6	&	1	&	\\
CCPC-z26-007	&	216.10	&	22.99	&	2.695	&	0.007	&	8	&	6	&	4.42	$\pm$	2.33		&	66.9	&	1	&	\\
CCPC-z27-006	&	214.36	&	52.60	&	2.710	&	0.008	&	5	&	2	&	6.32	$\pm$	4.21		&	73.1	&	1	&	\\
CCPC-z27-007	&	150.02	&	2.33	&	2.719	&	0.009	&	5	&	3	&	4.79	$\pm$	4.27		&	66.9	&	2	&	13\\
CCPC-z27-008	&	36.39	&	-4.51	&	2.729	&	0.006	&	6	&	2	&	7.27	$\pm$	4.83		&	85.6	&	1	&	\\
CCPC-z27-009	&	216.09	&	22.95	&	2.748	&	0.004	&	4	&	3	&	6.11	$\pm$	4.68		&	73.1	&	1	&	\\
CCPC-z27-010	&	214.39	&	52.60	&	2.747	&	0.009	&	9	&	4	&	2.91	$\pm$	1.73		&	18.1	&	1	&	\\
CCPC-z27-011	&	334.27	&	0.17	&	2.757	&	0.013	&	5	&	2	&	0.73	$\pm$	0.57		&	1.3	&	1	&	\\
CCPC-z27-012	&	16.48	&	-25.81	&	2.758	&	0.008	&	4	&	3	&	12.83	$\pm$	10.91		&	100.0	&	1	&	\\
CCPC-z27-013	&	36.40	&	-4.31	&	2.768	&	0.006	&	4	&	2	&	2.77	$\pm$	2.18		&	18.1	&	3	&	\\
CCPC-z27-014	&	143.33	&	28.80	&	2.792	&	0.011	&	4	&	2	&	1.29	$\pm$	0.90		&	10.6	&	1	&	\\
CCPC-z28-008	&	339.07	&	13.92	&	2.801	&	0.013	&	4	&	3	&	2.19	$\pm$	1.61		&	18.1	&	1	&	\\
CCPC-z28-009	&	13.34	&	12.51	&	2.803	&	0.004	&	4	&	2	&	2.38	$\pm$	1.83		&	18.1	&	1	&	\\
CCPC-z28-010	&	214.48	&	52.55	&	2.816	&	0.009	&	8	&	5	&	2.45	$\pm$	1.05	\tablenotemark{a}  	&	18.1	&	1	&	\\
CCPC-z28-011	&	36.36	&	-4.32	&	2.820	&	0.006	&	4	&	2	&	9.10	$\pm$	7.53		&	100.0	&	1	&	\\
CCPC-z28-012	&	339.93	&	11.86	&	2.836	&	0.009	&	4	&	1	&	2.06	$\pm$	1.59		&	18.1	&	1	&	\\
CCPC-z28-013	&	136.32	&	34.09	&	2.846	&	0.008	&	5	&	4	&	4.93	$\pm$	3.52		&	66.9	&	1	&	\\
CCPC-z28-014	&	334.37	&	0.24	&	2.863	&	0.006	&	5	&	4	&	4.97	$\pm$	2.71		&	66.9	&	1	&	\\
CCPC-z28-015	&	36.67	&	-4.36	&	2.859	&	0.010	&	4	&	1	&	2.34	$\pm$	1.60		&	18.1	&	1	&	\\
CCPC-z28-016	&	36.27	&	-4.28	&	2.866	&	0.006	&	5	&	1	&	15.46	$\pm$	11.42		&	100.0	&	1	&	\\
CCPC-z28-017	&	143.32	&	28.71	&	2.864	&	0.012	&	6	&	1	&	5.25	$\pm$	3.58		&	68.1	&	1	&	\\
CCPC-z28-018	&	255.19	&	64.17	&	2.889	&	0.012	&	9	&	6	&	1.52	$\pm$	1.05		&	10.6	&	1	&	\\
CCPC-z28-019	&	44.71	&	0.16	&	2.898	&	0.010	&	6	&	4	&	1.35	$\pm$	0.84		&	10.6	&	1	&	\\
CCPC-z29-008	&	36.84	&	-4.56	&	2.902	&	0.006	&	5	&	2	&	2.52	$\pm$	1.78		&	18.1	&	1	&	\\
CCPC-z29-009	&	136.34	&	34.14	&	2.905	&	0.010	&	5	&	5	&	13.33	$\pm$	9.13		&	100.0	&	1	&	\\
CCPC-z29-010	&	216.11	&	23.00	&	2.920	&	0.012	&	5	&	4	&	0.79	$\pm$	0.52		&	1.3	&	1	&	\\
CCPC-z29-011	&	339.95	&	11.87	&	2.925	&	0.008	&	13	&	3	&	9.62	$\pm$	5.14		&	100.0	&	1	&	\\
CCPC-z29-012	&	36.37	&	-4.41	&	2.932	&	0.012	&	5	&	2	&	1.04	$\pm$	0.76		&	10.6	&	1	&	\\
CCPC-z29-013	&	13.31	&	12.63	&	2.934	&	0.002	&	5	&	4	&	14.67	$\pm$	10.21		&	100.0	&	1	&	\\
CCPC-z29-014	&	214.30	&	52.49	&	2.965	&	0.012	&	14	&	6	&	0.58	$\pm$	0.25		&	1.3	&	1	&	\\
CCPC-z29-015	&	46.19	&	-0.11	&	2.965	&	0.009	&	4	&	3	&	4.45	$\pm$	3.24		&	66.9	&	1	&	\\
CCPC-z29-016	&	189.09	&	62.22	&	2.981	&	0.008	&	23	&	20	&	3.56	$\pm$	1.40		&	48.8	&	1	&	\\
CCPC-z30-004	&	255.25	&	64.16	&	3.000	&	0.010	&	4	&	3	&	0.61	$\pm$	0.50		&	1.3	&	1	&	\\
CCPC-z30-005	&	214.43	&	52.42	&	3.030	&	0.009	&	15	&	4	&	0.93	$\pm$	0.38		&	1.3	&	1	&	\\
CCPC-z30-006	&	339.06	&	13.95	&	3.050	&	0.010	&	7	&	6	&	3.53	$\pm$	2.21		&	48.8	&	1	&	\\
CCPC-z30-007	&	53.09	&	-27.67	&	3.074	&	0.009	&	11	&	2	&	1.77	$\pm$	0.86		&	10.6	&	1	&	\\
CCPC-z30-008	&	214.46	&	52.41	&	3.080	&	0.010	&	12	&	4	&	1.07	$\pm$	0.47		&	10.6	&	1	&	\\
CCPC-z30-009	&	44.74	&	0.21	&	3.088	&	0.004	&	4	&	4	&	4.90	$\pm$	3.96		&	66.9	&	1	&	\\
CCPC-z31-008	&	339.89	&	11.88	&	3.104	&	0.007	&	8	&	5	&	7.70	$\pm$	4.58		&	85.6	&	1	&	\\
CCPC-z31-009	&	339.07	&	14.00	&	3.107	&	0.007	&	5	&	5	&	4.93	$\pm$	3.46		&	66.9	&	1	&	\\
CCPC-z31-010	&	34.42	&	-4.53	&	3.119	&	0.010	&	4	&	4	&	0.60	$\pm$	0.74		&	1.3	&	2	&	\\
CCPC-z31-011	&	189.08	&	62.25	&	3.136	&	0.012	&	9	&	8	&	0.37	$\pm$	0.19		&	1.3	&	1	&	\\
CCPC-z31-012	&	34.22	&	-4.95	&	3.138	&	0.009	&	6	&	1	&	1.03	$\pm$	1.15		&	10.6	&	2	&	\\
CCPC-z31-013	&	216.10	&	22.89	&	3.139	&	0.006	&	6	&	5	&	1.49	$\pm$	1.01		&	10.6	&	1	&	\\
CCPC-z31-014	&	214.43	&	52.54	&	3.135	&	0.009	&	16	&	8	&	2.08	$\pm$	0.90		&	18.1	&	1	&	\\
CCPC-z31-015	&	339.87	&	11.88	&	3.148	&	0.008	&	9	&	2	&	8.15	$\pm$	4.56		&	100.0	&	1	&	\\
CCPC-z31-016	&	143.35	&	28.72	&	3.160	&	0.010	&	5	&	4	&	3.45	$\pm$	2.27		&	48.8	&	1	&	\\
CCPC-z31-017	&	36.76	&	-4.56	&	3.187	&	0.006	&	11	&	3	&	7.26	$\pm$	3.67		&	85.6	&	1	&	\\
CCPC-z32-004	&	13.40	&	12.41	&	3.209	&	0.009	&	4	&	3	&	1.40	$\pm$	1.02		&	10.6	&	1	&	\\
CCPC-z32-005	&	189.17	&	62.20	&	3.230	&	0.012	&	18	&	13	&	2.17	$\pm$	0.93		&	18.1	&	1	&	\\
CCPC-z32-006	&	143.34	&	28.75	&	3.231	&	0.008	&	4	&	4	&	4.25	$\pm$	3.41		&	66.9	&	1	&	\\
CCPC-z32-007	&	46.15	&	-0.19	&	3.233	&	0.007	&	5	&	4	&	11.47	$\pm$	9.32		&	100.0	&	1	&	\\
CCPC-z32-008	&	339.94	&	11.81	&	3.255	&	0.007	&	4	&	3	&	2.55	$\pm$	1.73		&	18.1	&	1	&	\\
CCPC-z32-009	&	16.47	&	-25.76	&	3.282	&	0.014	&	4	&	4	&	0.87	$\pm$	0.62		&	1.3	&	1	&	\\
CCPC-z32-010	&	214.41	&	52.46	&	3.281	&	0.010	&	10	&	5	&	1.65	$\pm$	0.83		&	10.6	&	1	&	\\
CCPC-z33-006	&	150.07	&	2.28	&	3.303	&	0.008	&	4	&	4	&	17.49	$\pm$	13.89		&	100.0	&	2	&	\\
CCPC-z33-007	&	334.35	&	0.07	&	3.310	&	0.012	&	7	&	5	&	11.52	$\pm$	9.15		&	100.0	&	1	&	\\
CCPC-z33-008	&	36.73	&	-4.74	&	3.312	&	0.007	&	4	&	2	&	1.68	$\pm$	1.42		&	10.6	&	1	&	17\\
CCPC-z33-009	&	214.32	&	52.42	&	3.351	&	0.012	&	7	&	4	&	0.69	$\pm$	0.44		&	1.3	&	1	&	\\
CCPC-z33-010	&	216.12	&	22.83	&	3.379	&	0.009	&	7	&	3	&	10.44	$\pm$	7.10		&	100.0	&	1	&	\\
CCPC-z34-003	&	214.40	&	52.45	&	3.401	&	0.013	&	7	&	4	&	1.02	$\pm$	0.64		&	10.6	&	1	&	\\
CCPC-z34-004	&	53.06	&	-27.88	&	3.395	&	0.012	&	6	&	2	&	0.46	$\pm$	0.24		&	1.3	&	1	&	\\
CCPC-z34-005	&	36.31	&	-4.57	&	3.423	&	0.010	&	8	&	2	&	6.41	$\pm$	4.52		&	73.1	&	1	&	\\
CCPC-z34-006	&	36.54	&	-4.63	&	3.472	&	0.013	&	6	&	1	&	7.82	$\pm$	5.40		&	85.6	&	2	&	\\
CCPC-z35-002	&	36.49	&	-4.16	&	3.508	&	0.012	&	4	&	1	&	4.10	$\pm$	4.23		&	41.3	&	1	&	\\
CCPC-z35-003	&	36.57	&	-4.58	&	3.525	&	0.011	&	7	&	2	&	1.92	$\pm$	1.20		&	13.1	&	1	&	\\
CCPC-z35-004	&	216.11	&	22.87	&	3.579	&	0.012	&	6	&	4	&	4.22	$\pm$	3.80		&	41.3	&	1	&	\\
CCPC-z35-005	&	36.53	&	-4.45	&	3.588	&	0.016	&	11	&	4	&	1.30	$\pm$	0.72		&	13.1	&	1	&	\\
CCPC-z36-003	&	53.01	&	-27.70	&	3.605	&	0.011	&	10	&	3	&	3.46	$\pm$	1.60		&	41.3	&	1	&	\\
CCPC-z36-004	&	36.39	&	-4.26	&	3.687	&	0.014	&	6	&	3	&	5.59	$\pm$	5.12		&	63.8	&	1	&	\\
CCPC-z36-005	&	36.55	&	-4.59	&	3.676	&	0.014	&	8	&	5	&	1.18	$\pm$	0.77		&	13.1	&	1	&	\\
CCPC-z36-006	&	34.17	&	-5.02	&	3.684	&	0.011	&	4	&	2	&	[3.62]	$\pm$	6.00		&	41.3	&	2	&	\\
CCPC-z36-007	&	34.54	&	-5.30	&	3.688	&	0.010	&	5	&	2	&	15.22	$\pm$	16.25		&	90.0	&	2	&	\\
CCPC-z37-002	&	36.77	&	-4.64	&	3.755	&	0.006	&	5	&	2	&	4.39	$\pm$	3.92		&	41.3	&	3	&	\\
CCPC-z37-003	&	53.01	&	-27.75	&	3.798	&	0.014	&	5	&	2	&	2.57	$\pm$	1.54		&	13.1	&	1	&	\\
CCPC-z38-001	&	36.34	&	-4.37	&	3.869	&	0.012	&	4	&	3	&	2.32	$\pm$	2.25		&	13.1	&	3	&	\\
CCPC-z40-001	&	189.14	&	62.24	&	4.050	&	0.011	&	11	&	6	&	5.61	$\pm$	2.99		&	63.8	&	1	&	21\\
CCPC-z40-002	&	204.59	&	-19.76	&	4.099	&	0.006	&	38	&	28	&	5.54	$\pm$	5.03		&	63.8	&	2	&	1,18,19\\
CCPC-z41-001	&	53.14	&	-27.82	&	4.125	&	0.011	&	5	&	4	&	3.28	$\pm$	2.38		&	41.3	&	2	&	\\
CCPC-z42-001	&	53.07	&	-27.76	&	4.287	&	0.010	&	5	&	3	&	2.90	$\pm$	1.92		&	13.1	&	3	&	\\
CCPC-z43-001	&	31.43	&	-5.09	&	4.387	&	0.007	&	8	&	2	&	1.58	$\pm$	1.17		&	13.1	&	2	&	\\
CCPC-z43-002	&	53.01	&	-27.70	&	4.398	&	0.009	&	4	&	2	&	5.76	$\pm$	4.75		&	63.8	&	3	&	\\
CCPC-z44-001	&	31.16	&	-4.74	&	4.415	&	0.015	&	19	&	8	&	1.73	$\pm$	1.20		&	13.1	&	2	&	\\
CCPC-z44-002	&	216.41	&	35.65	&	4.424	&	0.016	&	14	&	3	&	1.43	$\pm$	0.93		&	13.1	&	2	&	\\
CCPC-z44-003	&	189.15	&	62.23	&	4.424	&	0.010	&	5	&	3	&	13.79	$\pm$	12.03		&	90.0	&	1	&	\\
CCPC-z44-004	&	31.45	&	-4.96	&	4.462	&	0.019	&	12	&	1	&	1.00	$\pm$	0.61		&	1.9	&	1	&	\\
CCPC-z44-005	&	216.17	&	35.63	&	4.499	&	0.016	&	12	&	3	&	0.92	$\pm$	0.73		&	1.9	&	2	&	\\
CCPC-z45-001	&	53.04	&	-27.77	&	4.517	&	0.013	&	7	&	2	&	1.88	$\pm$	1.08		&	10.0	&	1	&	\\
CCPC-z45-002	&	53.33	&	-27.90	&	4.521	&	0.014	&	8	&	4	&	7.25	$\pm$	4.68		&	51.3	&	1	&	\\
CCPC-z45-003	&	31.43	&	-4.85	&	4.533	&	0.016	&	8	&	1	&	0.42	$\pm$	0.28		&	2.5	&	2	&	\\
CCPC-z48-001	&	53.02	&	-27.78	&	4.811	&	0.010	&	6	&	3	&	5.23	$\pm$	4.28		&	33.8	&	2	&	\\
CCPC-z48-002	&	240.97	&	43.38	&	4.839	&	0.014	&	6	&	3	&	4.20	$\pm$	4.70		&	21.3	&	2	&	11\\
CCPC-z49-001	&	163.64	&	-12.72	&	4.998	&	0.021	&	5	&	5	&	2.05	$\pm$	2.19		&	10.0	&	1	&	\\
CCPC-z50-001	&	54.58	&	0.40	&	5.070	&	0.014	&	4	&	2	&	16.75	$\pm$	18.30		&	68.1	&	1	&	15\\
CCPC-z51-001	&	141.03	&	-22.04	&	5.177	&	0.007	&	7	&	6	&	0.33	$\pm$	0.47		&	2.5	&	1	&	1\\
CCPC-z51-002	&	189.20	&	62.16	&	5.188	&	0.002	&	4	&	2	&	[2]	$\pm$	4.16		&	10.0	&	1	&	12\\
CCPC-z56-001	&	188.99	&	62.17	&	5.638	&	0.025	&	7	&	4	&	2.44	$\pm$	2.27		&	10.0	&	2	&	\\
CCPC-z56-002	&	200.95	&	27.40	&	5.683	&	0.017	&	12	&	4	&	3.00	$\pm$	3.18		&	10.0	&	2	&	\\
CCPC-z56-003	&	150.11	&	1.54	&	5.685	&	0.018	&	7	&	2	&	[1.5]	$\pm$	2.35		&	10.0	&	2	&	\\
CCPC-z56-004	&	200.97	&	27.73	&	5.686	&	0.014	&	8	&	3	&	[6]	$\pm$	12.21		&	33.8	&	2	&	\\
CCPC-z56-005	&	150.26	&	1.86	&	5.692	&	0.021	&	5	&	2	&	[0.5]	$\pm$	0.80		&	2.5	&	2	&	\\
CCPC-z56-006	&	40.07	&	-1.50	&	5.692	&	0.020	&	6	&	1	&	[4]	$\pm$	8.20		&	21.3	&	2	&	\\
CCPC-z56-007	&	334.44	&	0.67	&	5.698	&	0.020	&	4	&	2	&	17.62	$\pm$	20.35		&	68.1	&	2	&	\\
CCPC-z56-008	&	34.43	&	-5.47	&	5.693	&	0.009	&	7	&	7	&	17.79	$\pm$	16.31		&	68.1	&	1	&	14\\
CCPC-z57-001	&	201.24	&	27.27	&	5.709	&	0.017	&	8	&	3	&	2.45	$\pm$	1.91		&	10.0	&	2	&	\\
CCPC-z57-002	&	138.38	&	46.21	&	5.704	&	0.009	&	7	&	4	&	[1.98]	$\pm$	3.10		&	10.0	&	1	&	\\
CCPC-z57-003	&	190.27	&	62.36	&	5.710	&	0.021	&	8	&	5	&	1.33	$\pm$	1.22		&	10.0	&	3	&	\\
CCPC-z57-004	&	16.46	&	-25.78	&	5.772	&	0.011	&	5	&	3	&	4.21	$\pm$	4.16		&	21.3	&	1	&	\\
CCPC-z58-001	&	53.11	&	-27.93	&	5.811	&	0.020	&	6	&	1	&	1.08	$\pm$	0.79		&	10.0	&	1	&	\\
CCPC-z59-001	&	189.23	&	62.25	&	5.961	&	0.017	&	7	&	4	&	0.33	$\pm$	0.37		&	2.5	&	2	&	\\
CCPC-z60-001	&	201.09	&	27.22	&	6.013	&	0.021	&	10	&	6	&	3.19	$\pm$	2.49		&	21.3	&	1	&	16\\
CCPC-z65-001	&	334.60	&	0.78	&	6.520	&	0.028	&	5	&	2	&	[1.38]	$\pm$	2.22		&	10.0	&	2	&	\\
CCPC-z65-002	&	201.14	&	27.68	&	6.543	&	0.018	&	9	&	4	&	9.10	$\pm$	9.83		&	68.1	&	2	&	\\
CCPC-z65-003	&	39.91	&	-1.58	&	6.551	&	0.013	&	4	&	3	&	4.14	$\pm$	5.07		&	21.3	&	2	&	\\
CCPC-z65-004	&	200.91	&	27.49	&	6.562	&	0.019	&	7	&	4	&	2.50	$\pm$	2.89		&	10.0	&	2	&	\\
CCPC-z65-005	&	201.14	&	27.28	&	6.564	&	0.014	&	9	&	3	&	4.15	$\pm$	4.44		&	21.3	&	2	&	
\enddata
\tablenotetext{a}{Counts of field galaxies were limited to the surface area of the overdense region.}
\tablecomments{The names and positions of candidate protoclusters, with redshifts corresponding to the average value for the system with their dispersion $\sigma_z$. The number of galaxies within $R=10$ and 20 cMpc from the search center are included. The galaxy overdensity ($\delta_{gal}$) calculation is outlined in Section 2.3. Bracketed values should be considered highly uncertain, as they were computed based on $1\le N<3$ field galaxies. A probability is assigned to each system that it will collapse into a cluster by $z=0.$. These probability distribution functions at the relevant redshifts can be found in Figure 8 of \citet{chi13} from analysis of protocluster $\delta_{gal}$ values in the Millennium simulation. The `Q' column provides a rating of the source quality of redshifts as identified within NED. The best quality redshift systems are listed as a `1'. If the overdensity was previously identified, the discovery references are listed in the last column. References: (1) \citet{2012ApJ...749..169G}, (2) \citet{2013AnA...559A...2G}, (3) \citet{1980ApJ...242L..55B}, (4) \citet{die13}, (5) \citet{2014ApJ...795L..20Y}, (6) \citet{2005ApJ...626...44S}, (7) \citet{2007ApJ...668...23P}, (8) \citet{gob13}, (9) \citet{1997AnA...326..580P}, (10) \citet{2015ApJ...808...37C}, (11) \citet{2009ApJ...700...20L}, (12) \citet{2010MNRAS.408L..31D}, (13) \citet{2012MNRAS.423.2436S}, (14) \citet{2005ApJ...620L...1O}, (15) \citet{2013MNRAS.432.2869H}, (16) \citet{2012ApJ...750..137T}, (17) \citet{2014AnA...572A..41L}, (18) \citet{ven02}, (19) \citet{2004Natur.427...47M}, (20) \citet{1999AJ....118.2547K}, (21) \citet{2008ApJ...675.1171P}, (22) \citep{2009AnA...494..443P}, (23)  \citep{2003ApJ...592..721G} } \label{tab:tot_cat}
\end{deluxetable*}

\LongTables
\begin{deluxetable*}{c c}
\tablewidth{0pt}
\tablecolumns{2}
\tablecaption{CCPC: Member Redshift Reference List}
\tablehead{
\colhead{Candidate} & \colhead{Redshift} \\
\colhead{Name} & \colhead{References} 
} 
\startdata
CCPC-z20-001	 & 1,2,3,4,5,6,7,8,9,10,11,12,13,14 \\
CCPC-z20-002	 & 15,16 \\
CCPC-z20-003	 & 17,18 \\
CCPC-z20-004	 & 19,20,21 \\
CCPC-z20-005	 & 22,23,24,25,26 \\
CCPC-z20-006	 & 27,13,28,3 \\
CCPC-z20-007	 & 29,30 \\
CCPC-z20-008	 & 31,3,4,32 \\
CCPC-z20-009	 & 33,34,35,36 \\
CCPC-z21-001	 & 37,27,38,39 \\
CCPC-z21-002	 & 27,38 \\
CCPC-z21-003	 & 13,40,41,3,4 \\
CCPC-z21-004	 & 42,43,44,45,46 \\
CCPC-z21-005	 & 47,48,49,50 \\
CCPC-z21-006	 & 51 \\
CCPC-z21-007	 & 27,38 \\
CCPC-z21-008	 & 52,53,34,33 \\
CCPC-z21-009	 & 27,39,54,50,32 \\
CCPC-z21-010	 & 55,56 \\
CCPC-z21-011	 & 57,13,51,3,4,40,58,59 \\
CCPC-z22-001	 & 22,60,26 \\
CCPC-z22-002	 & 4,3,41,61,11,62,27,63,64,65 \\
CCPC-z22-003	 & 6,1,31 \\
CCPC-z22-004	 & 66,67,68 \\
CCPC-z22-005	 & 69 \\
CCPC-z22-006	 & 33,70,34,71 \\
CCPC-z22-007	 & 72,66,68,73,74,50 \\
CCPC-z23-001	 & 22,25,75,76,77,78,79,26 \\
CCPC-z23-002	 & 51,37,80 \\
CCPC-z23-003	 & 47 \\
CCPC-z23-004	 & 66,68 \\
CCPC-z23-005	 & 22,25,81,82,24 \\
CCPC-z23-006	 & 69 \\
CCPC-z23-007	 & 83,84,85,86 \\
CCPC-z24-001	 & 22,81,87,79,88,25 \\
CCPC-z24-002	 & 4,51,11,5,31,62,65,89,3 \\
CCPC-z24-003	 & 87,90 \\
CCPC-z24-004	 & 50,74,66 \\
CCPC-z24-005	 & 33,34,91,52 \\
CCPC-z24-006	 & 69 \\
CCPC-z24-007	 & 92,51,93 \\
CCPC-z24-008	 & 27,94,38,50 \\
CCPC-z24-009	 & 41,3,13,51,95,96,59,97,98,28,40 \\
CCPC-z24-010	 & 99,100 \\
CCPC-z25-001	 & 101,37,27,94 \\
CCPC-z25-002	 & 66,102 \\
CCPC-z25-003	 & 66 \\
CCPC-z25-004	 & 103,3 \\
CCPC-z25-005	 & 95,87,104,22,105,79,25,106,107,78,88,108,109 \\
CCPC-z25-006	 & 68,74,110,66 \\
CCPC-z25-007	 & 51 \\
CCPC-z25-008	 & 98,111,3,65,112,5,13,34 \\
CCPC-z26-001	 & 51 \\
CCPC-z26-002	 & 51 \\
CCPC-z26-003	 & 3,51 \\
CCPC-z26-004	 & 22 \\
CCPC-z26-005	 & 113,51 \\
CCPC-z26-006	 & 68,66,110 \\
CCPC-z26-007	 & 51 \\
CCPC-z27-006	 & 51,31 \\
CCPC-z27-007	 & 34,53 \\
CCPC-z27-008	 & 20 \\
CCPC-z27-009	 & 51 \\
CCPC-z27-010	 & 51 \\
CCPC-z27-011	 & 114,51 \\
CCPC-z27-012	 & 69 \\
CCPC-z27-013	 & 20 \\
CCPC-z27-014	 & 20 \\
CCPC-z28-008	 & 51 \\
CCPC-z28-009	 & 51 \\ 
CCPC-z28-010	 & 51,48 \\
CCPC-z28-011	 & 20 \\
CCPC-z28-012	 & 51 \\
CCPC-z28-013	 & 51 \\
CCPC-z28-014	 & 51,115 \\
CCPC-z28-015	 & 20 \\
CCPC-z28-016	 & 20,21 \\
CCPC-z28-017	 & 20 \\
CCPC-z28-018	 & 66,116,73,68 \\
CCPC-z28-019	 & 51 \\
CCPC-z29-008	 & 20 \\
CCPC-z29-009	 & 51 \\
CCPC-z29-010	 & 51 \\
CCPC-z29-011	 & 51 \\
CCPC-z29-012	 & 20 \\
CCPC-z29-013	 & 51 \\
CCPC-z29-014	 & 51,47 \\
CCPC-z29-015	 & 51,117,118,92\\ 
CCPC-z29-016	 & 51,61,40,3,62,103,65 \\
CCPC-z30-004	 & 68,116,73 \\
CCPC-z30-005	 & 51 \\
CCPC-z30-006	 & 51,38 \\
CCPC-z30-007	 & 22,78,119,87 \\
CCPC-z30-008	 & 51 \\
CCPC-z30-009	 & 51,38 \\
CCPC-z31-008	 & 51 \\
CCPC-z31-009	 & 51 \\
CCPC-z31-010	 & 120 \\
CCPC-z31-011	 & 120,51,13,34,3 \\
CCPC-z31-012	 & 120 \\
CCPC-z31-013	 & 51,118 \\
CCPC-z31-014	 & 51,47 \\
CCPC-z31-015	 & 51 \\
CCPC-z31-016	 & 51 \\
CCPC-z31-017	 & 20 \\
CCPC-z32-004	 & 20 \\
CCPC-z32-005	 & 112,121,62,51,40,3 \\
CCPC-z32-006	 & 3,122 \\
CCPC-z32-007	 & 51,117 \\
CCPC-z32-008	 & 51 \\
CCPC-z32-009	 & 69 \\
CCPC-z32-010	 & 51 \\
CCPC-z33-006	 & 34 \\
CCPC-z33-007	 & 34,51 \\
CCPC-z33-008	 & 20 \\
CCPC-z33-009	 & 49,51 \\
CCPC-z33-010	 & 51,123 \\
CCPC-z34-003	 & 49,47,51 \\
CCPC-z34-004	 & 22,107,124 \\
CCPC-z34-005	 & 125,20,21 \\
CCPC-z34-006	 & 20 \\
CCPC-z35-002	 & 20 \\
CCPC-z35-003	 & 20 \\
CCPC-z35-004	 & 51,123 \\
CCPC-z35-005	 & 20 \\
CCPC-z36-003	 & 126,127,124,22,88 \\
CCPC-z36-004	 & 20 \\
CCPC-z36-005	 & 20 \\
CCPC-z36-006	 & 120 \\
CCPC-z36-007	 & 120,128 \\
CCPC-z37-002	 & 20 \\
CCPC-z37-003	 & 22,129,40,127 \\
CCPC-z38-001	 & 20,125 \\
CCPC-z40-001	 & 130,131,11,132,40,133 \\
CCPC-z40-002	 & 134,135 \\
CCPC-z41-001	 & 22,136,40 \\
CCPC-z42-001	 & 22,127,131 \\
CCPC-z43-001	 & 137 \\
CCPC-z43-002	 & 22,40,131 \\
CCPC-z44-001	 & 137 \\
CCPC-z44-002	 & 138,139 \\
CCPC-z44-003	 & 140,141,131,142 \\
CCPC-z44-004	 & 137,139 \\
CCPC-z44-005	 & 139,138 \\
CCPC-z45-001	 & 143,144,129 \\
CCPC-z45-002	 & 22,144,143 \\
CCPC-z45-003	 & 137 \\
CCPC-z48-001	 & 22,127,126,145 \\
CCPC-z48-002	 & 146 \\
CCPC-z49-001	 & 147 \\
CCPC-z50-001	 & 148,149 \\
CCPC-z51-001	 & 150,151 \\
CCPC-z51-002	 & 5,152,140,153 \\
CCPC-z56-001	 & 154,155,156,157,158 \\
CCPC-z56-002	 & 159,160 \\
CCPC-z56-003	 & 161,162 \\
CCPC-z56-004	 & 160,159 \\
CCPC-z56-005	 & 162,161 \\
CCPC-z56-006	 & 154,155 \\
CCPC-z56-007	 & 154 \\
CCPC-z56-008	 & 163,120 \\
CCPC-z57-001	 & 160,159 \\
CCPC-z57-002	 & 164 \\
CCPC-z57-003	 & 154 \\
CCPC-z57-004	 & 165 \\
CCPC-z58-001	 & 127,166,167,168 \\
CCPC-z59-001	 & 156 \\
CCPC-z60-001	 & 169,170,171 \\
CCPC-z65-001	 & 154,155 \\
CCPC-z65-002	 & 159,172,173,174 \\
CCPC-z65-003	 & 154,168 \\
CCPC-z65-004	 & 159,174 \\
CCPC-z65-005	 & 174,173,159 \\
\enddata
\tablecomments{References for the spectroscopic measurements used in each CCPC system. Some of the above references have also utilized spectra from the catalogs of \citet{2004ApJS..155..271S,2004AnA...428.1043L,2005AnA...434...53V,2006AnA...454..423V,2007ApJS..172...70L,2008AnA...478...83V,2009ApJ...695.1163V}
References: (1) \citet{2004ApJ...617...64S}, (2) \citet{2010ApJ...719.1393D}, (3) \citet{2006ApJ...653.1004R}, (4) \citet{2010ApJ...718..112Y}, (5) \citet{2002AJ....124.1839B}, (6) \citet{2004ApJ...614..671C}, (7) \citet{2013MNRAS.429.3047B}, (8) \citet{2010MNRAS.405..219B}, (9) \citet{2009MNRAS.395L..67F}, (10) \citet{2005ApJ...631..101P}, (11) \citet{1999ApJ...513...34F}, (12) \citet{2009ApJ...691..560C}, (13) \citet{2002AJ....124.1886M}, (14) \citet{2008ApJ...675.1171P}, (15) \citet{gob13}, (16) \citet{2000AJ....120.2220H}, (17) \citet{2013AnA...559A...2G}, (18) \citet{1990AJ....100.1014M}, (19) \citet{2007AnA...467...73T}, (20) \citet{2005AnA...439..845L}, (21) \citet{2010MNRAS.401..294S}, (22) \citet{2010AnA...512A..12B}, (23) \citet{2008ApJS..179...19L}, (24) \citet{2008ApJ...677..219K}, (25) \citet{2011ApJ...729...48B}, (26) \citet{2011ApJ...743..144T}, (27) \citet{2006ApJ...646..107E}, (28) \citet{2003AJ....126.1183C}, (29) \citet{1980ApJ...242L..55B}, (30) \citet{2009MNRAS.392.1539T}, (31) \citet{2004ApJ...616...71S}, (32) \citet{2004ApJ...604..534S}, (33) \citet{die13}, (34) \citet{2011ApJS..192....5A}, (35) \citet{2012ApJ...761..139C}, (36) \citet{2010ApJ...716..348B}, (37) \citet{2010MNRAS.405.2302H}, (38) \citet{1989QSO...M...0000H}, (39) \citet{2004ApJ...612..108S}, (40) \citet{2013ApJ...772...48P}, (41) \citet{2004ApJ...612..122E}, (42) \citet{2000AnA...361L..25P}, (43) \citet{2005AJ....130..867C}, (44) \citet{2004AnA...428..817K}, (45) \citet{2002AnA...396..109P}, (46) \citet{2004AnA...428..793K}, (47) \citet{2011ApJ...735...86W}, (48) \citet{2006MNRAS.371..221G}, (49) \citet{2005ApJ...620..595W}, (50) \citet{2003ApJ...591..101E}, (51) \citet{2003ApJ...592..728S}, (52) \citet{2011ApJ...743...86M}, (53) \citet{2007ApJS..172...70L}, (54) \citet{1992NED11.R......1N}, (55) \citet{2000AnA...362....9M}, (56) \citet{1999MNRAS.305..685O}, (57) \citet{2011PASJ...63S.437T}, (58) \citet{2008ApJ...686..127W}, (59) \citet{2011MNRAS.412.1913I}, (60) \citet{2013ApJ...765L...2B}, (61) \citet{2004AJ....127.3137C}, (62) \citet{1997ApJ...481..673L}, (63) \citet{2002AJ....123.1163B}, (64) \citet{2008ApJS..179....1T}, (65) \citet{2004AJ....127.3121W}, (66) \citet{2005ApJ...626..698S}, (67) \citet{2012ApJ...745...85L}, (68) \citet{2007ApJ...668...23P}, (69) \citet{2004AnA...418..885N}, (70) \citet{2011ApJ...729..140F}, (71) \citet{2009ApJ...696.1195T}, (72) \citet{2011ApJ...740L..31E}, (73) \citet{2008PrivC.U..D...1L}, (74) \citet{1995AnA...294..377V}, (75) \citet{2010ApJS..191..124S}, (76) \citet{2009AnA...504..751S}, (77) \citet{2011AJ....141...14S}, (78) \citet{2013AnA...549A..63K}, (79) \citet{2004ApJS..155...73Z}, (80) \citet{2009MNRAS.400..299L}, (81) \citet{2008ApJ...682..985W}, (82) \citet{2011MNRAS.411.2739C}, (83) \citet{1998AJ....116.2659P}, (84) \citet{1996ApJ...456L..21P}, (85) \citet{2000ApJ...528L..81A}, (86) \citet{1999AJ....118.2547K}, (87) \citet{2009ApJ...706..885W}, (88) \citet{2010ApJ...720..368X}, (89) \citet{2003AJ....126..632B}, (90) \citet{2008MNRAS.384.1611K}, (91) \citet{2007ApJS..172..383T}, (92) \citet{2013MNRAS.430..425B}, (93) \citet{1995AJ....109.1522C}, (94) \citet{2005ApJ...629..636A}, (95) \citet{2011MNRAS.413...80C}, (96) \citet{1998AJ....115.1400F}, (97) \citet{2011ApJ...733L..11R}, (98) \citet{2006ApJ...647...74W}, (99) \citet{1997AnA...326..505R}, (100) \citet{2011PASJ...63S.415T}, (101) \citet{1987ApJ...314..111A}, (102) \citet{1989AnA...218...71R}, (103) \citet{2007ApJ...660..167D}, (104) \citet{2011AnA...526A..86G}, (105) \citet{2007ApJ...659..941T}, (106) \citet{2004ApJS..155..271S}, (107) \citet{2004AnA...428.1043L}, (108) \citet{2009ApJ...693.1713T}, (109) \citet{2012ApJS..203...15B}, (110) \citet{2006ApJ...637..648S}, (111) \citet{2004ApJ...611..732C}, (112) \citet{1996Natur.381..759L}, (113) \citet{2001ApJ...562...95S}, (114) \citet{2006AnA...457...79G}, (115) \citet{2009ApJ...699..667M}, (116) \citet{2010MNRAS.401.1521M}, (117) \citet{1991MNRAS.250...24Y}, (118) \citet{1998AJ....115.2184S}, (119) \citet{2000AnA...359..489C}, (120) \citet{2008ApJS..176..301O}, (121) \citet{2008ApJ...689..687B}, (122) \citet{1998AJ....115...55L}, (123) \citet{1998ApJ...494...60P}, (124) \citet{2011AnA...528A..88G}, (125) \citet{2008AnA...492...81P}, (126) \citet{2008AnA...478...83V}, (127) \citet{2006AnA...454..423V}, (128) \citet{2006ApJ...648...54S}, (129) \citet{2010ApJ...725.1011V}, (130) \citet{2009ApJ...695L.176D}, (131) \citet{2011ApJ...738...69S}, (132) \citet{2001ApJS..135...41F}, (133) \citet{2009ApJ...694.1517D}, (134) \citet{2007AnA...461..823V}, (135) \citet{2001AJ....121.1241D}, (136) \citet{2007AJ....134..169X}, (137) \citet{2009ApJ...706..762W}, (138) \citet{2007ApJ...671.1227D}, (139) \citet{2004ApJ...617..707D}, (140) \citet{2009ApJ...704..117K}, (141) \citet{1999PASP..111.1475S}, (142) \citet{1999ApJ...519....1S}, (143) \citet{2009MNRAS.393.1174F}, (144) \citet{2013MNRAS.431.3589Z}, (145) \citet{2005AnA...434...53V}, (146) \citet{2009ApJ...700...20L}, (147) \citet{2010MNRAS.408L..31D}, (148) \citet{2013MNRAS.432.2869H}, (149) \citet{2001SDSSe.1...0000:}, (150) \citet{2004AnA...424L..17V}, (151) \citet{1999ApJ...518L..61V}, (152) \citet{2014AnA...562A..35N}, (153) \citet{2002ApJ...570...92D}, (154) \citet{2010ApJ...725..394H}, (155) \citet{2007ApJ...668..853K}, (156) \citet{2011ApJ...728L...2S}, (157) \citet{1998ApJ...505L..95W}, (158) \citet{2001AJ....122..598D}, (159) \citet{2011ApJ...734..119K}, (160) \citet{2006PASJ...58..313S}, (161) \citet{2007ApJS..172..523M}, (162) \citet{2009ApJ...701..915T}, (163) \citet{2005ApJ...620L...1O}, (164) \citet{1998MNRAS.293L...6T}, (165) \citet{2006AnA...455..145T}, (166) \citet{2005ApJ...626..666M}, (167) \citet{2004MNRAS.355..374B}, (168) \citet{2007MNRAS.376.1861F}, (169) \citet{2012ApJ...750..137T}, (170) \citet{2005ApJ...634..142N}, (171) \citet{2011ApJ...743...65J}, (172) \citet{2013ApJ...772...99J}, (173) \citet{2006ApJ...648....7K}, (174) , \citet{2005PASJ...57..165T} } \label{tab:tot_ref}
\end{deluxetable*}

\LongTables
\begin{deluxetable*}{c c c c c c c}
\tablewidth{0pt}
\tablecolumns{7}
\tablecaption{CCPC: Mass Estimates}
\tablehead{
\colhead{Candidate} & \colhead{$R_e$}	&\colhead{$\sigma$}	&\colhead{Virial Mass} &	\colhead{$\delta_{m}$}	& \colhead{Overdensity}	&	\colhead{Overdensity Mass}	\\
\colhead{Name} & \colhead{(Mpc)}	&\colhead{(km s$^{-1}$)}	&\colhead{Estimate ($10^{14}$ $M_{\odot}$)}	&\colhead{($b=3$)}	&\colhead{Volume (cMpc$^3$)}	&\colhead{Estimate ($10^{14}$ $M_{\odot}$)}
} 
\startdata
CCPC-z20-001	&	4.2	&	563	&	7.7	&	1.69		&	2945	&	3.5	\\
CCPC-z20-002	&	4.5	&	749	&	14.7	&	3.13		&	5361	&	9.8	\\
CCPC-z20-003	&	2.3	&	367	&	1.8	&	6.48		&	125	&	0.4	\\
CCPC-z20-004	&	4.9	&	619	&	10.8	&	0.74		&	152	&	0.1	\\
CCPC-z20-005	&	5.0	&	773	&	17.4	&	0.33		&	11341	&	6.5	\\
CCPC-z20-006	&	3.4	&	827	&	13.3	&	0.16		&	1249	&	0.6	\\
CCPC-z20-007	&	2.6	&	565	&	4.8	&	0.67		&	40	&	 $<0.1$	\\
CCPC-z20-008	&	4.3	&	671	&	11.3	&	0.43		&	2126	&	1.4	\\
CCPC-z20-009	&	4.3	&	446	&	4.9	&	4.38		&	589	&	1.4	\\
CCPC-z21-001	&	3.4	&	689	&	9.3	&	1.39		&	690	&	0.7	\\
CCPC-z21-002	&	3.0	&	390	&	2.7	&	0.63		&	10	&	 $<0.1$	\\
CCPC-z21-003	&	4.5	&	791	&	16.4	&	0.64		&	969	&	0.7	\\
CCPC-z21-004	&	3.4	&	701	&	9.6	&	2.11		&	519	&	0.7	\\
CCPC-z21-005	&	5.0	&	689	&	13.8	&	3.02		&	799	&	1.5	\\
CCPC-z21-006	&	2.9	&	479	&	3.9	&	6.28		&	58	&	0.2	\\
CCPC-z21-007	&	1.2	&	156	&	0.2	&	5.76		&	13	&	 $<0.1$	\\
CCPC-z21-008	&	3.7	&	211	&	1.0	&	3.14		&	176	&	0.3	\\
CCPC-z21-009	&	4.0	&	500	&	5.7	&	0.98		&	26	&	 $<0.1$	\\
CCPC-z21-010	&	2.2	&	604	&	4.6	&	1.20		&	382	&	0.4	\\
CCPC-z21-011	&	4.4	&	903	&	20.9	&	0.23		&	5692	&	3.1	\\
CCPC-z22-001	&	4.7	&	718	&	14.1	&	0.21		&	7546	&	4.0	\\
CCPC-z22-002	&	5.5	&	845	&	22.6	&	0.11		&	7628	&	3.6	\\
CCPC-z22-003	&	3.3	&	583	&	6.5	&	1.94		&	107	&	0.1	\\
CCPC-z22-004	&	2.9	&	655	&	7.1	&	0.35		&	977	&	0.6	\\
CCPC-z22-005	&	1.3	&	394	&	1.2	&	1.51		&	15	&	 $<0.1$	\\
CCPC-z22-006	&	5.7	&	549	&	9.9	&	0.65		&	1973	&	1.4	\\
CCPC-z22-007	&	3.1	&	709	&	8.9	&	2.59		&	3032	&	4.7	\\
CCPC-z23-001	&	4.3	&	731	&	13.4	&	0.69		&	8267	&	6.0	\\
CCPC-z23-002	&	2.6	&	783	&	9.3	&	3.82		&	60	&	0.1	\\
CCPC-z23-003	&	2.3	&	711	&	6.8	&	3.91		&	171	&	0.4	\\
CCPC-z23-004	&	2.5	&	678	&	6.7	&	0.15		&	95	&	 $<0.1$	\\
CCPC-z23-005	&	3.9	&	661	&	9.9	&	0.23		&	5170	&	2.7	\\
CCPC-z23-006	&	2.7	&	658	&	6.7	&	1.55		&	225	&	0.2	\\
CCPC-z23-007	&	1.5	&	442	&	1.7	&	5.54		&	82	&	0.2	\\
CCPC-z24-001	&	5.1	&	644	&	12.3	&	0.19		&	7295	&	3.7	\\
CCPC-z24-002	&	3.4	&	805	&	12.8	&	0.24		&	3232	&	1.7	\\
CCPC-z24-003	&	2.4	&	399	&	2.2	&	5.02		&	327	&	0.8	\\
CCPC-z24-004	&	1.8	&	577	&	3.4	&	0.20		&	40	&	 $<0.1$	\\
CCPC-z24-005	&	4.1	&	770	&	14.1	&	3.09		&	9110	&	15.5	\\
CCPC-z24-006	&	3.1	&	487	&	4.2	&	1.23		&	192	&	0.2	\\
CCPC-z24-007	&	4.8	&	764	&	16.3	&	0.84		&	2157	&	1.8	\\
CCPC-z24-008	&	3.8	&	831	&	15.1	&	0.96		&	856	&	0.7	\\
CCPC-z24-009	&	2.5	&	571	&	4.6	&	1.87		&	2580	&	3.2	\\
CCPC-z24-010	&	0.9	&	162	&	0.1	&	0.67		&	7	&	 $<0.1$	\\
CCPC-z25-001	&	3.6	&	398	&	3.3	&	0.67		&	10	&	 $<0.1$	\\
CCPC-z25-002	&	1.3	&	208	&	0.3	&	6.62		&	28	&	0.1	\\
CCPC-z25-003	&	2.2	&	295	&	1.1	&	3.63		&	116	&	0.2	\\
CCPC-z25-004	&	3.0	&	486	&	4.1	&	0.83		&	22	&	 $<0.1$	\\
CCPC-z25-005	&	4.5	&	737	&	14.2	&	0.70		&	18669	&	14.0	\\
CCPC-z25-006	&	1.6	&	417	&	1.6	&	2.10		&	40	&	0.1	\\
CCPC-z25-007	&	3.1	&	620	&	6.9	&	3.63		&	215	&	0.4	\\
CCPC-z25-008	&	3.6	&	792	&	13.0	&	0.16		&	1286	&	0.6	\\
CCPC-z26-001	&	0.7	&	245	&	0.2	&	2.44		&	3	&	 $<0.1$	\\
CCPC-z26-002	&	1.5	&	355	&	1.1	&	0.88		&	42	&	 $<0.1$	\\
CCPC-z26-003	&	3.7	&	760	&	12.4	&	0.19		&	198	&	0.1	\\
CCPC-z26-004	&	2.7	&	825	&	10.5	&	0.37		&	8934	&	5.4	\\
CCPC-z26-005	&	3.8	&	849	&	15.9	&	1.54		&	2558	&	2.8	\\
CCPC-z26-006	&	2.2	&	366	&	1.7	&	2.41		&	139	&	0.2	\\
CCPC-z26-007	&	2.8	&	599	&	5.7	&	1.47		&	1484	&	1.6	\\
CCPC-z27-006	&	4.2	&	627	&	9.6	&	2.11		&	87	&	0.1	\\
CCPC-z27-007	&	3.3	&	730	&	10.2	&	1.60		&	894	&	1.0	\\
CCPC-z27-008	&	3.5	&	463	&	4.3	&	2.42		&	5570	&	7.9	\\
CCPC-z27-009	&	0.8	&	322	&	0.5	&	2.04		&	5	&	 $<0.1$	\\
CCPC-z27-010	&	3.4	&	732	&	10.6	&	0.97		&	1095	&	0.9	\\
CCPC-z27-011	&	4.8	&	1036	&	29.9	&	0.24		&	21	&	 $<0.1$	\\
CCPC-z27-012	&	1.3	&	632	&	2.9	&	4.28		&	217	&	0.5	\\
CCPC-z27-013	&	2.7	&	482	&	3.7	&	0.92		&	139	&	0.1	\\
CCPC-z27-014	&	3.1	&	853	&	12.9	&	0.43		&	1326	&	0.8	\\
CCPC-z28-008	&	3.5	&	1014	&	20.6	&	0.73		&	568	&	0.4	\\
CCPC-z28-009	&	1.6	&	337	&	1.1	&	0.79		&	14	&	 $<0.1$	\\
CCPC-z28-010	&	2.7	&	693	&	7.5	&	0.82		&	2796	&	2.1	\\
CCPC-z28-011	&	2.6	&	504	&	3.8	&	3.03		&	280	&	0.5	\\
CCPC-z28-012	&	4.3	&	685	&	11.7	&	0.69		&	2	&	 $<0.1$	\\
CCPC-z28-013	&	2.7	&	636	&	6.2	&	1.64		&	578	&	0.6	\\
CCPC-z28-014	&	2.7	&	438	&	3.0	&	1.66		&	3364	&	3.9	\\
CCPC-z28-015	&	3.6	&	772	&	12.3	&	0.78		&	959	&	0.7	\\
CCPC-z28-016	&	4.0	&	429	&	4.2	&	5.15		&	458	&	1.2	\\
CCPC-z28-017	&	4.6	&	927	&	22.7	&	1.75		&	8	&	 $<0.1$	\\
CCPC-z28-018	&	3.7	&	930	&	18.4	&	0.51		&	350	&	0.2	\\
CCPC-z28-019	&	2.8	&	792	&	10.2	&	0.45		&	1193	&	0.7	\\
CCPC-z29-008	&	3.3	&	487	&	4.5	&	0.84		&	1386	&	1.1	\\
CCPC-z29-009	&	2.2	&	757	&	7.2	&	4.44		&	27	&	0.1	\\
CCPC-z29-010	&	2.4	&	900	&	11.1	&	0.26		&	167	&	0.1	\\
CCPC-z29-011	&	4.1	&	628	&	9.3	&	3.21		&	4579	&	8.0	\\
CCPC-z29-012	&	4.5	&	917	&	21.8	&	0.35		&	3760	&	2.1	\\
CCPC-z29-013	&	1.7	&	189	&	0.3	&	4.89		&	56	&	0.1	\\
CCPC-z29-014	&	3.8	&	916	&	18.5	&	0.19		&	5527	&	2.9	\\
CCPC-z29-015	&	1.8	&	653	&	4.3	&	1.48		&	539	&	0.6	\\
CCPC-z29-016	&	2.5	&	601	&	5.1	&	1.19		&	3857	&	3.6	\\
CCPC-z30-004	&	2.6	&	713	&	7.7	&	0.20		&	409	&	0.2	\\
CCPC-z30-005	&	3.6	&	704	&	10.2	&	0.31		&	5825	&	3.2	\\
CCPC-z30-006	&	1.8	&	743	&	5.8	&	1.18		&	681	&	0.6	\\
CCPC-z30-007	&	4.1	&	657	&	10.3	&	0.59		&	1394	&	1.0	\\
CCPC-z30-008	&	2.9	&	726	&	8.9	&	0.36		&	6417	&	3.7	\\
CCPC-z30-009	&	1.3	&	267	&	0.5	&	1.63		&	33	&	 $<0.1$	\\
CCPC-z31-008	&	2.6	&	541	&	4.3	&	2.57		&	1038	&	1.6	\\
CCPC-z31-009	&	1.8	&	517	&	2.7	&	1.64		&	79	&	0.1	\\
CCPC-z31-010	&	2.3	&	711	&	6.6	&	0.20		&	157	&	0.1	\\
CCPC-z31-011	&	3.1	&	875	&	13.8	&	0.12		&	1898	&	0.9	\\
CCPC-z31-012	&	4.1	&	641	&	9.7	&	0.34		&	4055	&	2.2	\\
CCPC-z31-013	&	2.1	&	467	&	2.6	&	0.50		&	666	&	0.4	\\
CCPC-z31-014	&	3.2	&	630	&	7.4	&	0.69		&	7991	&	5.6	\\
CCPC-z31-015	&	3.7	&	558	&	6.6	&	2.72		&	2011	&	3.0	\\
CCPC-z31-016	&	1.8	&	699	&	5.0	&	1.15		&	292	&	0.3	\\
CCPC-z31-017	&	3.7	&	462	&	4.5	&	2.42		&	10025	&	14.6	\\
CCPC-z32-004	&	2.6	&	650	&	6.3	&	0.47		&	4118	&	2.5	\\
CCPC-z32-005	&	2.7	&	833	&	10.7	&	0.72		&	3469	&	2.5	\\
CCPC-z32-006	&	2.0	&	575	&	3.8	&	1.42		&	1	&	 $<0.1$	\\
CCPC-z32-007	&	2.6	&	504	&	3.8	&	3.82		&	280	&	0.6	\\
CCPC-z32-008	&	2.2	&	463	&	2.7	&	0.85		&	418	&	0.3	\\
CCPC-z32-009	&	2.6	&	1007	&	15.3	&	0.29		&	528	&	0.3	\\
CCPC-z32-010	&	3.0	&	670	&	7.7	&	0.55		&	4068	&	2.7	\\
CCPC-z33-006	&	1.8	&	578	&	3.5	&	5.83		&	115	&	0.3	\\
CCPC-z33-007	&	3.2	&	818	&	12.3	&	3.84		&	101	&	0.2	\\
CCPC-z33-008	&	2.3	&	452	&	2.7	&	0.56		&	29	&	 $<0.1$	\\
CCPC-z33-009	&	3.3	&	813	&	12.5	&	0.23		&	3019	&	1.6	\\
CCPC-z33-010	&	3.4	&	611	&	7.3	&	3.48		&	730	&	1.4	\\
CCPC-z34-003	&	3.6	&	890	&	16.4	&	0.34		&	3174	&	1.8	\\
CCPC-z34-004	&	4.0	&	785	&	14.1	&	0.15		&	8601	&	4.2	\\
CCPC-z34-005	&	3.9	&	706	&	11.3	&	2.14		&	1437	&	1.9	\\
CCPC-z34-006	&	4.0	&	878	&	17.7	&	2.61		&	270	&	0.4	\\
CCPC-z35-002	&	3.9	&	785	&	14.0	&	1.37		&	3108	&	3.1	\\
CCPC-z35-003	&	3.8	&	723	&	11.4	&	0.64		&	1557	&	1.1	\\
CCPC-z35-004	&	2.6	&	758	&	8.5	&	1.41		&	1237	&	1.3	\\
CCPC-z35-005	&	4.3	&	1054	&	27.8	&	0.43		&	10638	&	6.4	\\
CCPC-z36-003	&	3.3	&	710	&	9.7	&	1.15		&	5269	&	4.7	\\
CCPC-z36-004	&	3.5	&	905	&	16.7	&	1.86		&	8092	&	9.7	\\
CCPC-z36-005	&	3.4	&	891	&	15.7	&	0.39		&	2381	&	1.4	\\
CCPC-z36-006	&	3.0	&	728	&	9.3	&	1.21		&	160	&	0.1	\\
CCPC-z36-007	&	3.3	&	634	&	7.7	&	5.07		&	165	&	0.4	\\
CCPC-z37-002	&	2.9	&	372	&	2.3	&	1.46		&	25	&	 $<0.1$	\\
CCPC-z37-003	&	3.4	&	856	&	14.3	&	0.86		&	1295	&	1.0	\\
CCPC-z38-001	&	2.6	&	722	&	7.7	&	0.77		&	66	&	 $<0.1$	\\
CCPC-z40-001	&	2.8	&	628	&	6.3	&	1.87		&	2412	&	2.8	\\
CCPC-z40-002	&	1.6	&	359	&	1.2	&	1.85		&	1504	&	1.8	\\
CCPC-z41-001	&	2.6	&	651	&	6.3	&	1.09		&	776	&	0.7	\\
CCPC-z42-001	&	1.6	&	580	&	3.0	&	0.97		&	137	&	0.1	\\
CCPC-z43-001	&	2.3	&	389	&	2.0	&	0.53		&	1717	&	1.1	\\
CCPC-z43-002	&	1.7	&	505	&	2.5	&	1.92		&	47	&	0.1	\\
CCPC-z44-001	&	3.3	&	843	&	13.6	&	0.58		&	15375	&	10.0	\\
CCPC-z44-002	&	2.9	&	905	&	13.6	&	0.48		&	10887	&	6.6	\\
CCPC-z44-003	&	1.4	&	575	&	2.6	&	4.60		&	456	&	1.0	\\
CCPC-z44-004	&	3.1	&	1022	&	18.8	&	0.33		&	11586	&	6.5	\\
CCPC-z44-005	&	3.2	&	870	&	13.9	&	0.31		&	5413	&	2.9	\\
CCPC-z45-001	&	2.7	&	720	&	8.1	&	0.63		&	5331	&	3.6	\\
CCPC-z45-002	&	2.6	&	766	&	8.7	&	2.42		&	4367	&	6.1	\\
CCPC-z45-003	&	3.5	&	878	&	15.7	&	0.14		&	3857	&	1.8	\\
CCPC-z48-001	&	1.6	&	493	&	2.2	&	1.74		&	13	&	 $<0.1$	\\
CCPC-z48-002	&	2.2	&	736	&	6.9	&	1.40		&	563	&	0.5	\\
CCPC-z49-001	&	2.1	&	1052	&	13.5	&	0.68		&	2	&	 $<0.1$	\\
CCPC-z50-001	&	1.8	&	699	&	5.0	&	5.58		&	17	&	 $<0.1$	\\
CCPC-z51-001	&	1.5	&	356	&	1.1	&	0.11		&	203	&	0.1	\\
CCPC-z51-002	&	1.1	&	110	&	0.1	&	0.67		&	55	&	 $<0.1$	\\
CCPC-z56-001	&	2.4	&	1137	&	17.7	&	0.81		&	324	&	0.2	\\
CCPC-z56-002	&	2.5	&	763	&	8.3	&	1.00		&	4720	&	3.9	\\
CCPC-z56-003	&	2.5	&	820	&	9.8	&	0.50		&	3728	&	2.3	\\
CCPC-z56-004	&	1.8	&	614	&	3.9	&	2.00		&	2766	&	3.4	\\
CCPC-z56-005	&	2.5	&	932	&	12.6	&	0.17		&	2128	&	1.0	\\
CCPC-z56-006	&	2.6	&	909	&	12.3	&	1.33		&	3625	&	3.5	\\
CCPC-z56-007	&	1.5	&	907	&	7.2	&	5.87		&	311	&	0.9	\\
CCPC-z56-008	&	0.7	&	391	&	0.6	&	5.93		&	95	&	0.3	\\
CCPC-z57-001	&	1.9	&	752	&	6.1	&	0.82		&	3942	&	2.9	\\
CCPC-z57-002	&	1.2	&	404	&	1.1	&	0.66		&	583	&	0.4	\\
CCPC-z57-003	&	1.7	&	947	&	8.9	&	0.44		&	3484	&	2.1	\\
CCPC-z57-004	&	1.3	&	465	&	1.6	&	1.40		&	309	&	0.3	\\
CCPC-z58-001	&	3.0	&	881	&	13.3	&	0.36		&	269	&	0.1	\\
CCPC-z59-001	&	1.8	&	743	&	5.6	&	0.11		&	2938	&	1.4	\\
CCPC-z60-001	&	1.7	&	879	&	7.6	&	1.06		&	998	&	0.8	\\
CCPC-z65-001	&	2.0	&	1113	&	14.0	&	0.46		&	2378	&	1.4	\\
CCPC-z65-002	&	1.6	&	703	&	4.6	&	3.03		&	2541	&	4.1	\\
CCPC-z65-003	&	1.0	&	510	&	1.5	&	1.38		&	99	&	0.1	\\
CCPC-z65-004	&	1.6	&	733	&	5.0	&	0.83		&	429	&	0.3	\\
CCPC-z65-005	&	2.0	&	566	&	3.6	&	1.38		&	6918	&	6.9	
\enddata
\tablenotetext{a}{Counts of field galaxies were limited to the surface area of the overdense region.}
\tablecomments{The mass estimates for the entire CCPC2 sample. The $R_e$ (in physical units) and velocity dispersions $\sigma$ are used to compute a virial mass estimate, with the caveat that these systems are not expected to be in equilibrium at the relevant redshifts. We obtain the mass overdensity $\delta_{m}$ by assuming galaxies are linearly biased tracers of mass with a slope of $b=3$. From this and the volume contained within the galaxy distribution we can compute an expected mass of the cluster \citep{1998ApJ...492..428S}. This assumes that the total volume will collapse into a single halo at $z=0$.
} \label{tab:tot_mass}
\end{deluxetable*}

\begin{deluxetable*}{c c c c c c c c}
\tablewidth{0pt}
\tablecolumns{8}
\tablecaption{CCPC II: Objects of Interest (OOIs)}
\tablehead{
\colhead{Candidate}	&	\colhead{RA}	&	\colhead{DEC}	&	\colhead{Redshift}	&	\colhead{$N$}	&	 \colhead{$N_{R\le10}$} 	&	\colhead{$R_{e}$}	&	\colhead{Reference}	\\
\colhead{Name}	&	\colhead{($deg$)}	&	\colhead{($deg$)}	&	\colhead{($z_{avg}$)}	&	\colhead{}	&	\colhead{cMpc}	&	\colhead{$(Mpc)$}	&	\colhead{}	
} 
\startdata
OOI-z22-001	&	30.199	&	1.749	&	2.2	&	11	&	10	&	2.65	&	1	\\
OOI-z22-002	&	30.569	&	1.488	&	2.2	&	14	&	4	&	4.45	&	1	\\
OOI-z22-003	&	30.581	&	1.928	&	2.2	&	10	&	2	&	5.1	&	1	\\
OOI-z22-004	&	150.852	&	0.181	&	2.2	&	17	&	11	&	2.4	&	1	\\
OOI-z22-005	&	151.318	&	0.697	&	2.2	&	19	&	9	&	3.2	&	1	\\
OOI-z22-006	&	197.999	&	42.715	&	2.2	&	4	&	3	&	1	&	1	\\
OOI-z23-001	&	325.525	&	-44.580	&	2.38	&	10	&	4	&	3.15	&	2	\\
OOI-z23-002	&	325.843	&	-44.218	&	2.38	&	10	&	2	&	4.1	&	2	\\
OOI-z23-001	&	325.525	&	-44.580	&	2.38	&	10	&	4	&	3.15	&	2	\\
OOI-z23-002	&	325.843	&	-44.218	&	2.38	&	10	&	2	&	4.1	&	2	\\
OOI-z57-001	&	149.654	&	1.539	&	5.7	&	6	&	1	&	1.8	&	3	\\
OOI-z57-002	&	149.746	&	2.752	&	5.7	&	7	&	2	&	1.75	&	3	\\
OOI-z57-003	&	150.398	&	2.773	&	5.7	&	8	&	4	&	1.35	&	3	\\
OOI-z57-004	&	230.905	&	-0.144	&	5.7	&	5	&	3	&	1.45	&	4	\\
OOI-z57-005	&	150.168	&	2.318	&	5.7	&	5	&	3	&	1.4	&	5	\\
OOI-z65-001	&	314.069	&	-4.630	&	6.50	&	4	&	4	&	0.75	&	6 \\
OOI-z65-002	&	34.4896	&	-5.146 &	6.57	&	4	&	1	&	1.85	&	7
\enddata
\tablecomments{Candidate systems found within NED as spectroscopic sources. Further inspection revealed that these galaxies are mainly from NB imaging and do not have spectroscopic information, but fulfill all other CCPC criteria. \emph{The last two systems are spectroscopic sources, but did not have any field galaxies with which to compute an overdensity.} The references of their estimated redshifts are included. References: (1) \citet{2011MNRAS.416.2041M} (2) \citet{2004ApJ...614...75F} (3) \citet{2007ApJS..172..523M} (4) \citet{2008ApJ...679..942M} (5) \citet{2012ApJ...760..128M},
(6) \citet{2013AJ....146...96D}, (7) \citet{2013ApJ...772...99J}.
} \label{tab:ooi}
\end{deluxetable*}

\begin{figure*}
\centering
\begin{subfigure}
\centering
\includegraphics[height=7.5cm,width=7.5cm]{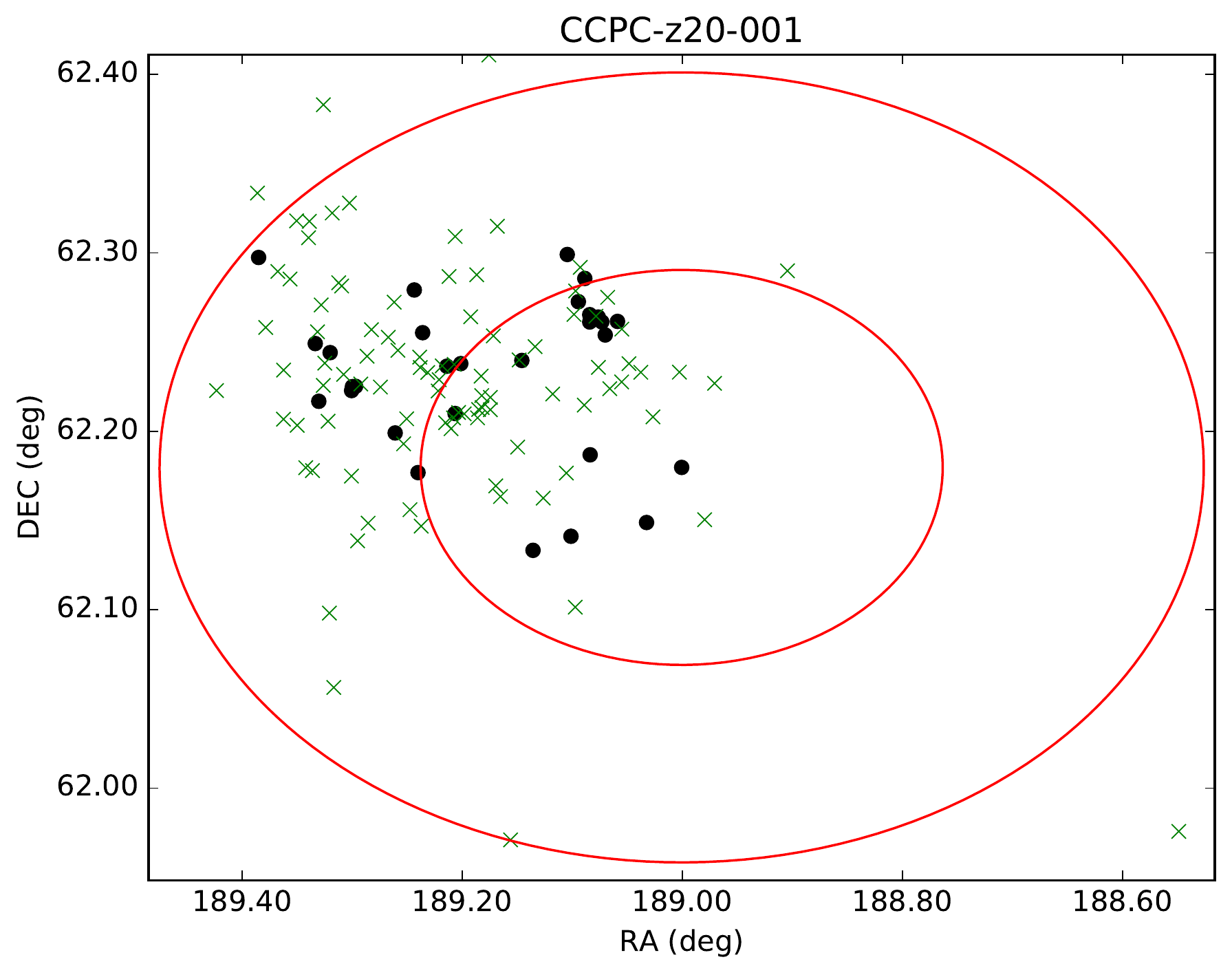}
\label{fig:CCPC-z20-001_sky}
\end{subfigure}
\hfill
\begin{subfigure}
\centering
\includegraphics[scale=0.52]{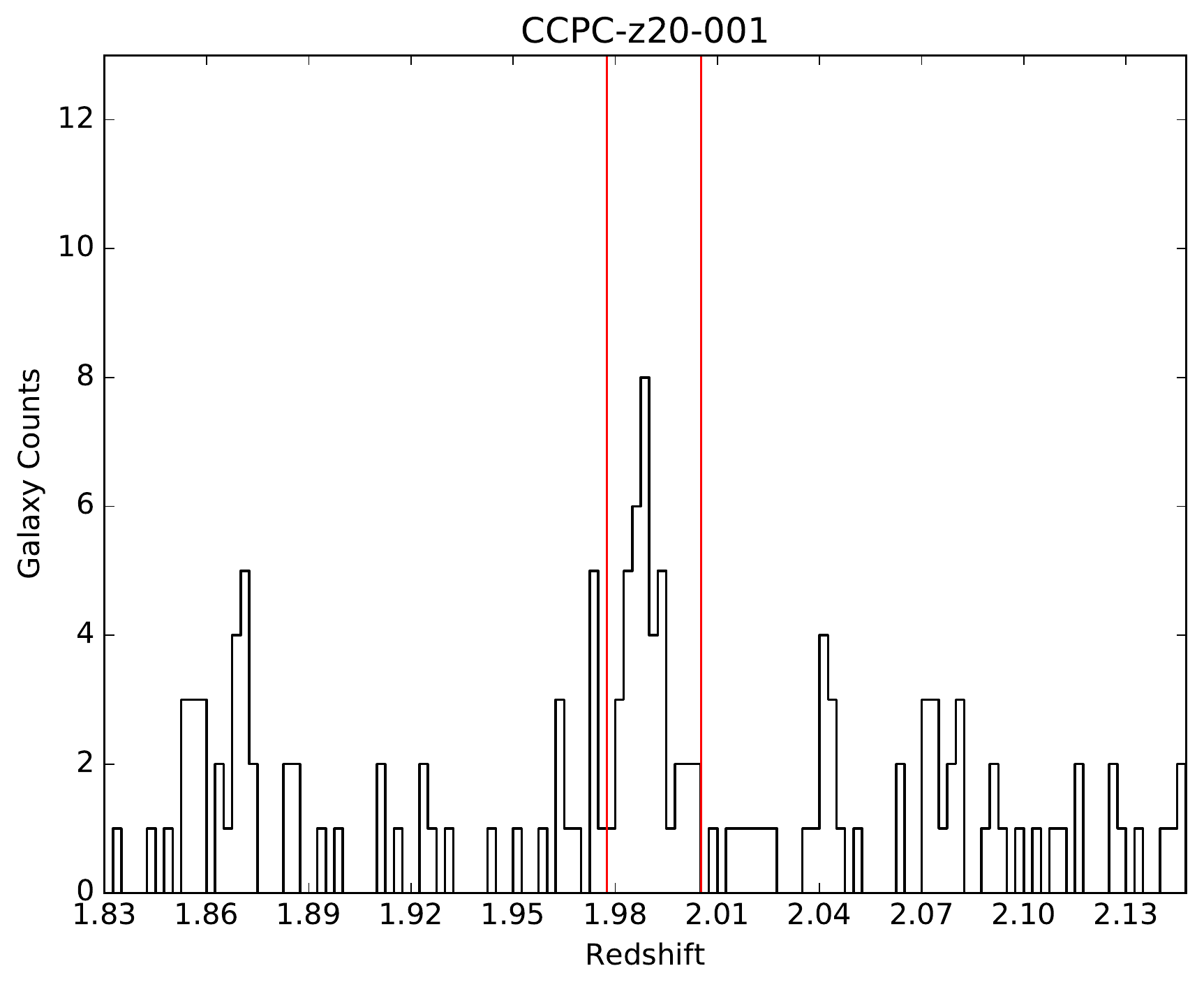}
\label{fig:CCPC-z20-001}
\end{subfigure}
\hfill
\caption{\emph{Left:} The outer and inner red circles represents 20 and 10 cMpc from the center of the protocluster, respectively. Black points are the protocluster galaxies and green $\times$ symbols indicate `field' sources within $\Delta z \pm 0.15$ of the redshift of the structure. This serves as a proxy of the FOV of sources in the combined surveys. \emph{Right:} Number of galaxies plotted as a function of redshift ($N(z)$ distribution). The red vertical lines represent $\pm$20 comoving Mpc (the search criteria) from the mean redshift of the system. The calculation of the overdensity $\delta_{gal}$ is restricted to $z\pm\sigma$ and is typically a factor of a few smaller.}. 
\end{figure*}

\begin{figure*}
\centering
\begin{subfigure}
\centering
\includegraphics[height=7.5cm,width=7.5cm]{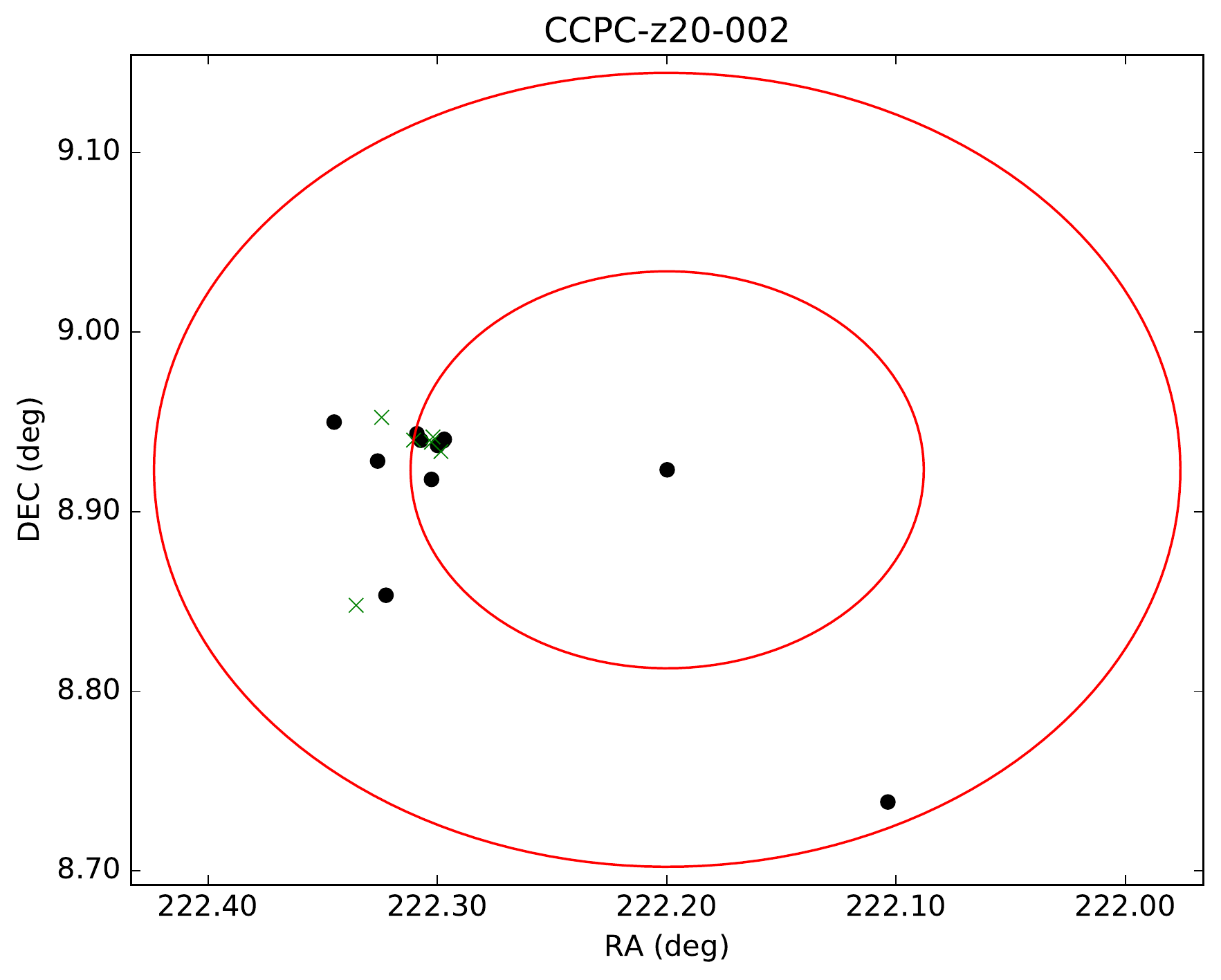}
\label{fig:CCPC-z20-002_sky}
\end{subfigure}
\hfill
\begin{subfigure}
\centering
\includegraphics[scale=0.52]{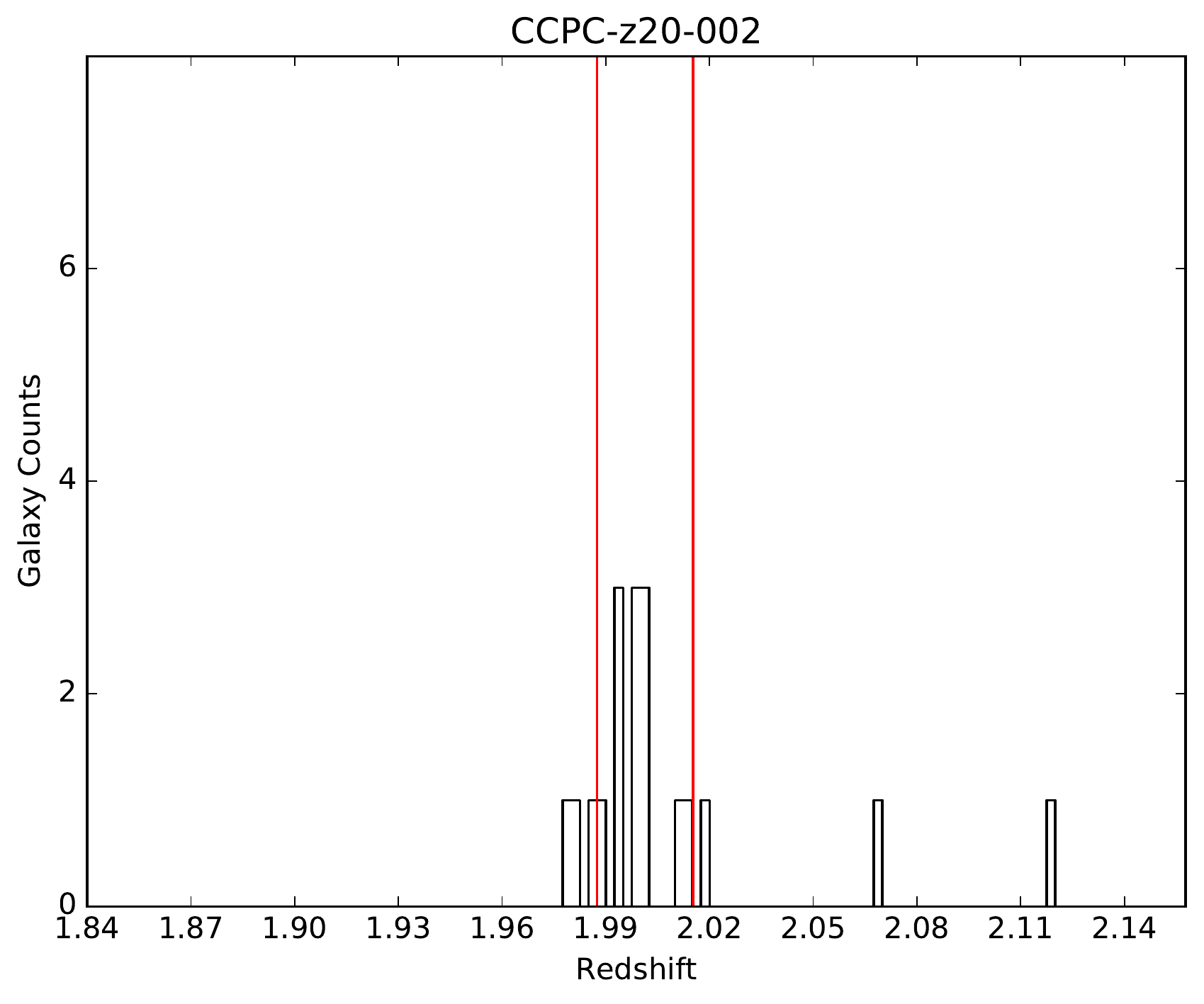}
\label{fig:CCPC-z20-002}
\end{subfigure}
\hfill
\end{figure*}

\begin{figure*}
\centering
\begin{subfigure}
\centering
\includegraphics[height=7.5cm,width=7.5cm]{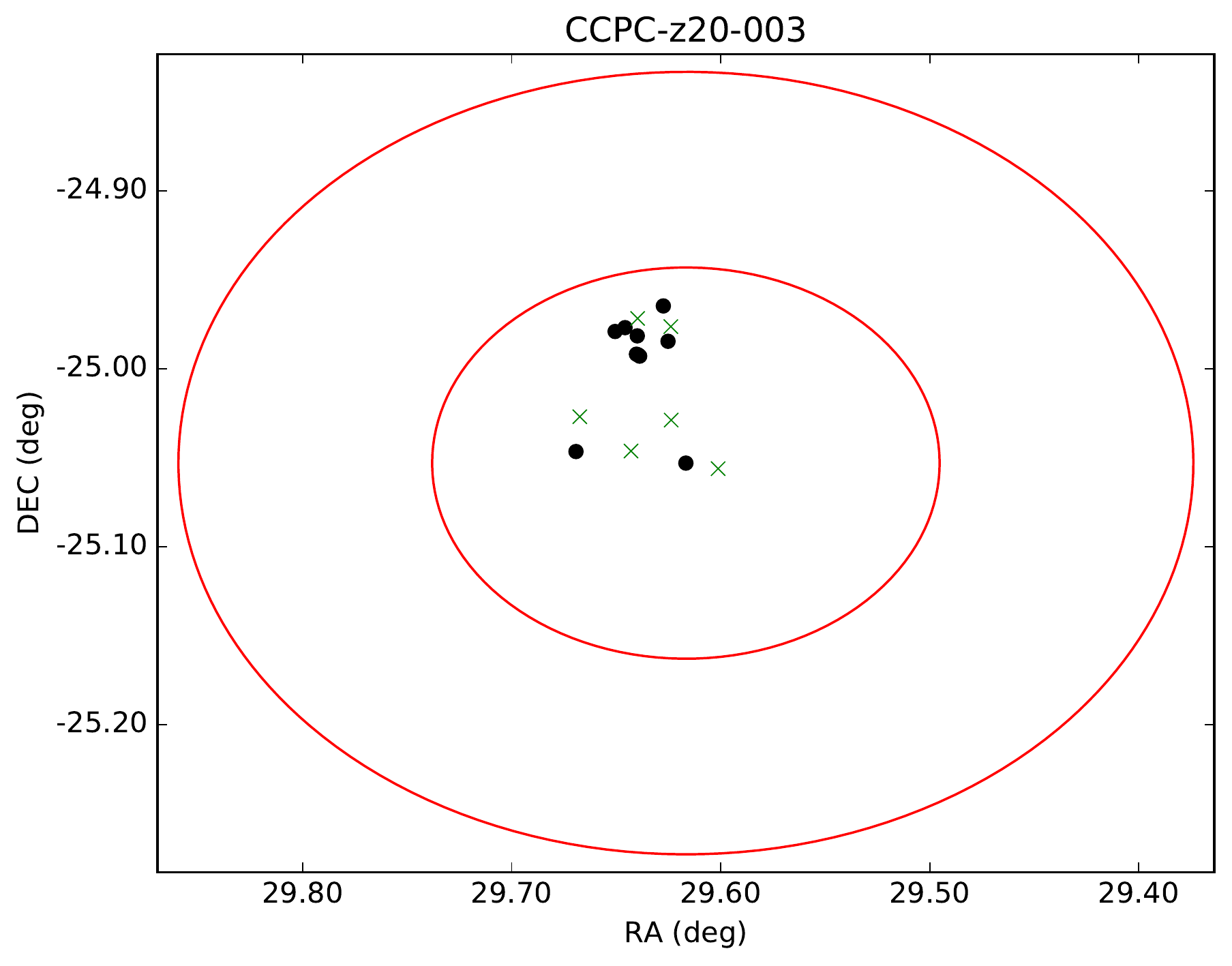}
\label{fig:CCPC-z20-003_sky}
\end{subfigure}
\hfill
\begin{subfigure}
\centering
\includegraphics[scale=0.52]{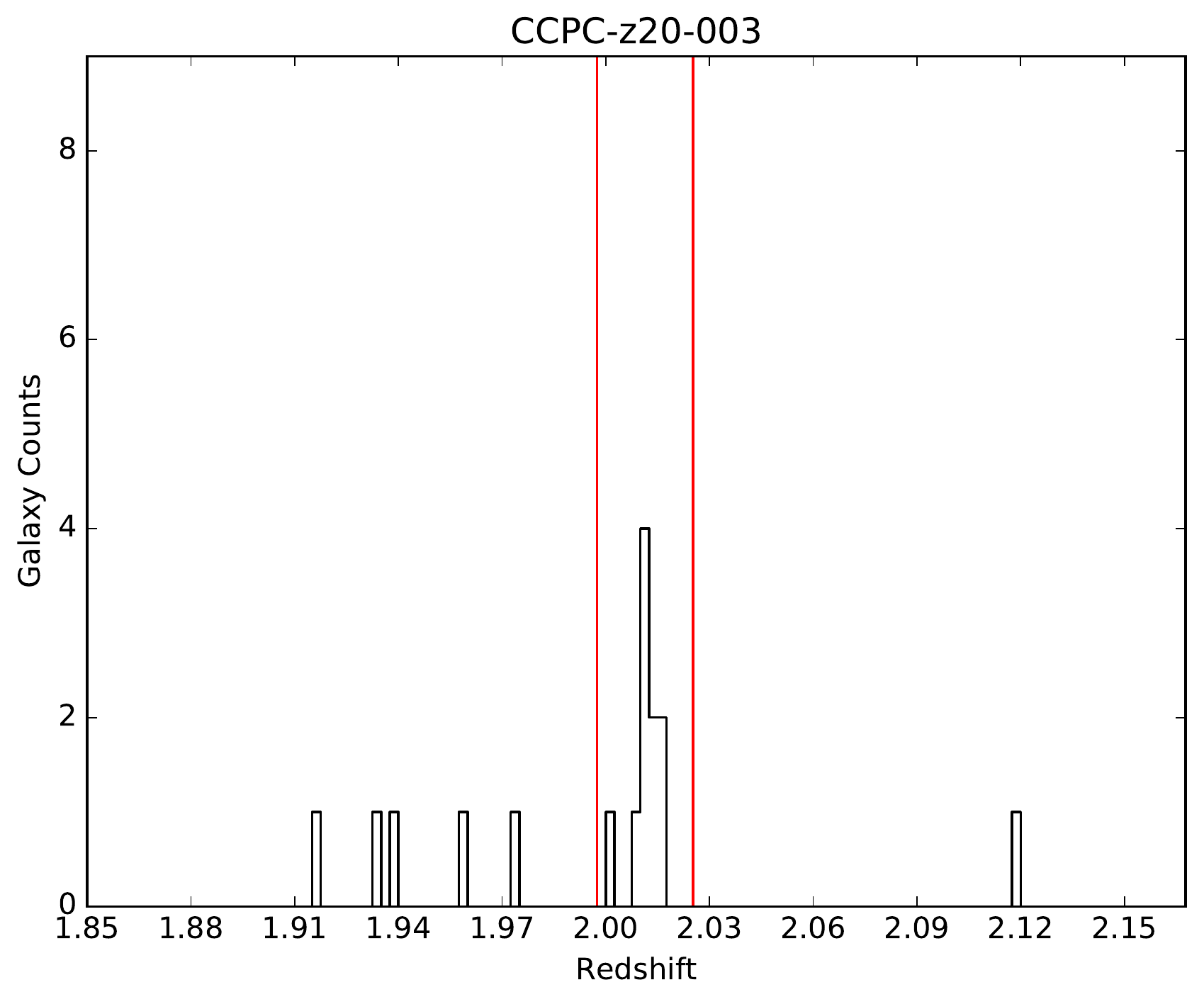}
\label{fig:CCPC-z20-003}
\end{subfigure}
\hfill
\end{figure*}
\clearpage 

\begin{figure*}
\centering
\begin{subfigure}
\centering
\includegraphics[height=7.5cm,width=7.5cm]{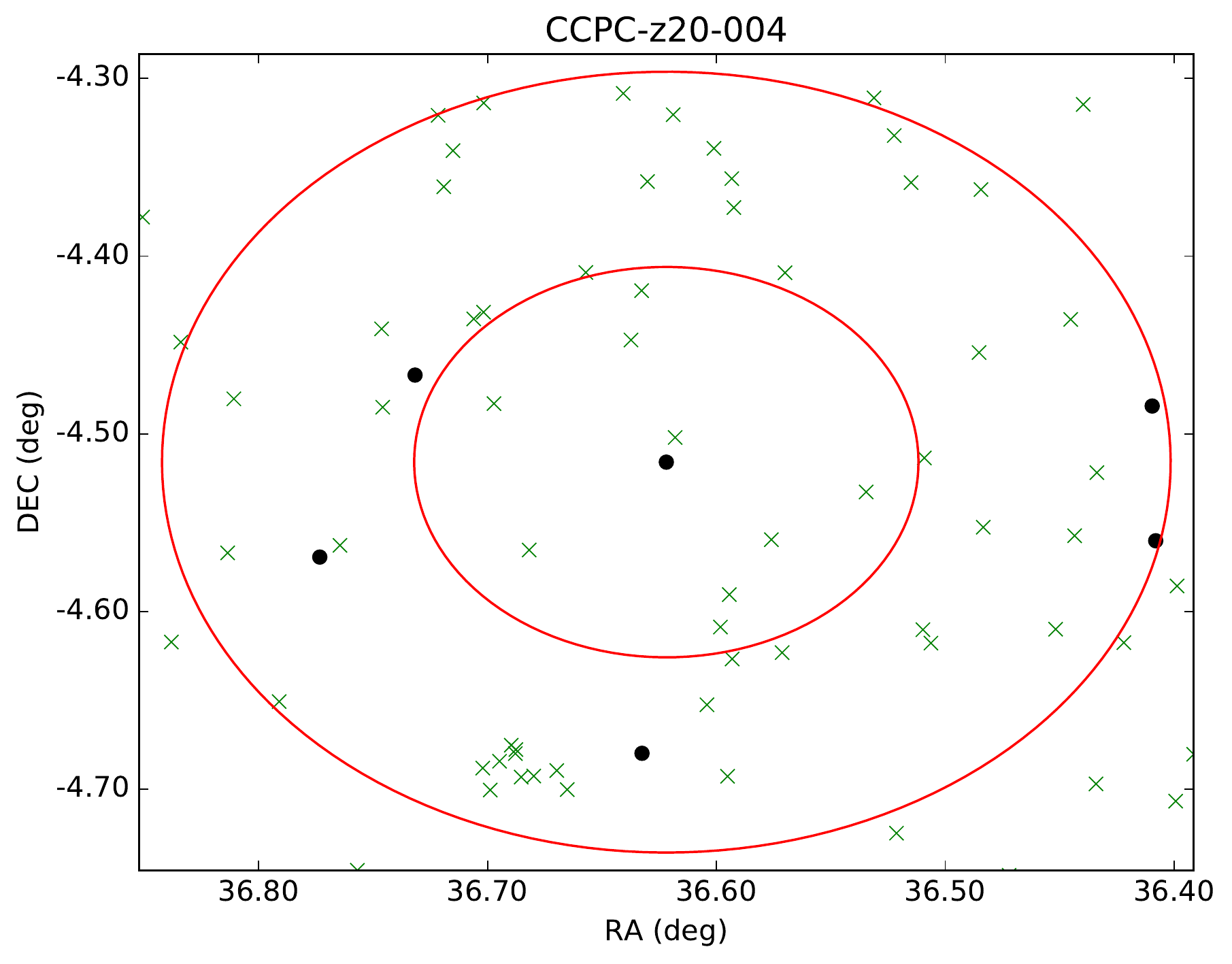}
\label{fig:CCPC-z20-004_sky}
\end{subfigure}
\hfill
\begin{subfigure}
\centering
\includegraphics[scale=0.52]{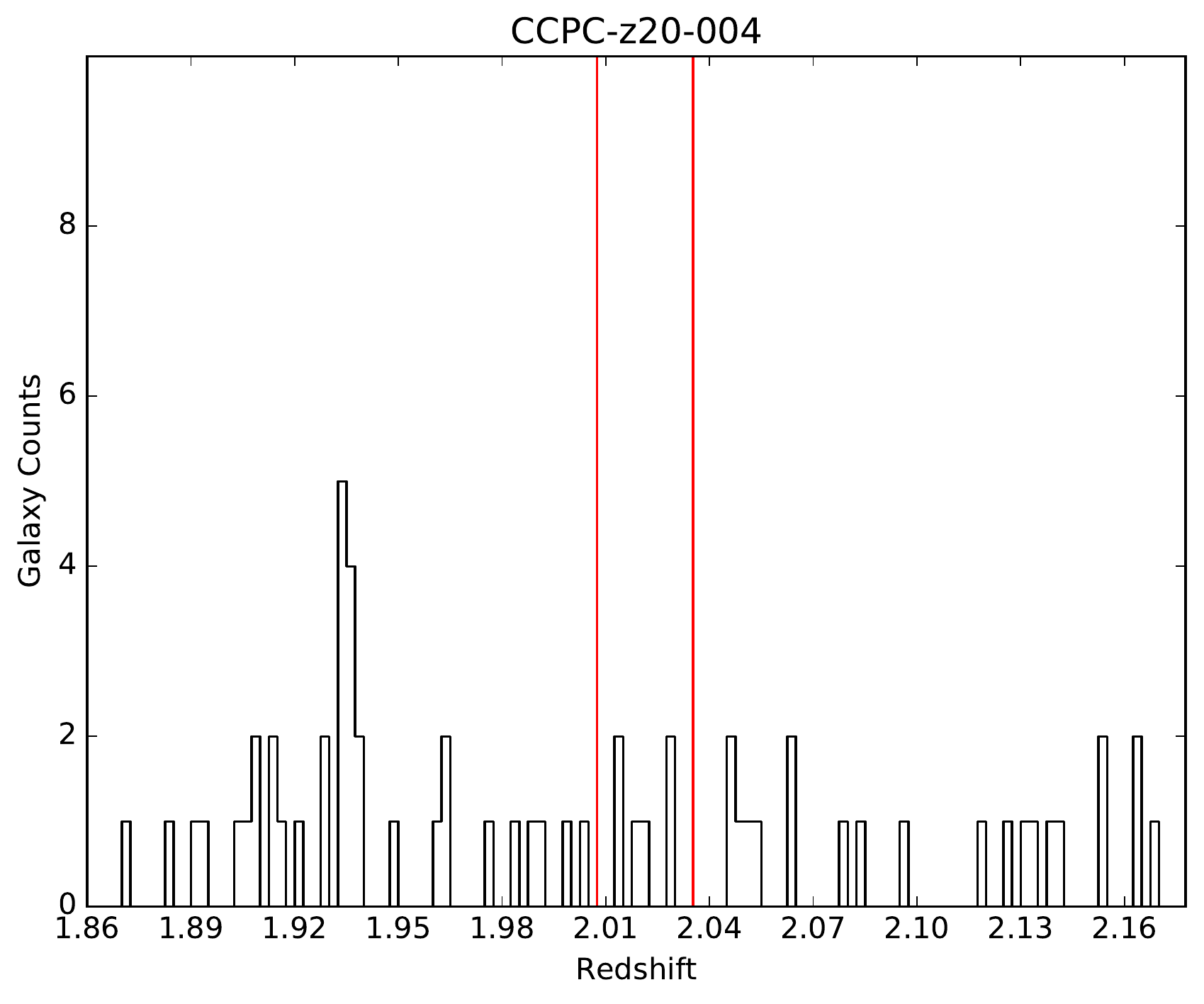}
\label{fig:CCPC-z20-004}
\end{subfigure}
\hfill
\end{figure*}

\begin{figure*}
\centering
\begin{subfigure}
\centering
\includegraphics[height=7.5cm,width=7.5cm]{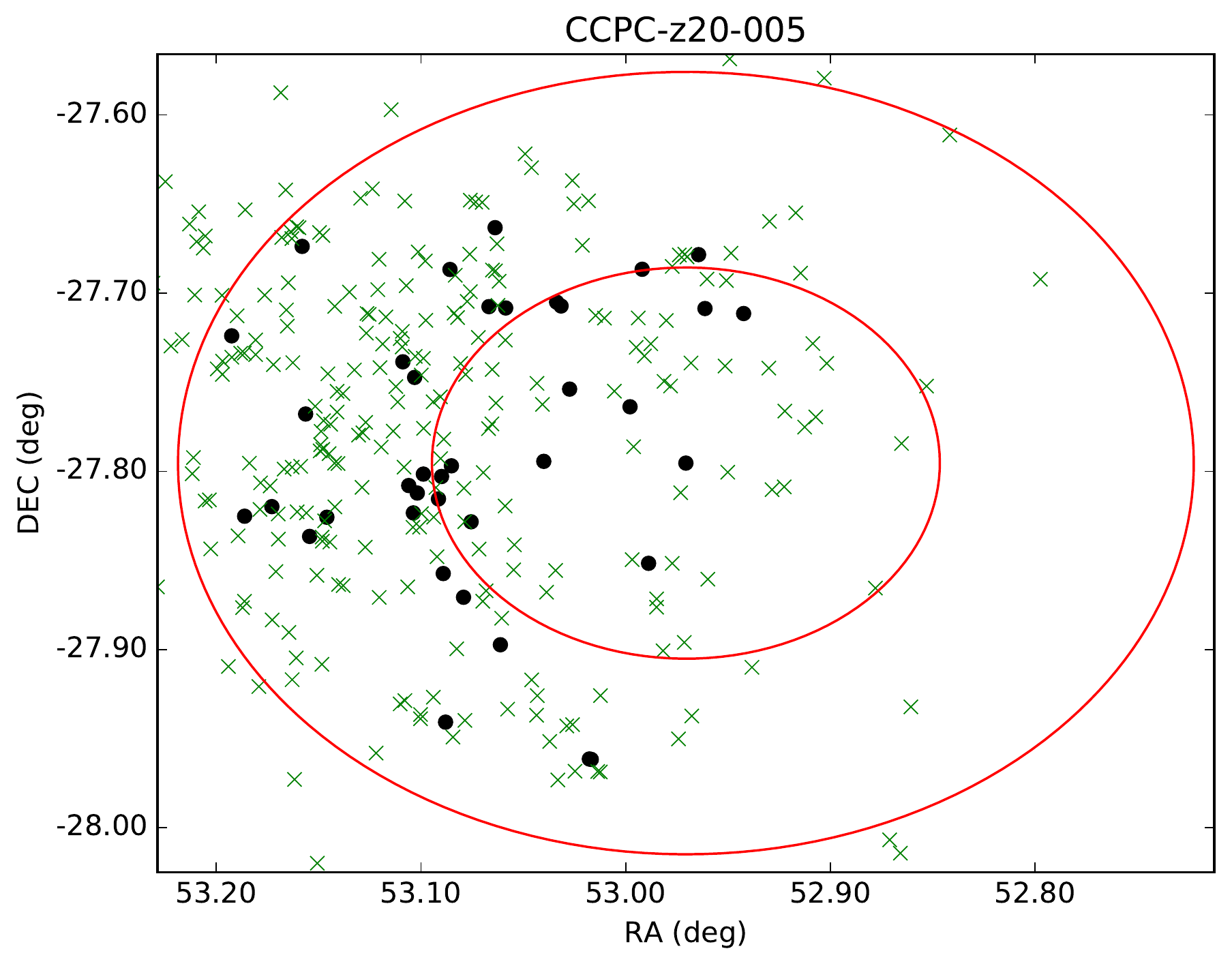}
\label{fig:CCPC-z20-005_sky}
\end{subfigure}
\hfill
\begin{subfigure}
\centering
\includegraphics[scale=0.52]{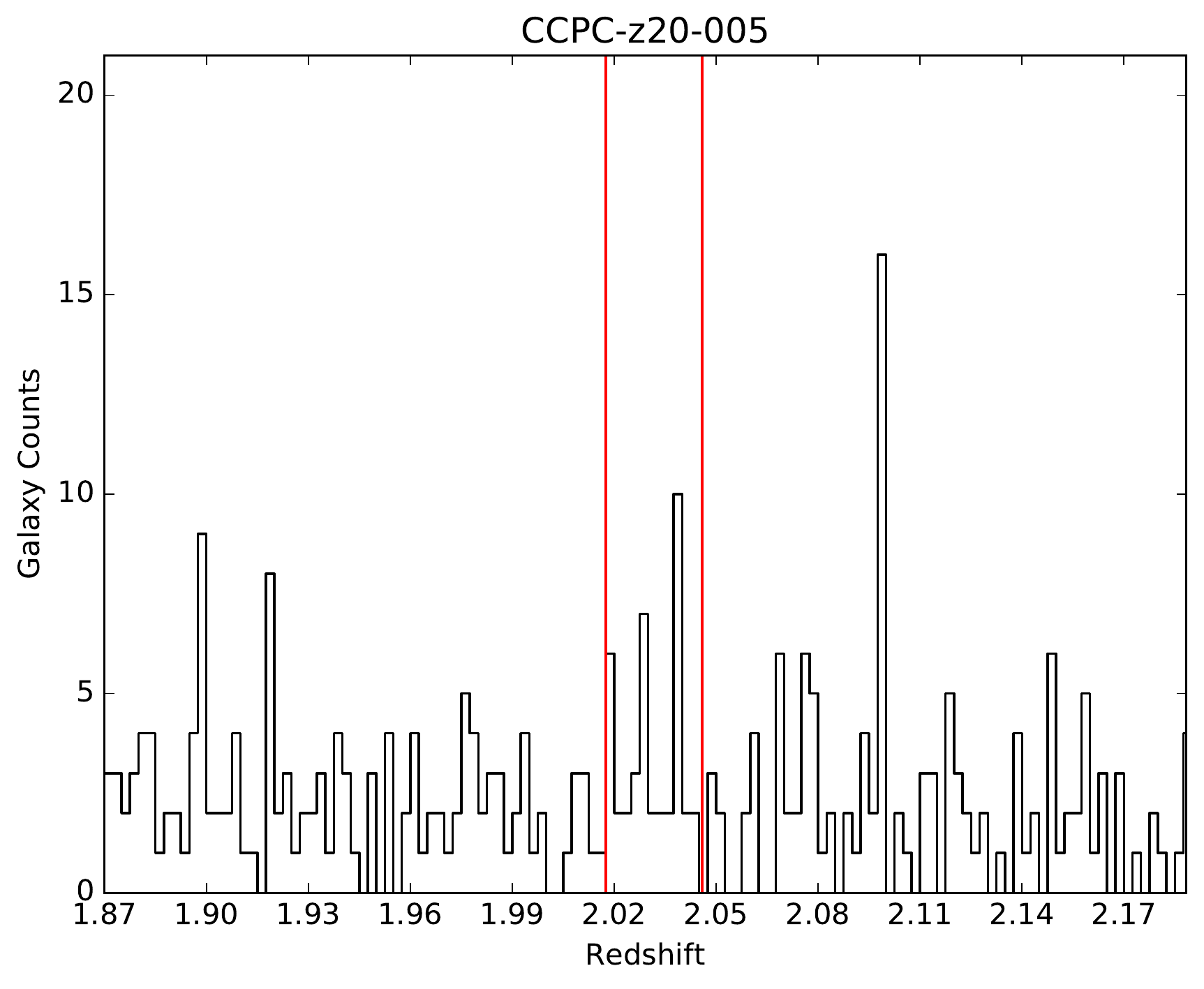}
\label{fig:CCPC-z20-005}
\end{subfigure}
\hfill
\end{figure*}

\begin{figure*}
\centering
\begin{subfigure}
\centering
\includegraphics[height=7.5cm,width=7.5cm]{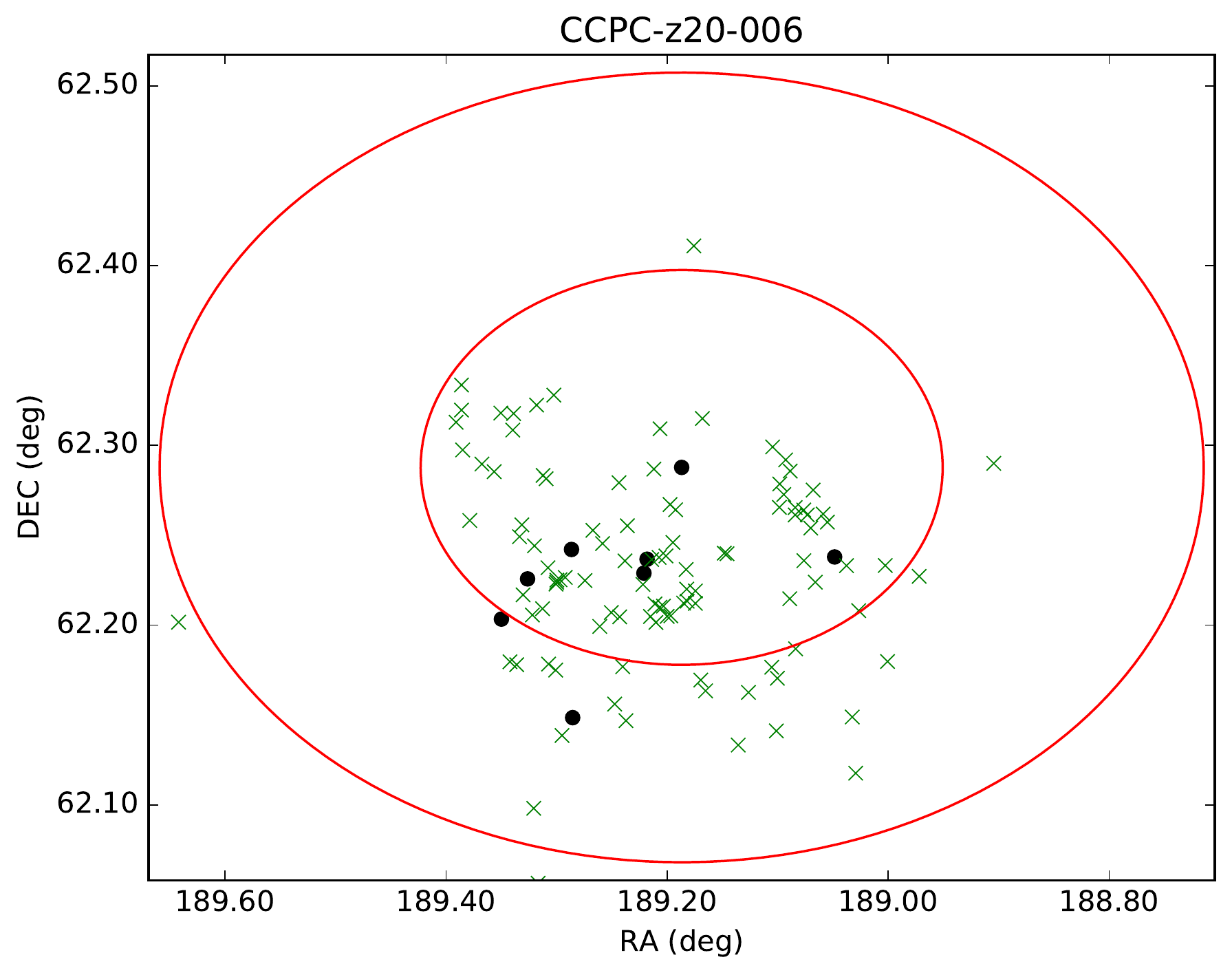}
\label{fig:CCPC-z20-006_sky}
\end{subfigure}
\hfill
\begin{subfigure}
\centering
\includegraphics[scale=0.52]{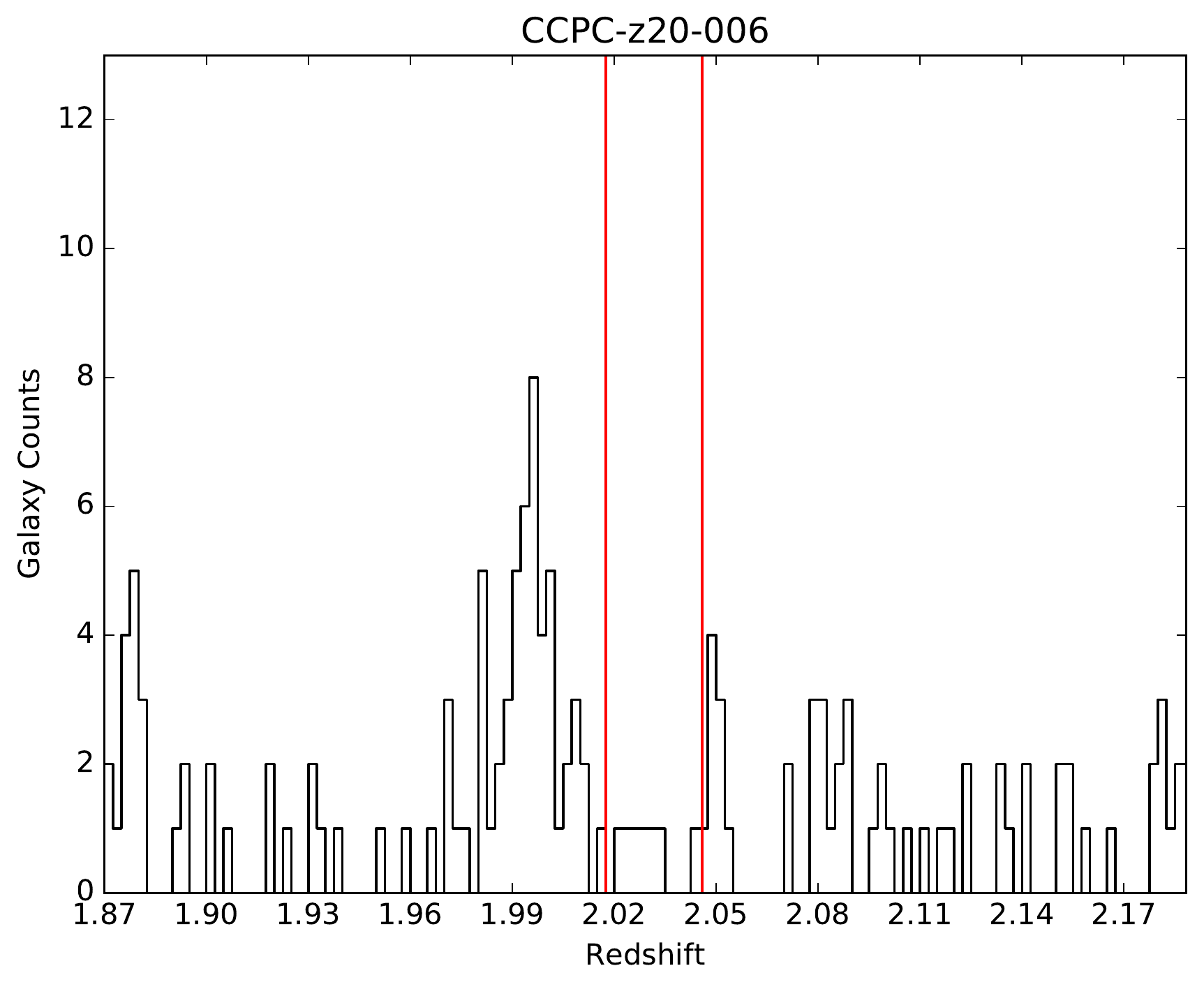}
\label{fig:CCPC-z20-006}
\end{subfigure}
\hfill
\end{figure*}
\clearpage 

\begin{figure*}
\centering
\begin{subfigure}
\centering
\includegraphics[height=7.5cm,width=7.5cm]{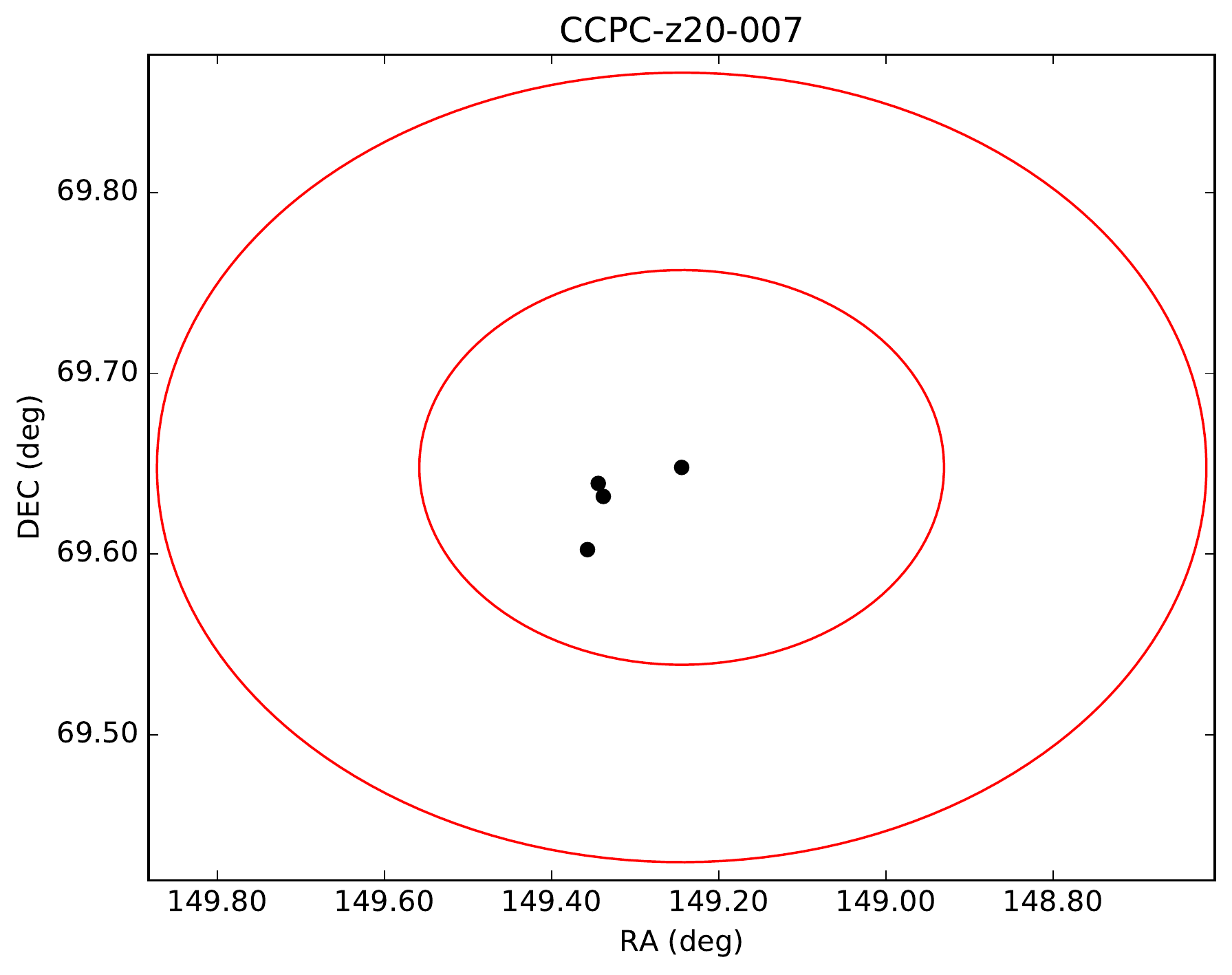}
\label{fig:CCPC-z20-007_sky}
\end{subfigure}
\hfill
\begin{subfigure}
\centering
\includegraphics[scale=0.52]{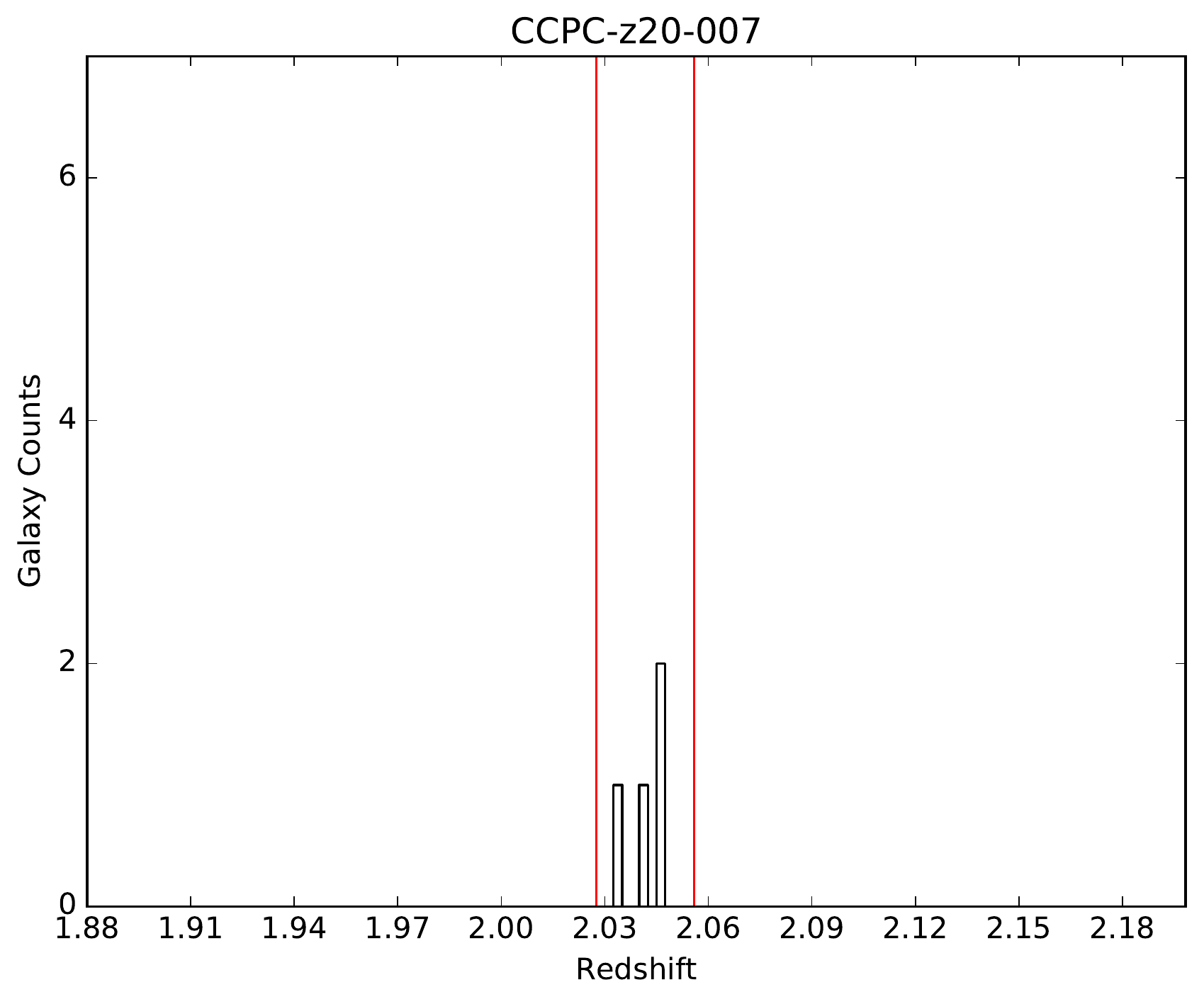}
\label{fig:CCPC-z20-007}
\end{subfigure}
\hfill
\end{figure*}

\begin{figure*}
\centering
\begin{subfigure}
\centering
\includegraphics[height=7.5cm,width=7.5cm]{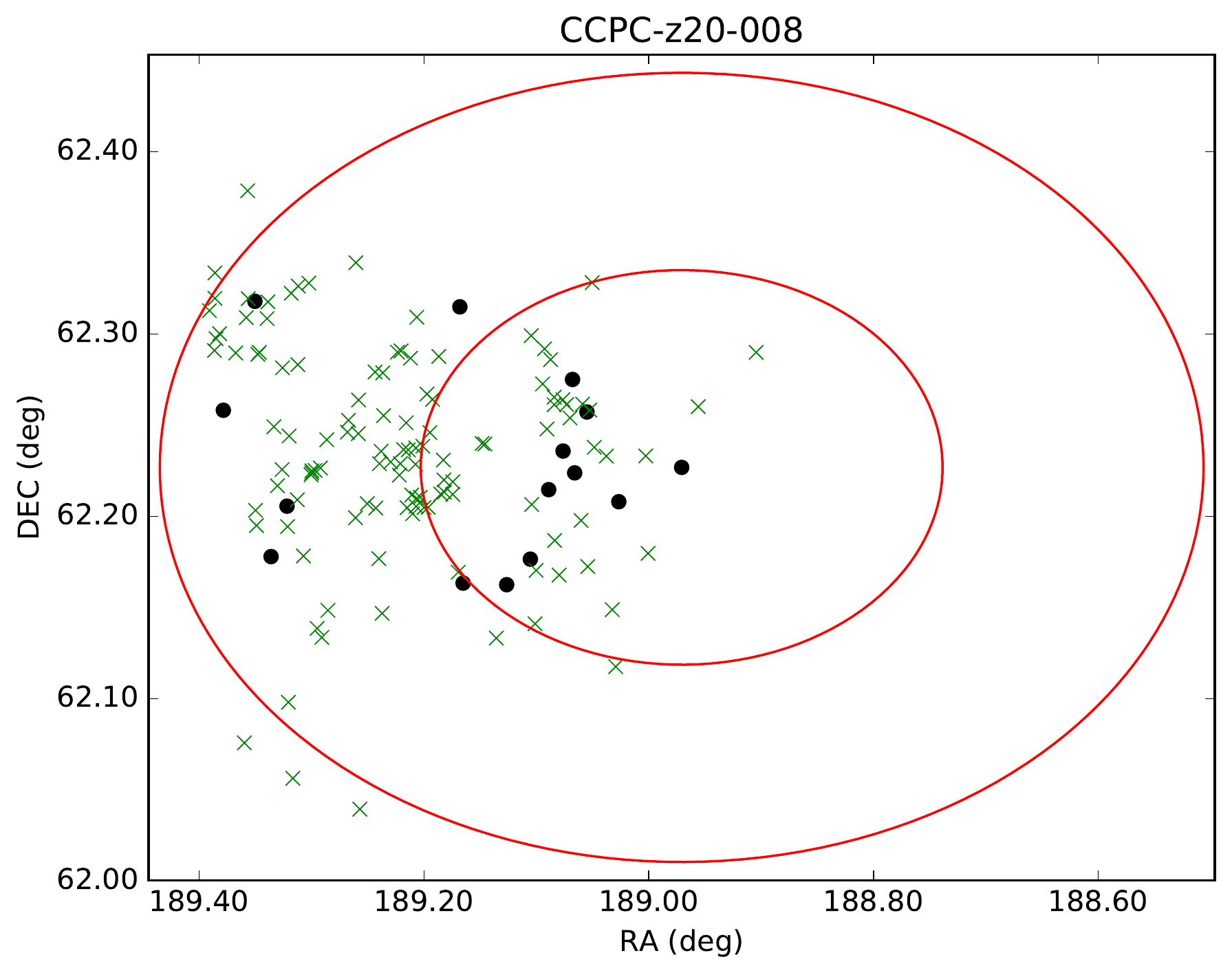}
\label{fig:CCPC-z20-008_sky}
\end{subfigure}
\hfill
\begin{subfigure}
\centering
\includegraphics[scale=0.52]{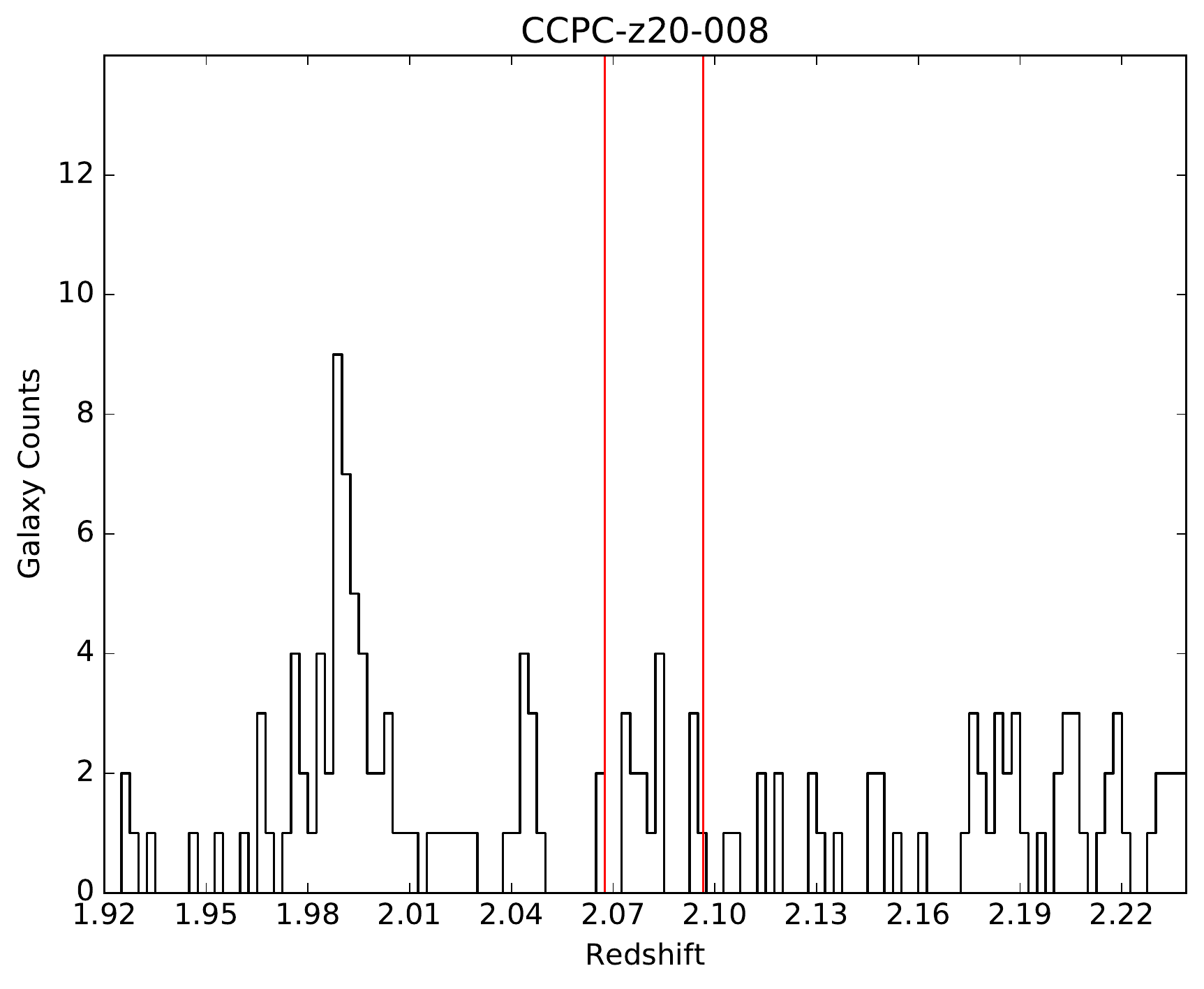}
\label{fig:CCPC-z20-008}
\end{subfigure}
\hfill
\end{figure*}

\begin{figure*}
\centering
\begin{subfigure}
\centering
\includegraphics[height=7.5cm,width=7.5cm]{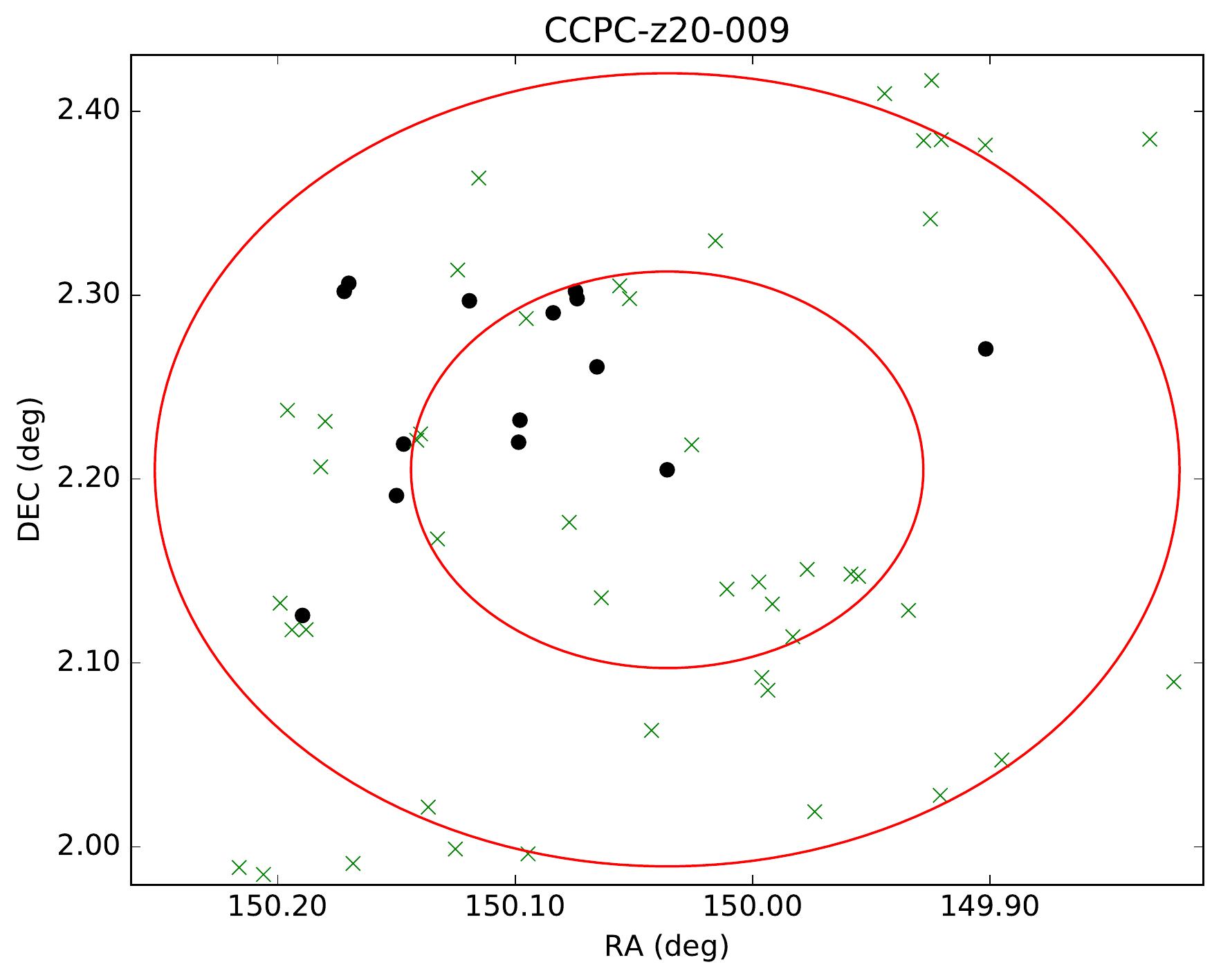}
\label{fig:CCPC-z20-009_sky}
\end{subfigure}
\hfill
\begin{subfigure}
\centering
\includegraphics[scale=0.52]{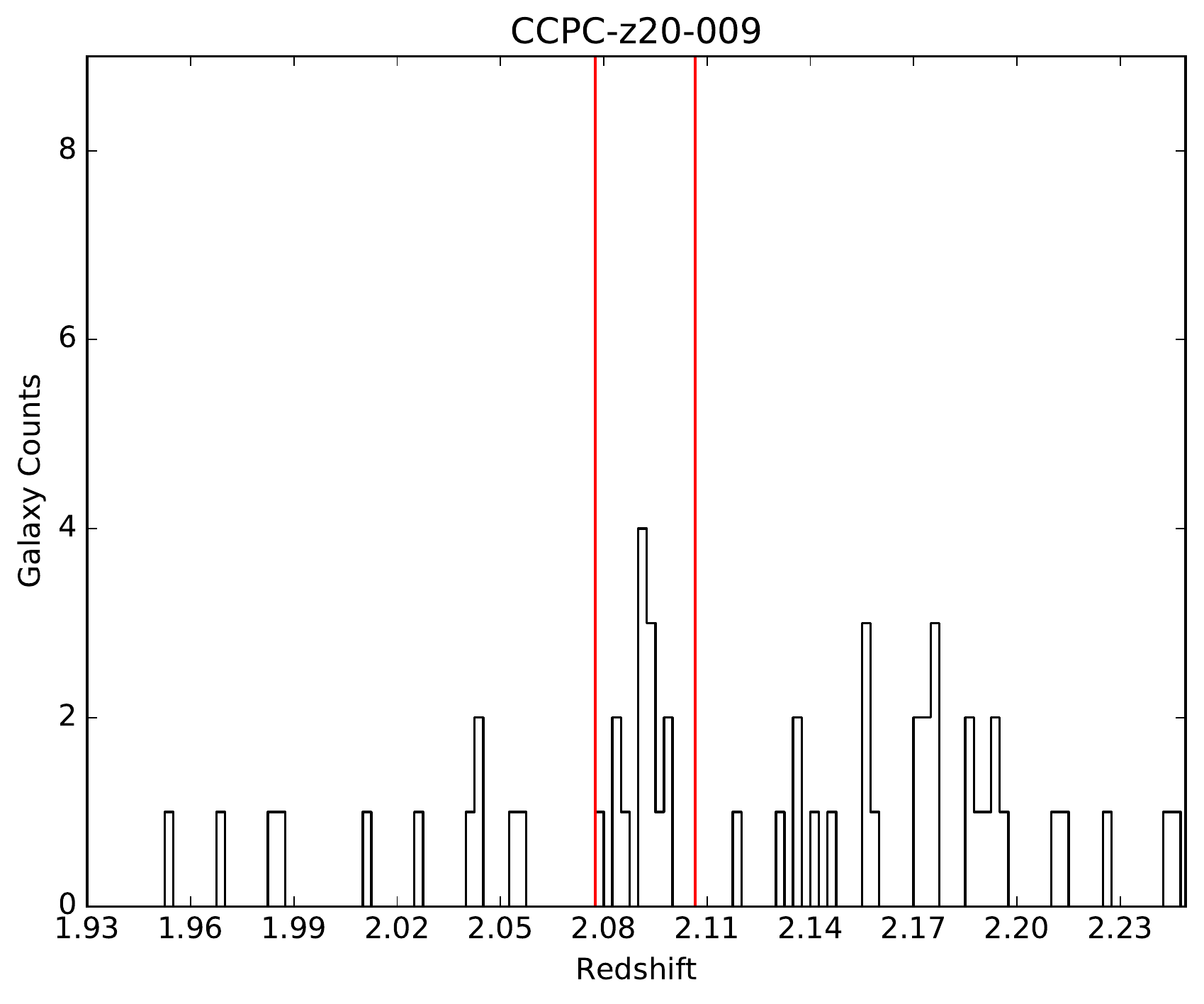}
\label{fig:CCPC-z20-009}
\end{subfigure}
\hfill
\end{figure*}
\clearpage 

\begin{figure*}
\centering
\begin{subfigure}
\centering
\includegraphics[height=7.5cm,width=7.5cm]{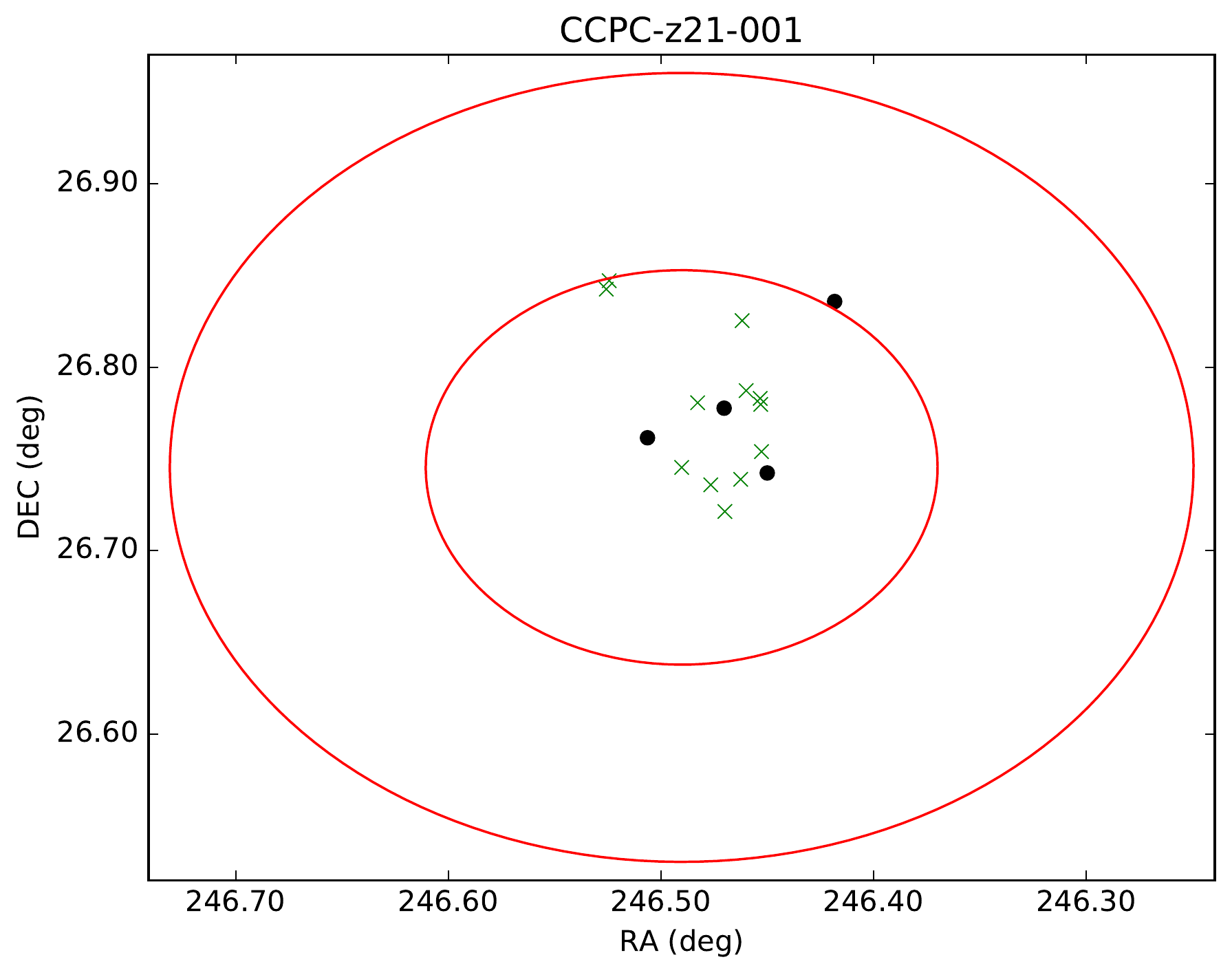}
\label{fig:CCPC-z21-001_sky}
\end{subfigure}
\hfill
\begin{subfigure}
\centering
\includegraphics[scale=0.52]{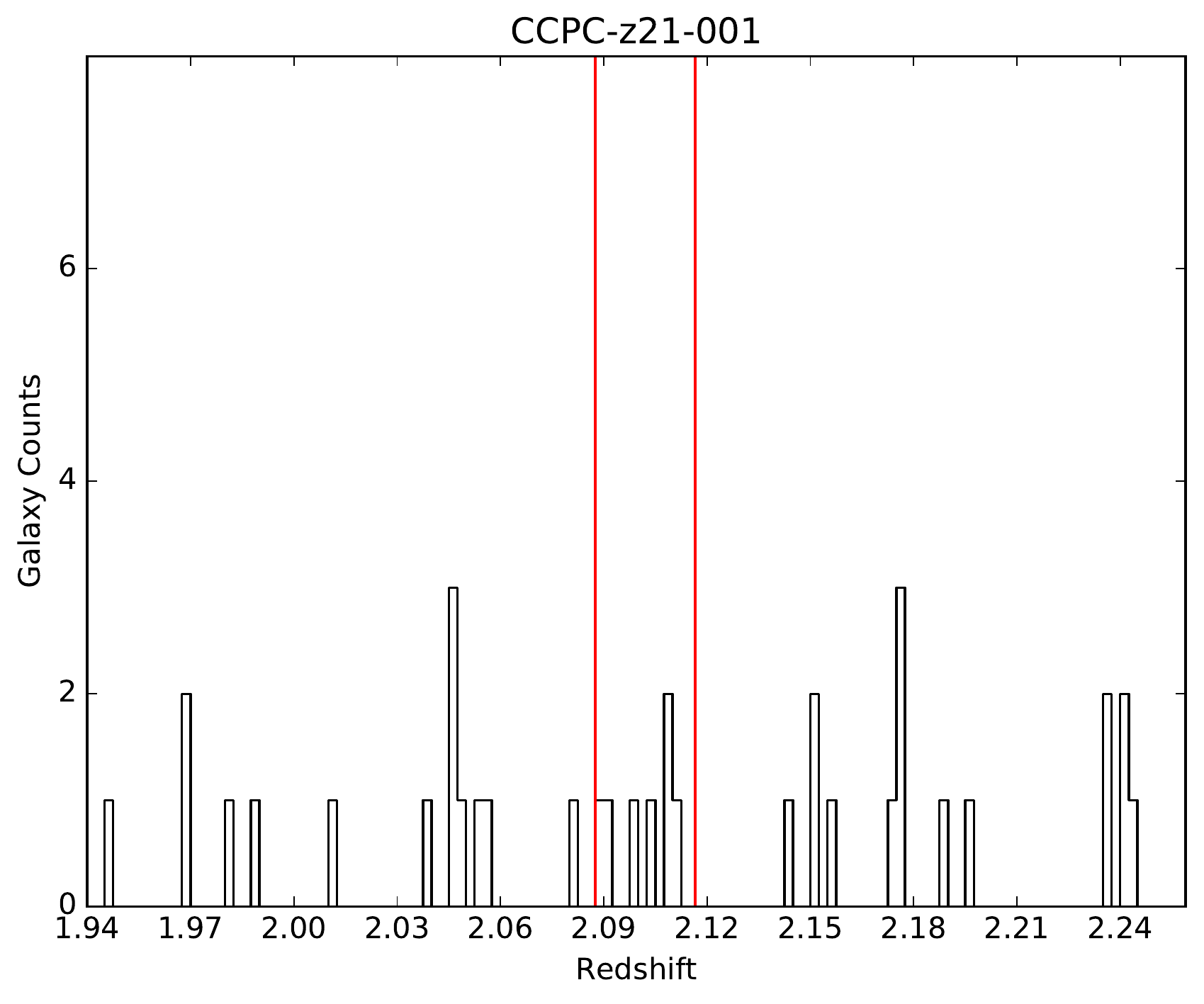}
\label{fig:CCPC-z21-001}
\end{subfigure}
\hfill
\end{figure*}

\begin{figure*}
\centering
\begin{subfigure}
\centering
\includegraphics[height=7.5cm,width=7.5cm]{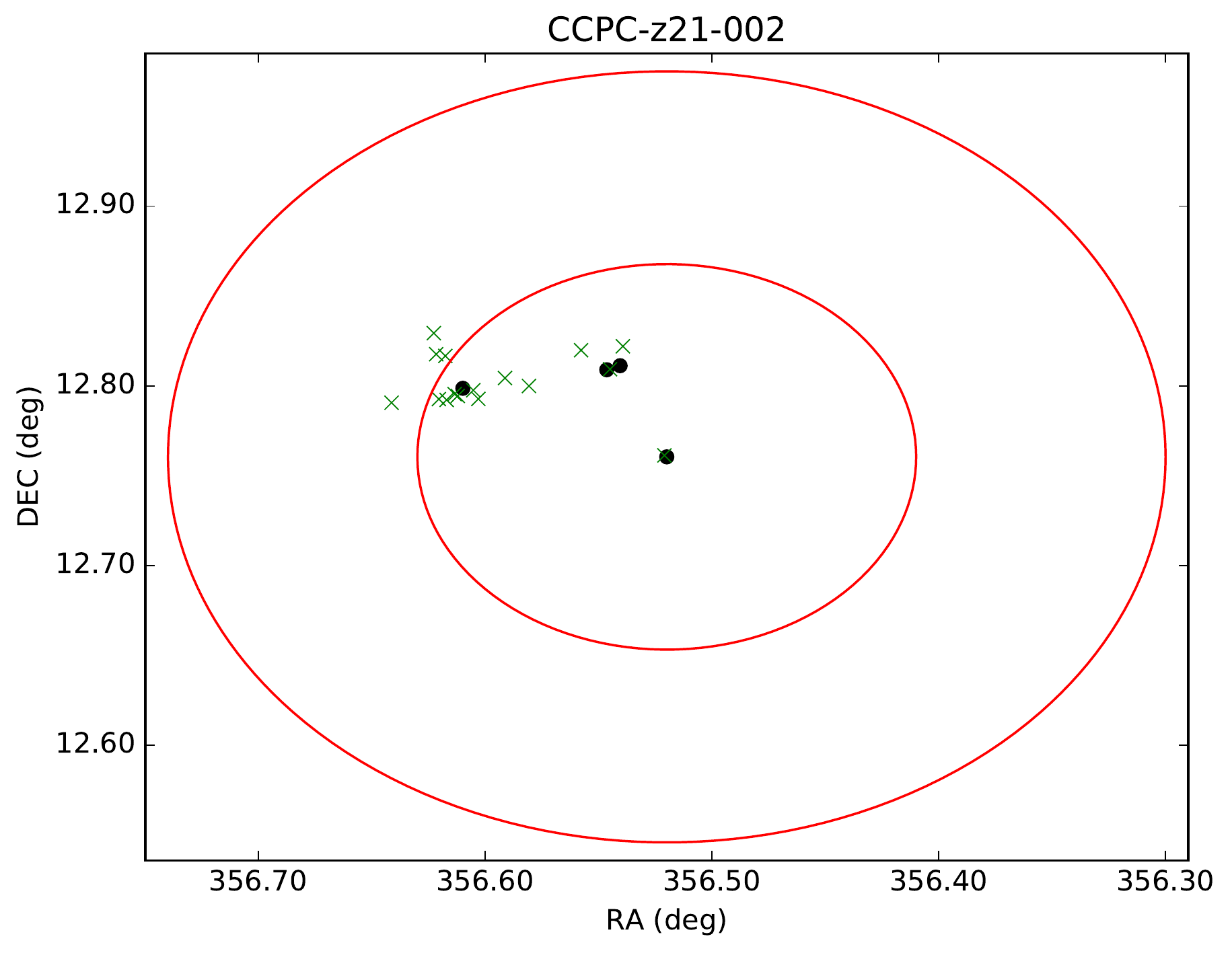}
\label{fig:CCPC-z21-002_sky}
\end{subfigure}
\hfill
\begin{subfigure}
\centering
\includegraphics[scale=0.52]{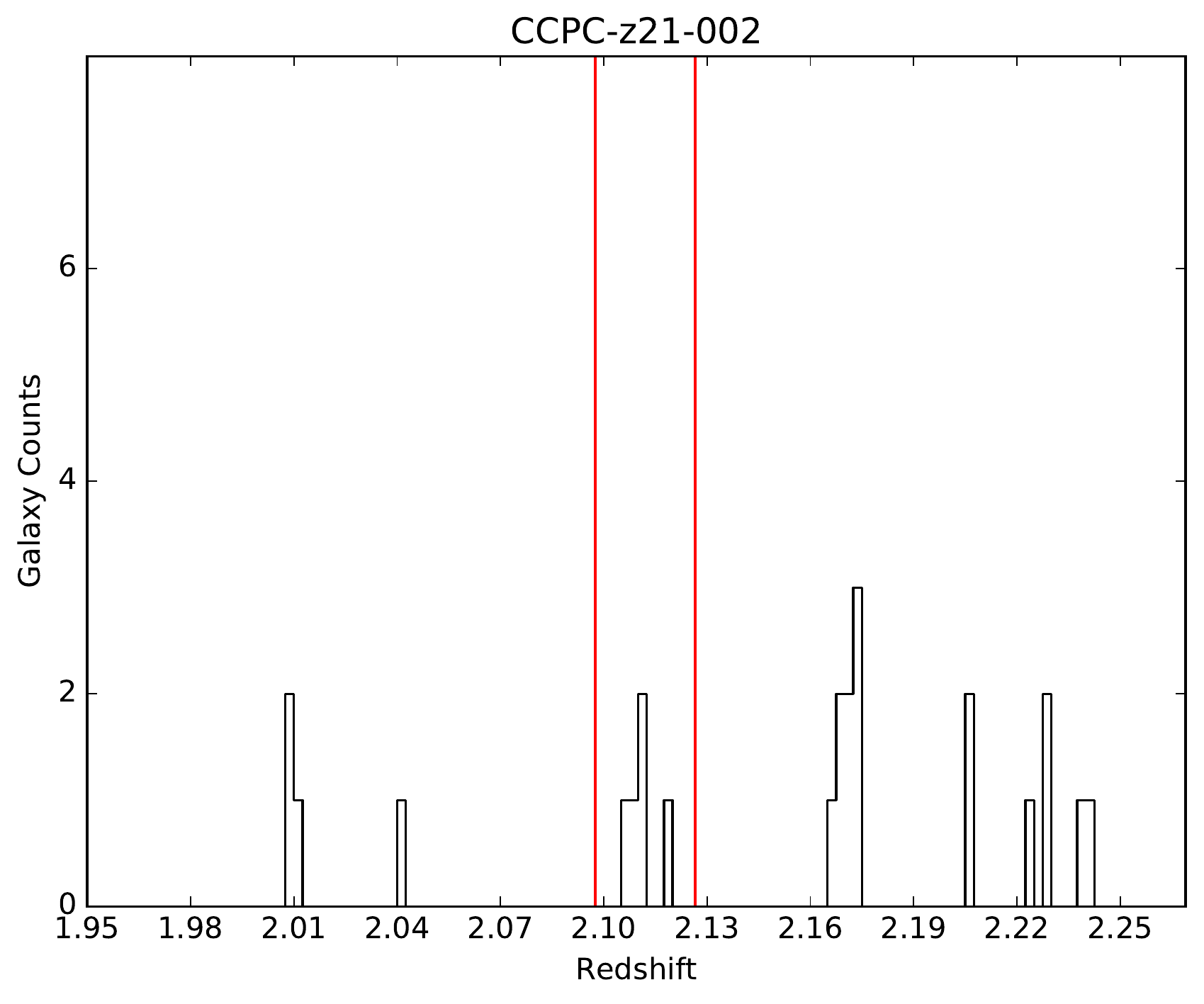}
\label{fig:CCPC-z21-002}
\end{subfigure}
\hfill
\end{figure*}

\begin{figure*}
\centering
\begin{subfigure}
\centering
\includegraphics[height=7.5cm,width=7.5cm]{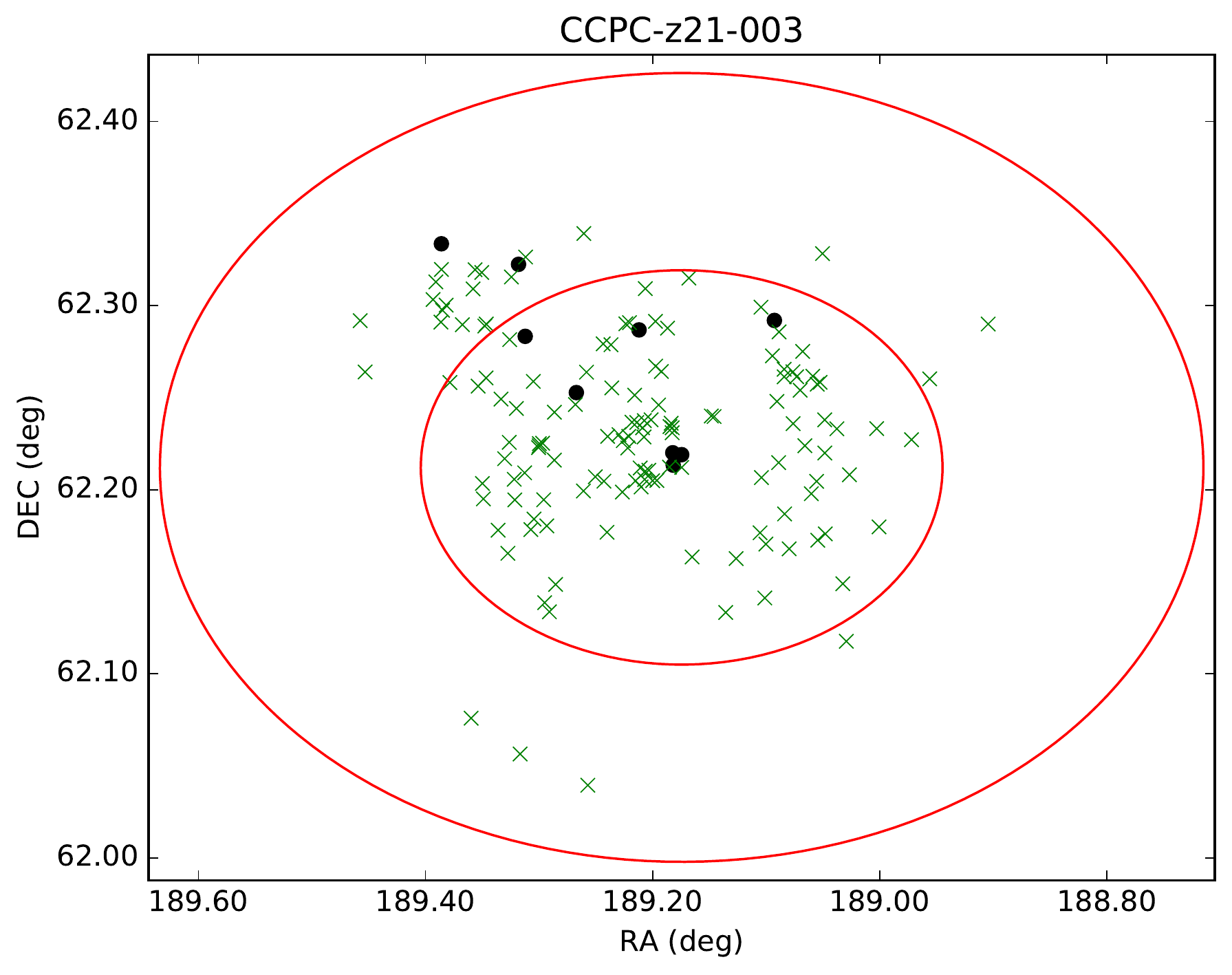}
\label{fig:CCPC-z21-003_sky}
\end{subfigure}
\hfill
\begin{subfigure}
\centering
\includegraphics[scale=0.52]{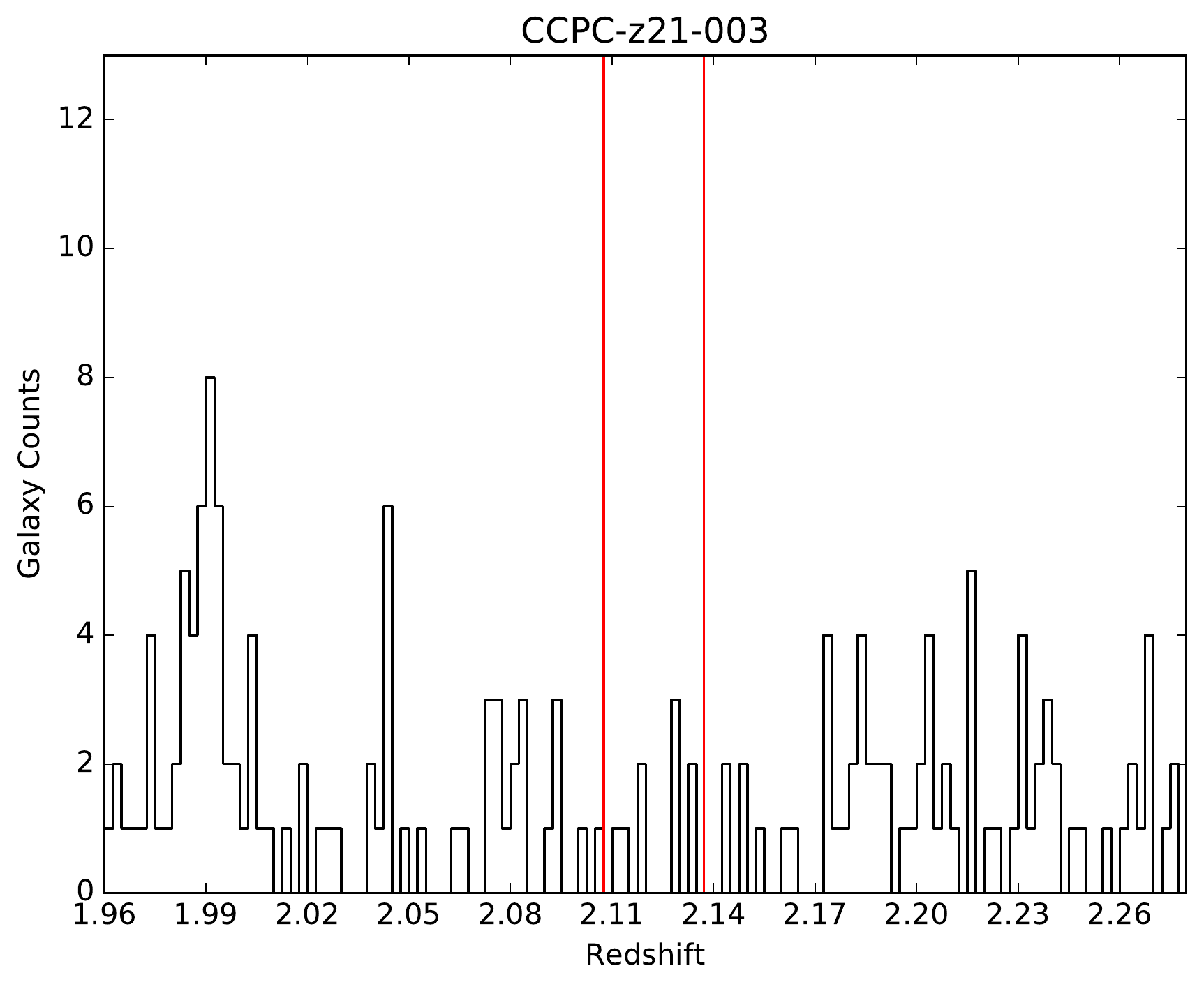}
\label{fig:CCPC-z21-003}
\end{subfigure}
\hfill
\end{figure*}
\clearpage 

\begin{figure*}
\centering
\begin{subfigure}
\centering
\includegraphics[height=7.5cm,width=7.5cm]{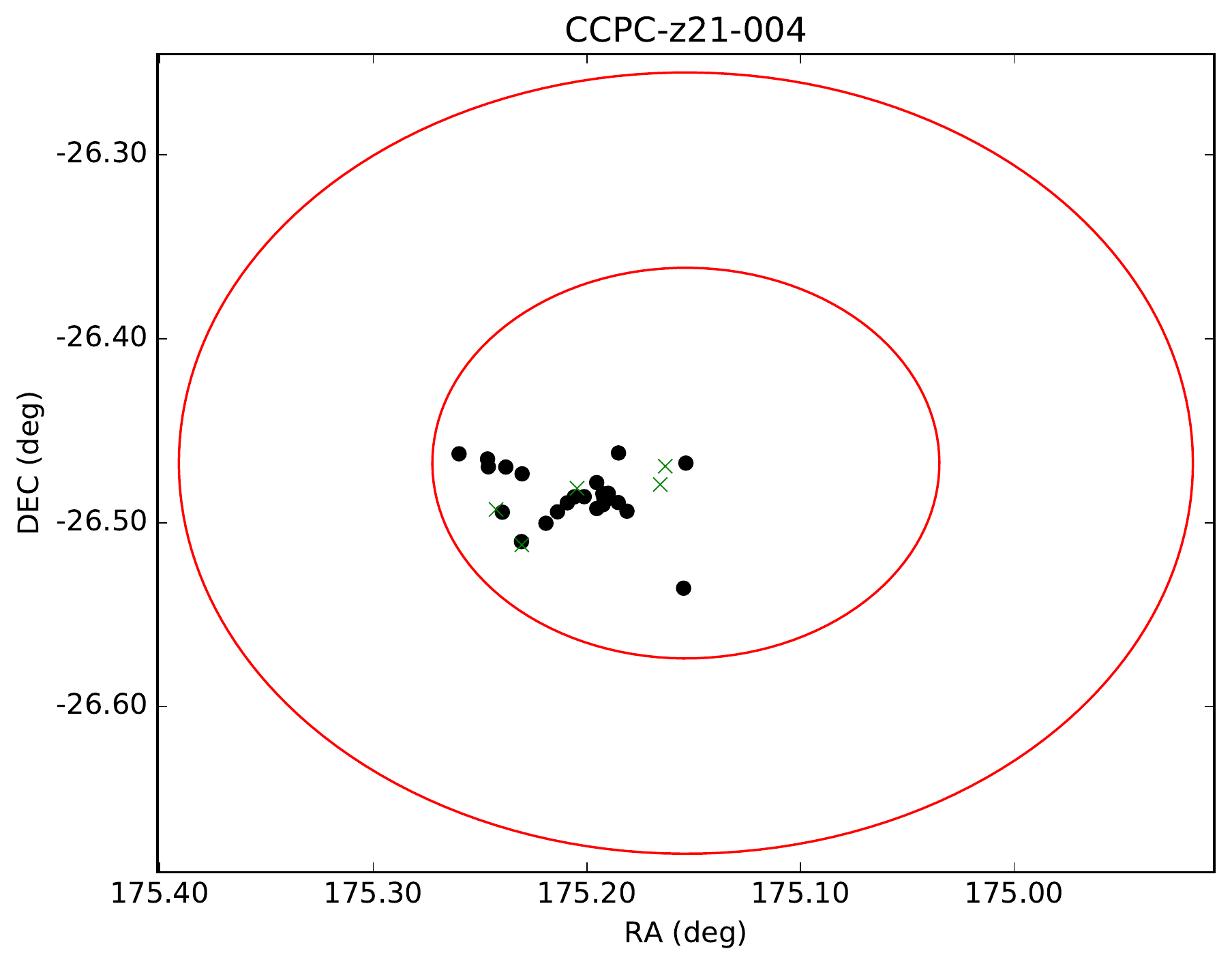}
\label{fig:CCPC-z21-004_sky}
\end{subfigure}
\hfill
\begin{subfigure}
\centering
\includegraphics[scale=0.52]{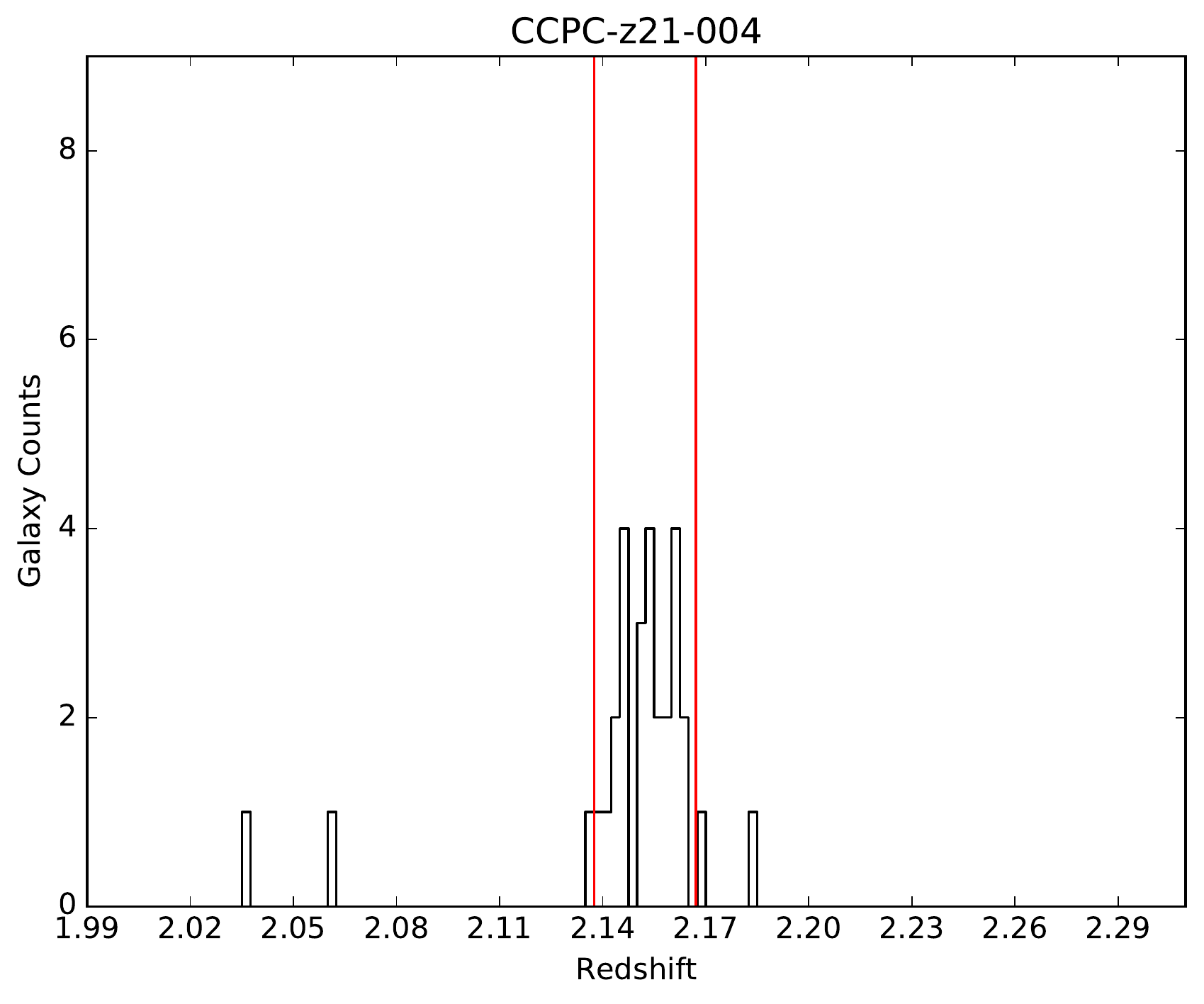}
\label{fig:CCPC-z21-004}
\end{subfigure}
\hfill
\end{figure*}

\begin{figure*}
\centering
\begin{subfigure}
\centering
\includegraphics[height=7.5cm,width=7.5cm]{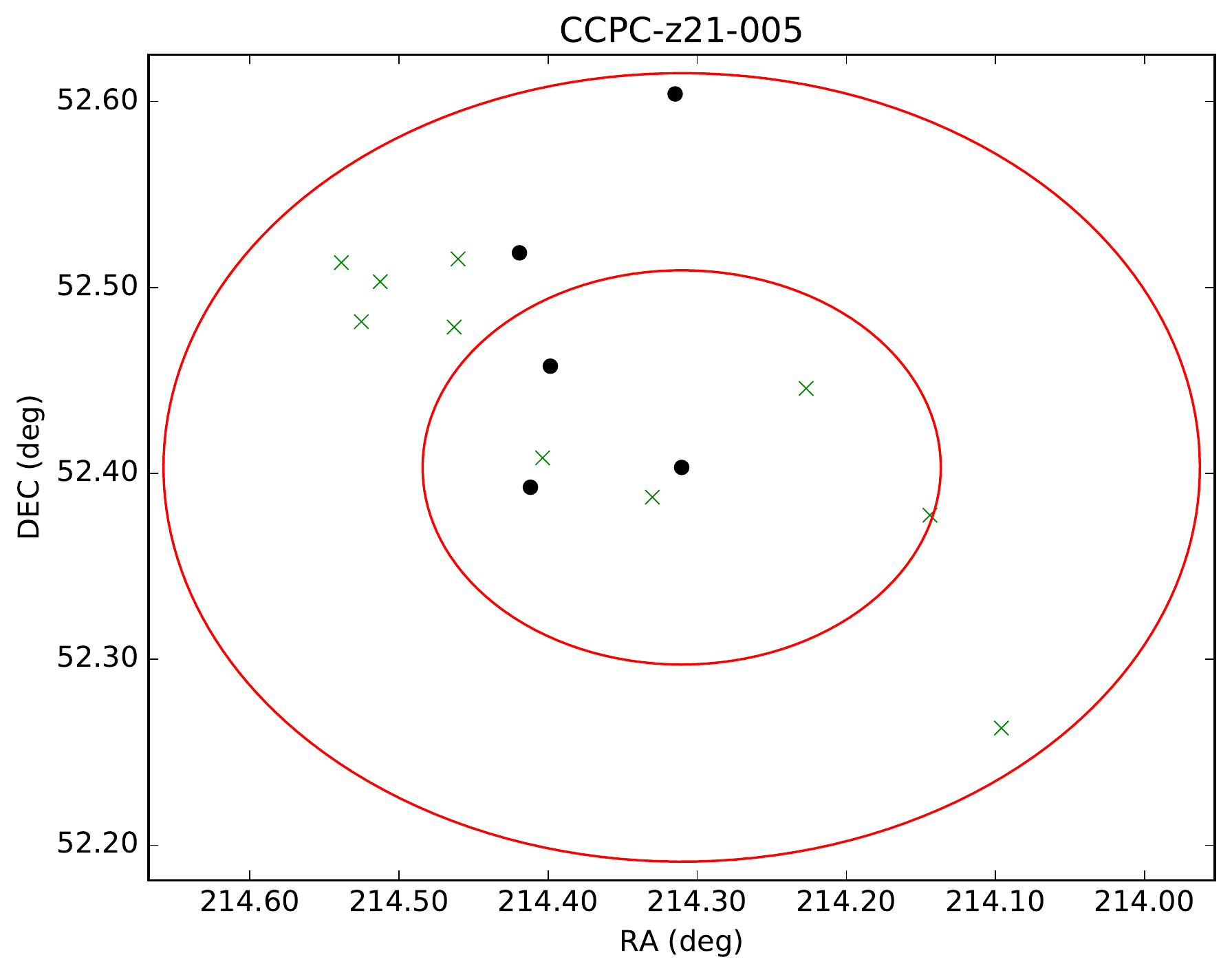}
\label{fig:CCPC-z21-005_sky}
\end{subfigure}
\hfill
\begin{subfigure}
\centering
\includegraphics[scale=0.52]{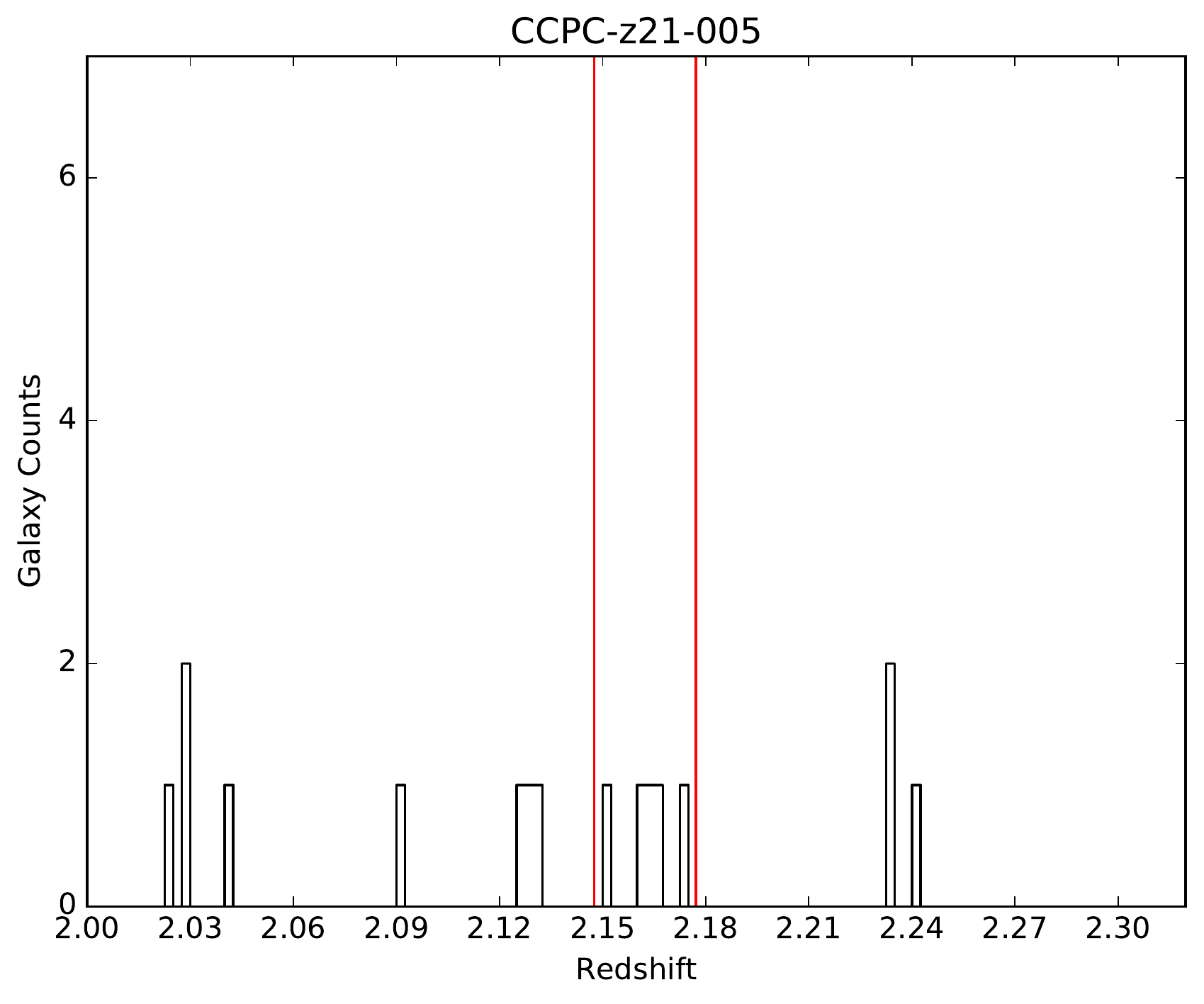}
\label{fig:CCPC-z21-005}
\end{subfigure}
\hfill
\end{figure*}

\begin{figure*}
\centering
\begin{subfigure}
\centering
\includegraphics[height=7.5cm,width=7.5cm]{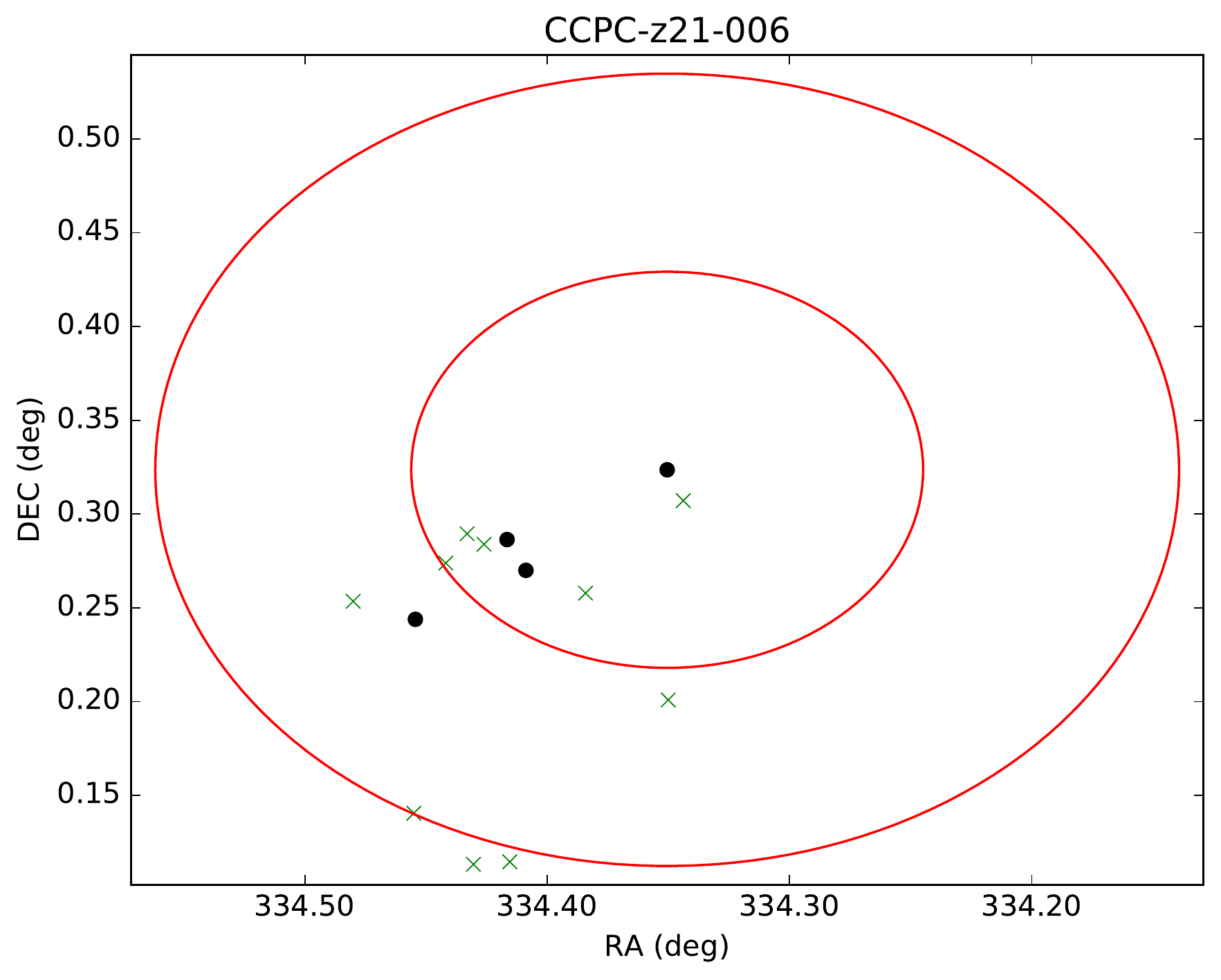}
\label{fig:CCPC-z21-006_sky}
\end{subfigure}
\hfill
\begin{subfigure}
\centering
\includegraphics[scale=0.52]{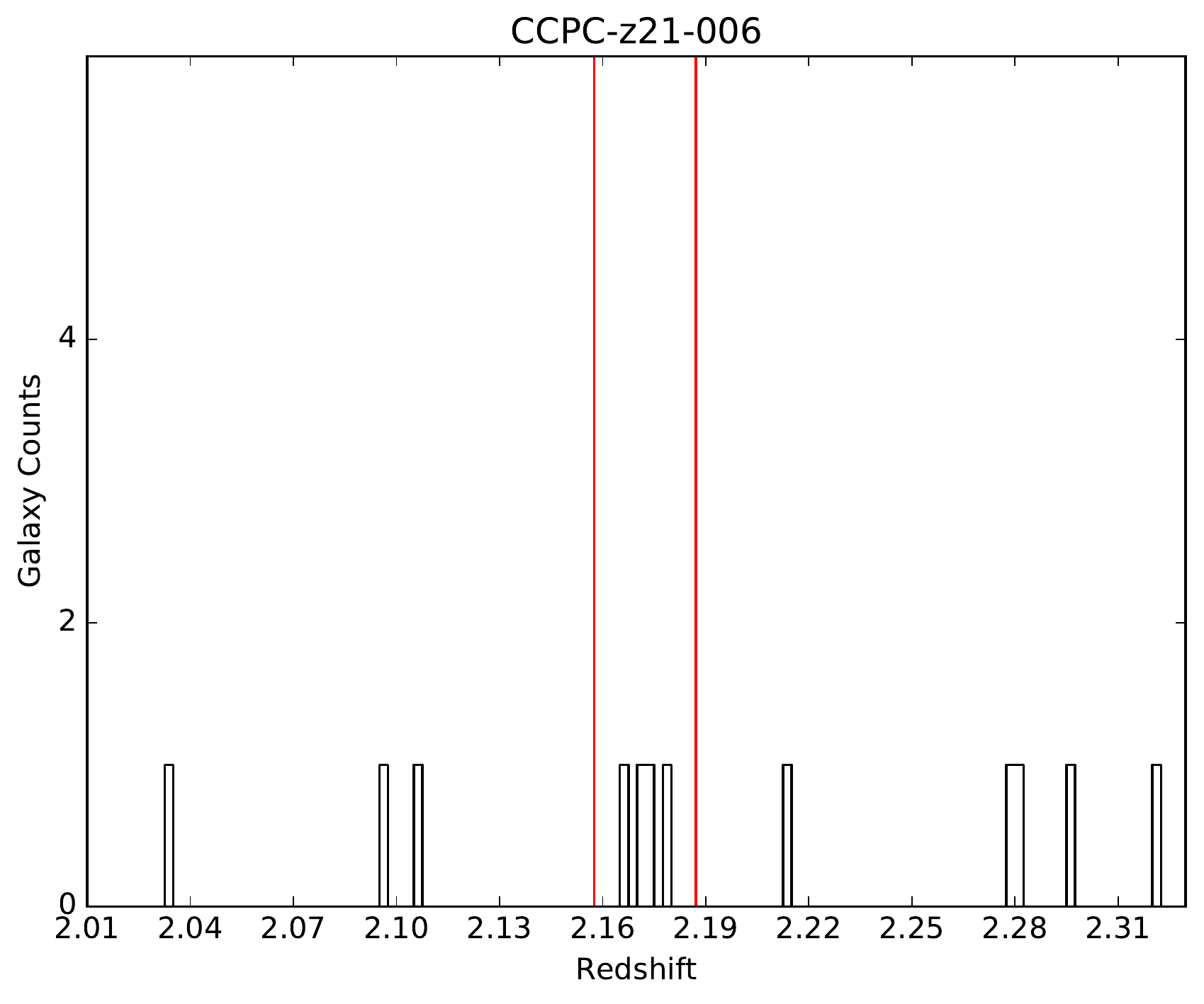}
\label{fig:CCPC-z21-006}
\end{subfigure}
\hfill
\end{figure*}
\clearpage 

\begin{figure*}
\centering
\begin{subfigure}
\centering
\includegraphics[height=7.5cm,width=7.5cm]{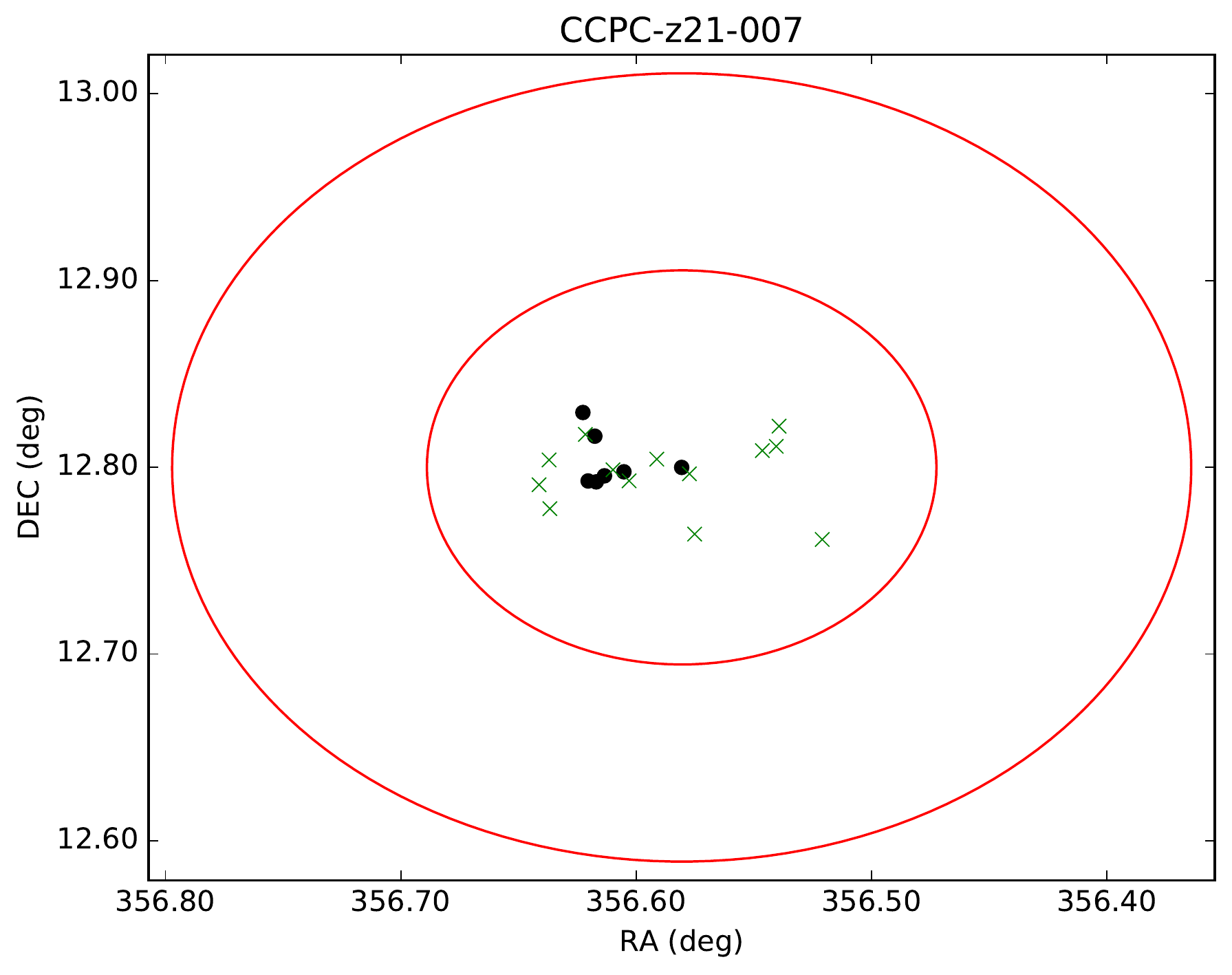}
\label{fig:CCPC-z21-007_sky}
\end{subfigure}
\hfill
\begin{subfigure}
\centering
\includegraphics[scale=0.52]{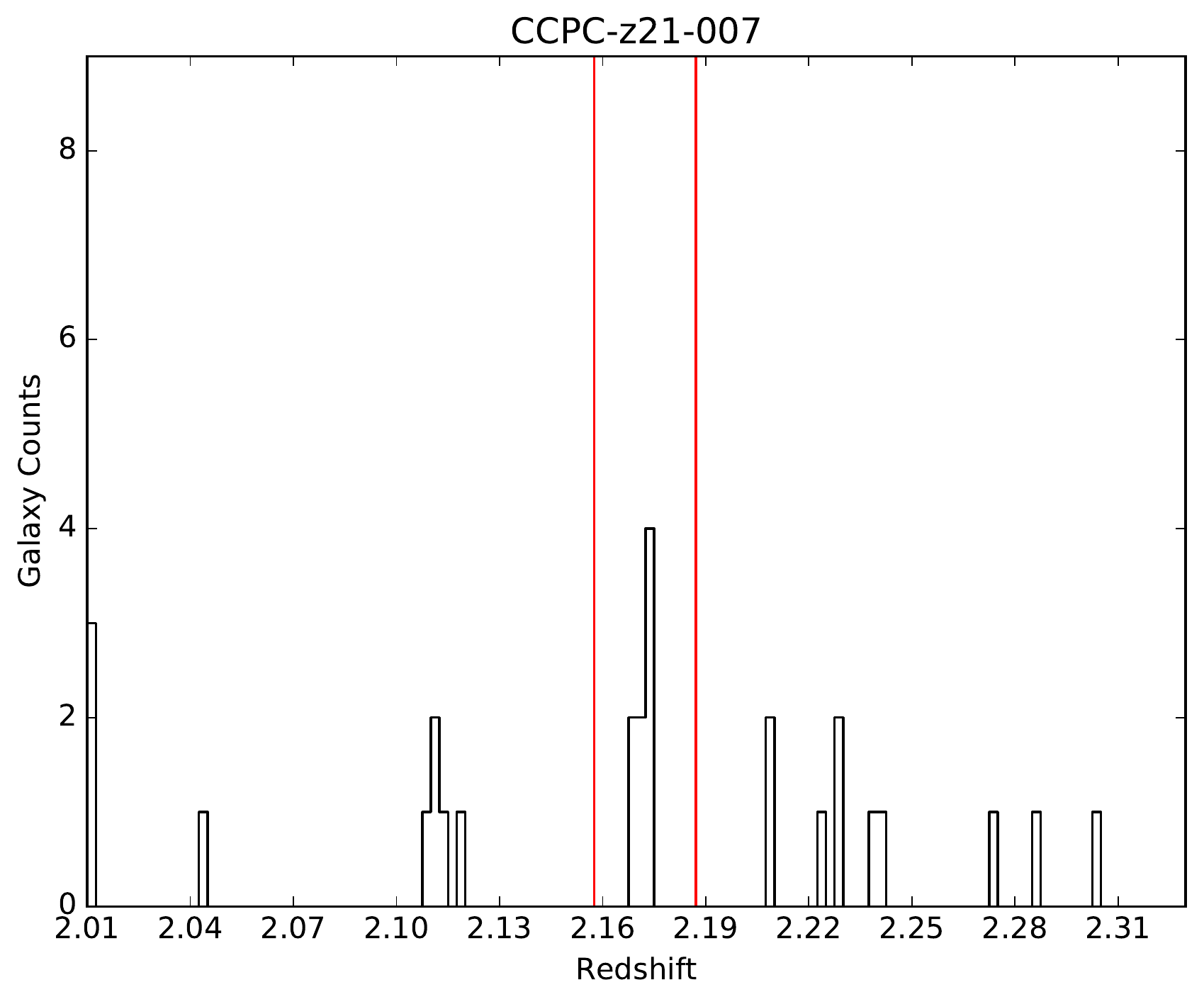}
\label{fig:CCPC-z21-007}
\end{subfigure}
\hfill
\end{figure*}

\begin{figure*}
\centering
\begin{subfigure}
\centering
\includegraphics[height=7.5cm,width=7.5cm]{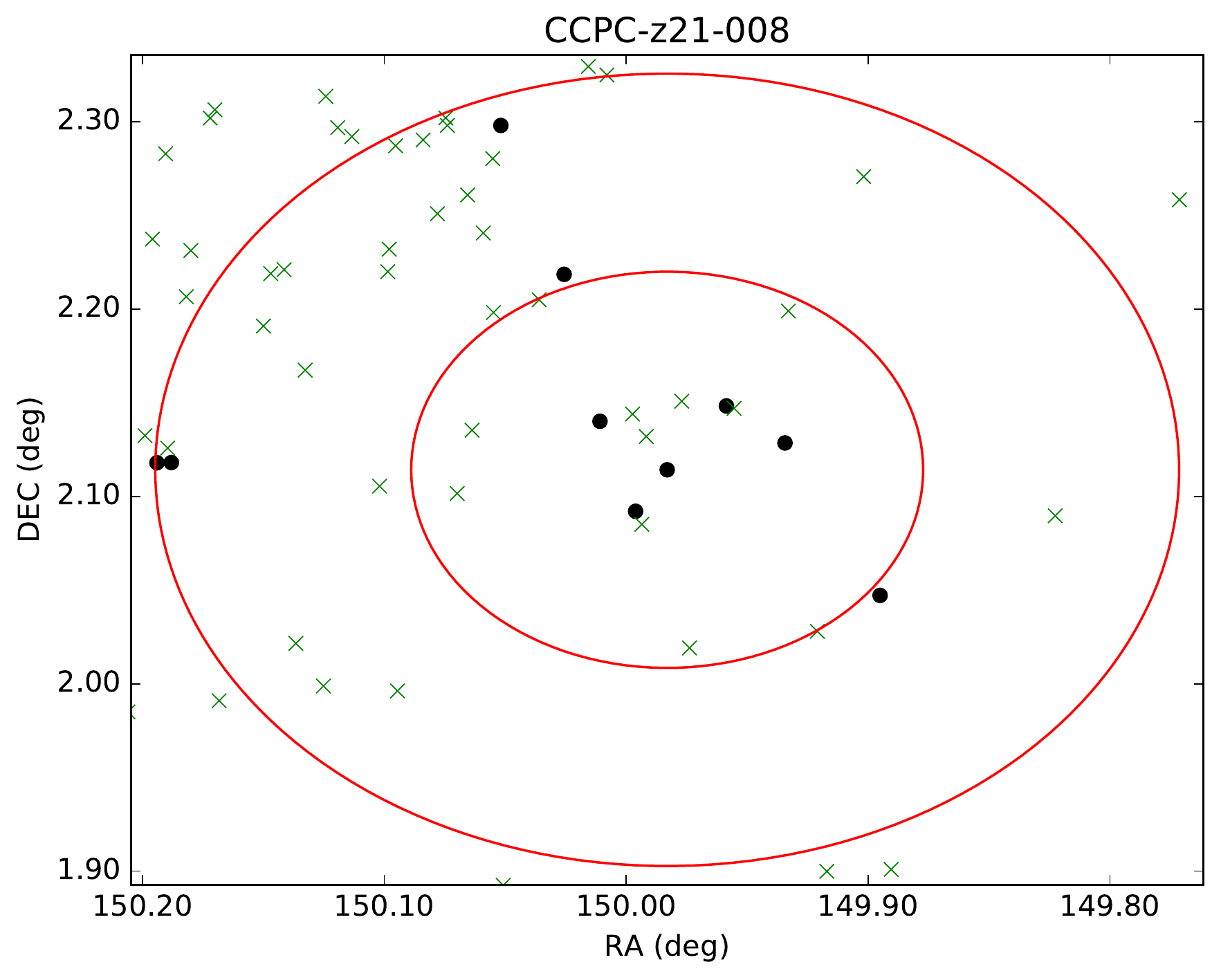}
\label{fig:CCPC-z21-008_sky}
\end{subfigure}
\hfill
\begin{subfigure}
\centering
\includegraphics[scale=0.52]{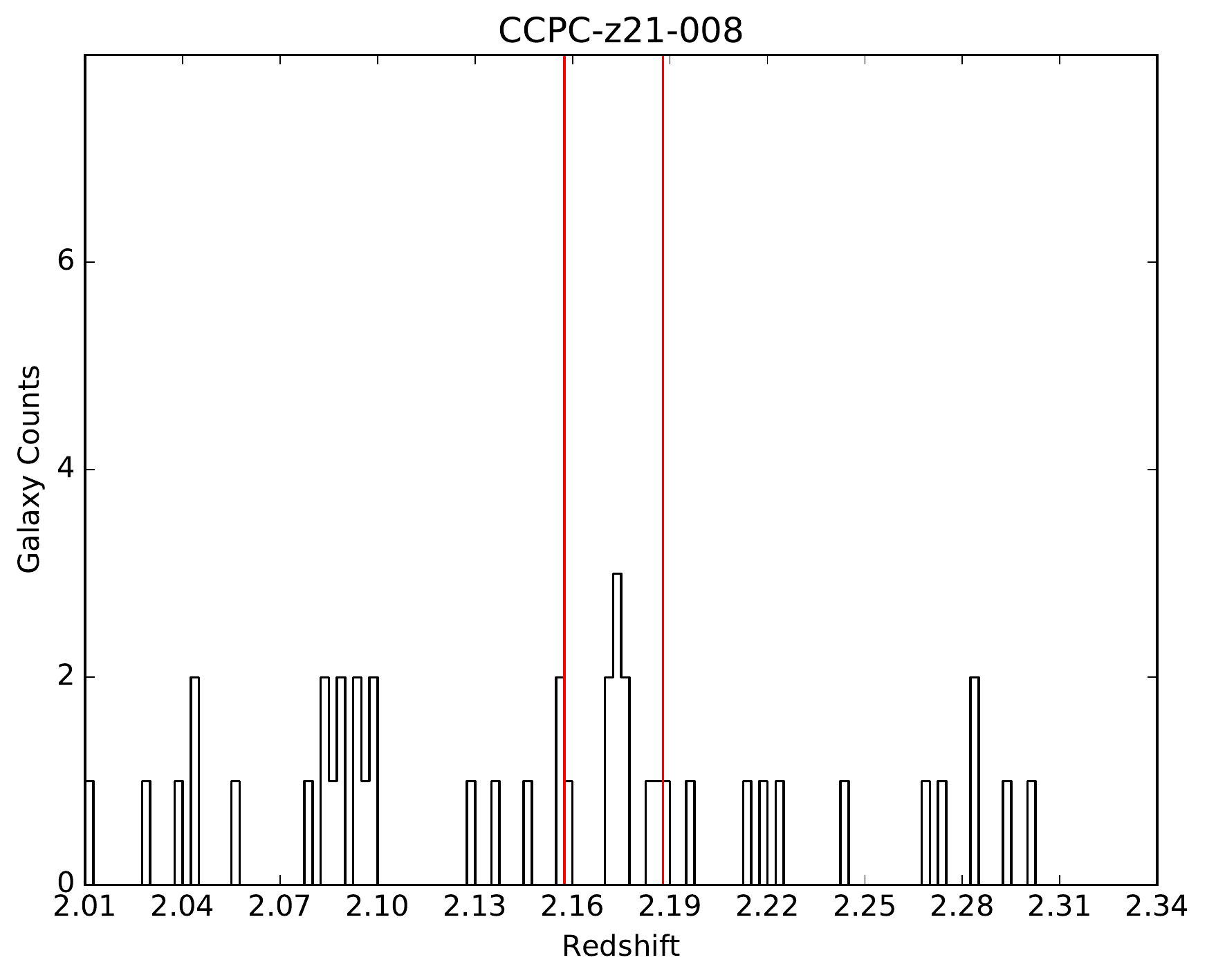}
\label{fig:CCPC-z21-008}
\end{subfigure}
\hfill
\end{figure*}

\begin{figure*}
\centering
\begin{subfigure}
\centering
\includegraphics[height=7.5cm,width=7.5cm]{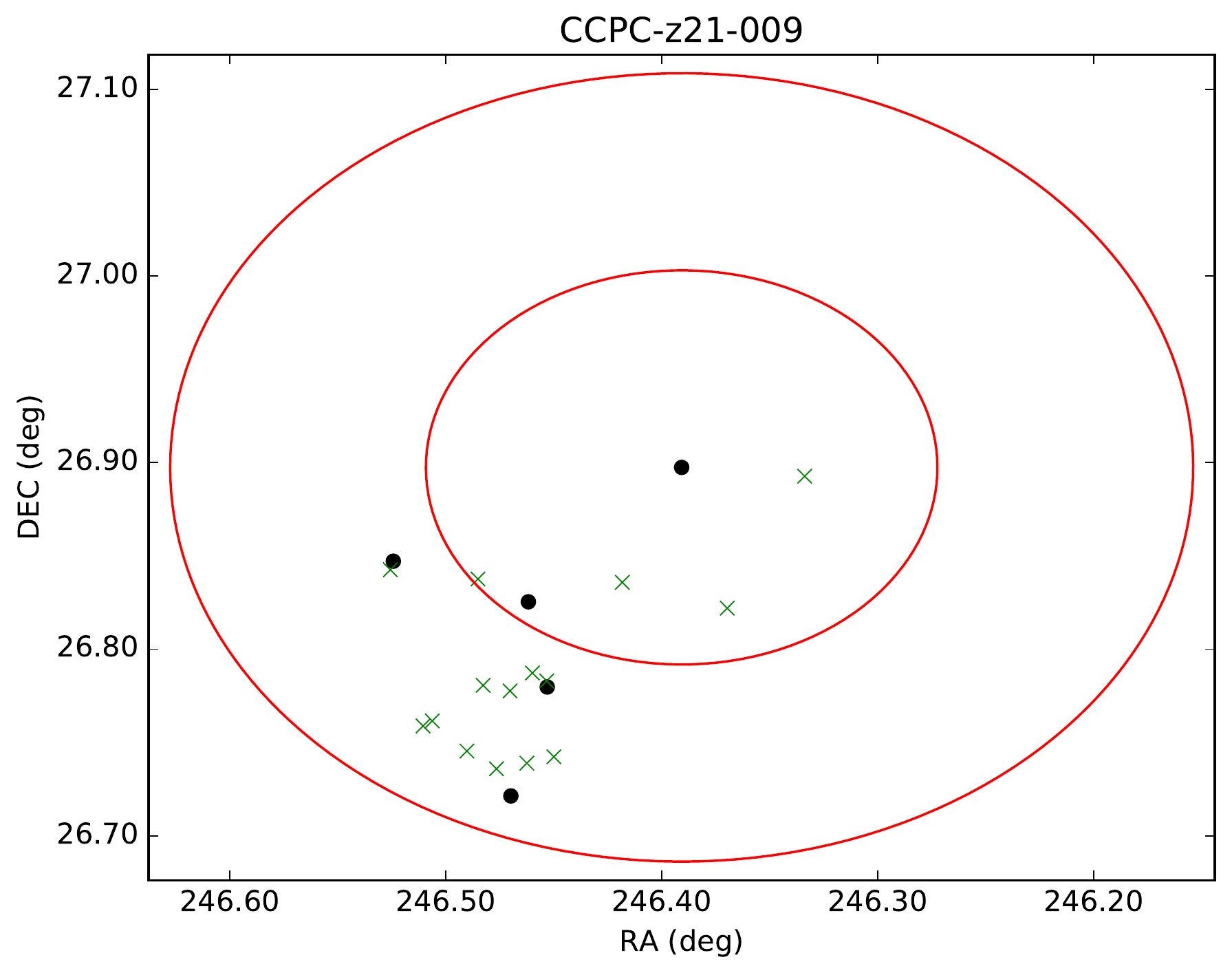}
\label{fig:CCPC-z21-009_sky}
\end{subfigure}
\hfill
\begin{subfigure}
\centering
\includegraphics[scale=0.52]{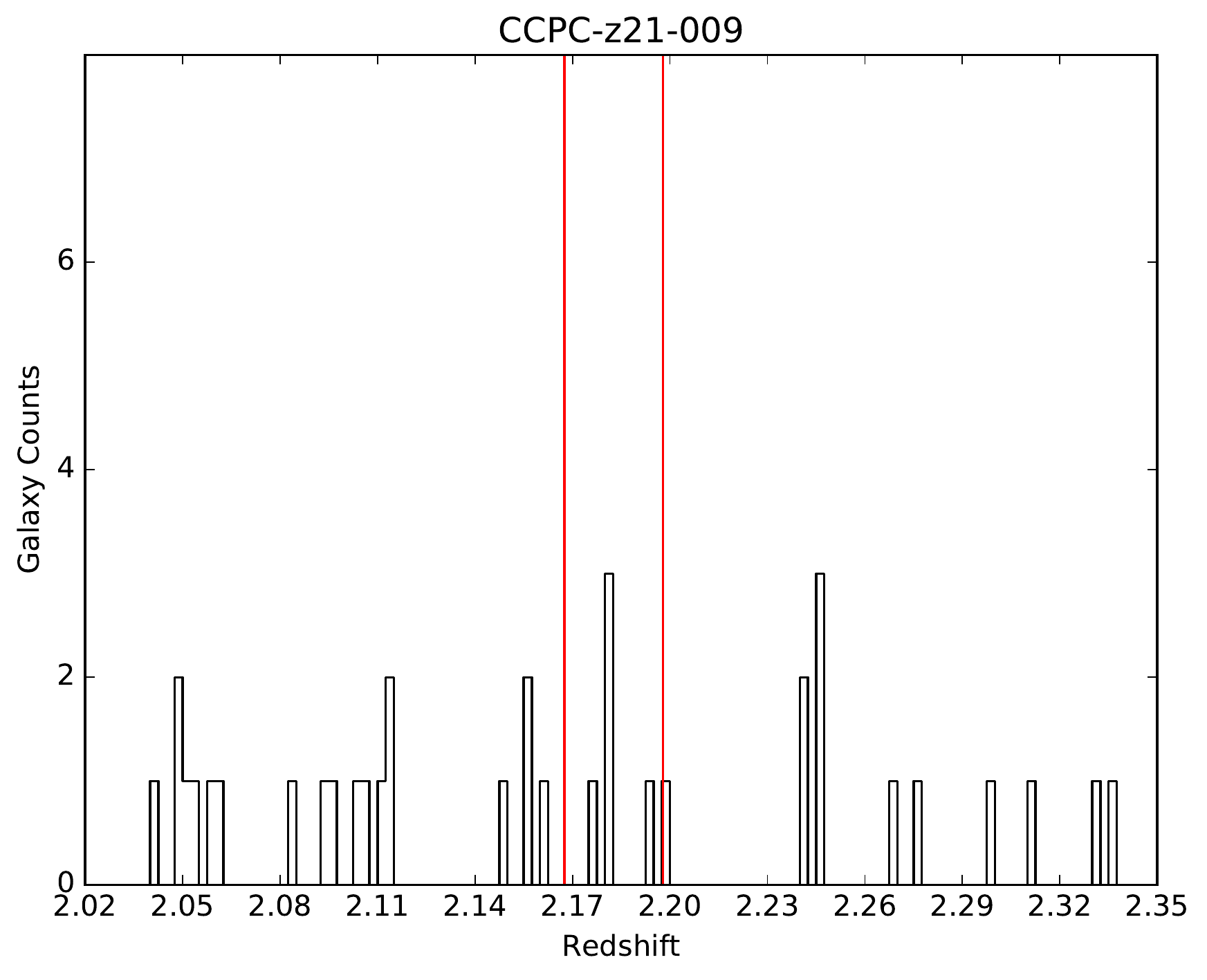}
\label{fig:CCPC-z21-009}
\end{subfigure}
\hfill
\end{figure*}
\clearpage 

\begin{figure*}
\centering
\begin{subfigure}
\centering
\includegraphics[height=7.5cm,width=7.5cm]{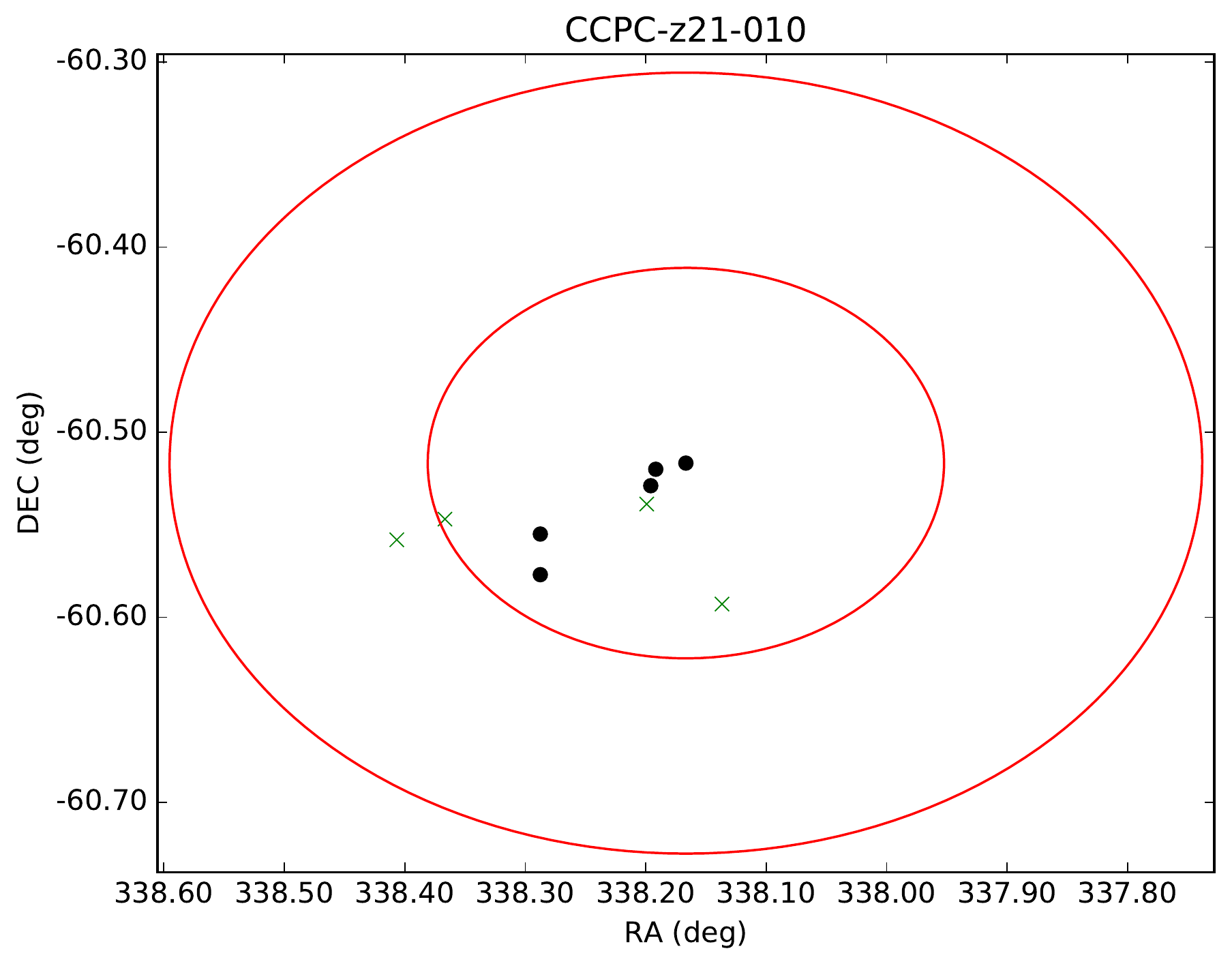}
\label{fig:CCPC-z21-010_sky}
\end{subfigure}
\hfill
\begin{subfigure}
\centering
\includegraphics[scale=0.52]{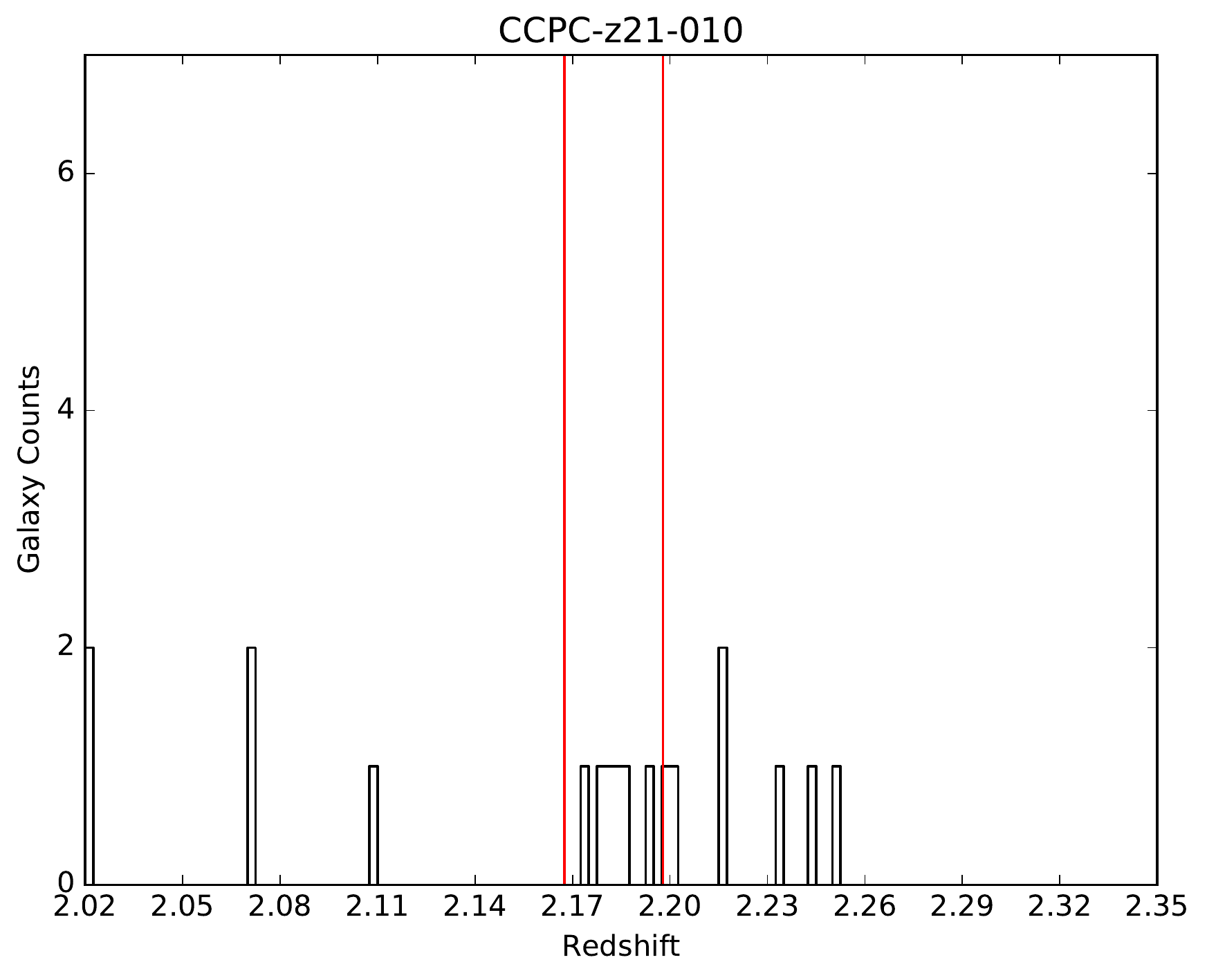}
\label{fig:CCPC-z21-010}
\end{subfigure}
\hfill
\end{figure*}

\begin{figure*}
\centering
\begin{subfigure}
\centering
\includegraphics[height=7.5cm,width=7.5cm]{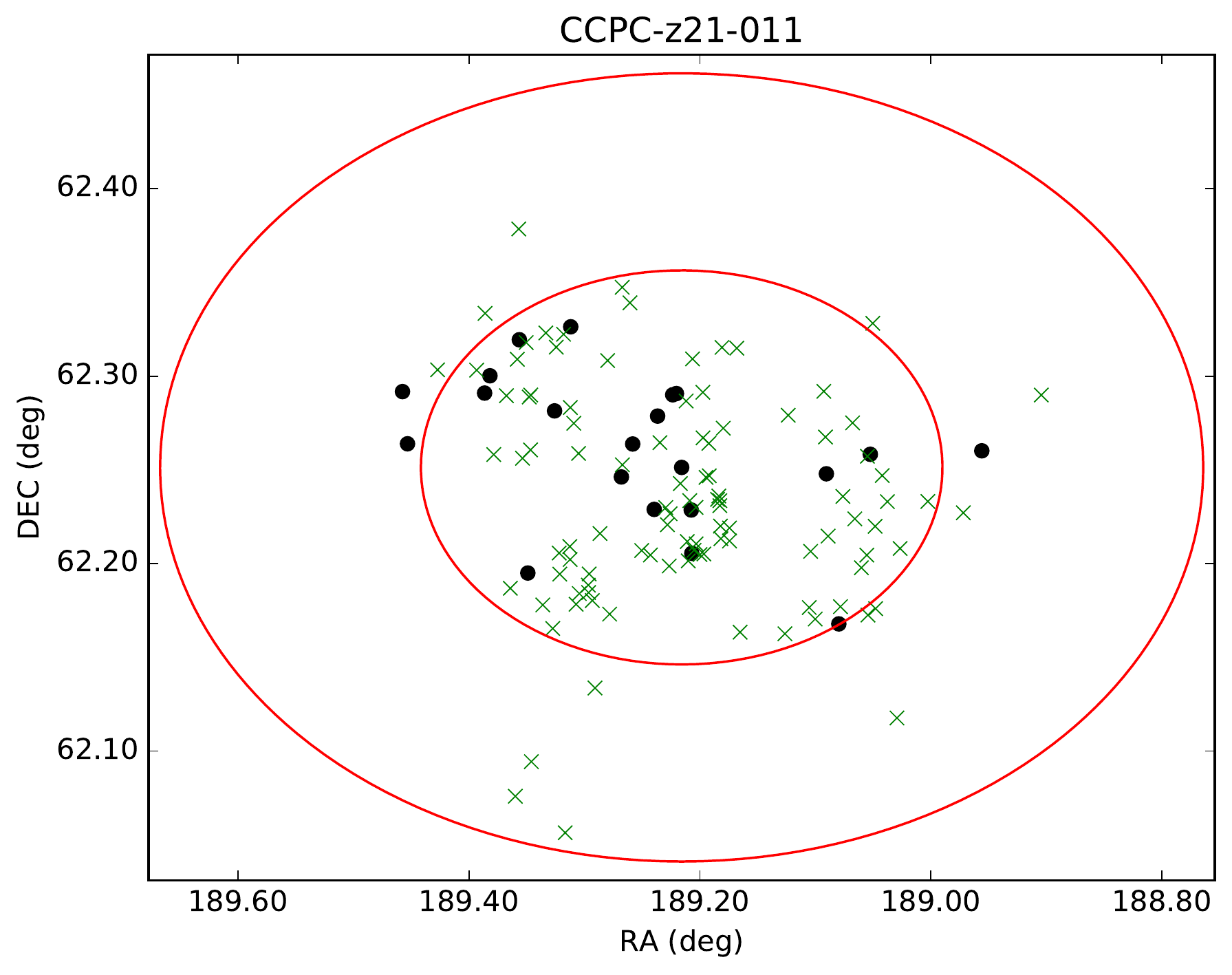}
\label{fig:CCPC-z21-011_sky}
\end{subfigure}
\hfill
\begin{subfigure}
\centering
\includegraphics[scale=0.52]{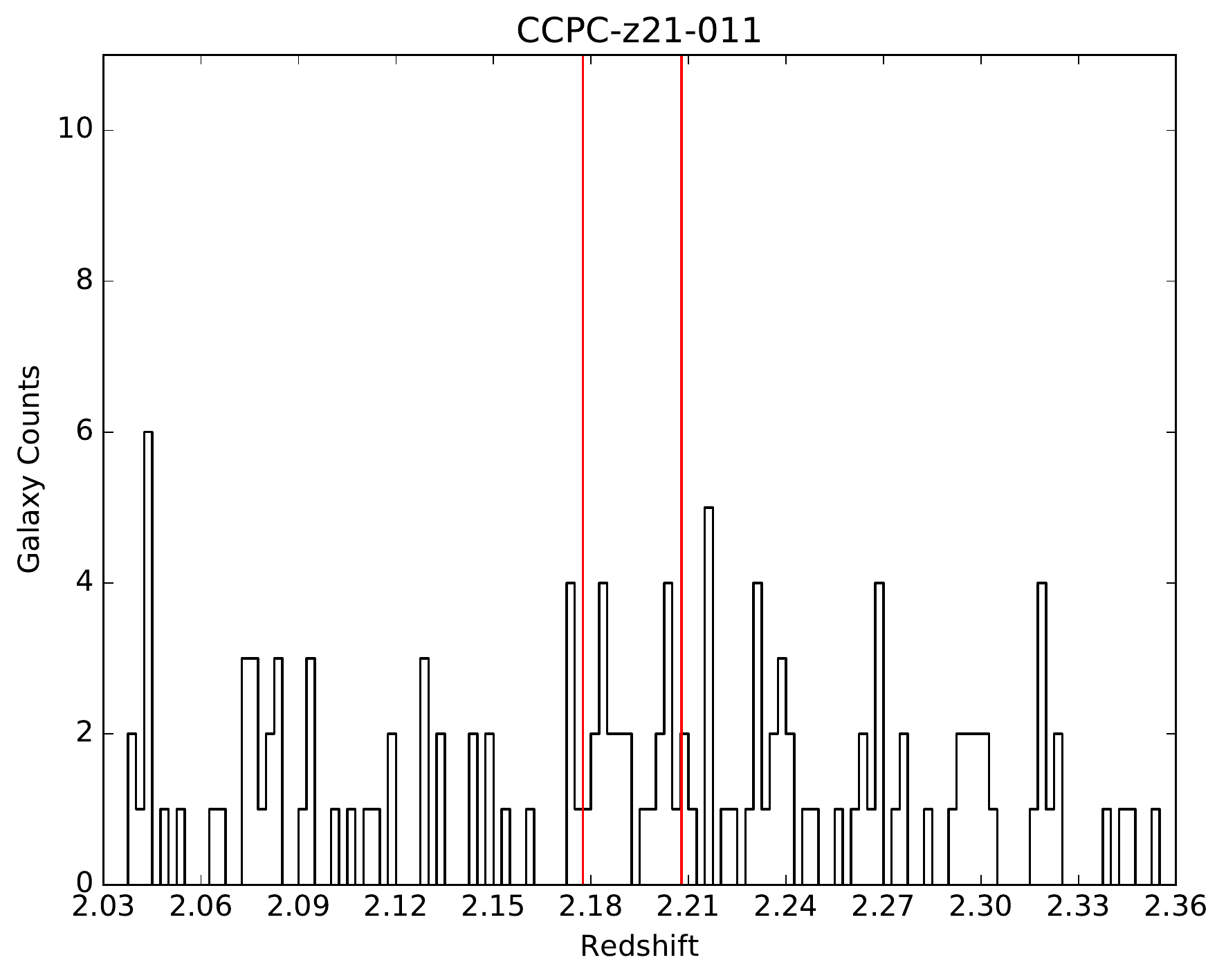}
\label{fig:CCPC-z21-011}
\end{subfigure}
\hfill
\end{figure*}

\begin{figure*}
\centering
\begin{subfigure}
\centering
\includegraphics[height=7.5cm,width=7.5cm]{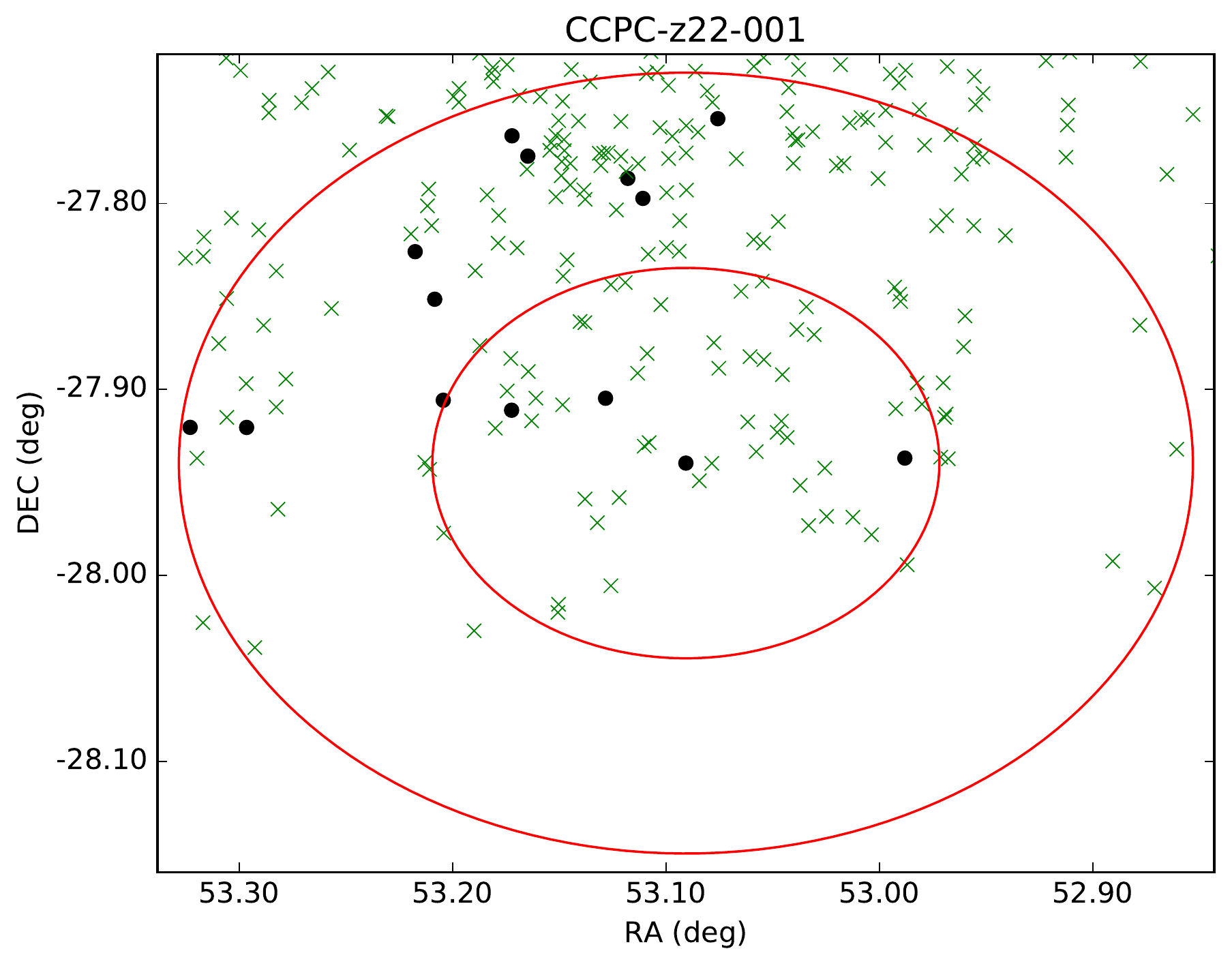}
\label{fig:CCPC-z22-001_sky}
\end{subfigure}
\hfill
\begin{subfigure}
\centering
\includegraphics[scale=0.52]{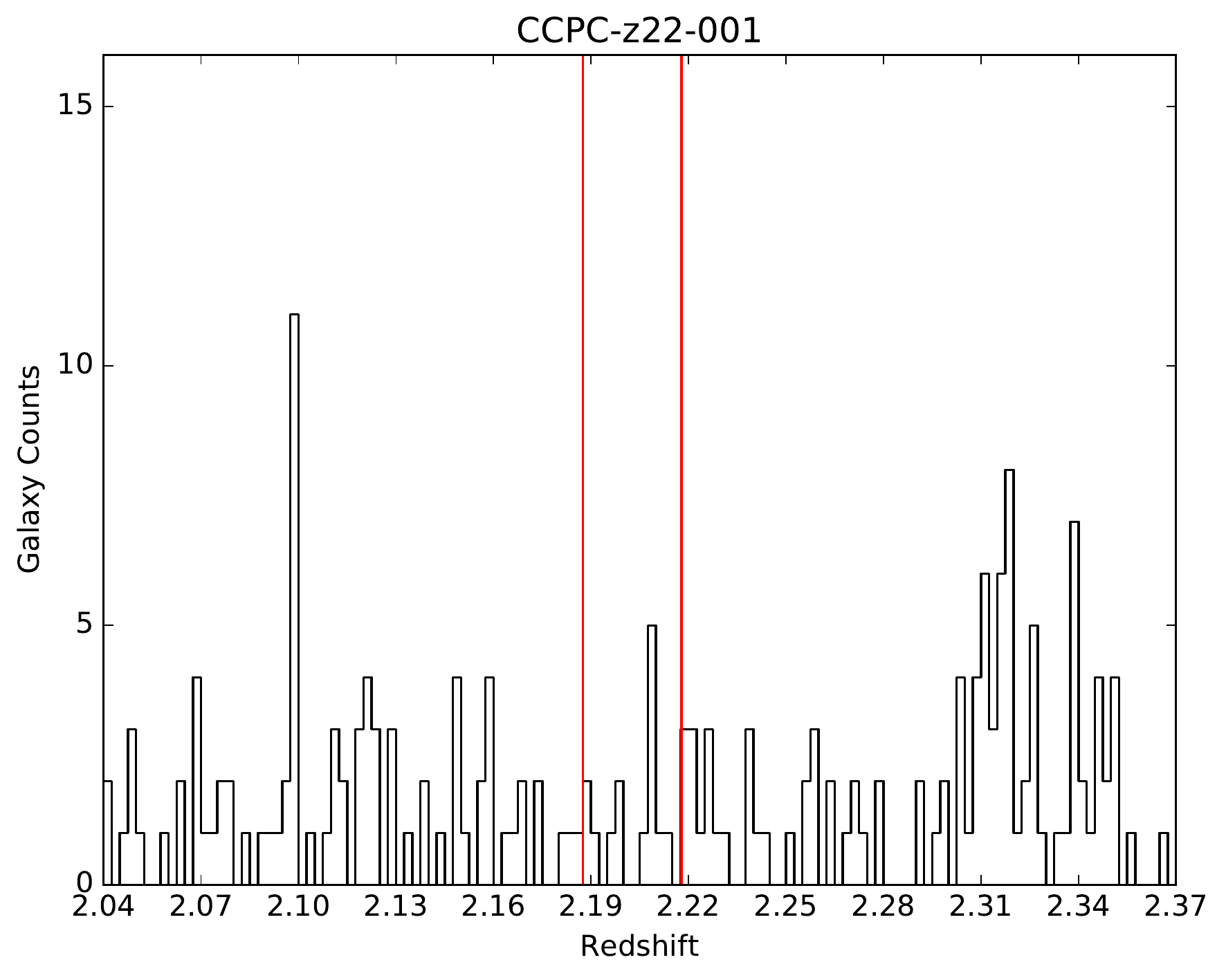}
\label{fig:CCPC-z22-001}
\end{subfigure}
\hfill
\end{figure*}
\clearpage 

\begin{figure*}
\centering
\begin{subfigure}
\centering
\includegraphics[height=7.5cm,width=7.5cm]{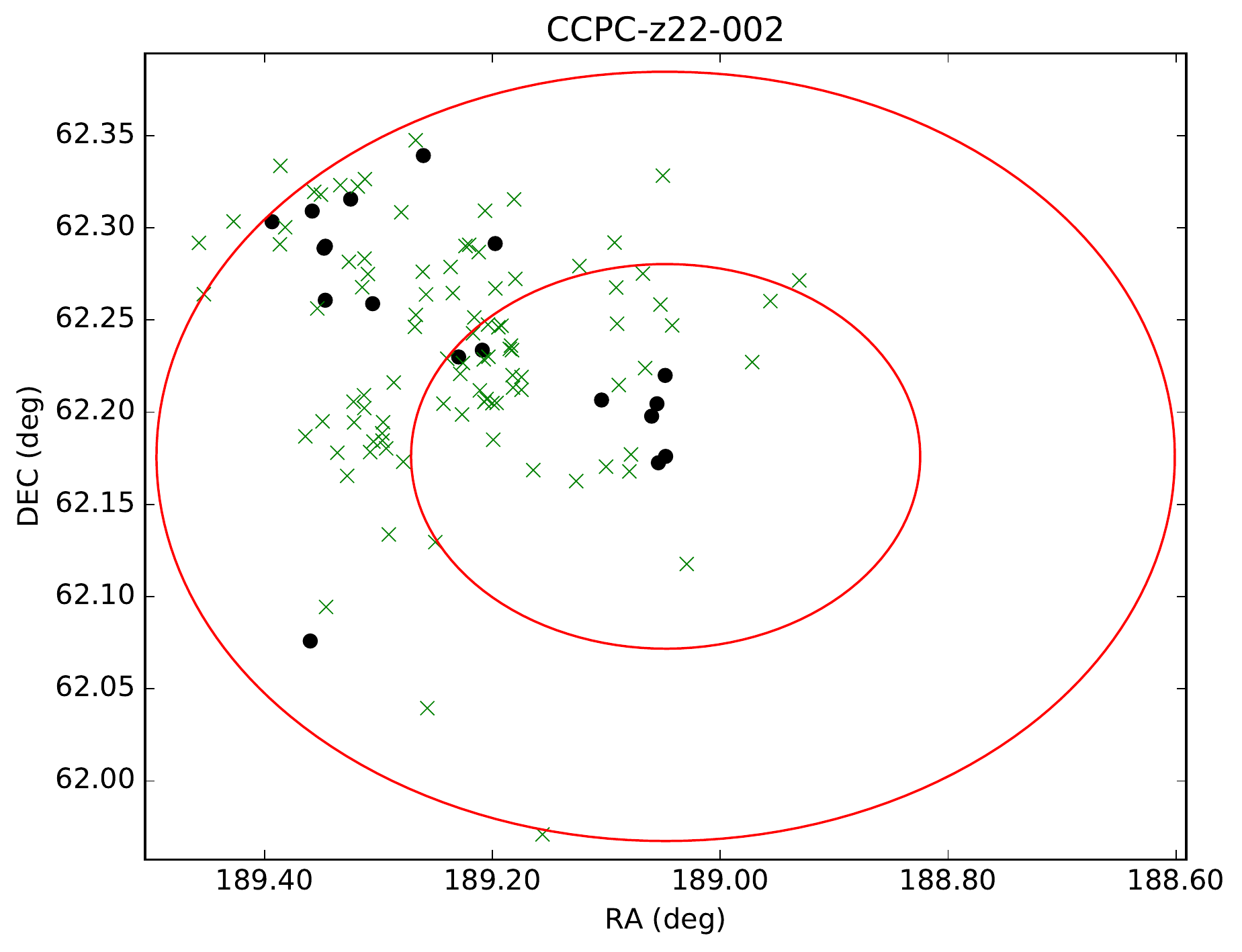}
\label{fig:CCPC-z22-002_sky}
\end{subfigure}
\hfill
\begin{subfigure}
\centering
\includegraphics[scale=0.52]{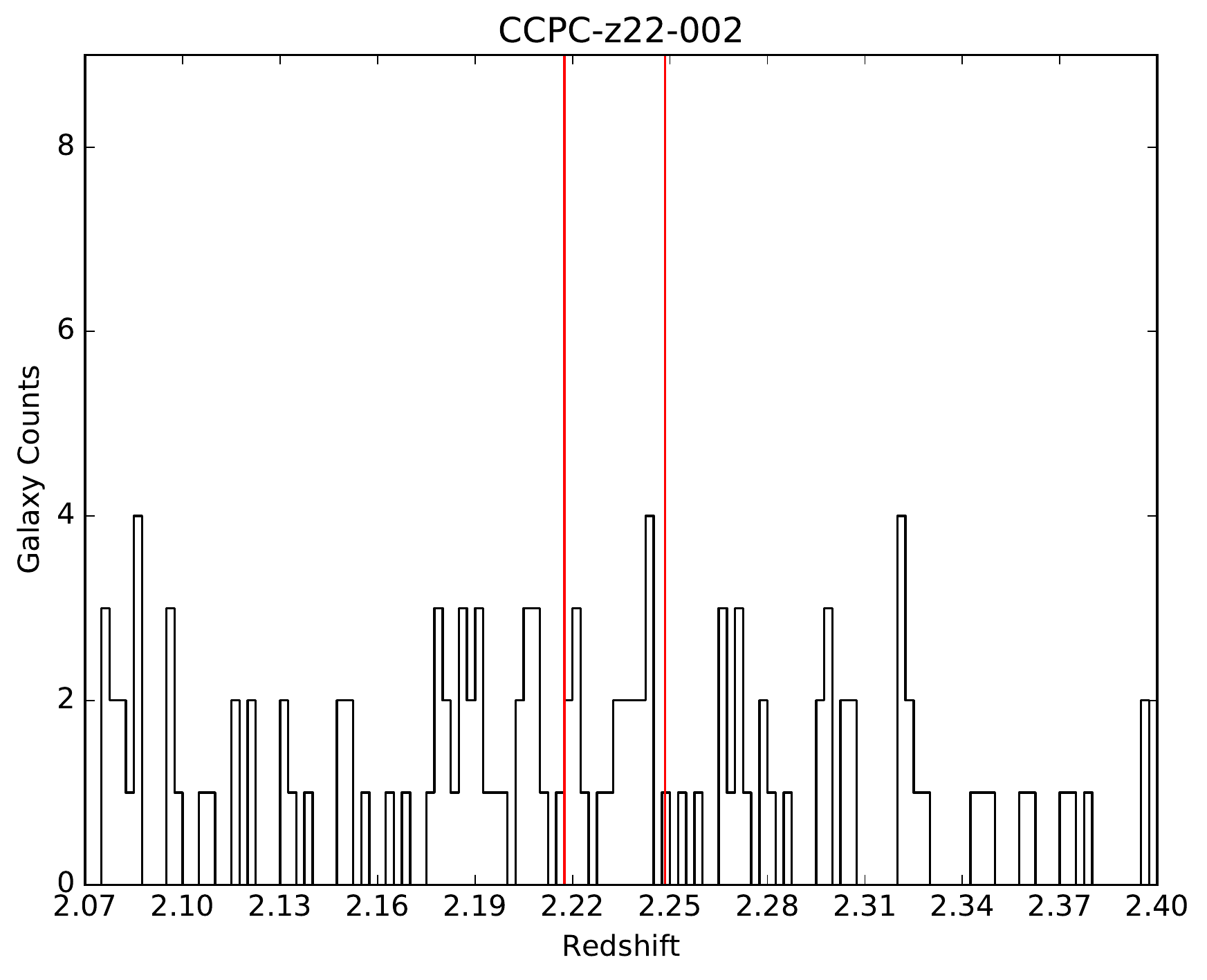}
\label{fig:CCPC-z22-002}
\end{subfigure}
\hfill
\end{figure*}

\begin{figure*}
\centering
\begin{subfigure}
\centering
\includegraphics[height=7.5cm,width=7.5cm]{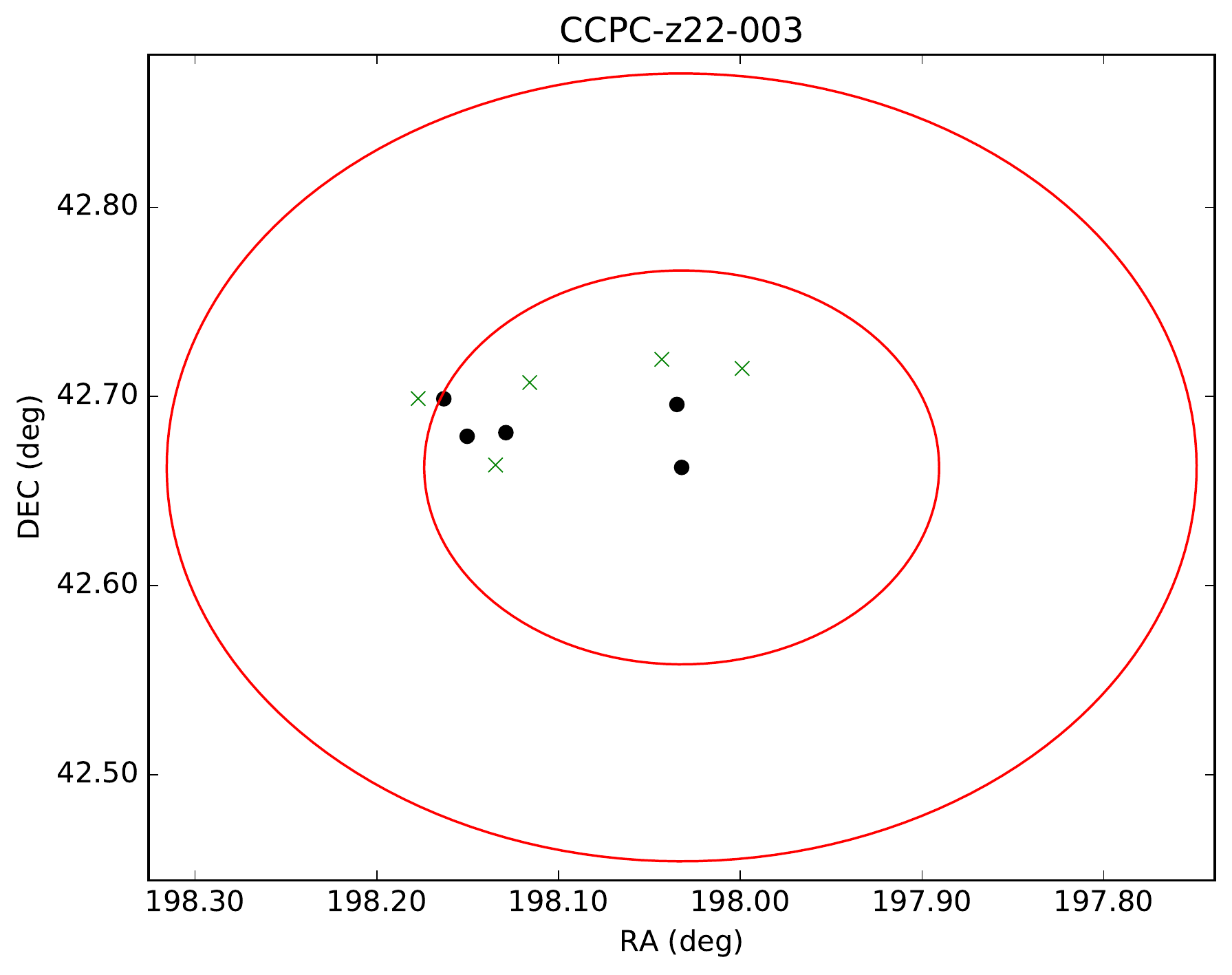}
\label{fig:CCPC-z22-003_sky}
\end{subfigure}
\hfill
\begin{subfigure}
\centering
\includegraphics[scale=0.52]{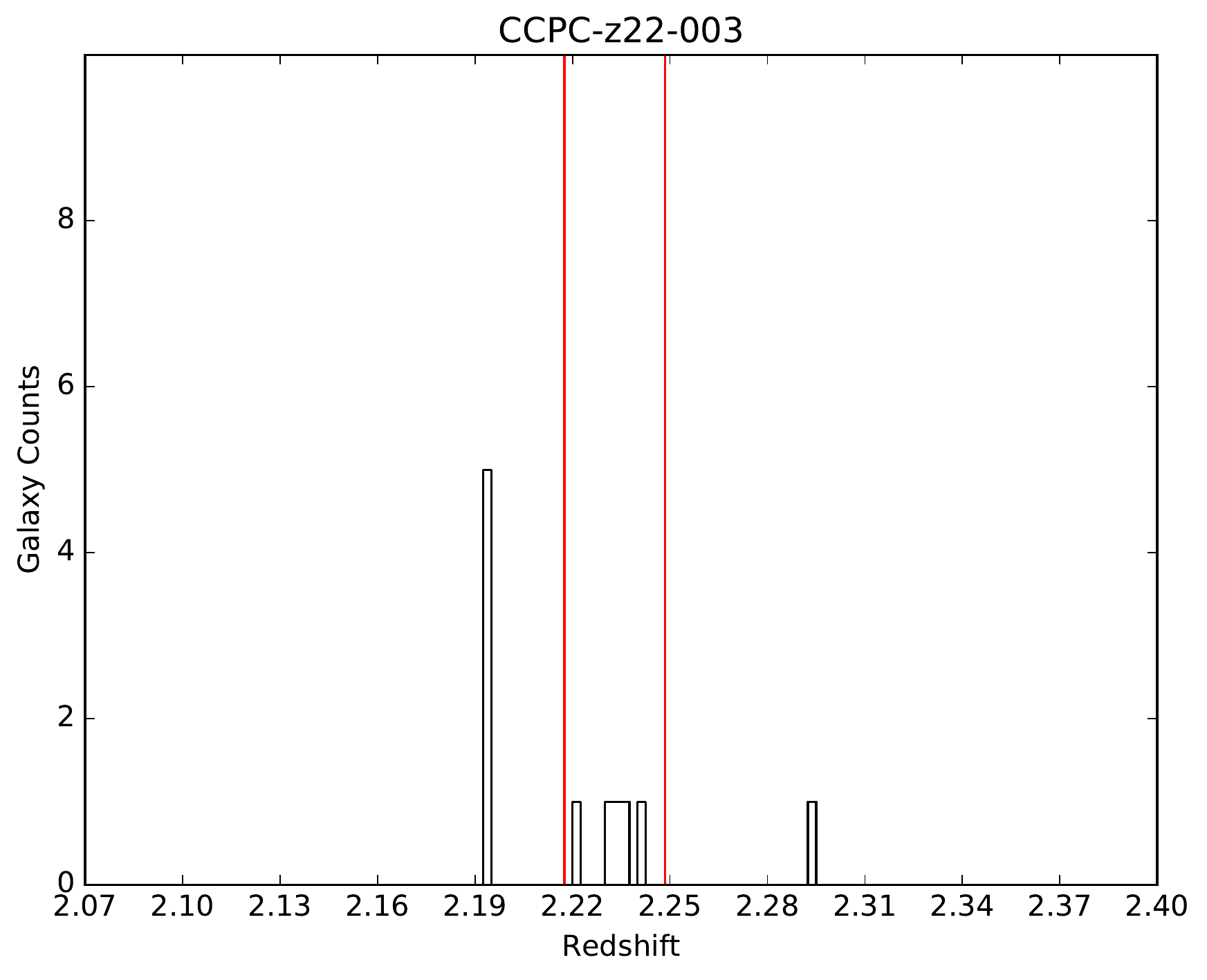}
\label{fig:CCPC-z22-003}
\end{subfigure}
\hfill
\end{figure*}

\begin{figure*}
\centering
\begin{subfigure}
\centering
\includegraphics[height=7.5cm,width=7.5cm]{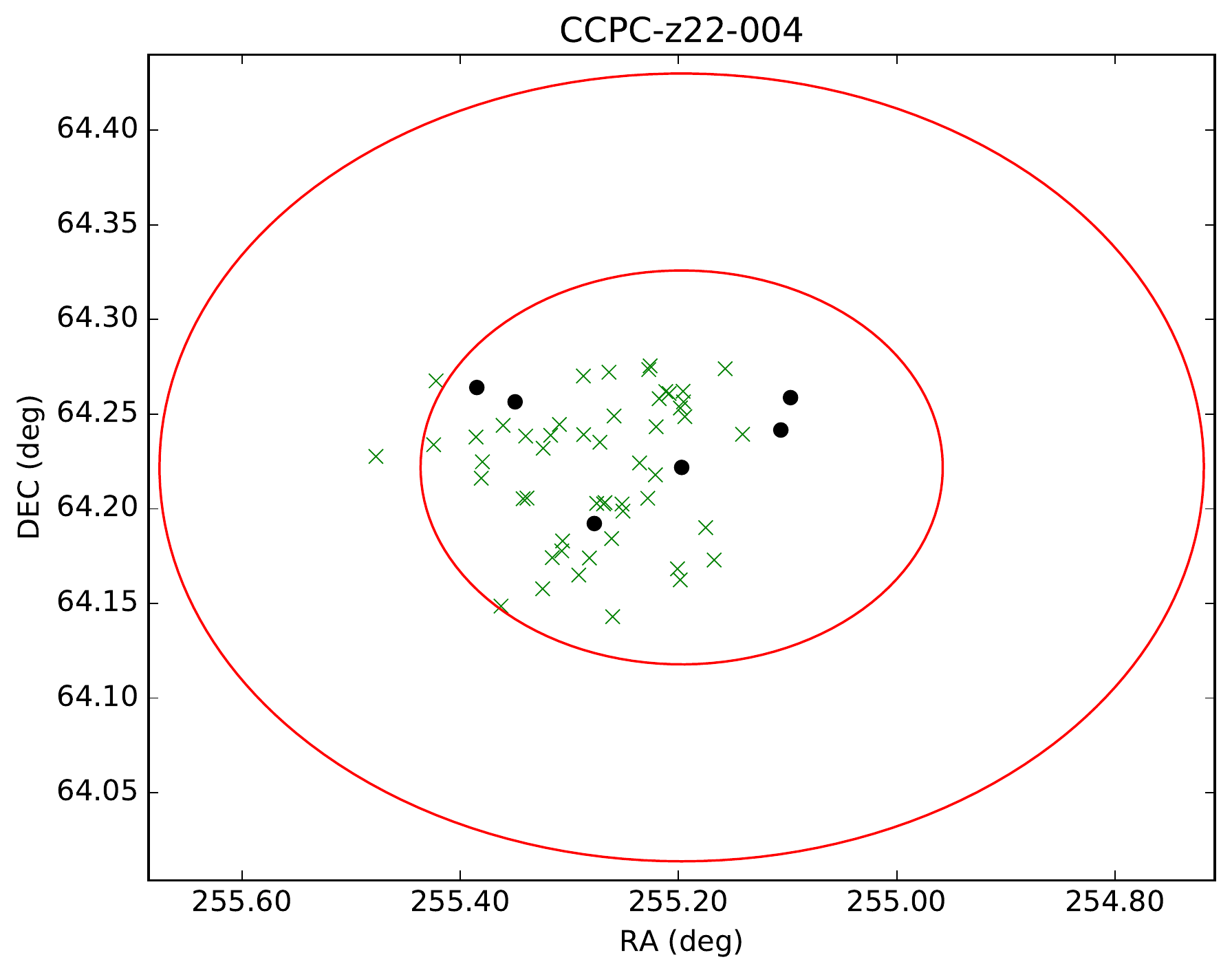}
\label{fig:CCPC-z22-004_sky}
\end{subfigure}
\hfill
\begin{subfigure}
\centering
\includegraphics[scale=0.52]{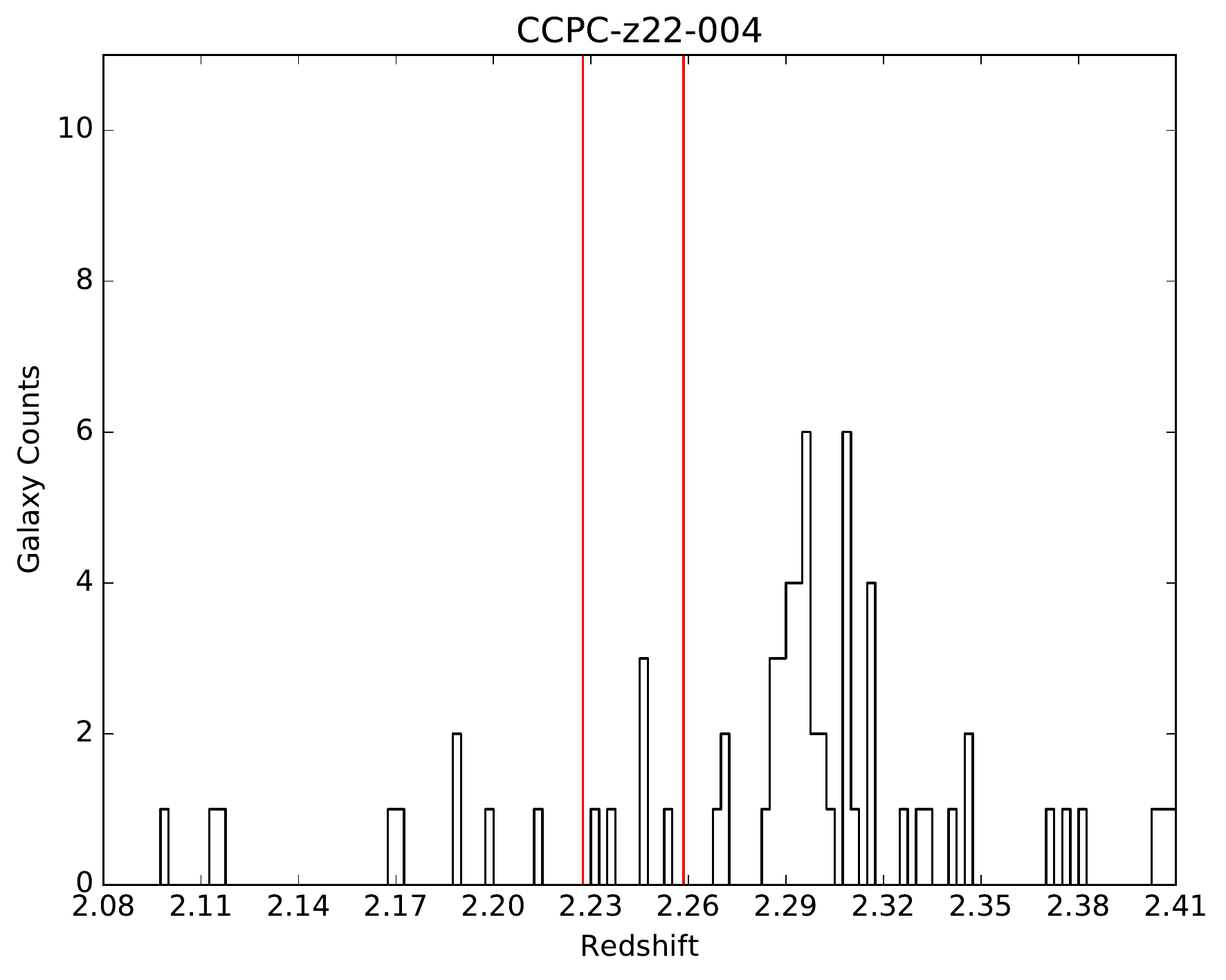}
\label{fig:CCPC-z22-004}
\end{subfigure}
\hfill
\end{figure*}
\clearpage 

\begin{figure*}
\centering
\begin{subfigure}
\centering
\includegraphics[height=7.5cm,width=7.5cm]{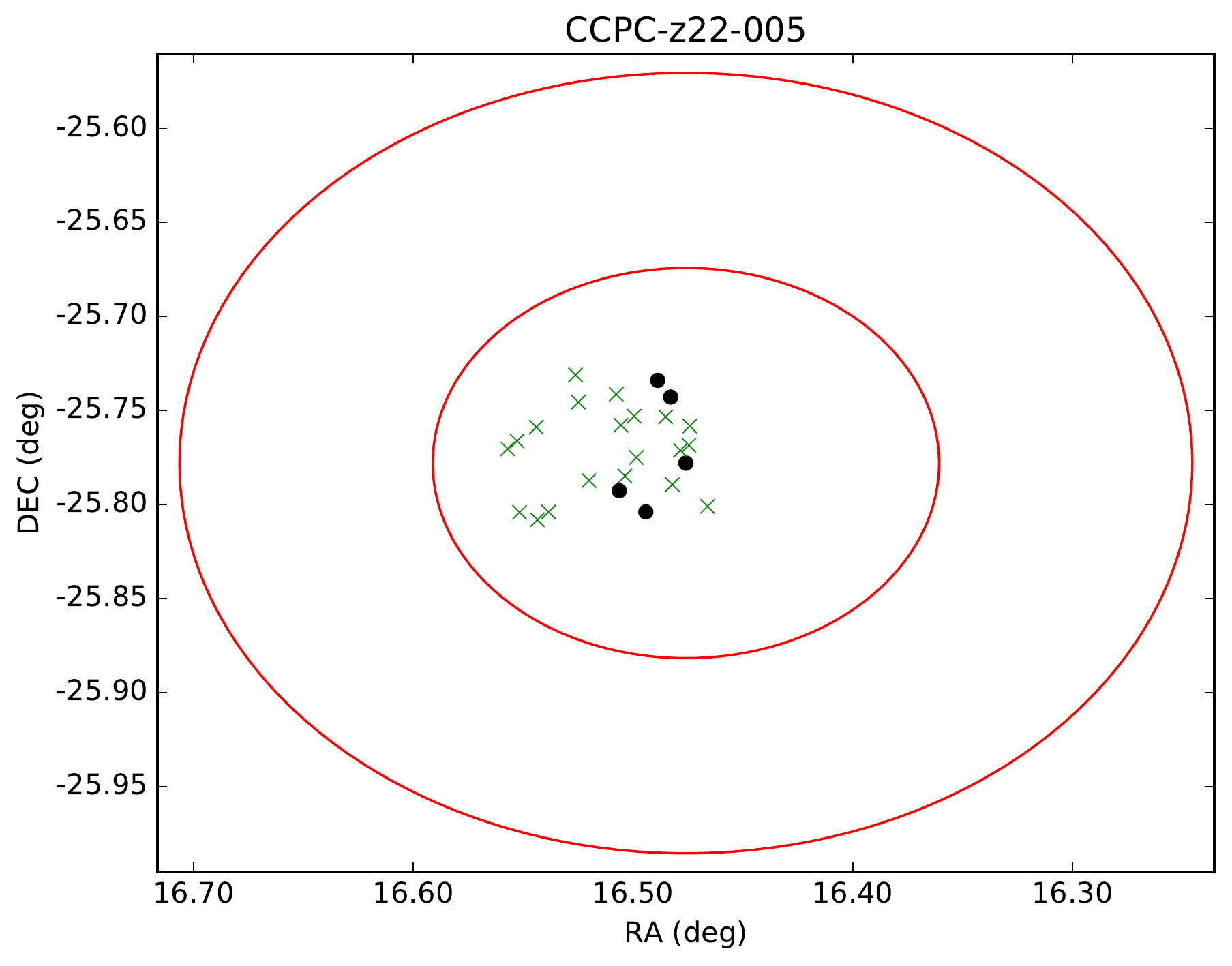}
\label{fig:CCPC-z22-005_sky}
\end{subfigure}
\hfill
\begin{subfigure}
\centering
\includegraphics[scale=0.52]{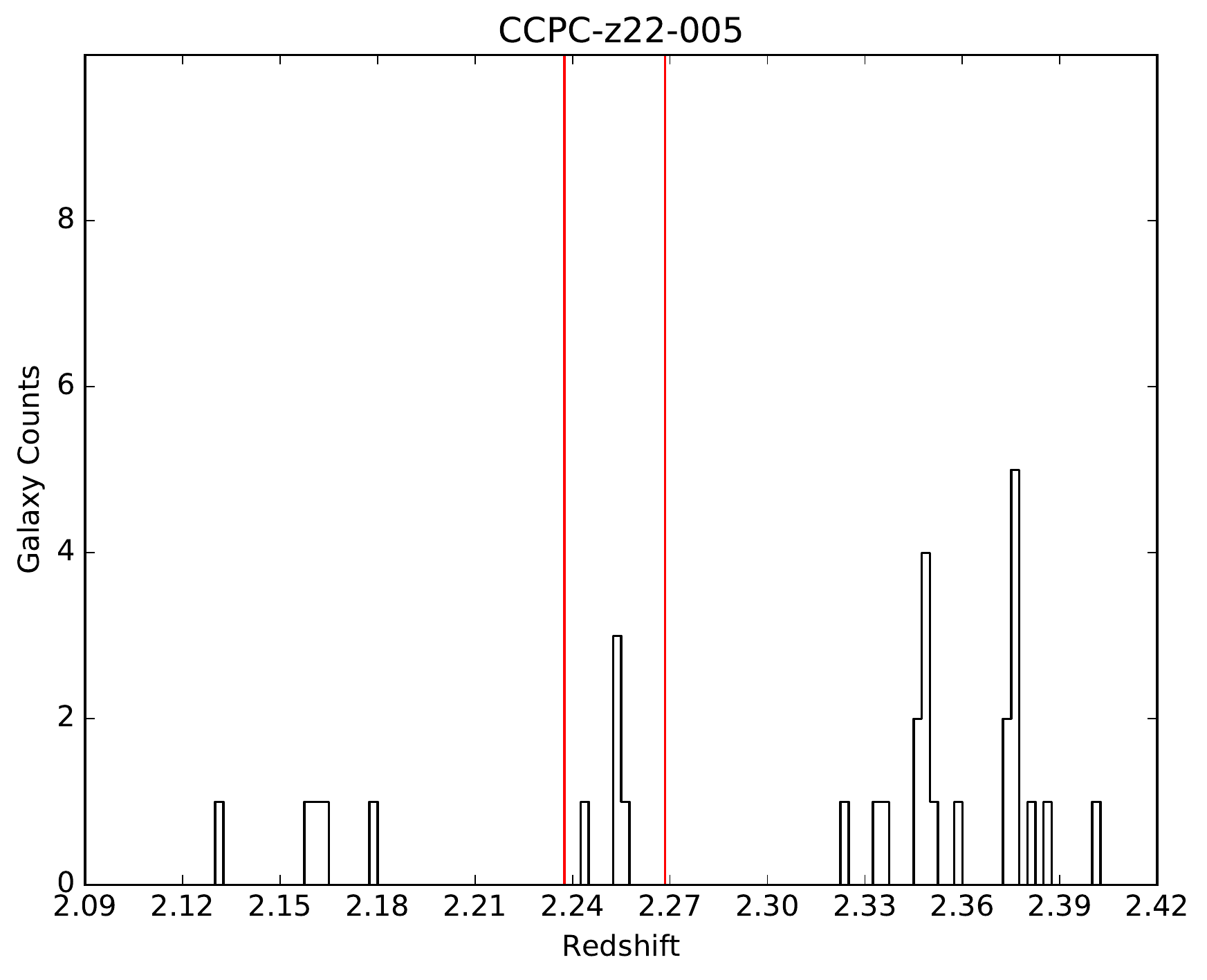}
\label{fig:CCPC-z22-005}
\end{subfigure}
\hfill
\end{figure*}

\begin{figure*}
\centering
\begin{subfigure}
\centering
\includegraphics[height=7.5cm,width=7.5cm]{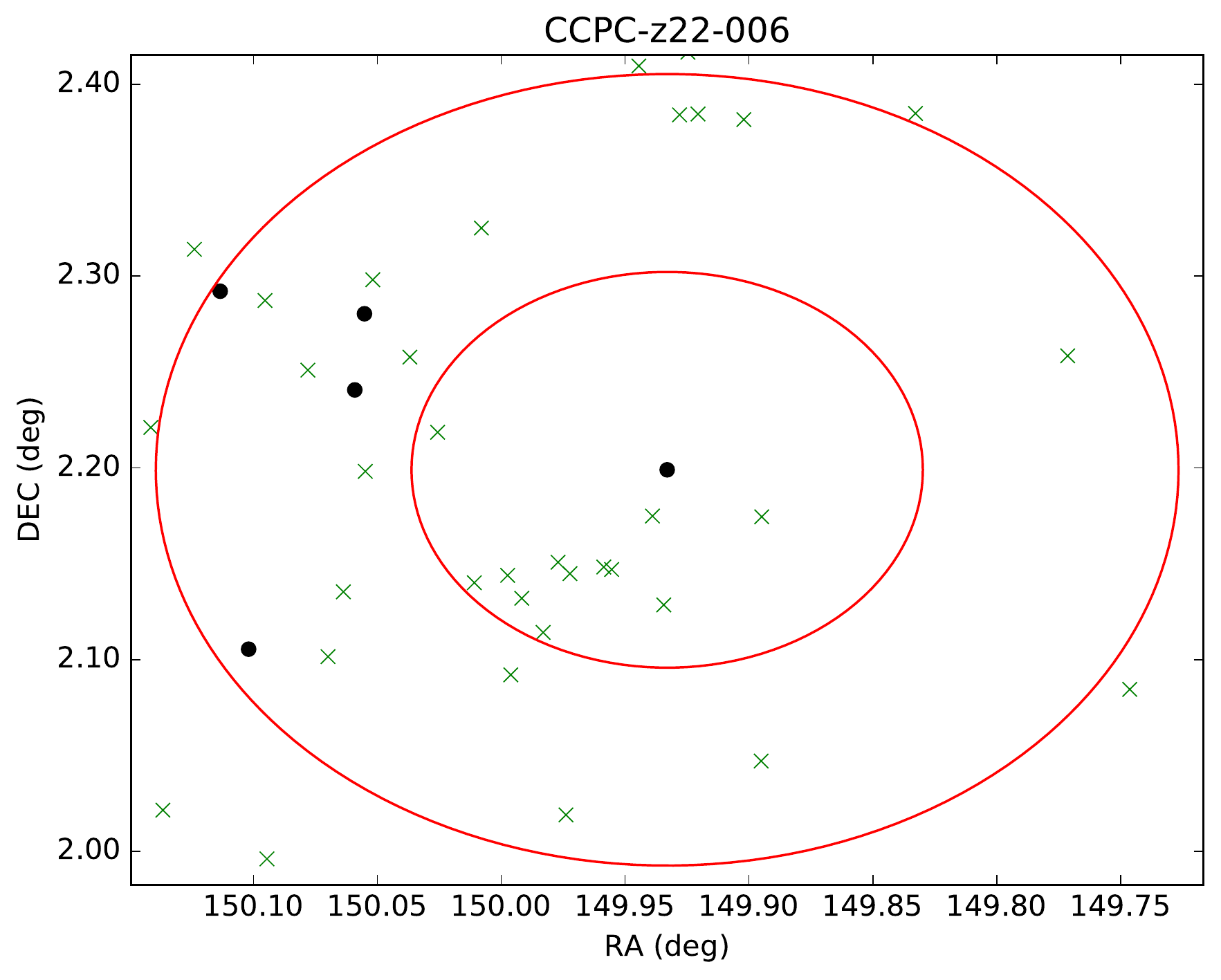}
\label{fig:CCPC-z22-006_sky}
\end{subfigure}
\hfill
\begin{subfigure}
\centering
\includegraphics[scale=0.52]{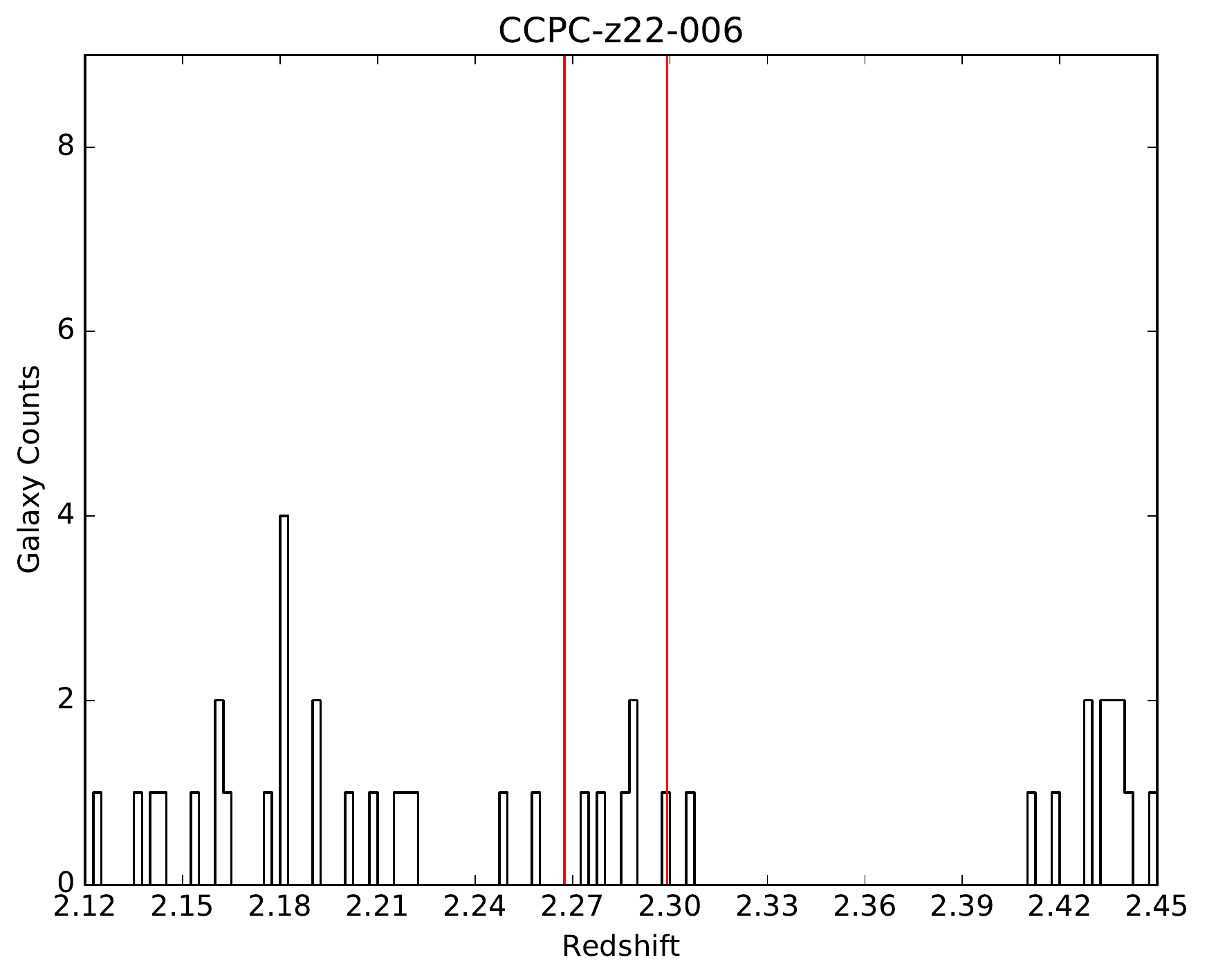}
\label{fig:CCPC-z22-006}
\end{subfigure}
\hfill
\end{figure*}

\begin{figure*}
\centering
\begin{subfigure}
\centering
\includegraphics[height=7.5cm,width=7.5cm]{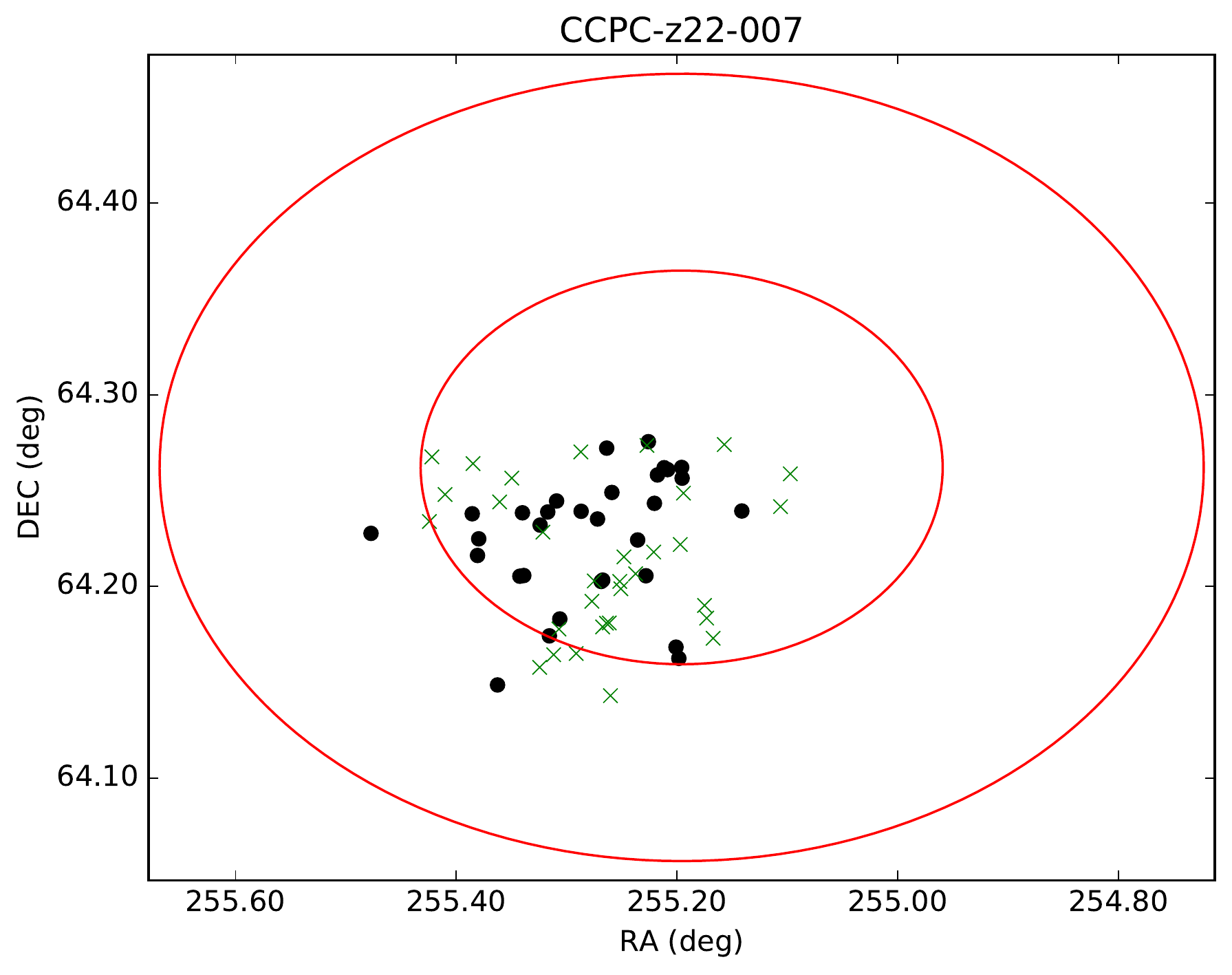}
\label{fig:CCPC-z22-007_sky}
\end{subfigure}
\hfill
\begin{subfigure}
\centering
\includegraphics[scale=0.52]{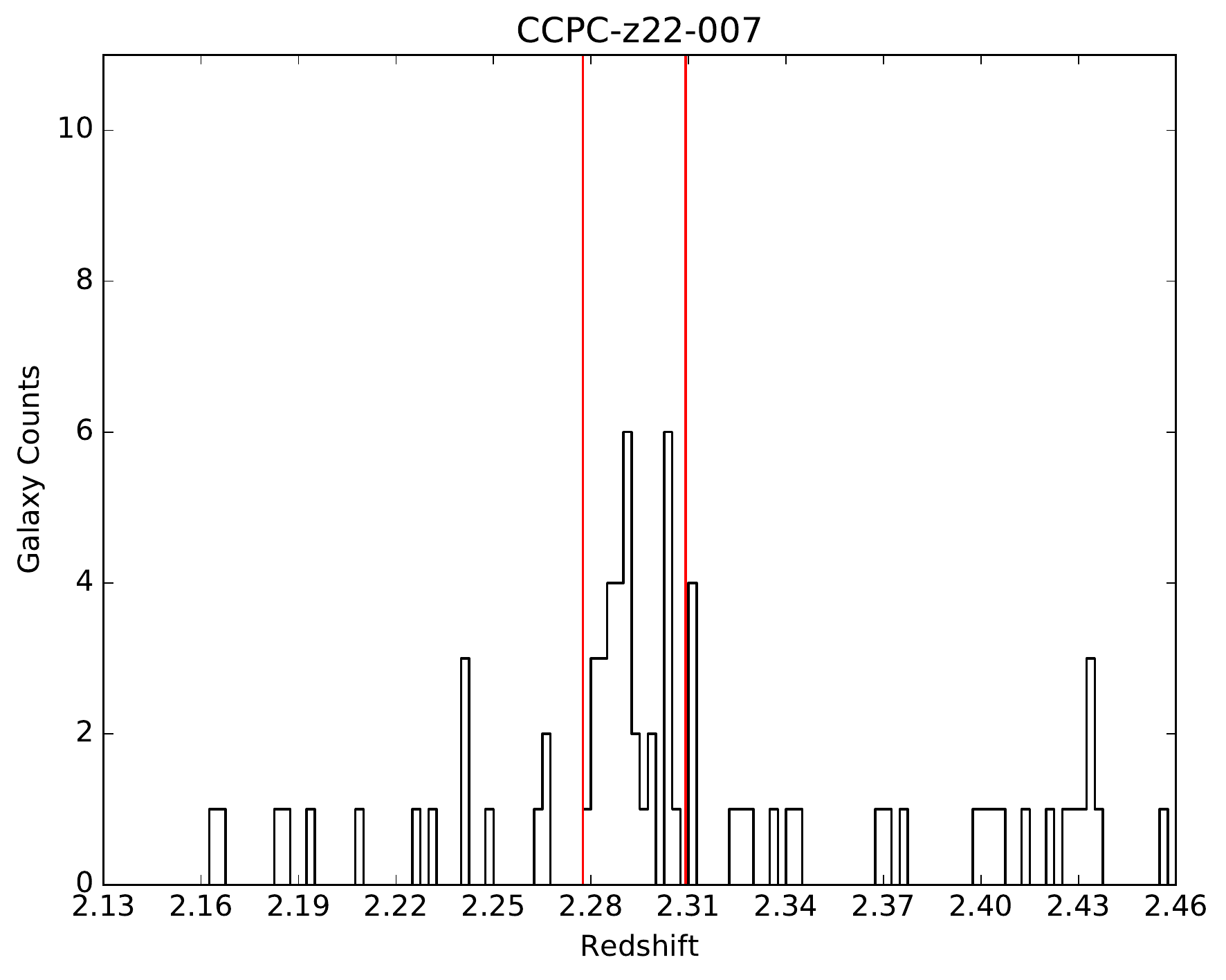}
\label{fig:CCPC-z22-007}
\end{subfigure}
\hfill
\end{figure*}
\clearpage 

\begin{figure*}
\centering
\begin{subfigure}
\centering
\includegraphics[height=7.5cm,width=7.5cm]{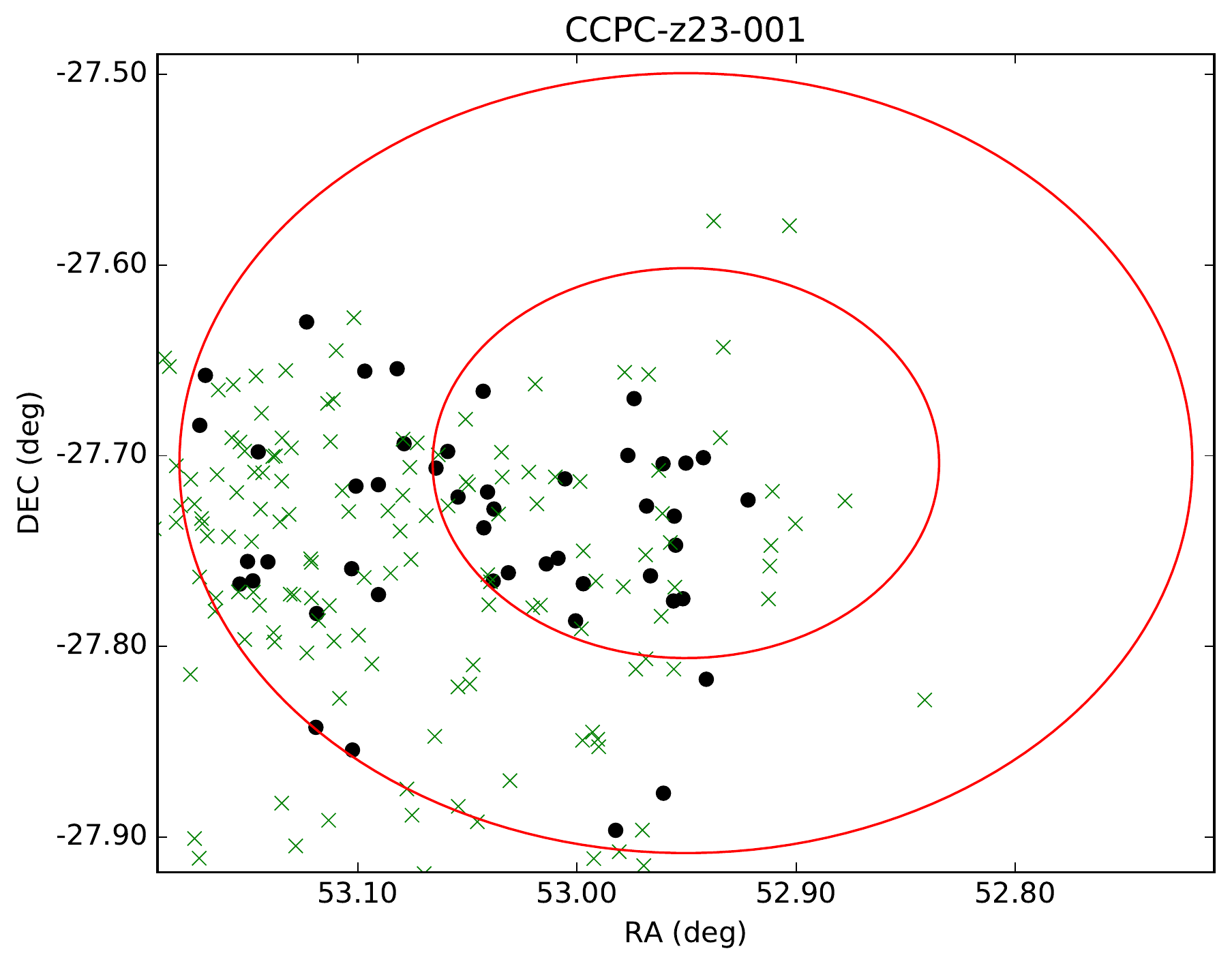}
\label{fig:CCPC-z23-001_sky}
\end{subfigure}
\hfill
\begin{subfigure}
\centering
\includegraphics[scale=0.52]{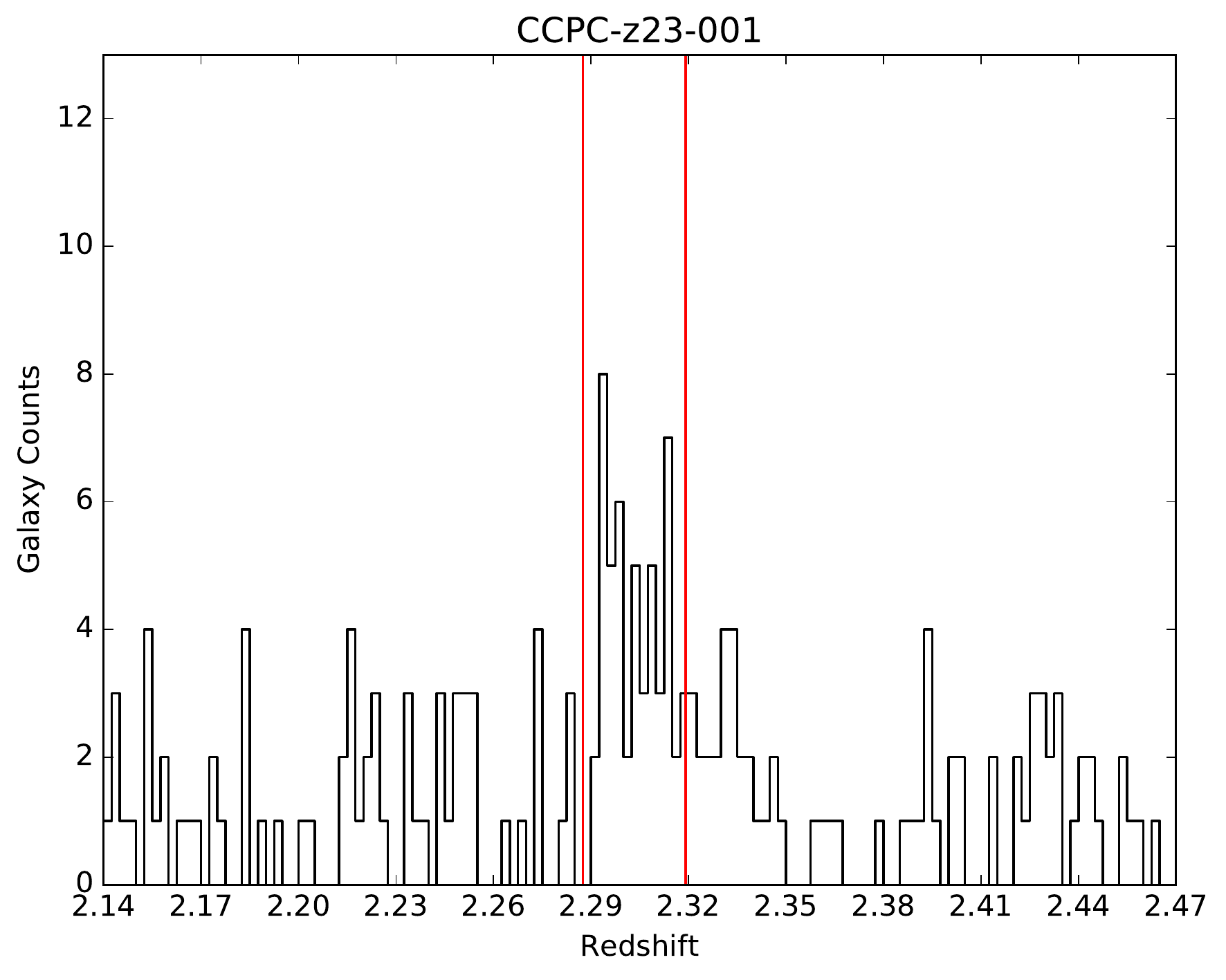}
\label{fig:CCPC-z23-001}
\end{subfigure}
\hfill
\end{figure*}

\begin{figure*}
\centering
\begin{subfigure}
\centering
\includegraphics[height=7.5cm,width=7.5cm]{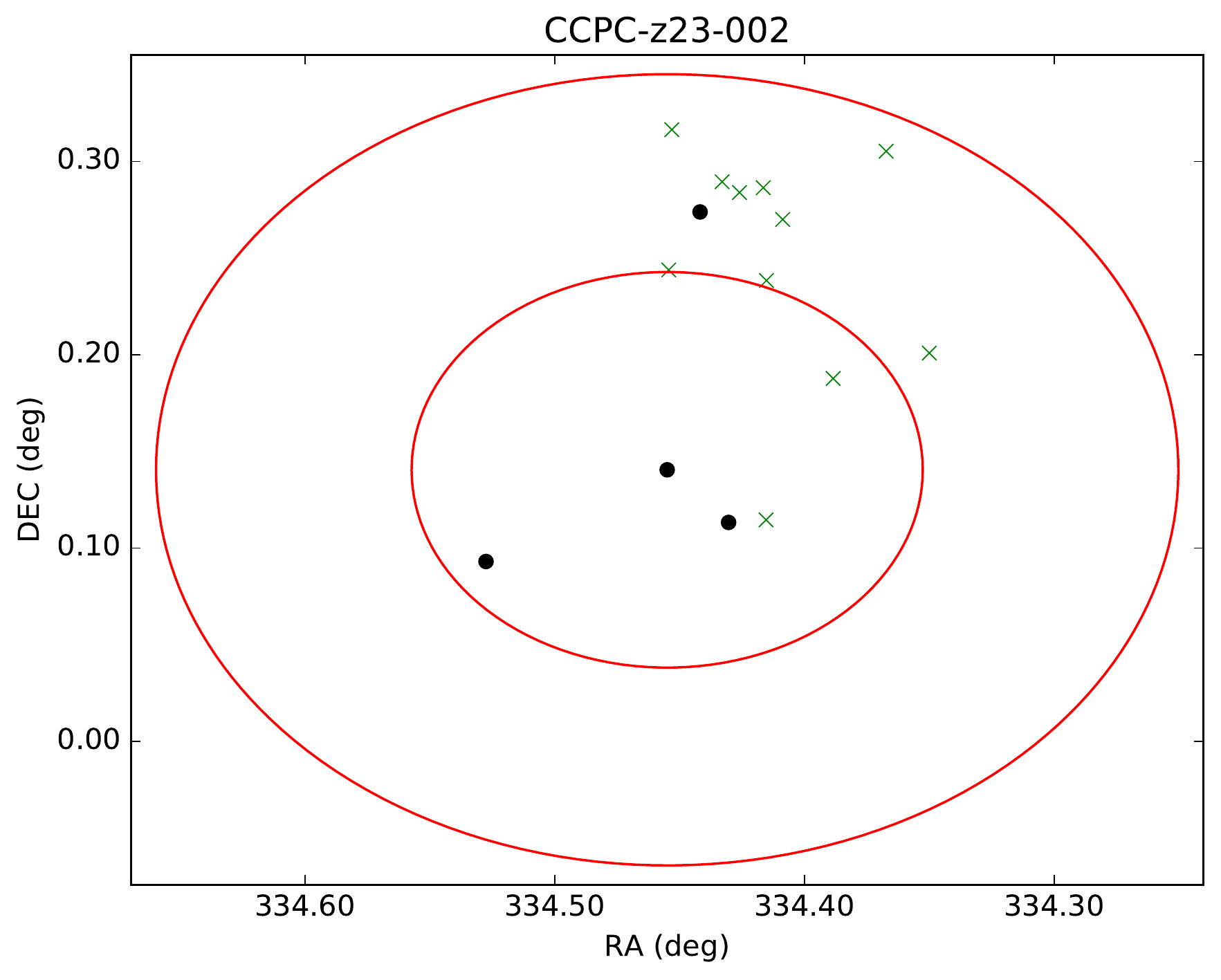}
\label{fig:CCPC-z23-002_sky}
\end{subfigure}
\hfill
\begin{subfigure}
\centering
\includegraphics[scale=0.52]{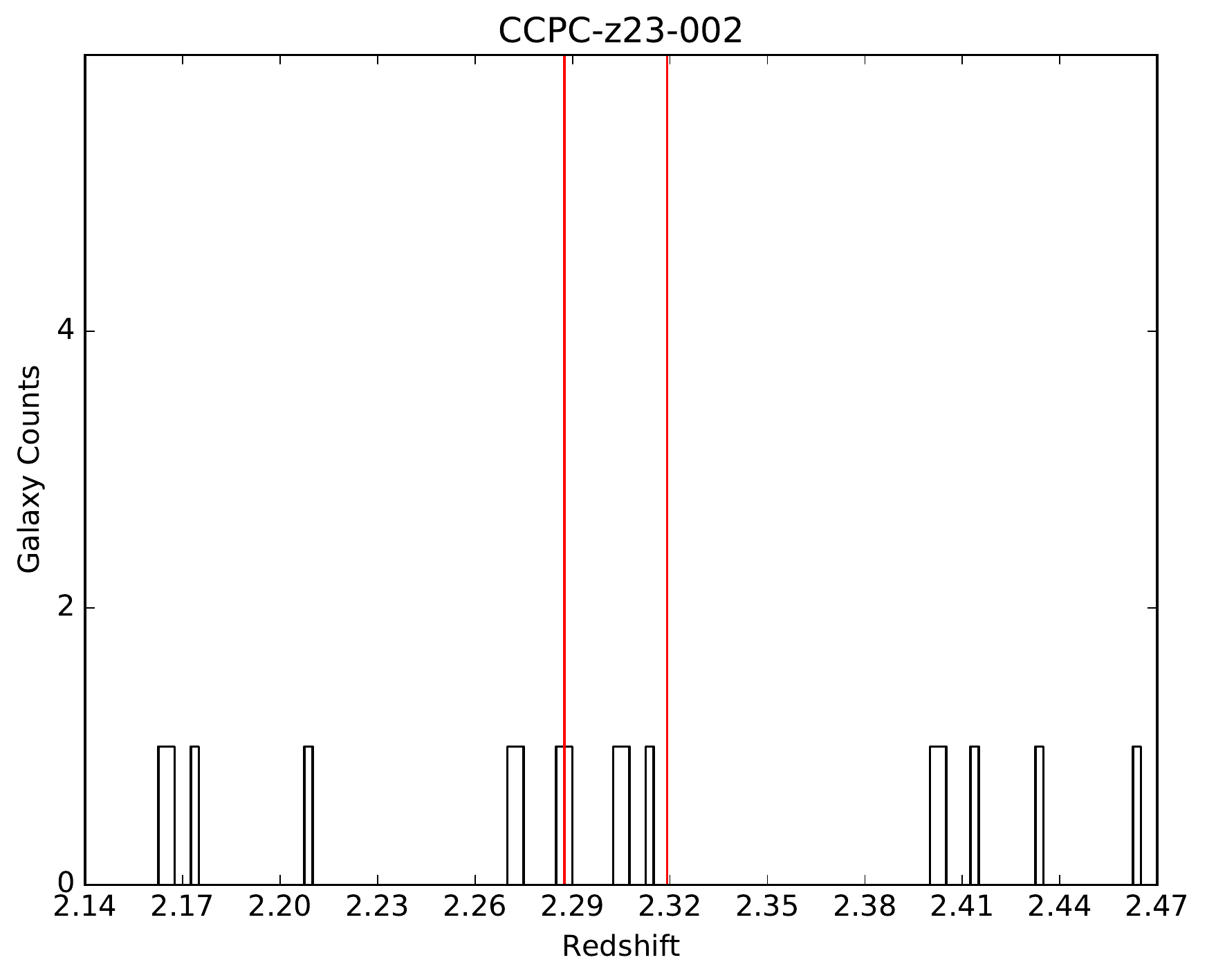}
\label{fig:CCPC-z23-002}
\end{subfigure}
\hfill
\end{figure*}

\begin{figure*}
\centering
\begin{subfigure}
\centering
\includegraphics[height=7.5cm,width=7.5cm]{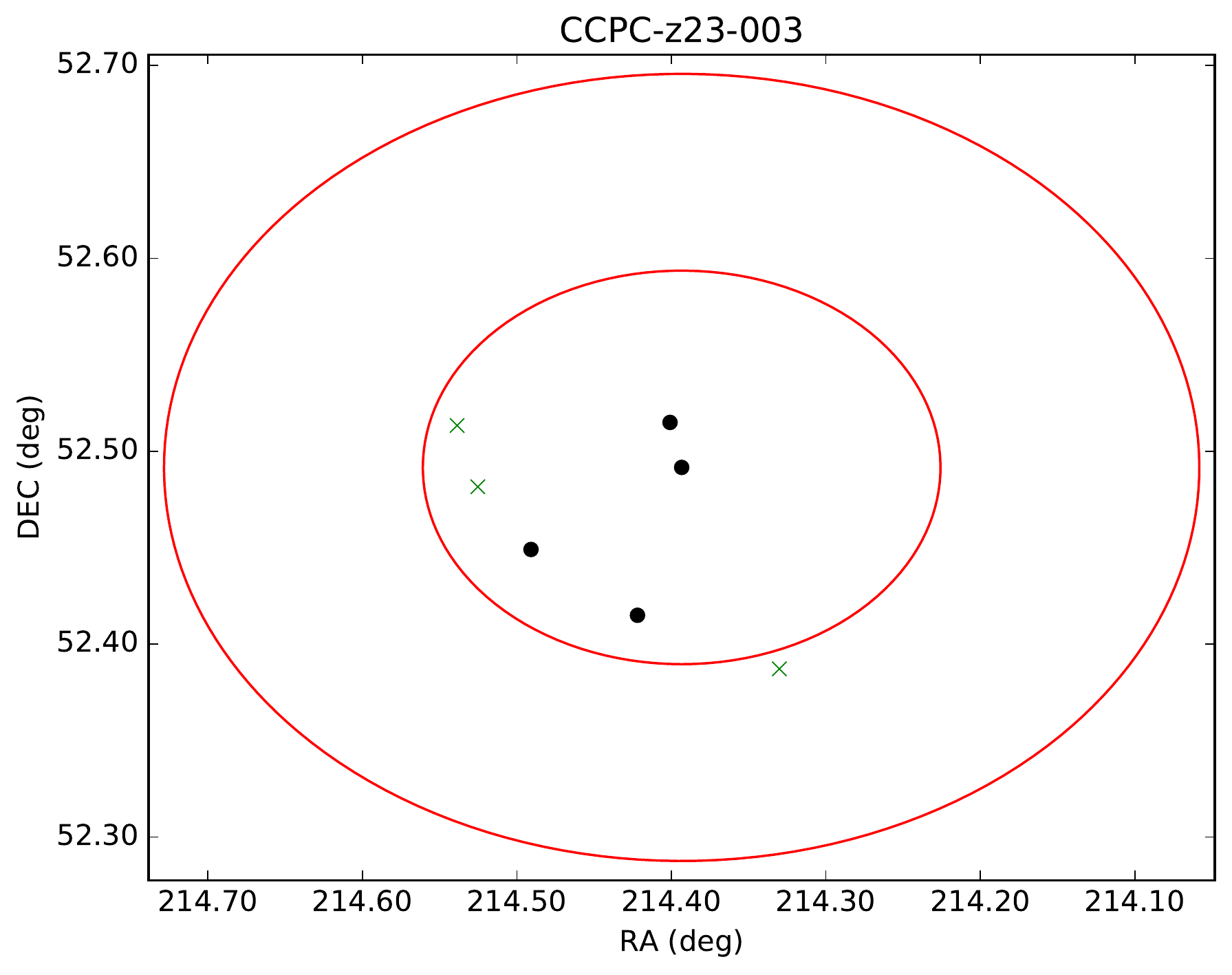}
\label{fig:CCPC-z23-003_sky}
\end{subfigure}
\hfill
\begin{subfigure}
\centering
\includegraphics[scale=0.52]{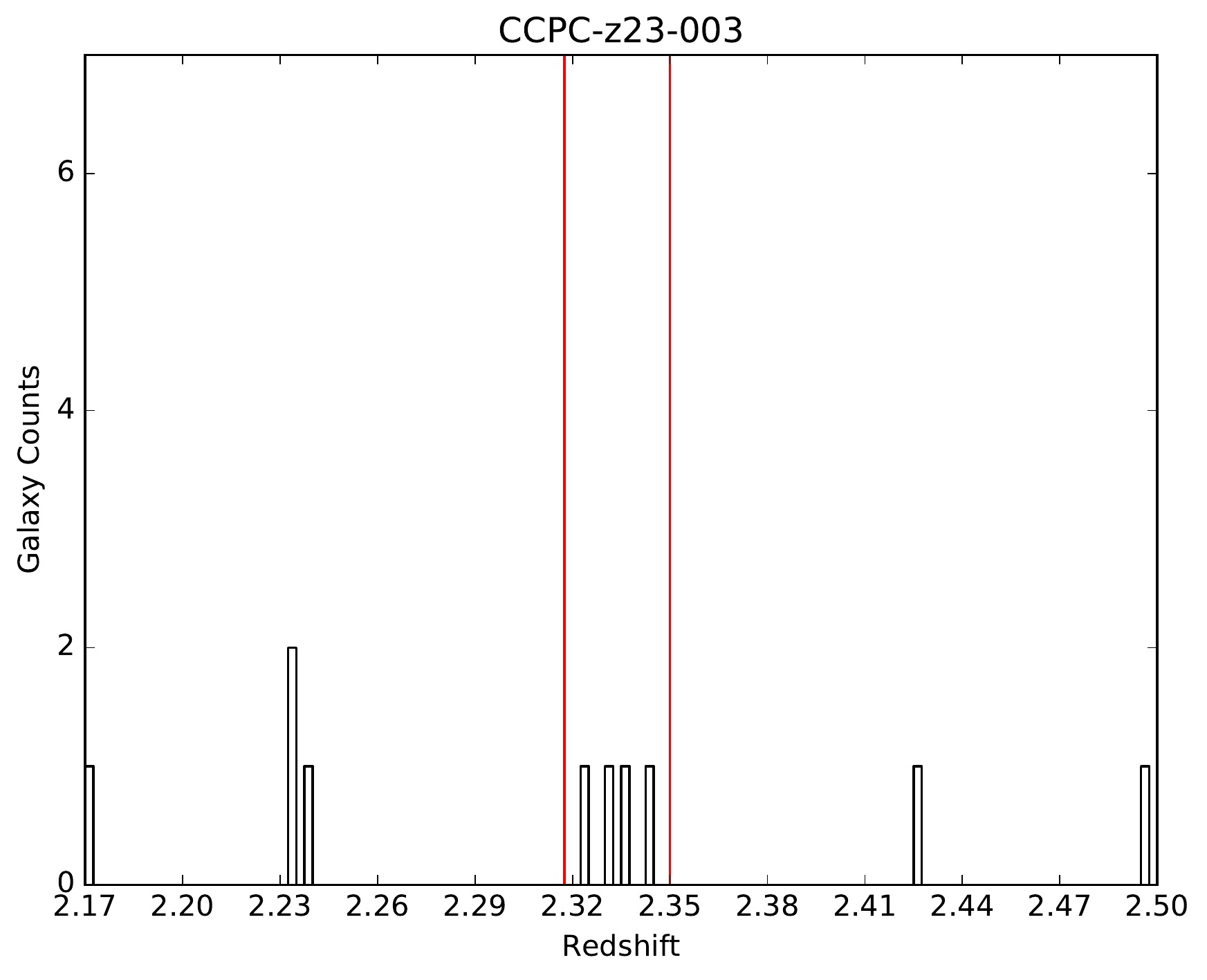}
\label{fig:CCPC-z23-003}
\end{subfigure}
\hfill
\end{figure*}
\clearpage 

\begin{figure*}
\centering
\begin{subfigure}
\centering
\includegraphics[height=7.5cm,width=7.5cm]{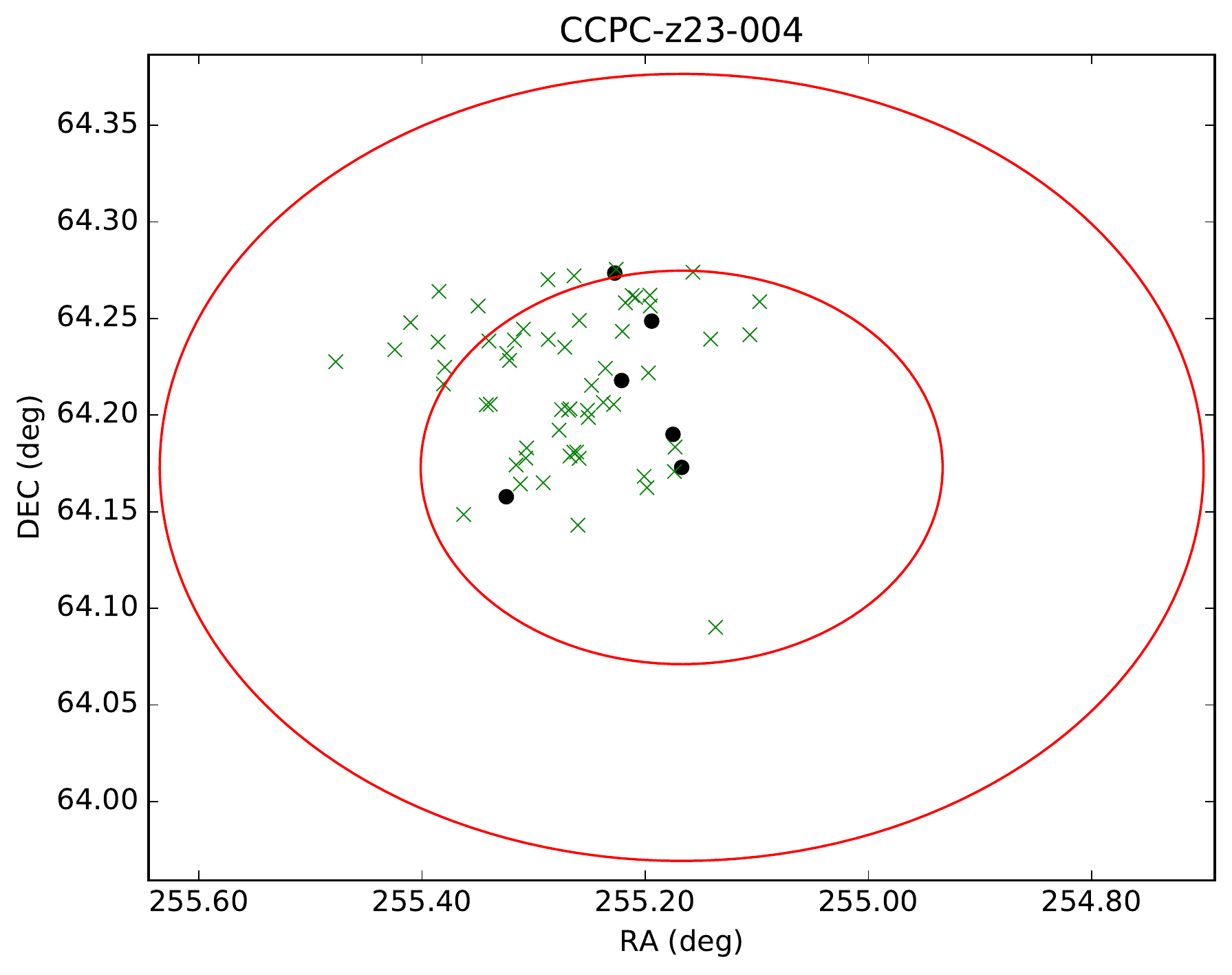}
\label{fig:CCPC-z23-004_sky}
\end{subfigure}
\hfill
\begin{subfigure}
\centering
\includegraphics[scale=0.52]{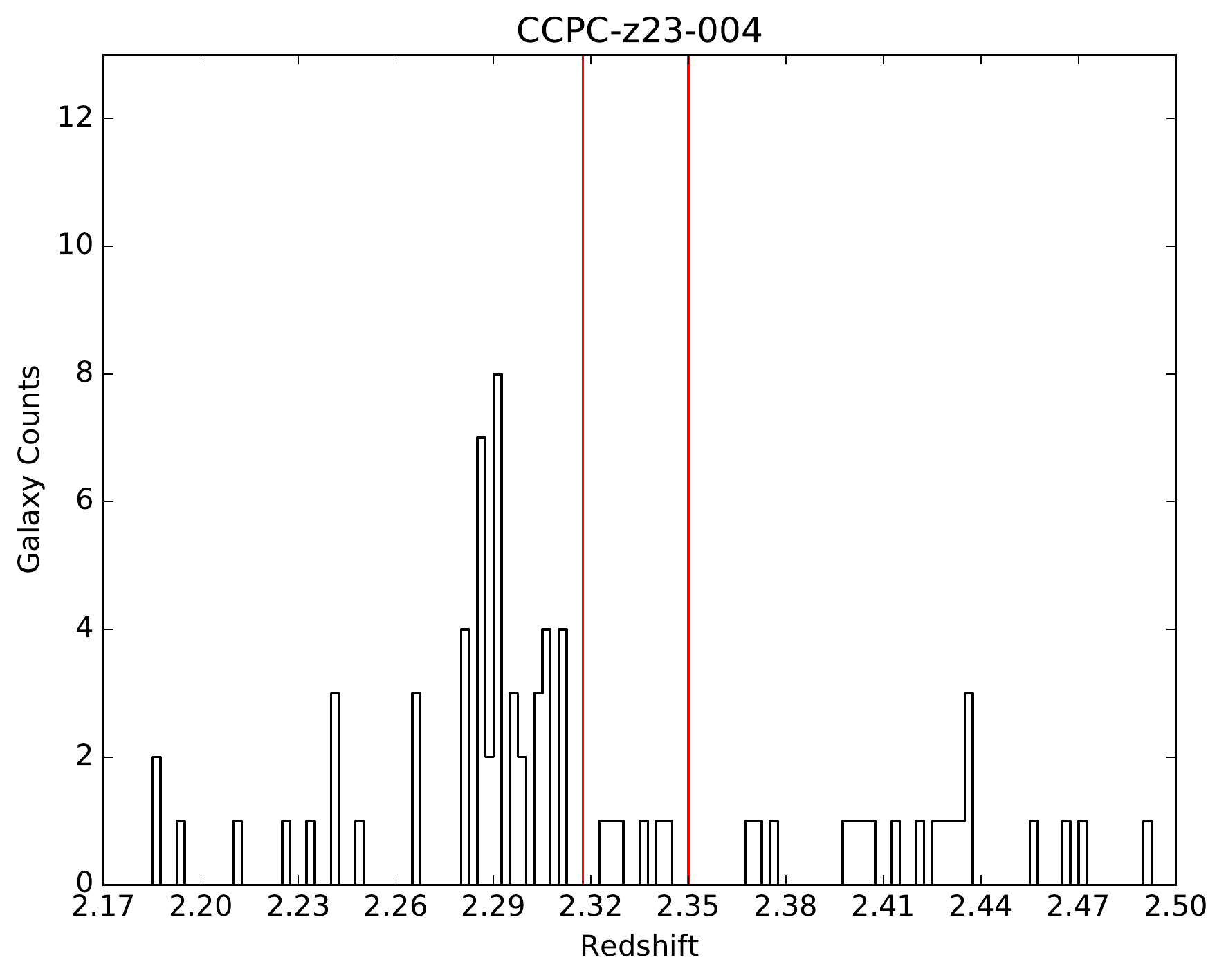}
\label{fig:CCPC-z23-004}
\end{subfigure}
\hfill
\end{figure*}

\begin{figure*}
\centering
\begin{subfigure}
\centering
\includegraphics[height=7.5cm,width=7.5cm]{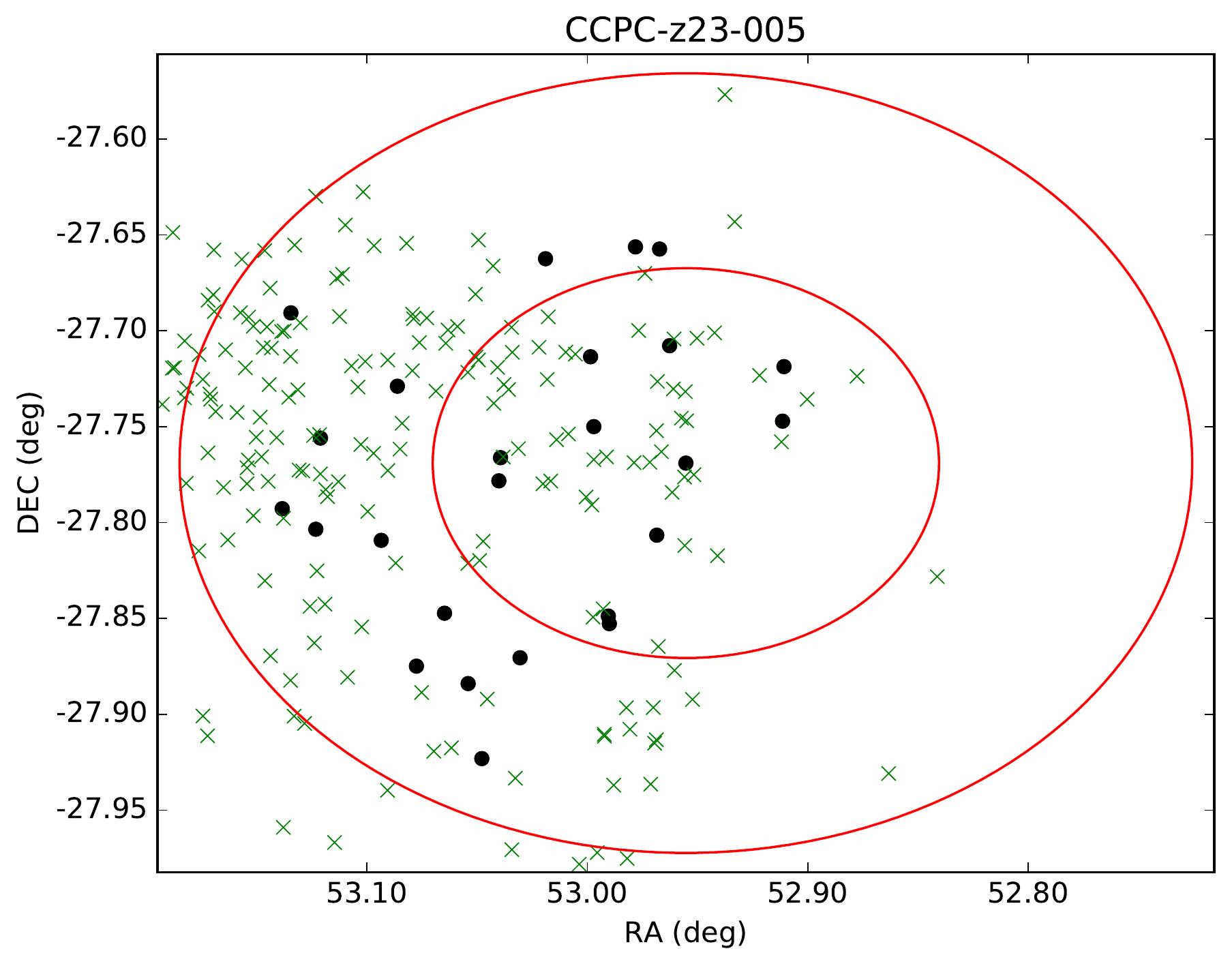}
\label{fig:CCPC-z23-005_sky}
\end{subfigure}
\hfill
\begin{subfigure}
\centering
\includegraphics[scale=0.52]{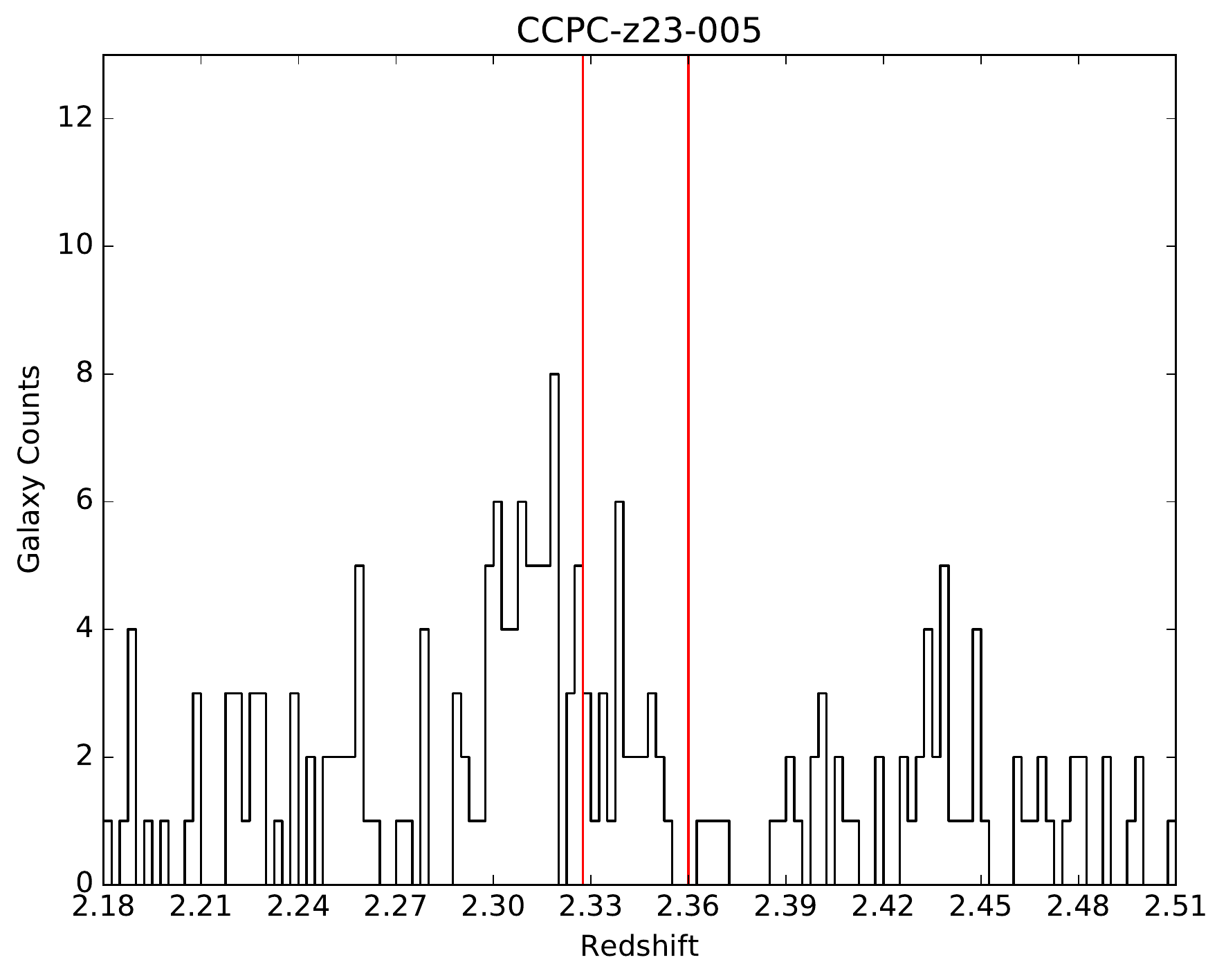}
\label{fig:CCPC-z23-005}
\end{subfigure}
\hfill
\end{figure*}

\begin{figure*}
\centering
\begin{subfigure}
\centering
\includegraphics[height=7.5cm,width=7.5cm]{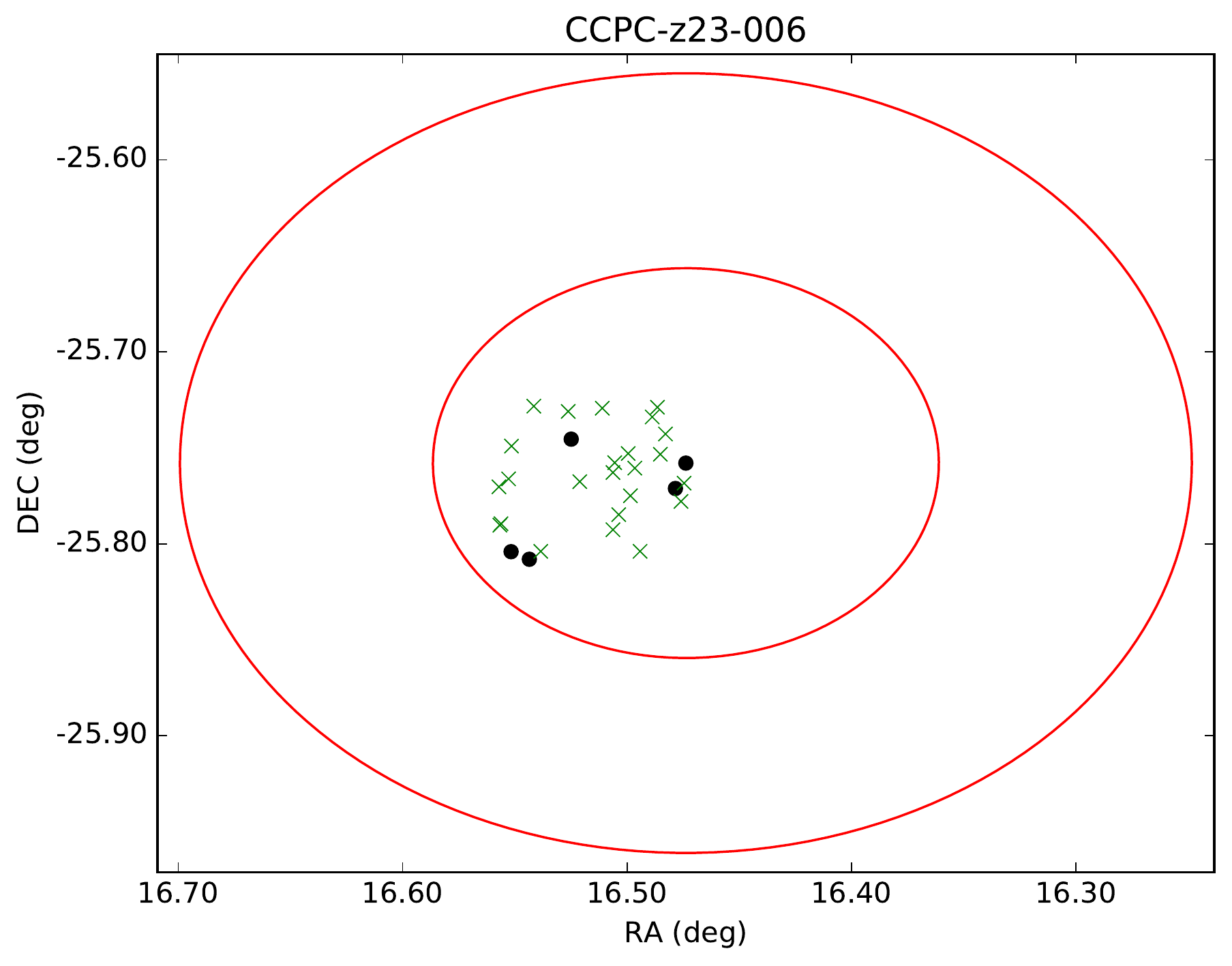}
\label{fig:CCPC-z23-006_sky}
\end{subfigure}
\hfill
\begin{subfigure}
\centering
\includegraphics[scale=0.52]{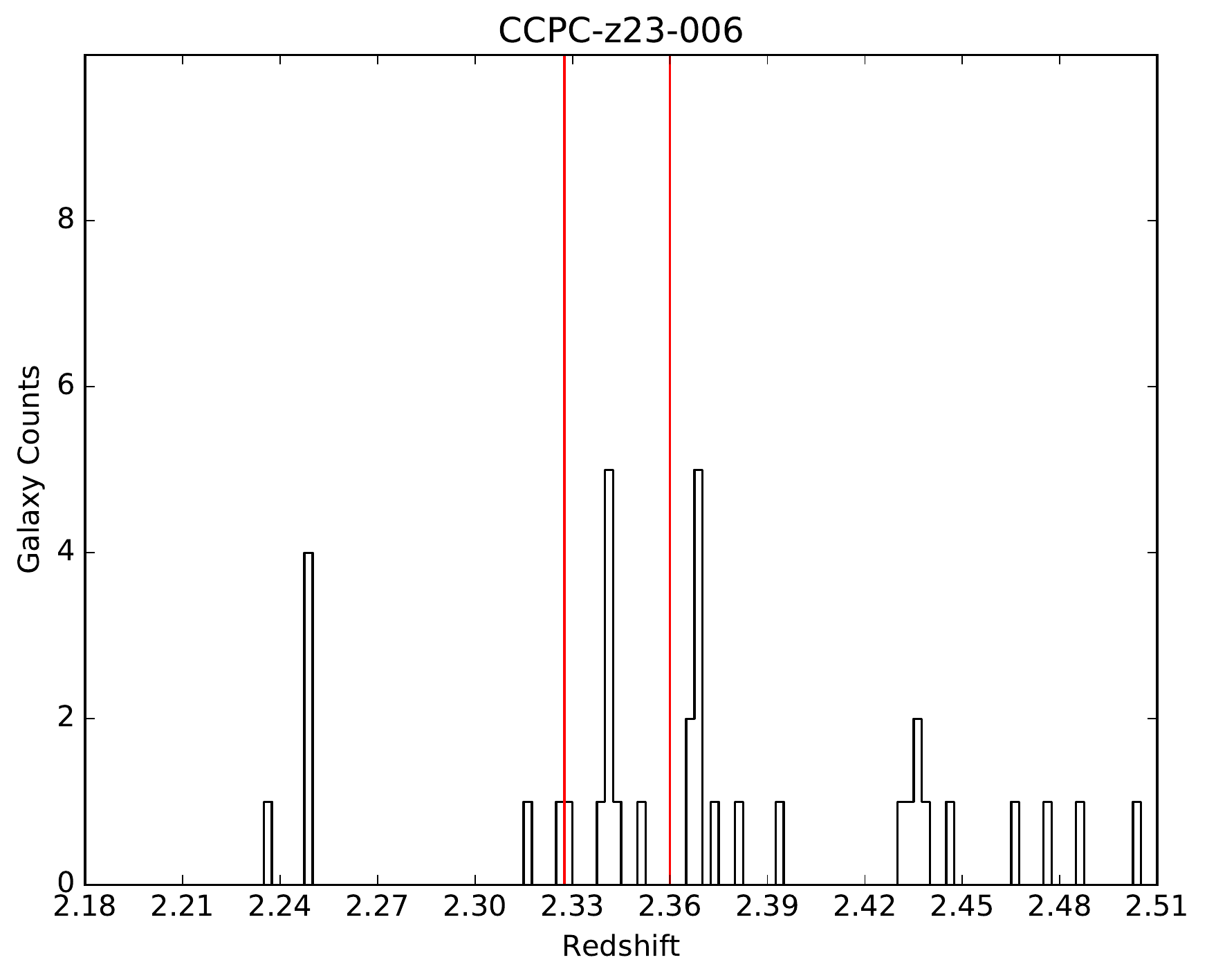}
\label{fig:CCPC-z23-006}
\end{subfigure}
\hfill
\end{figure*}
\clearpage 

\begin{figure*}
\centering
\begin{subfigure}
\centering
\includegraphics[height=7.5cm,width=7.5cm]{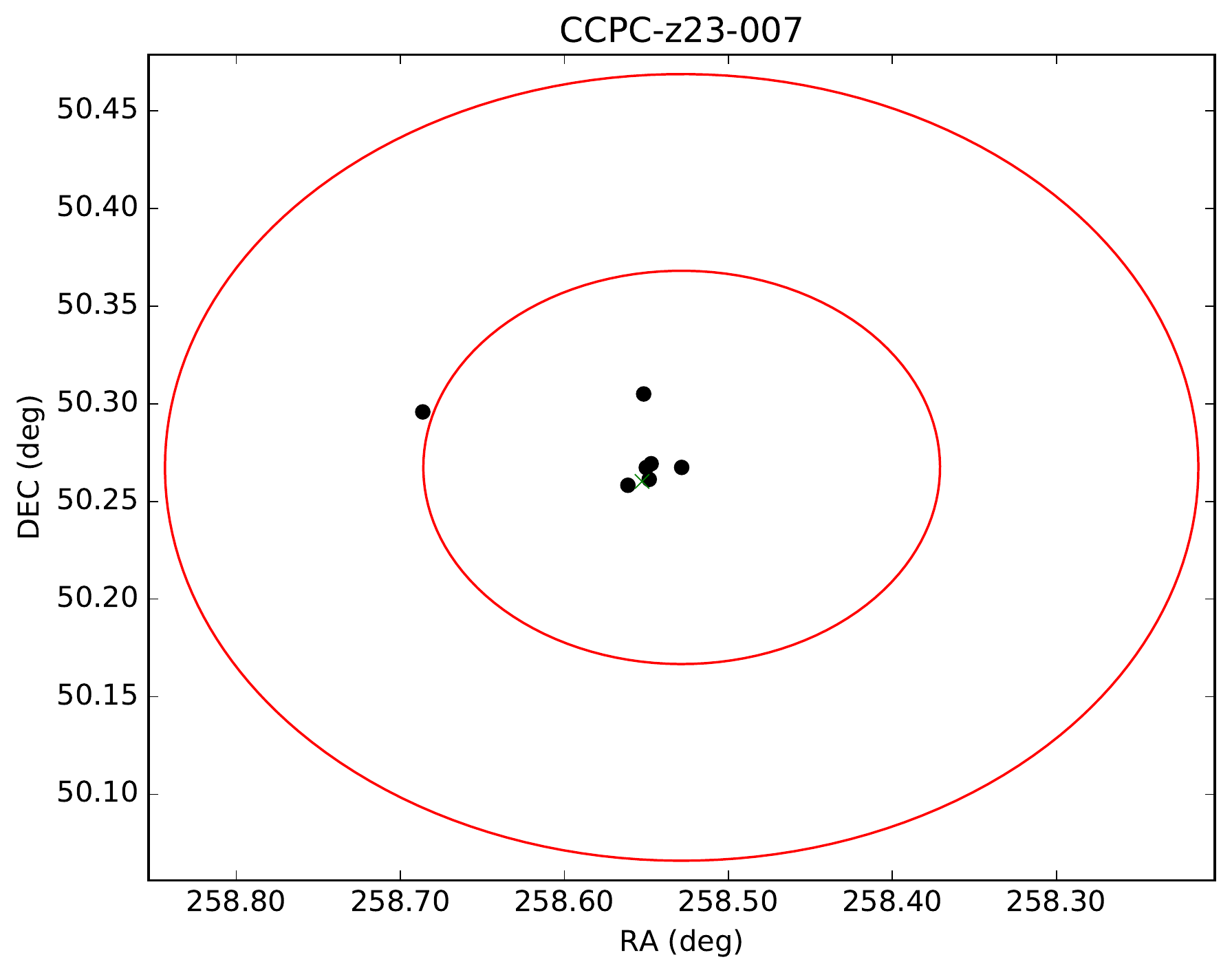}
\label{fig:CCPC-z23-007_sky}
\end{subfigure}
\hfill
\begin{subfigure}
\centering
\includegraphics[scale=0.52]{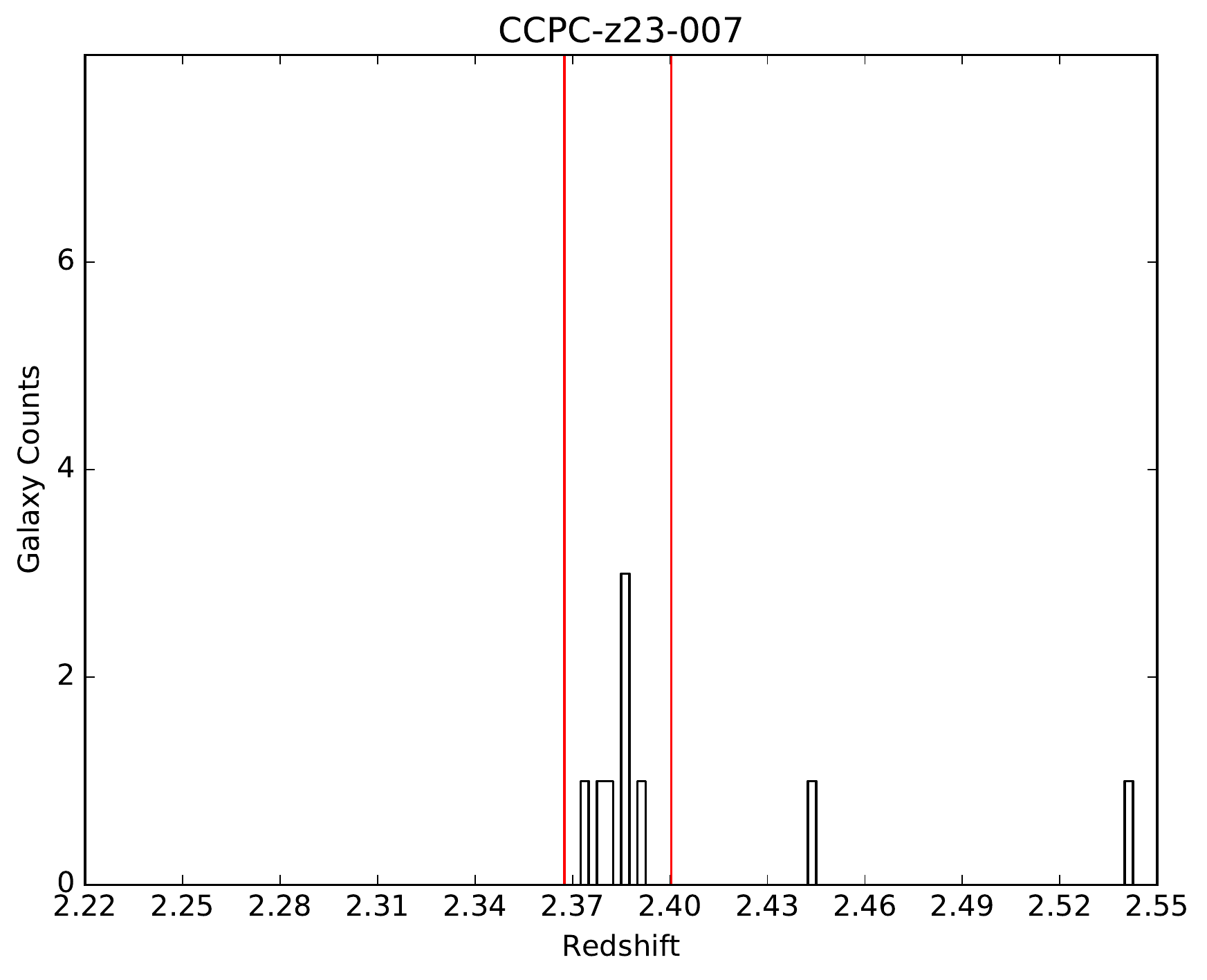}
\label{fig:CCPC-z23-007}
\end{subfigure}
\hfill
\end{figure*}

\begin{figure*}
\centering
\begin{subfigure}
\centering
\includegraphics[height=7.5cm,width=7.5cm]{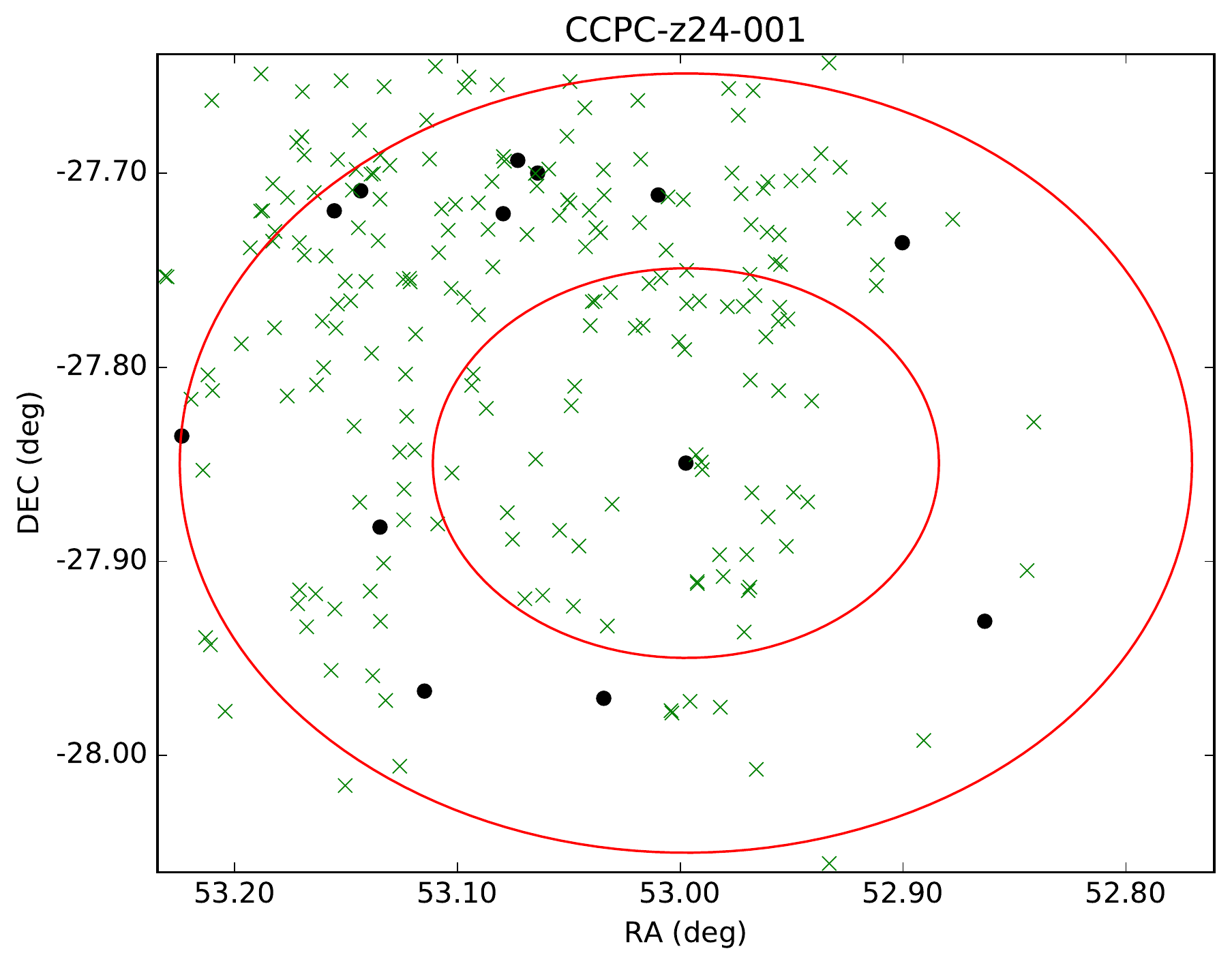}
\label{fig:CCPC-z24-001_sky}
\end{subfigure}
\hfill
\begin{subfigure}
\centering
\includegraphics[scale=0.52]{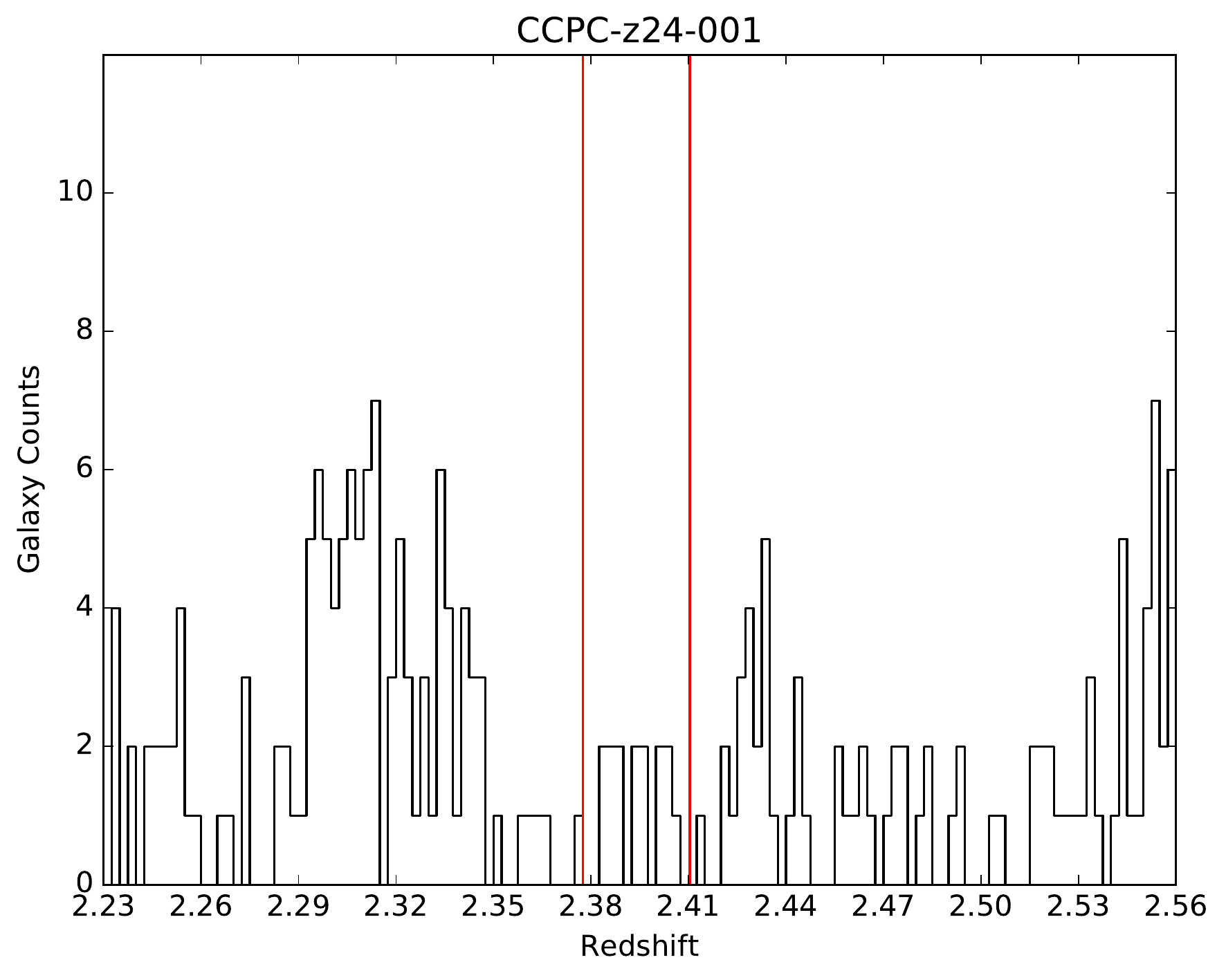}
\label{fig:CCPC-z24-001}
\end{subfigure}
\hfill
\end{figure*}

\begin{figure*}
\centering
\begin{subfigure}
\centering
\includegraphics[height=7.5cm,width=7.5cm]{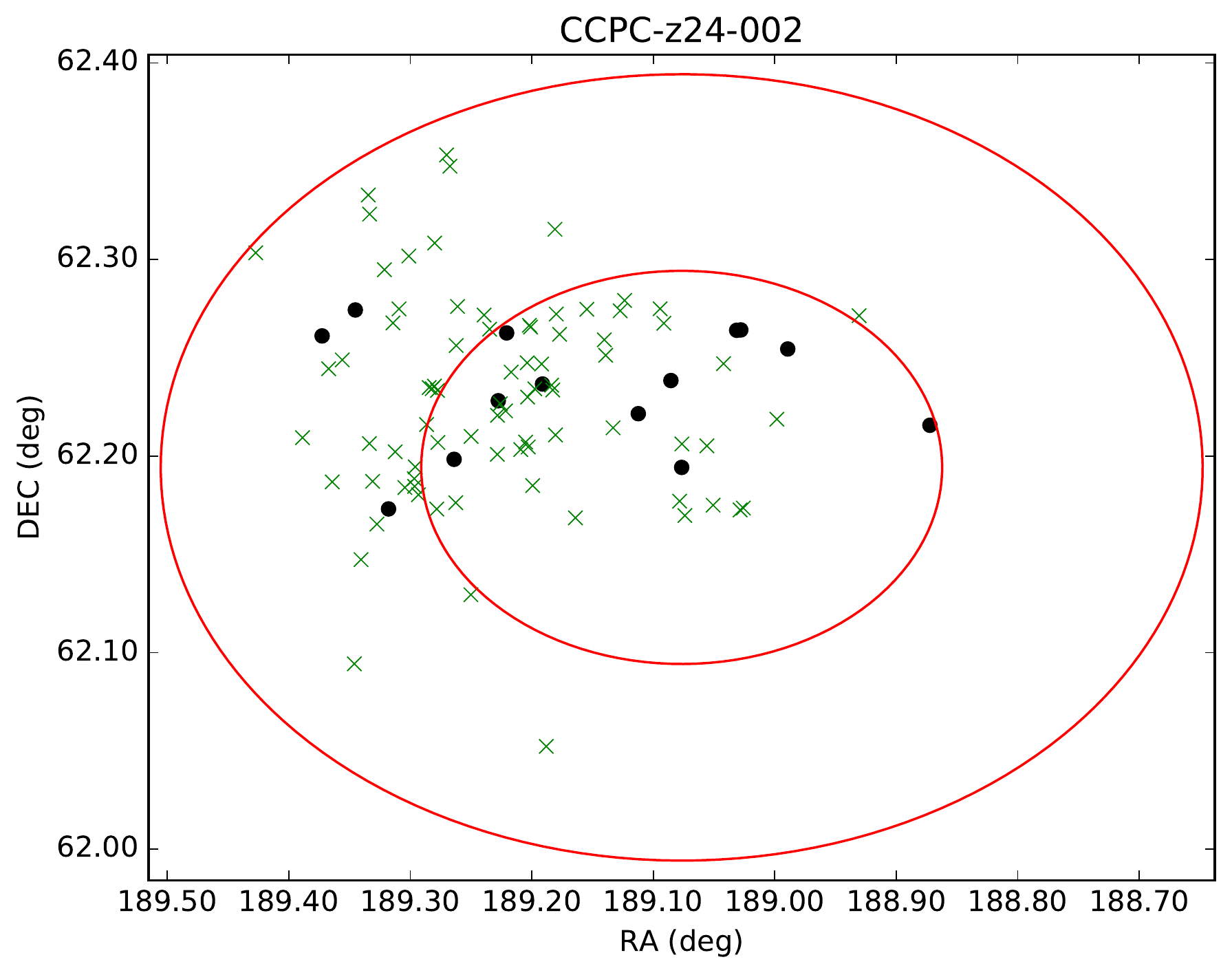}
\label{fig:CCPC-z24-002_sky}
\end{subfigure}
\hfill
\begin{subfigure}
\centering
\includegraphics[scale=0.52]{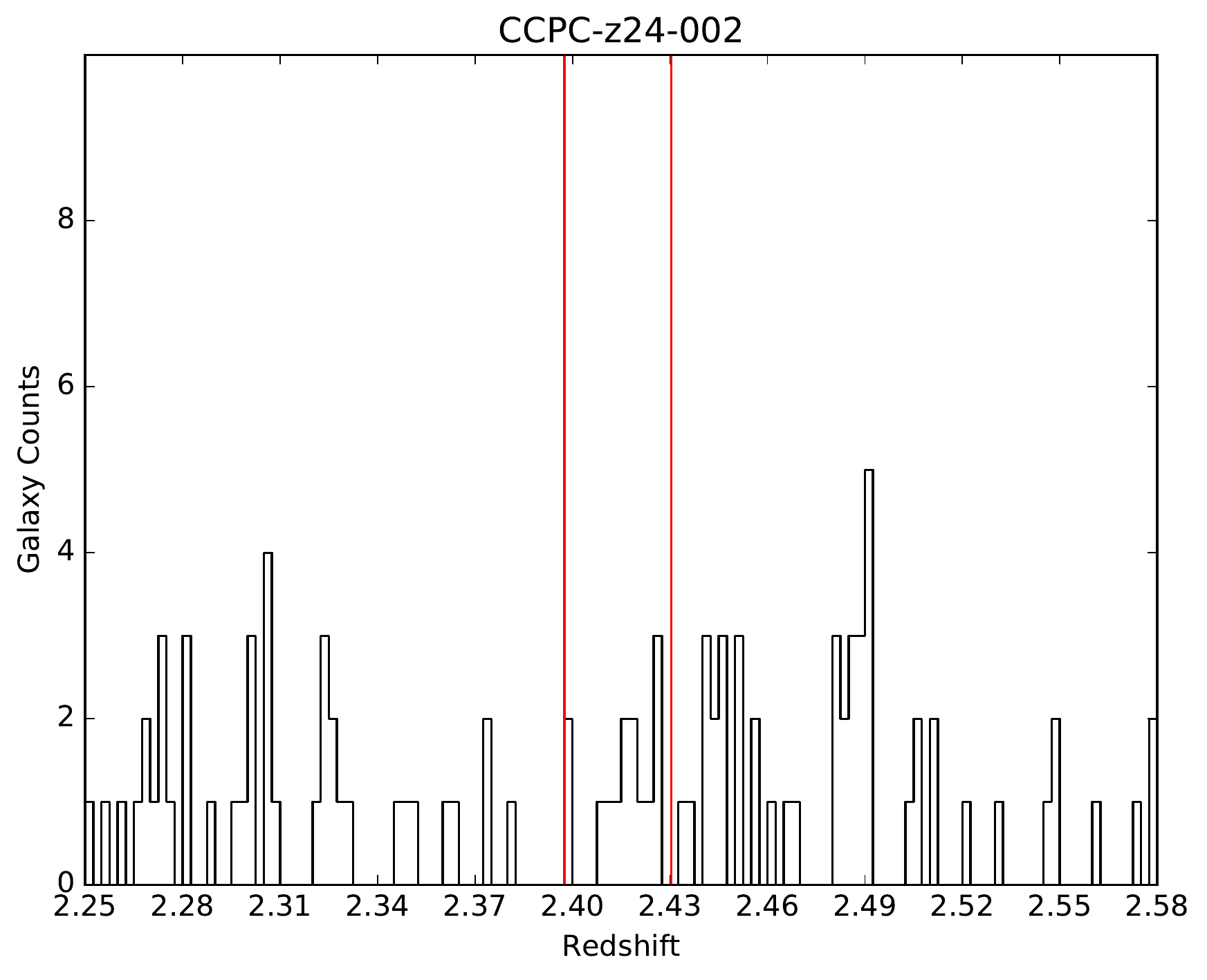}
\label{fig:CCPC-z24-002}
\end{subfigure}
\hfill
\end{figure*}
\clearpage 

\begin{figure*}
\centering
\begin{subfigure}
\centering
\includegraphics[height=7.5cm,width=7.5cm]{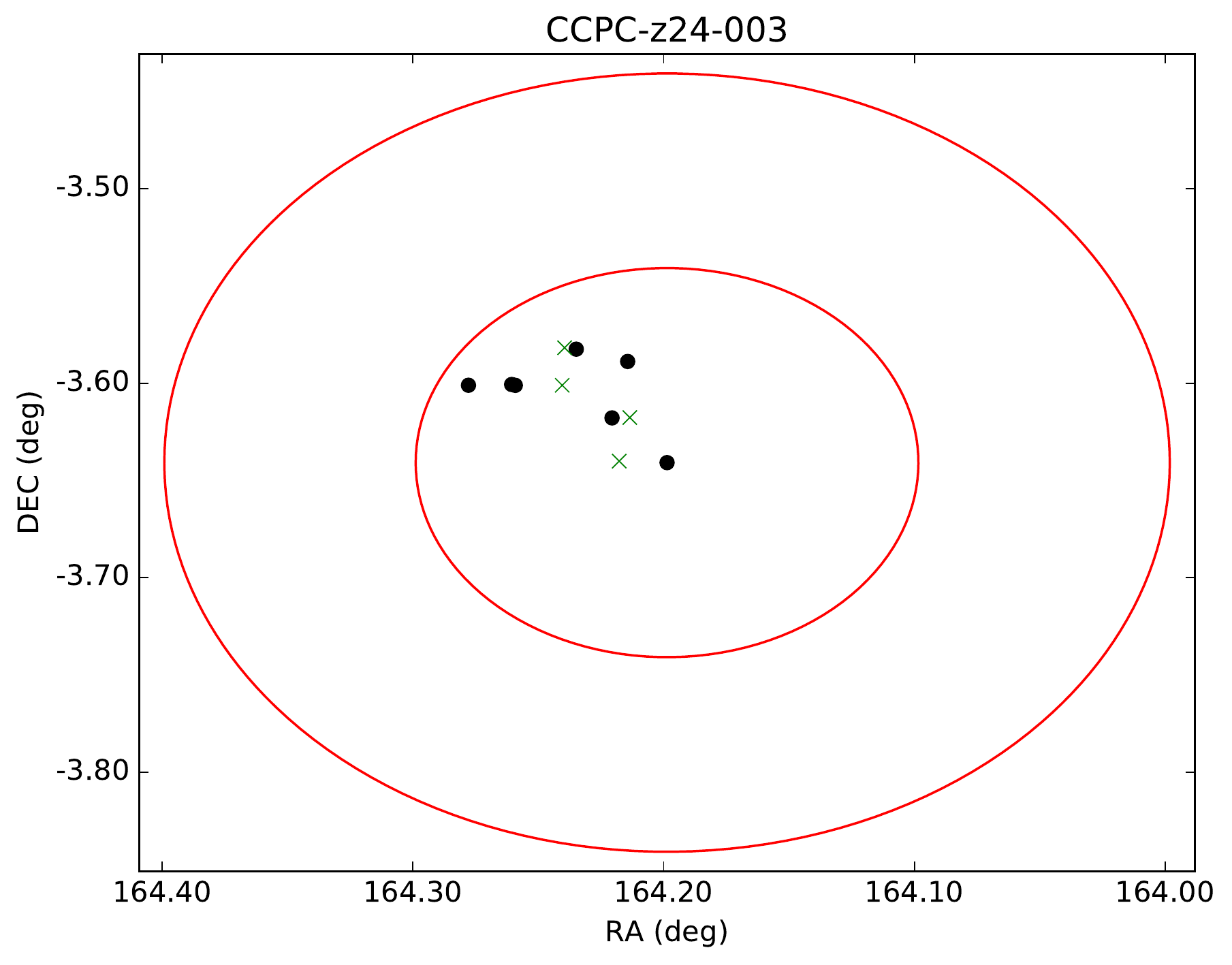}
\label{fig:CCPC-z24-003_sky}
\end{subfigure}
\hfill
\begin{subfigure}
\centering
\includegraphics[scale=0.52]{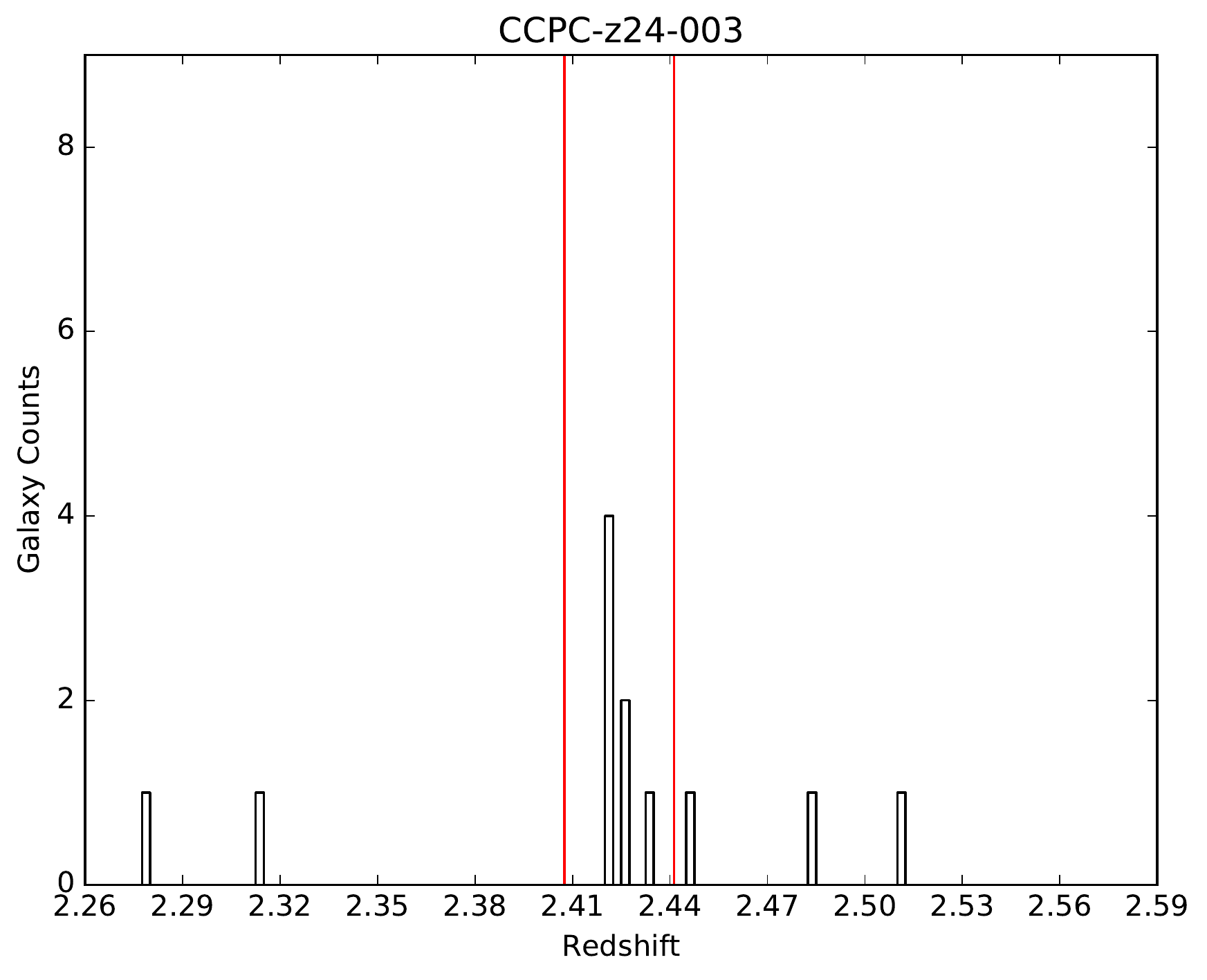}
\label{fig:CCPC-z24-003}
\end{subfigure}
\hfill
\end{figure*}

\begin{figure*}
\centering
\begin{subfigure}
\centering
\includegraphics[height=7.5cm,width=7.5cm]{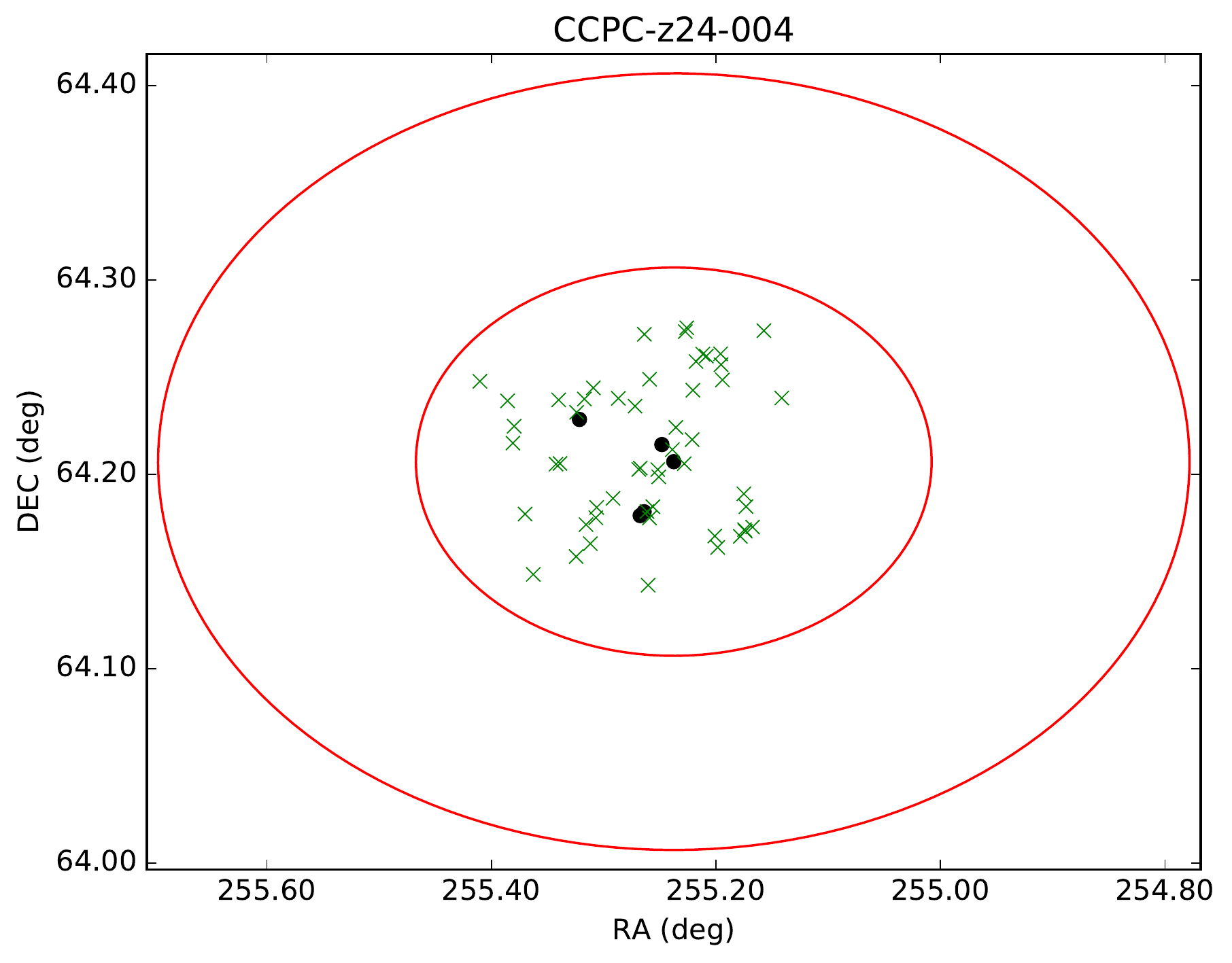}
\label{fig:CCPC-z24-004_sky}
\end{subfigure}
\hfill
\begin{subfigure}
\centering
\includegraphics[scale=0.52]{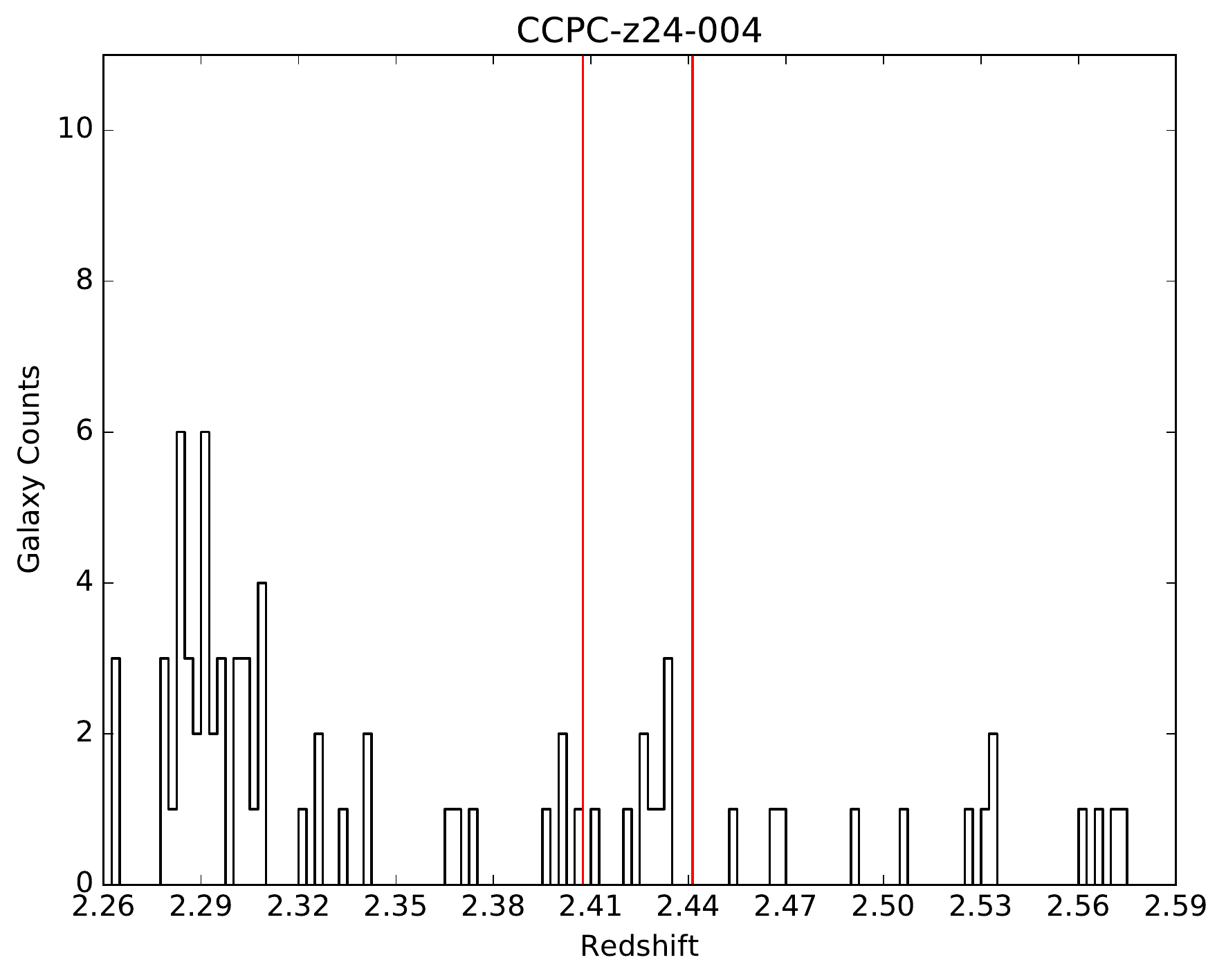}
\label{fig:CCPC-z24-004}
\end{subfigure}
\hfill
\end{figure*}

\begin{figure*}
\centering
\begin{subfigure}
\centering
\includegraphics[height=7.5cm,width=7.5cm]{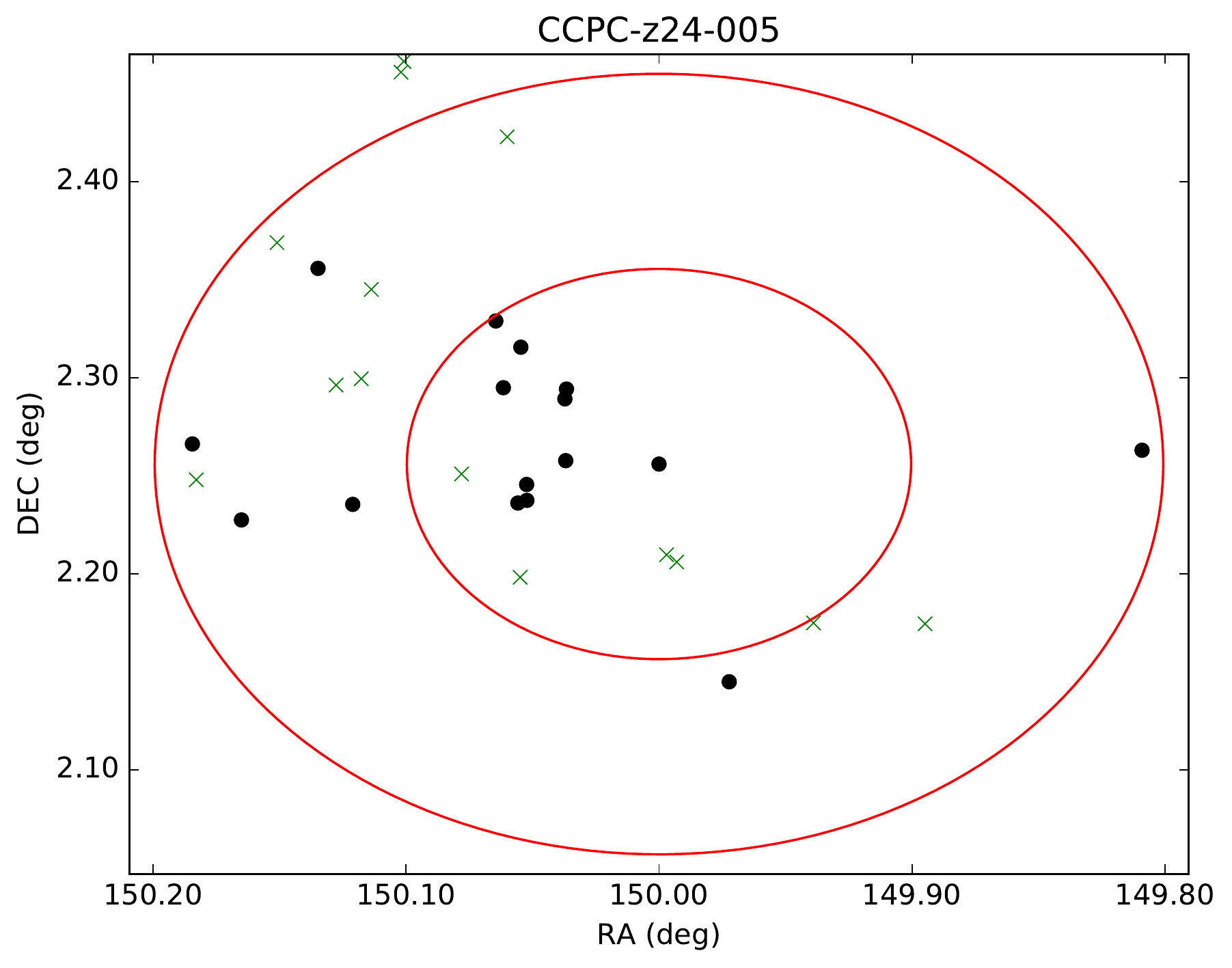}
\label{fig:CCPC-z24-005_sky}
\end{subfigure}
\hfill
\begin{subfigure}
\centering
\includegraphics[scale=0.52]{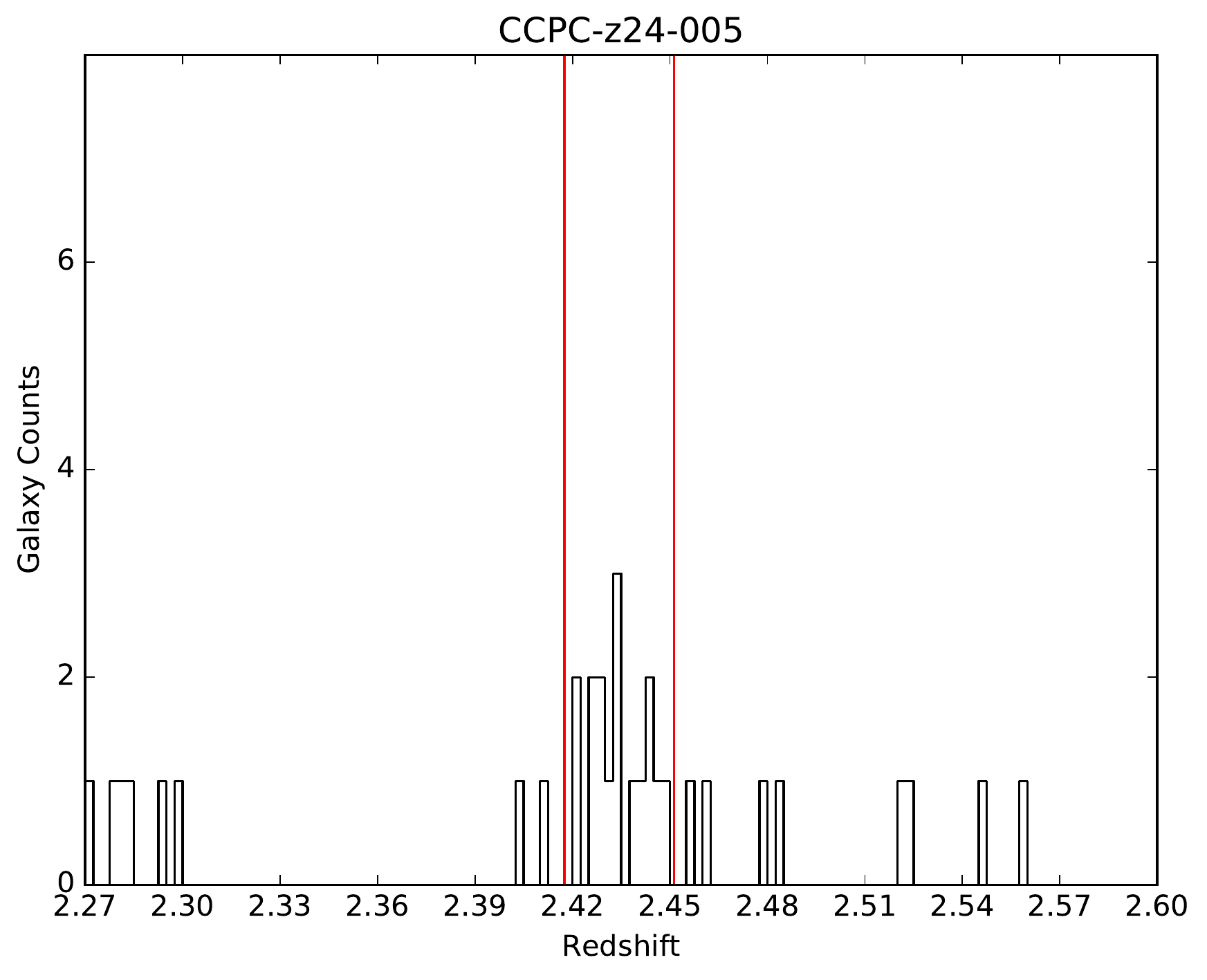}
\label{fig:CCPC-z24-005}
\end{subfigure}
\hfill
\end{figure*}
\clearpage 

\begin{figure*}
\centering
\begin{subfigure}
\centering
\includegraphics[height=7.5cm,width=7.5cm]{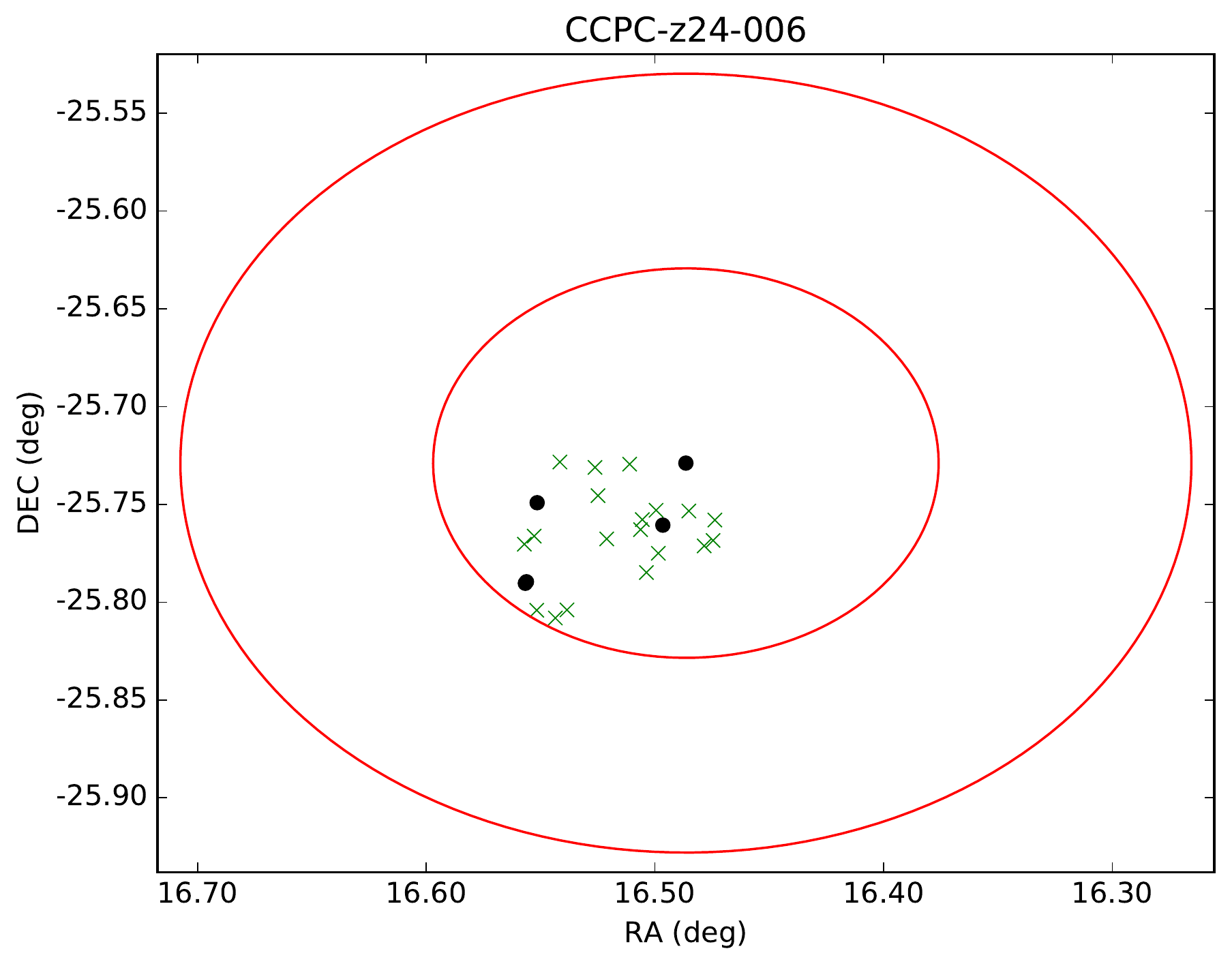}
\label{fig:CCPC-z24-006_sky}
\end{subfigure}
\hfill
\begin{subfigure}
\centering
\includegraphics[scale=0.52]{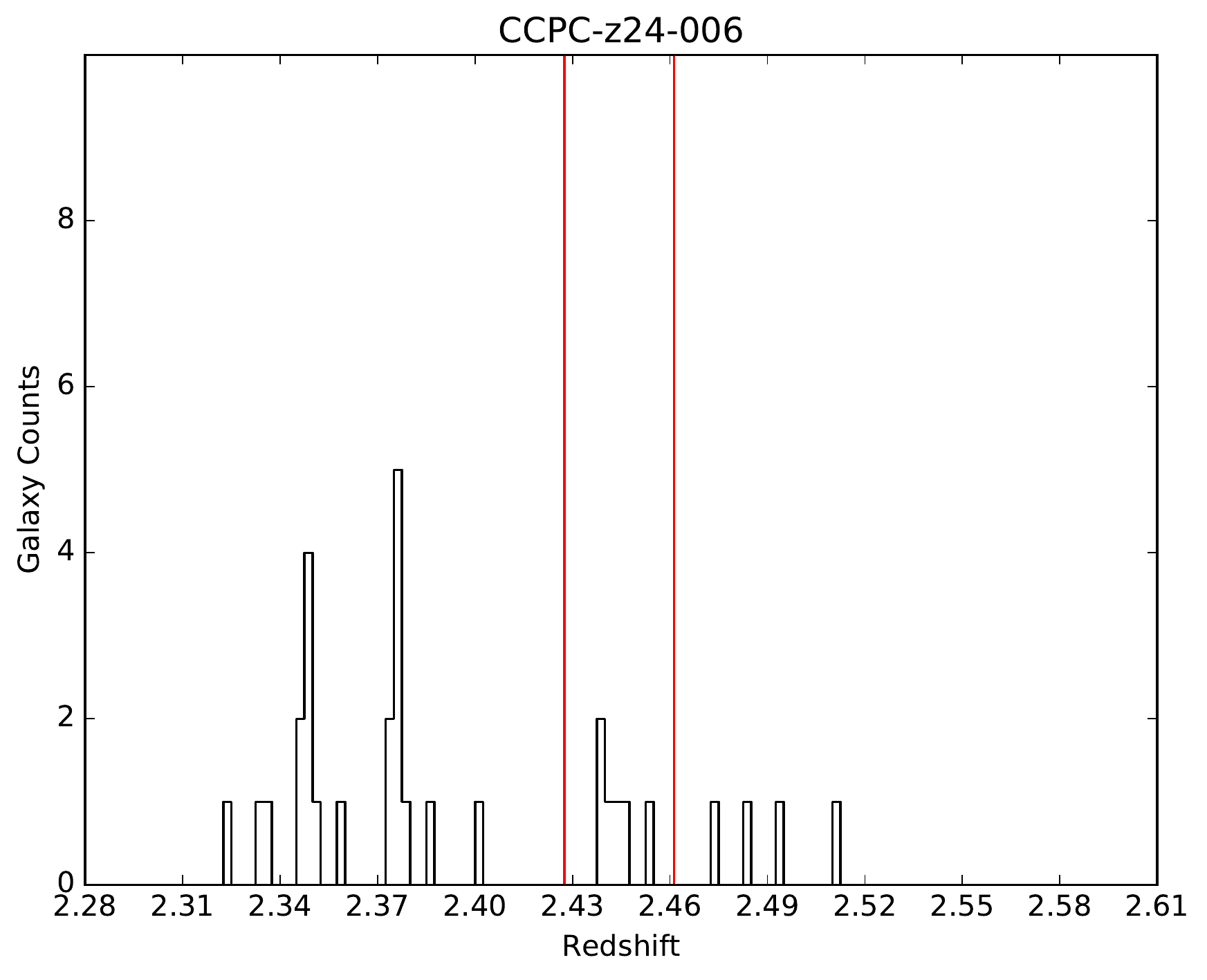}
\label{fig:CCPC-z24-006}
\end{subfigure}
\hfill
\end{figure*}

\begin{figure*}
\centering
\begin{subfigure}
\centering
\includegraphics[height=7.5cm,width=7.5cm]{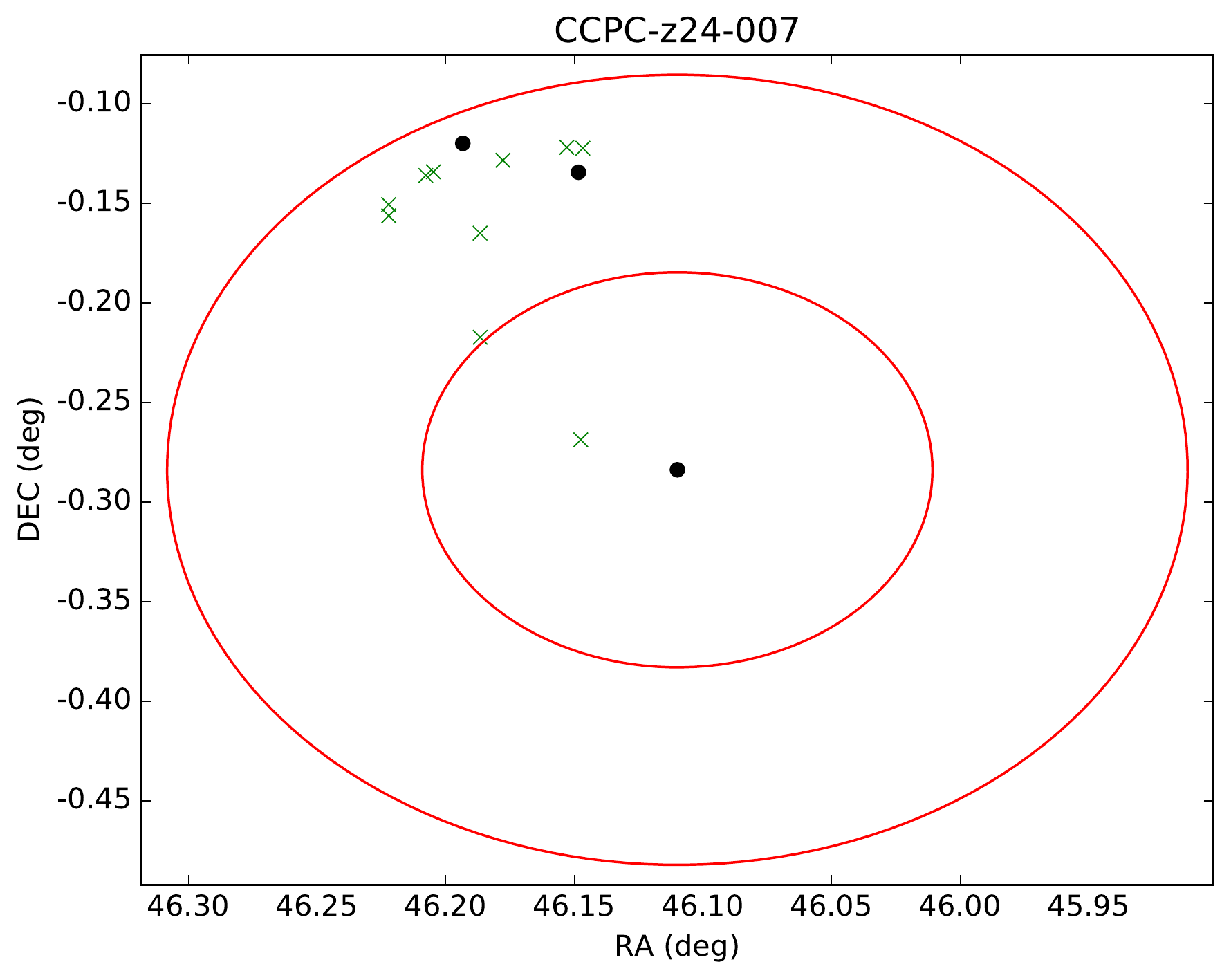}
\label{fig:CCPC-z24-007_sky}
\end{subfigure}
\hfill
\begin{subfigure}
\centering
\includegraphics[scale=0.52]{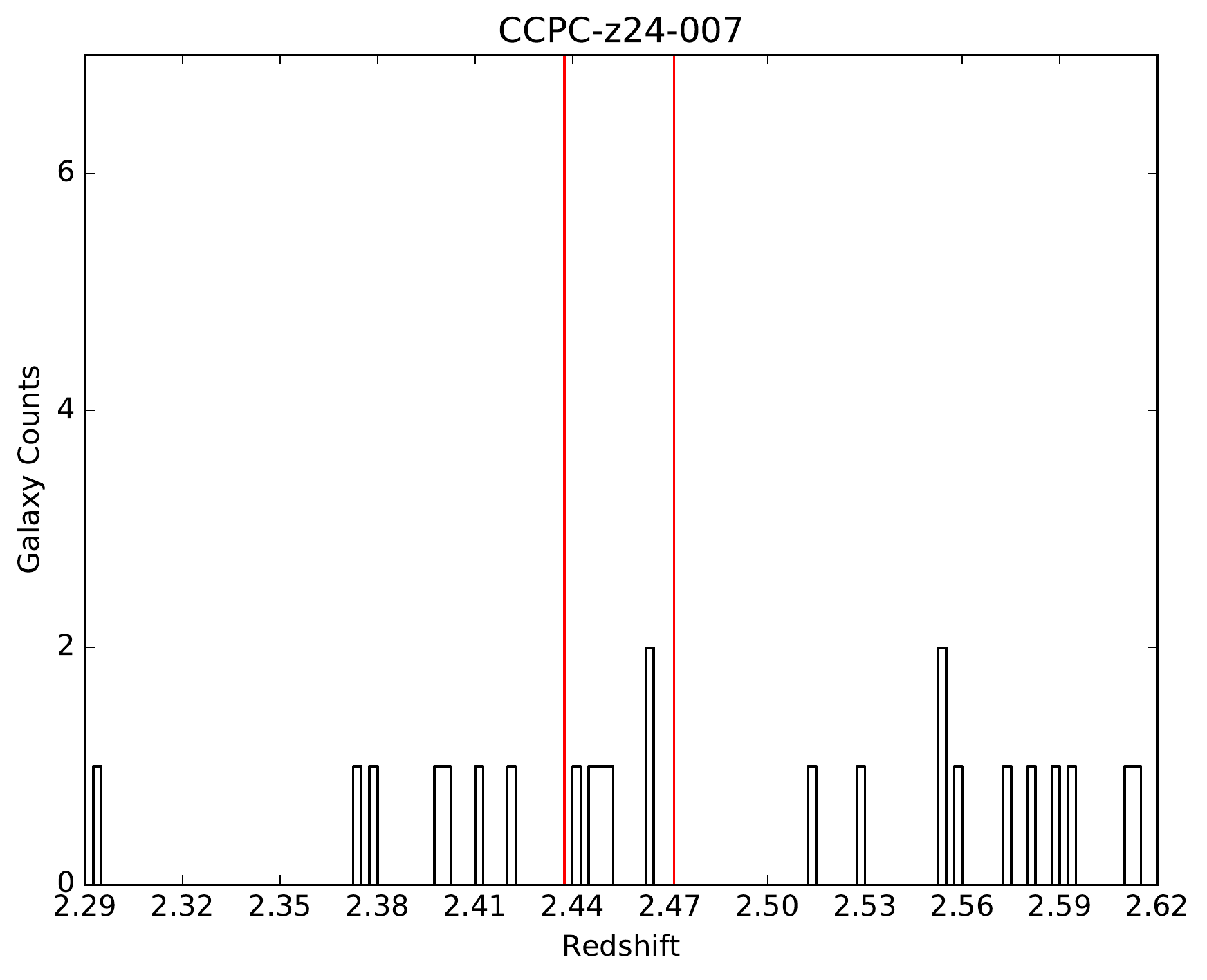}
\label{fig:CCPC-z24-007}
\end{subfigure}
\hfill
\end{figure*}

\begin{figure*}
\centering
\begin{subfigure}
\centering
\includegraphics[height=7.5cm,width=7.5cm]{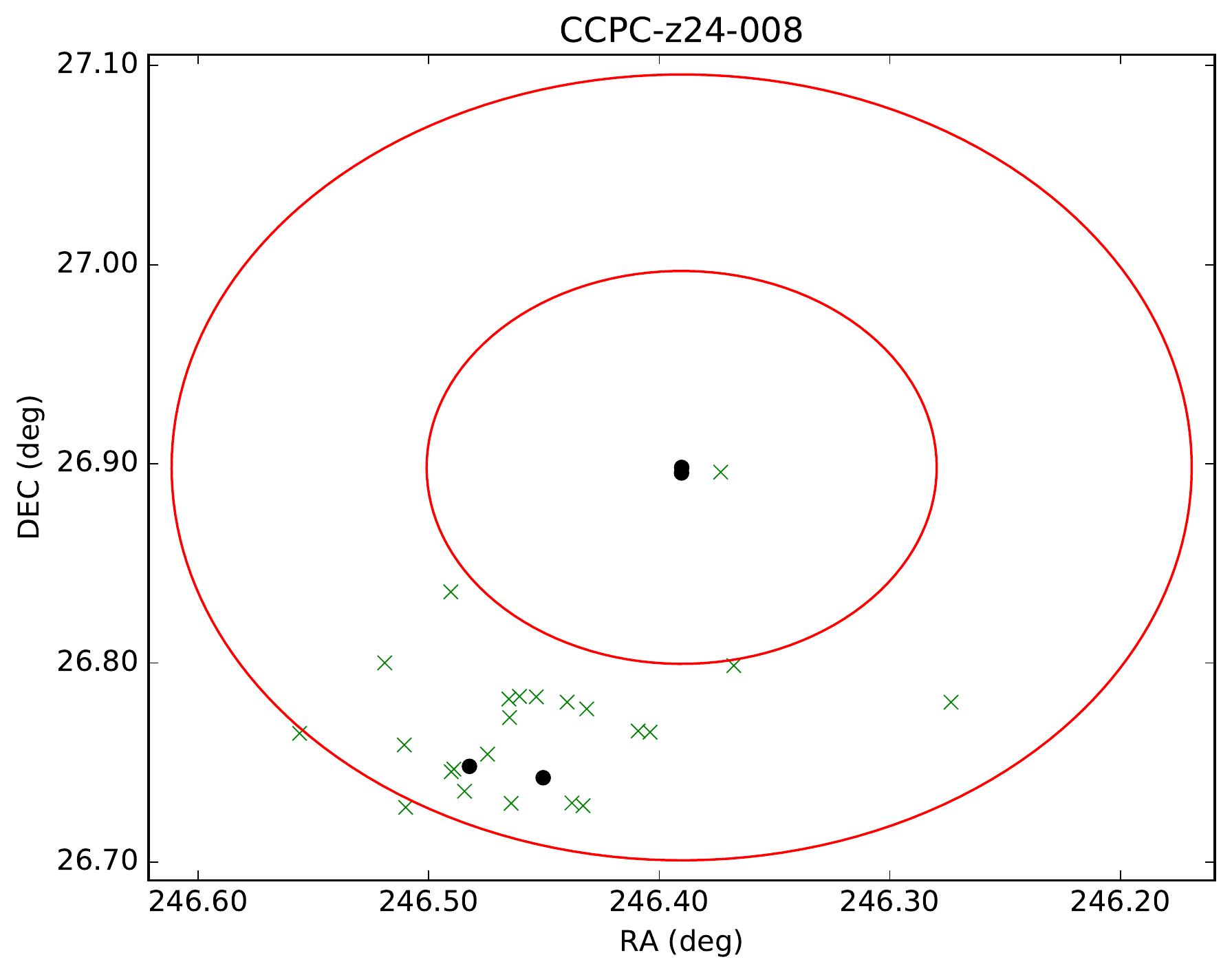}
\label{fig:CCPC-z24-008_sky}
\end{subfigure}
\hfill
\begin{subfigure}
\centering
\includegraphics[scale=0.52]{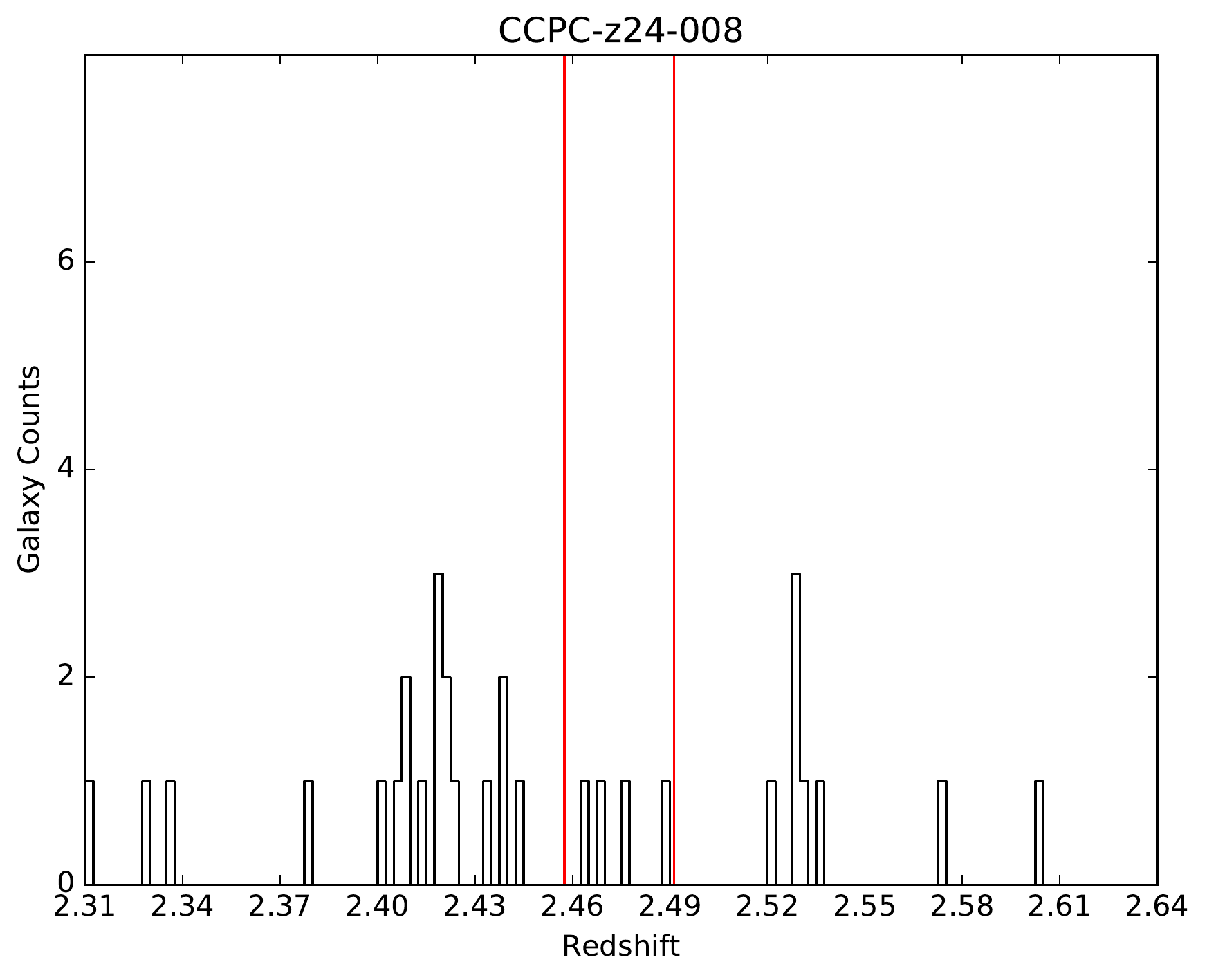}
\label{fig:CCPC-z24-008}
\end{subfigure}
\hfill
\end{figure*}
\clearpage 

\begin{figure*}
\centering
\begin{subfigure}
\centering
\includegraphics[height=7.5cm,width=7.5cm]{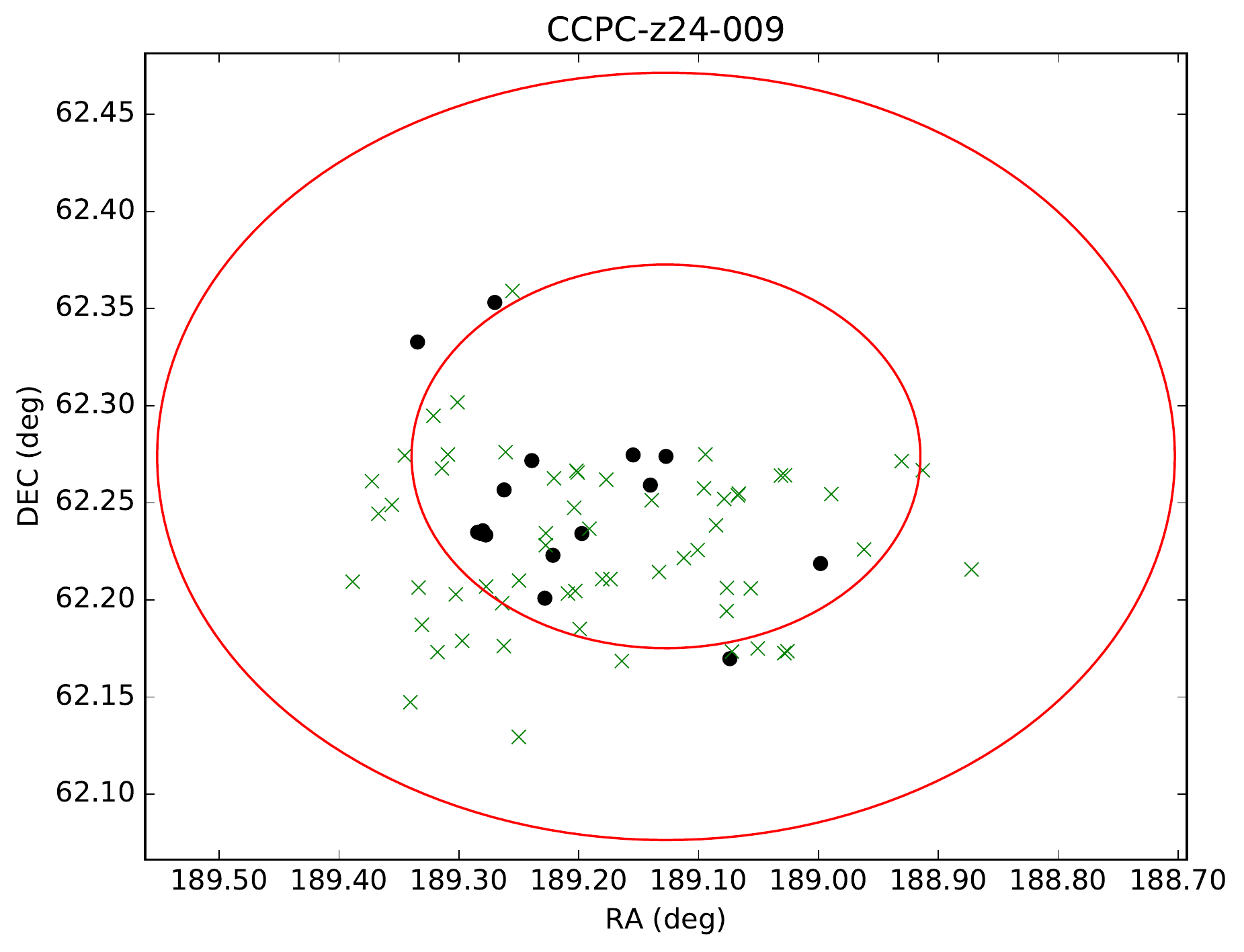}
\label{fig:CCPC-z24-009_sky}
\end{subfigure}
\hfill
\begin{subfigure}
\centering
\includegraphics[scale=0.52]{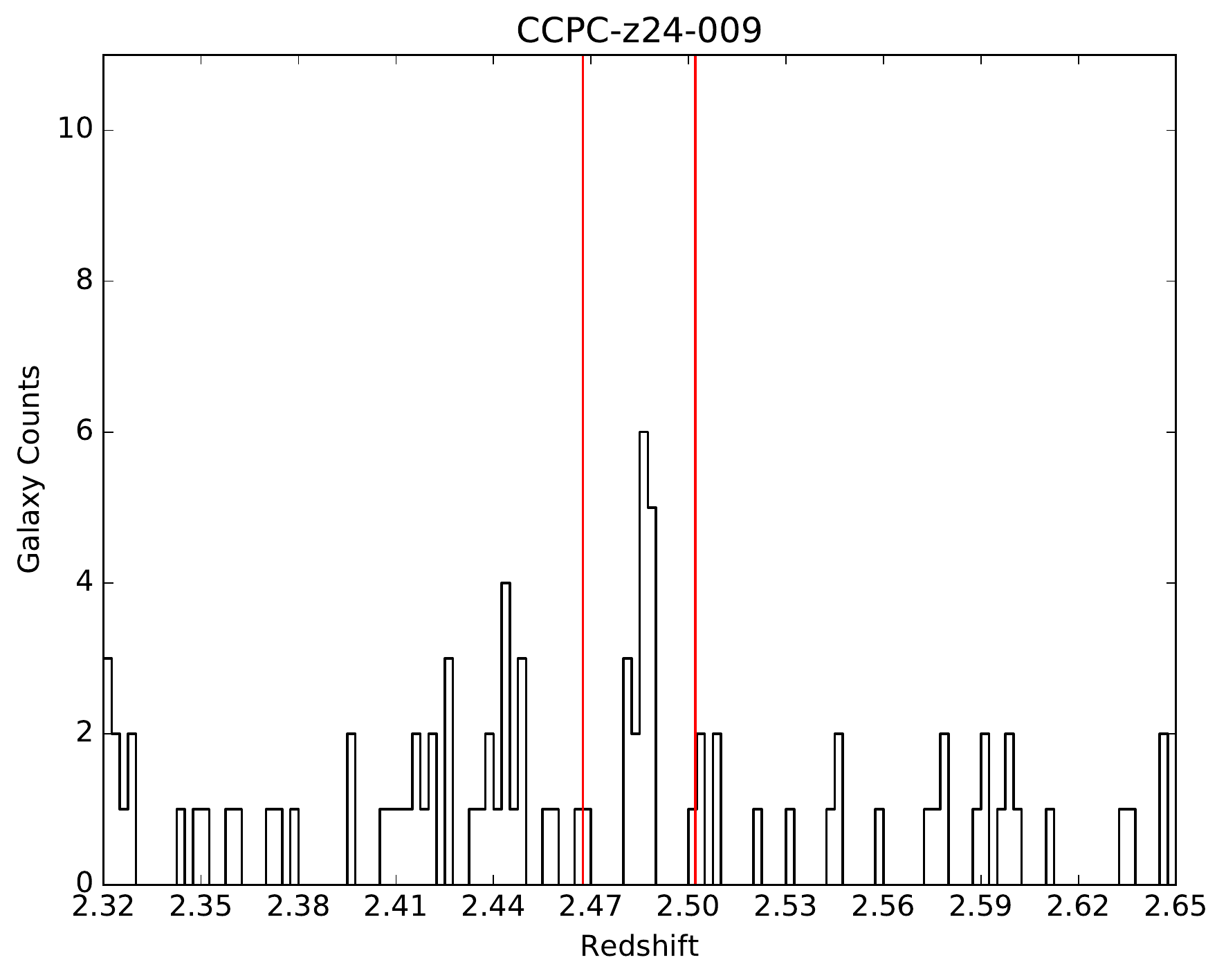}
\label{fig:CCPC-z24-009}
\end{subfigure}
\hfill
\end{figure*}

\begin{figure*}
\centering
\begin{subfigure}
\centering
\includegraphics[height=7.5cm,width=7.5cm]{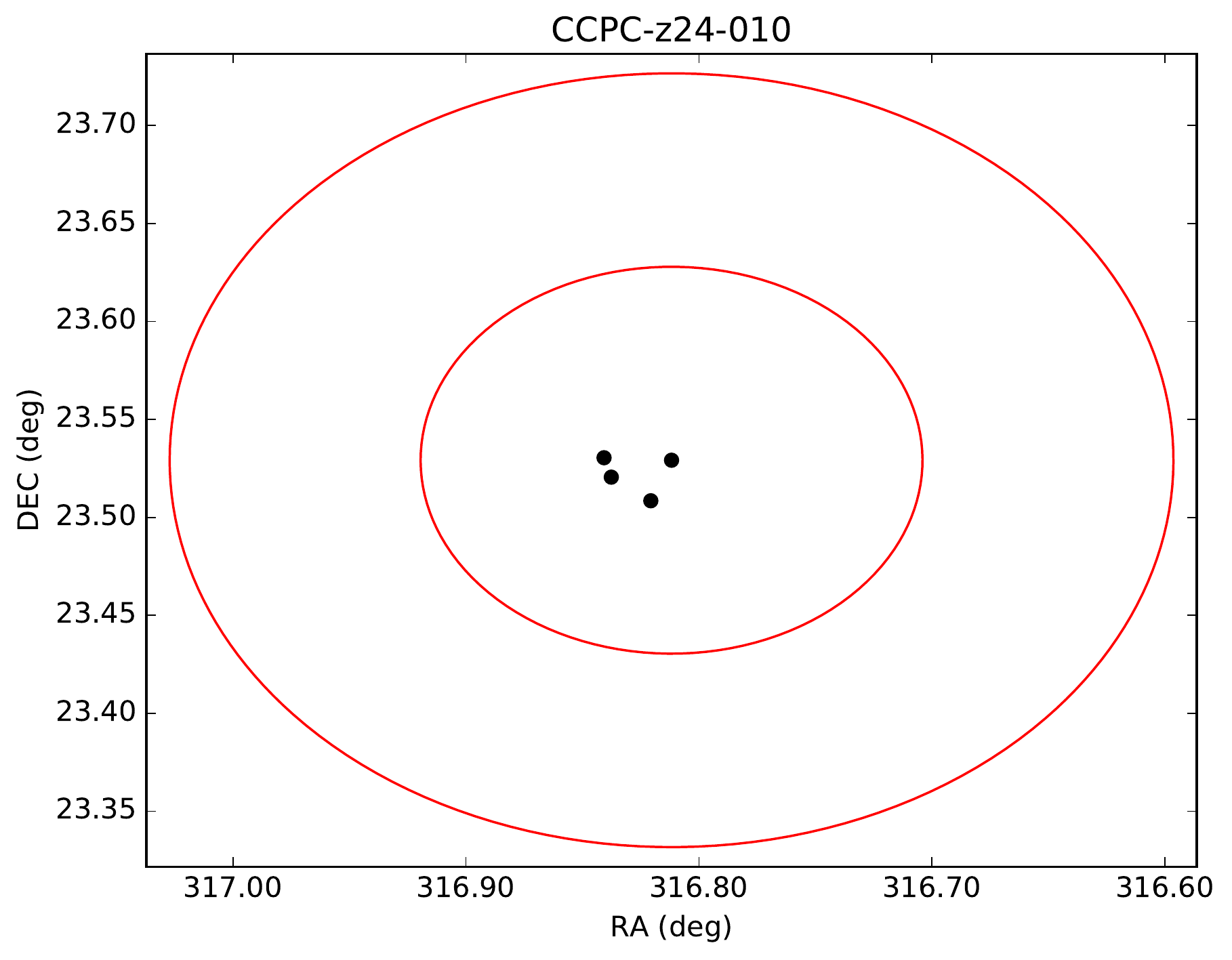}
\label{fig:CCPC-z24-010_sky}
\end{subfigure}
\hfill
\begin{subfigure}
\centering
\includegraphics[scale=0.52]{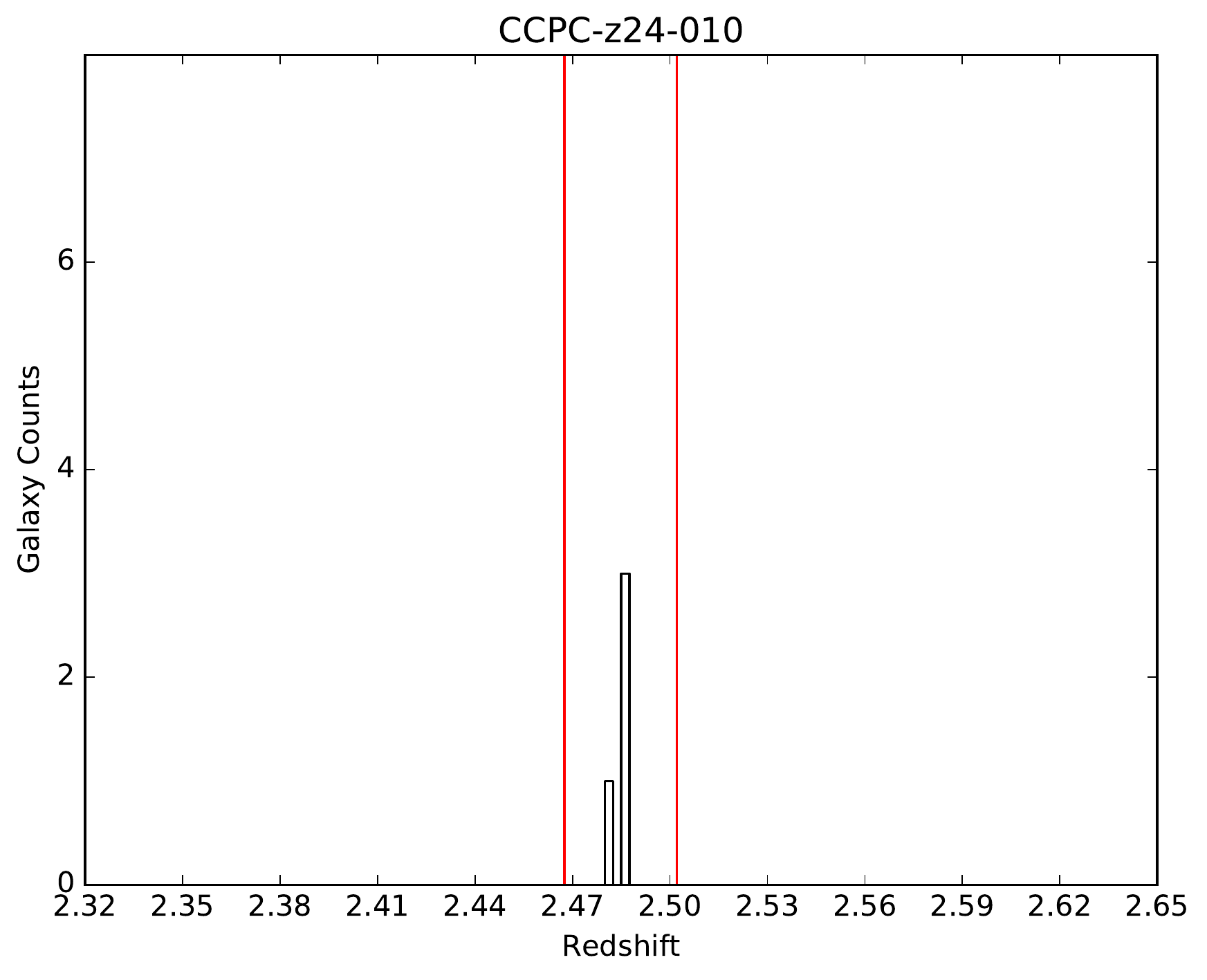}
\label{fig:CCPC-z24-010}
\end{subfigure}
\hfill
\end{figure*}

\begin{figure*}
\centering
\begin{subfigure}
\centering
\includegraphics[height=7.5cm,width=7.5cm]{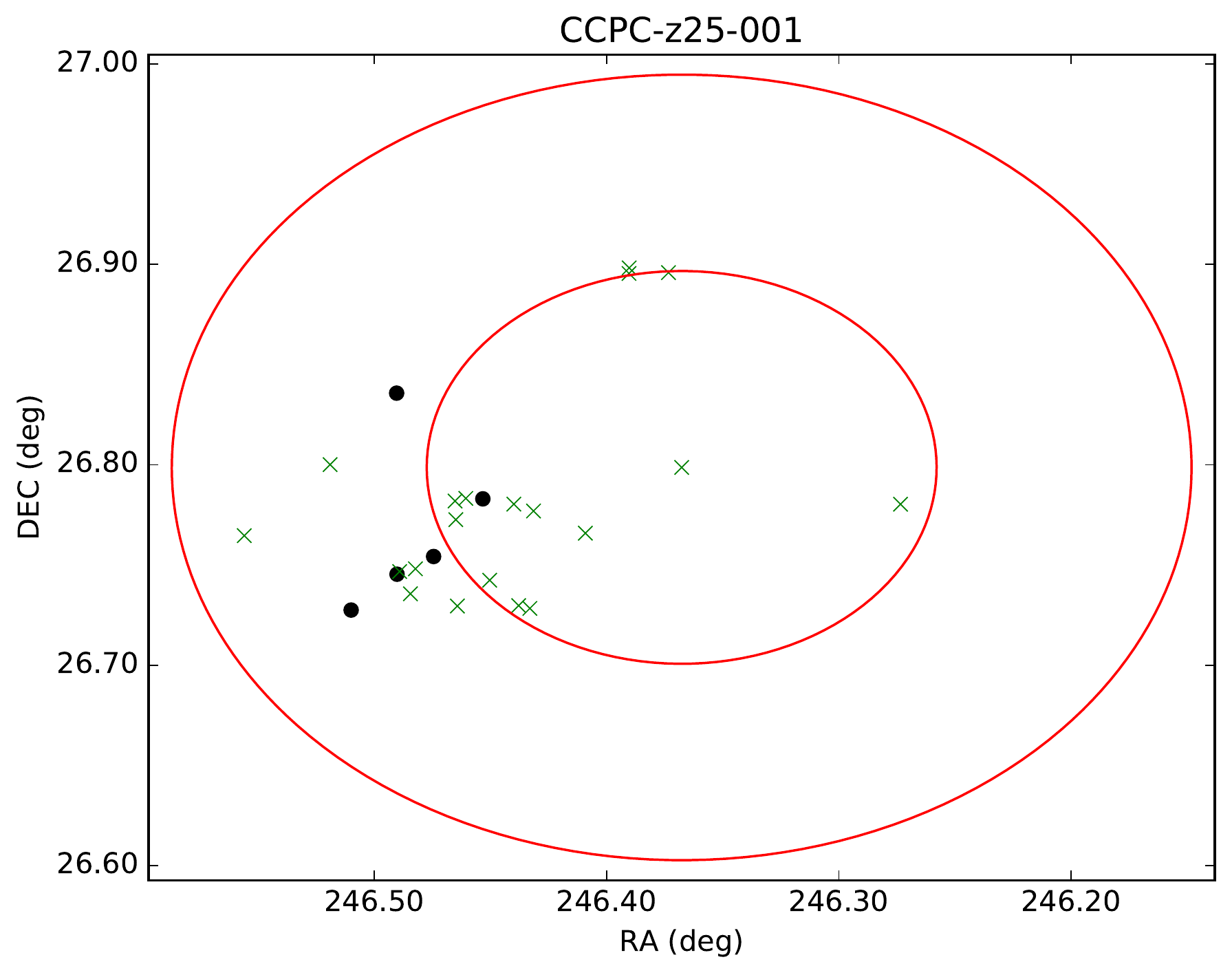}
\label{fig:CCPC-z25-001_sky}
\end{subfigure}
\hfill
\begin{subfigure}
\centering
\includegraphics[scale=0.52]{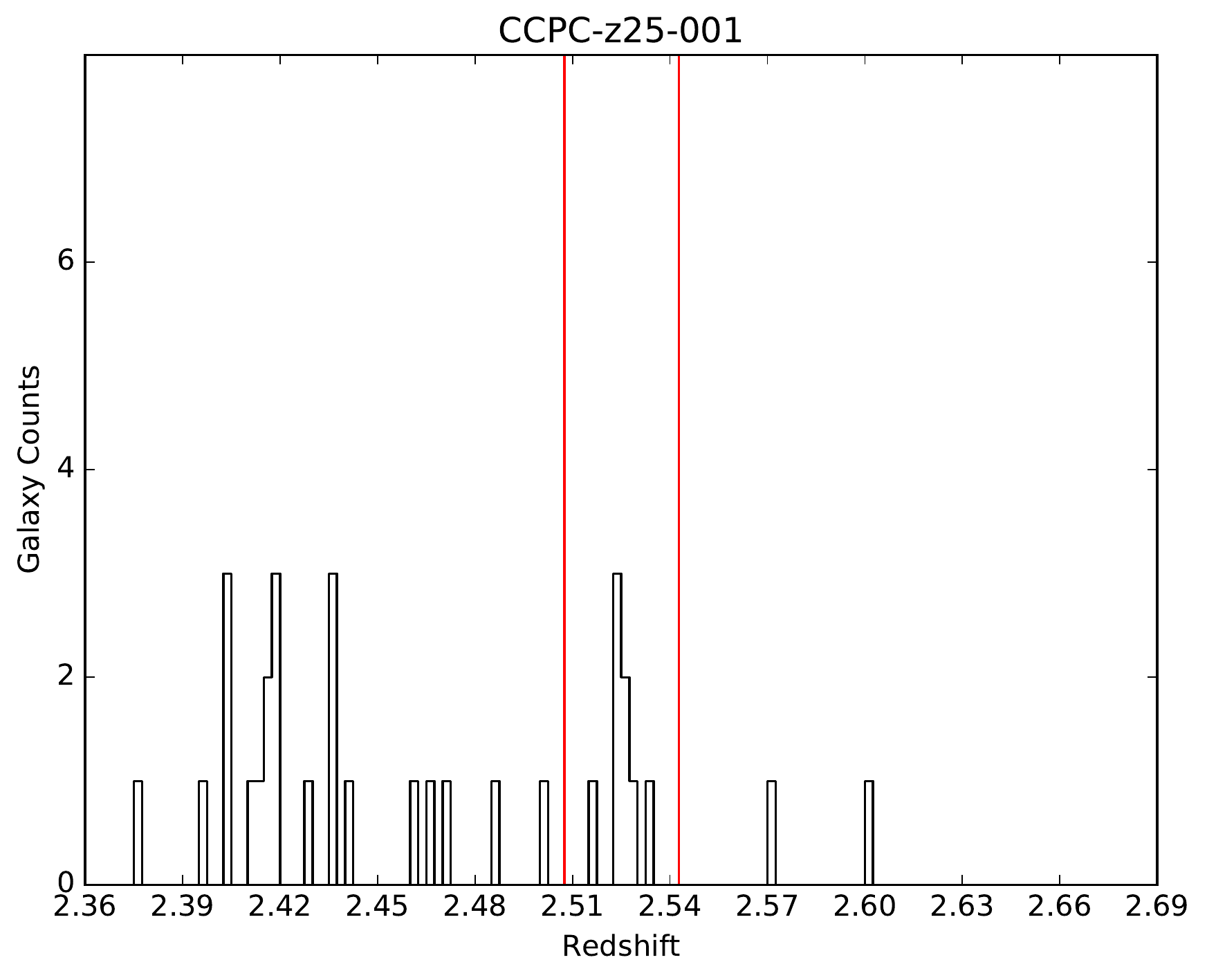}
\label{fig:CCPC-z25-001}
\end{subfigure}
\hfill
\end{figure*}
\clearpage 

\begin{figure*}
\centering
\begin{subfigure}
\centering
\includegraphics[height=7.5cm,width=7.5cm]{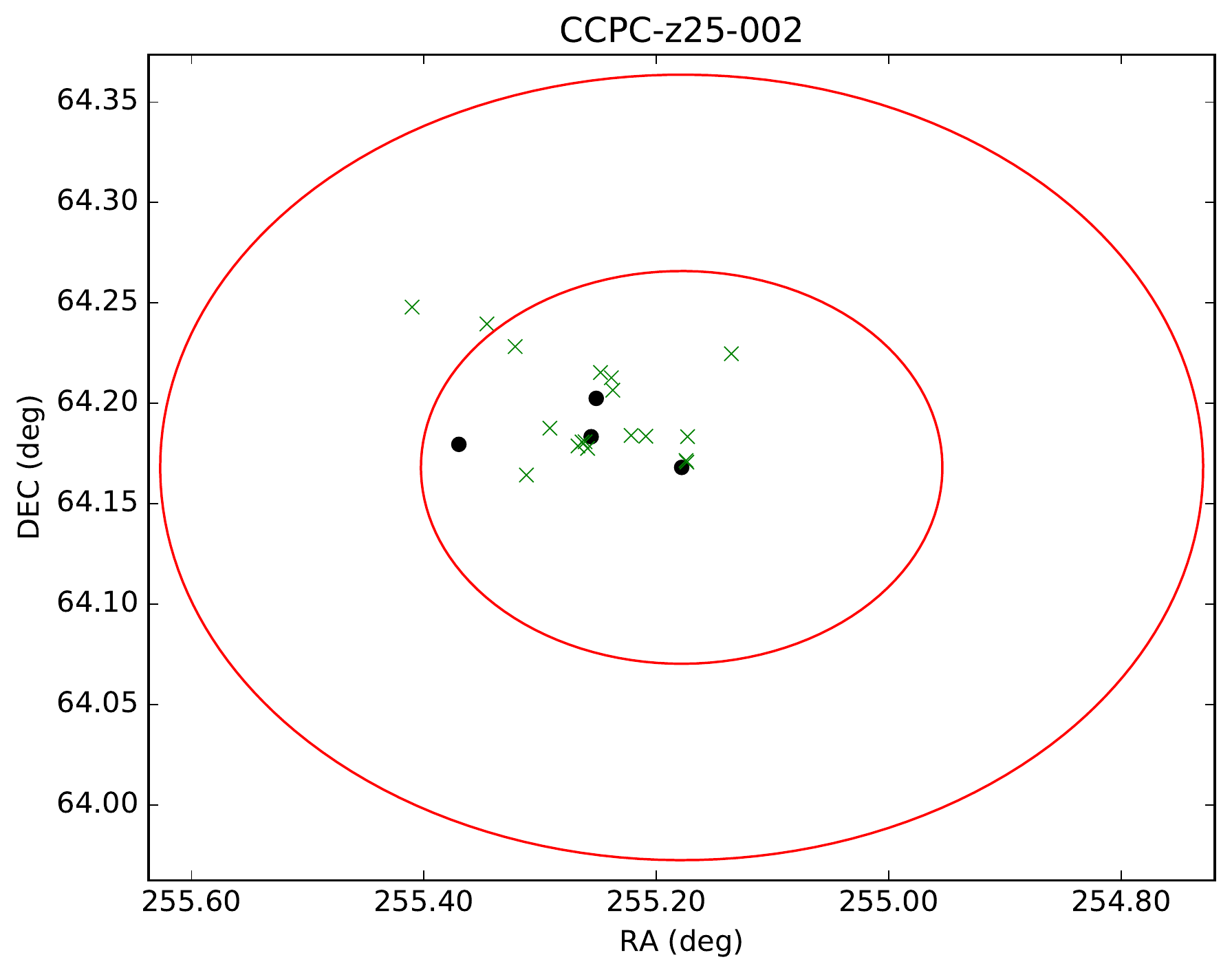}
\label{fig:CCPC-z25-002_sky}
\end{subfigure}
\hfill
\begin{subfigure}
\centering
\includegraphics[scale=0.52]{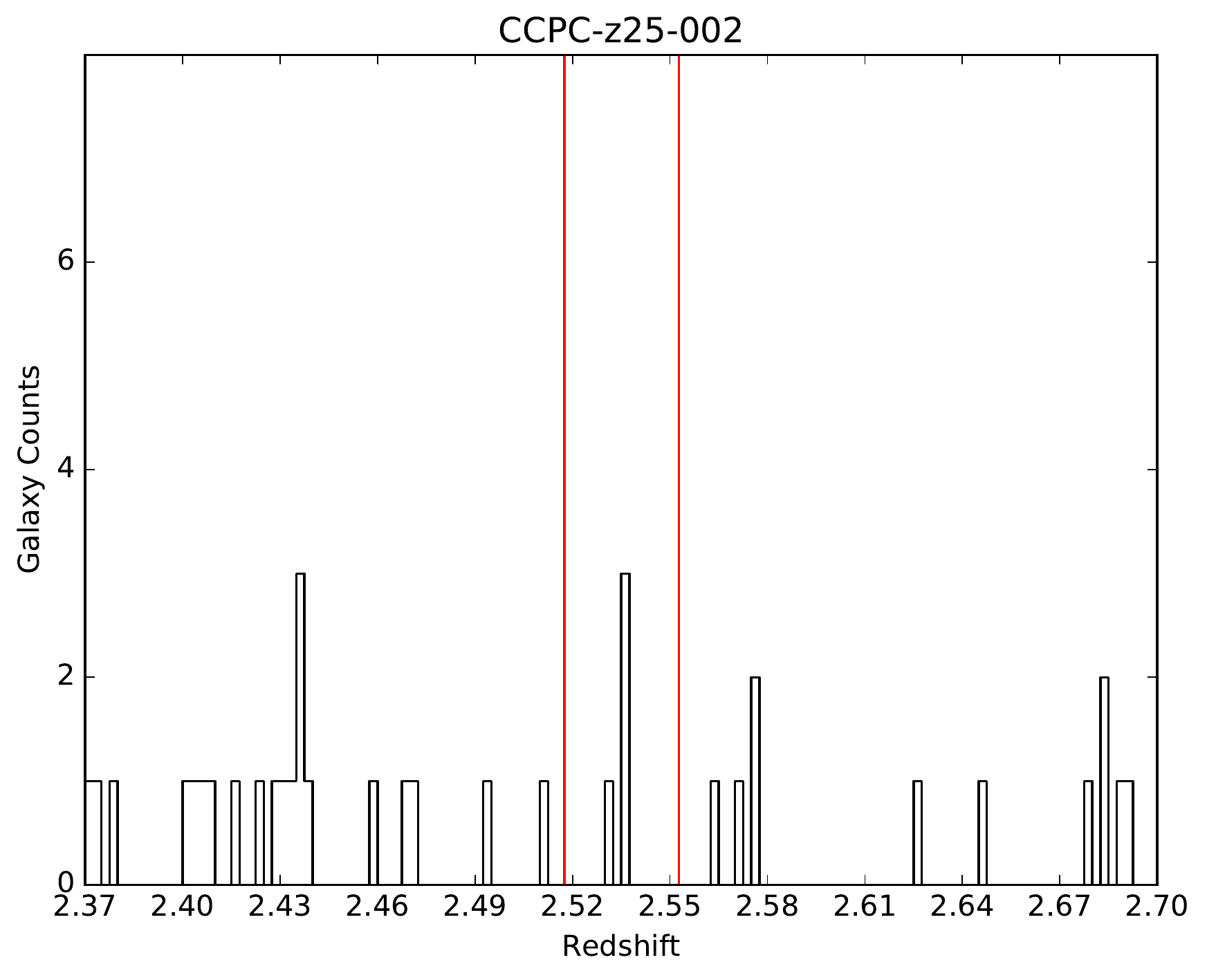}
\label{fig:CCPC-z25-002}
\end{subfigure}
\hfill
\end{figure*}

\begin{figure*}
\centering
\begin{subfigure}
\centering
\includegraphics[height=7.5cm,width=7.5cm]{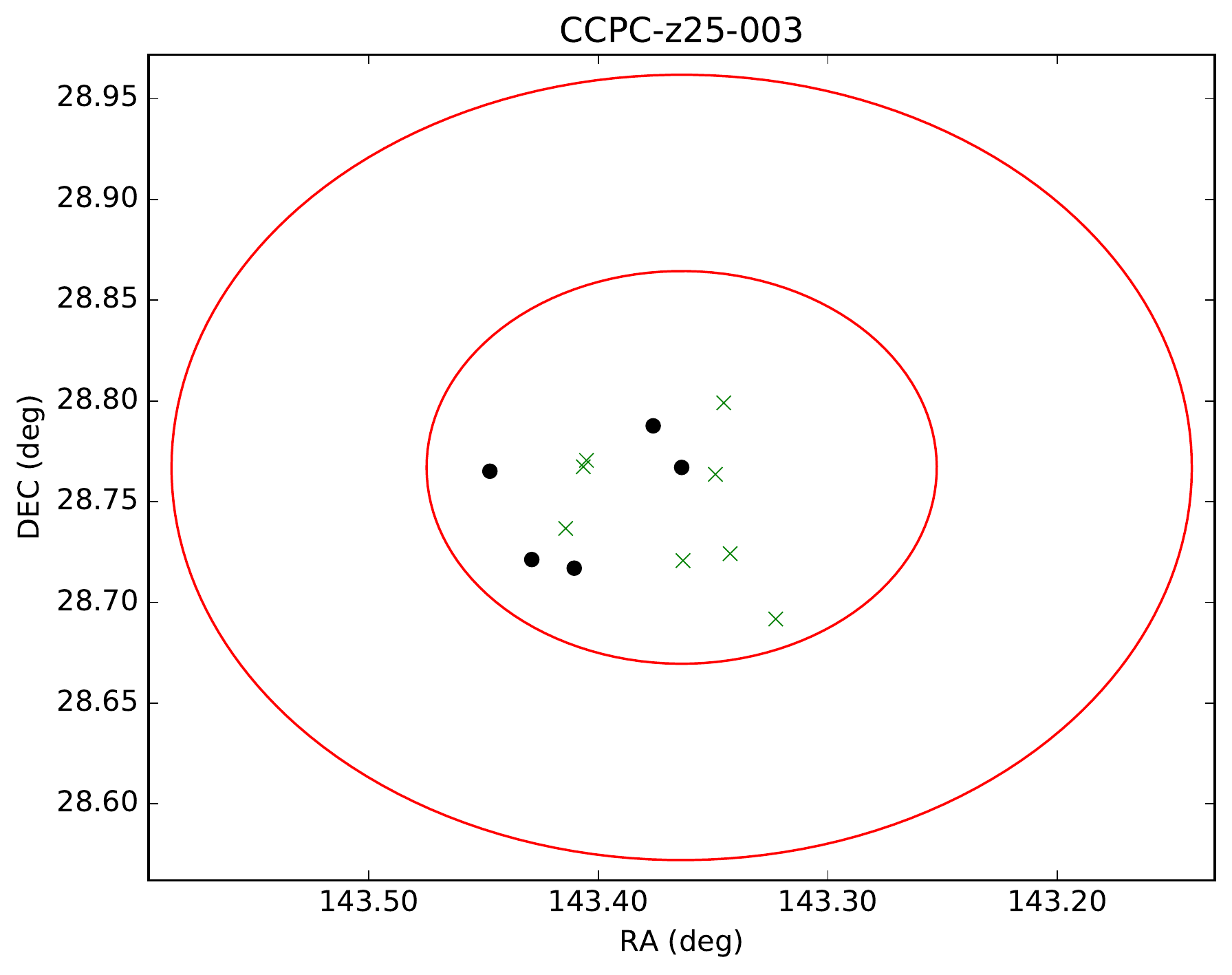}
\label{fig:CCPC-z25-003_sky}
\end{subfigure}
\hfill
\begin{subfigure}
\centering
\includegraphics[scale=0.52]{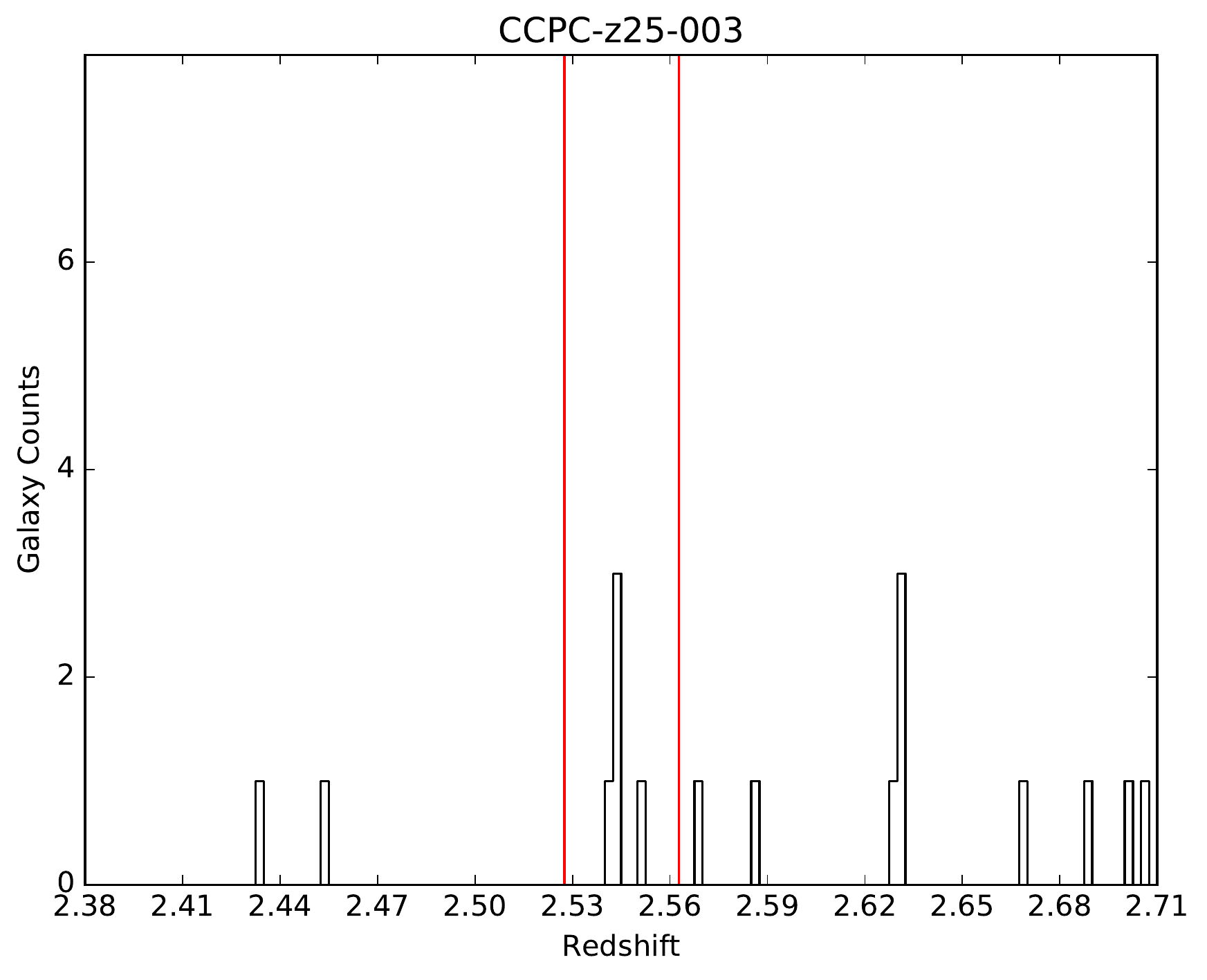}
\label{fig:CCPC-z25-003}
\end{subfigure}
\hfill
\end{figure*}

\begin{figure*}
\centering
\begin{subfigure}
\centering
\includegraphics[height=7.5cm,width=7.5cm]{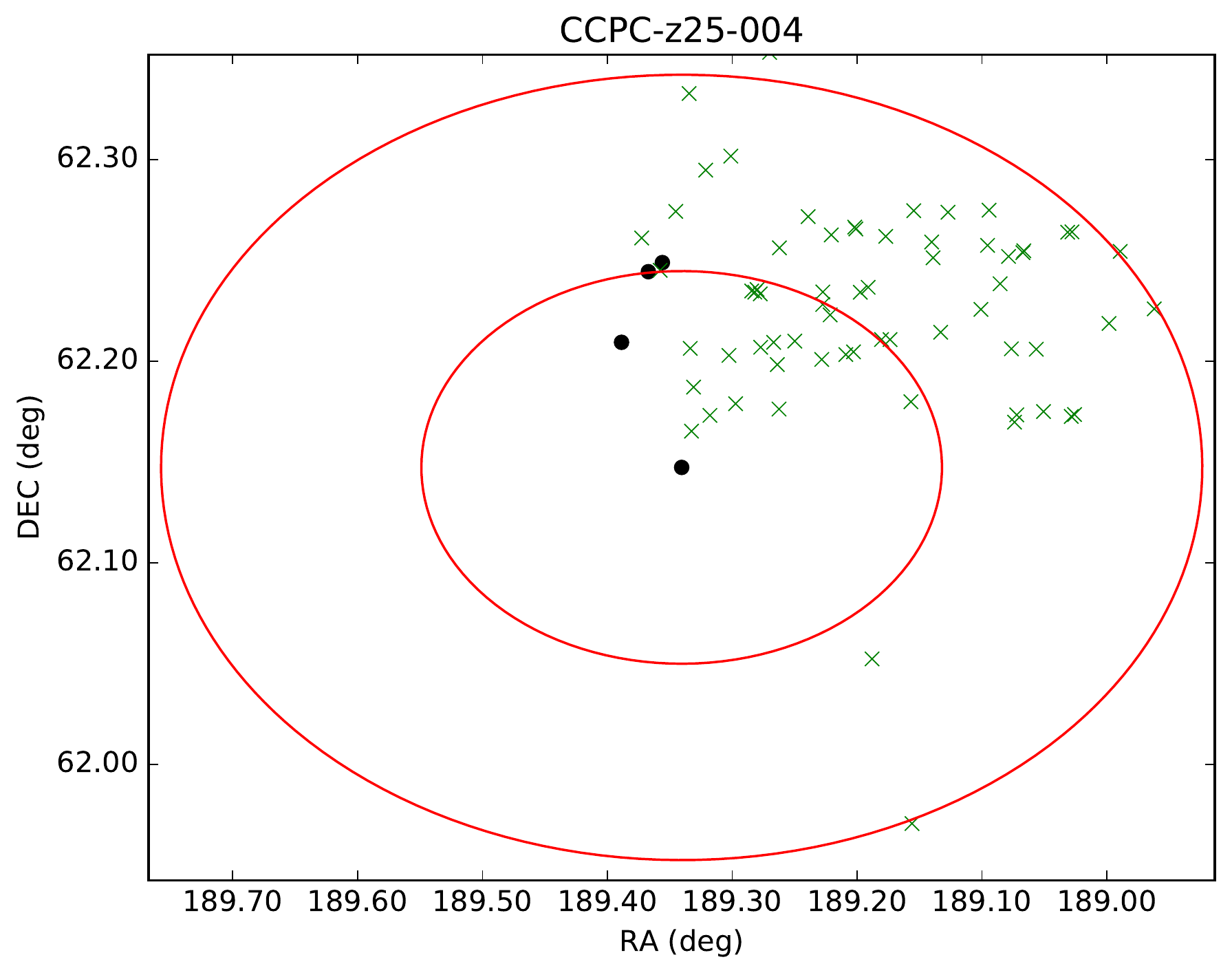}
\label{fig:CCPC-z25-004_sky}
\end{subfigure}
\hfill
\begin{subfigure}
\centering
\includegraphics[scale=0.52]{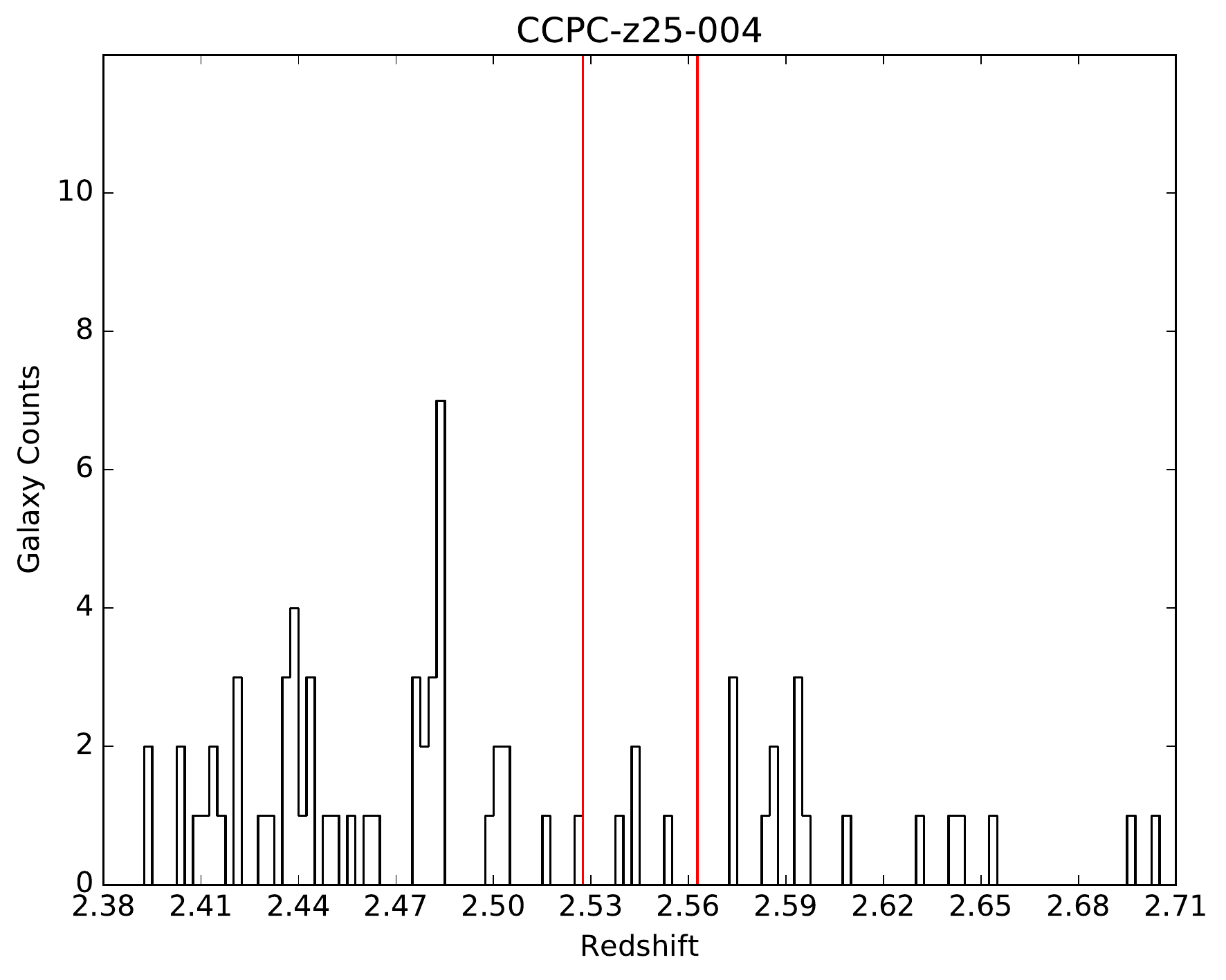}
\label{fig:CCPC-z25-004}
\end{subfigure}
\hfill
\end{figure*}
\clearpage 

\begin{figure*}
\centering
\begin{subfigure}
\centering
\includegraphics[height=7.5cm,width=7.5cm]{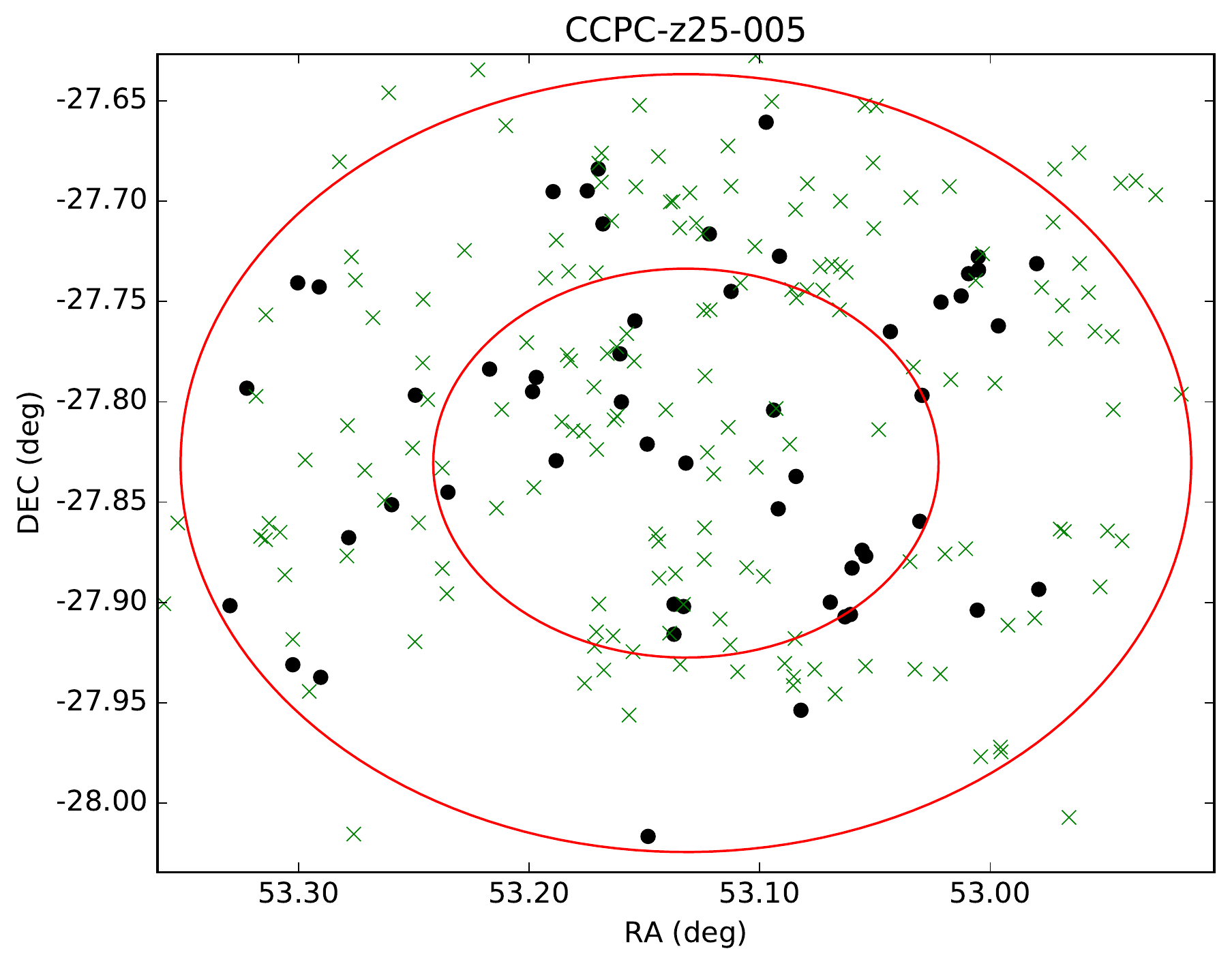}
\label{fig:CCPC-z25-005_sky}
\end{subfigure}
\hfill
\begin{subfigure}
\centering
\includegraphics[scale=0.52]{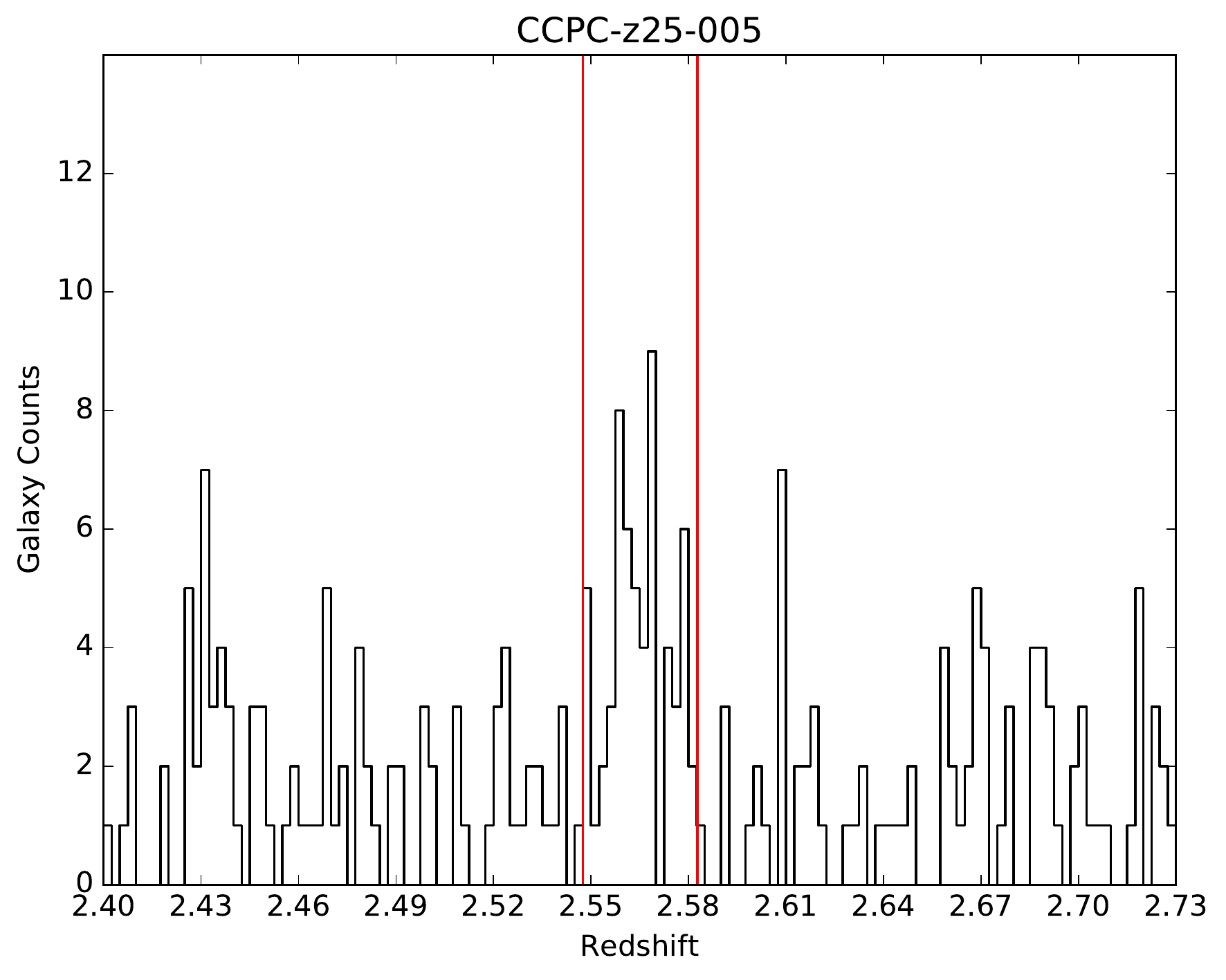}
\label{fig:CCPC-z25-005}
\end{subfigure}
\hfill
\end{figure*}

\begin{figure*}
\centering
\begin{subfigure}
\centering
\includegraphics[height=7.5cm,width=7.5cm]{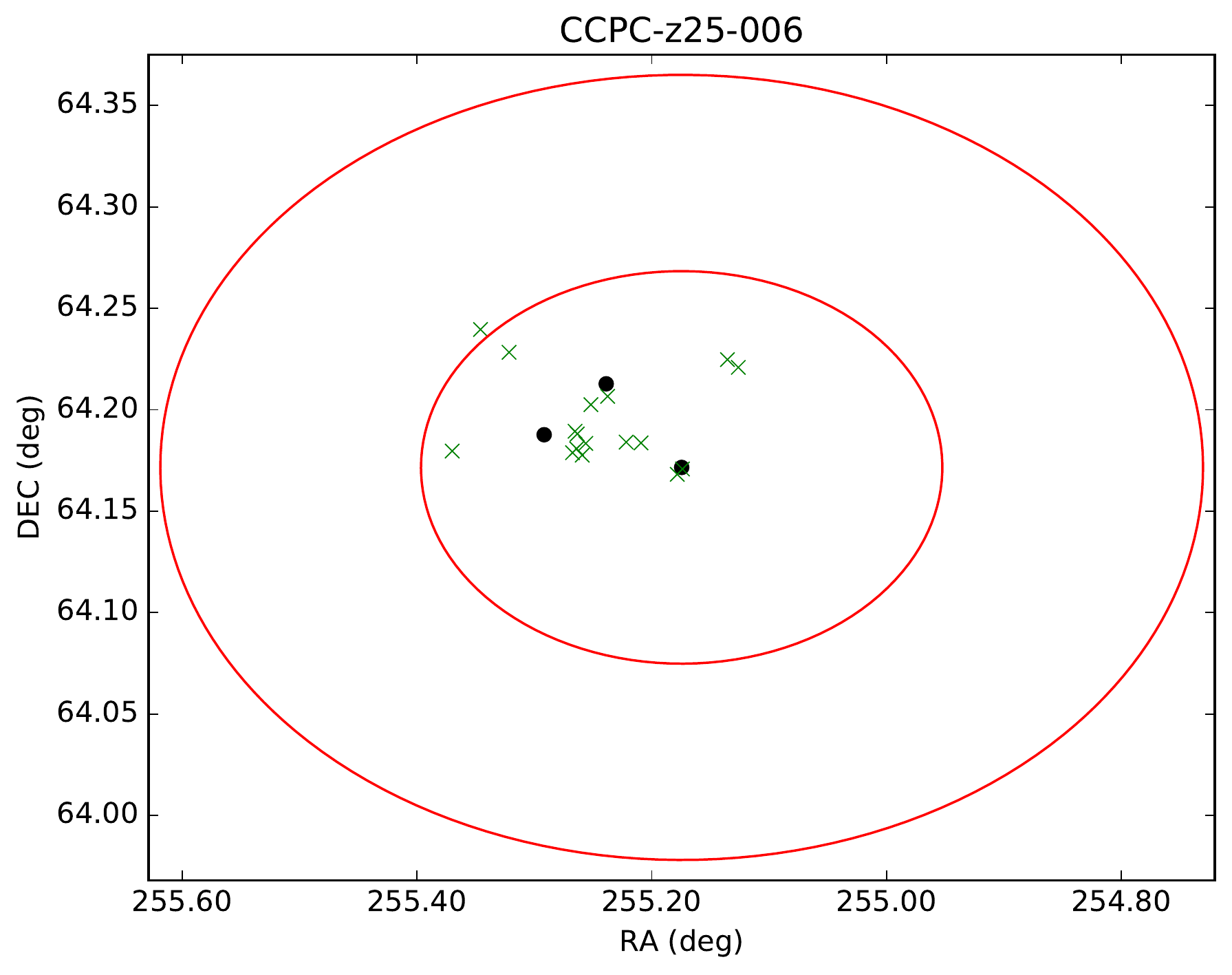}
\label{fig:CCPC-z25-006_sky}
\end{subfigure}
\hfill
\begin{subfigure}
\centering
\includegraphics[scale=0.52]{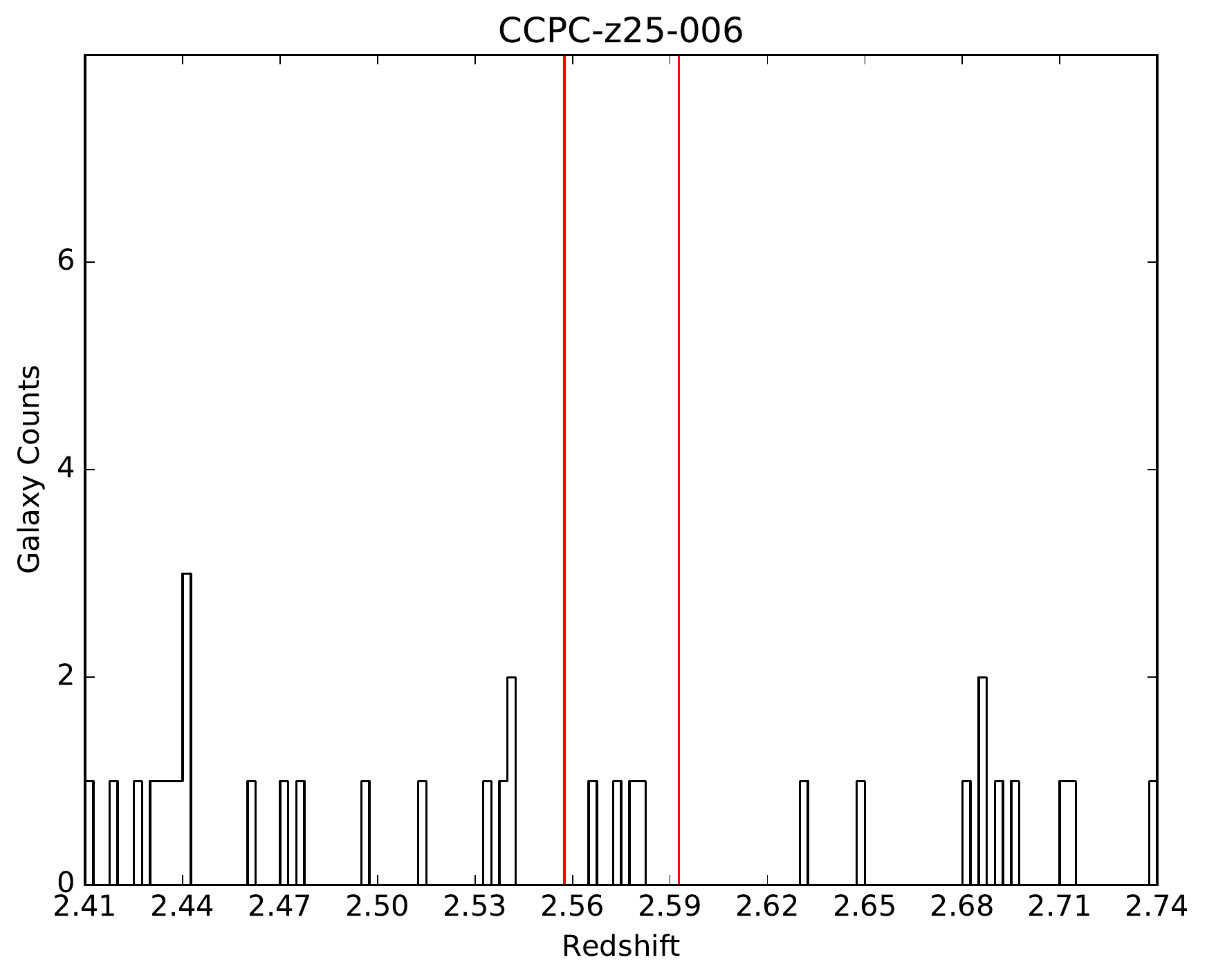}
\label{fig:CCPC-z25-006}
\end{subfigure}
\hfill
\end{figure*}

\begin{figure*}
\centering
\begin{subfigure}
\centering
\includegraphics[height=7.5cm,width=7.5cm]{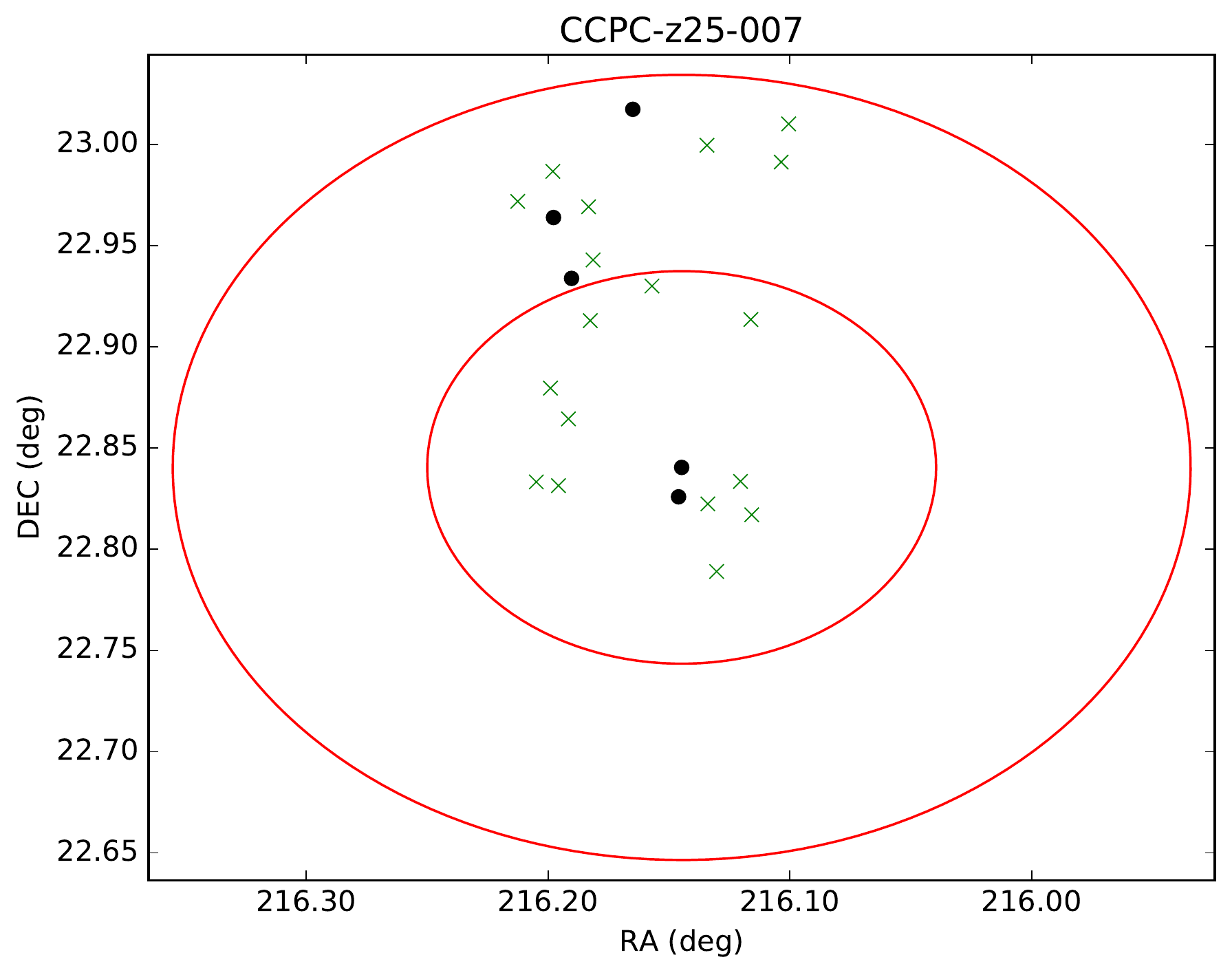}
\label{fig:CCPC-z25-007_sky}
\end{subfigure}
\hfill
\begin{subfigure}
\centering
\includegraphics[scale=0.52]{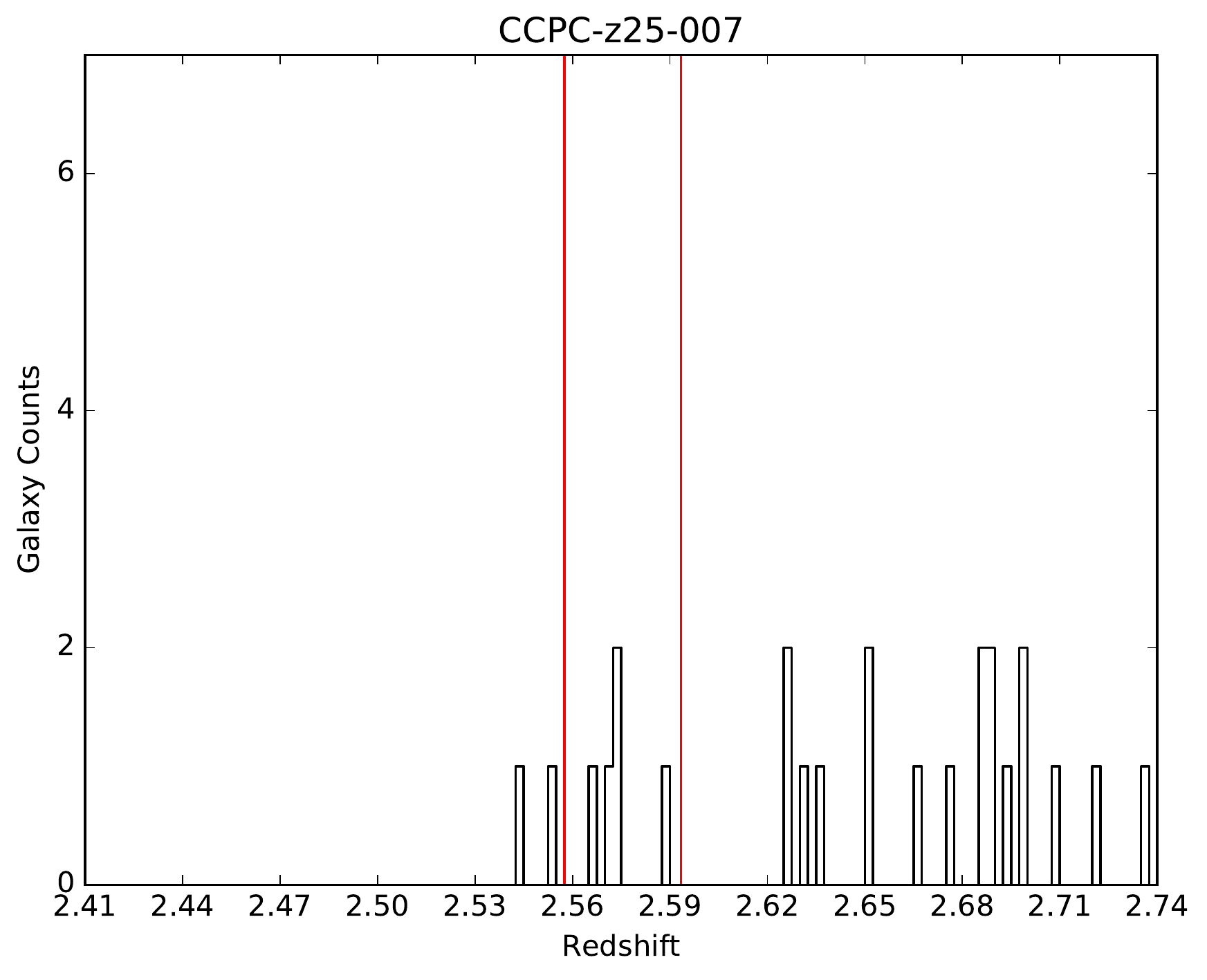}
\label{fig:CCPC-z25-007}
\end{subfigure}
\hfill
\end{figure*}
\clearpage 

\begin{figure*}
\centering
\begin{subfigure}
\centering
\includegraphics[height=7.5cm,width=7.5cm]{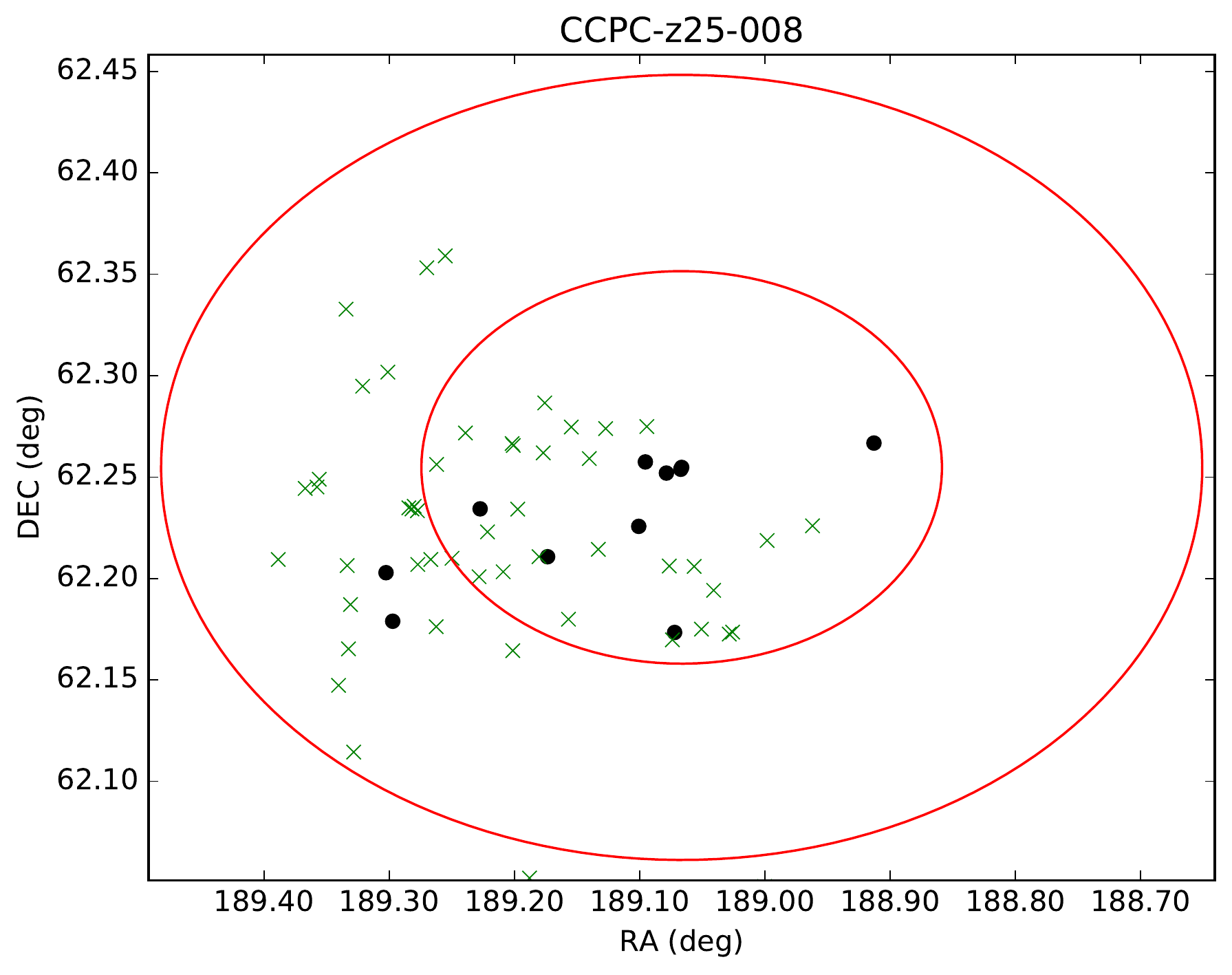}
\label{fig:CCPC-z25-008_sky}
\end{subfigure}
\hfill
\begin{subfigure}
\centering
\includegraphics[scale=0.52]{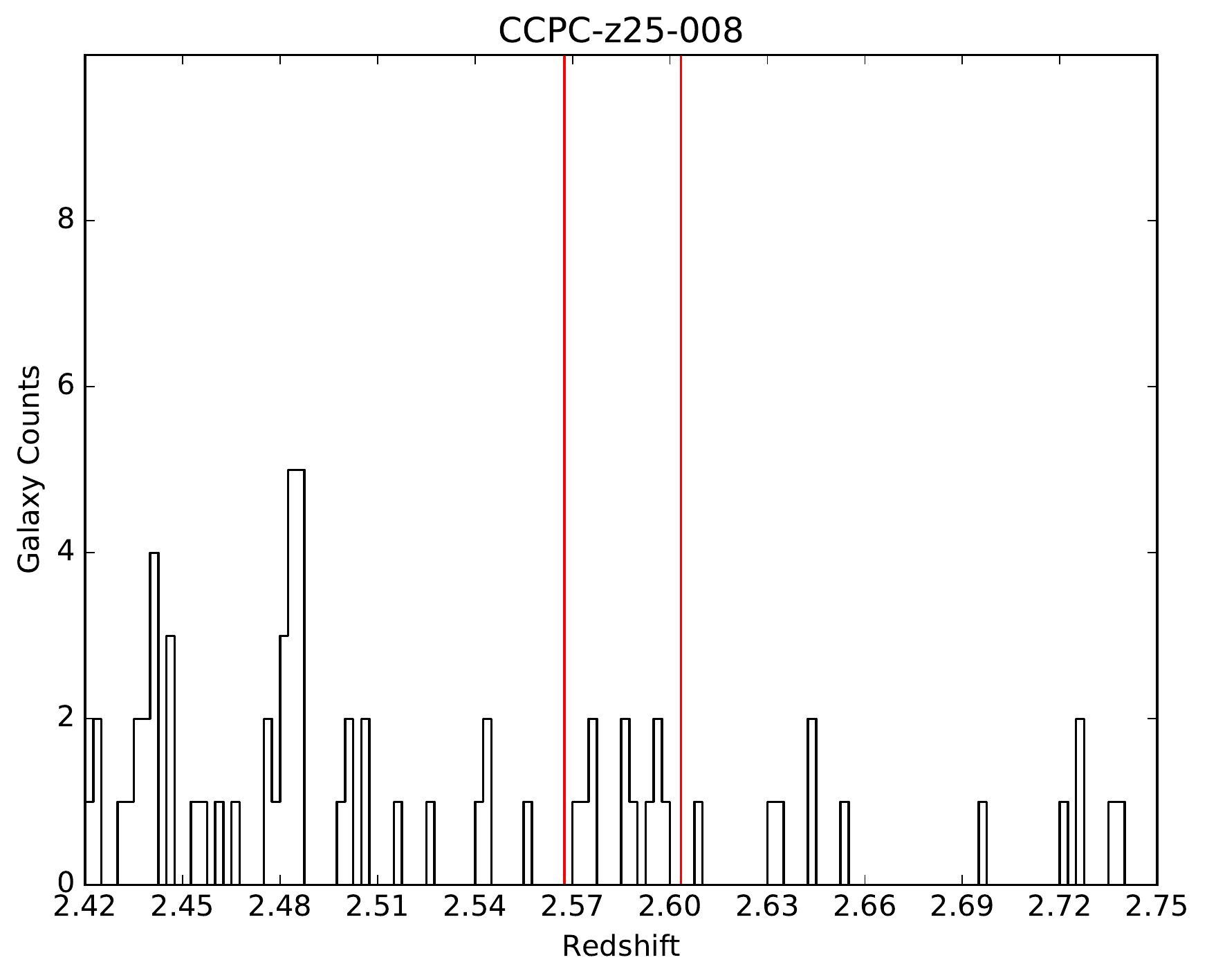}
\label{fig:CCPC-z25-008}
\end{subfigure}
\hfill
\end{figure*}

\begin{figure*}
\centering
\begin{subfigure}
\centering
\includegraphics[height=7.5cm,width=7.5cm]{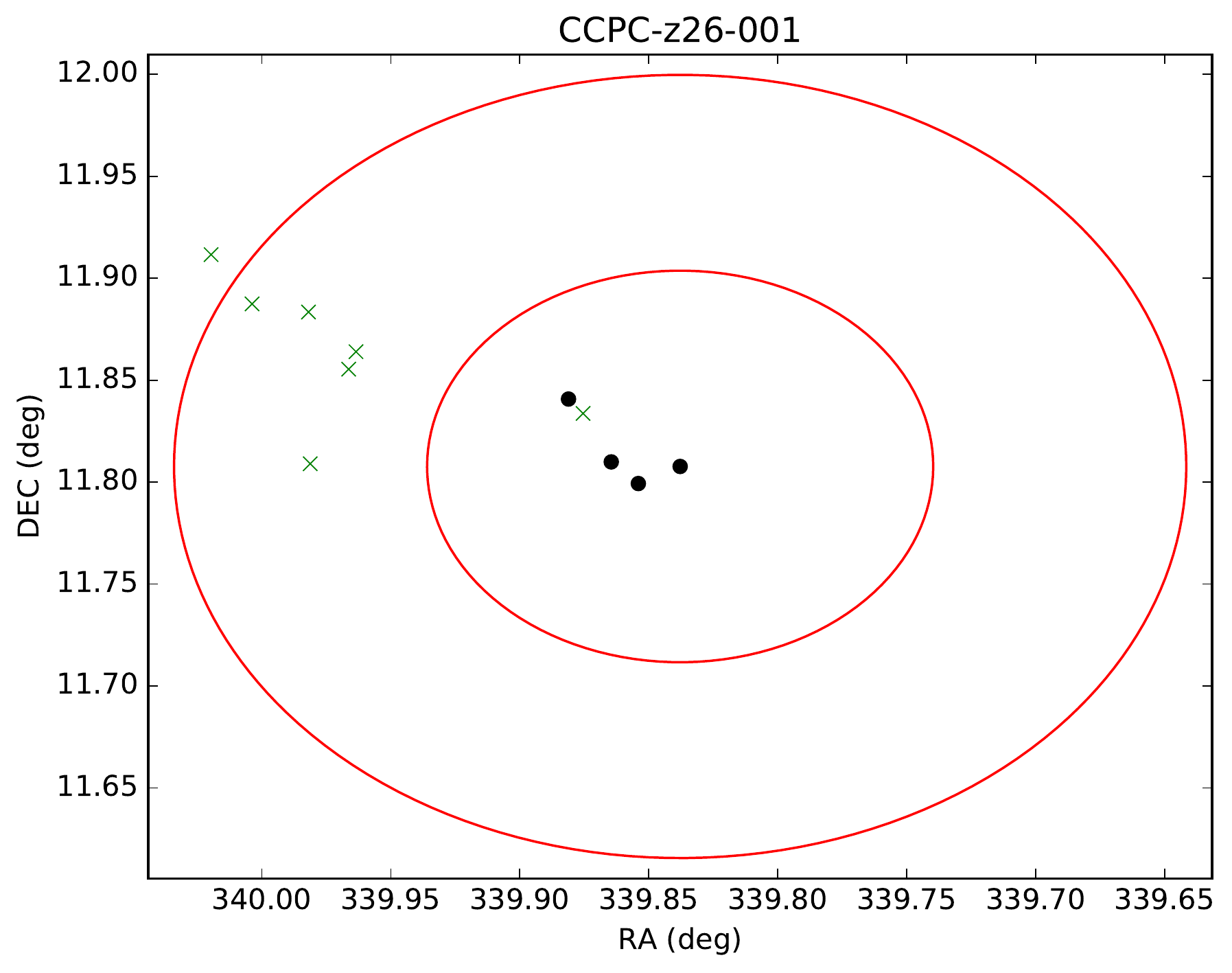}
\label{fig:CCPC-z26-001_sky}
\end{subfigure}
\hfill
\begin{subfigure}
\centering
\includegraphics[scale=0.52]{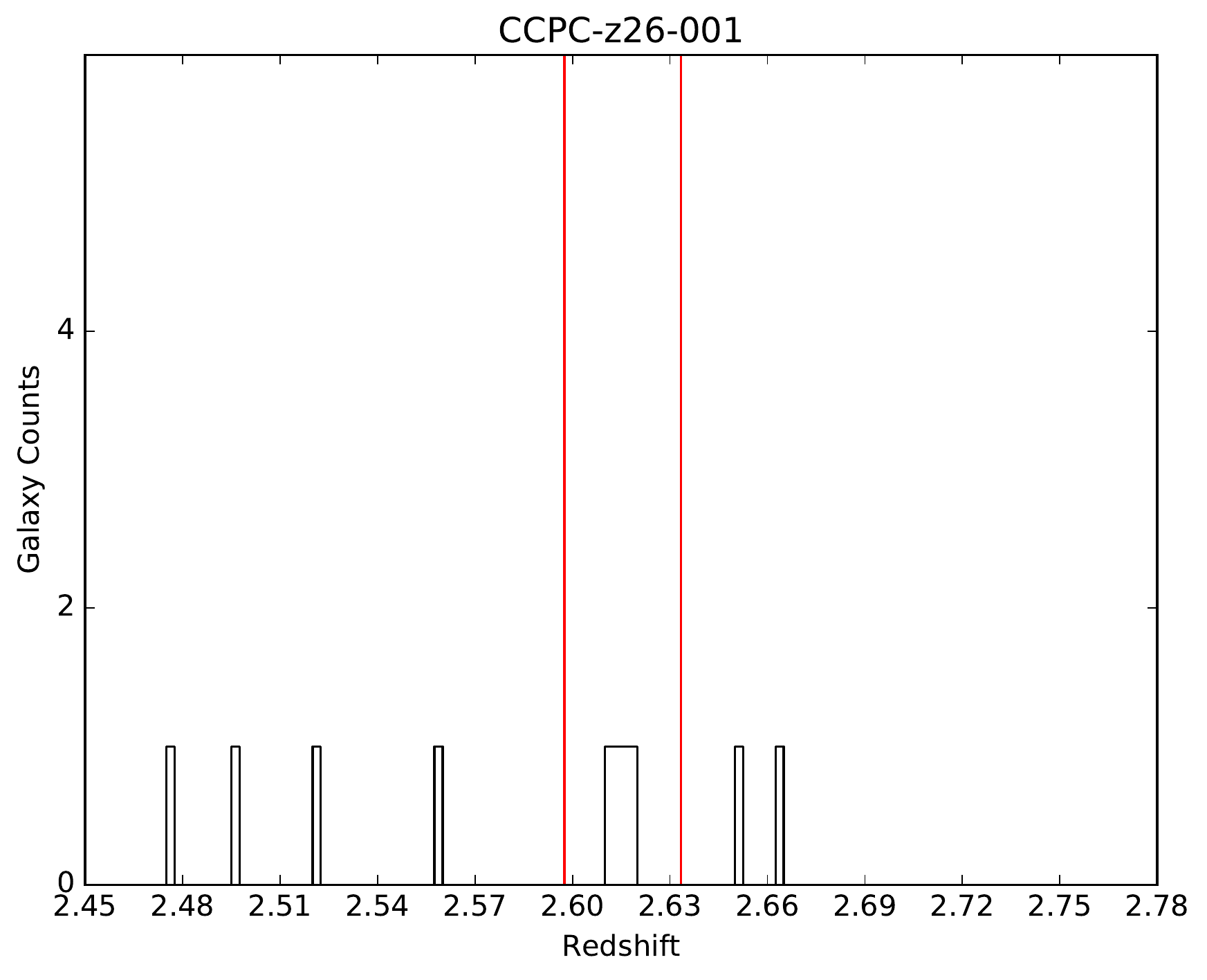}
\label{fig:CCPC-z26-001}
\end{subfigure}
\hfill
\end{figure*}

\begin{figure*}
\centering
\begin{subfigure}
\centering
\includegraphics[height=7.5cm,width=7.5cm]{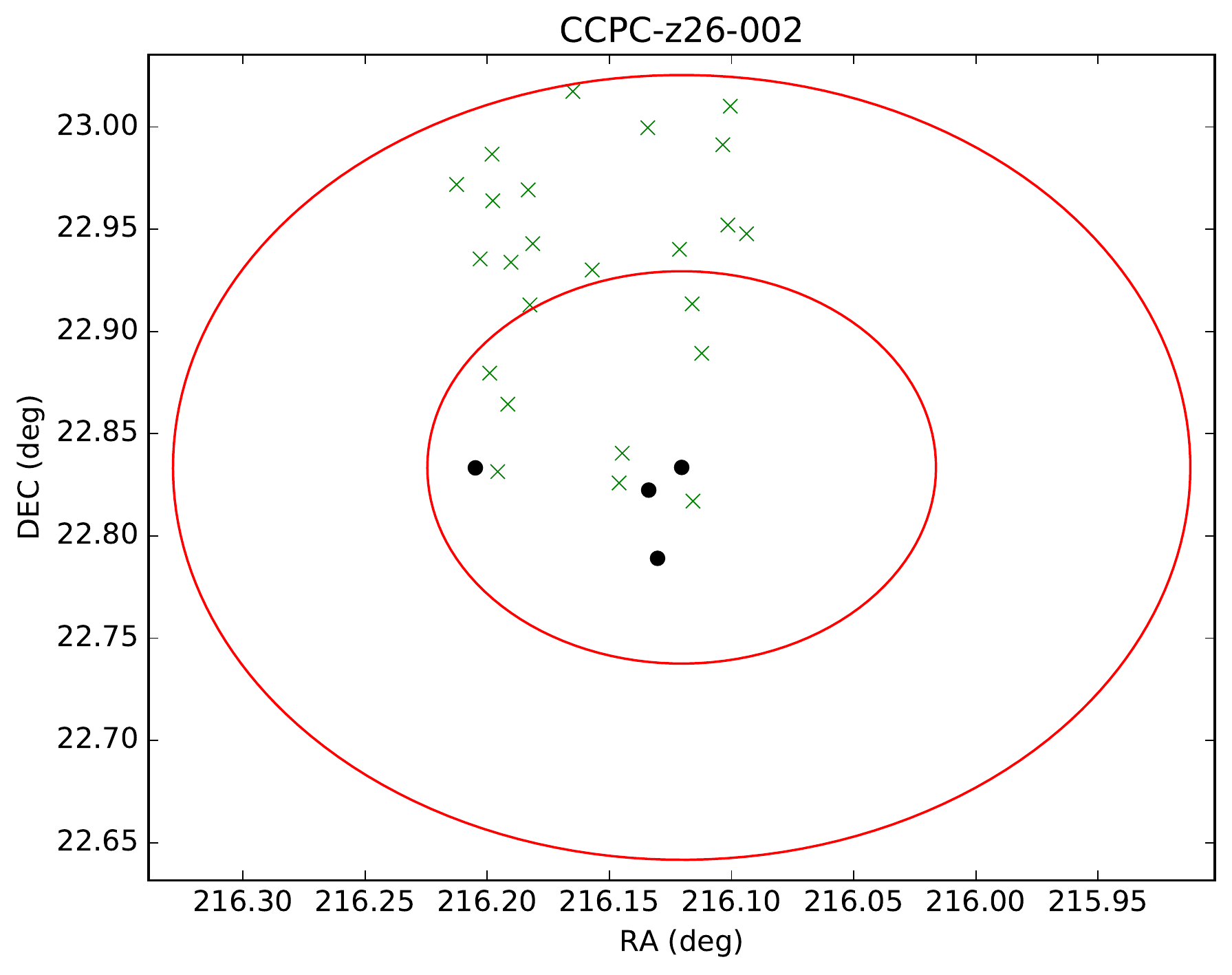}
\label{fig:CCPC-z26-002_sky}
\end{subfigure}
\hfill
\begin{subfigure}
\centering
\includegraphics[scale=0.52]{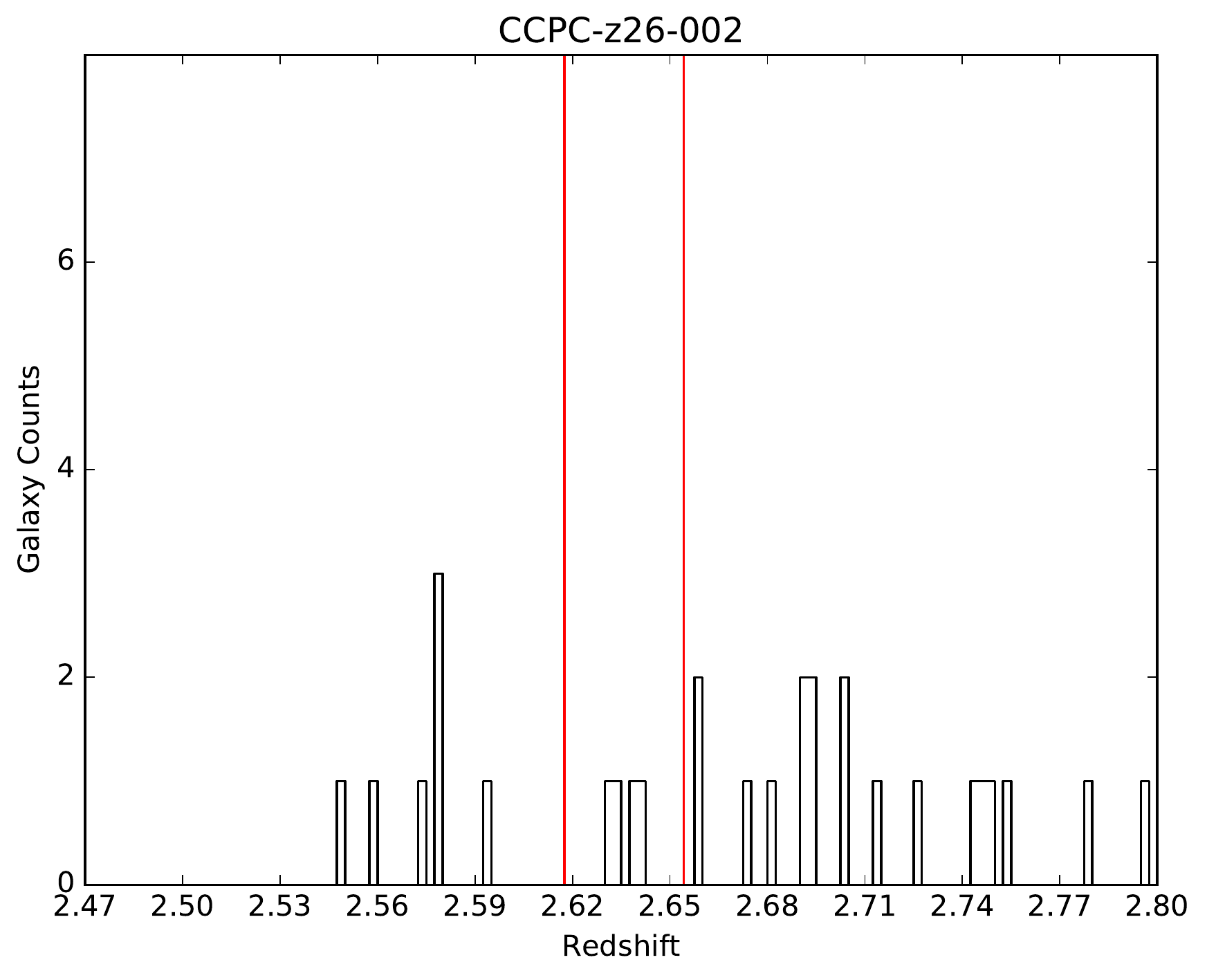}
\label{fig:CCPC-z26-002}
\end{subfigure}
\hfill
\end{figure*}
\clearpage 

\begin{figure*}
\centering
\begin{subfigure}
\centering
\includegraphics[height=7.5cm,width=7.5cm]{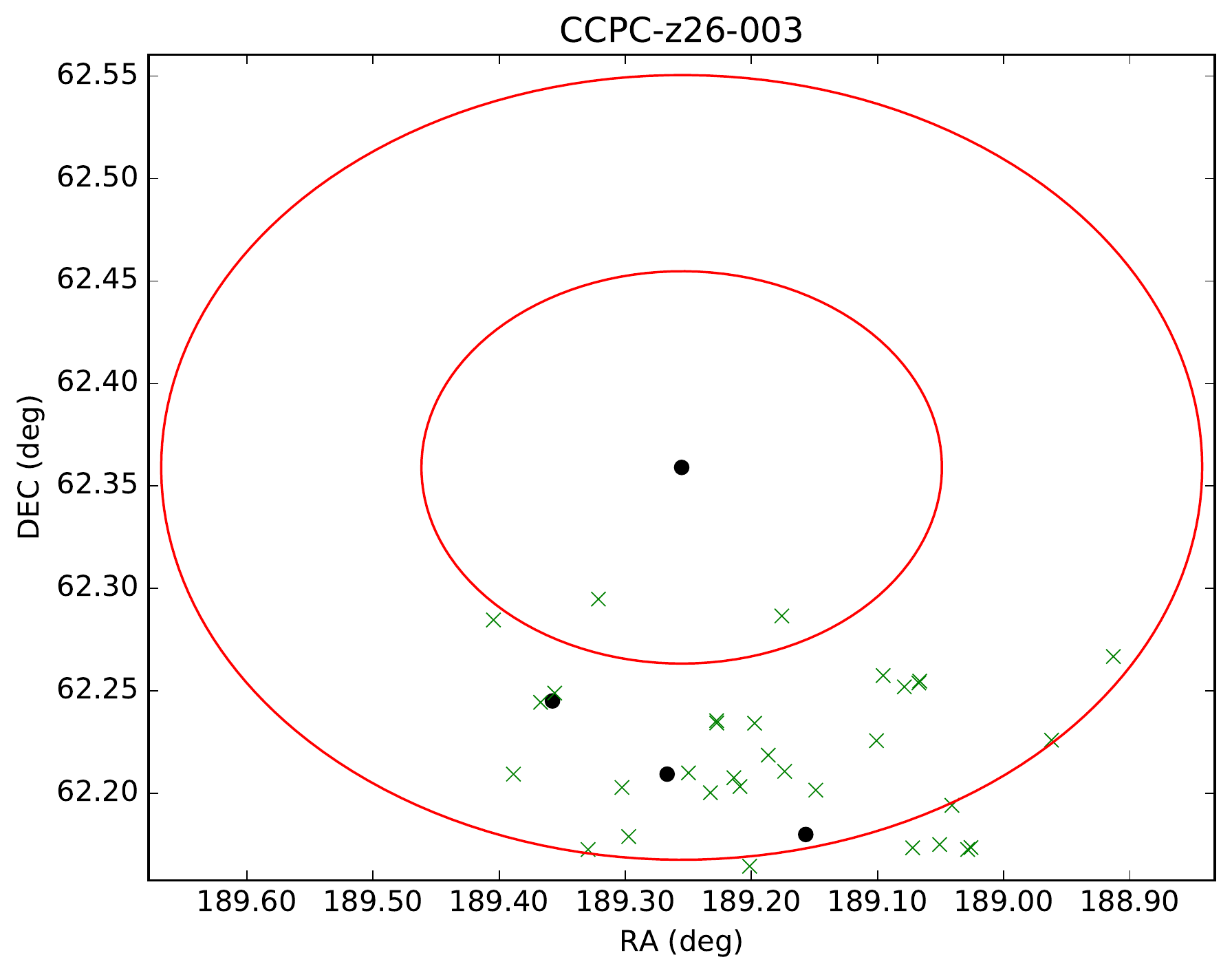}
\label{fig:CCPC-z26-003_sky}
\end{subfigure}
\hfill
\begin{subfigure}
\centering
\includegraphics[scale=0.52]{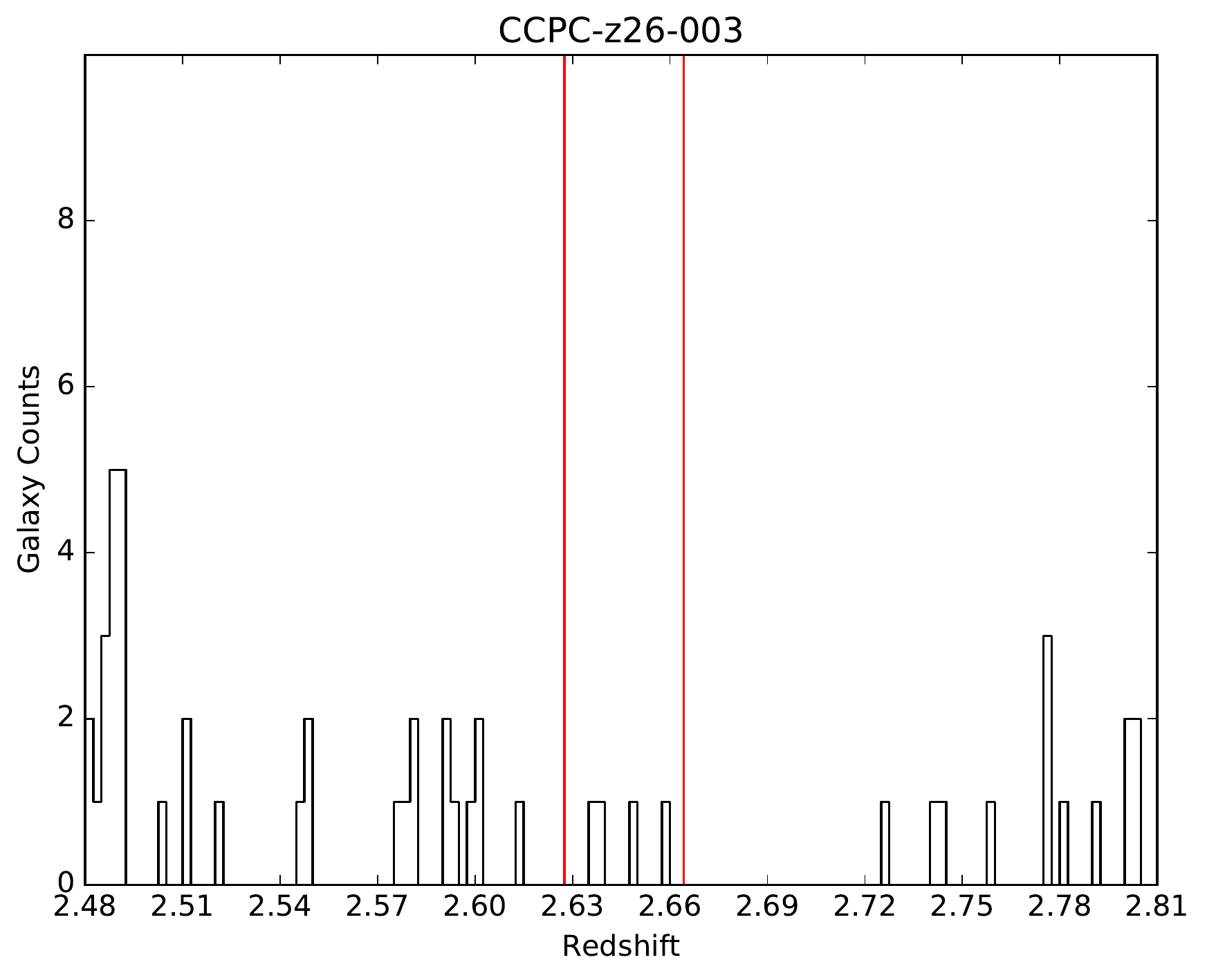}
\label{fig:CCPC-z26-003}
\end{subfigure}
\hfill
\end{figure*}

\begin{figure*}
\centering
\begin{subfigure}
\centering
\includegraphics[height=7.5cm,width=7.5cm]{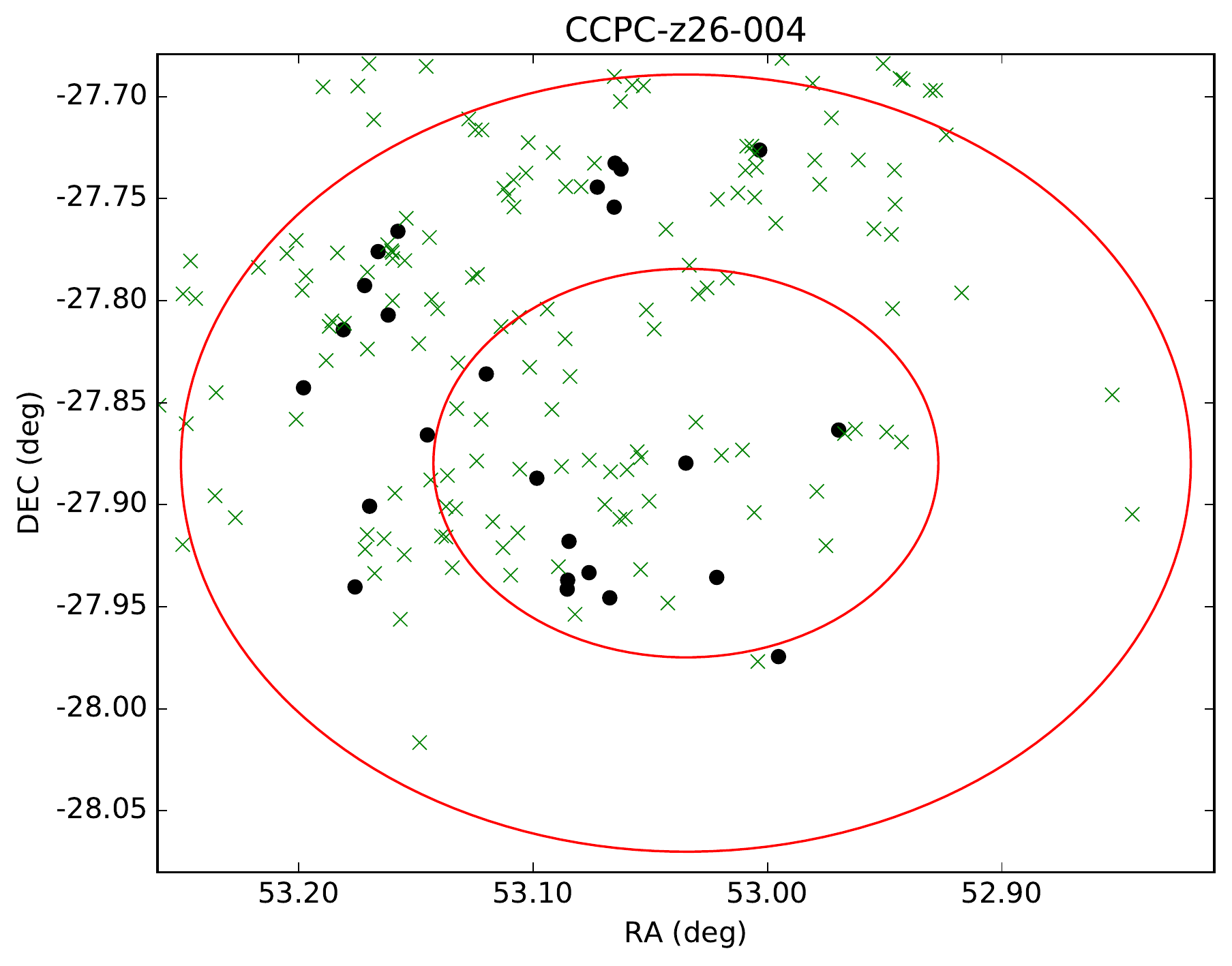}
\label{fig:CCPC-z26-004_sky}
\end{subfigure}
\hfill
\begin{subfigure}
\centering
\includegraphics[scale=0.52]{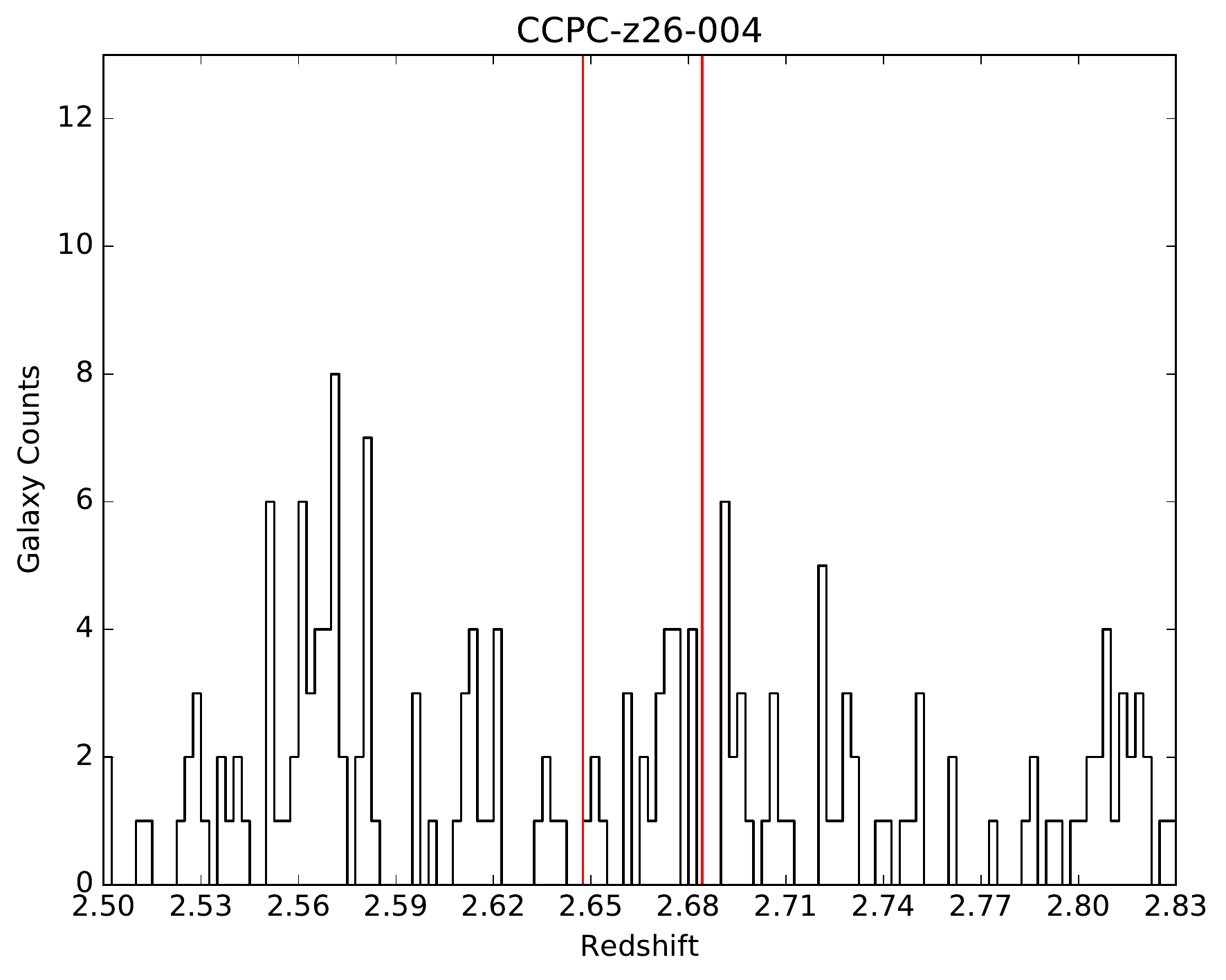}
\label{fig:CCPC-z26-004}
\end{subfigure}
\hfill
\end{figure*}

\begin{figure*}
\centering
\begin{subfigure}
\centering
\includegraphics[height=7.5cm,width=7.5cm]{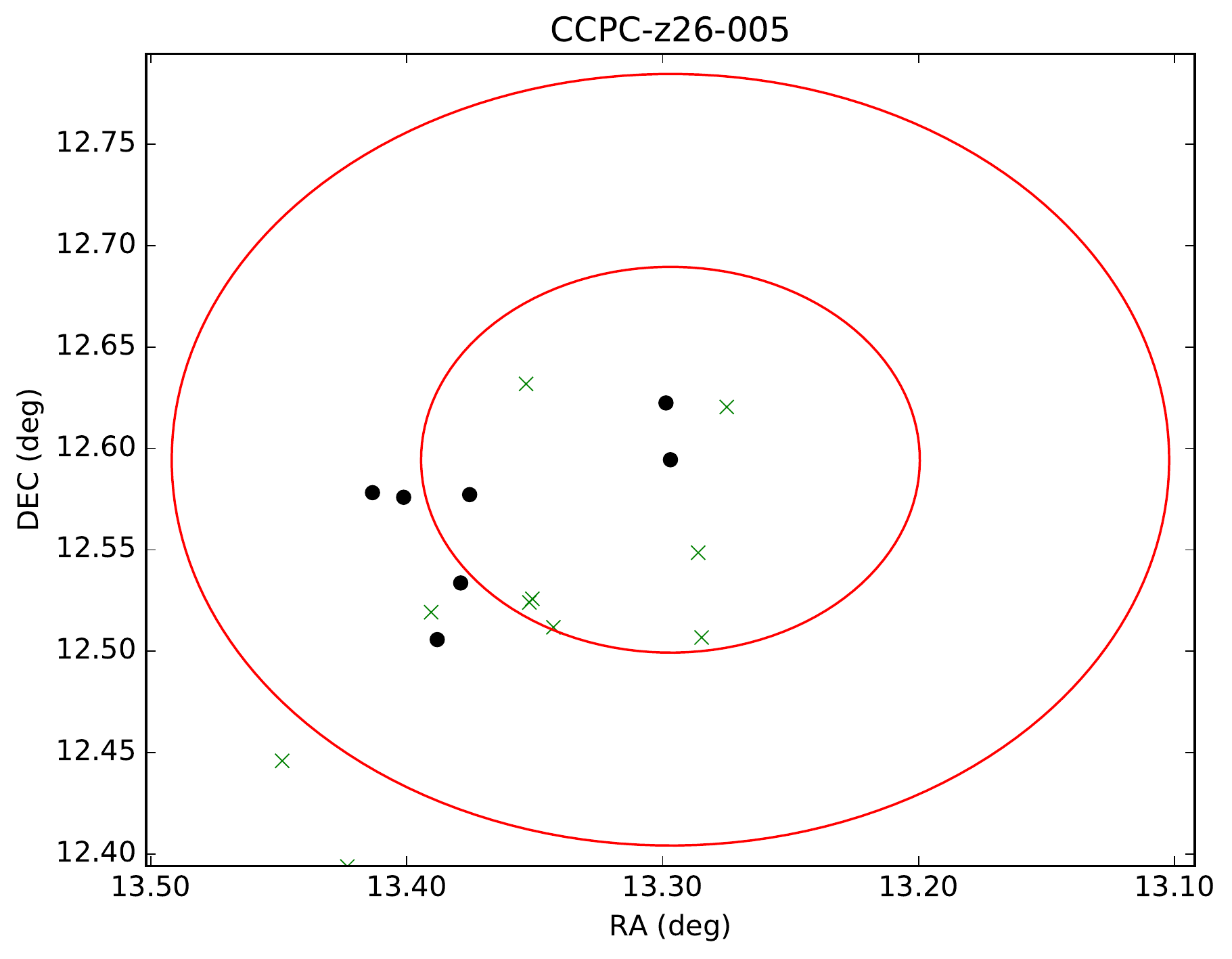}
\label{fig:CCPC-z26-005_sky}
\end{subfigure}
\hfill
\begin{subfigure}
\centering
\includegraphics[scale=0.52]{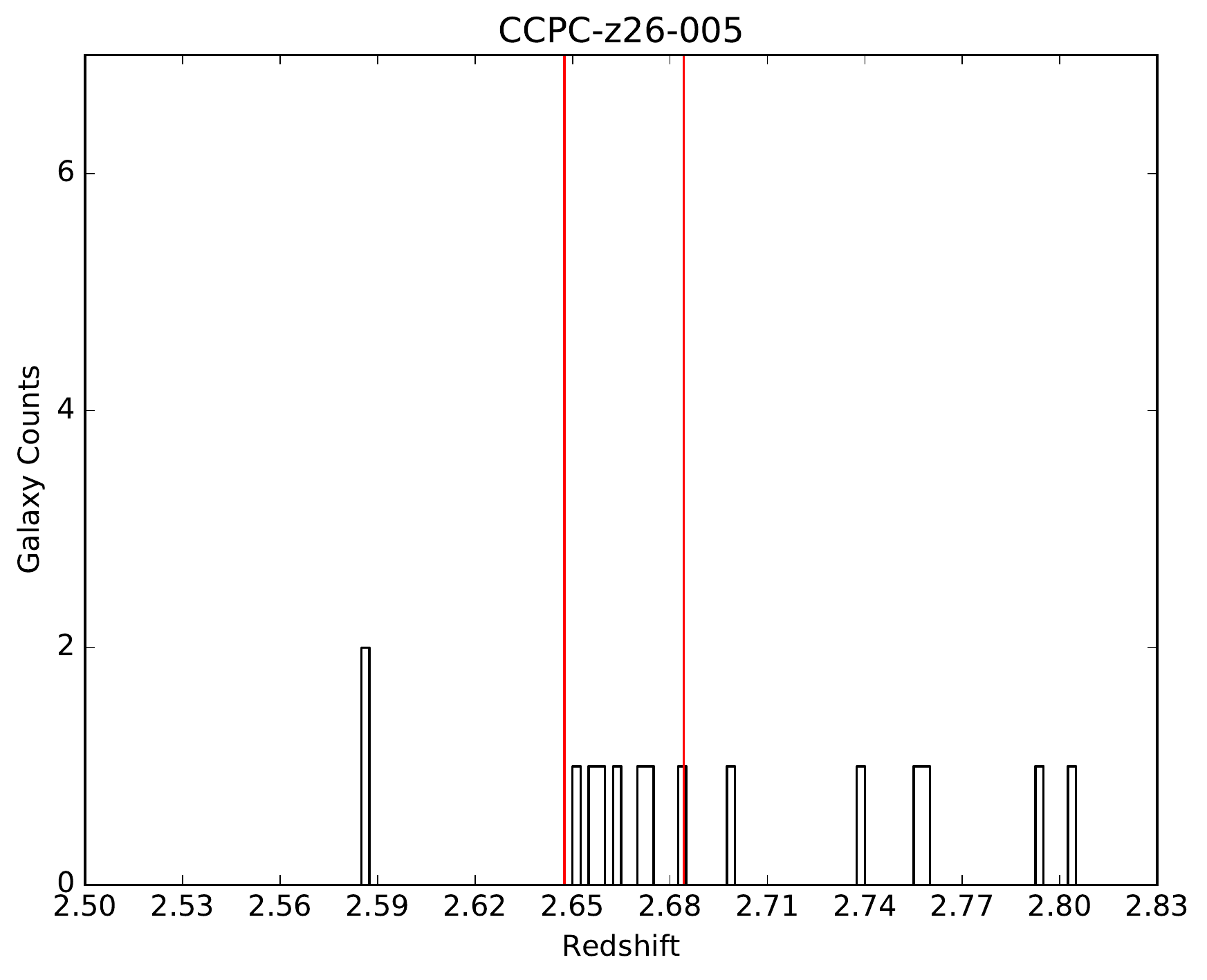}
\label{fig:CCPC-z26-005}
\end{subfigure}
\hfill
\end{figure*}
\clearpage 

\begin{figure*}
\centering
\begin{subfigure}
\centering
\includegraphics[height=7.5cm,width=7.5cm]{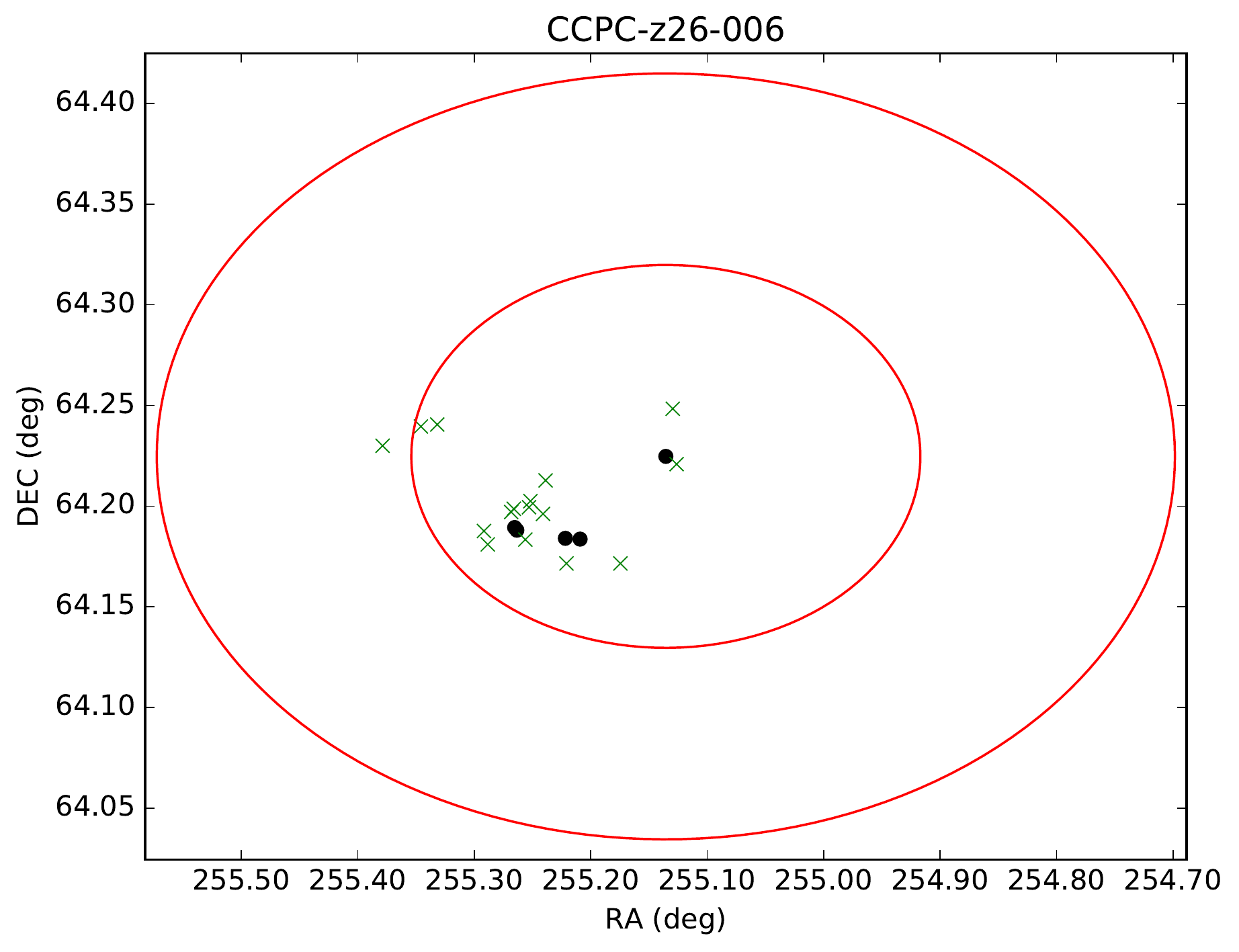}
\label{fig:CCPC-z26-006_sky}
\end{subfigure}
\hfill
\begin{subfigure}
\centering
\includegraphics[scale=0.52]{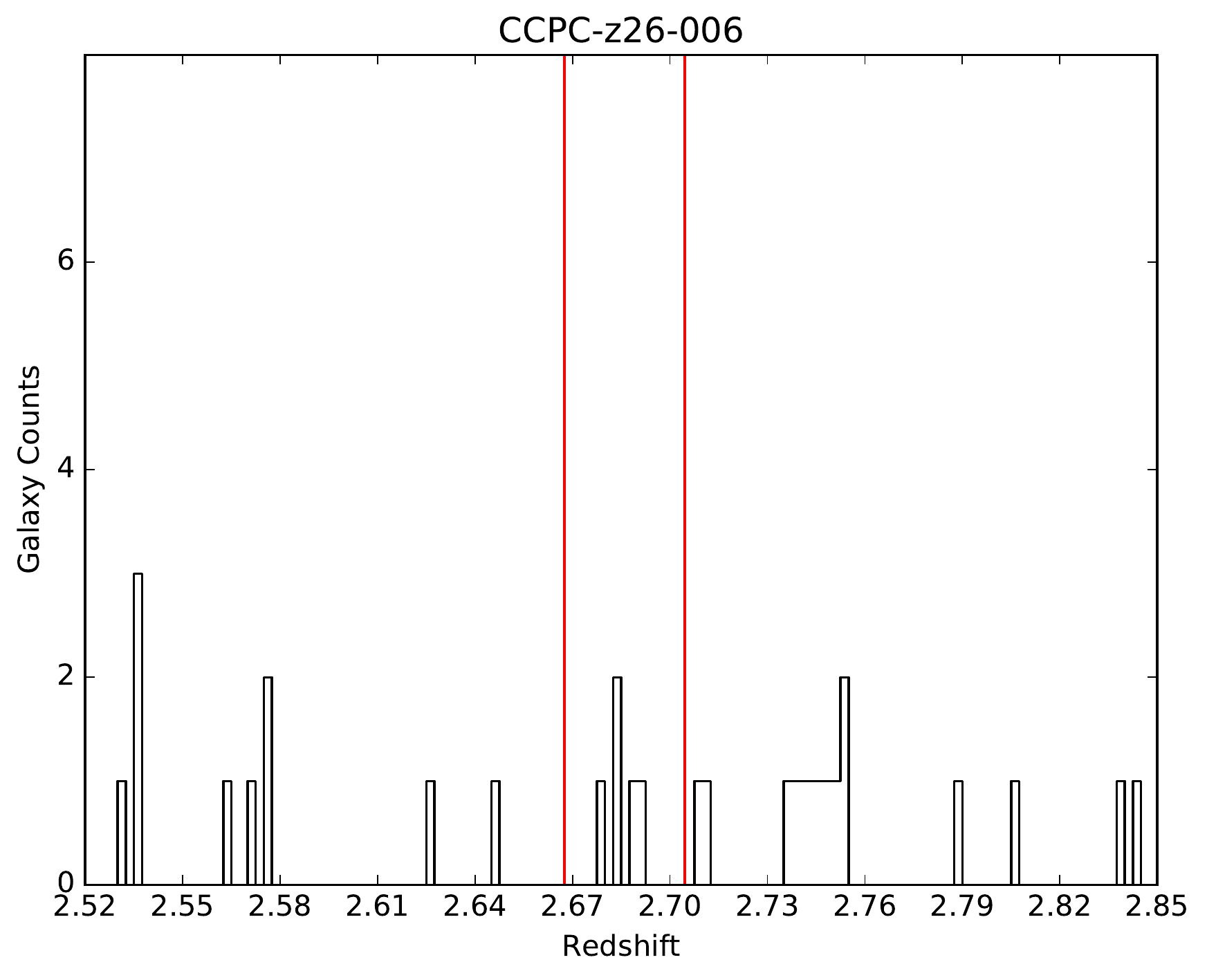}
\label{fig:CCPC-z26-006}
\end{subfigure}
\hfill
\end{figure*}

\begin{figure*}
\centering
\begin{subfigure}
\centering
\includegraphics[height=7.5cm,width=7.5cm]{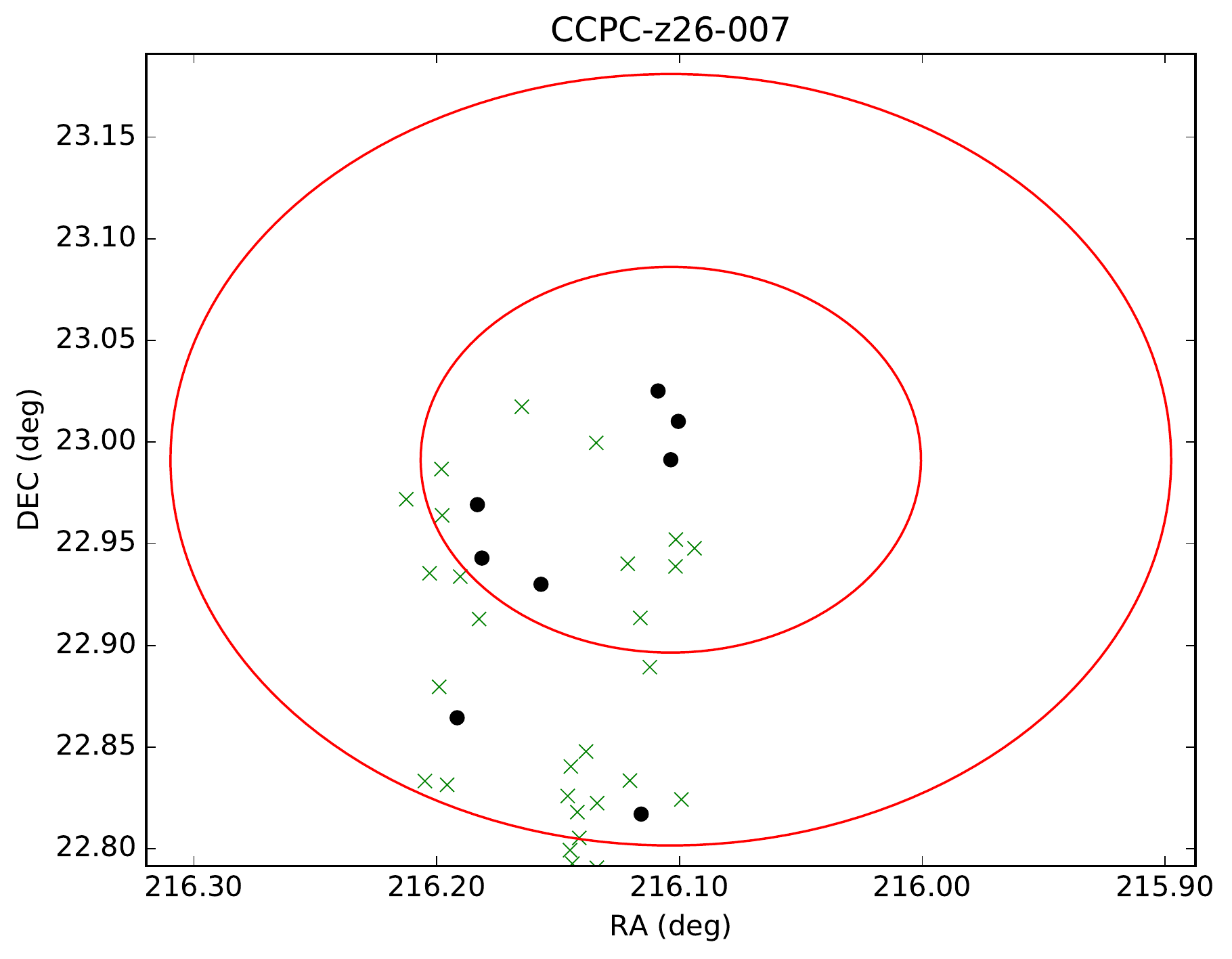}
\label{fig:CCPC-z26-007_sky}
\end{subfigure}
\hfill
\begin{subfigure}
\centering
\includegraphics[scale=0.52]{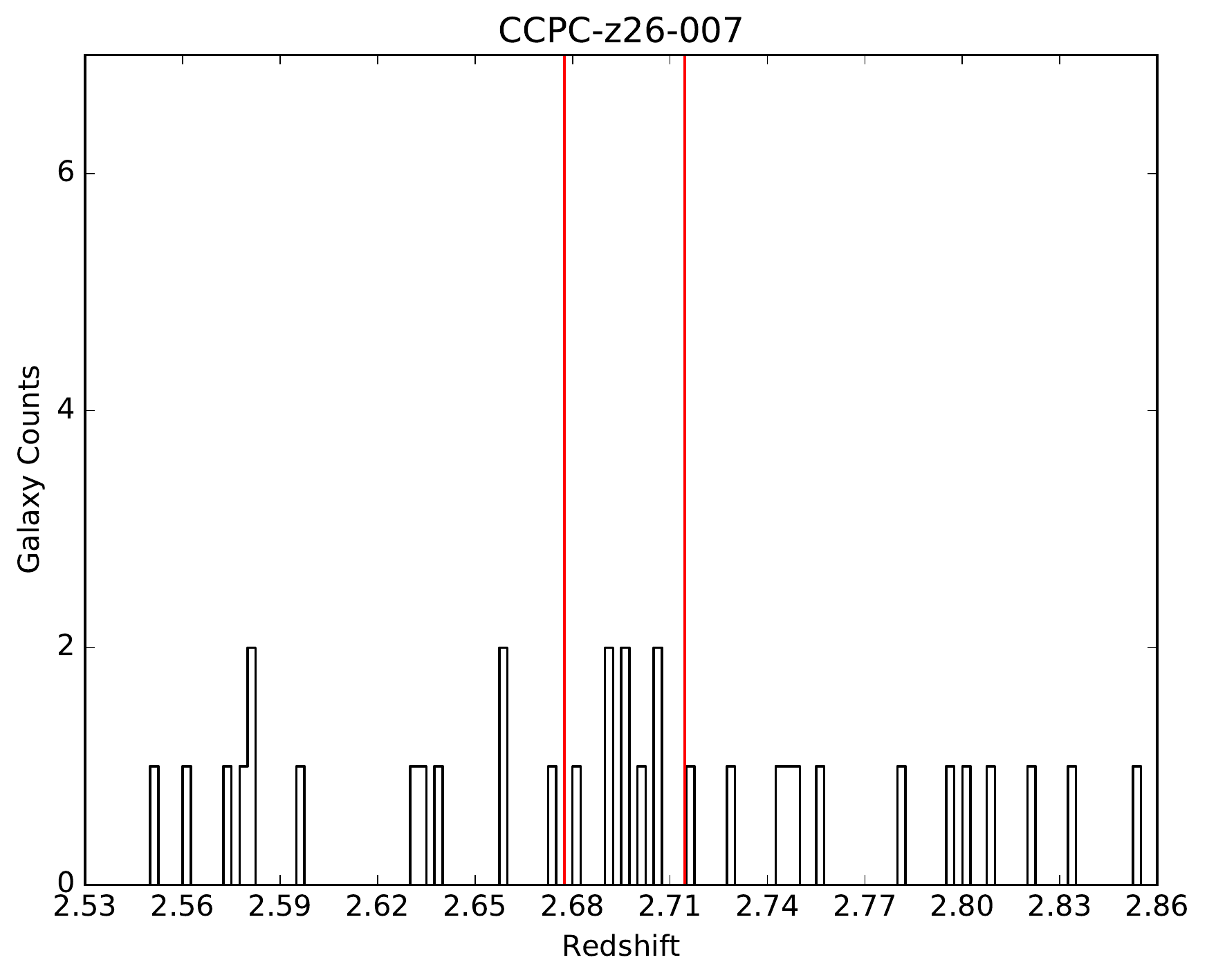}
\label{fig:CCPC-z26-007}
\end{subfigure}
\hfill
\end{figure*}

\begin{figure*}
\centering
\begin{subfigure}
\centering
\includegraphics[height=7.5cm,width=7.5cm]{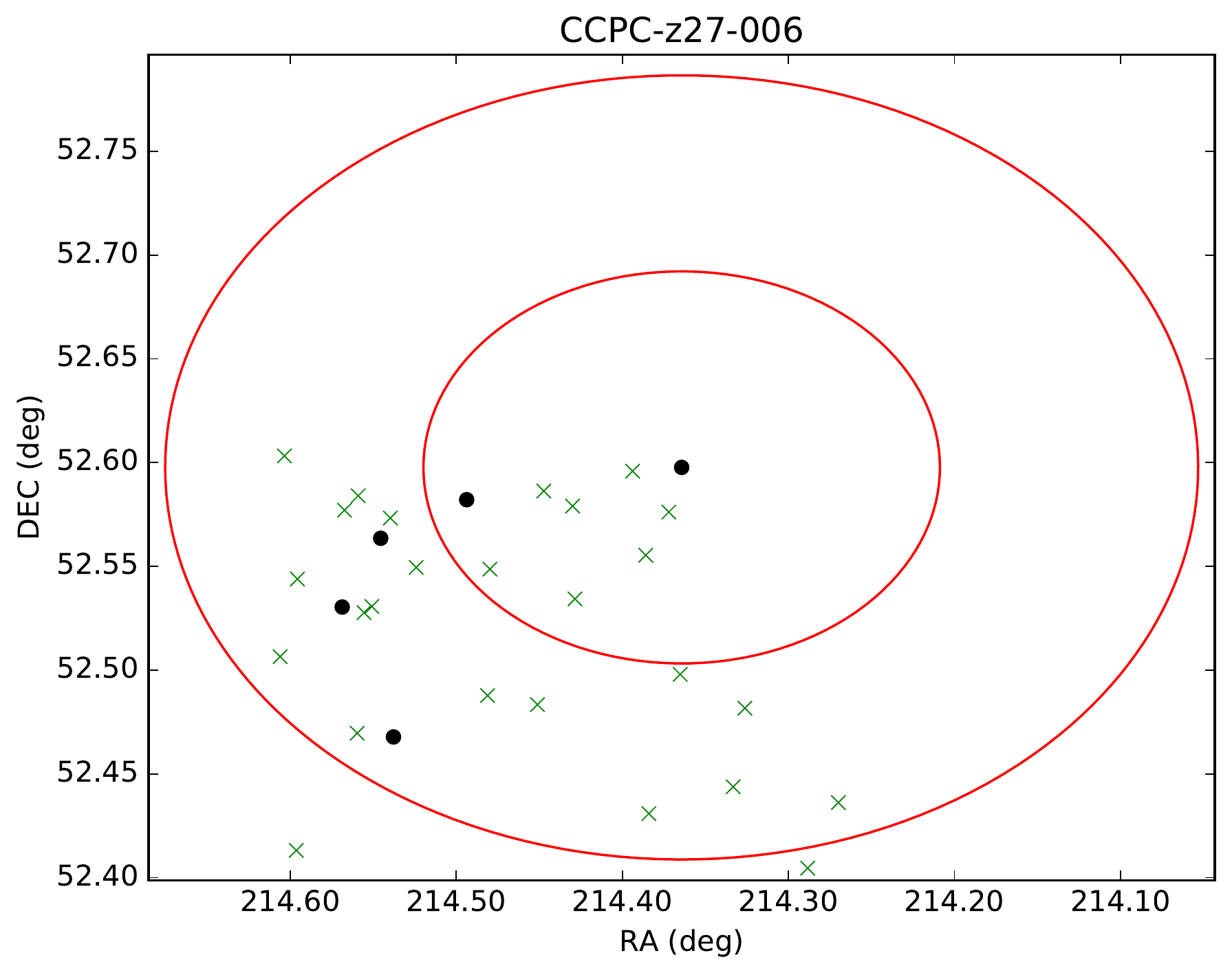}
\label{fig:CCPC-z27-006_sky}
\end{subfigure}
\hfill
\begin{subfigure}
\centering
\includegraphics[scale=0.52]{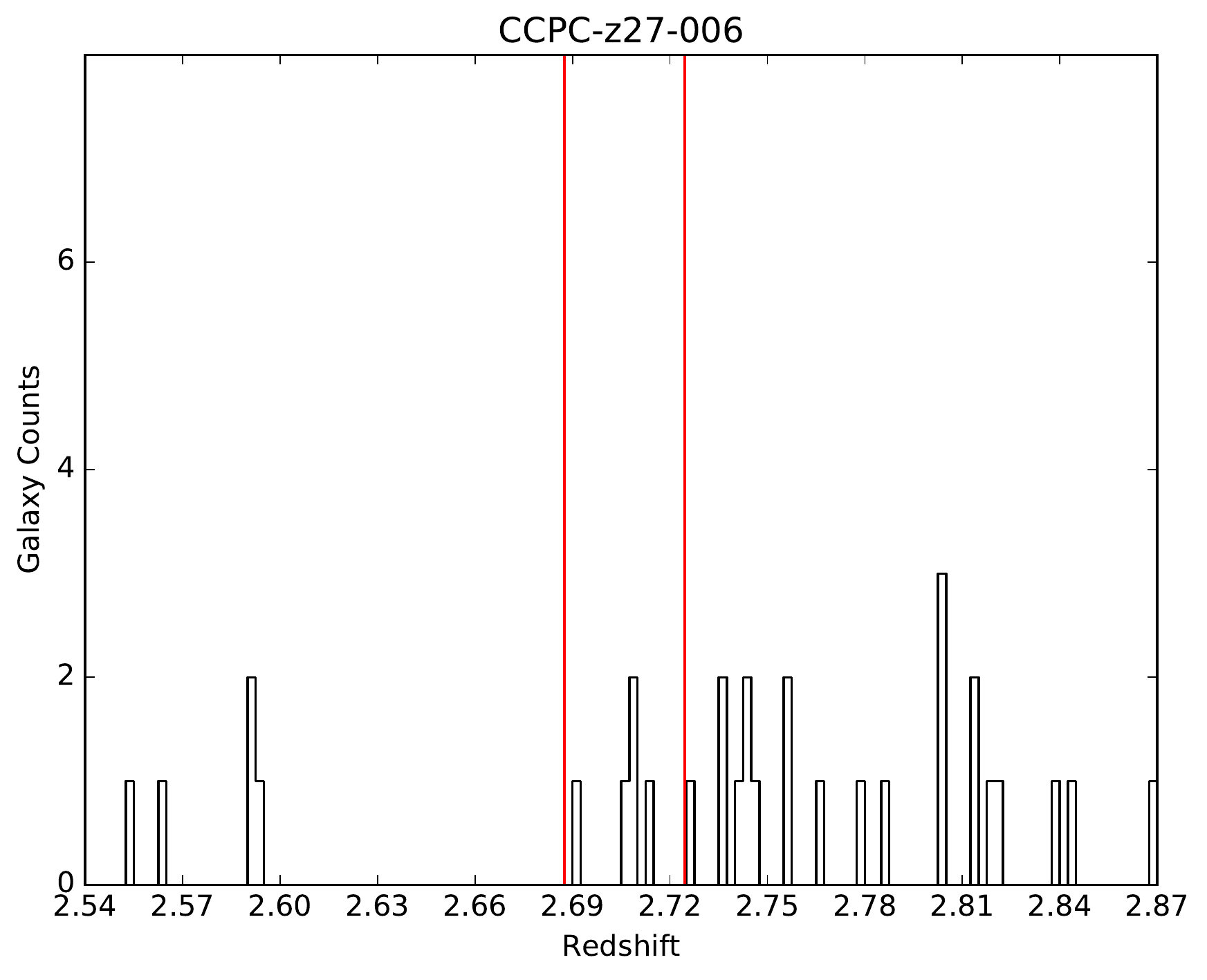}
\label{fig:CCPC-z27-006}
\end{subfigure}
\hfill
\end{figure*}
\clearpage 

\begin{figure*}
\centering
\begin{subfigure}
\centering
\includegraphics[height=7.5cm,width=7.5cm]{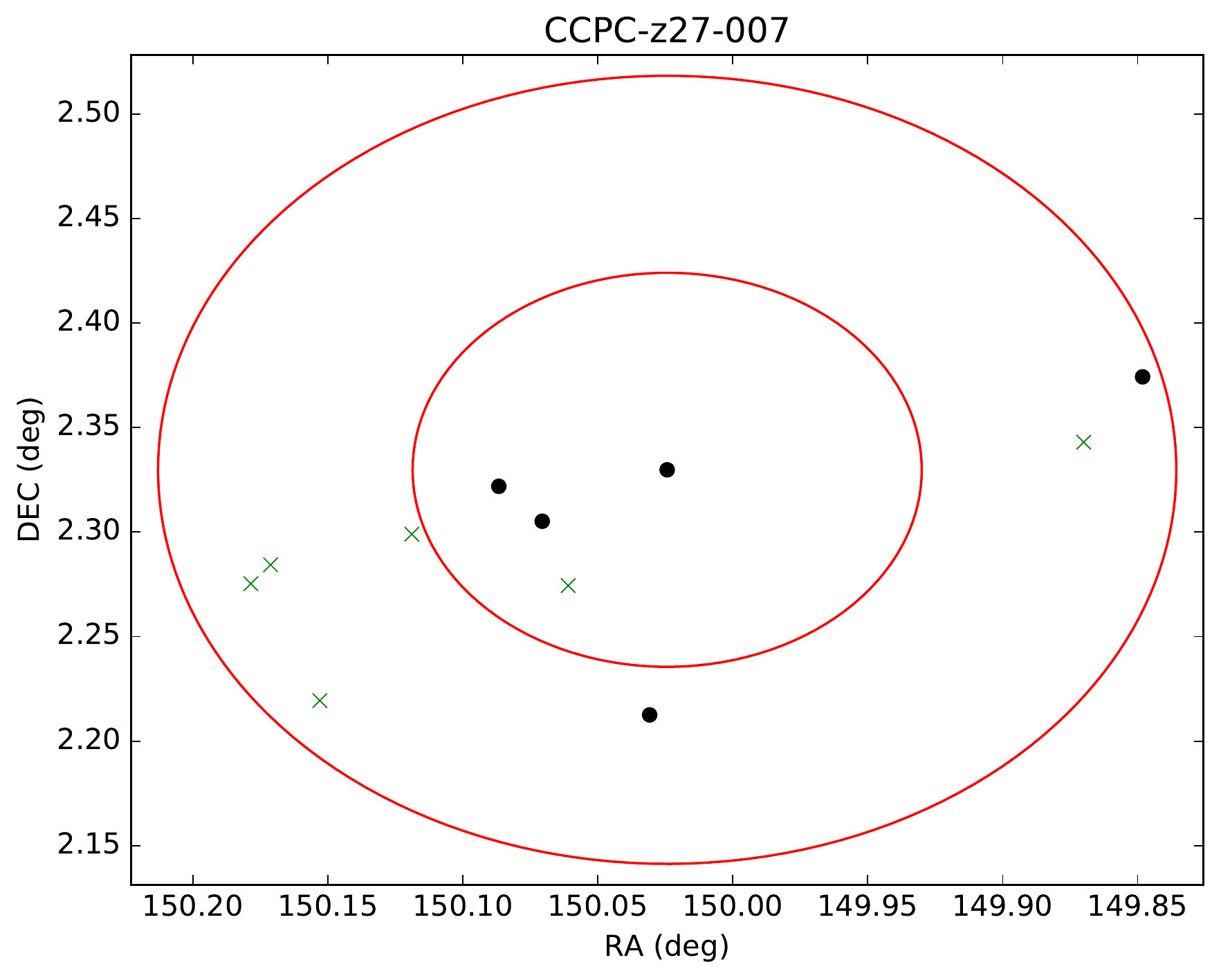}
\label{fig:CCPC-z27-007_sky}
\end{subfigure}
\hfill
\begin{subfigure}
\centering
\includegraphics[scale=0.52]{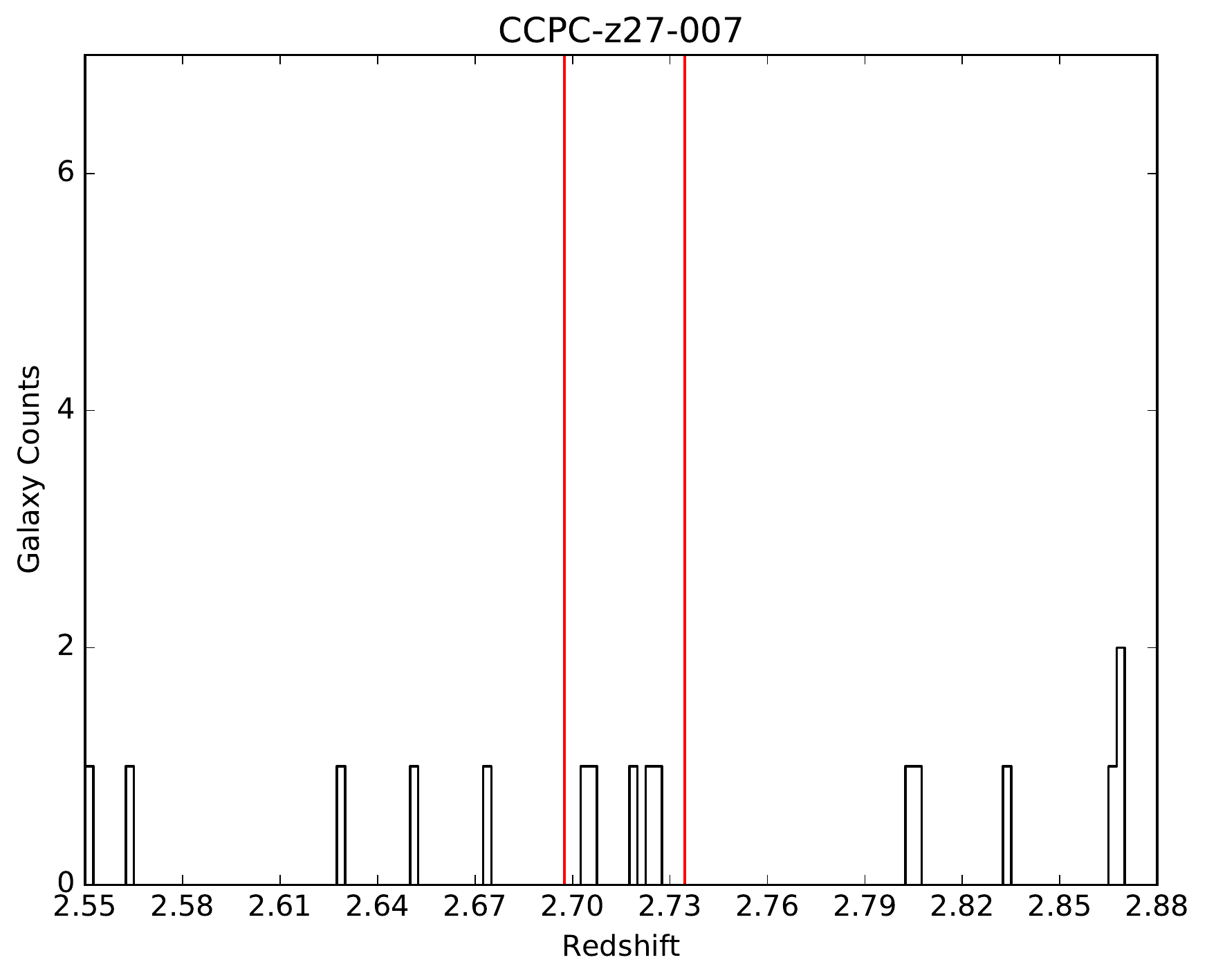}
\label{fig:CCPC-z27-007}
\end{subfigure}
\hfill
\end{figure*}

\begin{figure*}
\centering
\begin{subfigure}
\centering
\includegraphics[height=7.5cm,width=7.5cm]{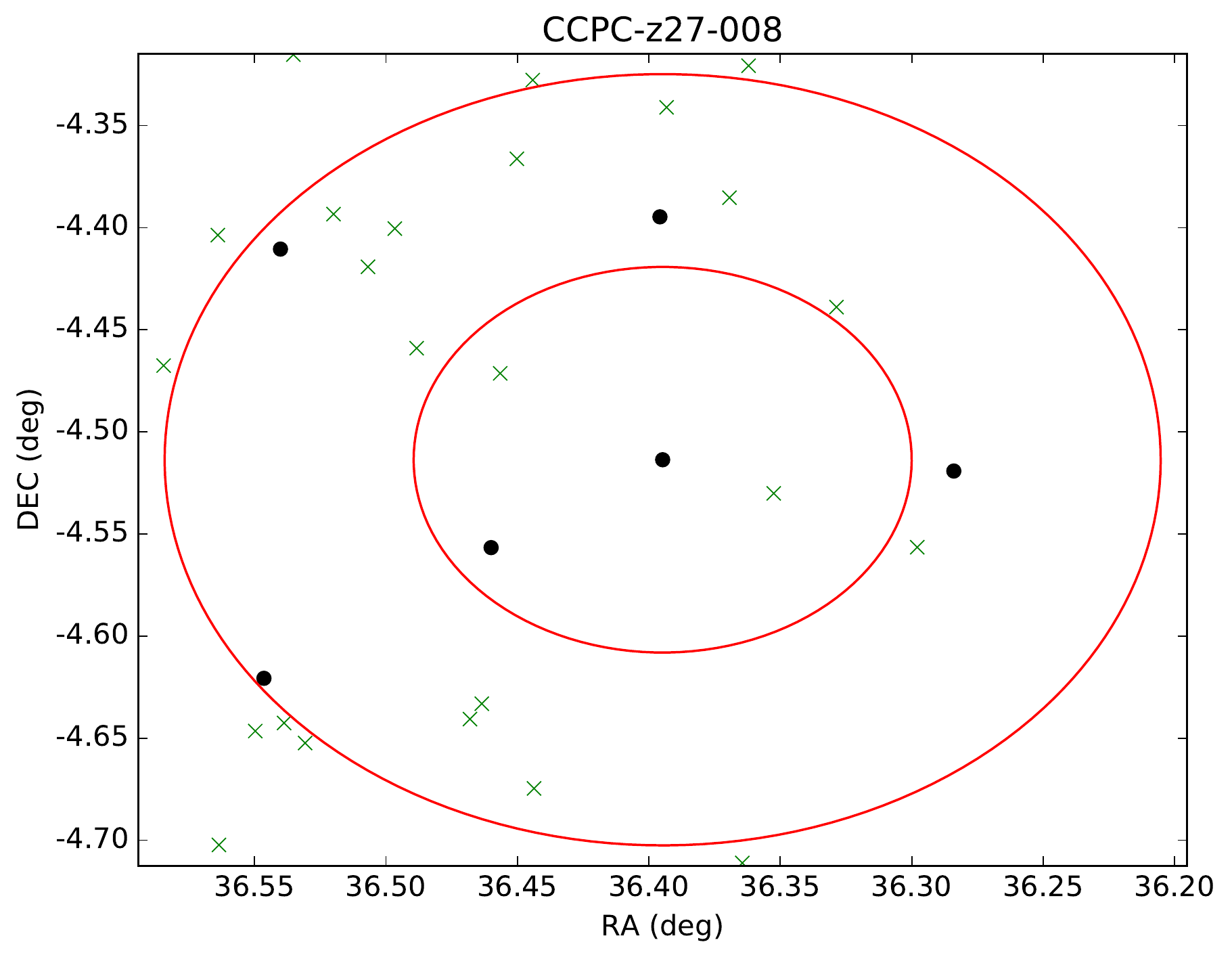}
\label{fig:CCPC-z27-008_sky}
\end{subfigure}
\hfill
\begin{subfigure}
\centering
\includegraphics[scale=0.52]{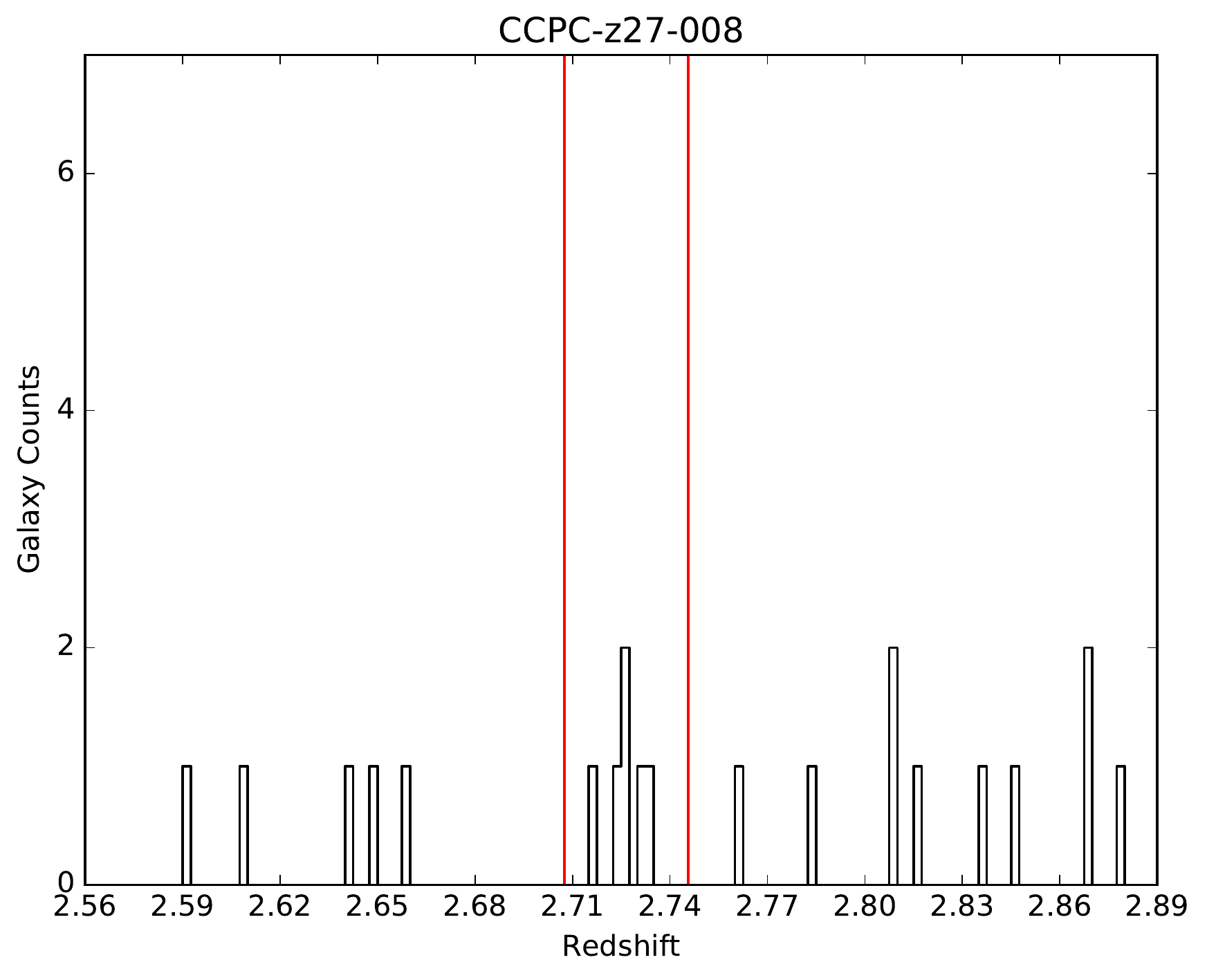}
\label{fig:CCPC-z27-008}
\end{subfigure}
\hfill
\end{figure*}

\begin{figure*}
\centering
\begin{subfigure}
\centering
\includegraphics[height=7.5cm,width=7.5cm]{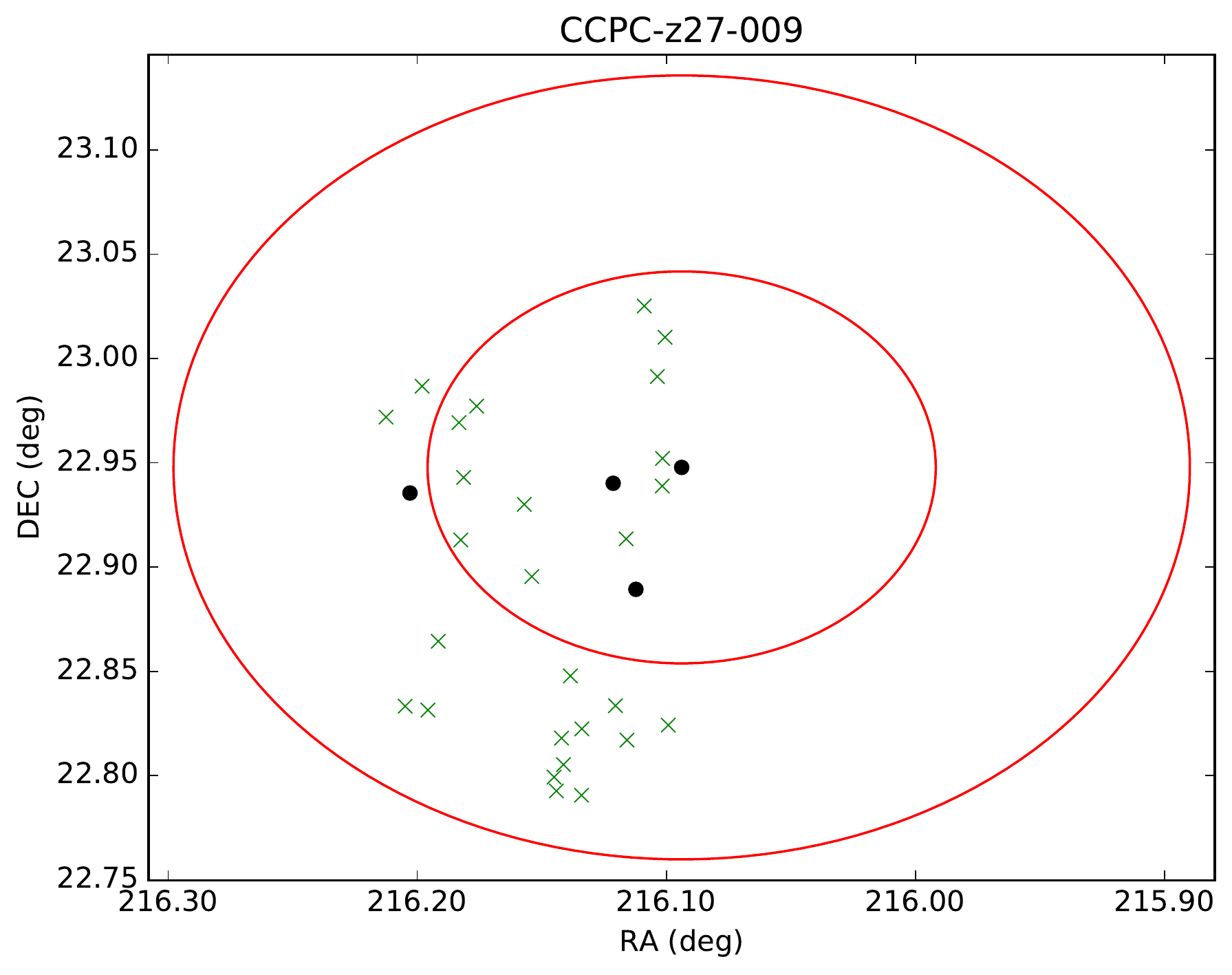}
\label{fig:CCPC-z27-009_sky}
\end{subfigure}
\hfill
\begin{subfigure}
\centering
\includegraphics[scale=0.52]{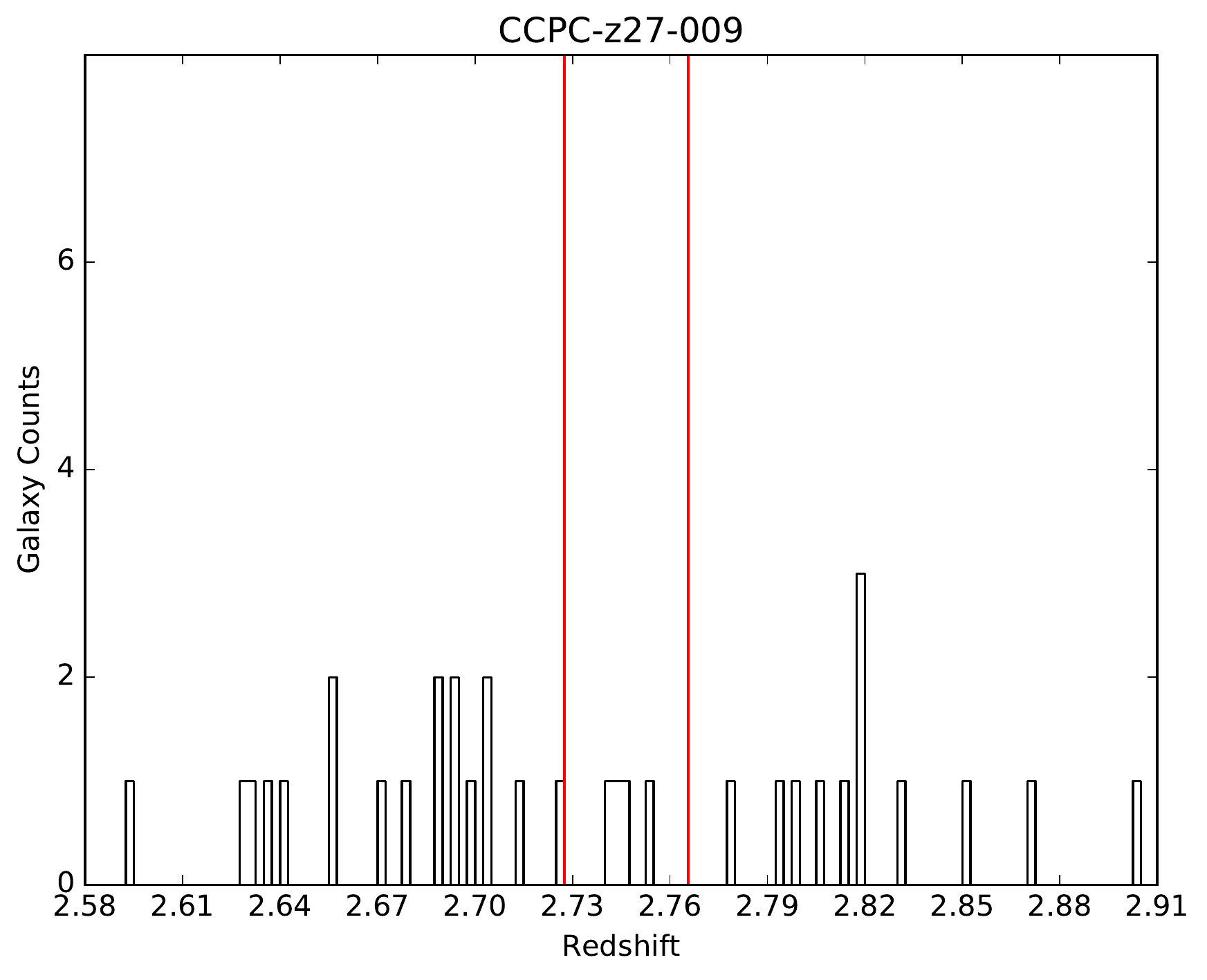}
\label{fig:CCPC-z27-009}
\end{subfigure}
\hfill
\end{figure*}
\clearpage 

\begin{figure*}
\centering
\begin{subfigure}
\centering
\includegraphics[height=7.5cm,width=7.5cm]{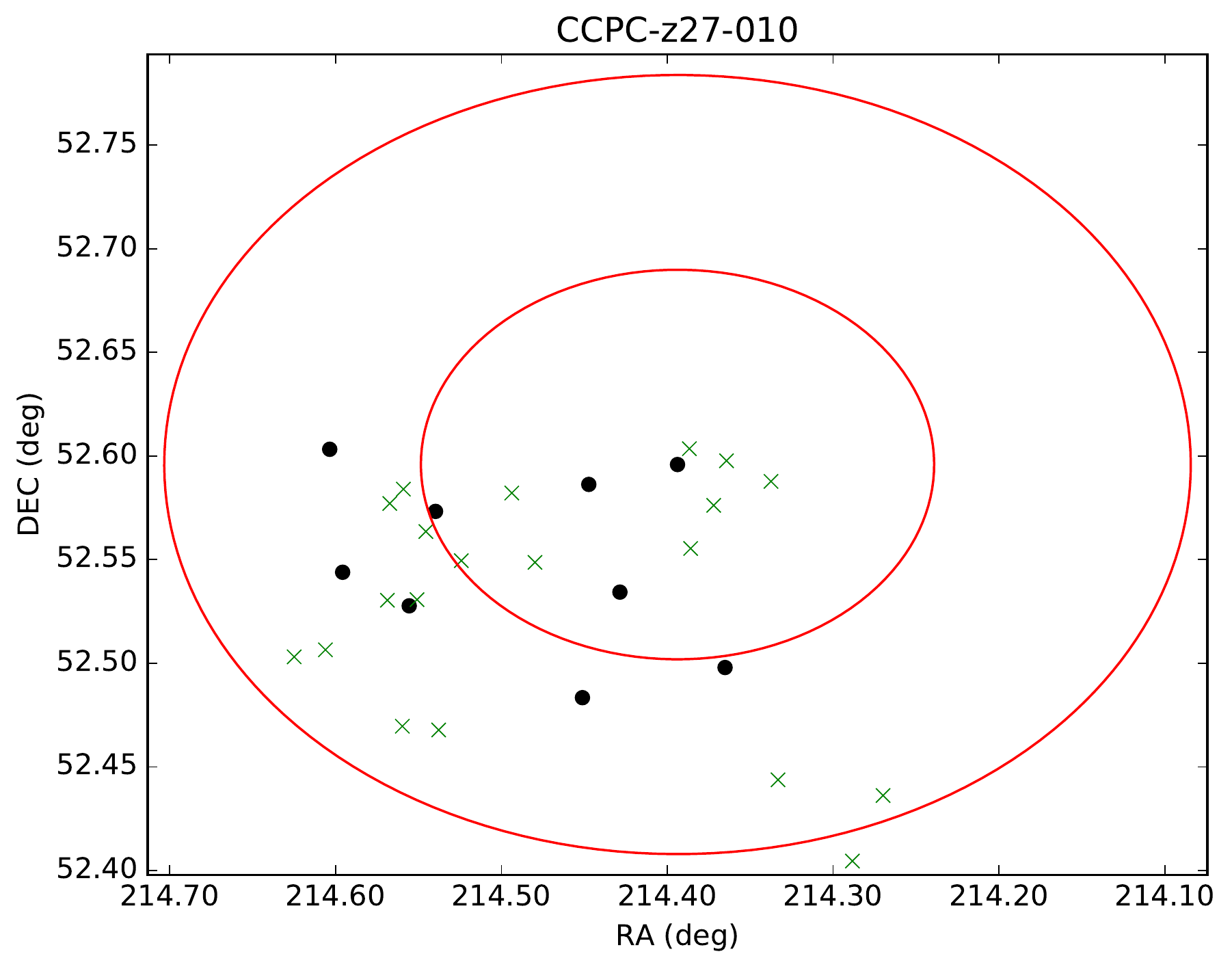}
\label{fig:CCPC-z27-010_sky}
\end{subfigure}
\hfill
\begin{subfigure}
\centering
\includegraphics[scale=0.52]{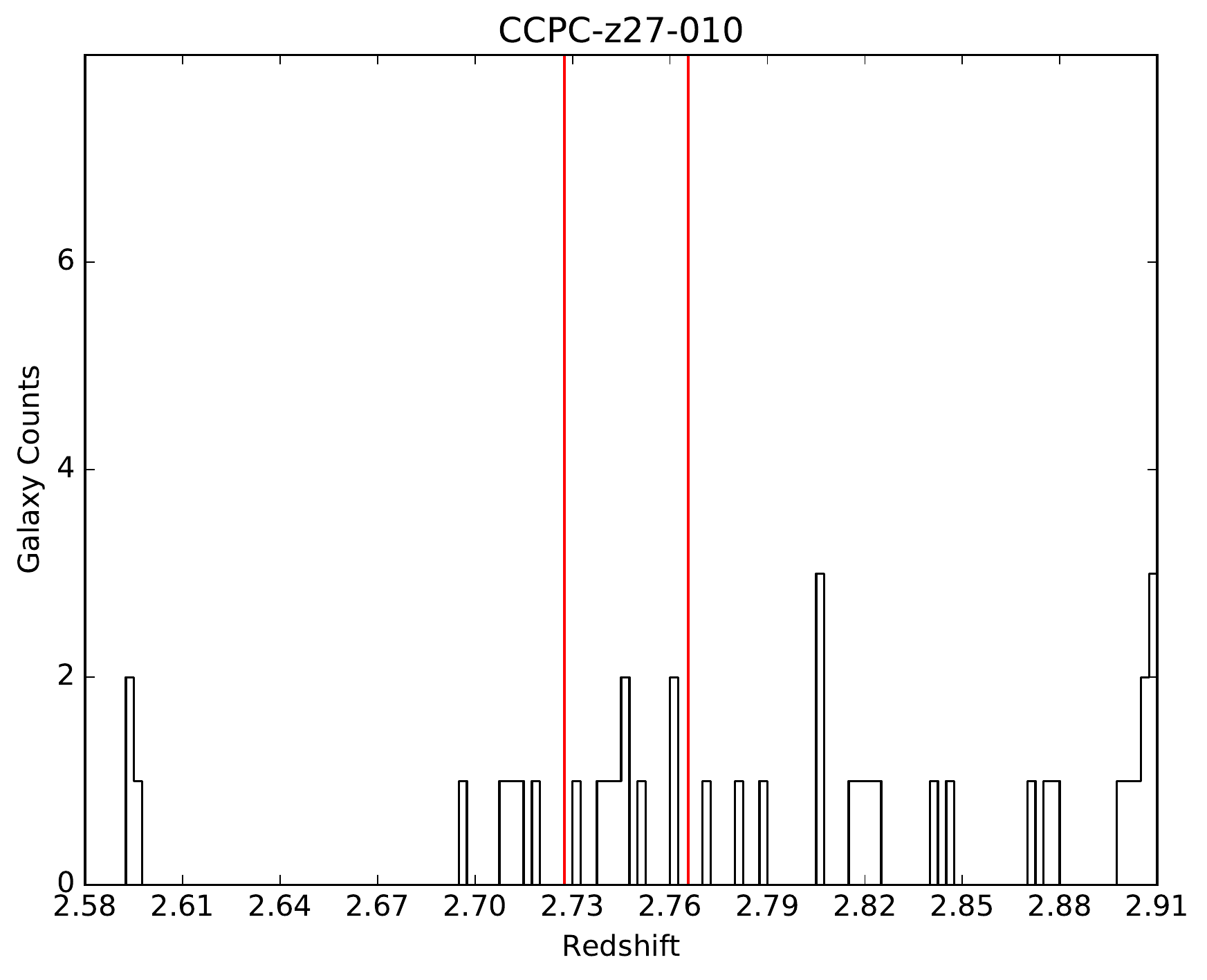}
\label{fig:CCPC-z27-010}
\end{subfigure}
\hfill
\end{figure*}

\begin{figure*}
\centering
\begin{subfigure}
\centering
\includegraphics[height=7.5cm,width=7.5cm]{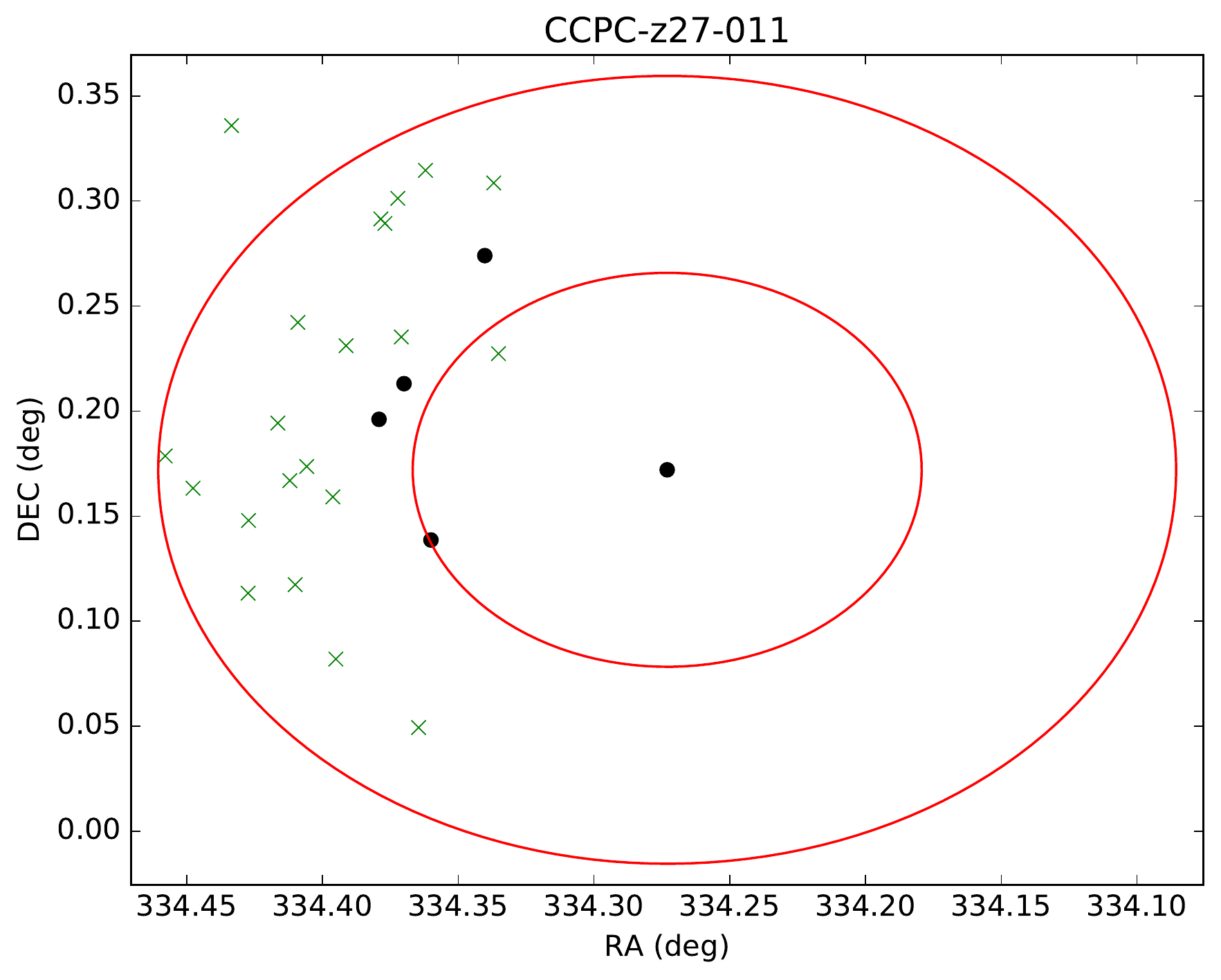}
\label{fig:CCPC-z27-011_sky}
\end{subfigure}
\hfill
\begin{subfigure}
\centering
\includegraphics[scale=0.52]{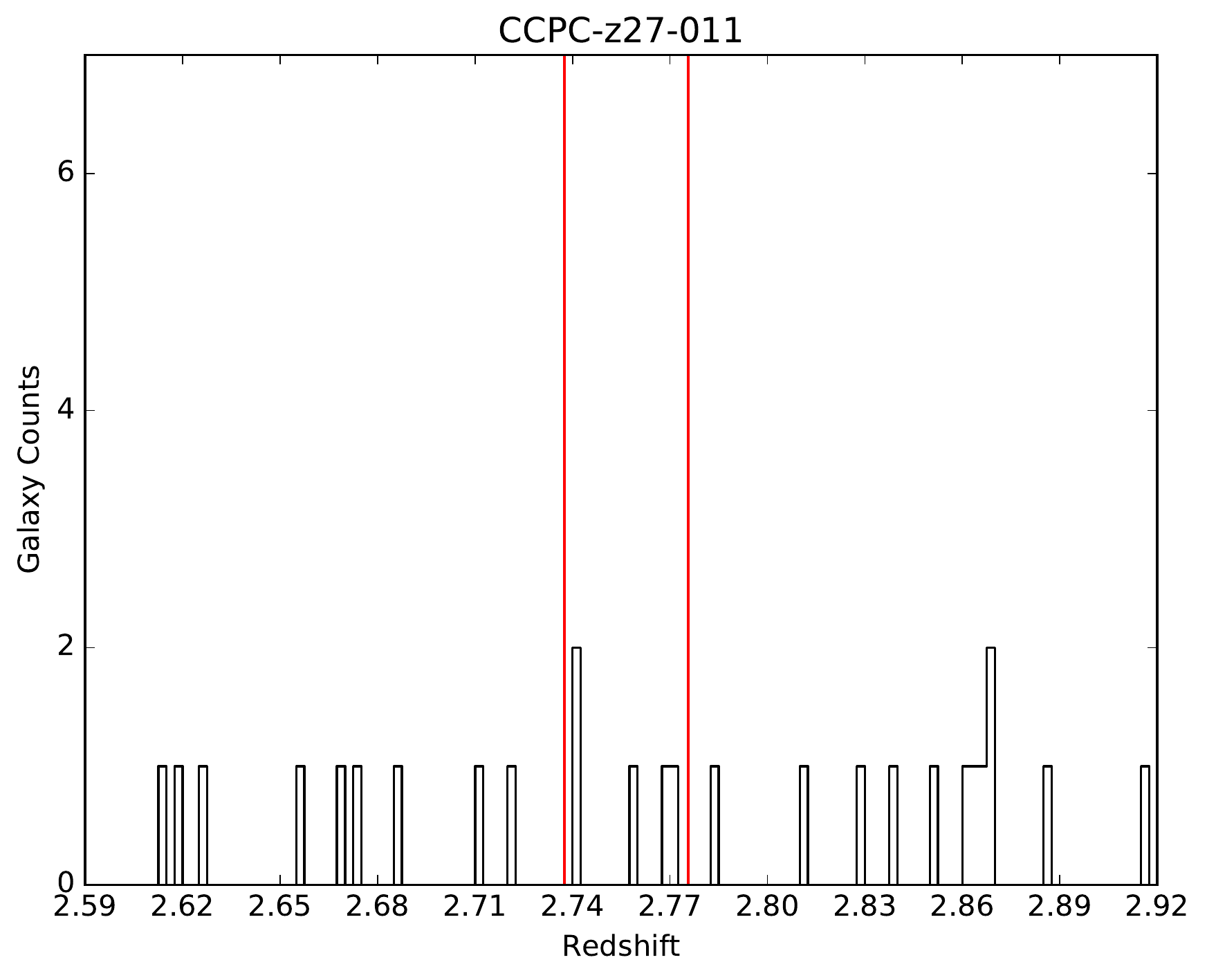}
\label{fig:CCPC-z27-011}
\end{subfigure}
\hfill
\end{figure*}

\begin{figure*}
\centering
\begin{subfigure}
\centering
\includegraphics[height=7.5cm,width=7.5cm]{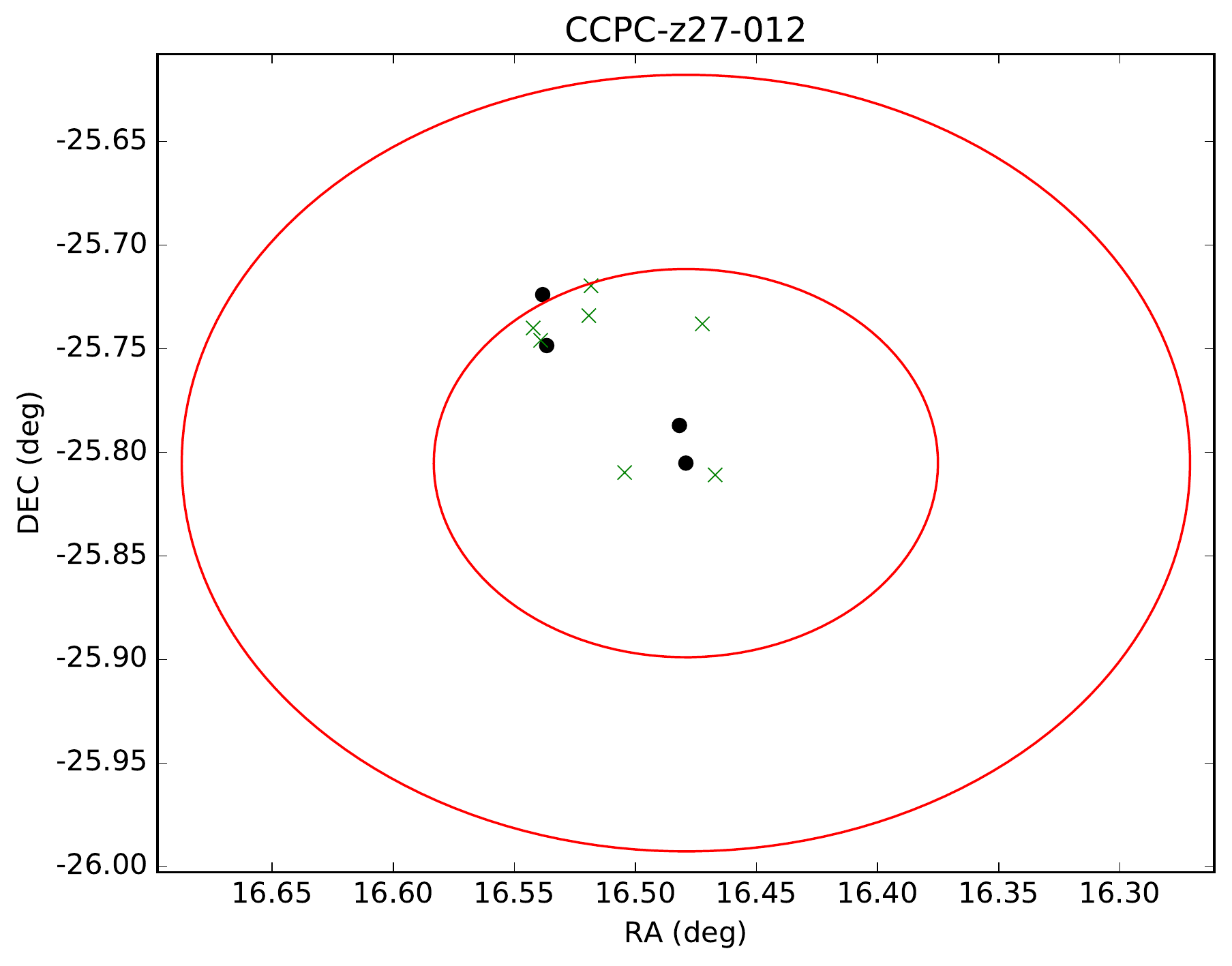}
\label{fig:CCPC-z27-012_sky}
\end{subfigure}
\hfill
\begin{subfigure}
\centering
\includegraphics[scale=0.52]{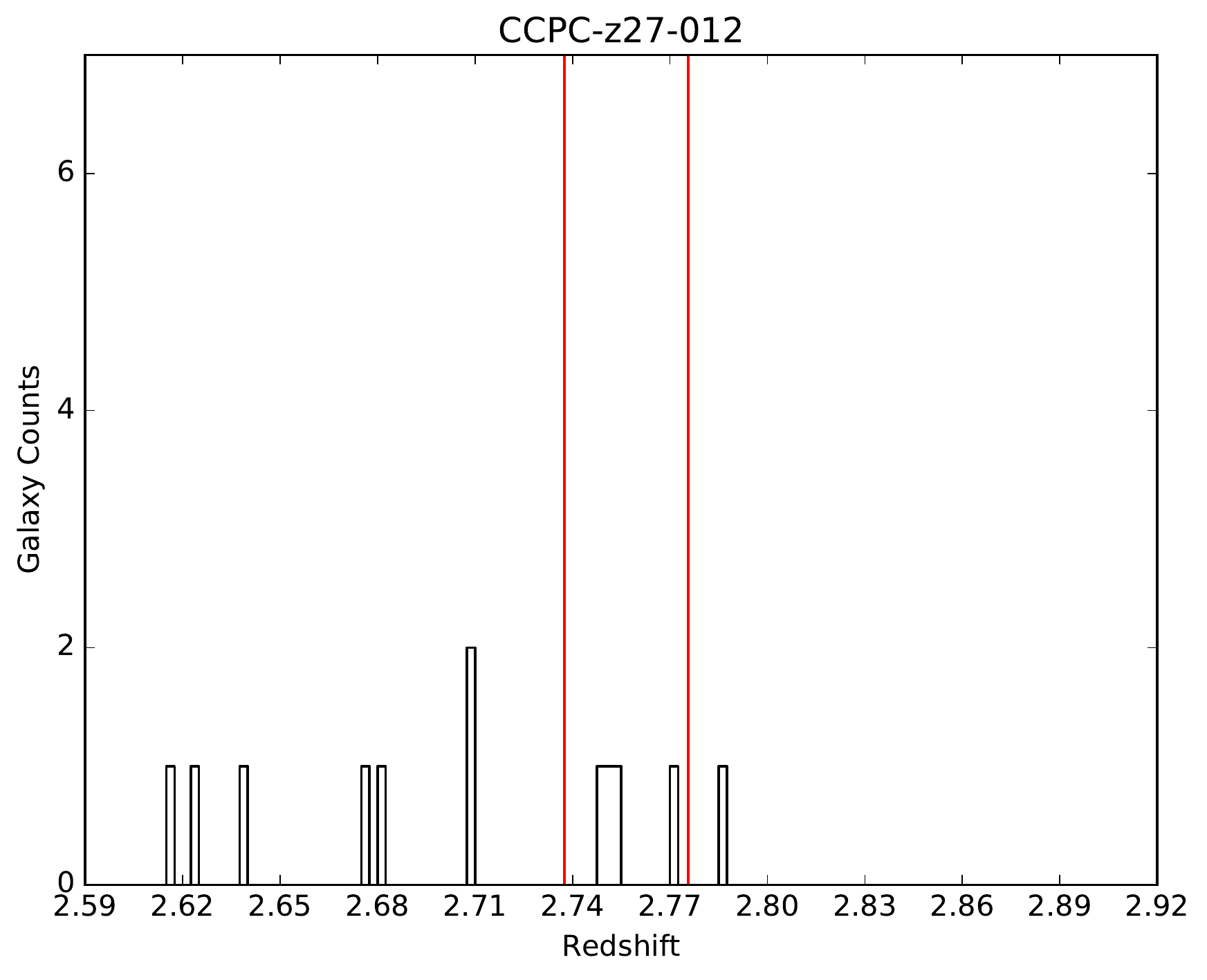}
\label{fig:CCPC-z27-012}
\end{subfigure}
\hfill
\end{figure*}
\clearpage 

\begin{figure*}
\centering
\begin{subfigure}
\centering
\includegraphics[height=7.5cm,width=7.5cm]{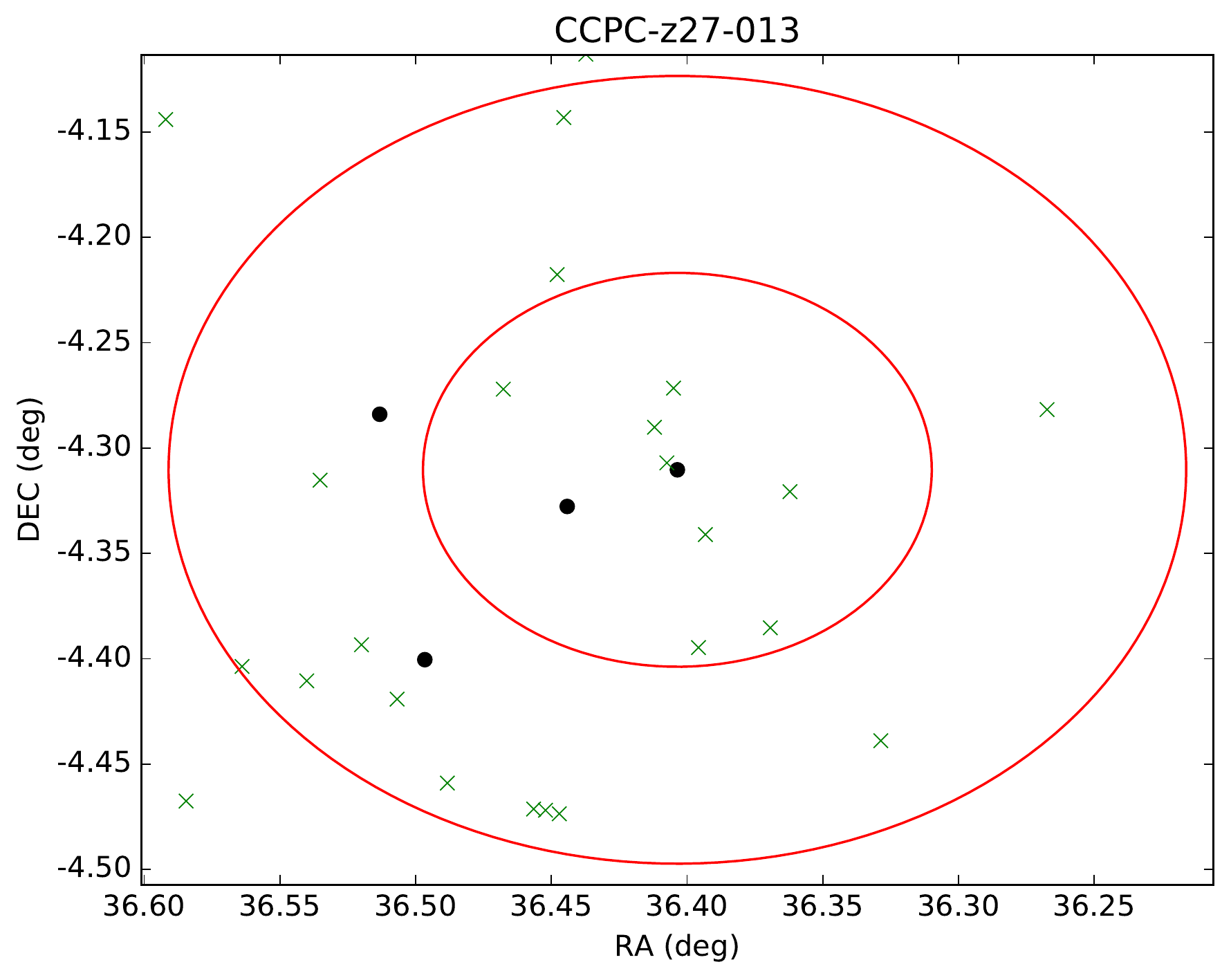}
\label{fig:CCPC-z27-013_sky}
\end{subfigure}
\hfill
\begin{subfigure}
\centering
\includegraphics[scale=0.52]{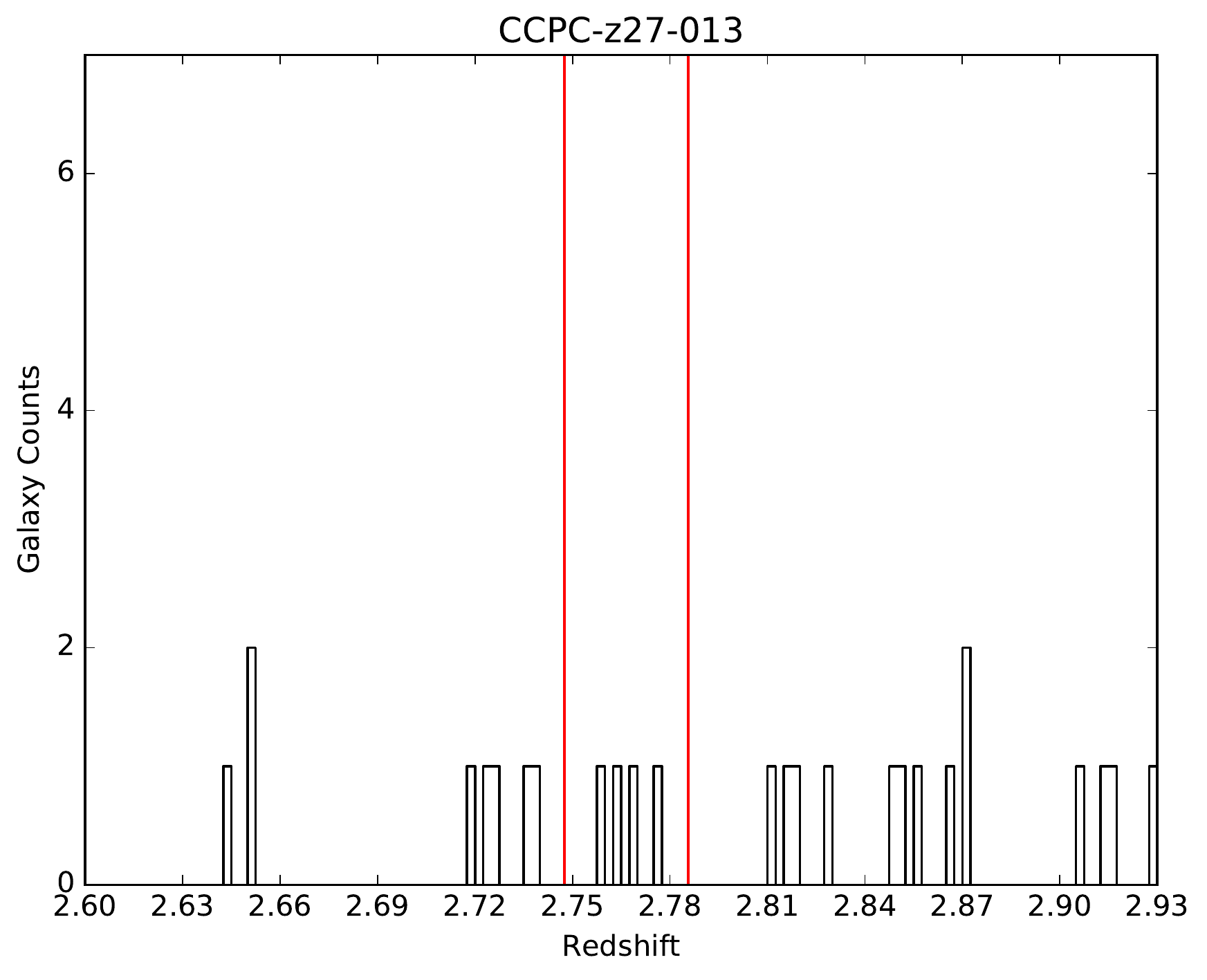}
\label{fig:CCPC-z27-013}
\end{subfigure}
\hfill
\end{figure*}

\begin{figure*}
\centering
\begin{subfigure}
\centering
\includegraphics[height=7.5cm,width=7.5cm]{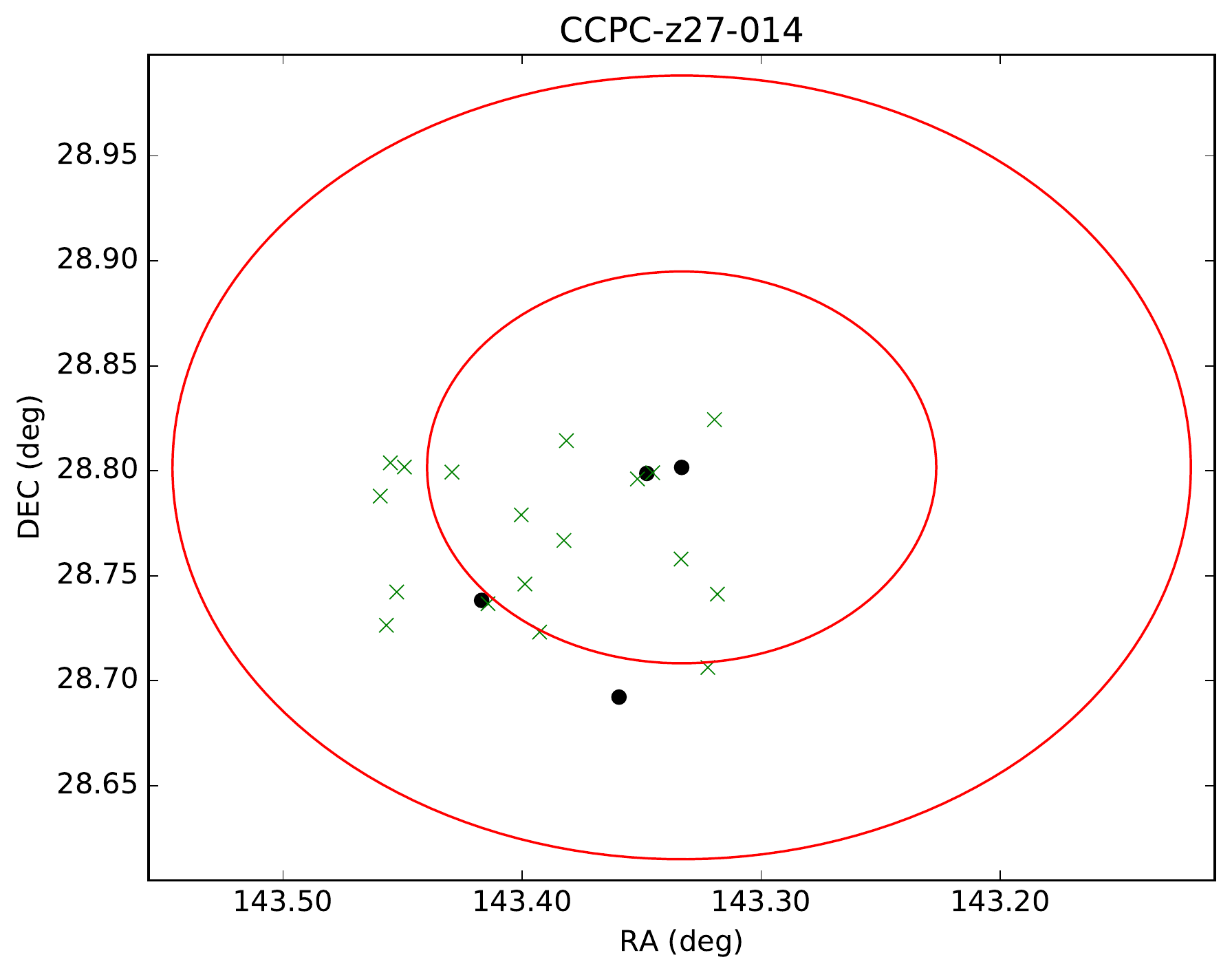}
\label{fig:CCPC-z27-014_sky}
\end{subfigure}
\hfill
\begin{subfigure}
\centering
\includegraphics[scale=0.52]{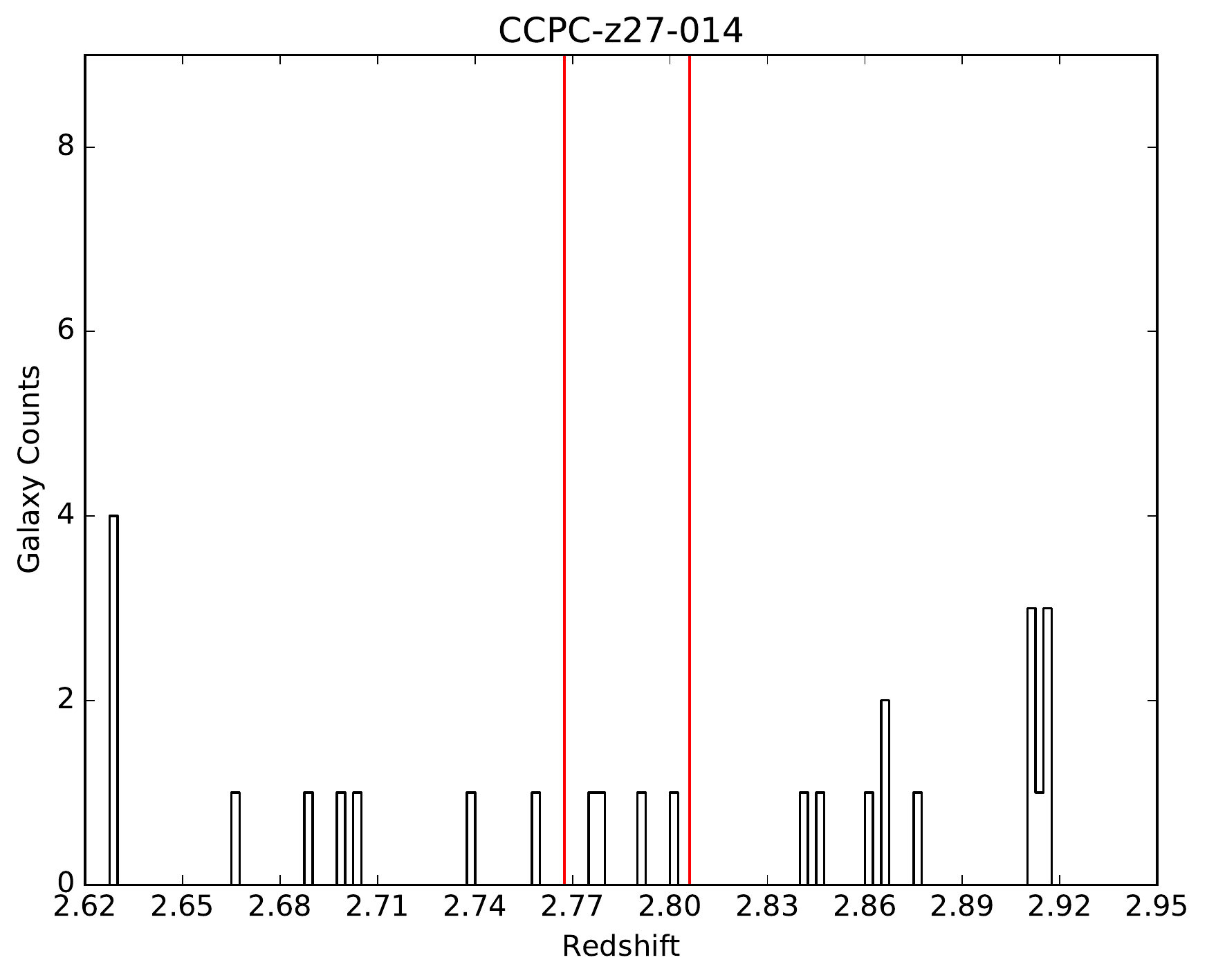}
\label{fig:CCPC-z27-014}
\end{subfigure}
\hfill
\end{figure*}

\begin{figure*}
\centering
\begin{subfigure}
\centering
\includegraphics[height=7.5cm,width=7.5cm]{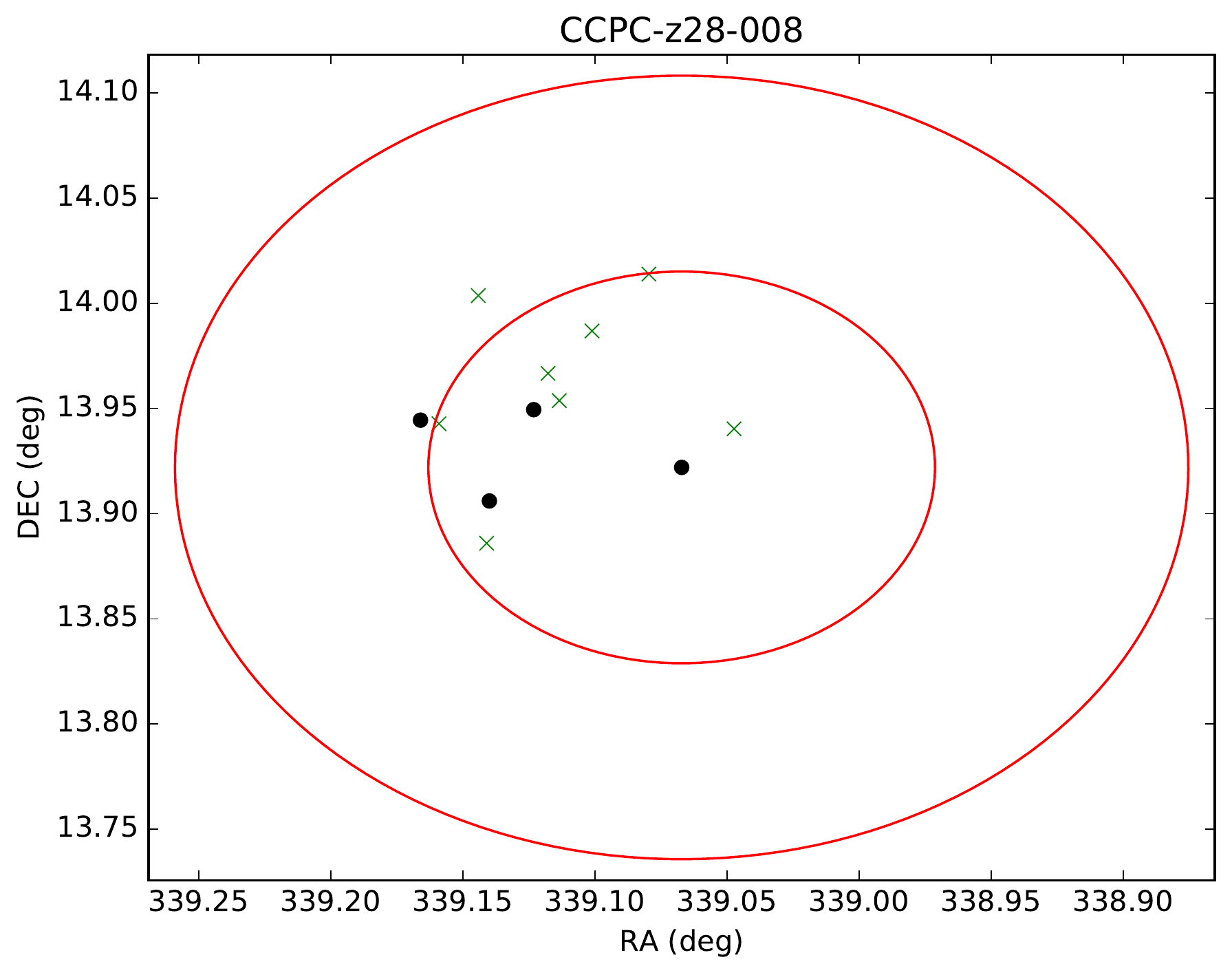}
\label{fig:CCPC-z28-008_sky}
\end{subfigure}
\hfill
\begin{subfigure}
\centering
\includegraphics[scale=0.52]{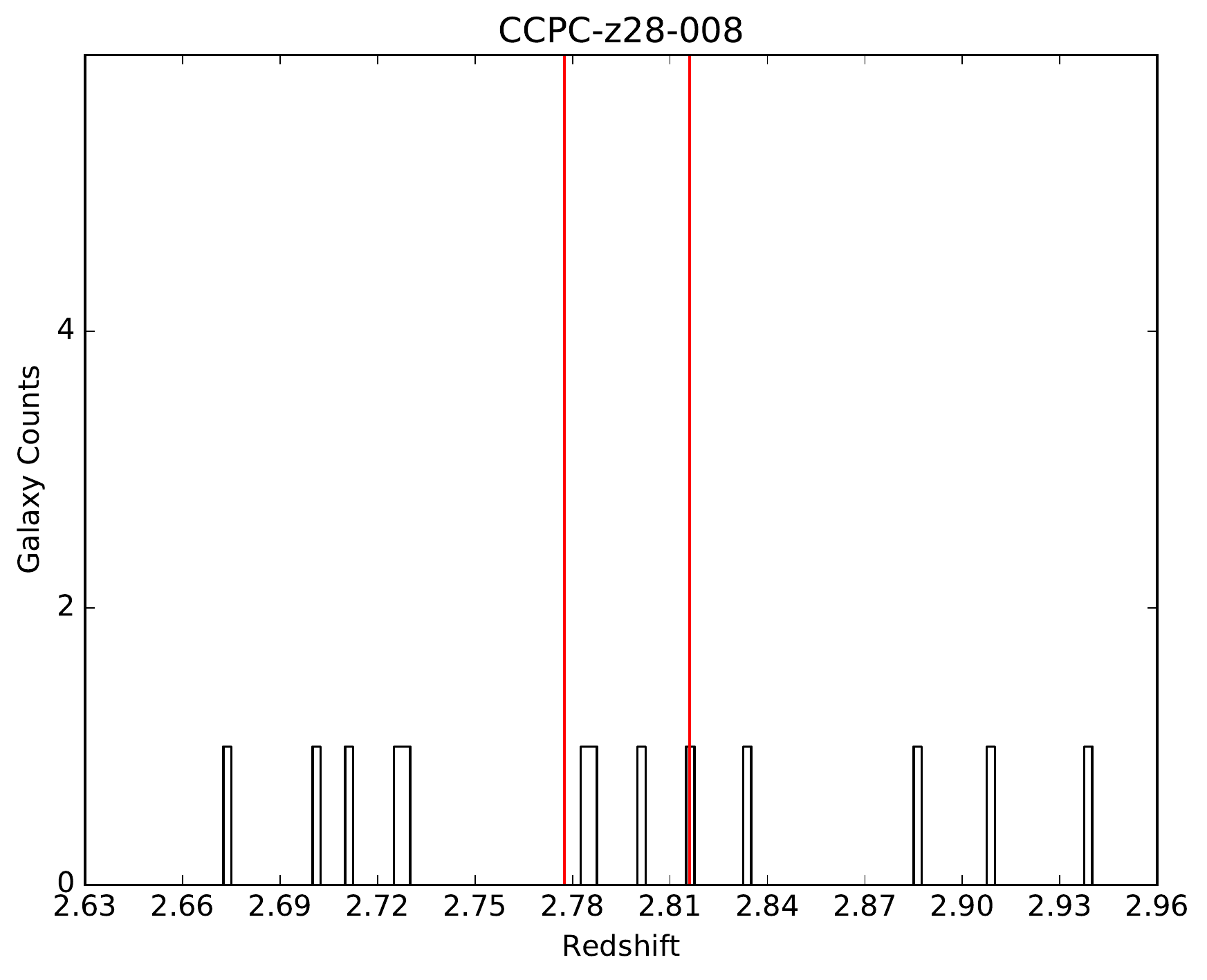}
\label{fig:CCPC-z28-008}
\end{subfigure}
\hfill
\end{figure*}
\clearpage 

\begin{figure*}
\centering
\begin{subfigure}
\centering
\includegraphics[height=7.5cm,width=7.5cm]{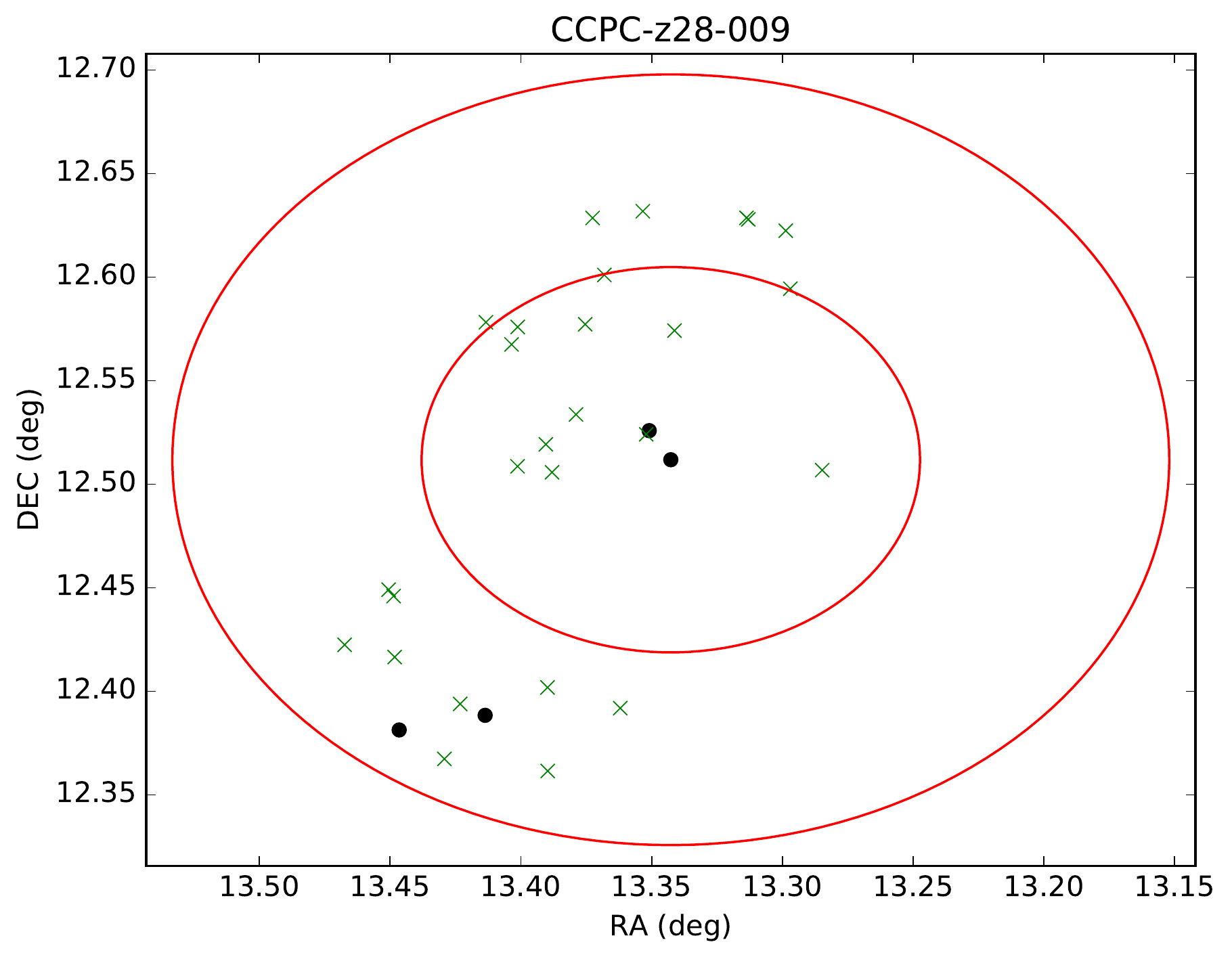}
\label{fig:CCPC-z28-009_sky}
\end{subfigure}
\hfill
\begin{subfigure}
\centering
\includegraphics[scale=0.52]{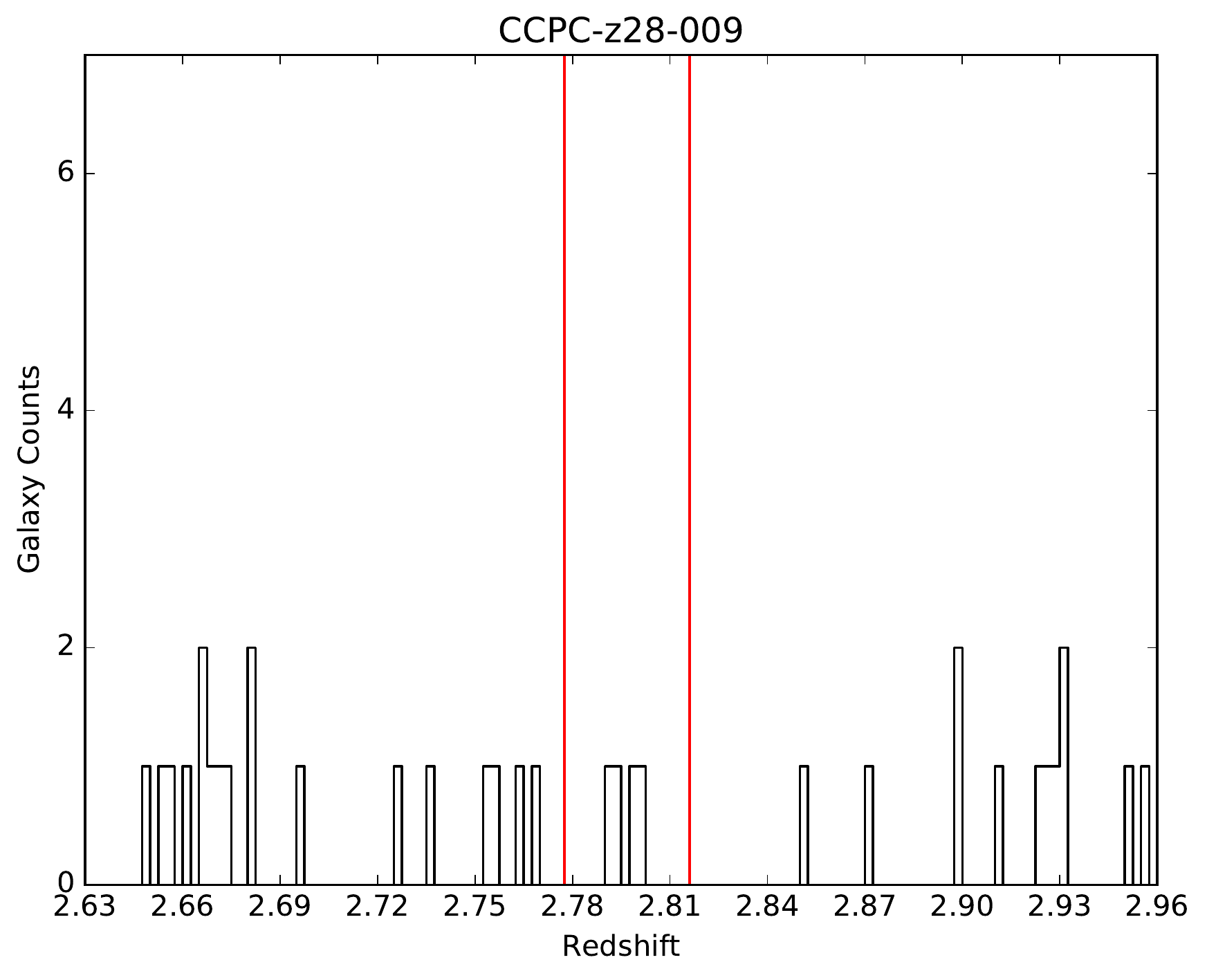}
\label{fig:CCPC-z28-009}
\end{subfigure}
\hfill
\end{figure*}

\begin{figure*}
\centering
\begin{subfigure}
\centering
\includegraphics[height=7.5cm,width=7.5cm]{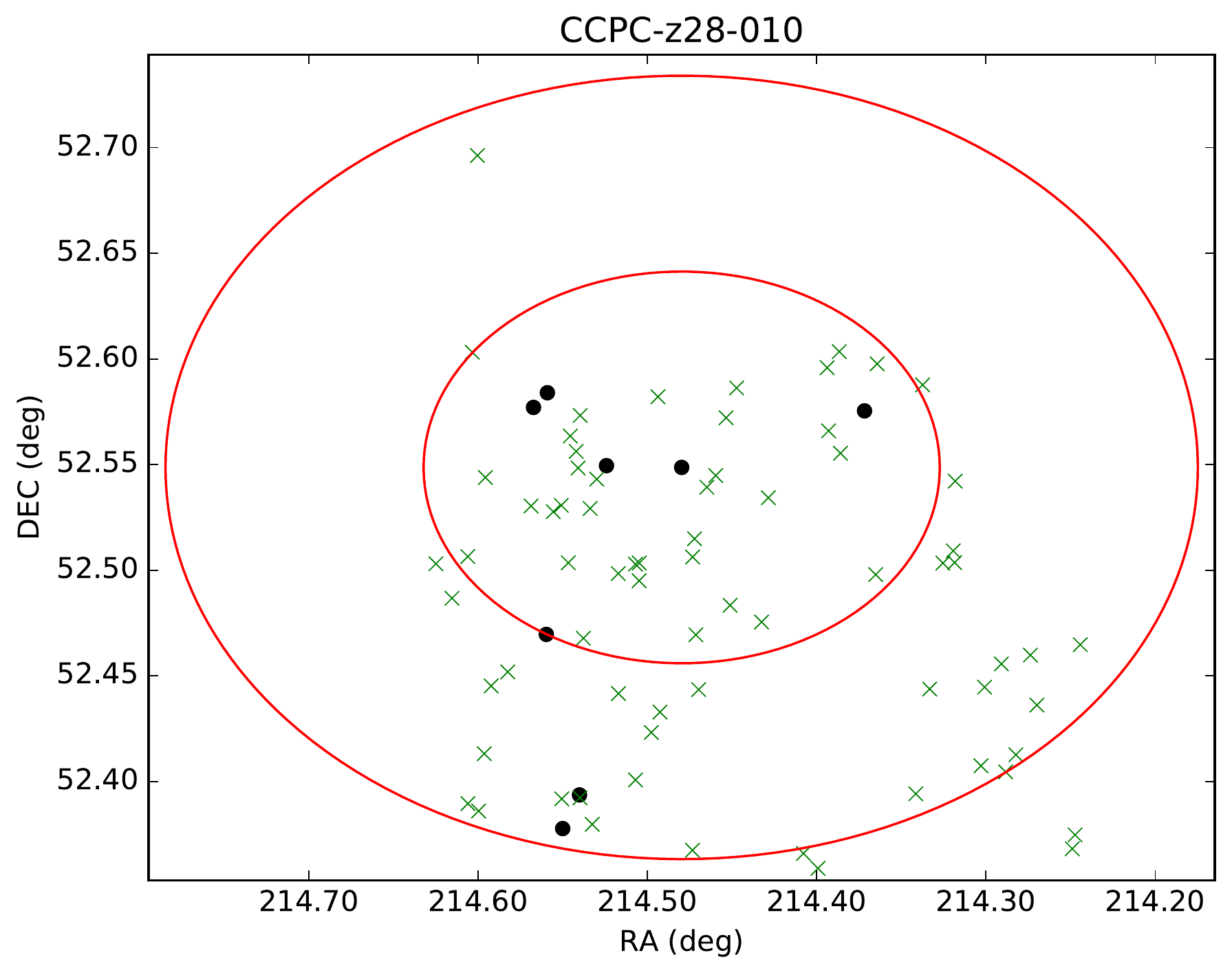}
\label{fig:CCPC-z28-010_sky}
\end{subfigure}
\hfill
\begin{subfigure}
\centering
\includegraphics[scale=0.52]{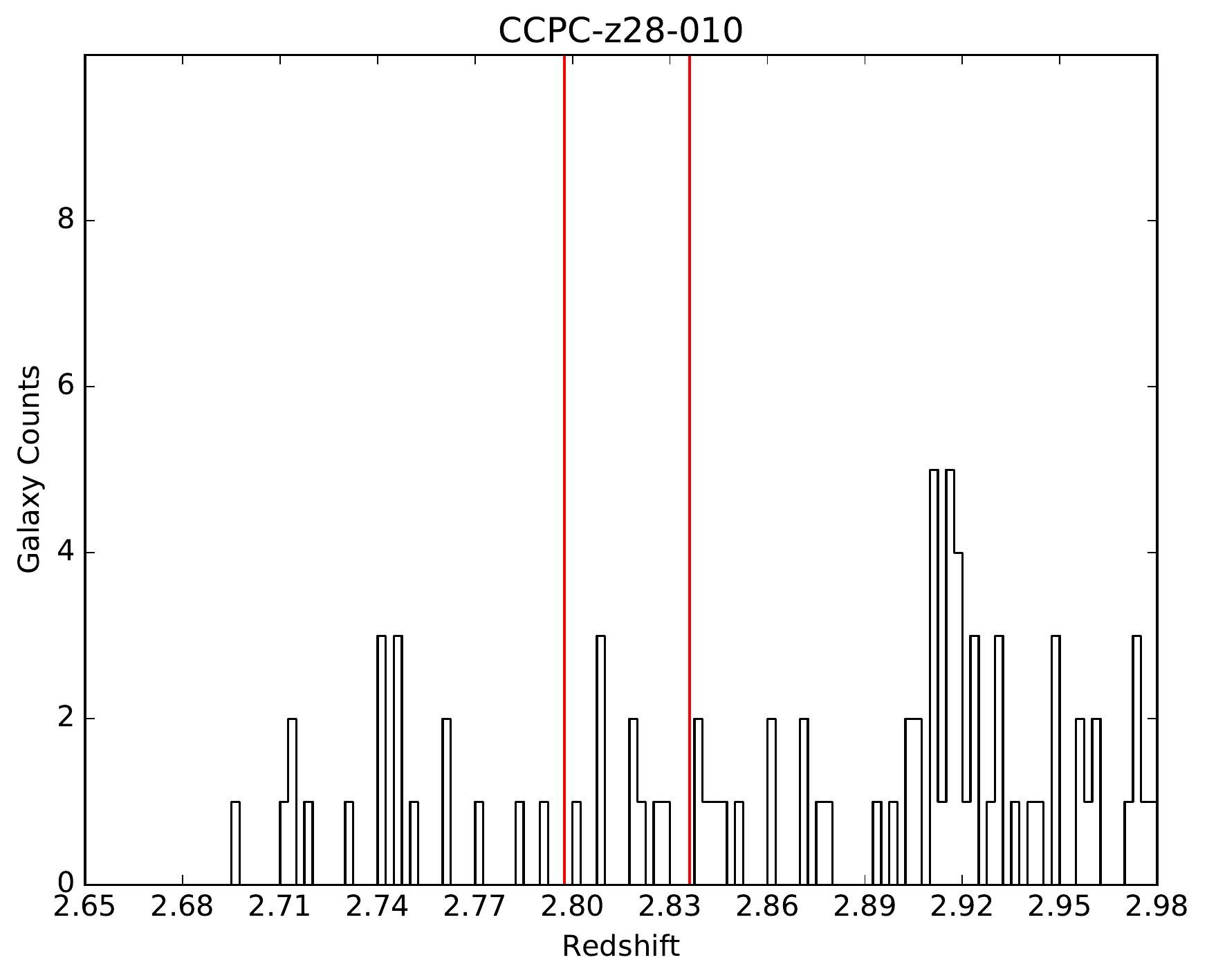}
\label{fig:CCPC-z28-010}
\end{subfigure}
\hfill
\end{figure*}

\begin{figure*}
\centering
\begin{subfigure}
\centering
\includegraphics[height=7.5cm,width=7.5cm]{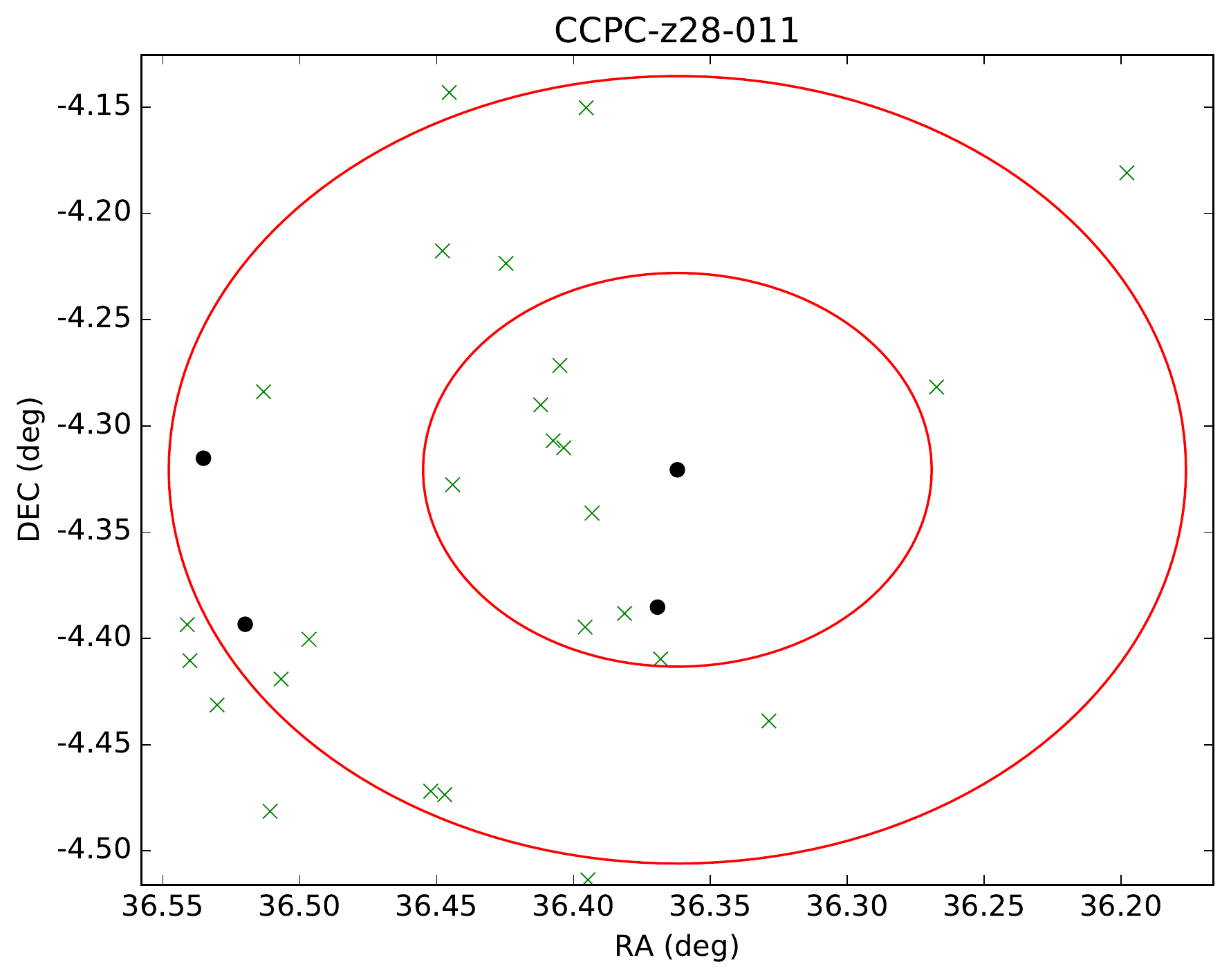}
\label{fig:CCPC-z28-011_sky}
\end{subfigure}
\hfill
\begin{subfigure}
\centering
\includegraphics[scale=0.52]{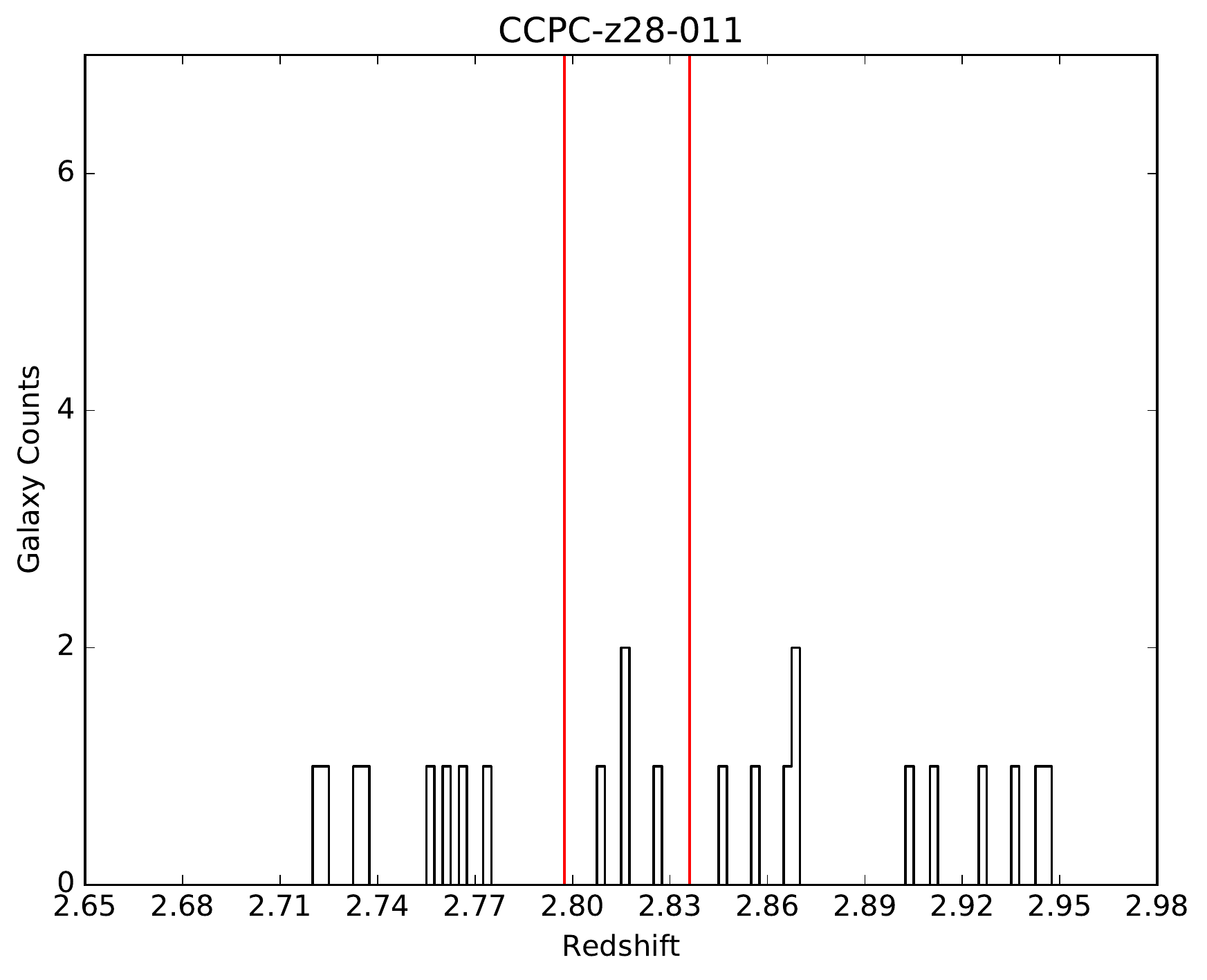}
\label{fig:CCPC-z28-011}
\end{subfigure}
\hfill
\end{figure*}
\clearpage 

\begin{figure*}
\centering
\begin{subfigure}
\centering
\includegraphics[height=7.5cm,width=7.5cm]{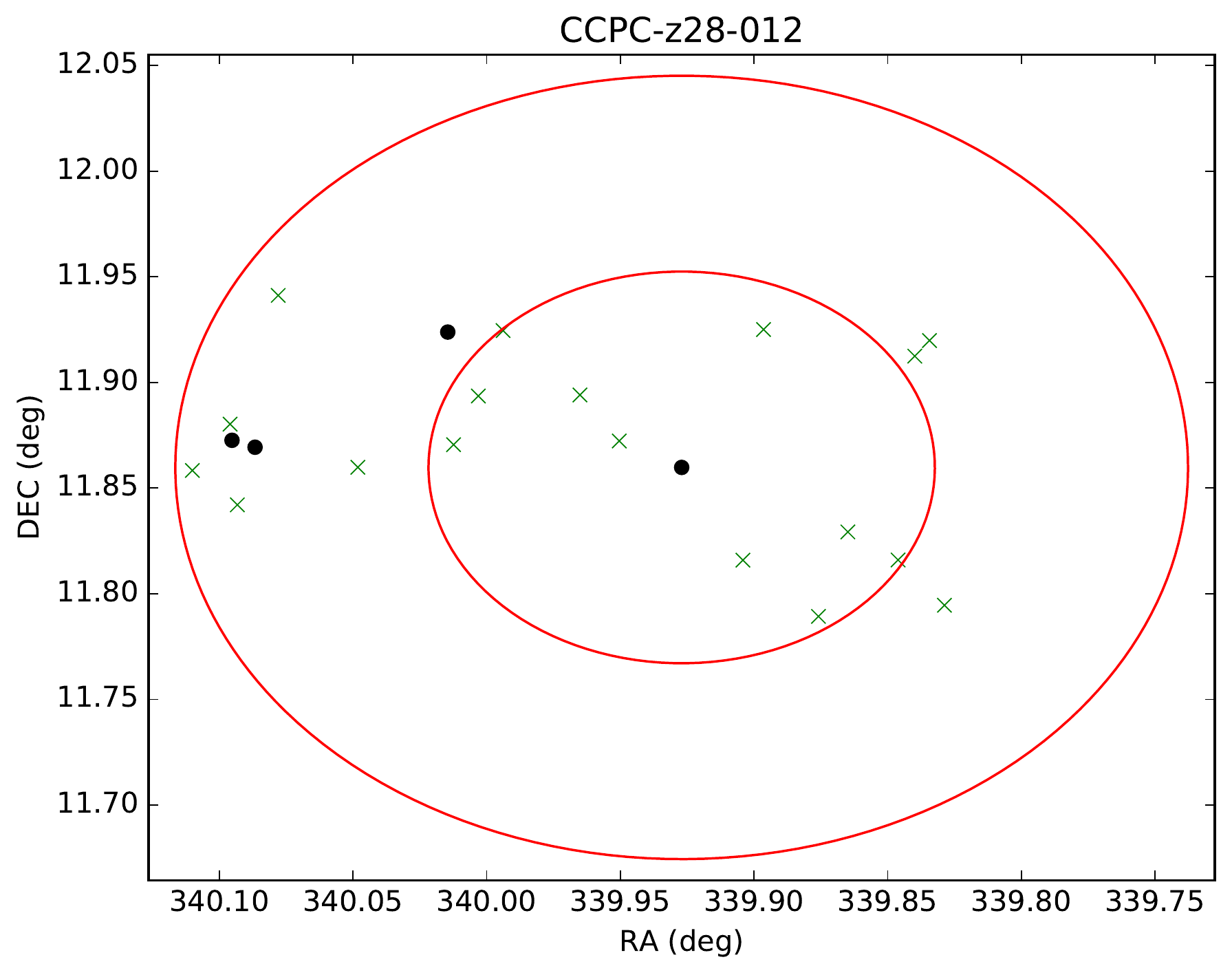}
\label{fig:CCPC-z28-012_sky}
\end{subfigure}
\hfill
\begin{subfigure}
\centering
\includegraphics[scale=0.52]{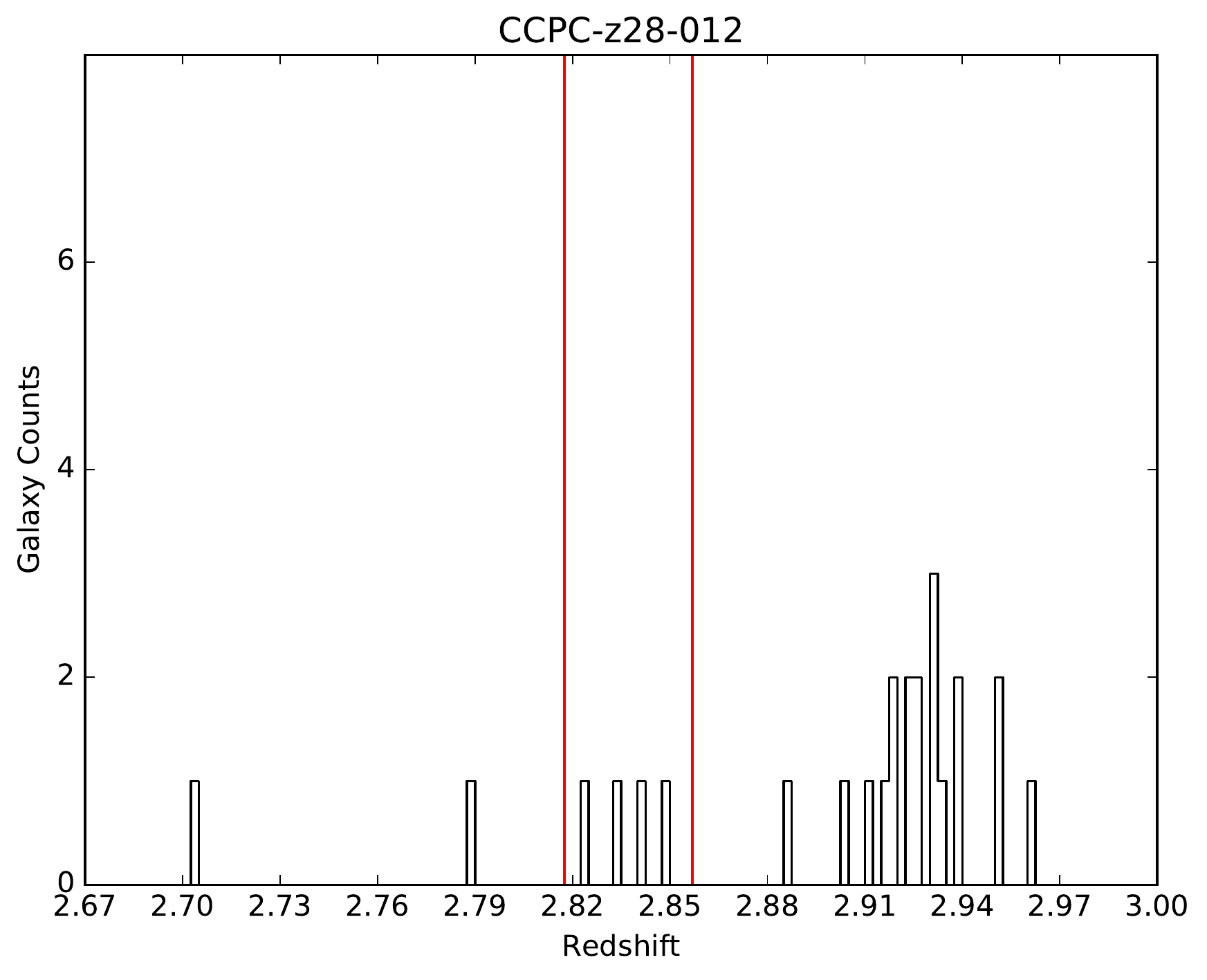}
\label{fig:CCPC-z28-012}
\end{subfigure}
\hfill
\end{figure*}

\begin{figure*}
\centering
\begin{subfigure}
\centering
\includegraphics[height=7.5cm,width=7.5cm]{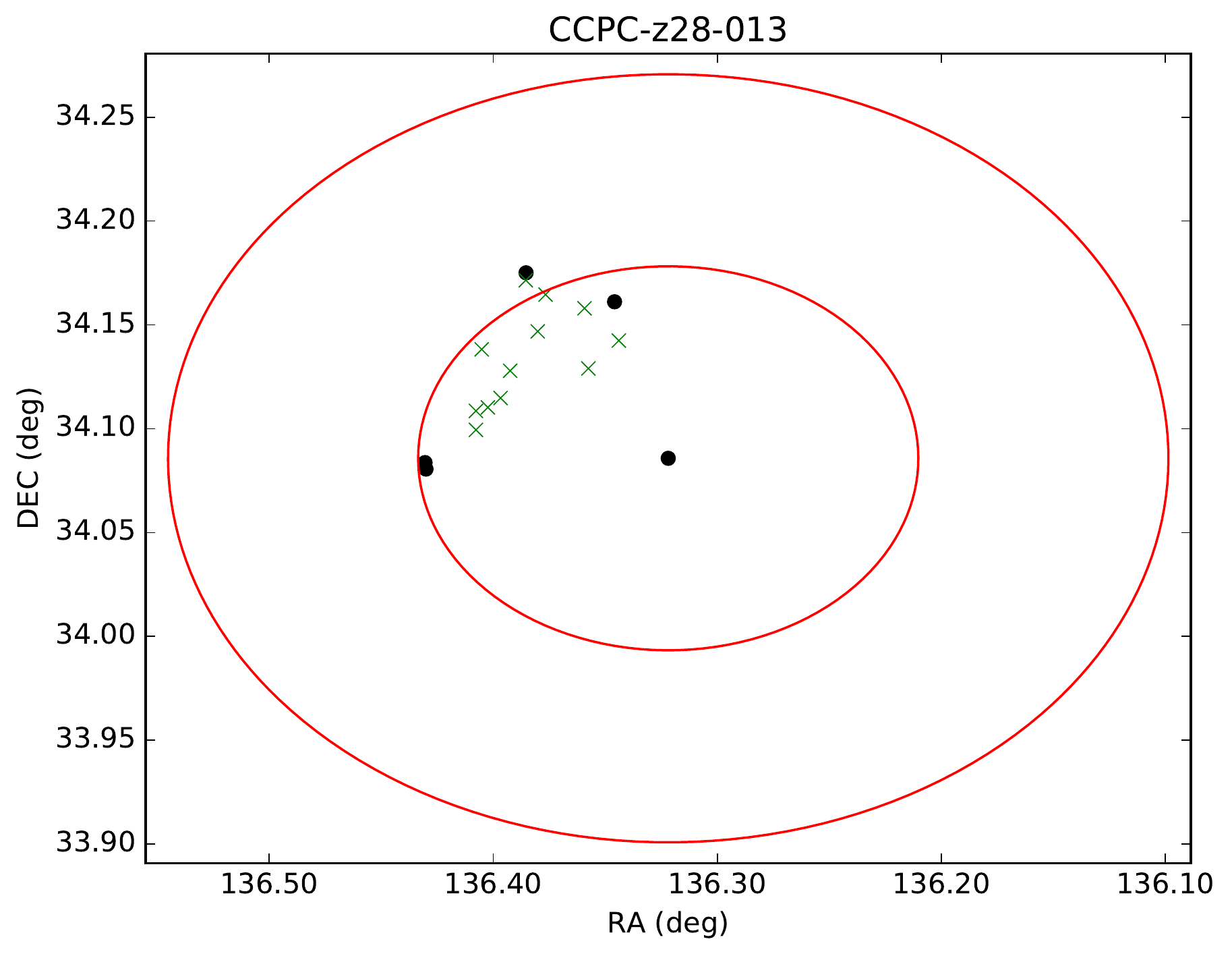}
\label{fig:CCPC-z28-013_sky}
\end{subfigure}
\hfill
\begin{subfigure}
\centering
\includegraphics[scale=0.52]{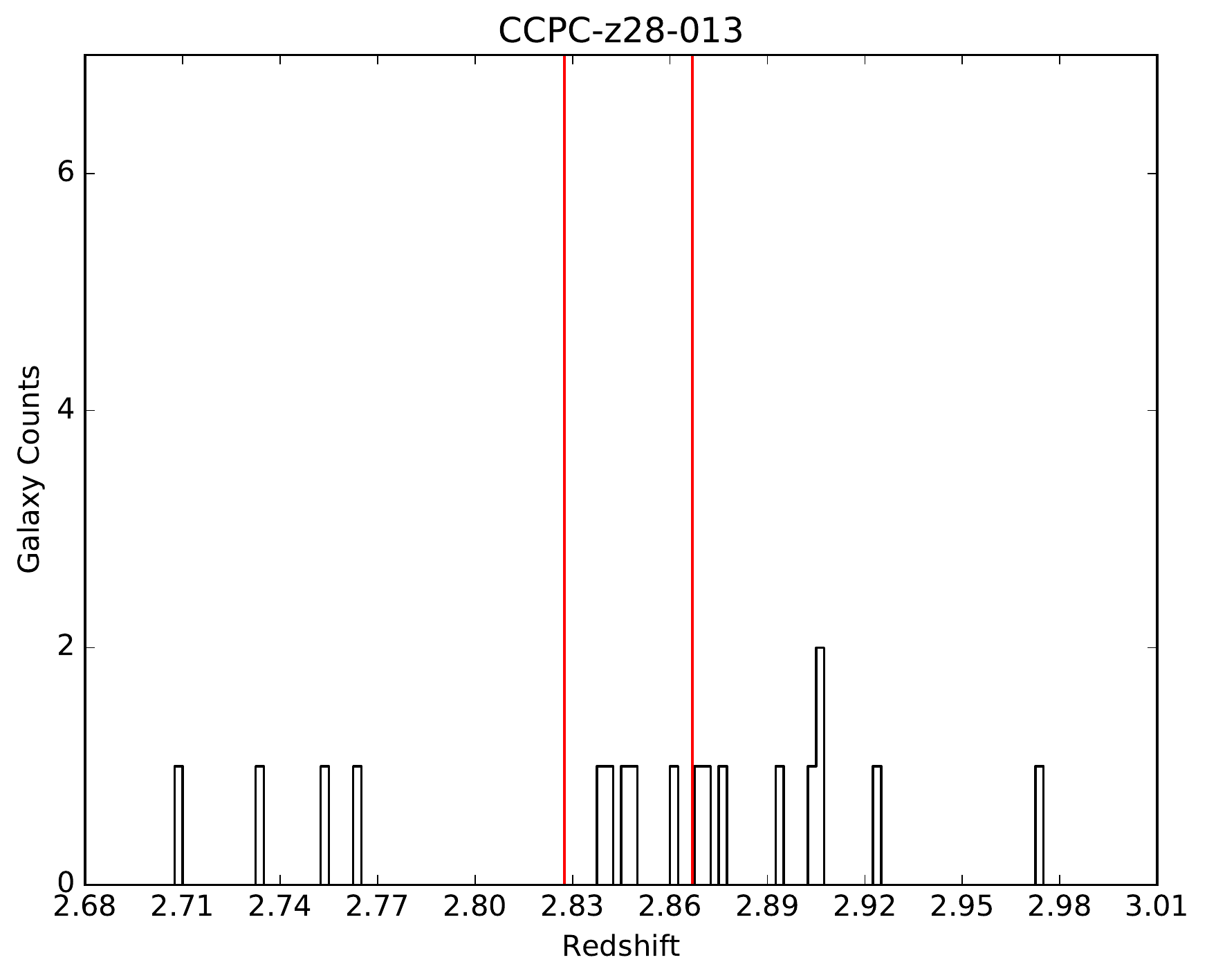}
\label{fig:CCPC-z28-013}
\end{subfigure}
\hfill
\end{figure*}

\begin{figure*}
\centering
\begin{subfigure}
\centering
\includegraphics[height=7.5cm,width=7.5cm]{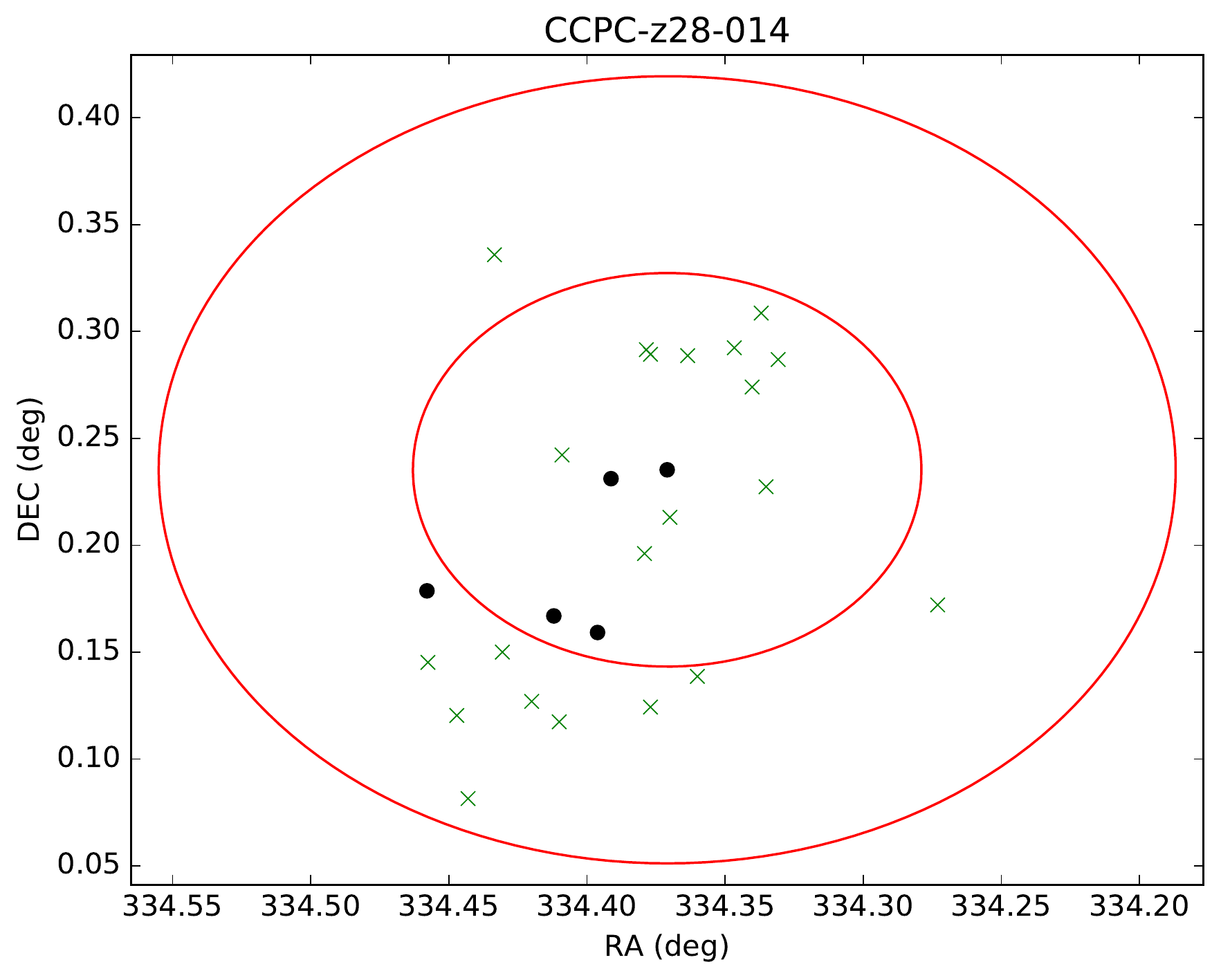}
\label{fig:CCPC-z28-014_sky}
\end{subfigure}
\hfill
\begin{subfigure}
\centering
\includegraphics[scale=0.52]{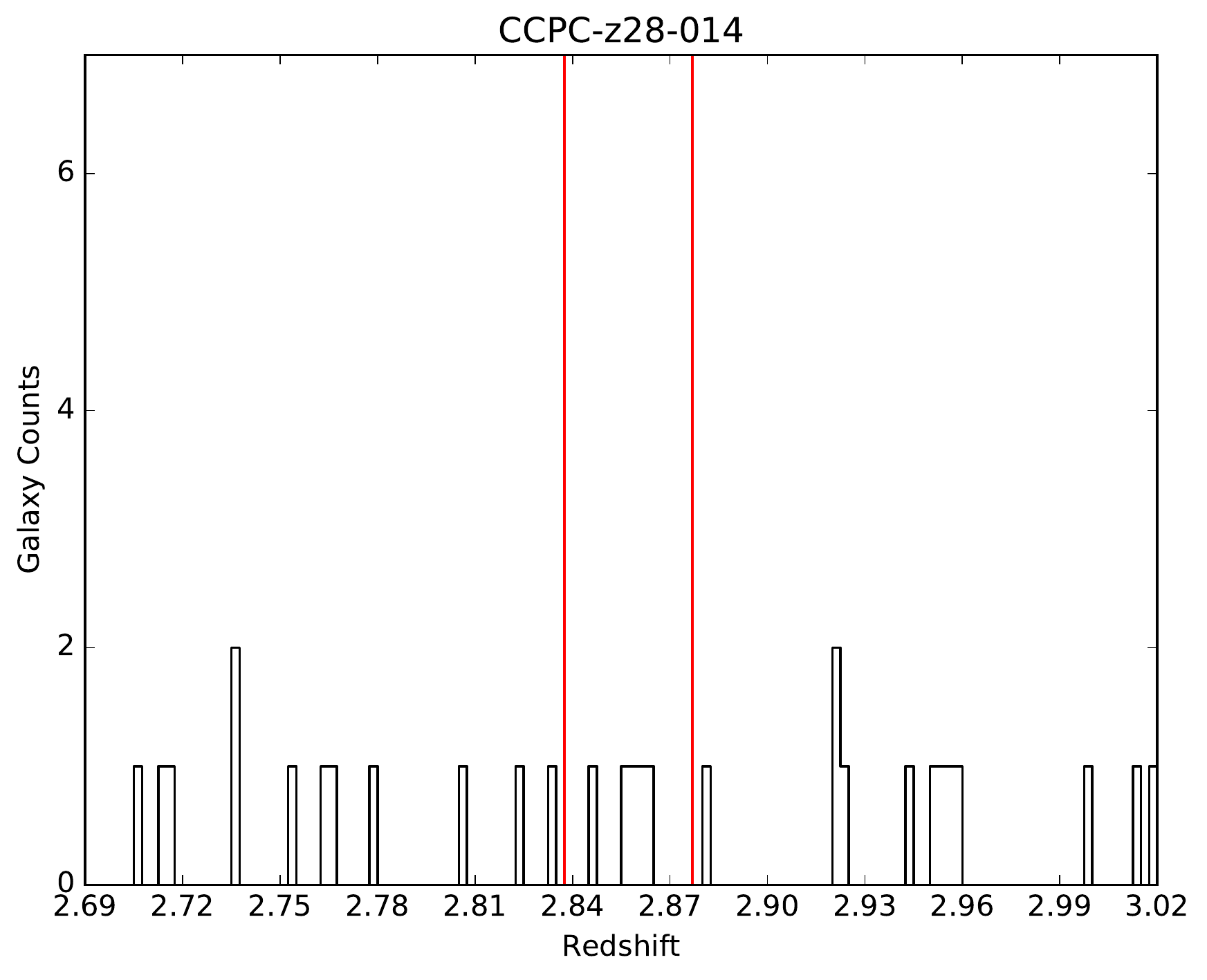}
\label{fig:CCPC-z28-014}
\end{subfigure}
\hfill
\end{figure*}
\clearpage 

\begin{figure*}
\centering
\begin{subfigure}
\centering
\includegraphics[height=7.5cm,width=7.5cm]{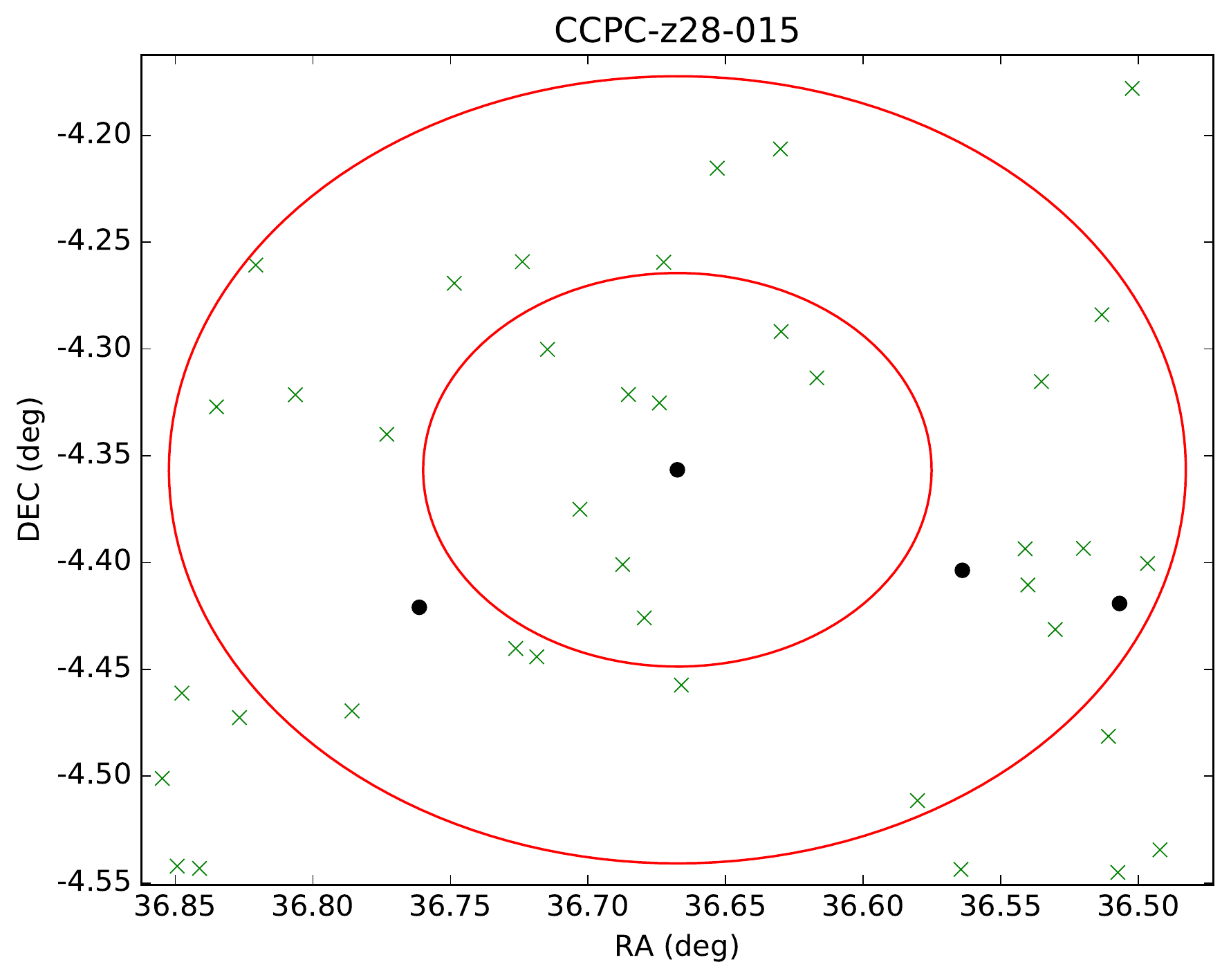}
\label{fig:CCPC-z28-015_sky}
\end{subfigure}
\hfill
\begin{subfigure}
\centering
\includegraphics[scale=0.52]{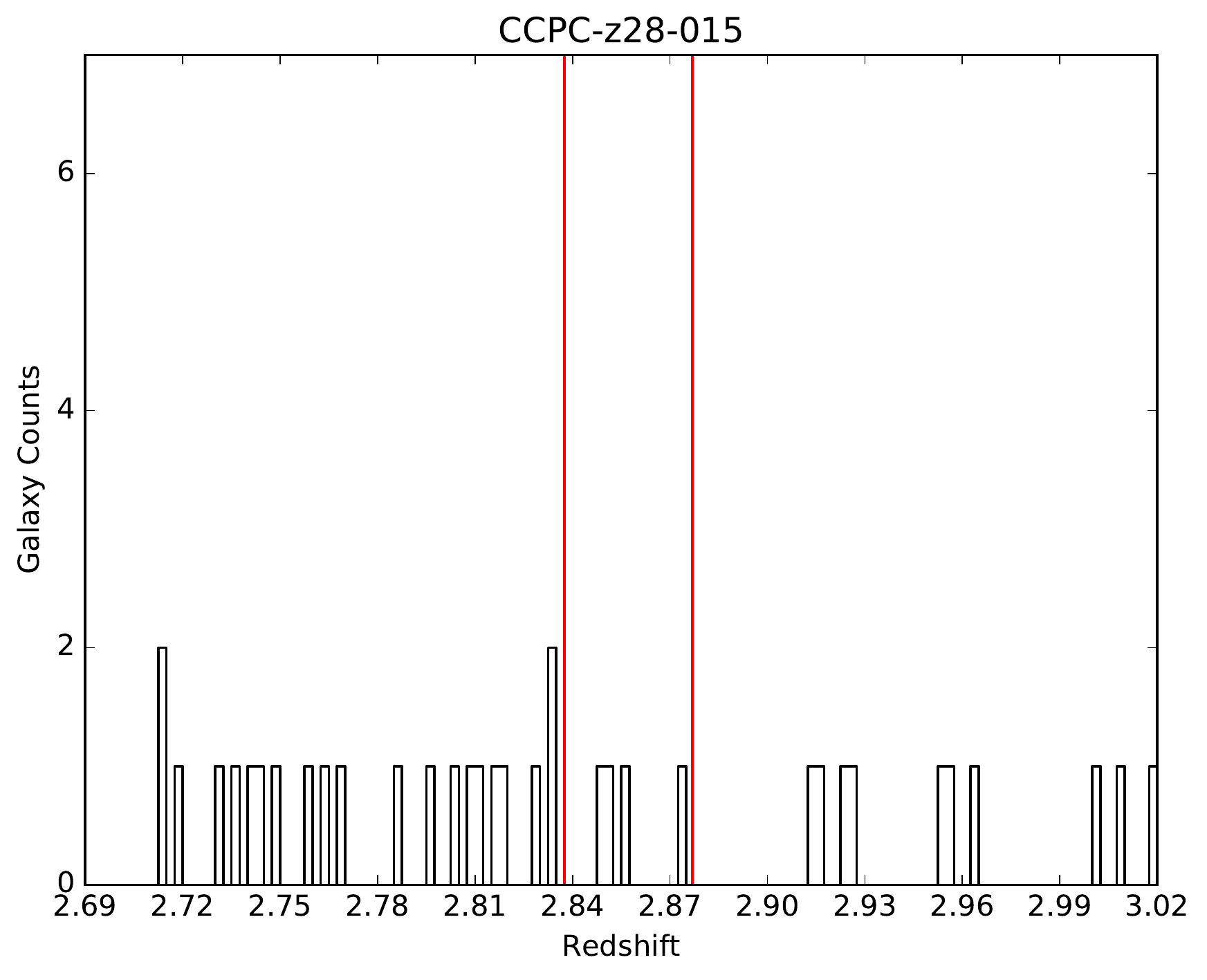}
\label{fig:CCPC-z28-015}
\end{subfigure}
\hfill
\end{figure*}

\begin{figure*}
\centering
\begin{subfigure}
\centering
\includegraphics[height=7.5cm,width=7.5cm]{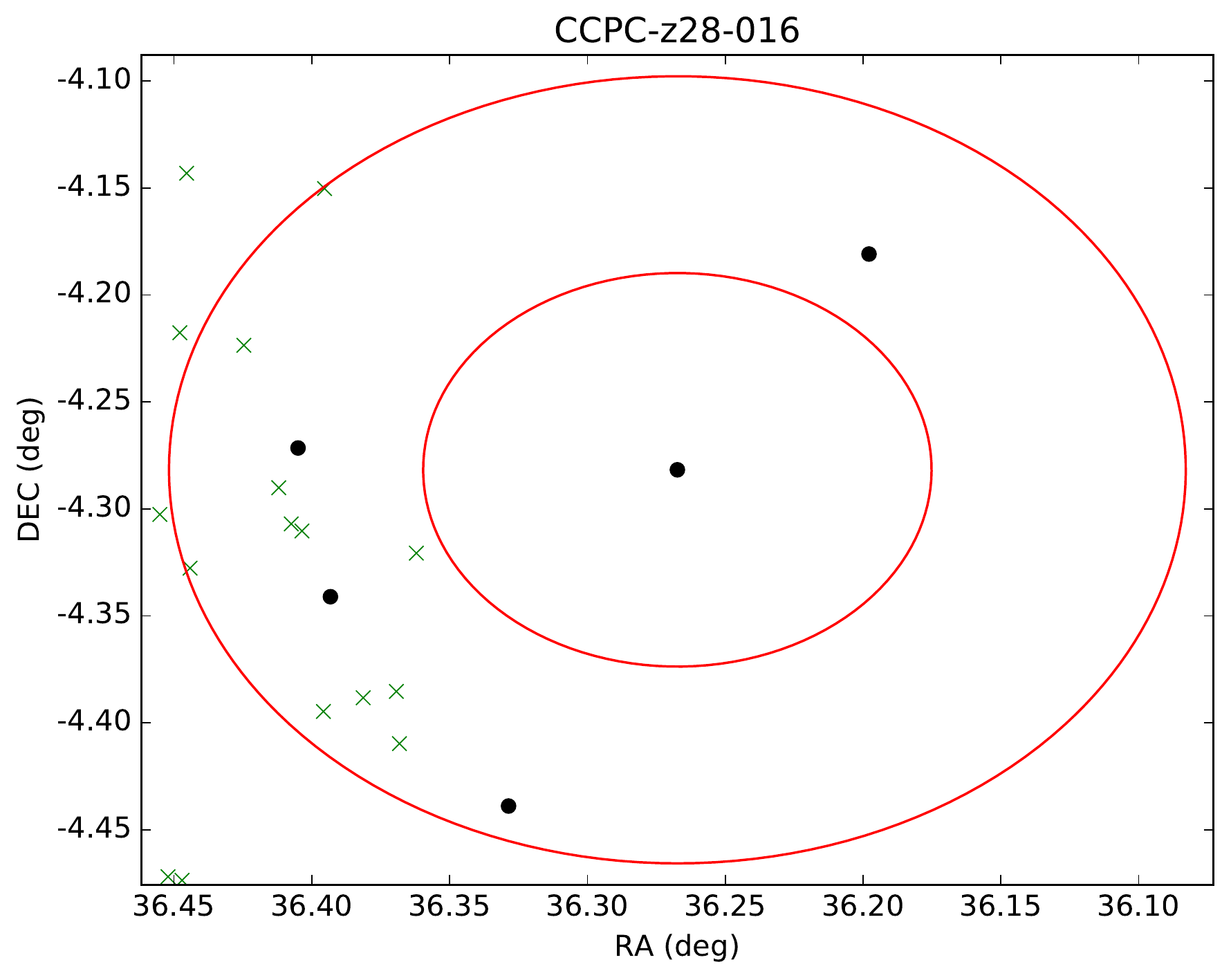}
\label{fig:CCPC-z28-016_sky}
\end{subfigure}
\hfill
\begin{subfigure}
\centering
\includegraphics[scale=0.52]{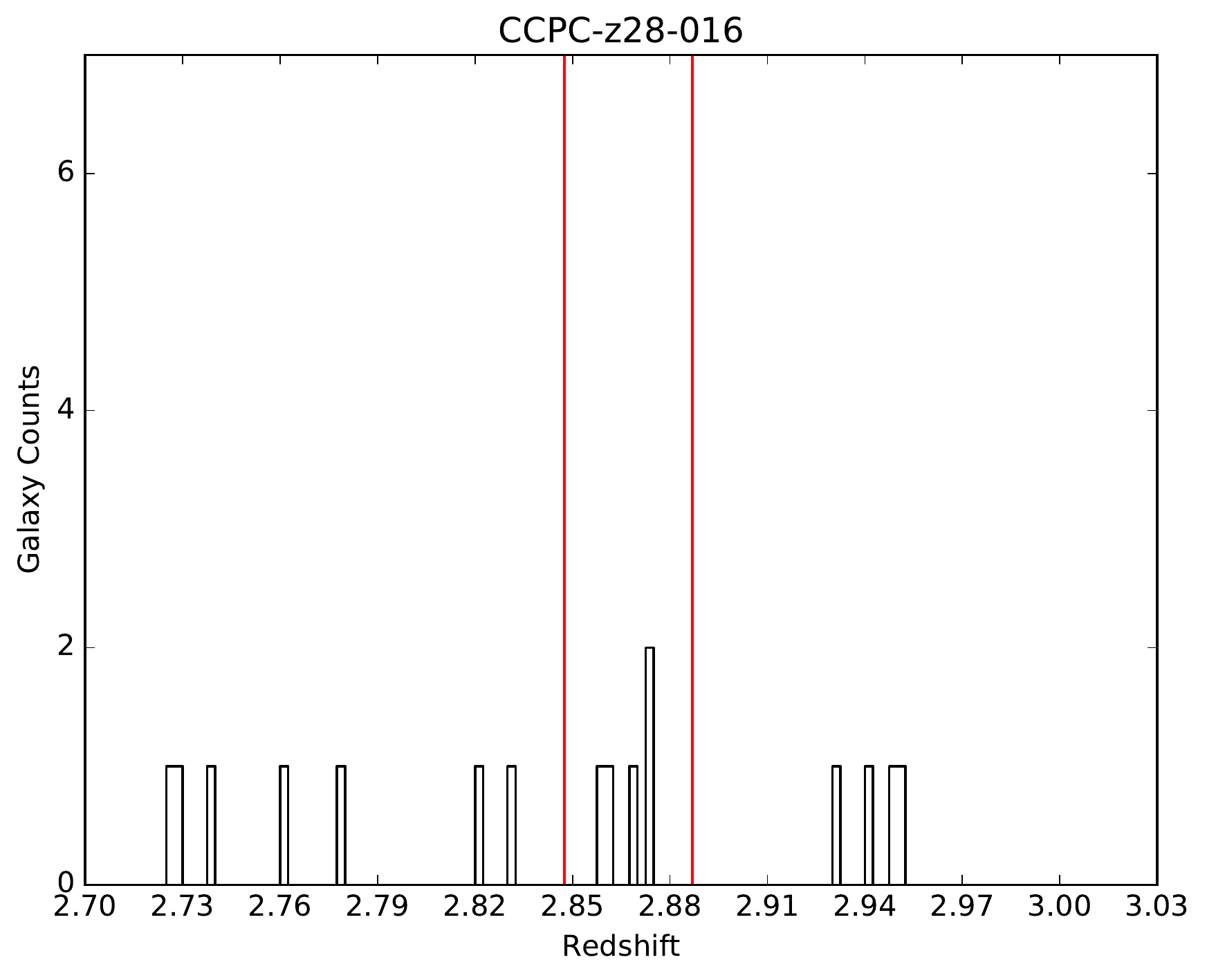}
\label{fig:CCPC-z28-016}
\end{subfigure}
\hfill
\end{figure*}

\begin{figure*}
\centering
\begin{subfigure}
\centering
\includegraphics[height=7.5cm,width=7.5cm]{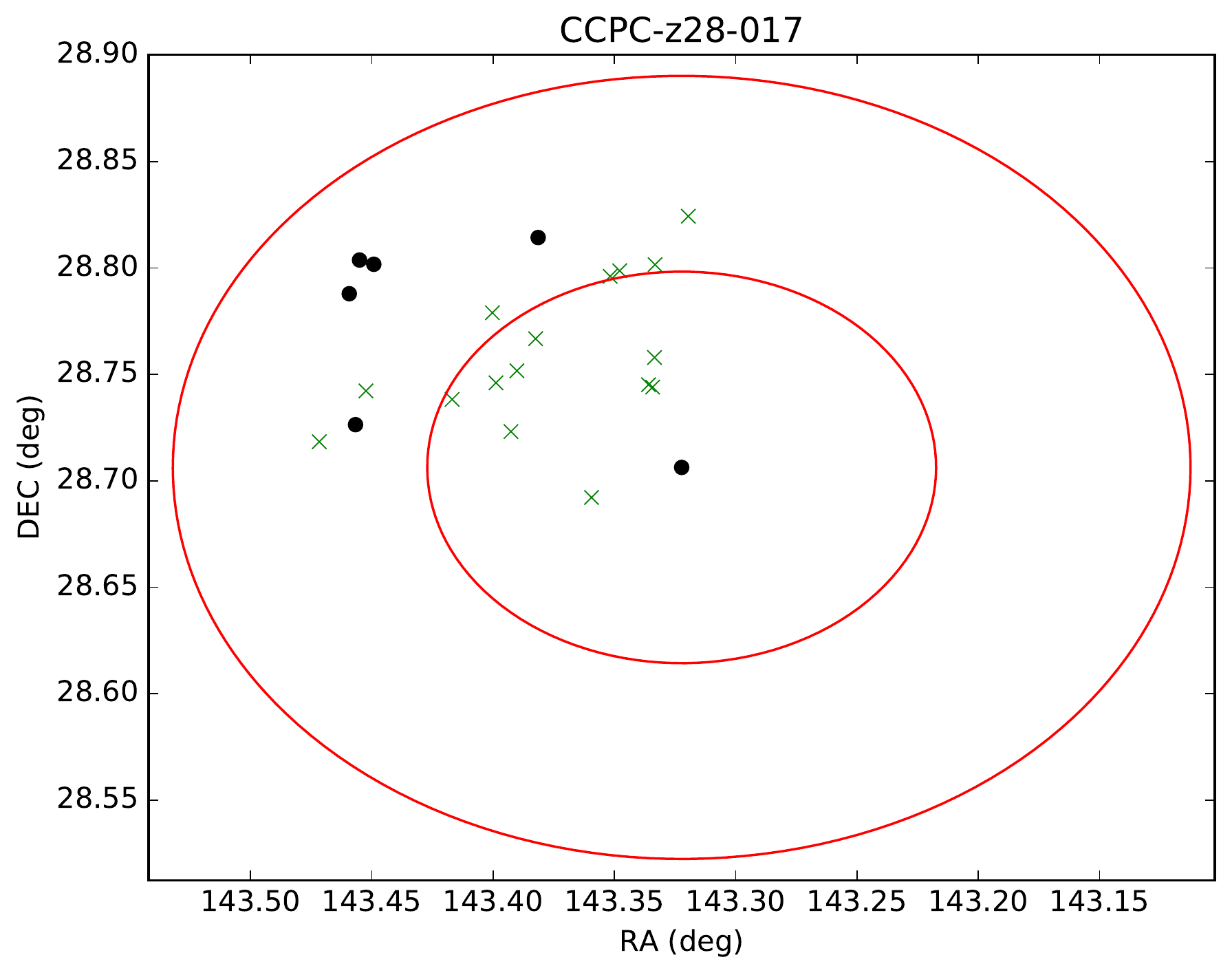}
\label{fig:CCPC-z28-017_sky}
\end{subfigure}
\hfill
\begin{subfigure}
\centering
\includegraphics[scale=0.52]{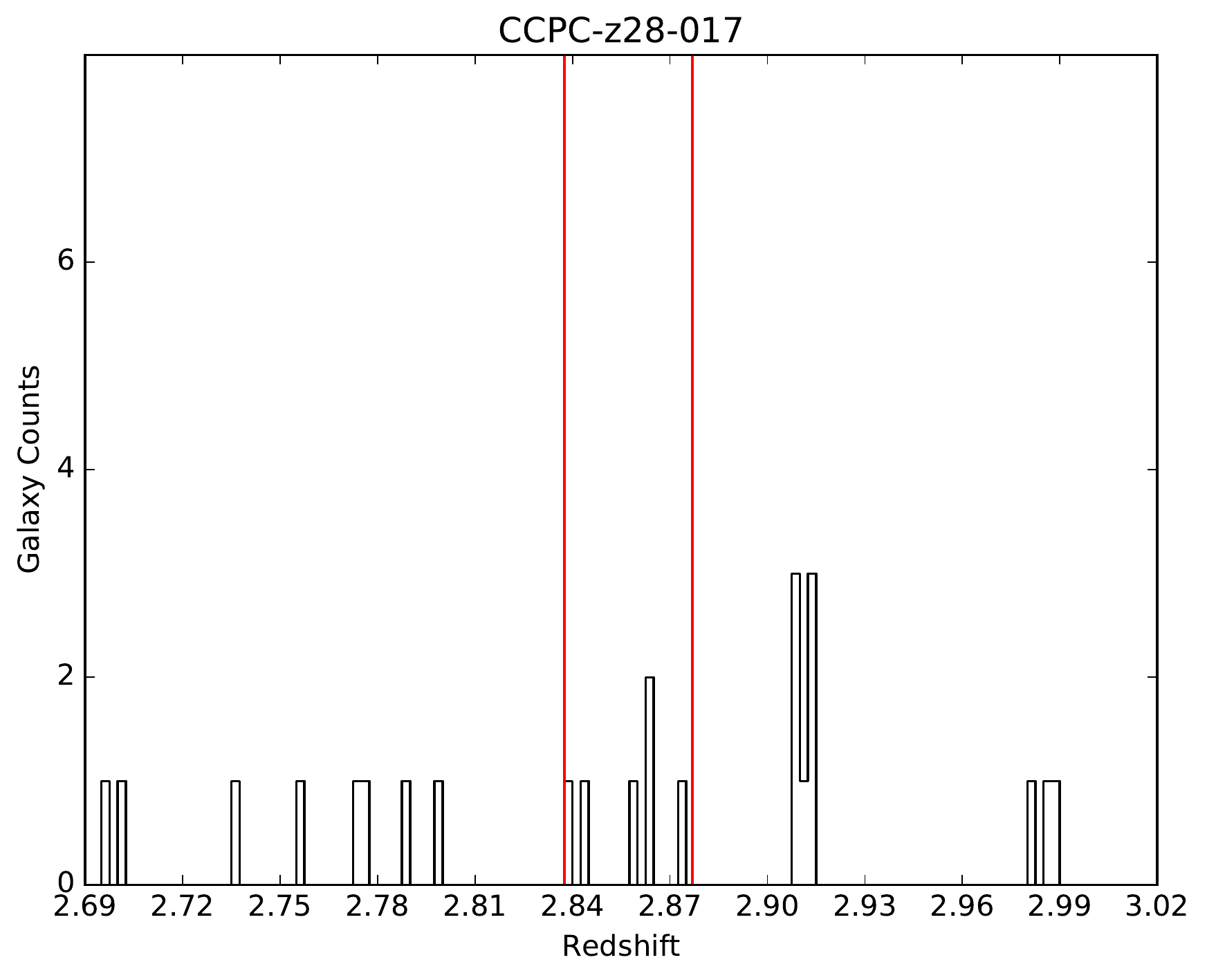}
\label{fig:CCPC-z28-017}
\end{subfigure}
\hfill
\end{figure*}
\clearpage 

\begin{figure*}
\centering
\begin{subfigure}
\centering
\includegraphics[height=7.5cm,width=7.5cm]{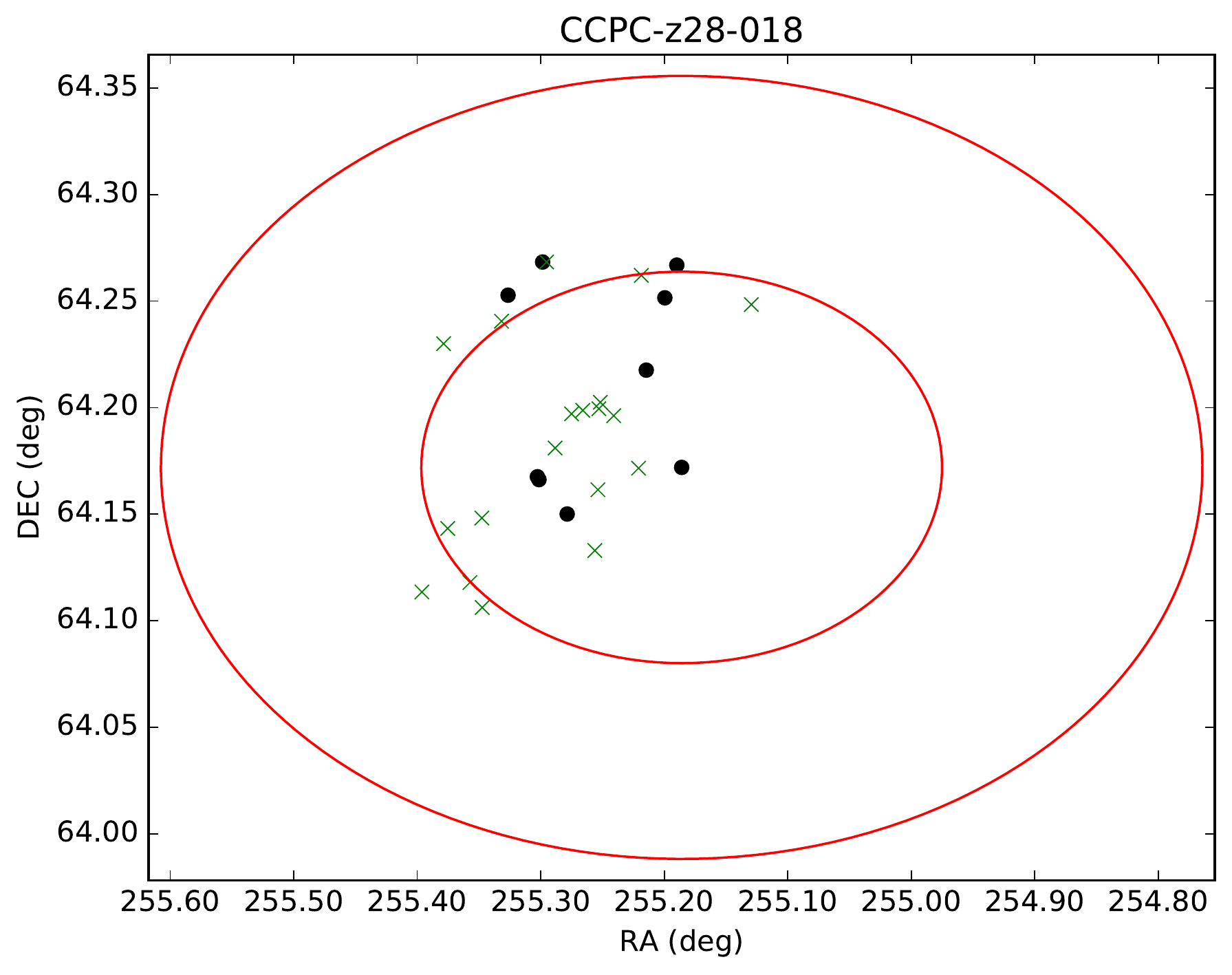}
\label{fig:CCPC-z28-018_sky}
\end{subfigure}
\hfill
\begin{subfigure}
\centering
\includegraphics[scale=0.52]{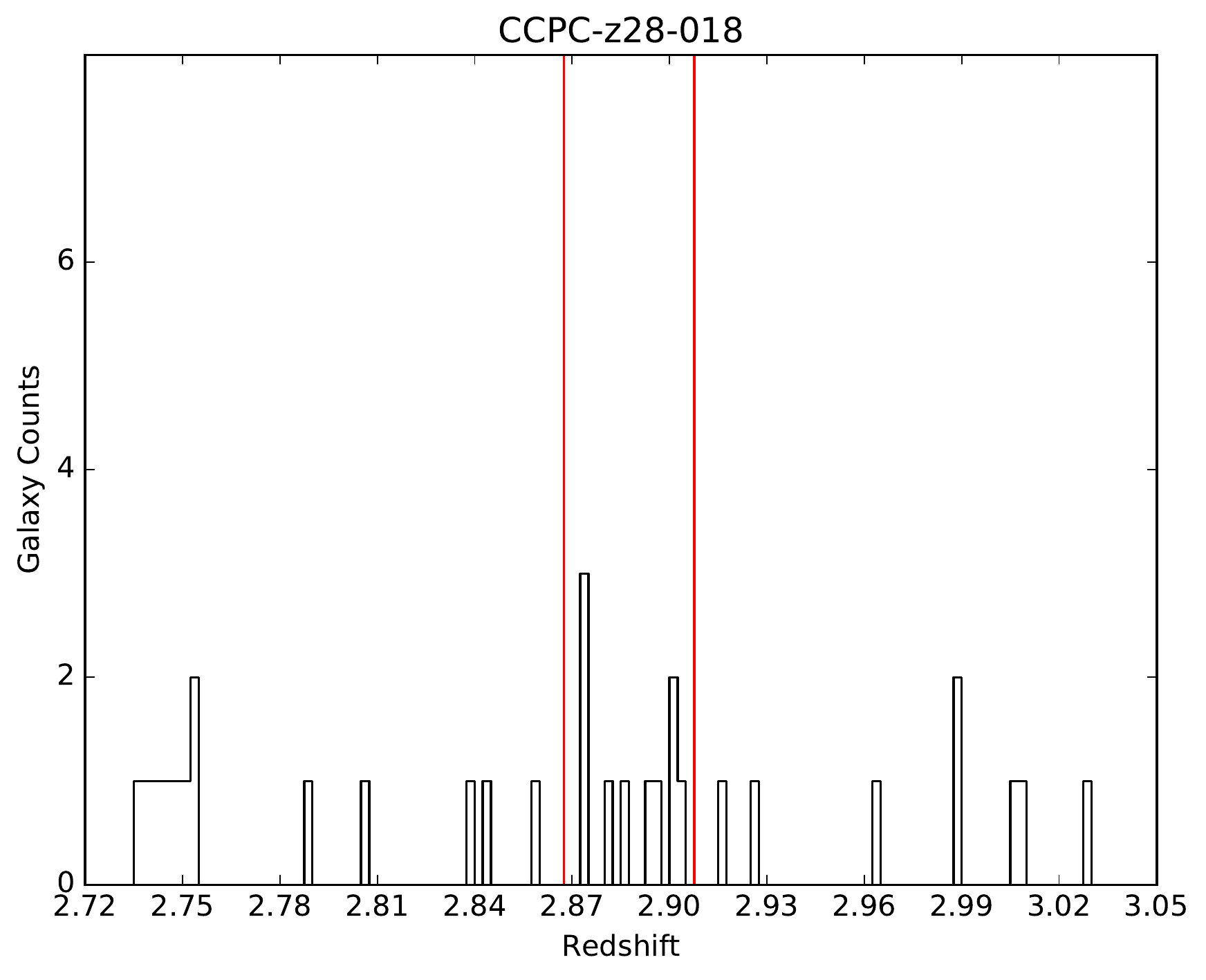}
\label{fig:CCPC-z28-018}
\end{subfigure}
\hfill
\end{figure*}

\begin{figure*}
\centering
\begin{subfigure}
\centering
\includegraphics[height=7.5cm,width=7.5cm]{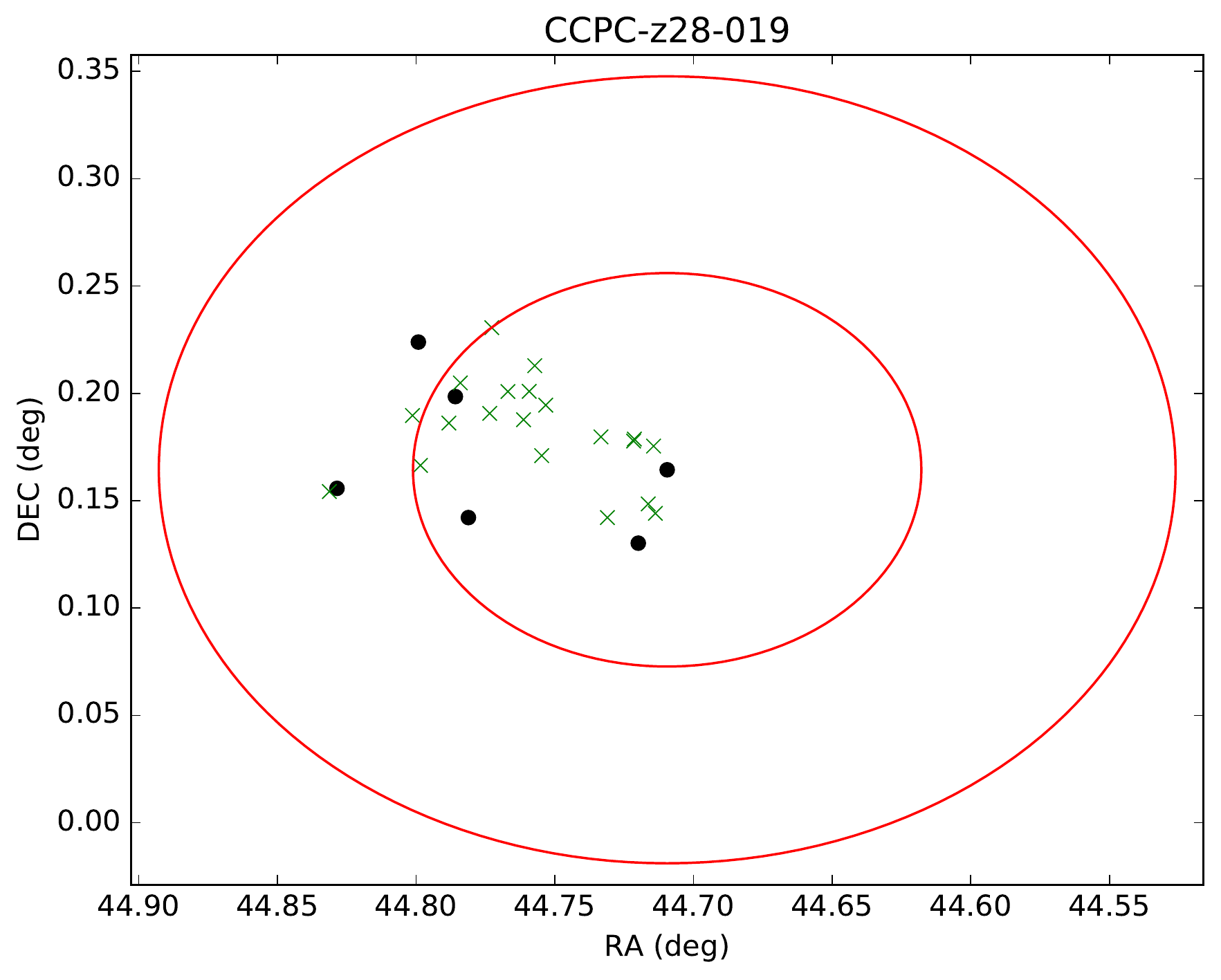}
\label{fig:CCPC-z28-019_sky}
\end{subfigure}
\hfill
\begin{subfigure}
\centering
\includegraphics[scale=0.52]{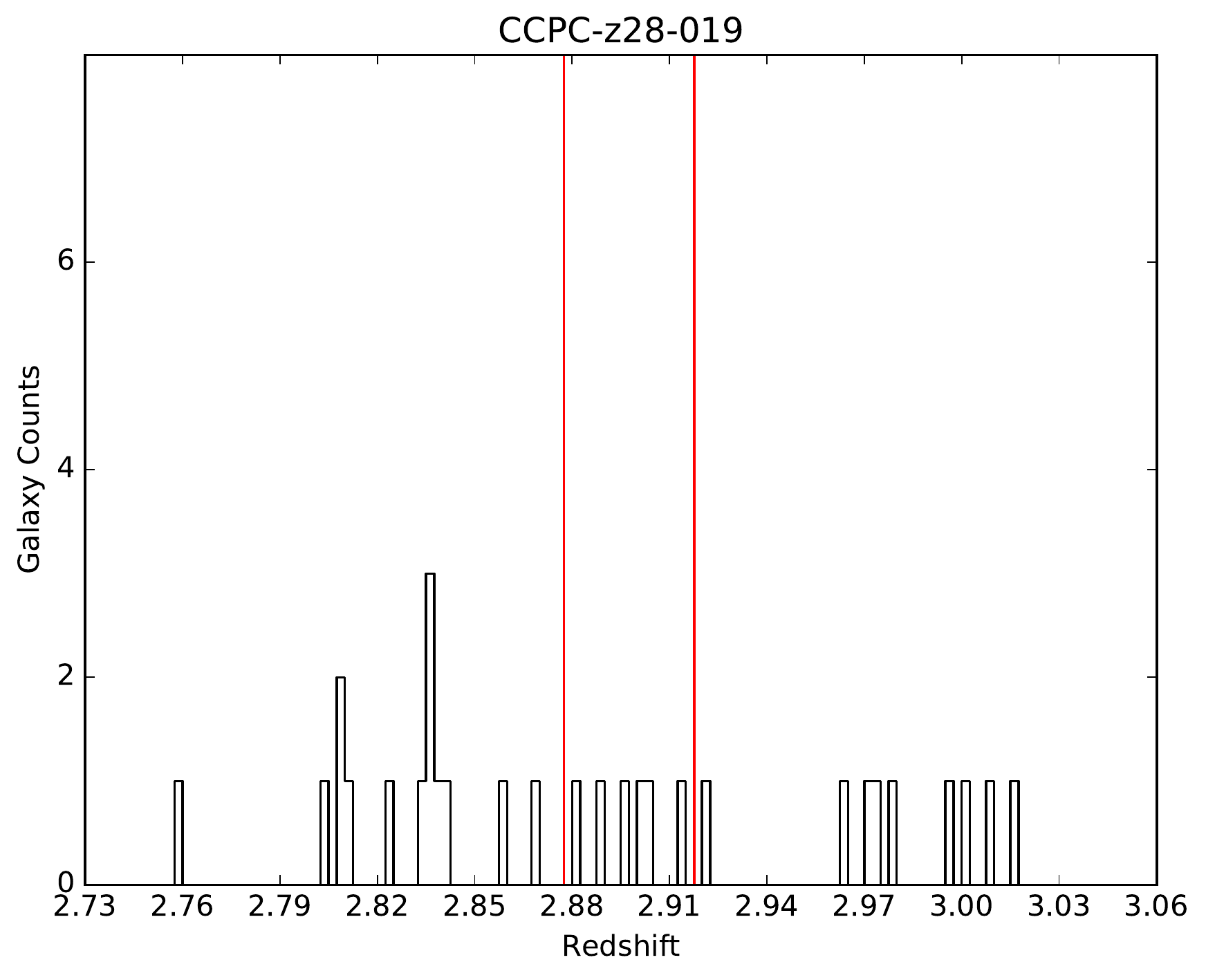}
\label{fig:CCPC-z28-019}
\end{subfigure}
\hfill
\end{figure*}

\begin{figure*}
\centering
\begin{subfigure}
\centering
\includegraphics[height=7.5cm,width=7.5cm]{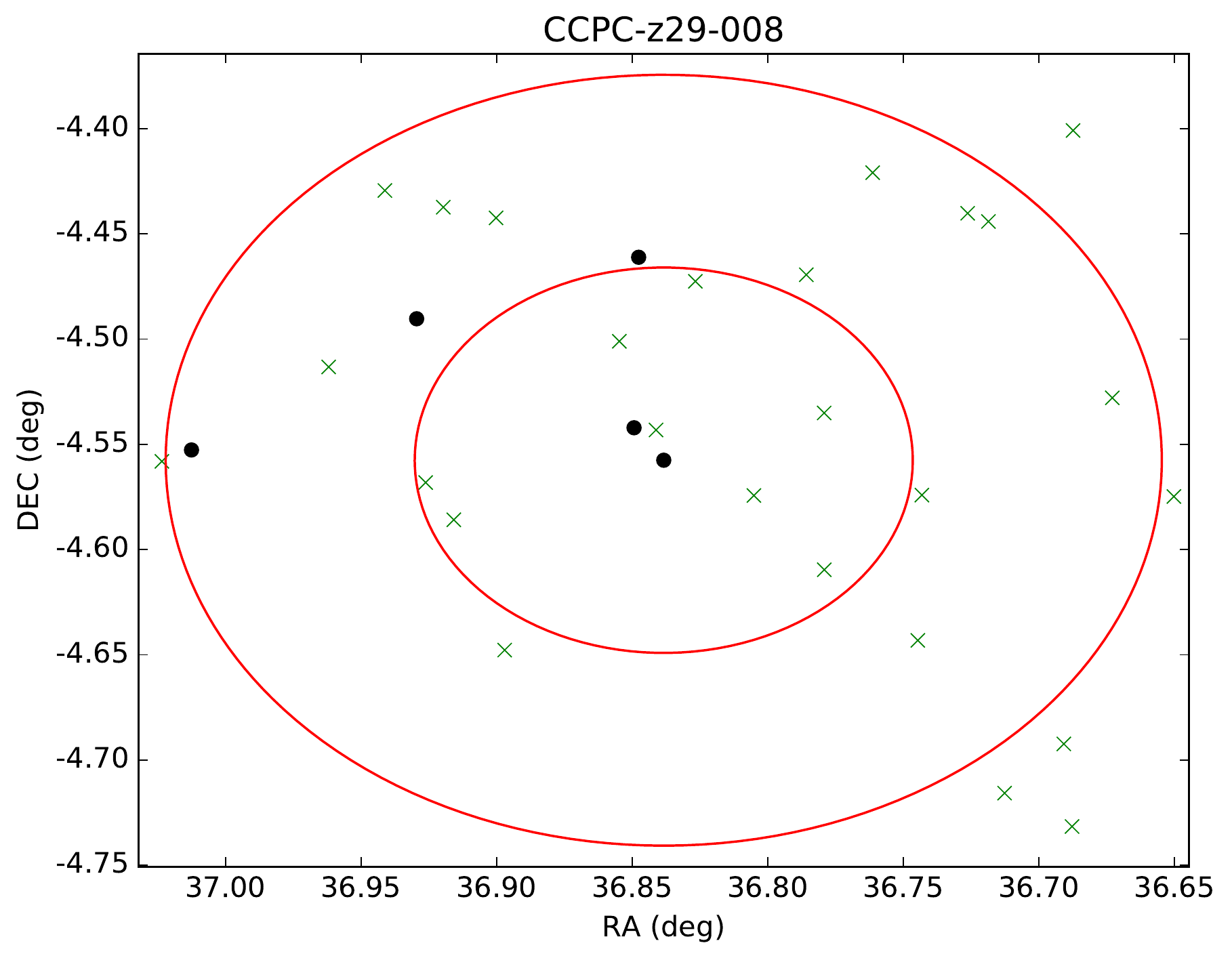}
\label{fig:CCPC-z29-008_sky}
\end{subfigure}
\hfill
\begin{subfigure}
\centering
\includegraphics[scale=0.52]{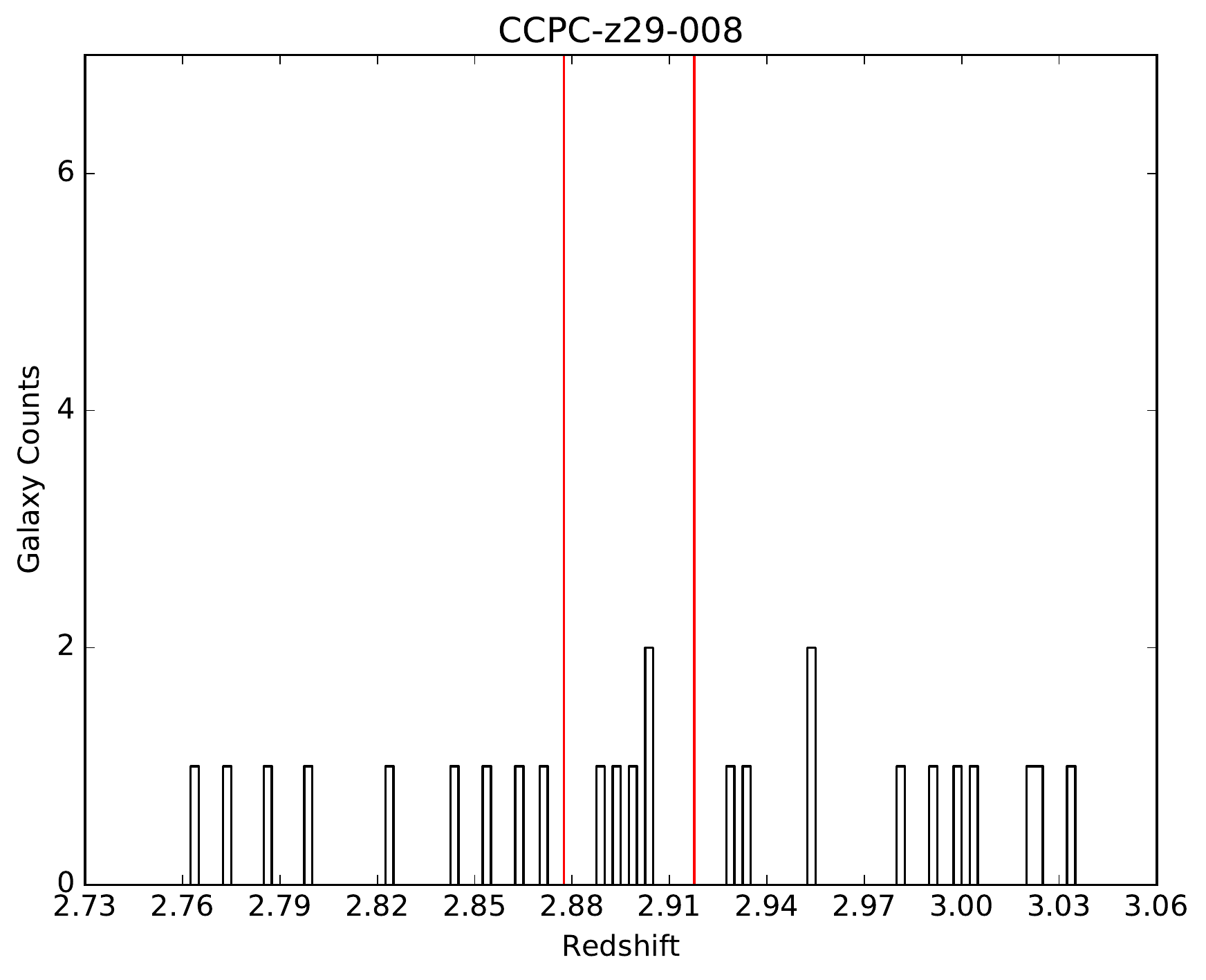}
\label{fig:CCPC-z29-008}
\end{subfigure}
\hfill
\end{figure*}
\clearpage 

\begin{figure*}
\centering
\begin{subfigure}
\centering
\includegraphics[height=7.5cm,width=7.5cm]{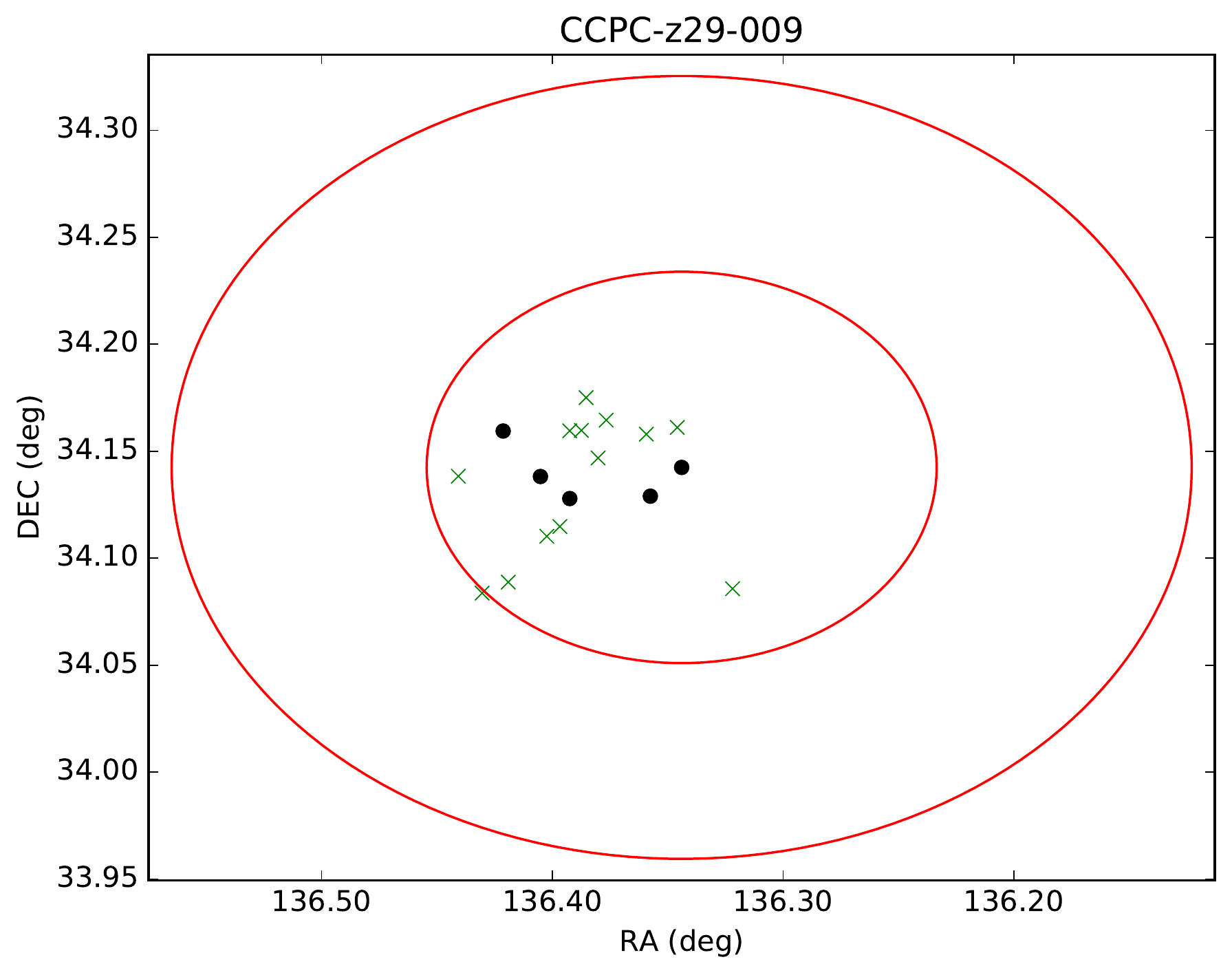}
\label{fig:CCPC-z29-009_sky}
\end{subfigure}
\hfill
\begin{subfigure}
\centering
\includegraphics[scale=0.52]{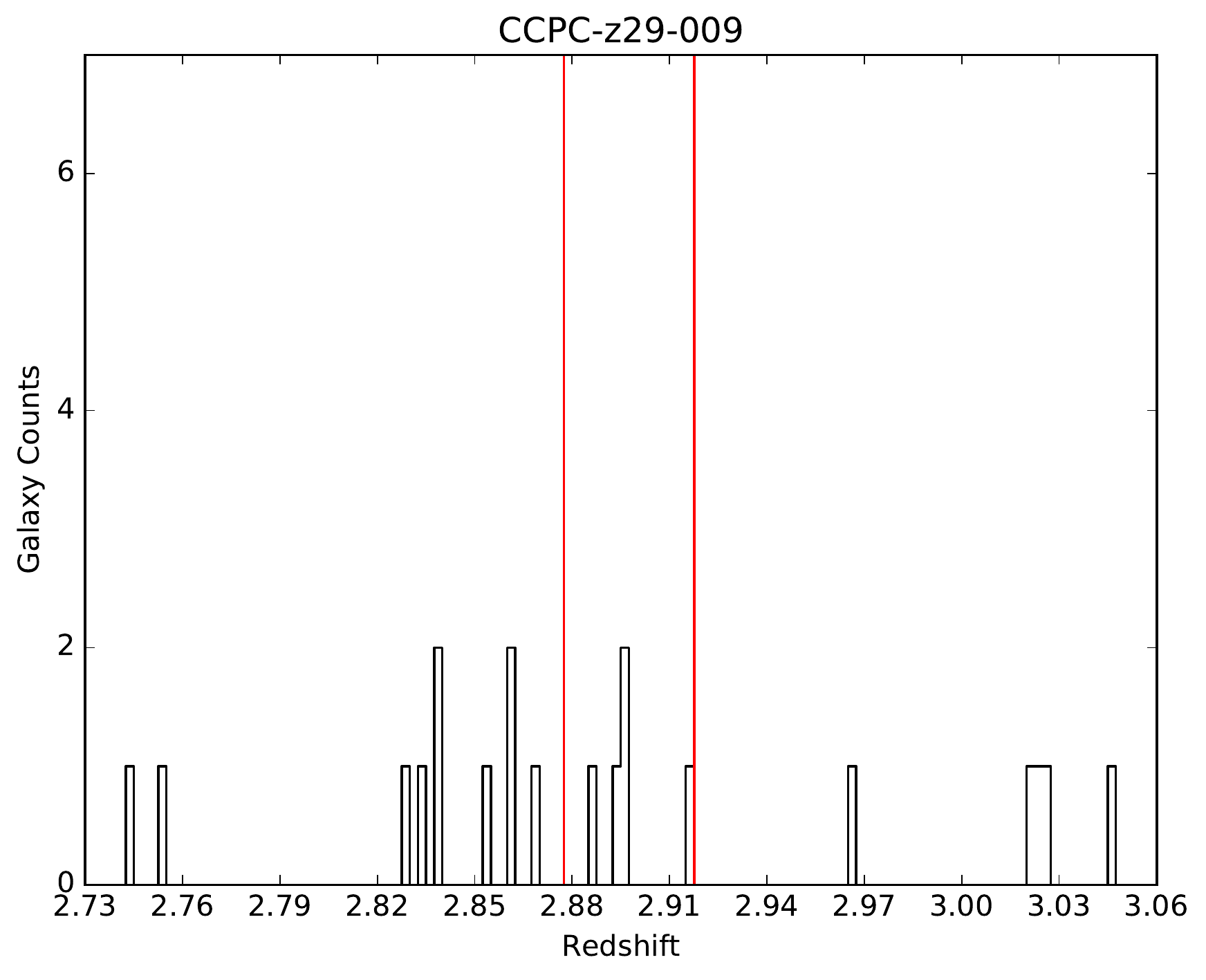}
\label{fig:CCPC-z29-009}
\end{subfigure}
\hfill
\end{figure*}

\begin{figure*}
\centering
\begin{subfigure}
\centering
\includegraphics[height=7.5cm,width=7.5cm]{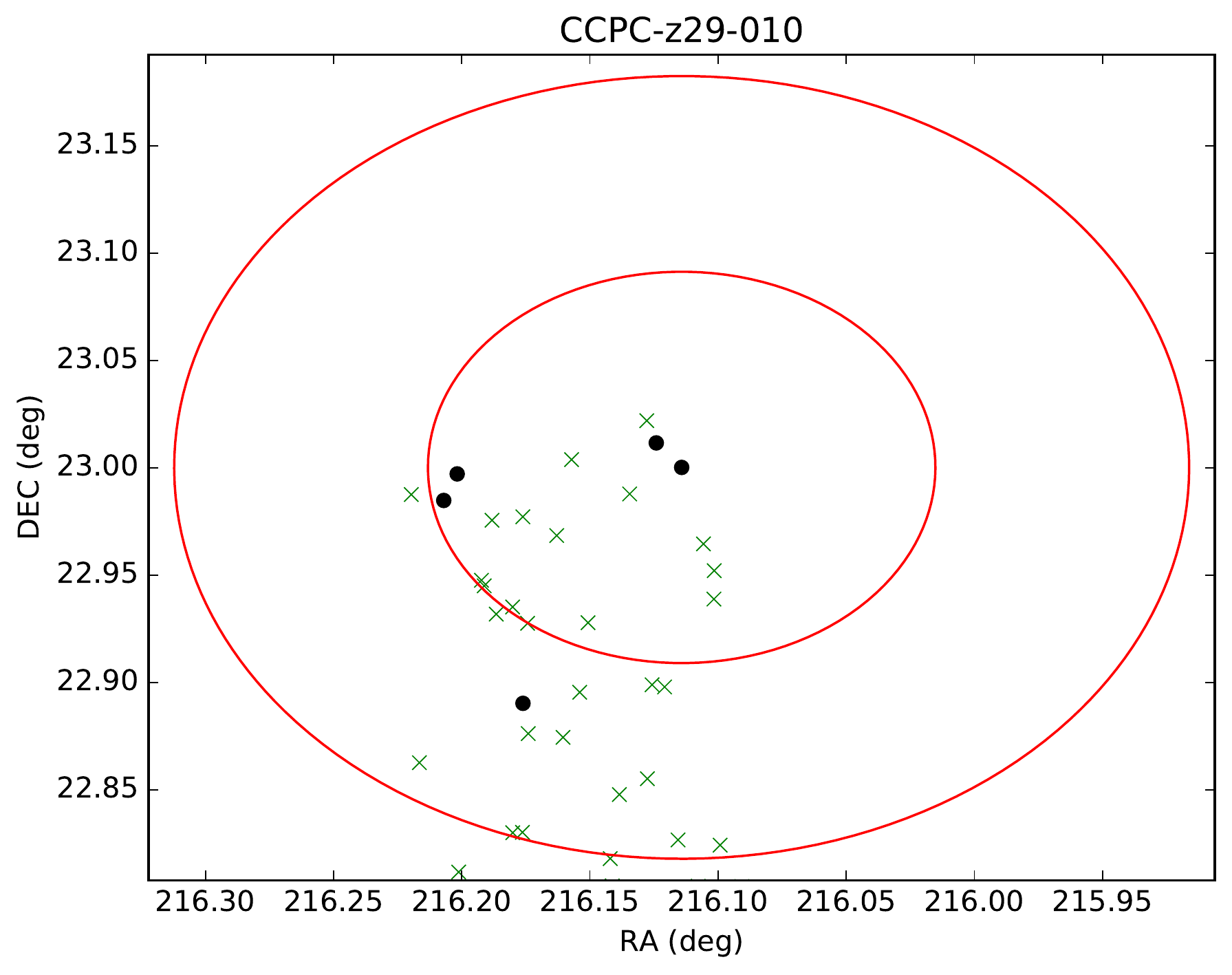}
\label{fig:CCPC-z29-010_sky}
\end{subfigure}
\hfill
\begin{subfigure}
\centering
\includegraphics[scale=0.52]{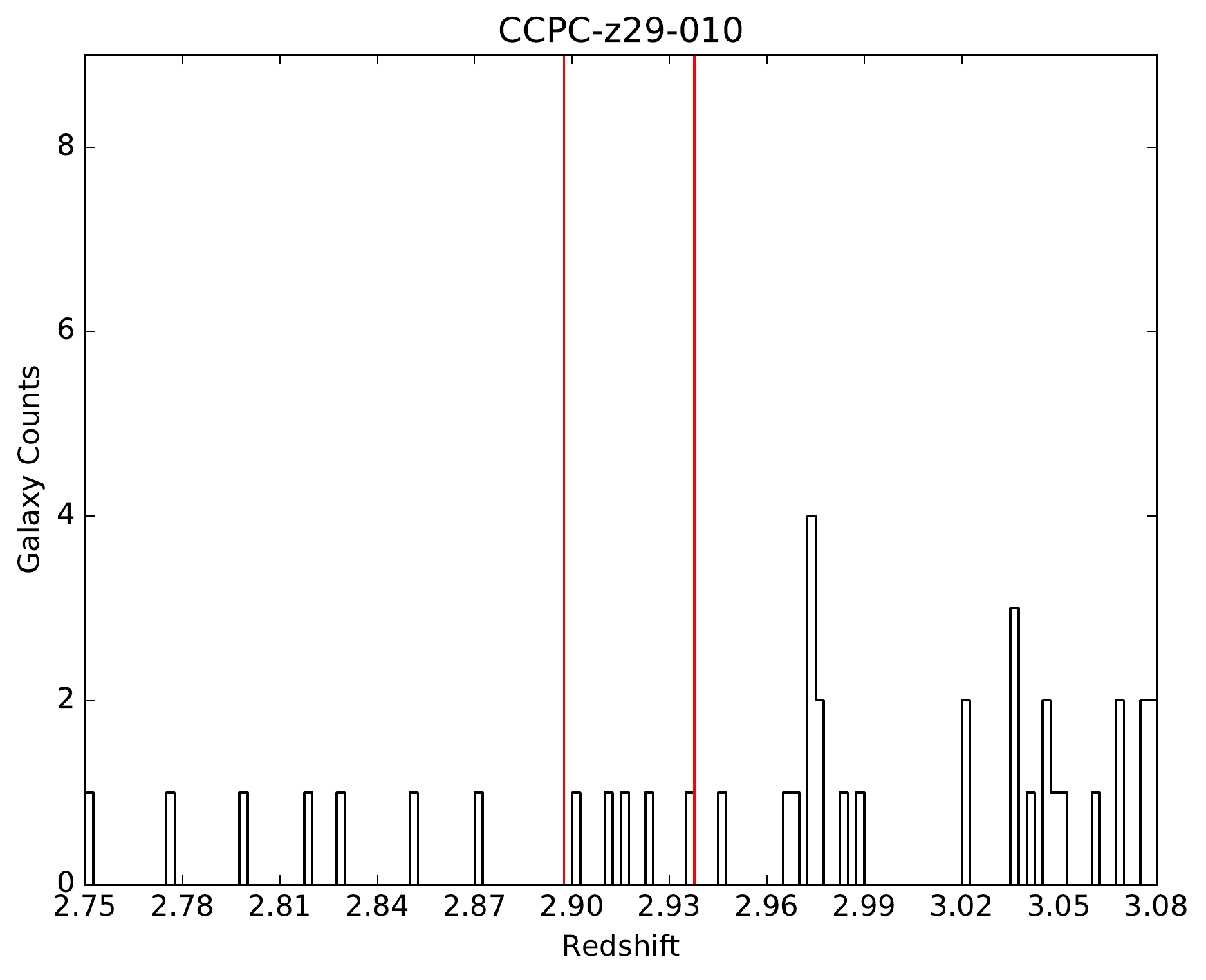}
\label{fig:CCPC-z29-010}
\end{subfigure}
\hfill
\end{figure*}

\begin{figure*}
\centering
\begin{subfigure}
\centering
\includegraphics[height=7.5cm,width=7.5cm]{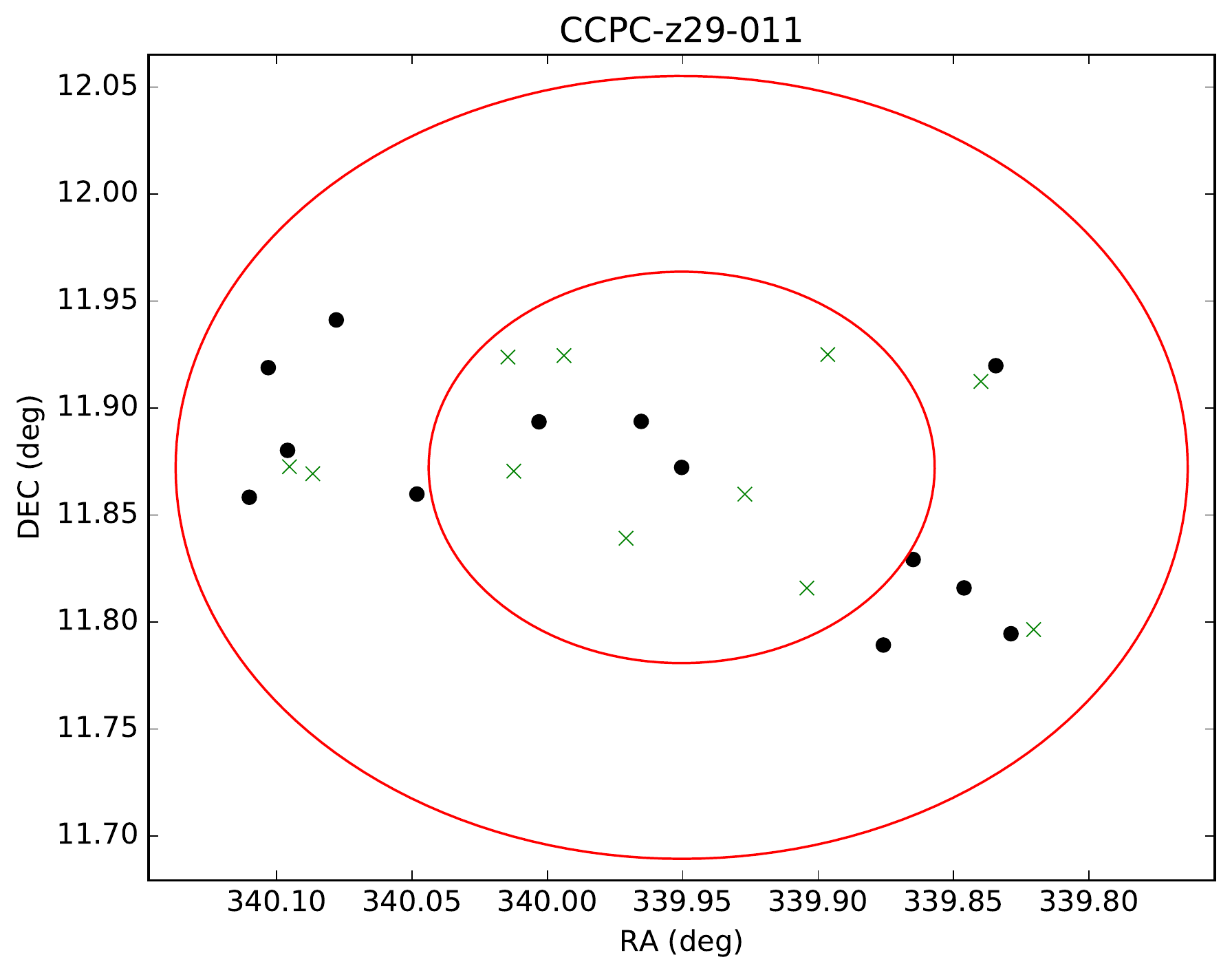}
\label{fig:CCPC-z29-011_sky}
\end{subfigure}
\hfill
\begin{subfigure}
\centering
\includegraphics[scale=0.52]{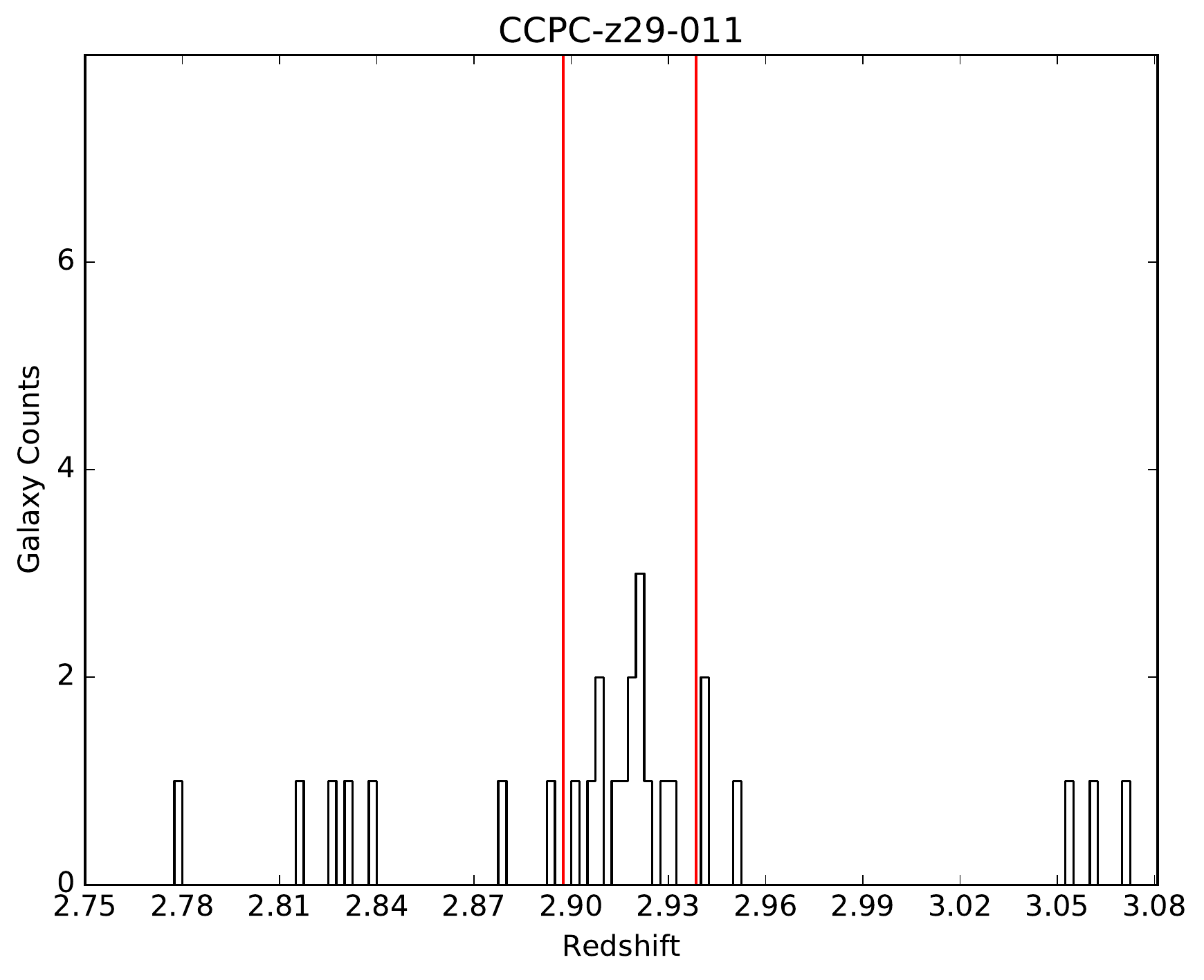}
\label{fig:CCPC-z29-011}
\end{subfigure}
\hfill
\end{figure*}
\clearpage 

\begin{figure*}
\centering
\begin{subfigure}
\centering
\includegraphics[height=7.5cm,width=7.5cm]{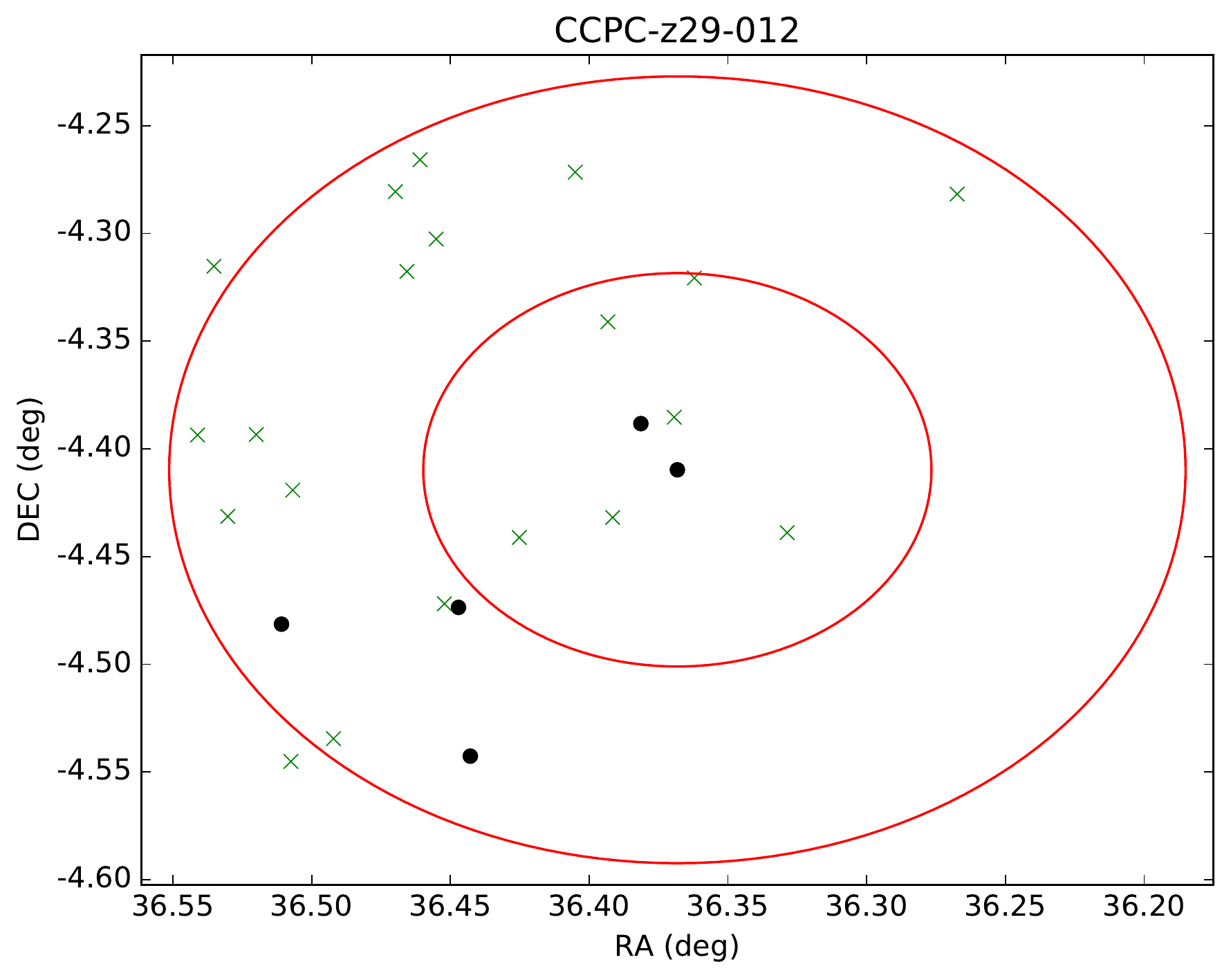}
\label{fig:CCPC-z29-012_sky}
\end{subfigure}
\hfill
\begin{subfigure}
\centering
\includegraphics[scale=0.52]{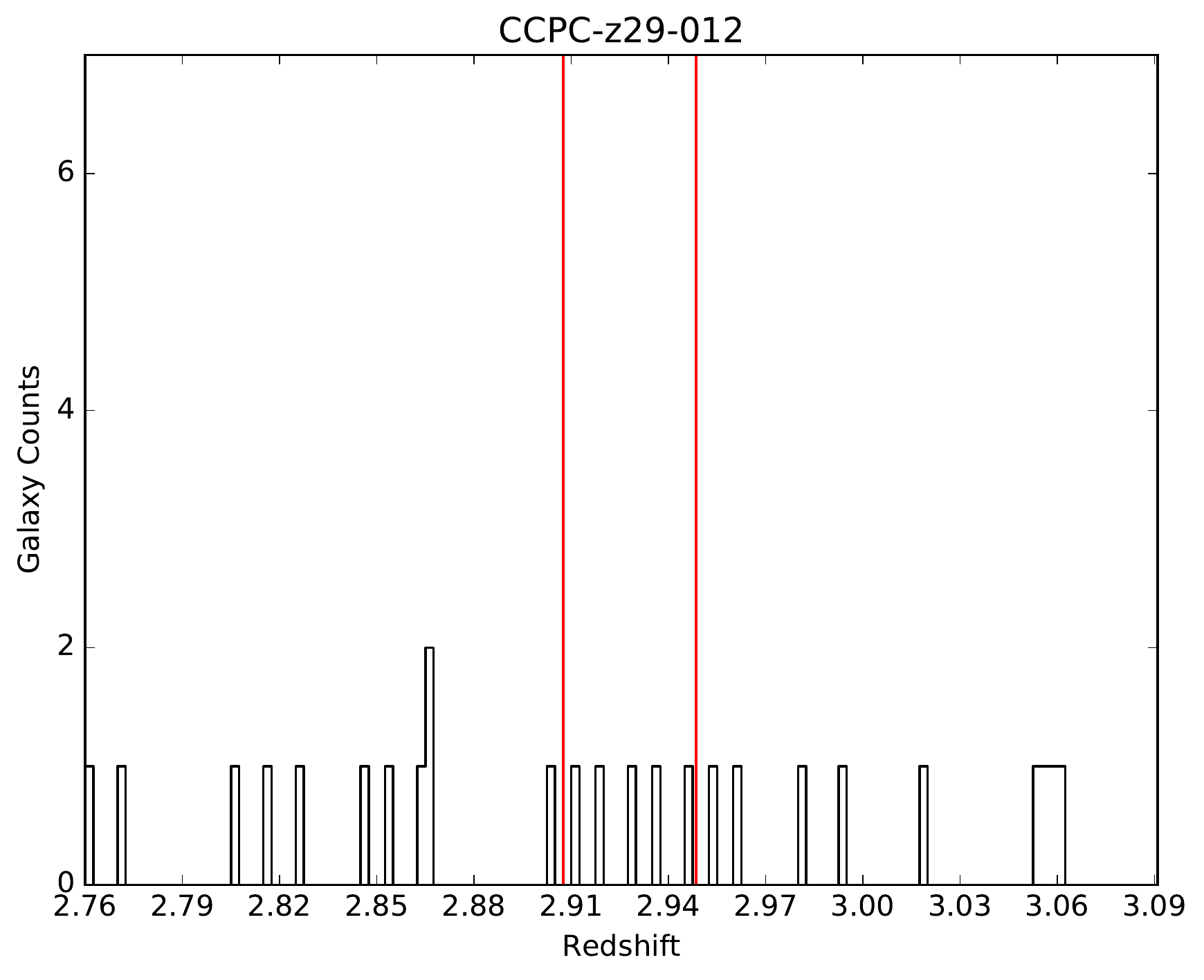}
\label{fig:CCPC-z29-012}
\end{subfigure}
\hfill
\end{figure*}

\begin{figure*}
\centering
\begin{subfigure}
\centering
\includegraphics[height=7.5cm,width=7.5cm]{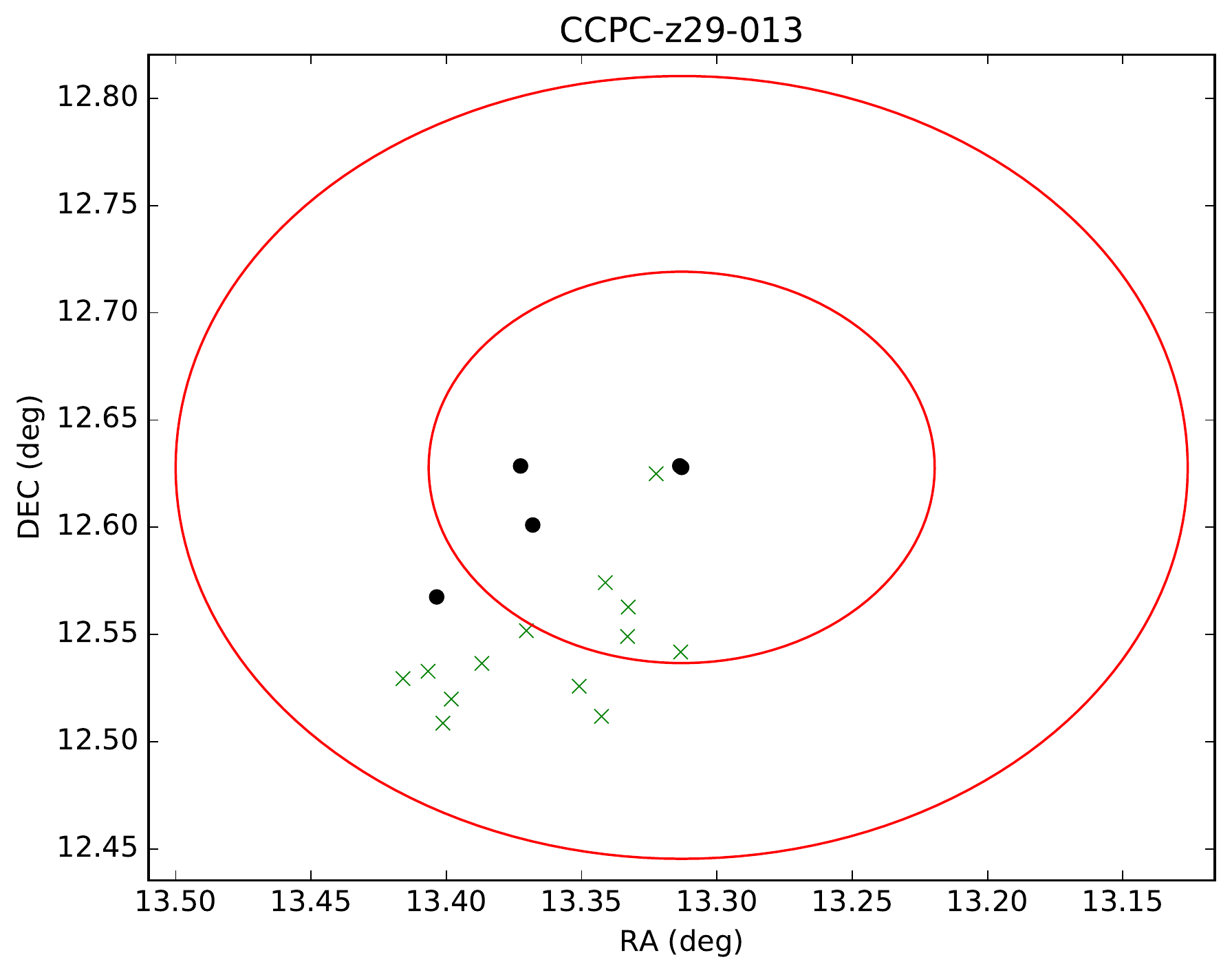}
\label{fig:CCPC-z29-013_sky}
\end{subfigure}
\hfill
\begin{subfigure}
\centering
\includegraphics[scale=0.52]{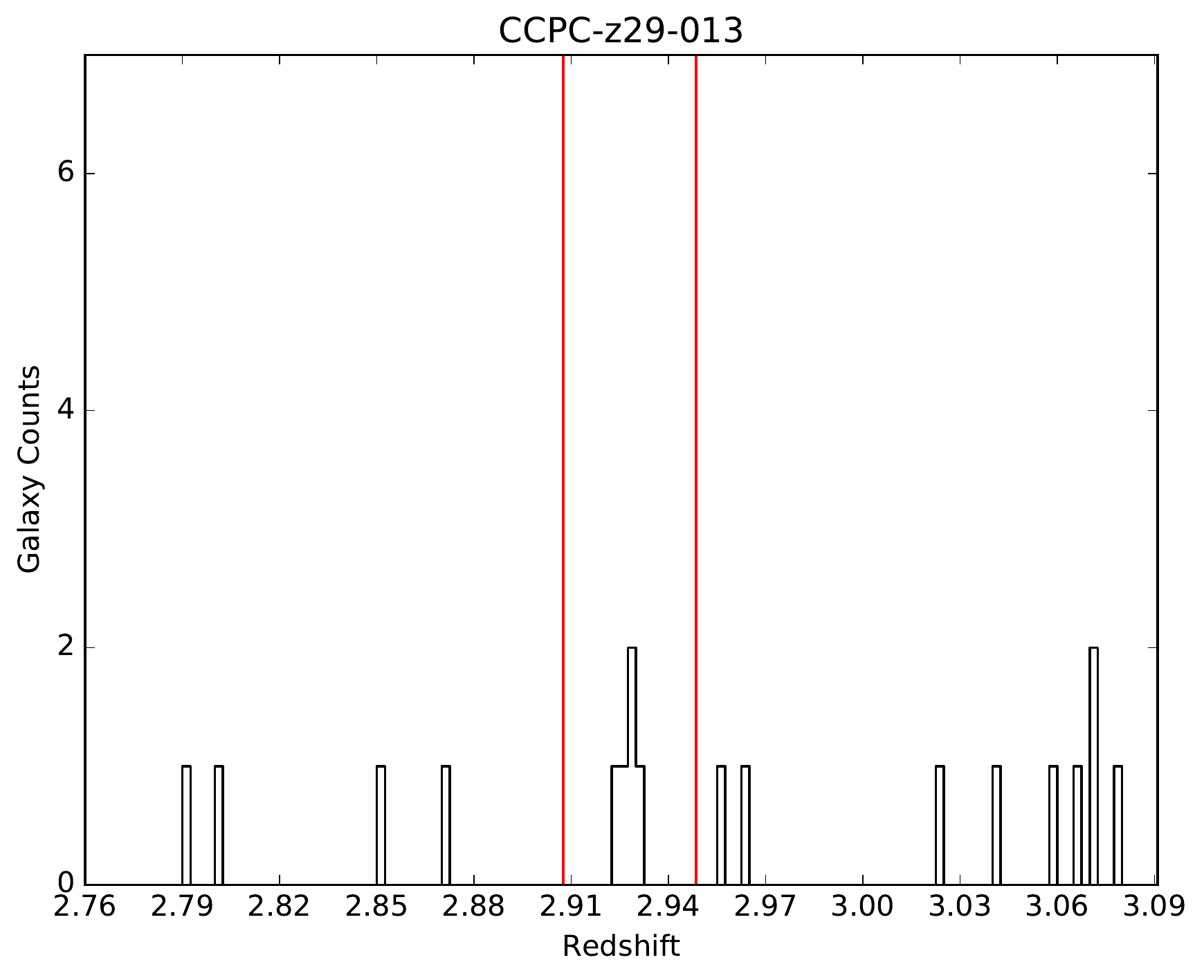}
\label{fig:CCPC-z29-013}
\end{subfigure}
\hfill
\end{figure*}

\begin{figure*}
\centering
\begin{subfigure}
\centering
\includegraphics[height=7.5cm,width=7.5cm]{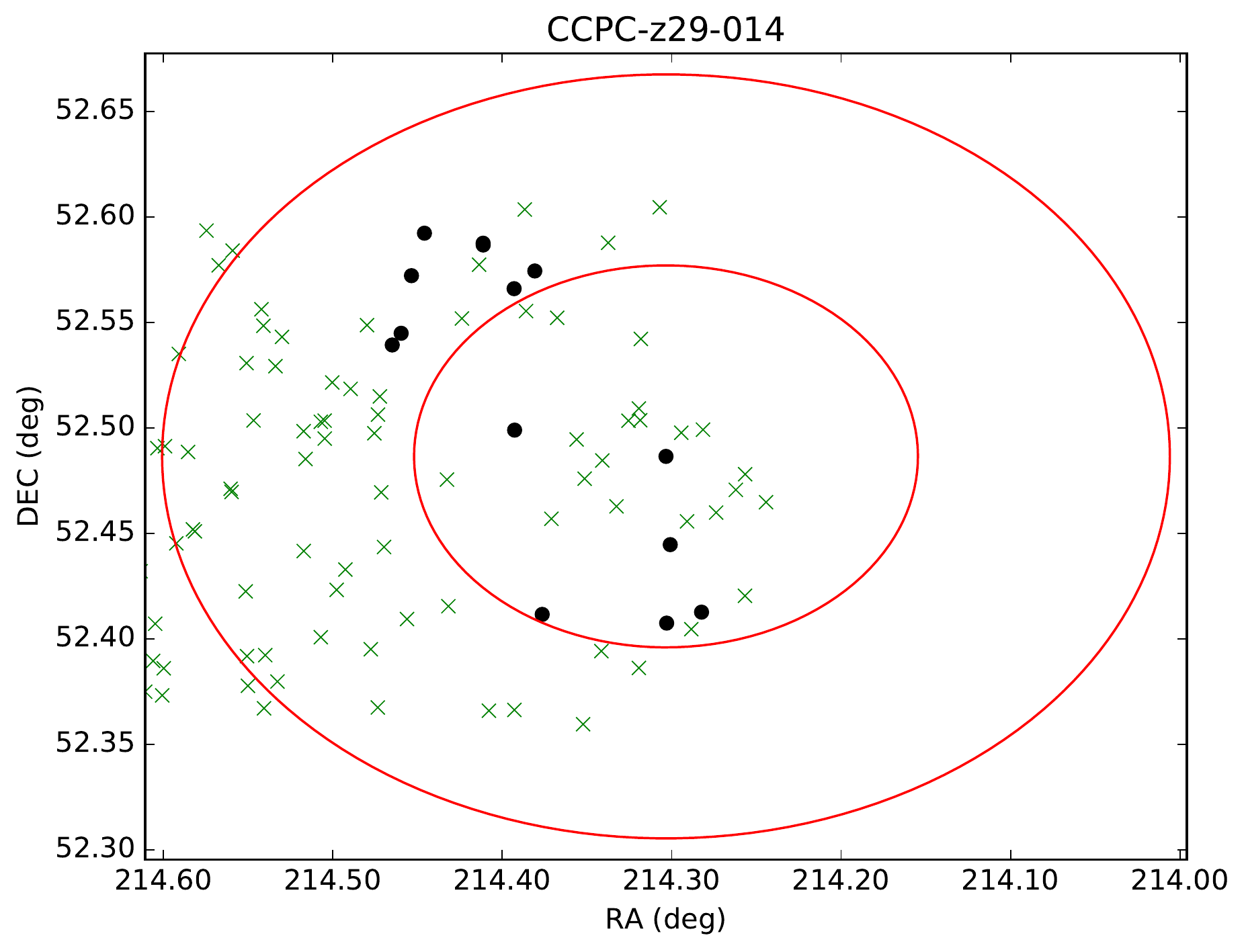}
\label{fig:CCPC-z29-014_sky}
\end{subfigure}
\hfill
\begin{subfigure}
\centering
\includegraphics[scale=0.52]{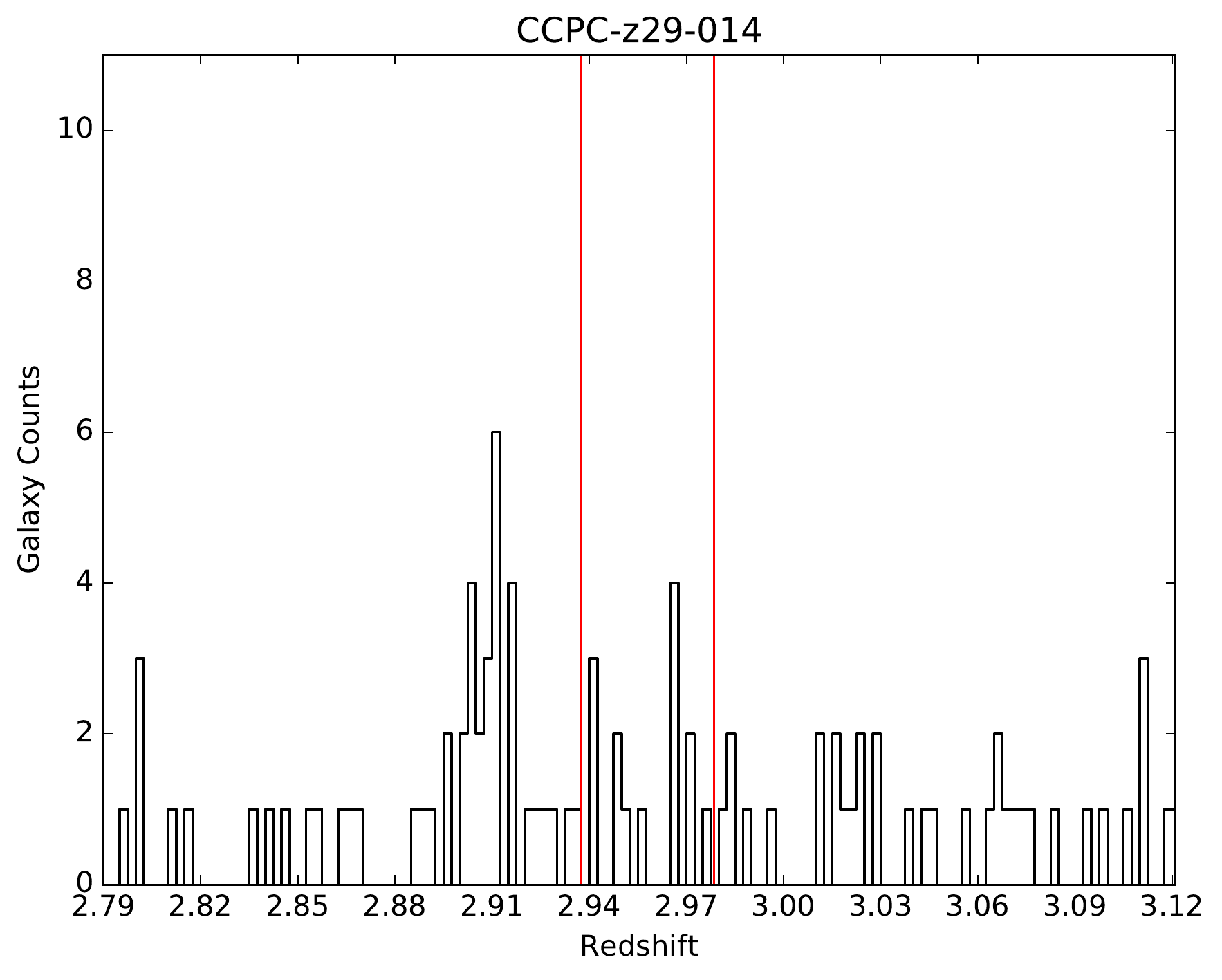}
\label{fig:CCPC-z29-014}
\end{subfigure}
\hfill
\end{figure*}
\clearpage 

\begin{figure*}
\centering
\begin{subfigure}
\centering
\includegraphics[height=7.5cm,width=7.5cm]{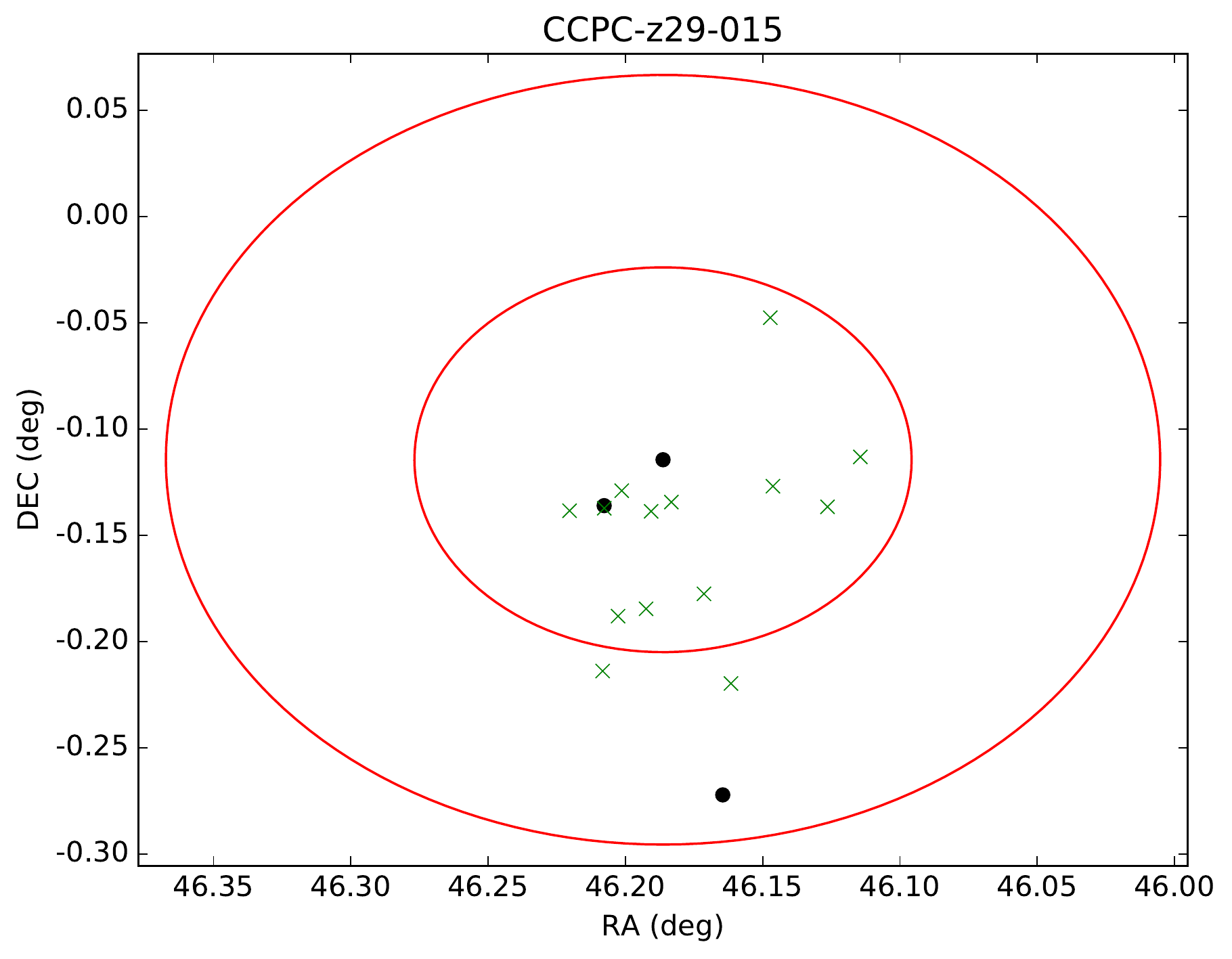}
\label{fig:CCPC-z29-015_sky}
\end{subfigure}
\hfill
\begin{subfigure}
\centering
\includegraphics[scale=0.52]{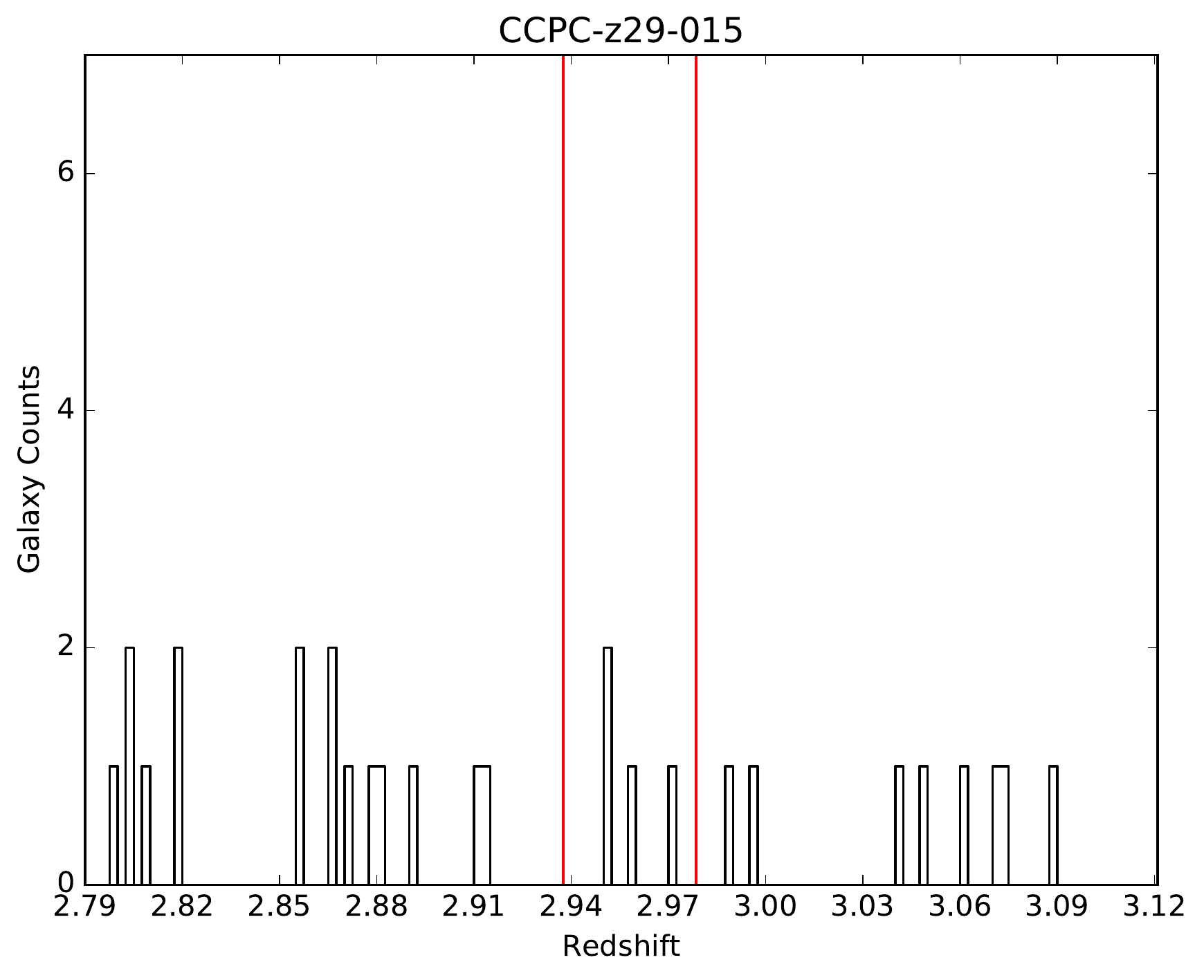}
\label{fig:CCPC-z29-015}
\end{subfigure}
\hfill
\end{figure*}

\begin{figure*}
\centering
\begin{subfigure}
\centering
\includegraphics[height=7.5cm,width=7.5cm]{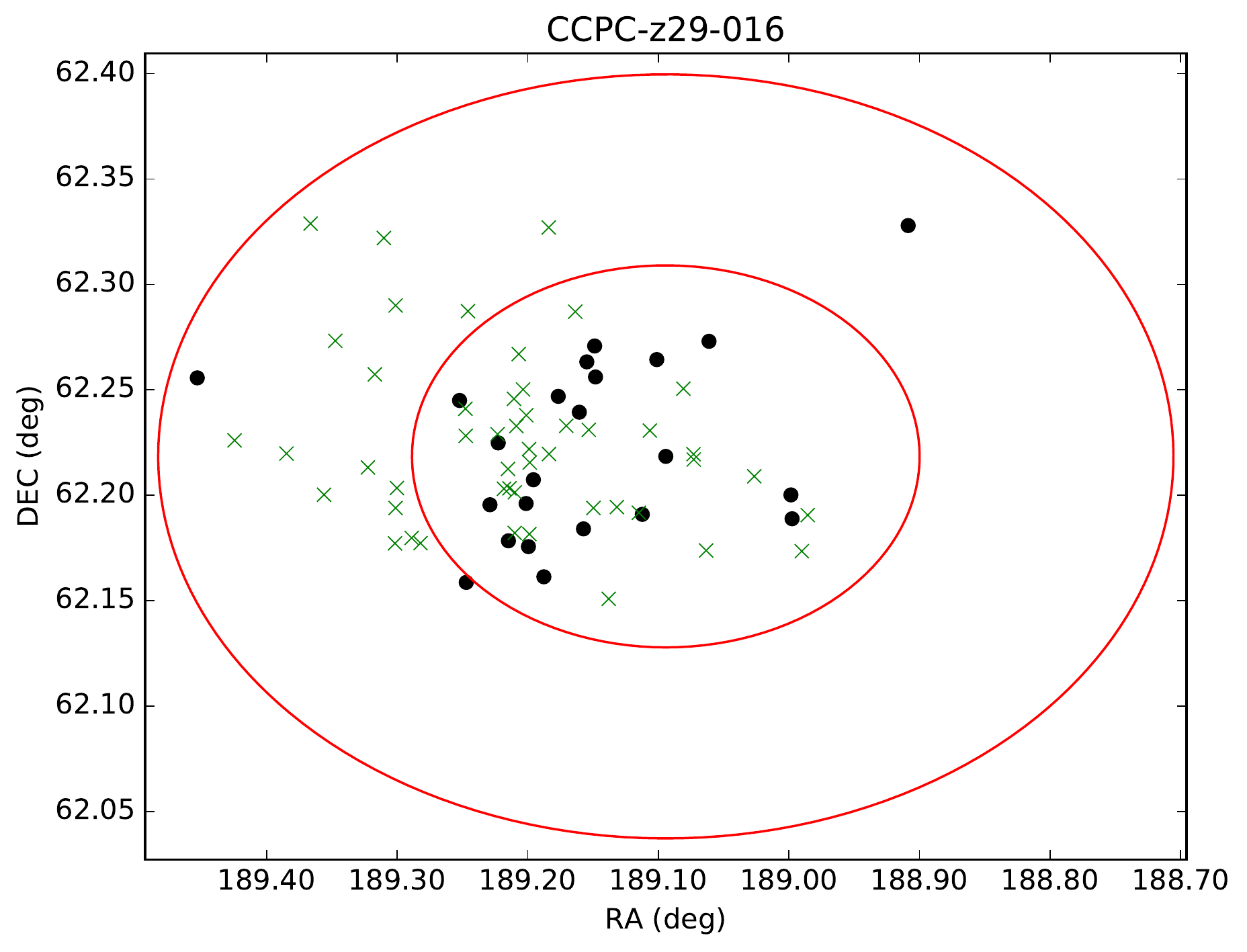}
\label{fig:CCPC-z29-016_sky}
\end{subfigure}
\hfill
\begin{subfigure}
\centering
\includegraphics[scale=0.52]{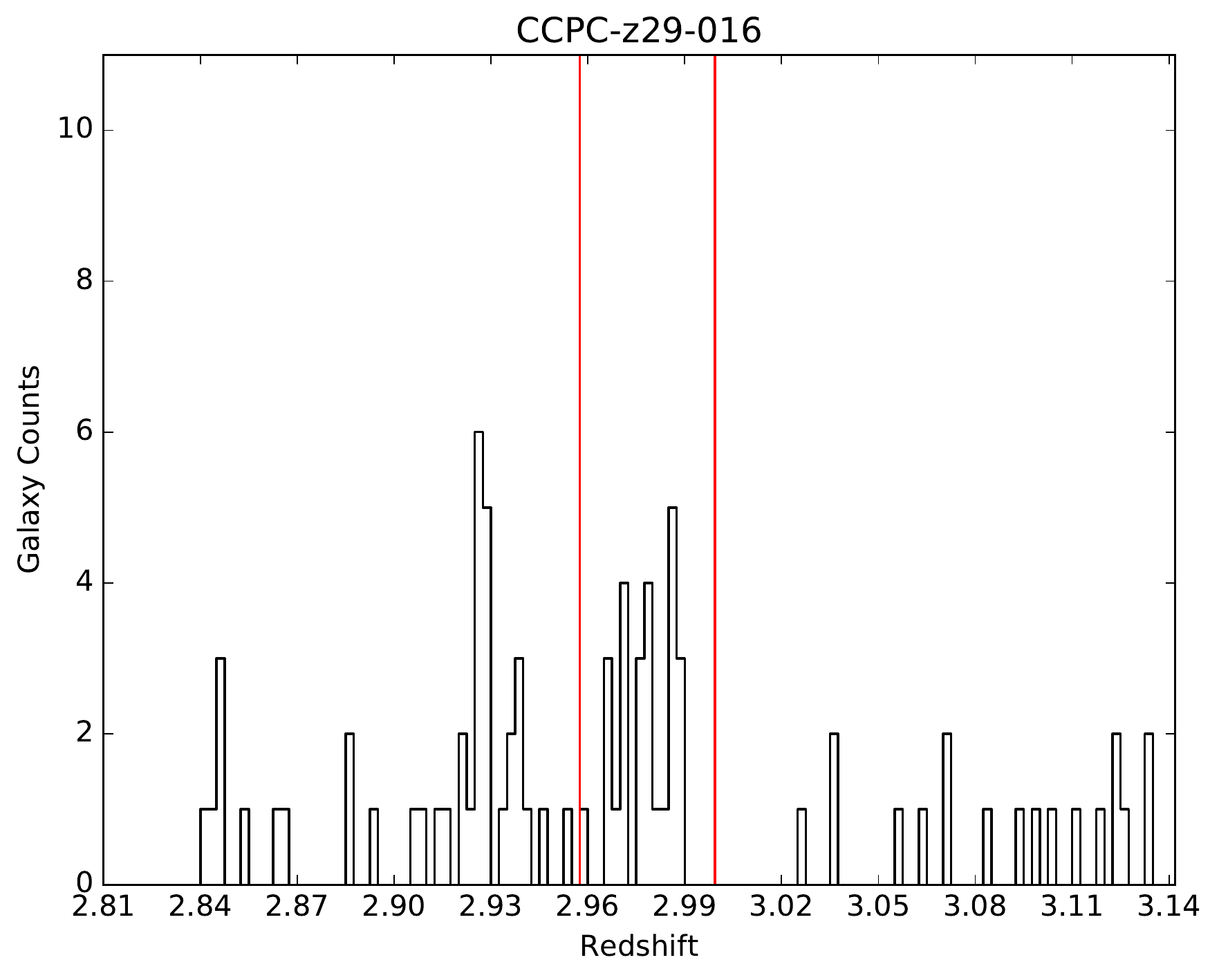}
\label{fig:CCPC-z29-016}
\end{subfigure}
\hfill
\end{figure*}

\begin{figure*}
\centering
\begin{subfigure}
\centering
\includegraphics[height=7.5cm,width=7.5cm]{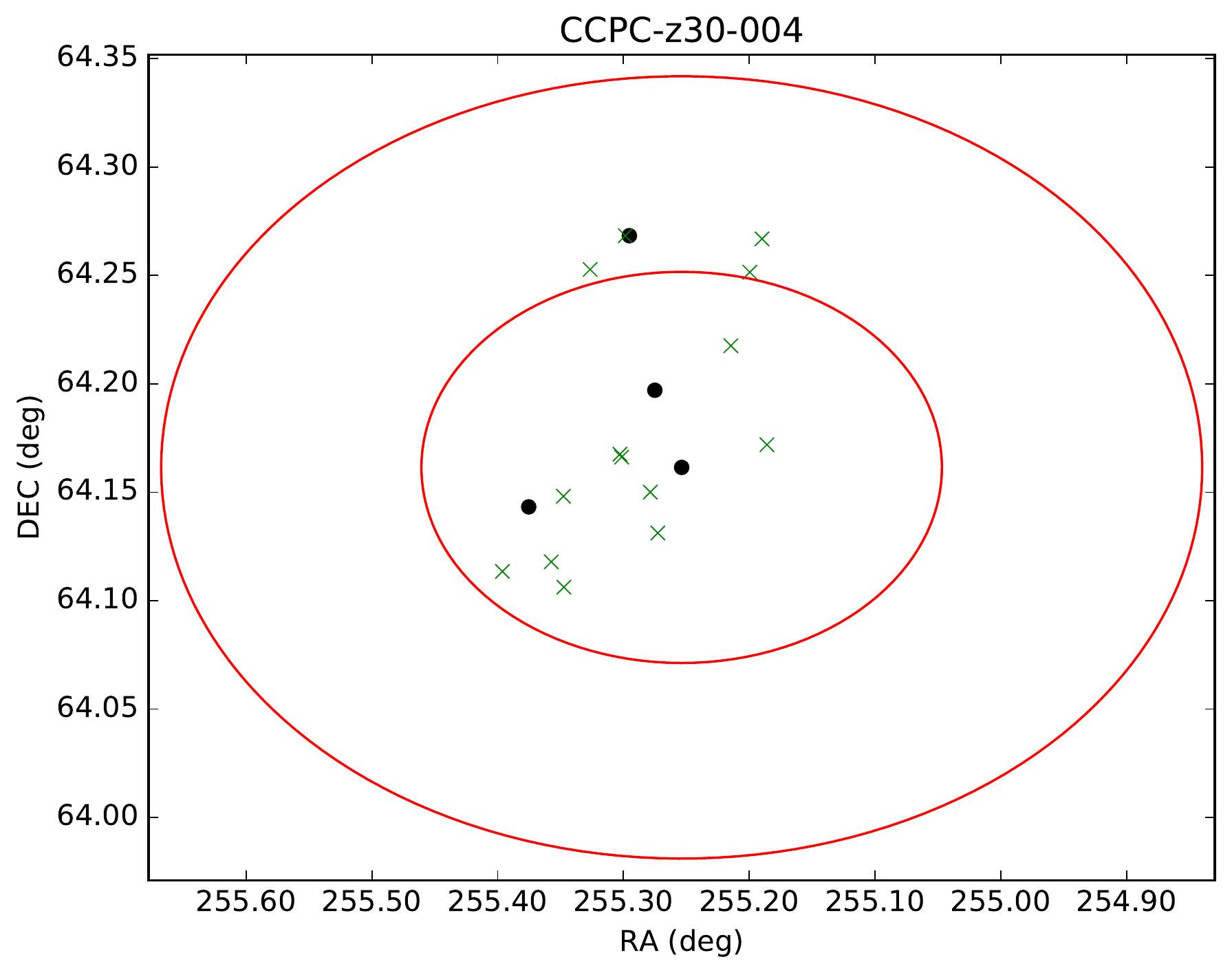}
\label{fig:CCPC-z30-004_sky}
\end{subfigure}
\hfill
\begin{subfigure}
\centering
\includegraphics[scale=0.52]{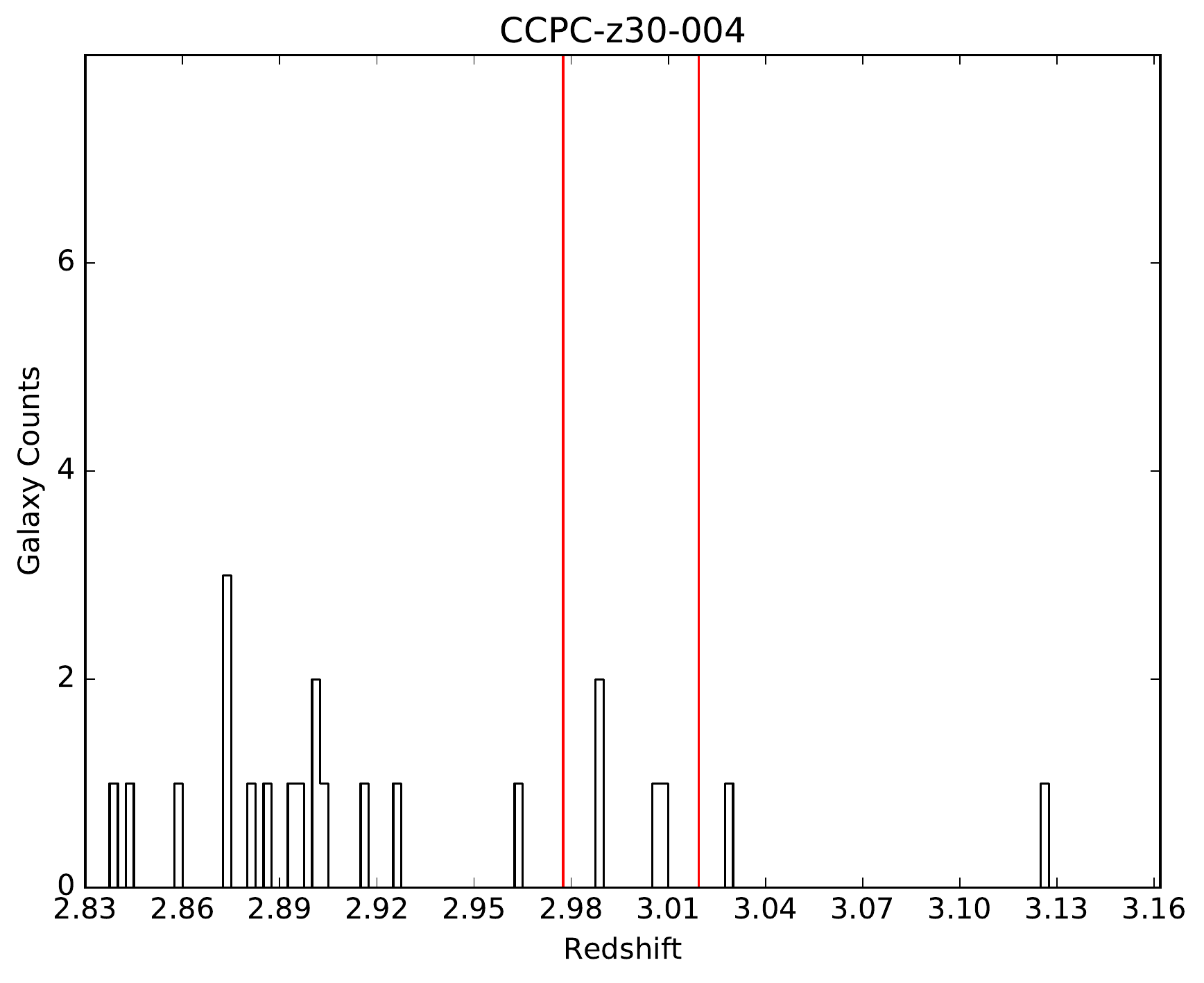}
\label{fig:CCPC-z30-004}
\end{subfigure}
\hfill
\end{figure*}
\clearpage 

\begin{figure*}
\centering
\begin{subfigure}
\centering
\includegraphics[height=7.5cm,width=7.5cm]{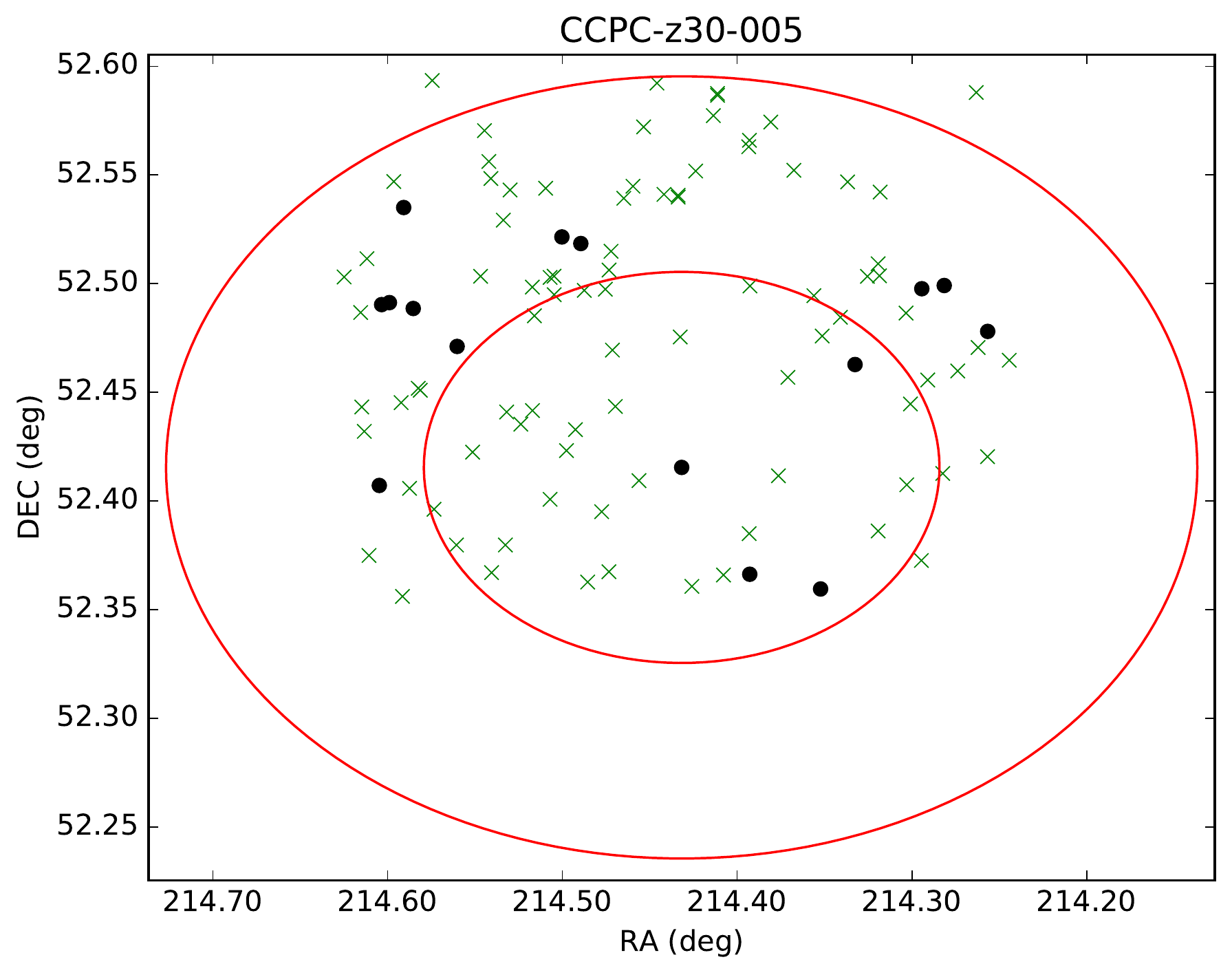}
\label{fig:CCPC-z30-005_sky}
\end{subfigure}
\hfill
\begin{subfigure}
\centering
\includegraphics[scale=0.52]{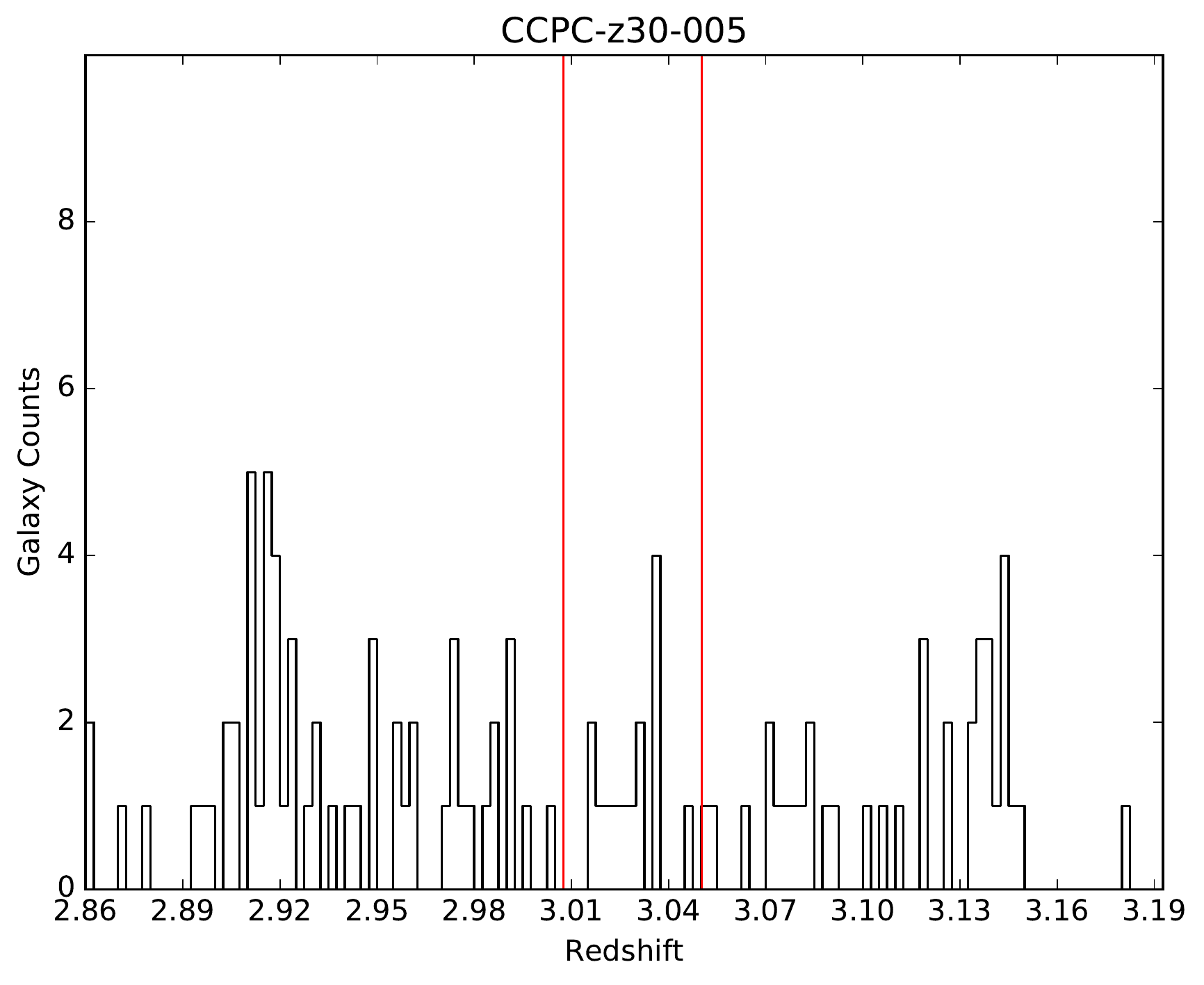}
\label{fig:CCPC-z30-005}
\end{subfigure}
\hfill
\end{figure*}

\begin{figure*}
\centering
\begin{subfigure}
\centering
\includegraphics[height=7.5cm,width=7.5cm]{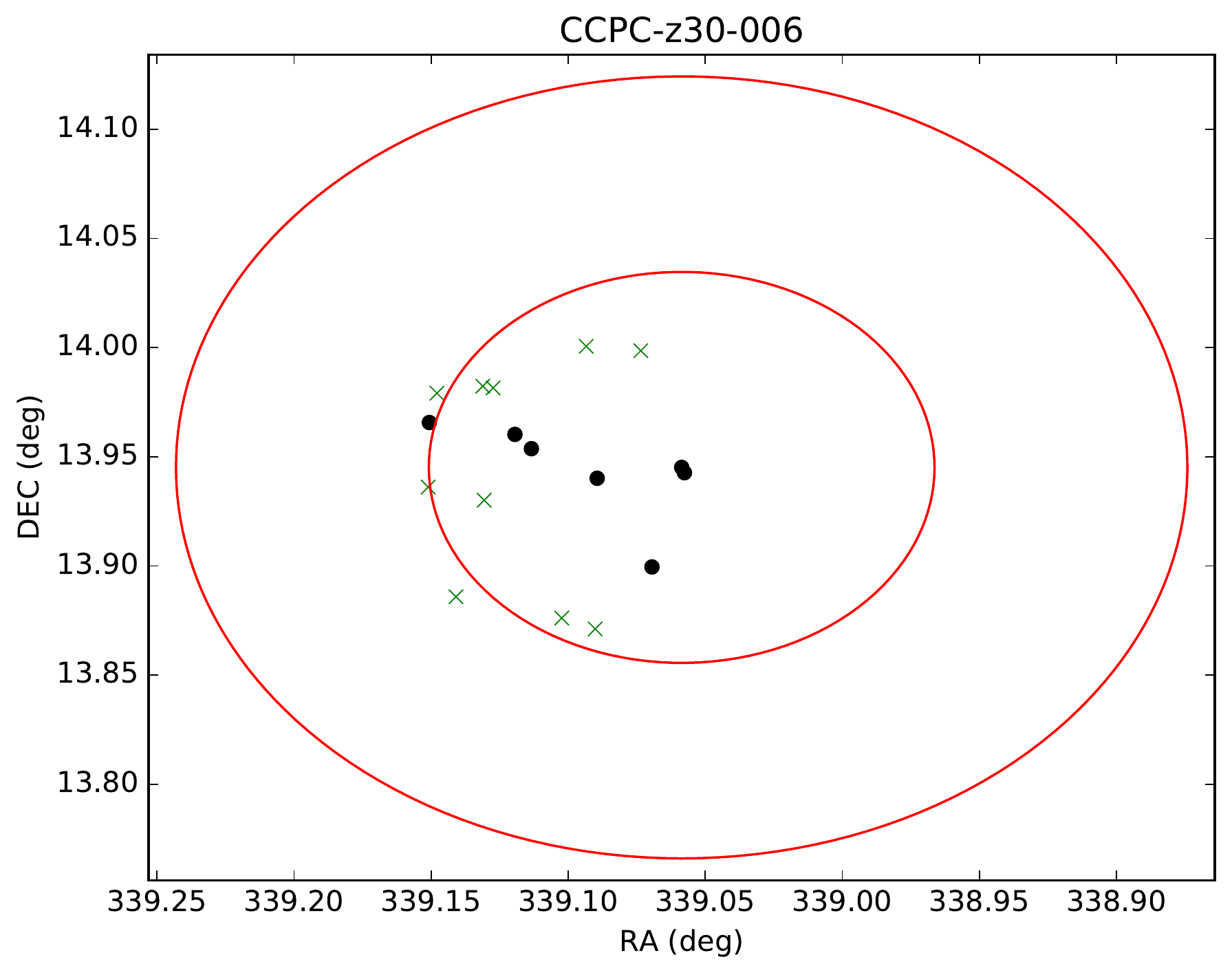}
\label{fig:CCPC-z30-006_sky}
\end{subfigure}
\hfill
\begin{subfigure}
\centering
\includegraphics[scale=0.52]{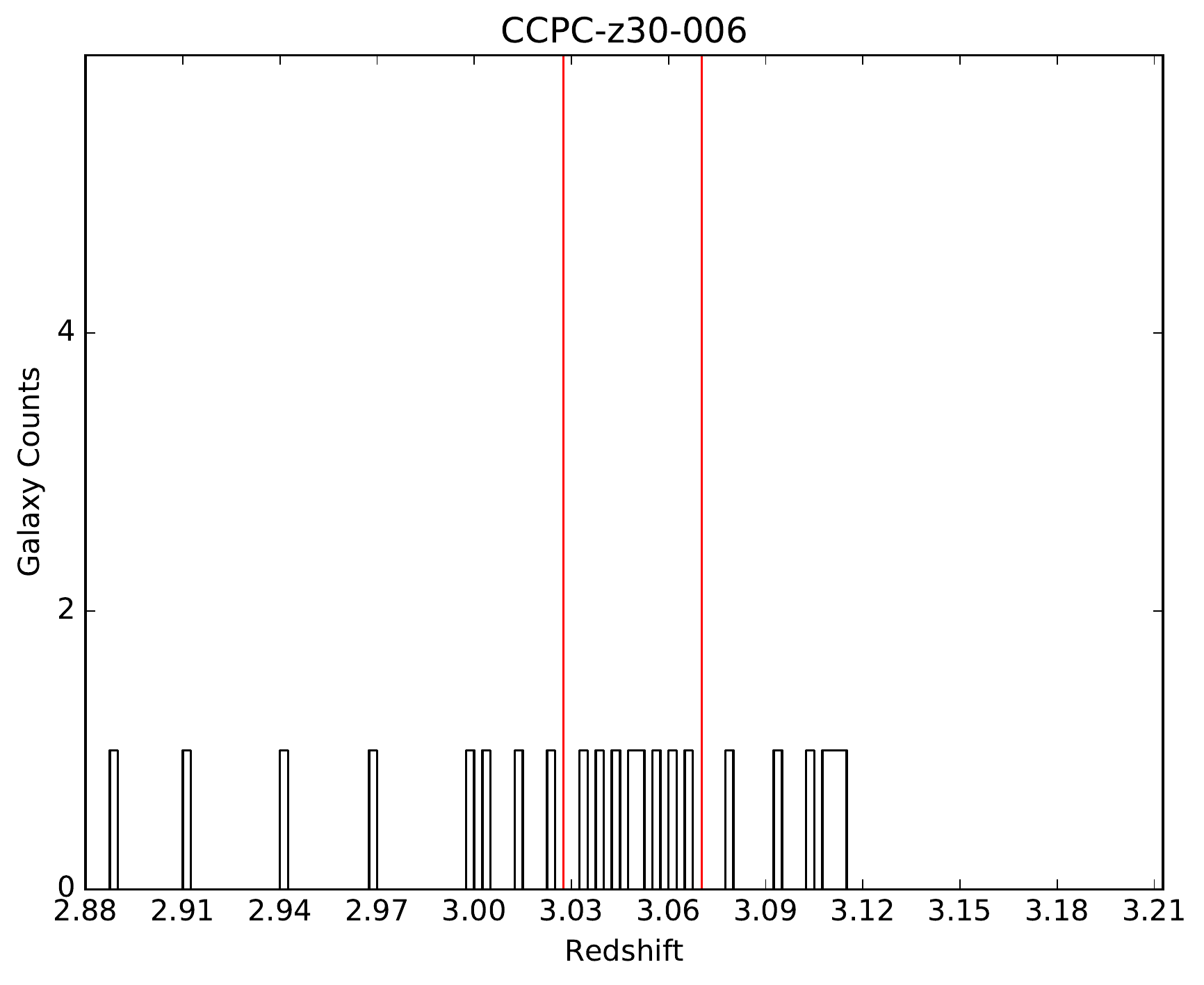}
\label{fig:CCPC-z30-006}
\end{subfigure}
\hfill
\end{figure*}

\begin{figure*}
\centering
\begin{subfigure}
\centering
\includegraphics[height=7.5cm,width=7.5cm]{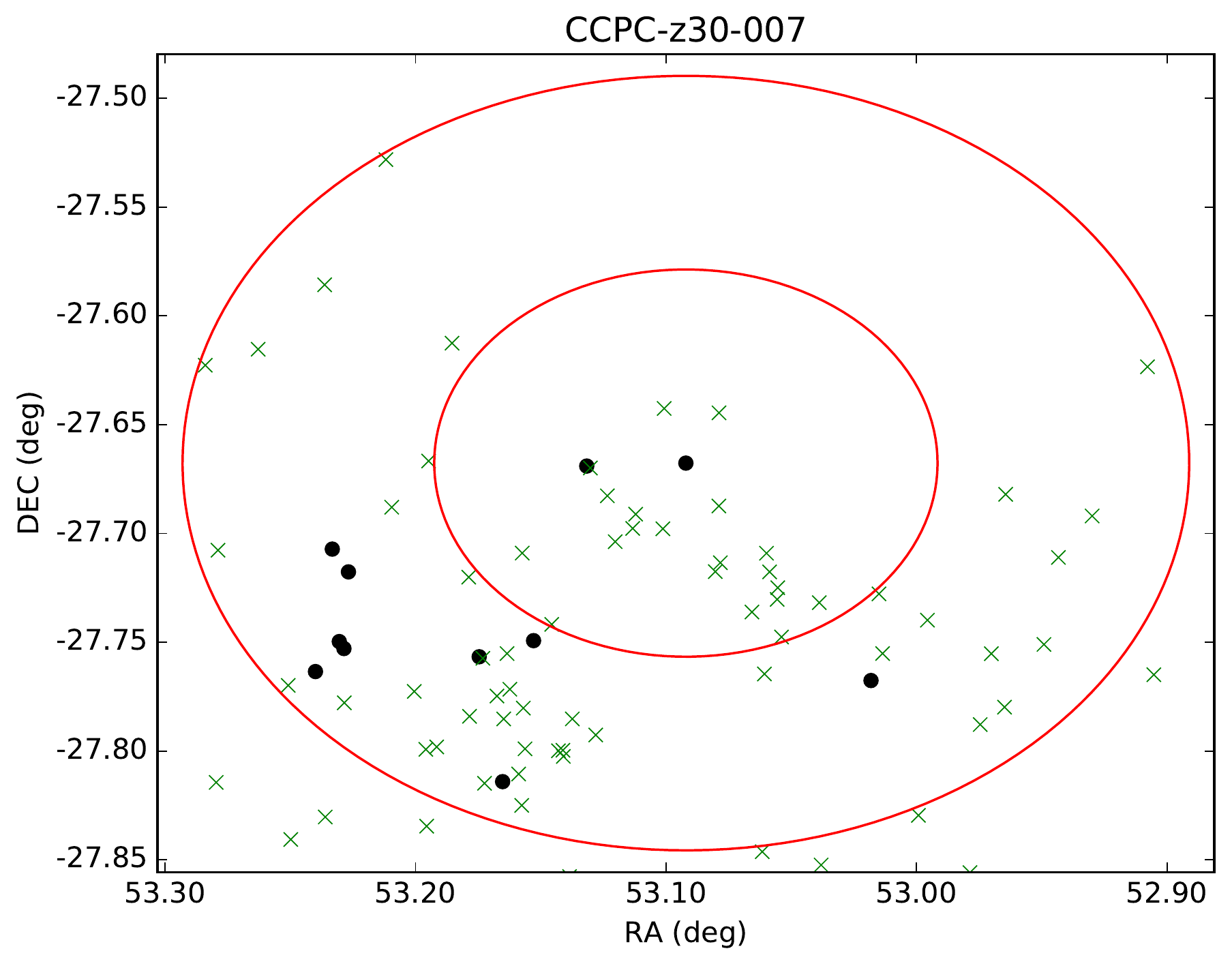}
\label{fig:CCPC-z30-007_sky}
\end{subfigure}
\hfill
\begin{subfigure}
\centering
\includegraphics[scale=0.52]{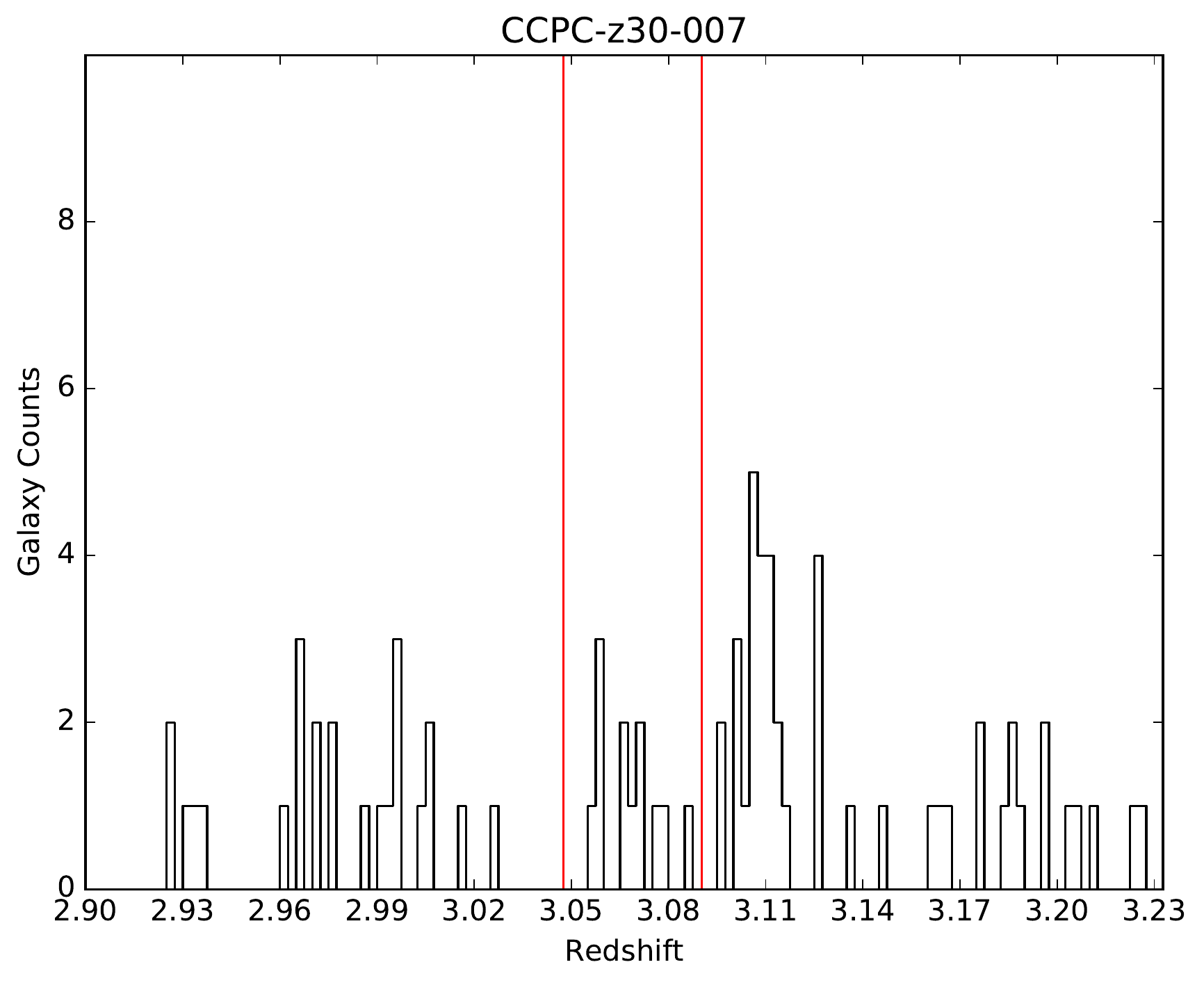}
\label{fig:CCPC-z30-007}
\end{subfigure}
\hfill
\end{figure*}
\clearpage 

\begin{figure*}
\centering
\begin{subfigure}
\centering
\includegraphics[height=7.5cm,width=7.5cm]{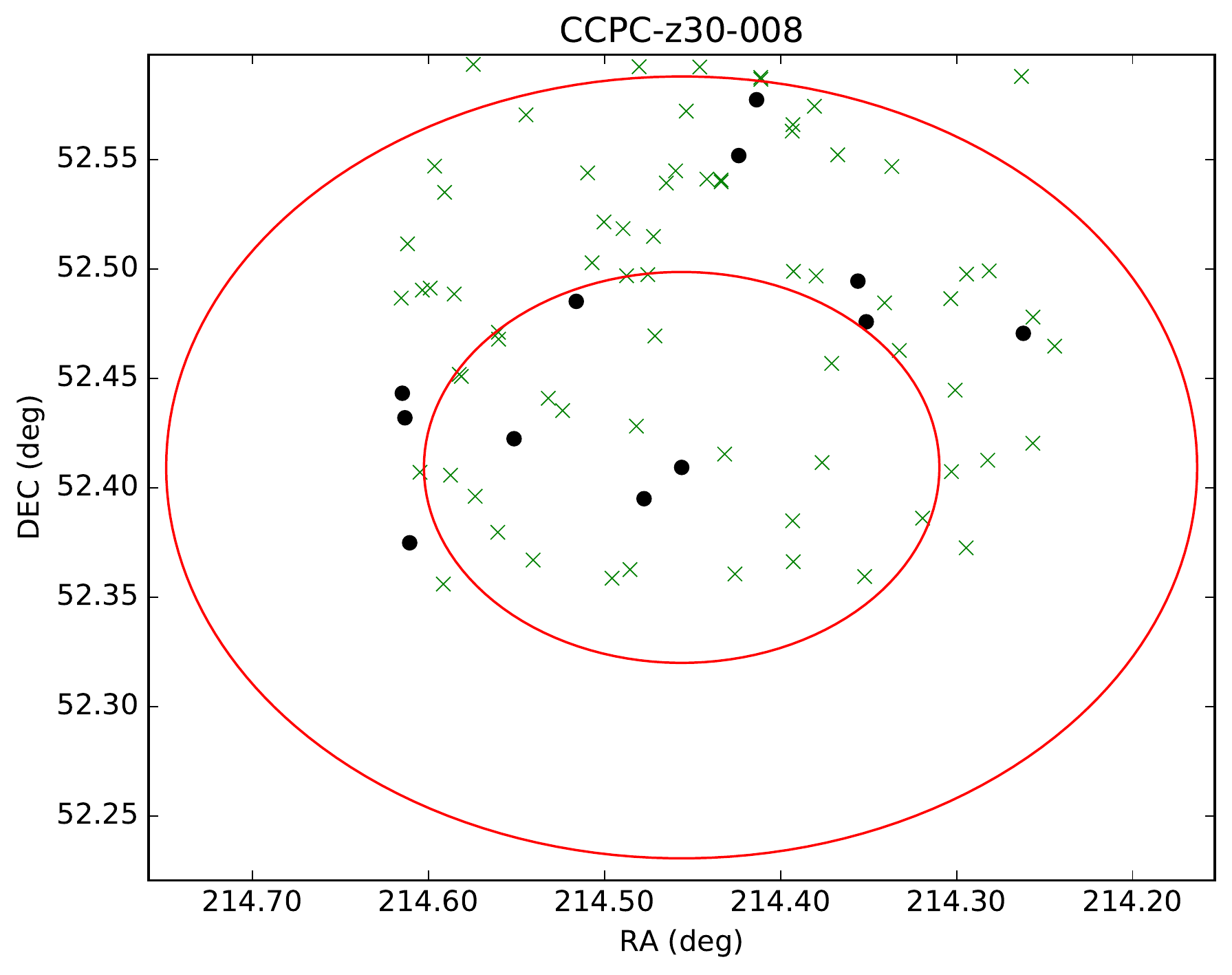}
\label{fig:CCPC-z30-008_sky}
\end{subfigure}
\hfill
\begin{subfigure}
\centering
\includegraphics[scale=0.52]{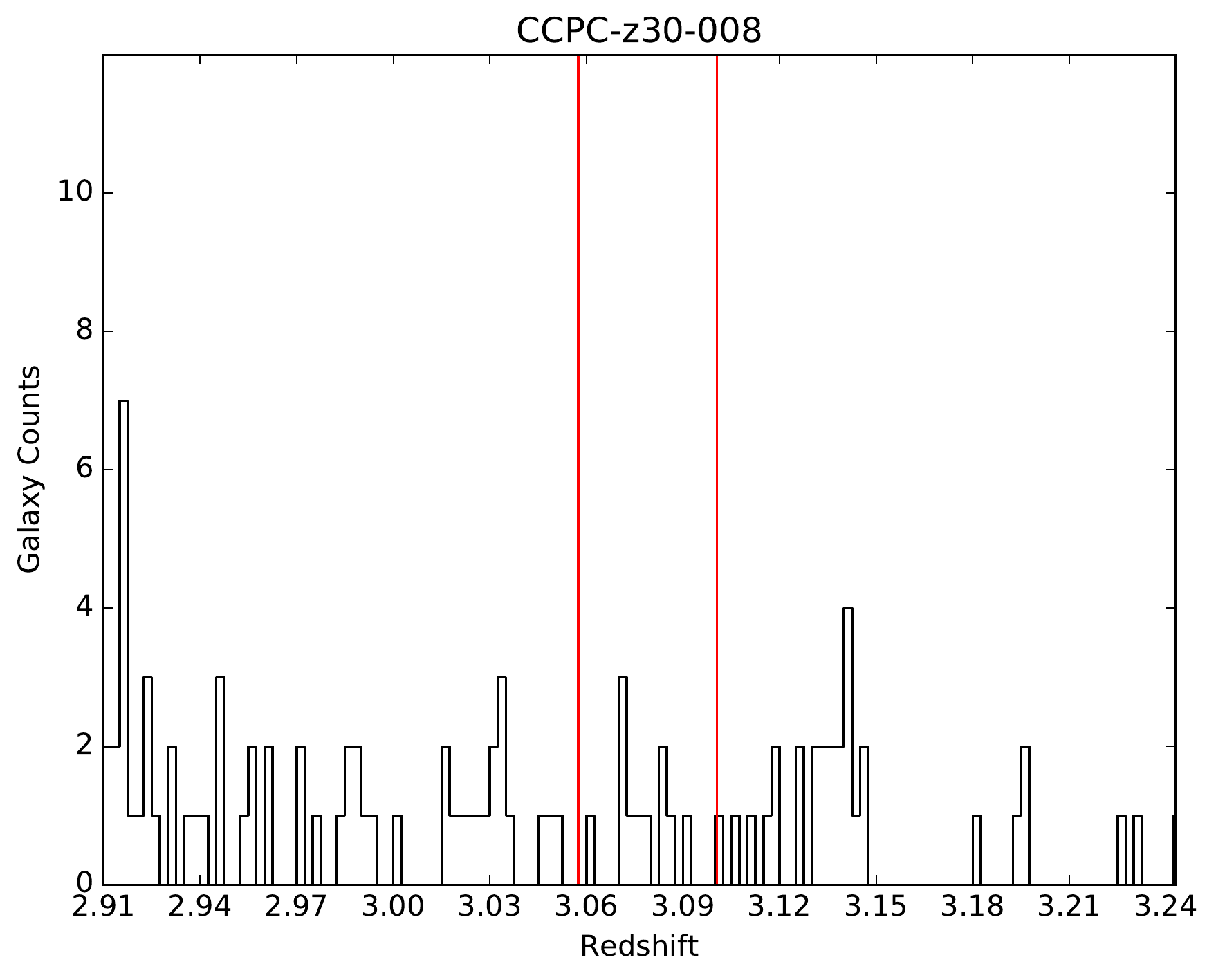}
\label{fig:CCPC-z30-008}
\end{subfigure}
\hfill
\end{figure*}

\begin{figure*}
\centering
\begin{subfigure}
\centering
\includegraphics[height=7.5cm,width=7.5cm]{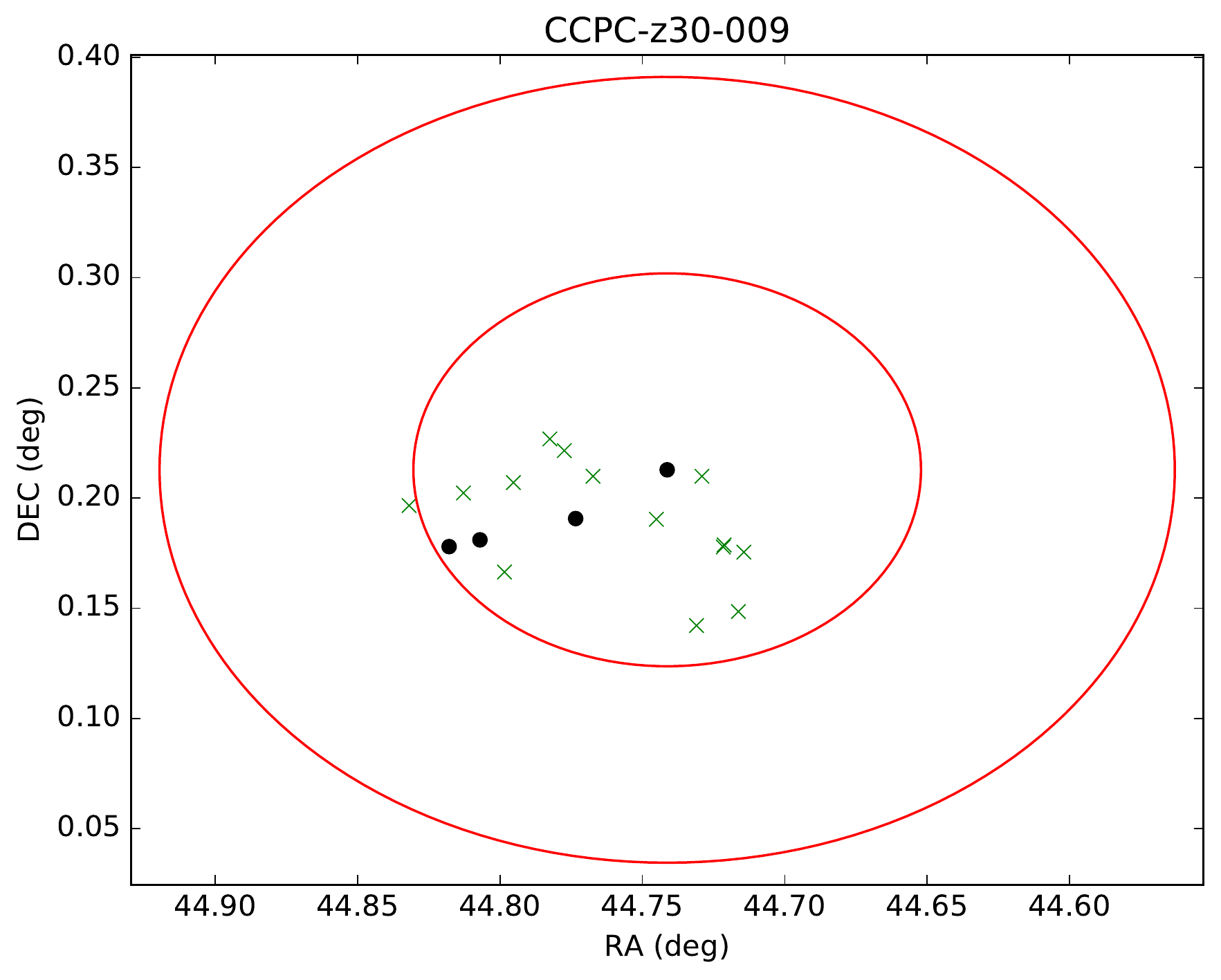}
\label{fig:CCPC-z30-009_sky}
\end{subfigure}
\hfill
\begin{subfigure}
\centering
\includegraphics[scale=0.52]{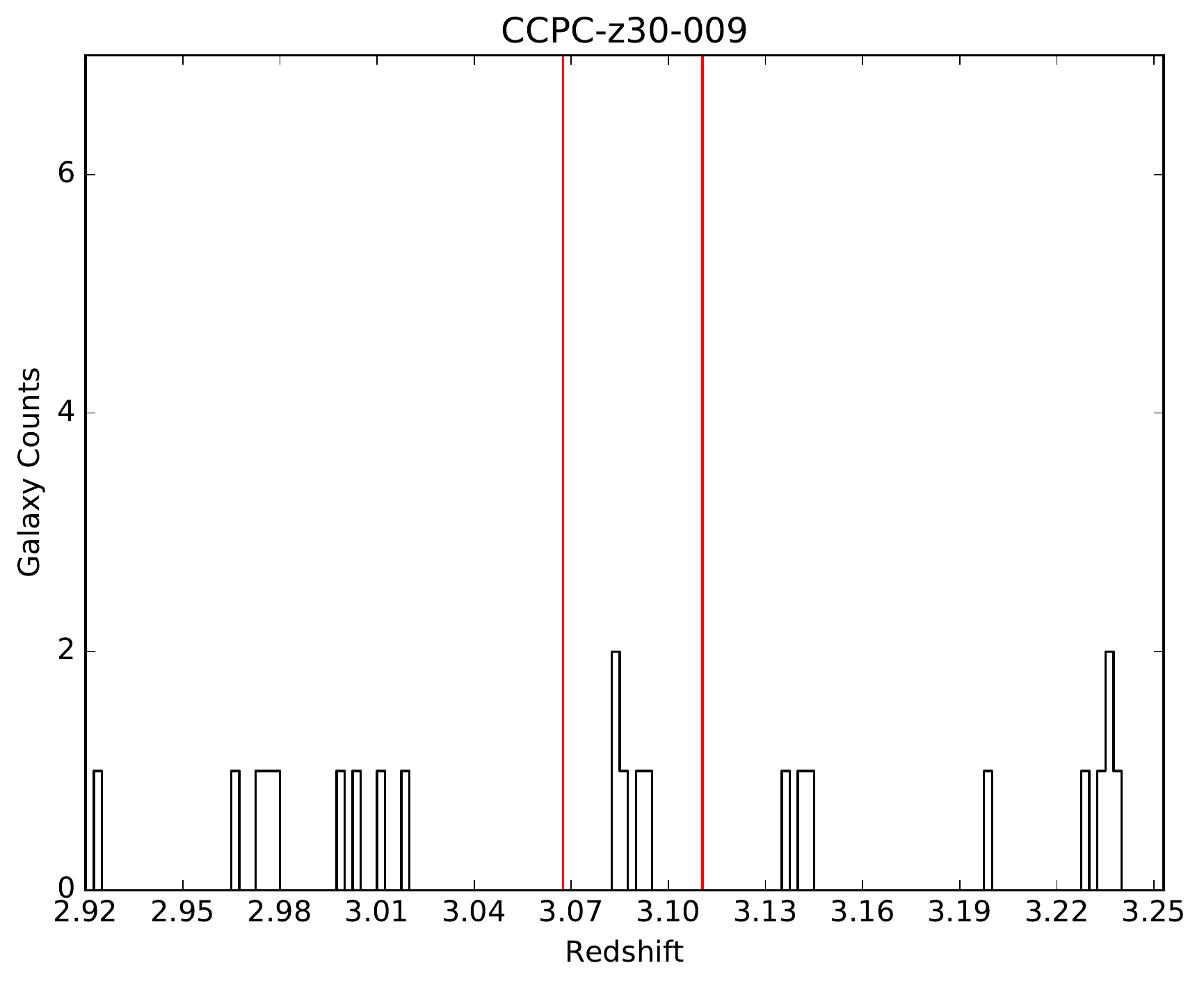}
\label{fig:CCPC-z30-009}
\end{subfigure}
\hfill
\end{figure*}

\begin{figure*}
\centering
\begin{subfigure}
\centering
\includegraphics[height=7.5cm,width=7.5cm]{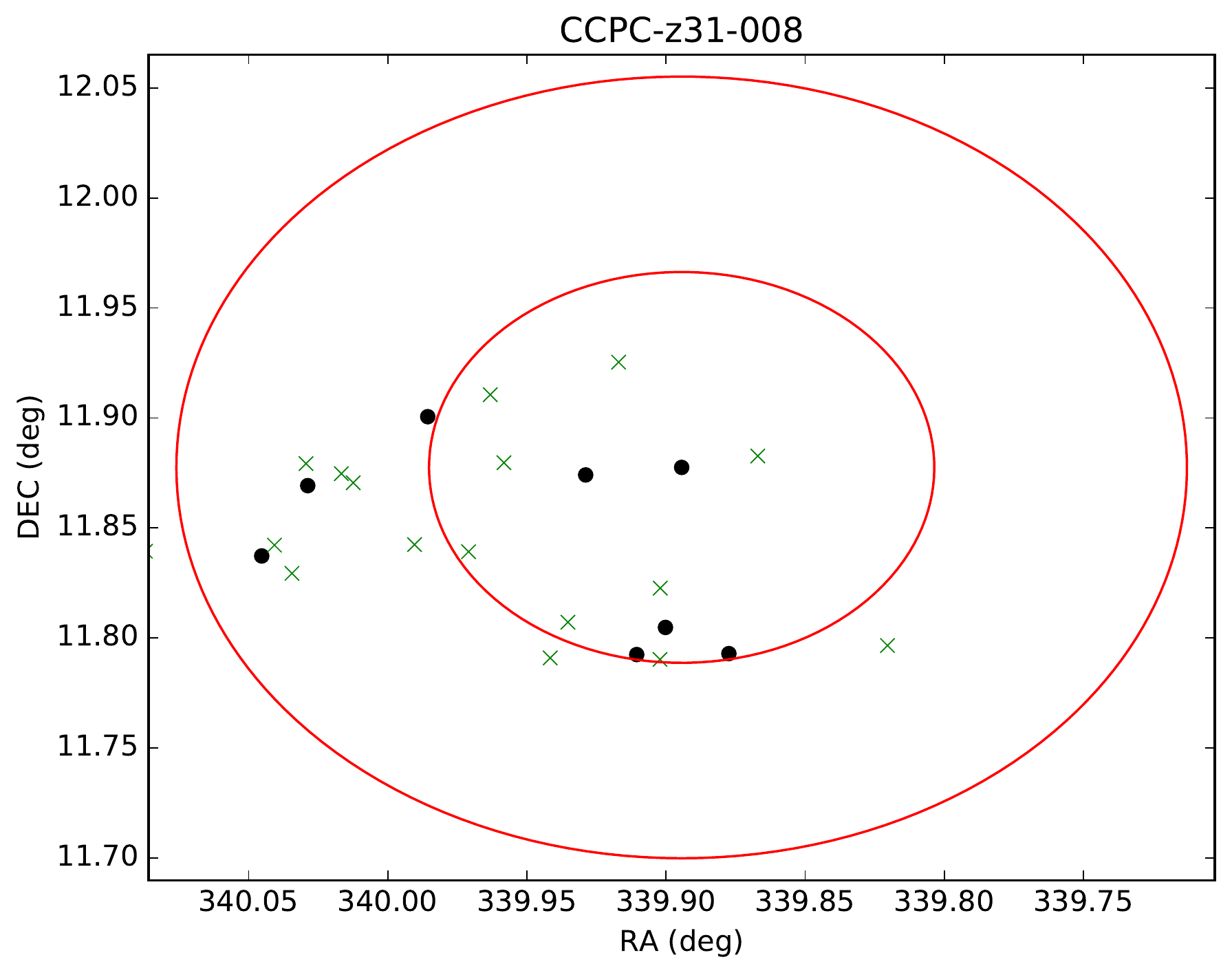}
\label{fig:CCPC-z31-008_sky}
\end{subfigure}
\hfill
\begin{subfigure}
\centering
\includegraphics[scale=0.52]{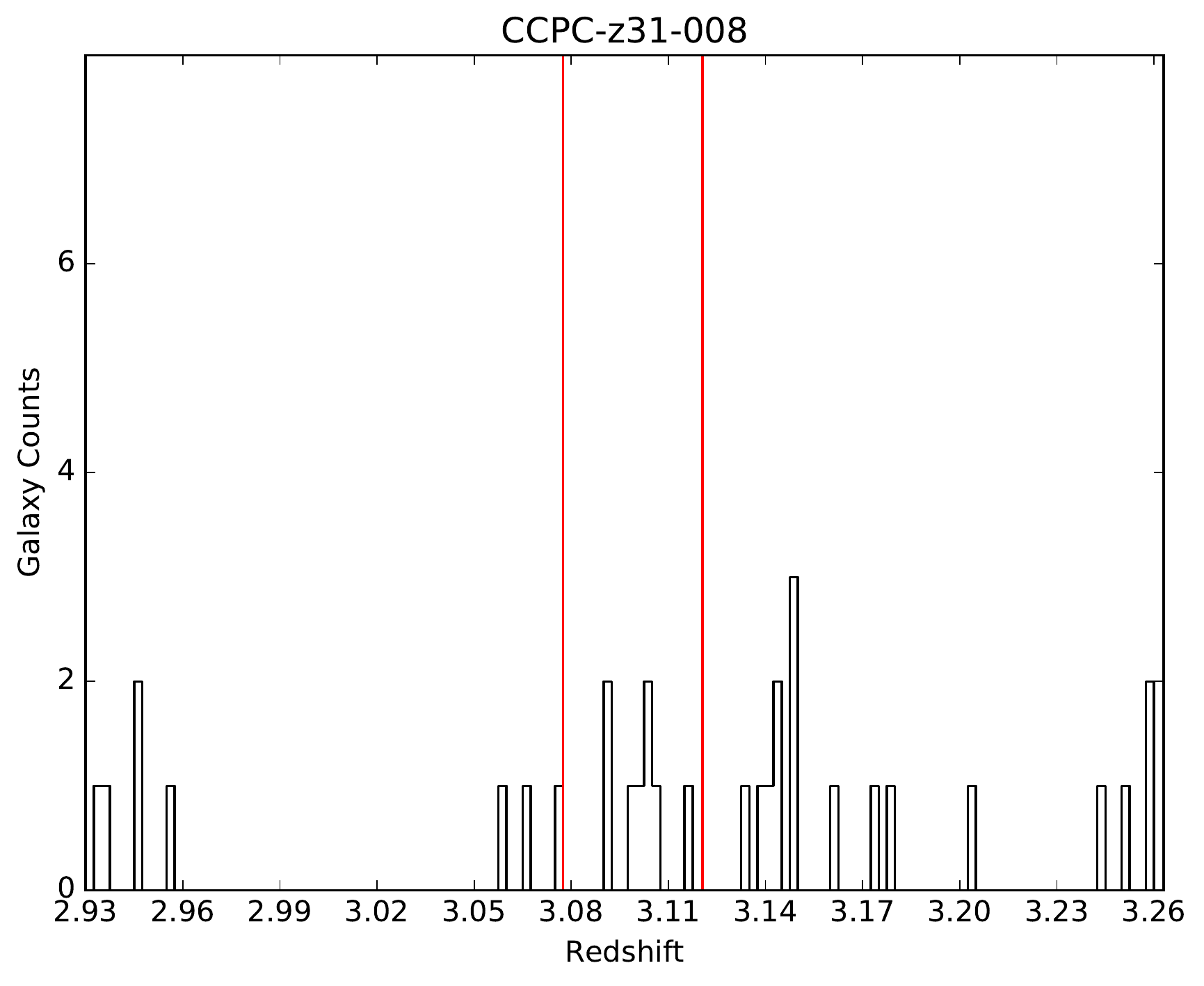}
\label{fig:CCPC-z31-008}
\end{subfigure}
\hfill
\end{figure*}
\clearpage 

\begin{figure*}
\centering
\begin{subfigure}
\centering
\includegraphics[height=7.5cm,width=7.5cm]{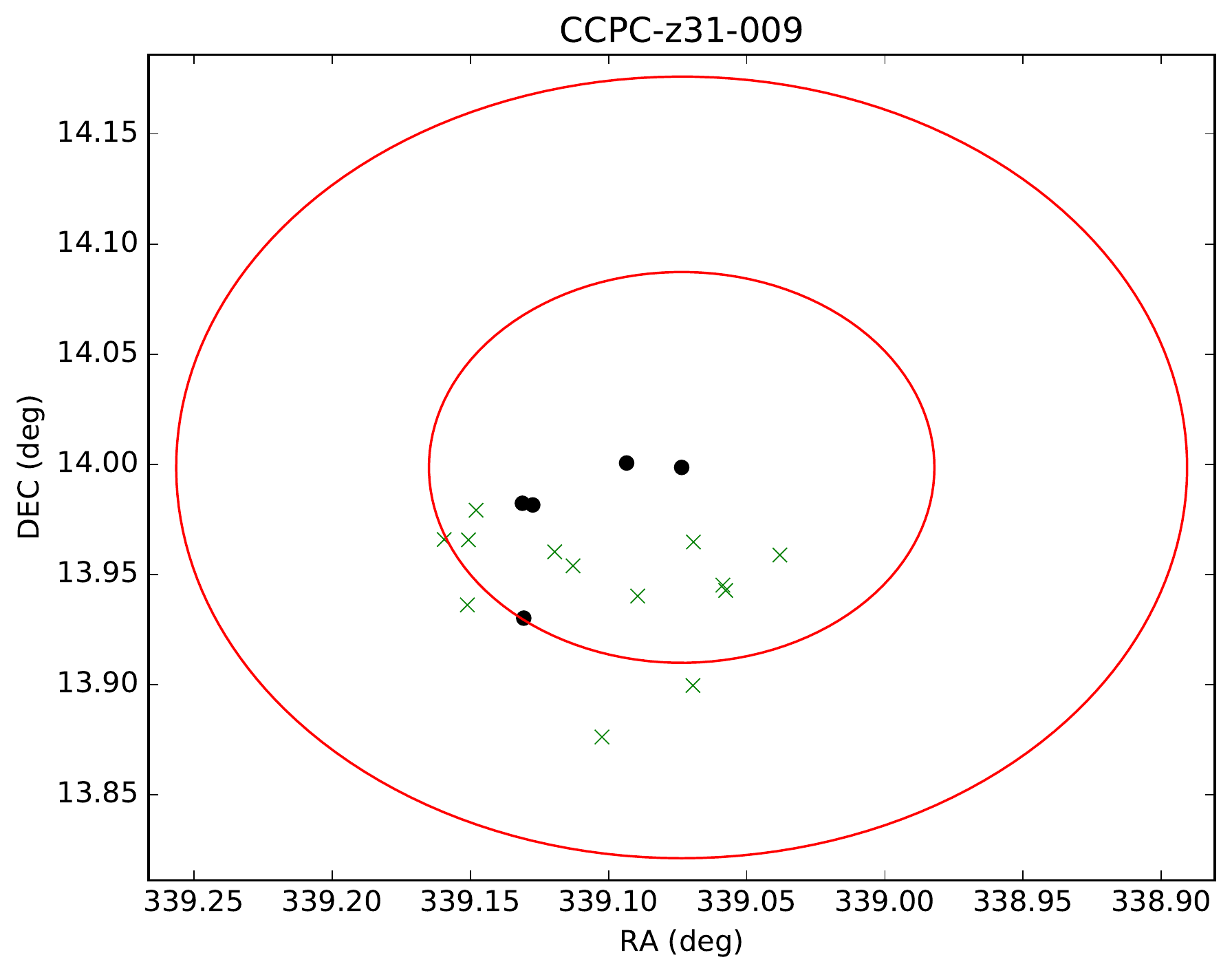}
\label{fig:CCPC-z31-009_sky}
\end{subfigure}
\hfill
\begin{subfigure}
\centering
\includegraphics[scale=0.52]{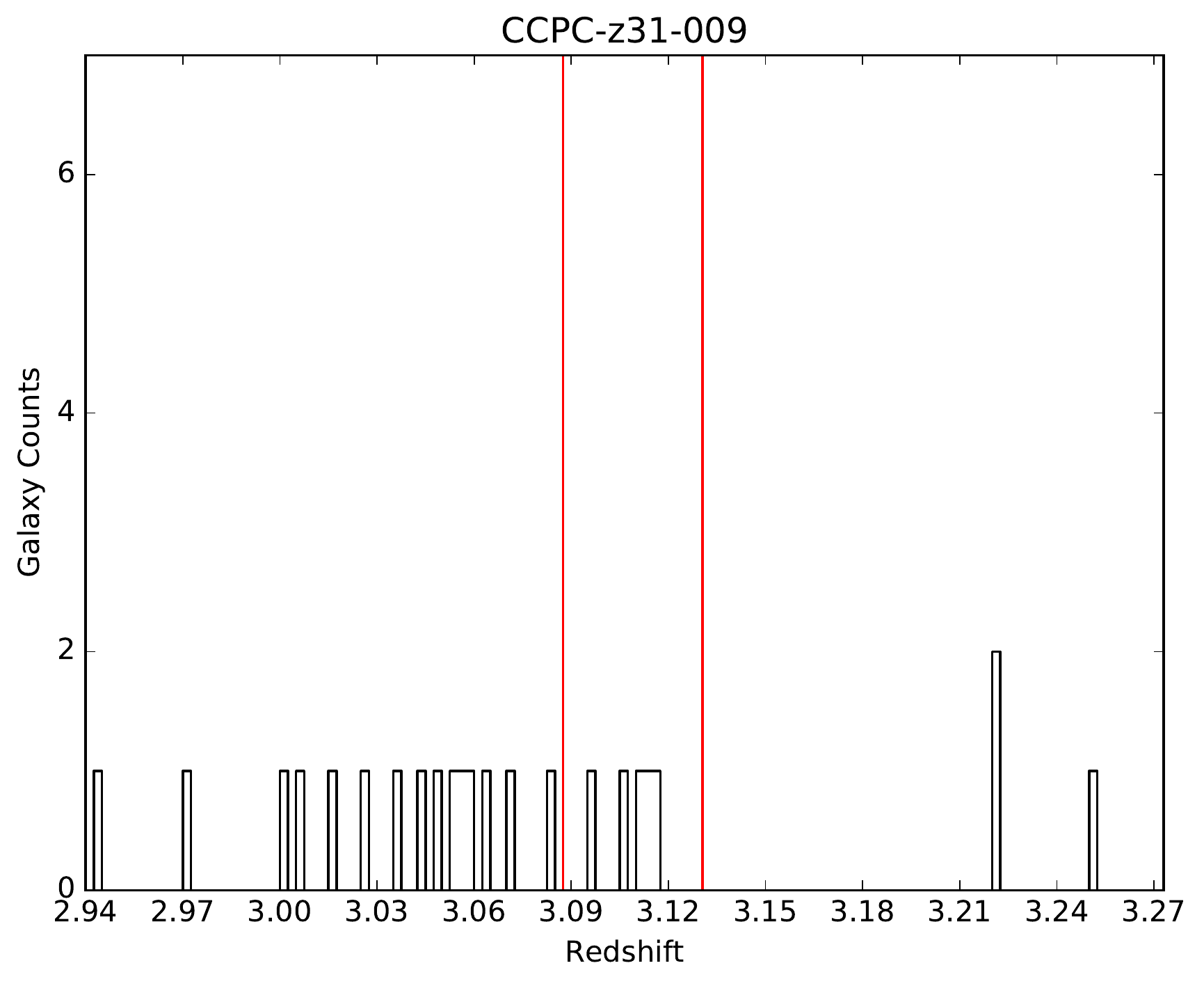}
\label{fig:CCPC-z31-009}
\end{subfigure}
\hfill
\end{figure*}

\begin{figure*}
\centering
\begin{subfigure}
\centering
\includegraphics[height=7.5cm,width=7.5cm]{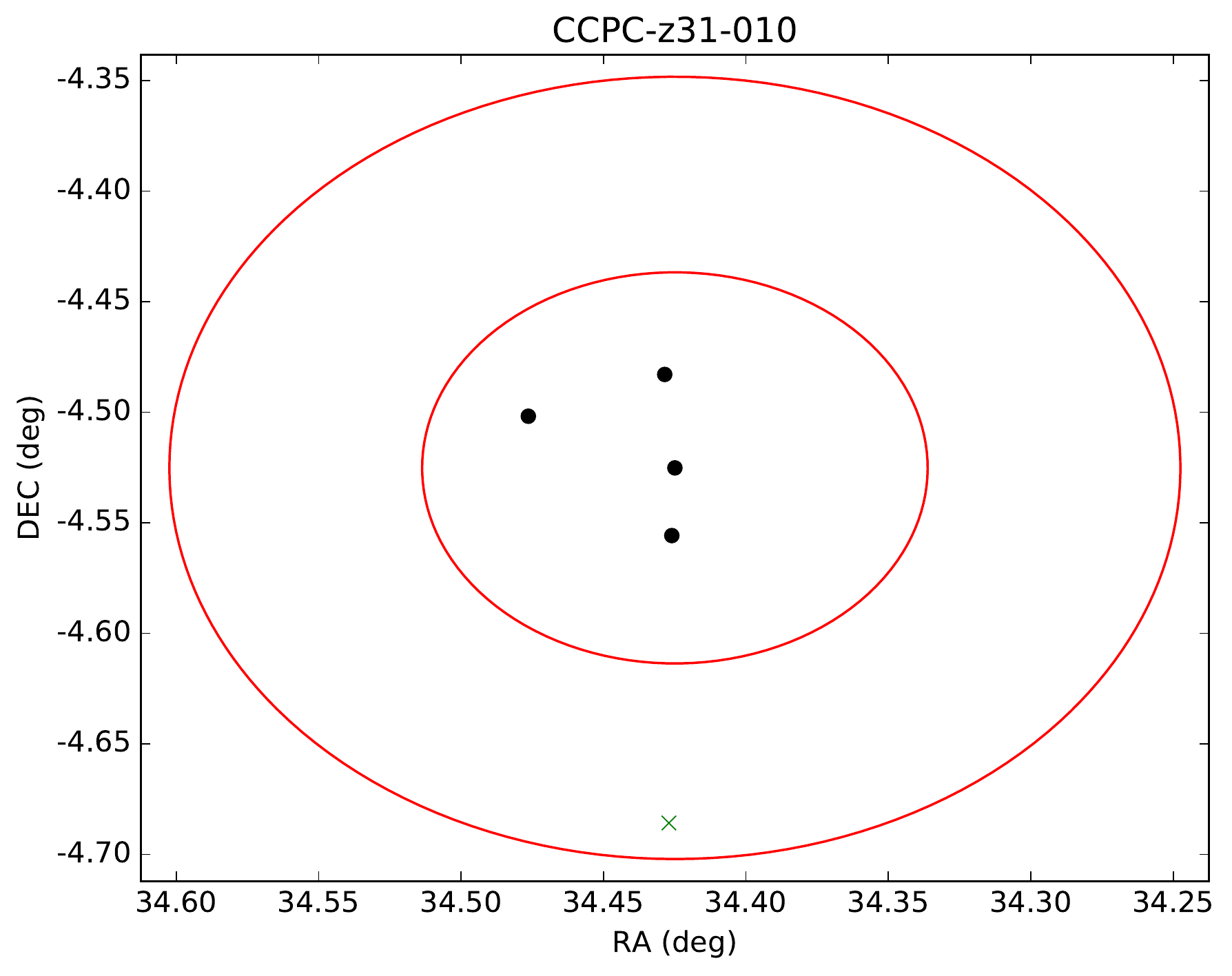}
\label{fig:CCPC-z31-010_sky}
\end{subfigure}
\hfill
\begin{subfigure}
\centering
\includegraphics[scale=0.52]{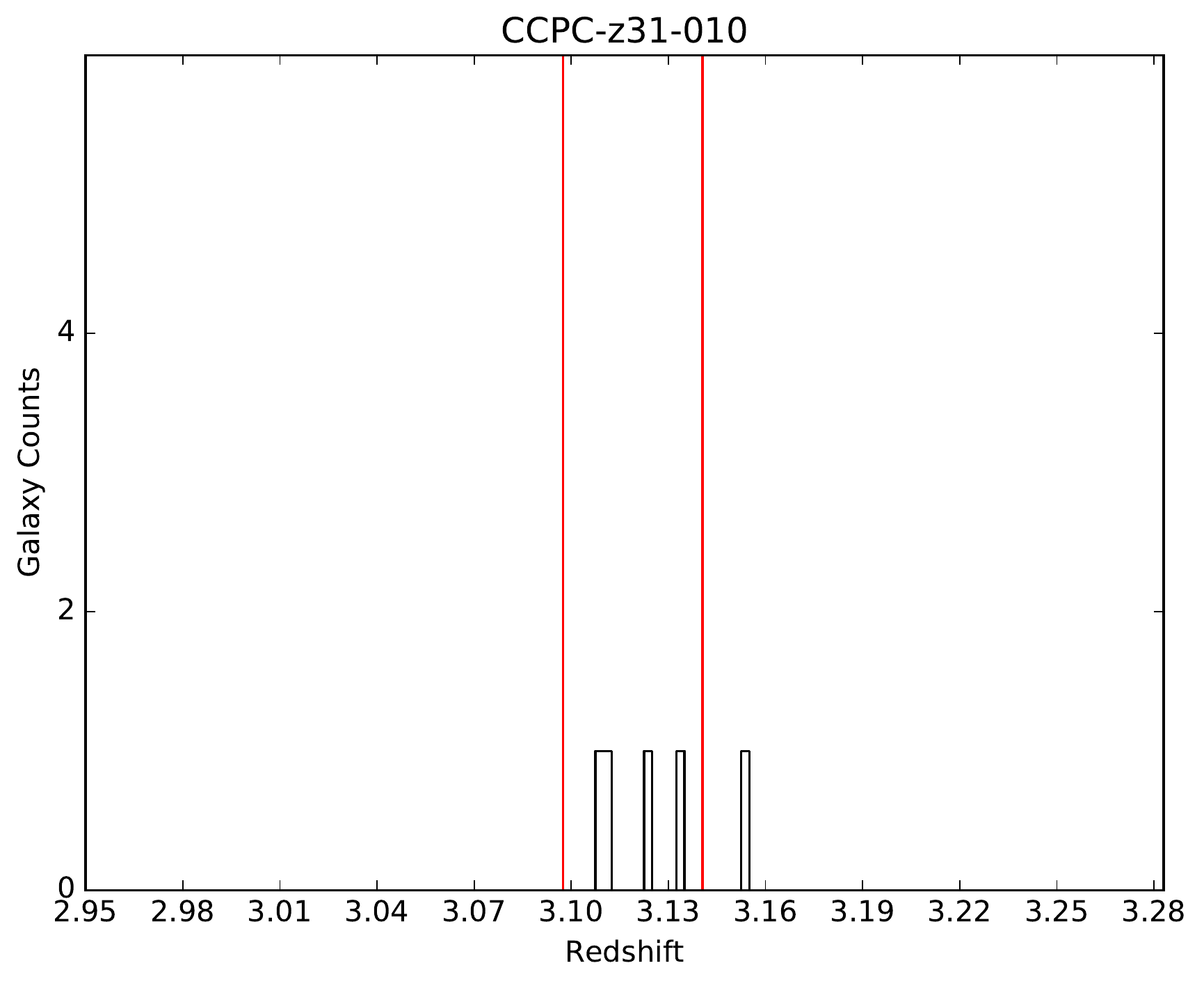}
\label{fig:CCPC-z31-010}
\end{subfigure}
\hfill
\end{figure*}

\begin{figure*}
\centering
\begin{subfigure}
\centering
\includegraphics[height=7.5cm,width=7.5cm]{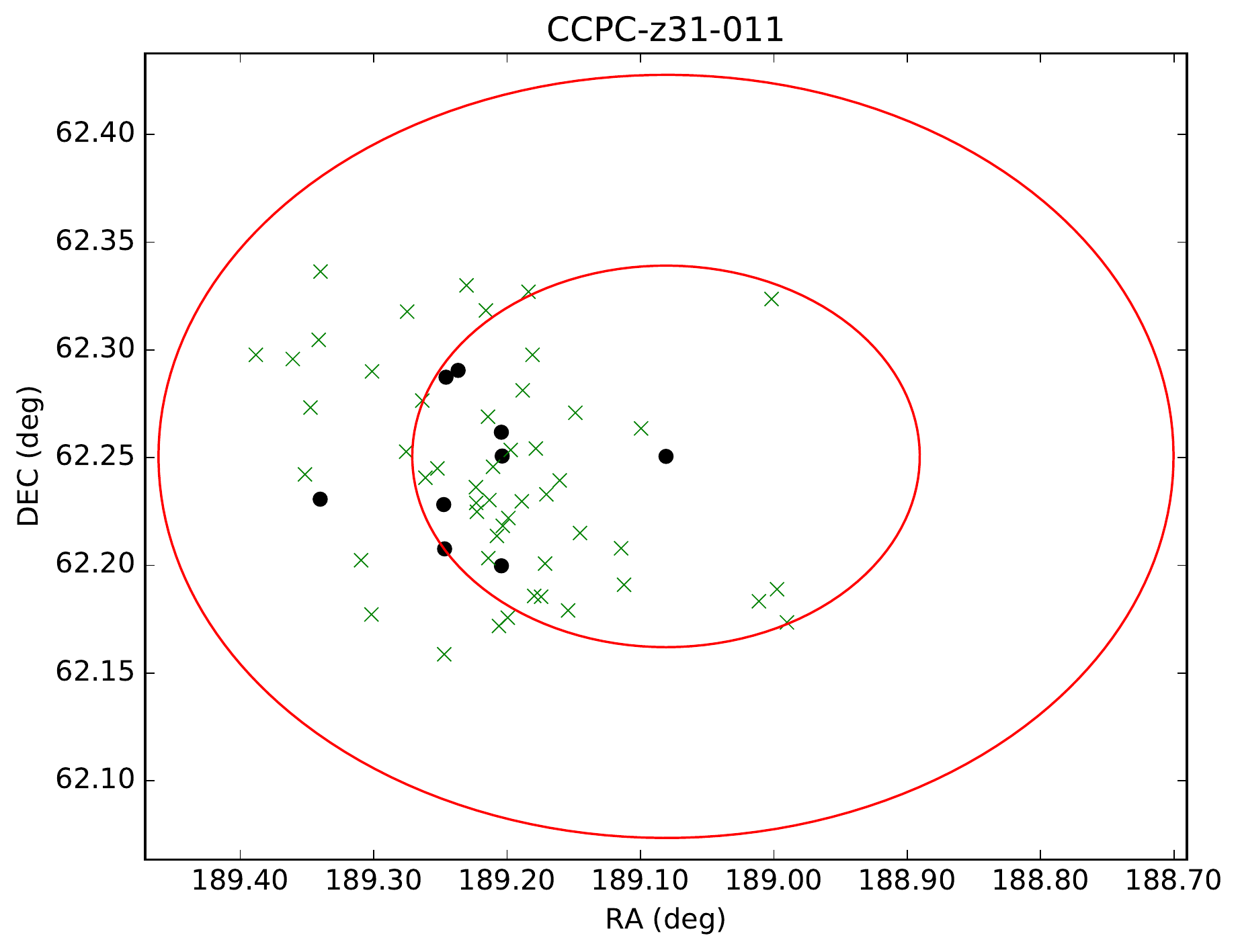}
\label{fig:CCPC-z31-011_sky}
\end{subfigure}
\hfill
\begin{subfigure}
\centering
\includegraphics[scale=0.52]{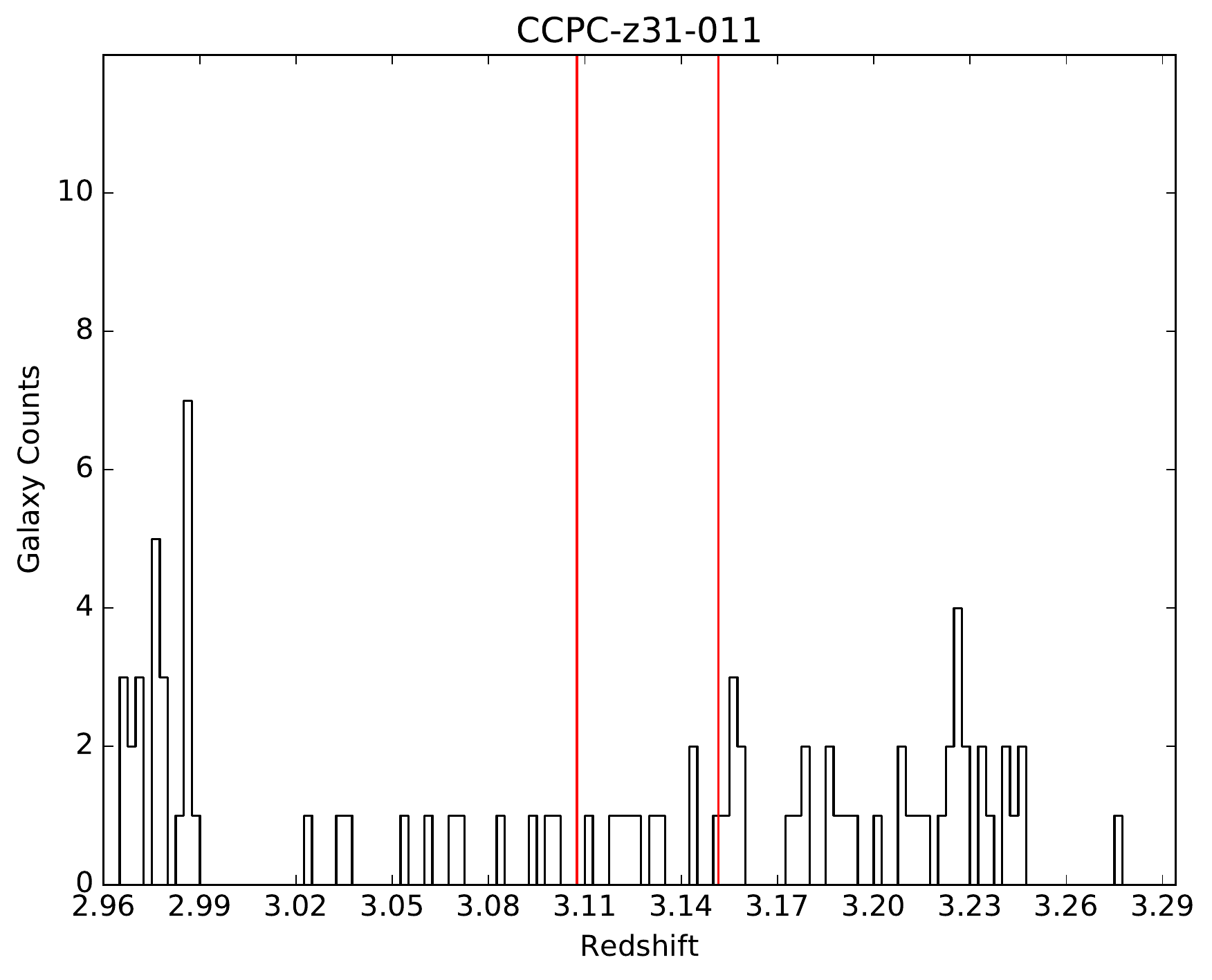}
\label{fig:CCPC-z31-011}
\end{subfigure}
\hfill
\end{figure*}
\clearpage 

\begin{figure*}
\centering
\begin{subfigure}
\centering
\includegraphics[height=7.5cm,width=7.5cm]{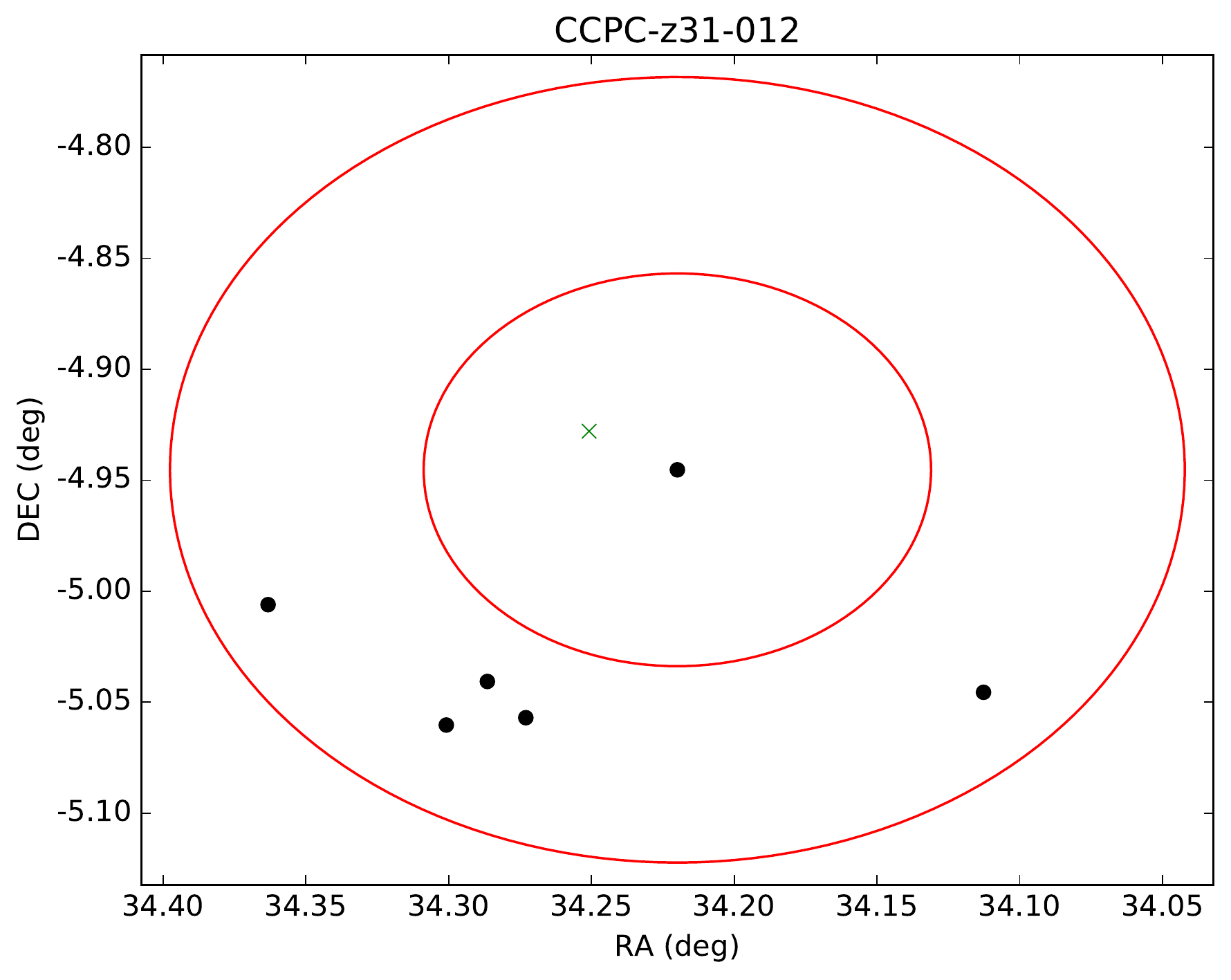}
\label{fig:CCPC-z31-012_sky}
\end{subfigure}
\hfill
\begin{subfigure}
\centering
\includegraphics[scale=0.52]{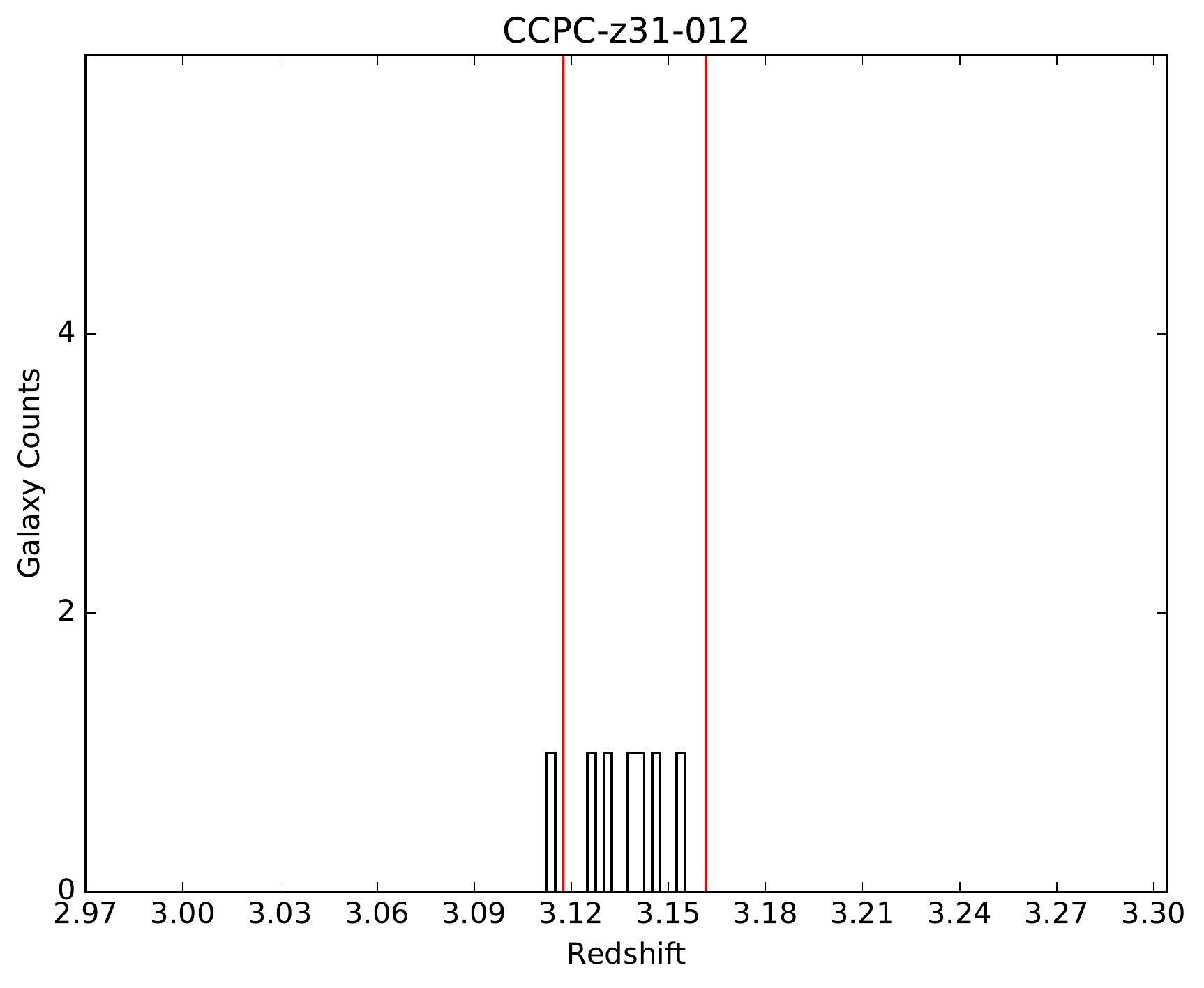}
\label{fig:CCPC-z31-012}
\end{subfigure}
\hfill
\end{figure*}

\begin{figure*}
\centering
\begin{subfigure}
\centering
\includegraphics[height=7.5cm,width=7.5cm]{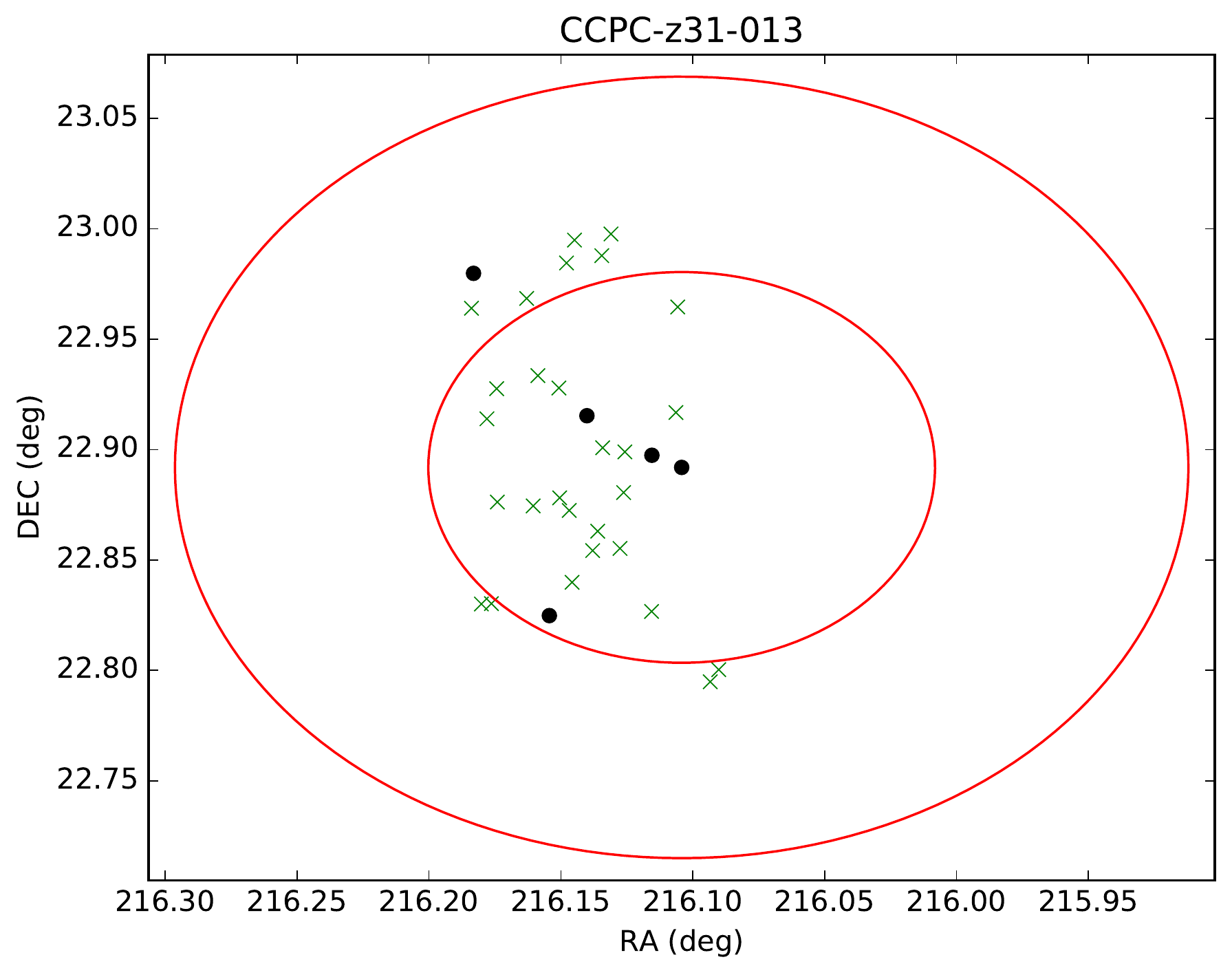}
\label{fig:CCPC-z31-013_sky}
\end{subfigure}
\hfill
\begin{subfigure}
\centering
\includegraphics[scale=0.52]{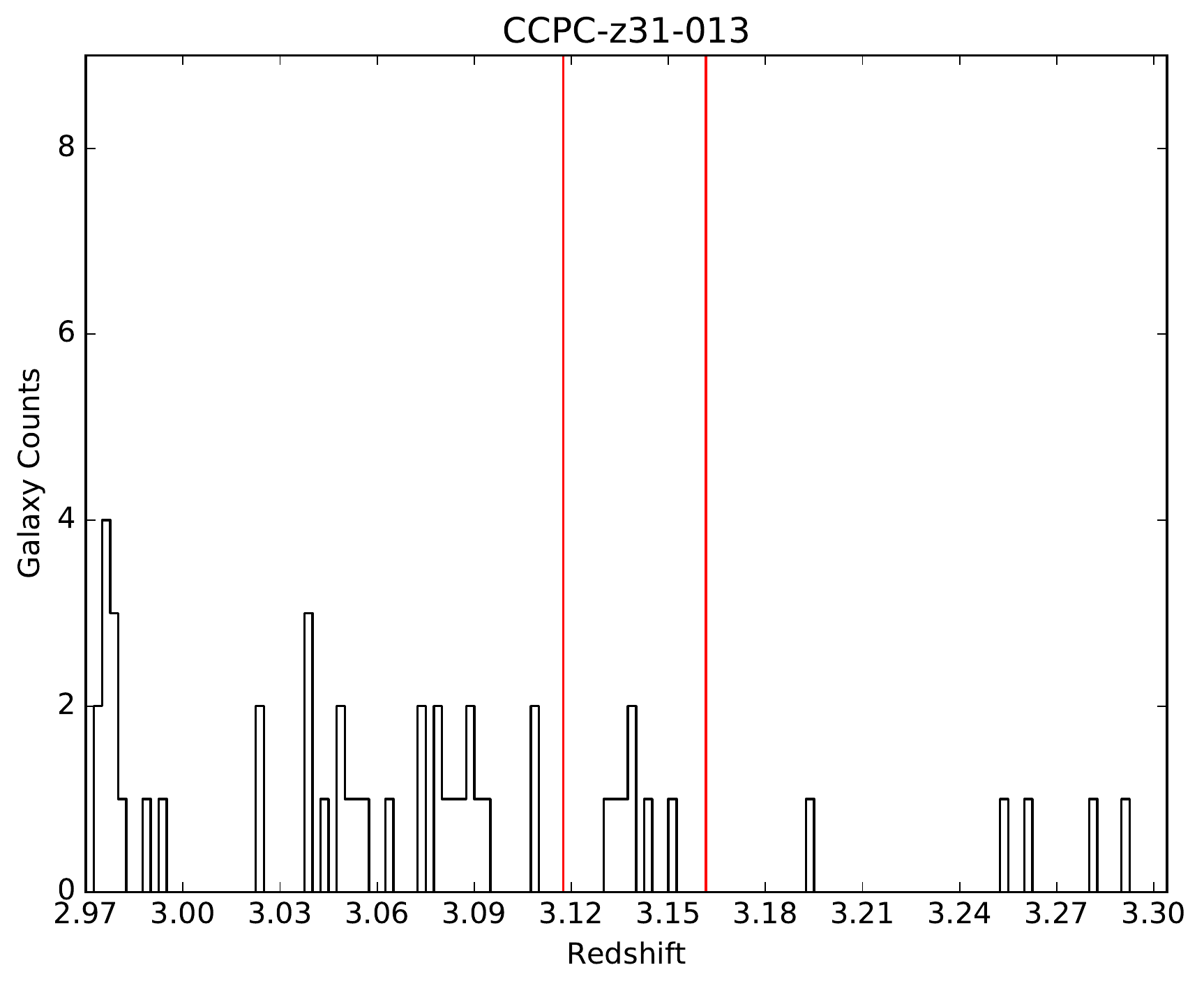}
\label{fig:CCPC-z31-013}
\end{subfigure}
\hfill
\end{figure*}

\begin{figure*}
\centering
\begin{subfigure}
\centering
\includegraphics[height=7.5cm,width=7.5cm]{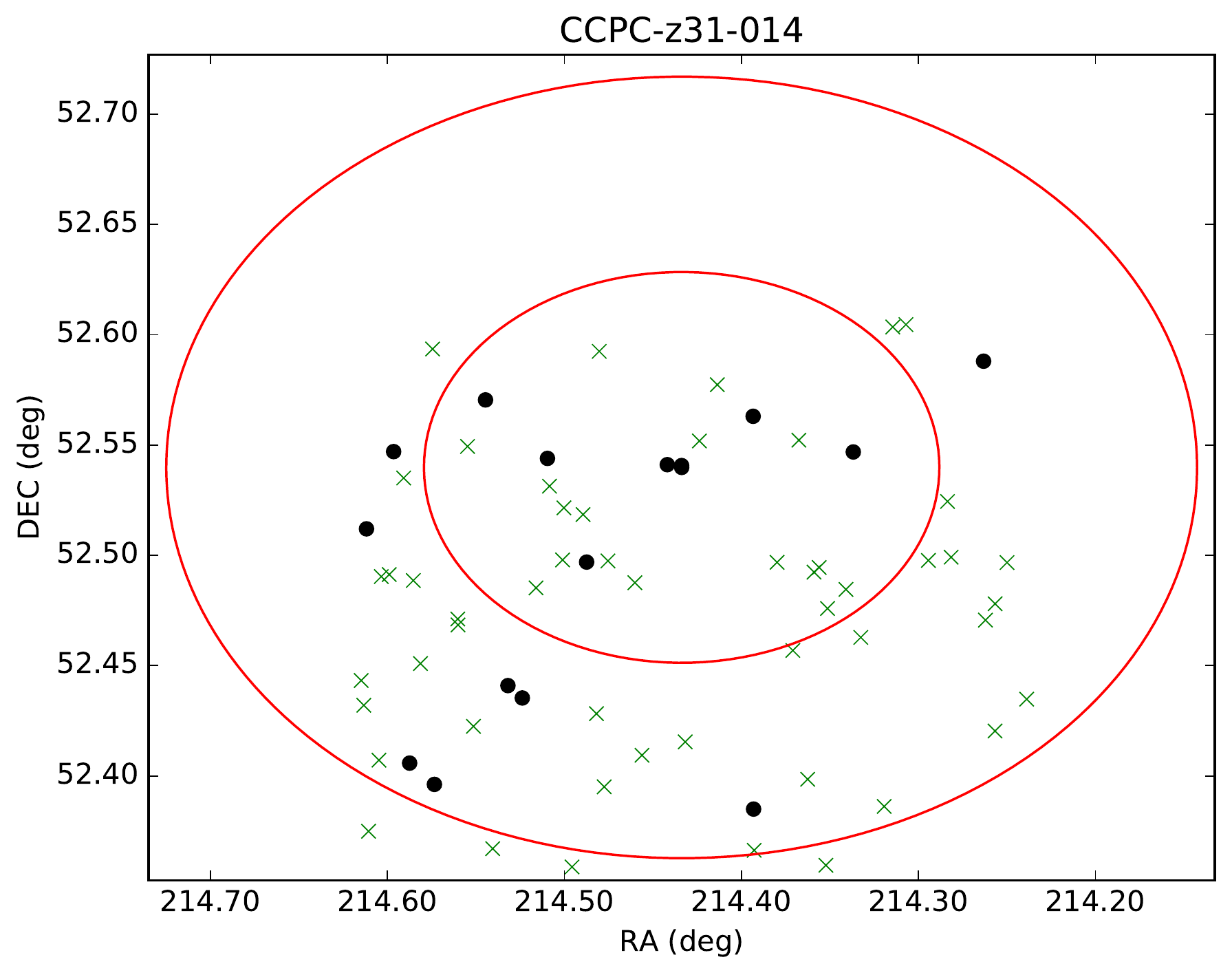}
\label{fig:CCPC-z31-014_sky}
\end{subfigure}
\hfill
\begin{subfigure}
\centering
\includegraphics[scale=0.52]{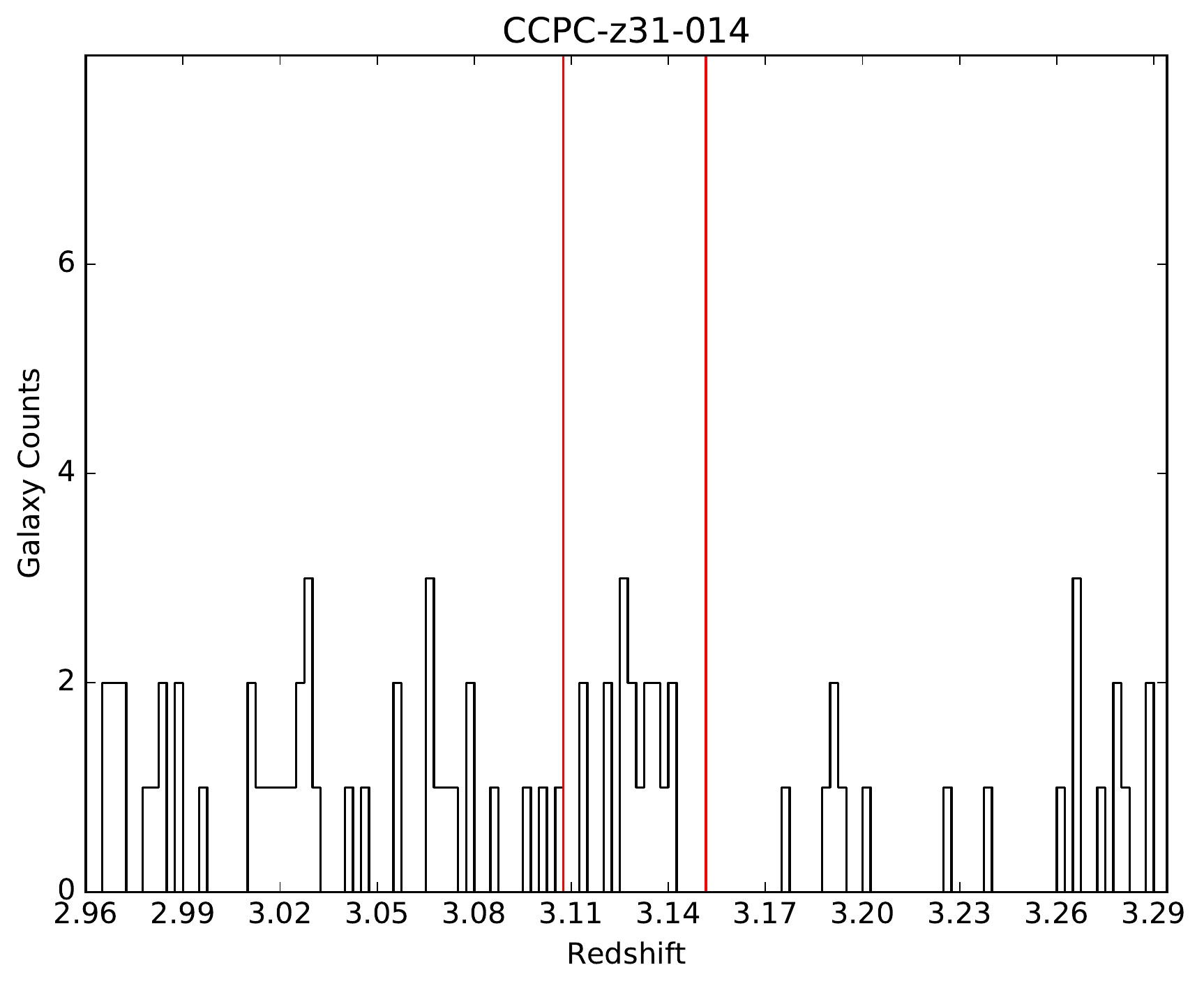}
\label{fig:CCPC-z31-014}
\end{subfigure}
\hfill
\end{figure*}
\clearpage 

\begin{figure*}
\centering
\begin{subfigure}
\centering
\includegraphics[height=7.5cm,width=7.5cm]{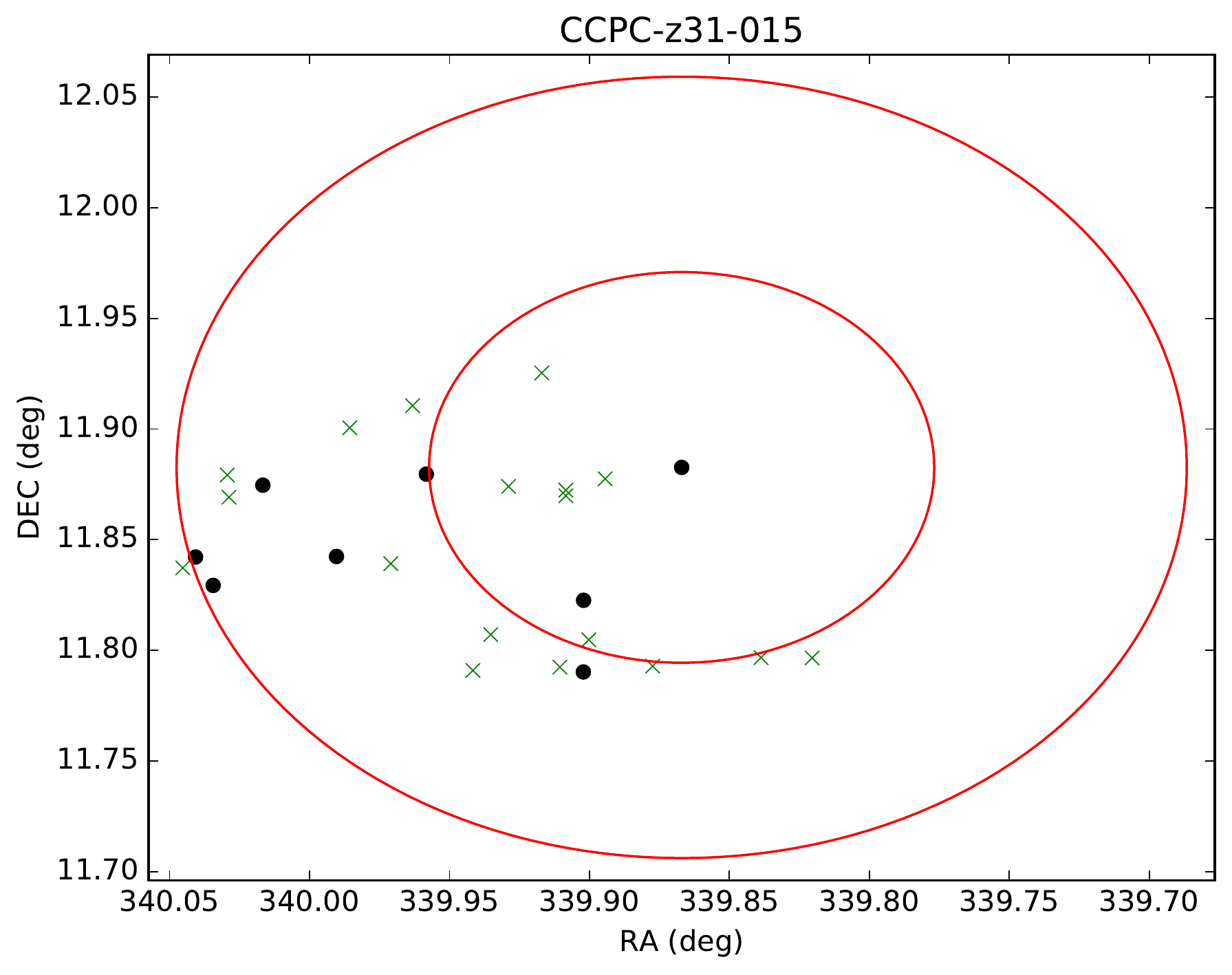}
\label{fig:CCPC-z31-015_sky}
\end{subfigure}
\hfill
\begin{subfigure}
\centering
\includegraphics[scale=0.52]{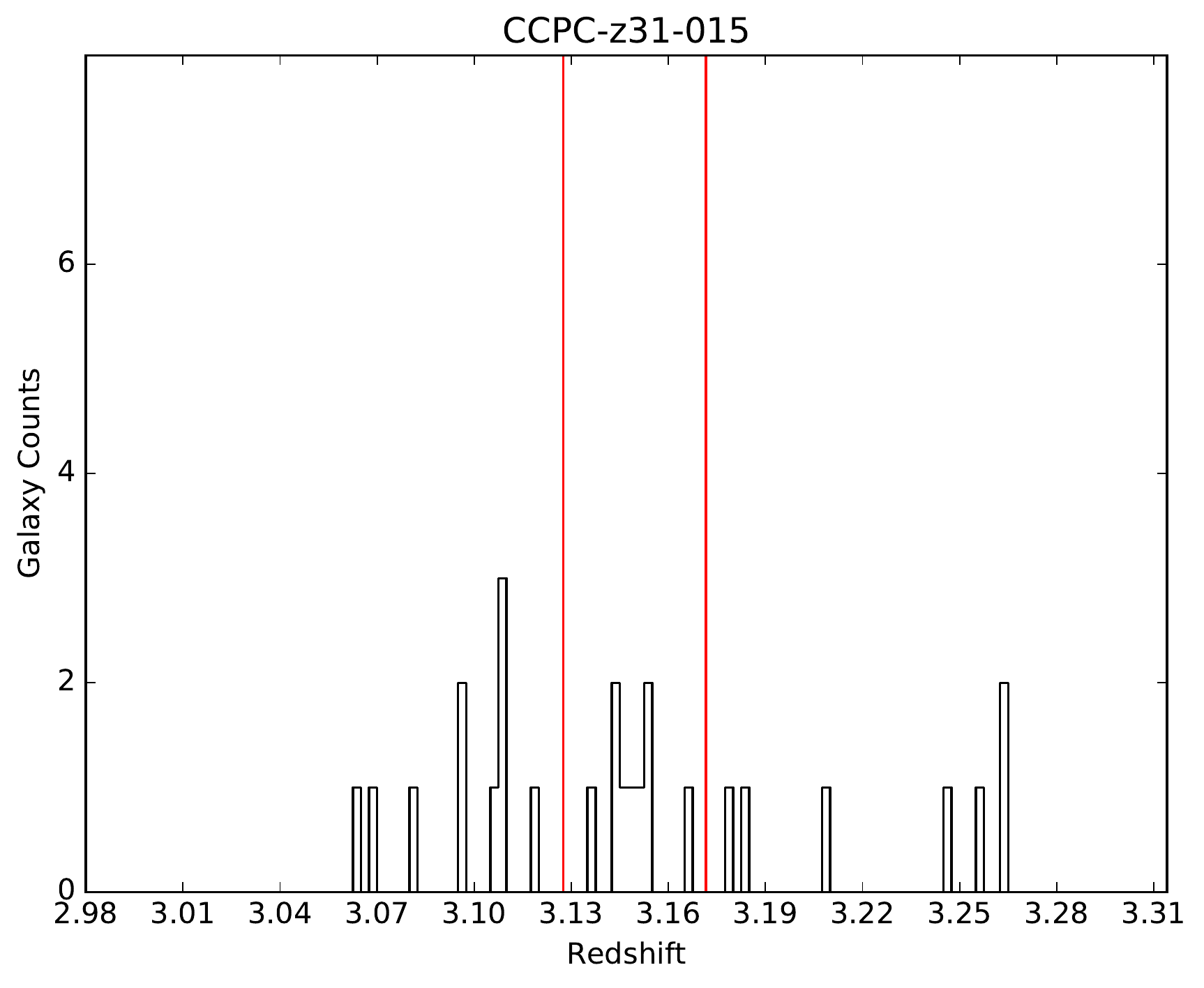}
\label{fig:CCPC-z31-015}
\end{subfigure}
\hfill
\end{figure*}

\begin{figure*}
\centering
\begin{subfigure}
\centering
\includegraphics[height=7.5cm,width=7.5cm]{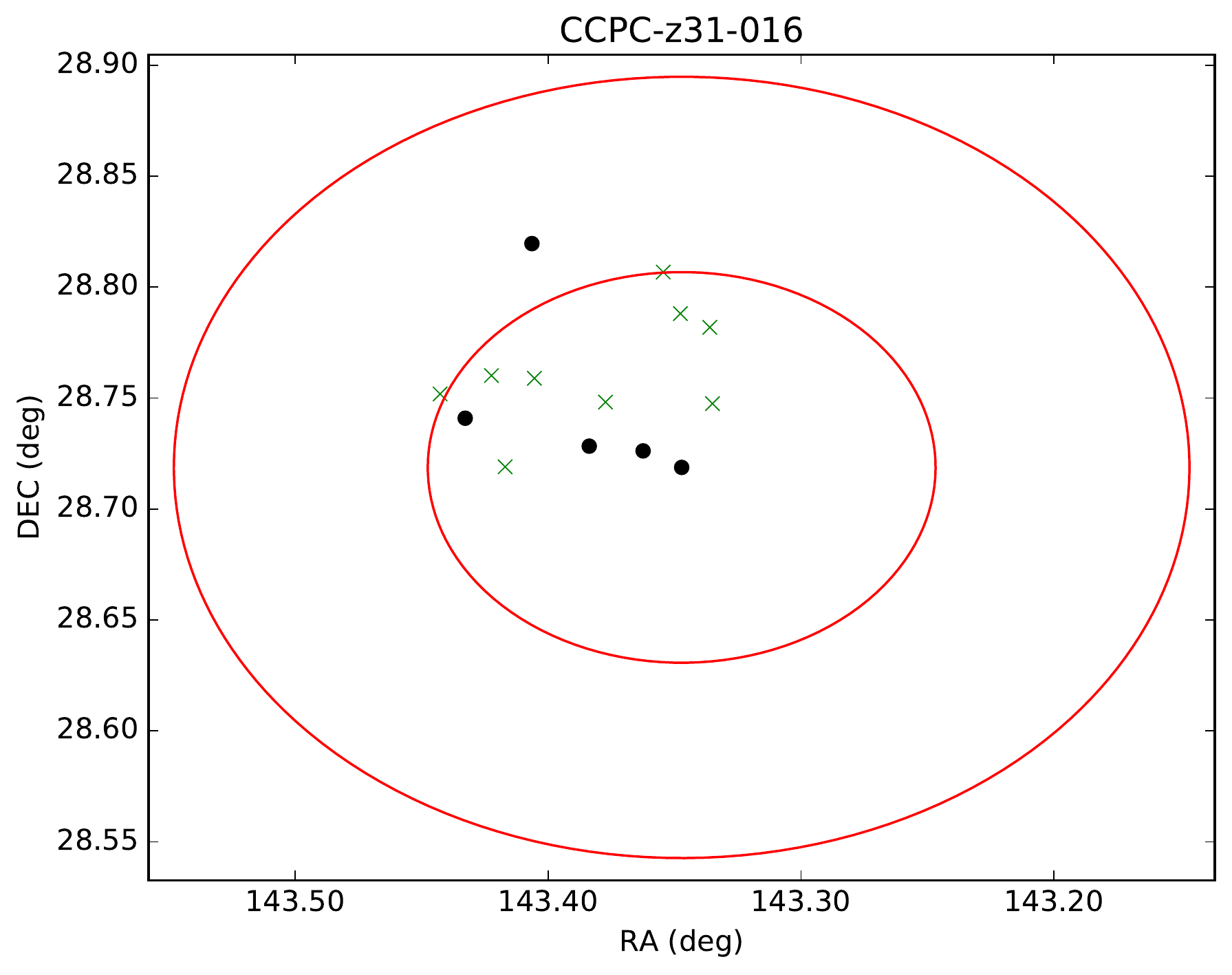}
\label{fig:CCPC-z31-016_sky}
\end{subfigure}
\hfill
\begin{subfigure}
\centering
\includegraphics[scale=0.52]{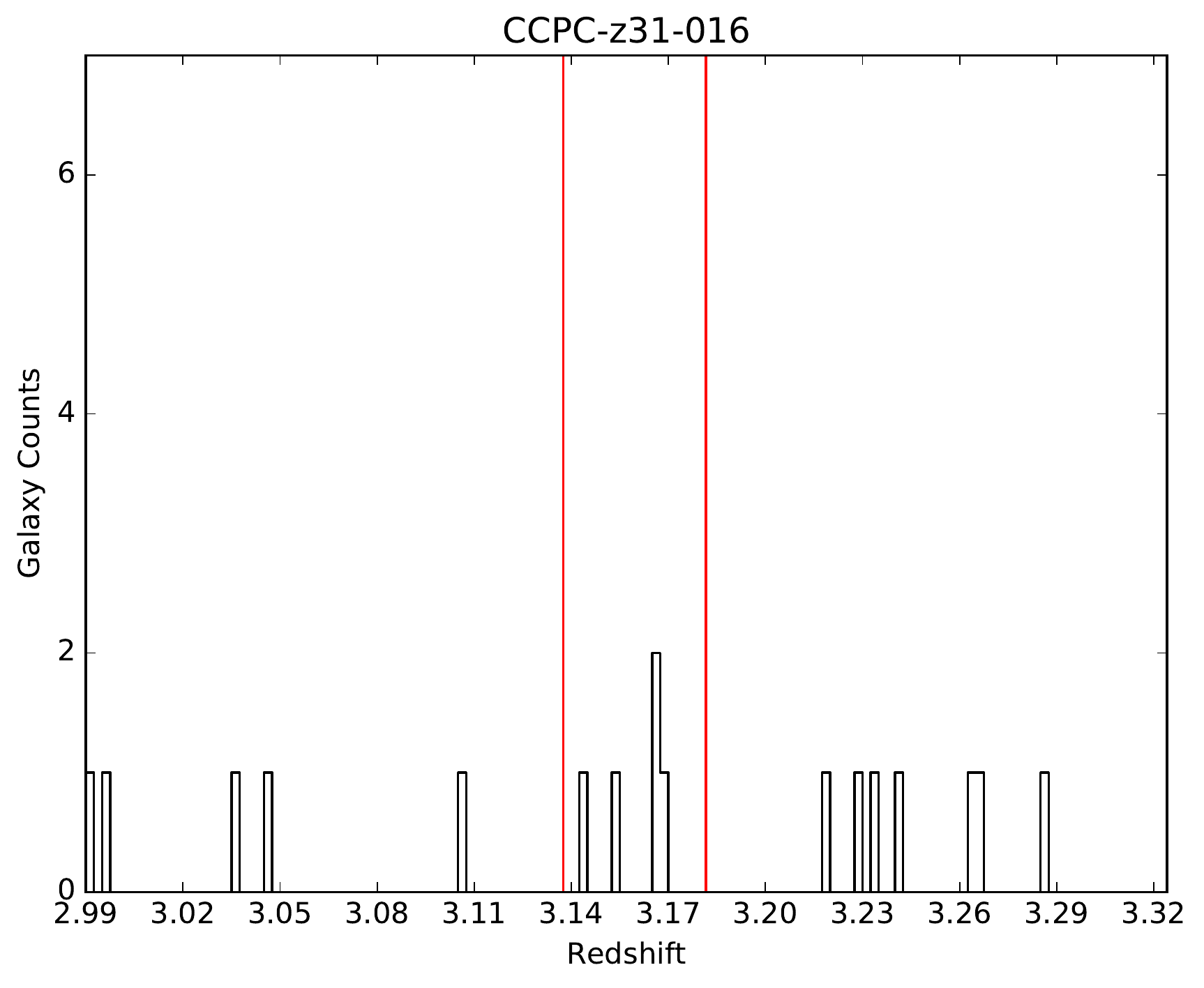}
\label{fig:CCPC-z31-016}
\end{subfigure}
\hfill
\end{figure*}

\begin{figure*}
\centering
\begin{subfigure}
\centering
\includegraphics[height=7.5cm,width=7.5cm]{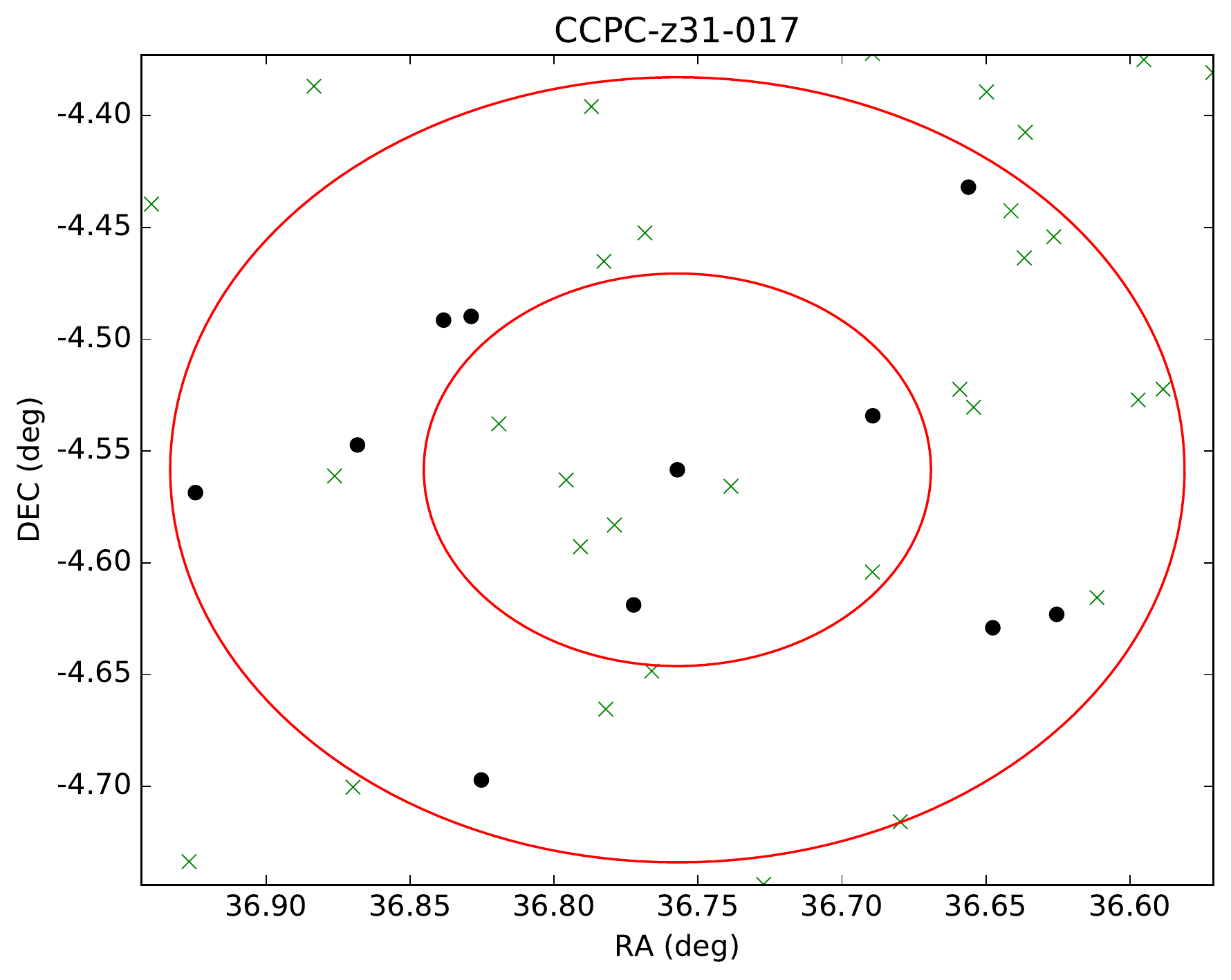}
\label{fig:CCPC-z31-017_sky}
\end{subfigure}
\hfill
\begin{subfigure}
\centering
\includegraphics[scale=0.52]{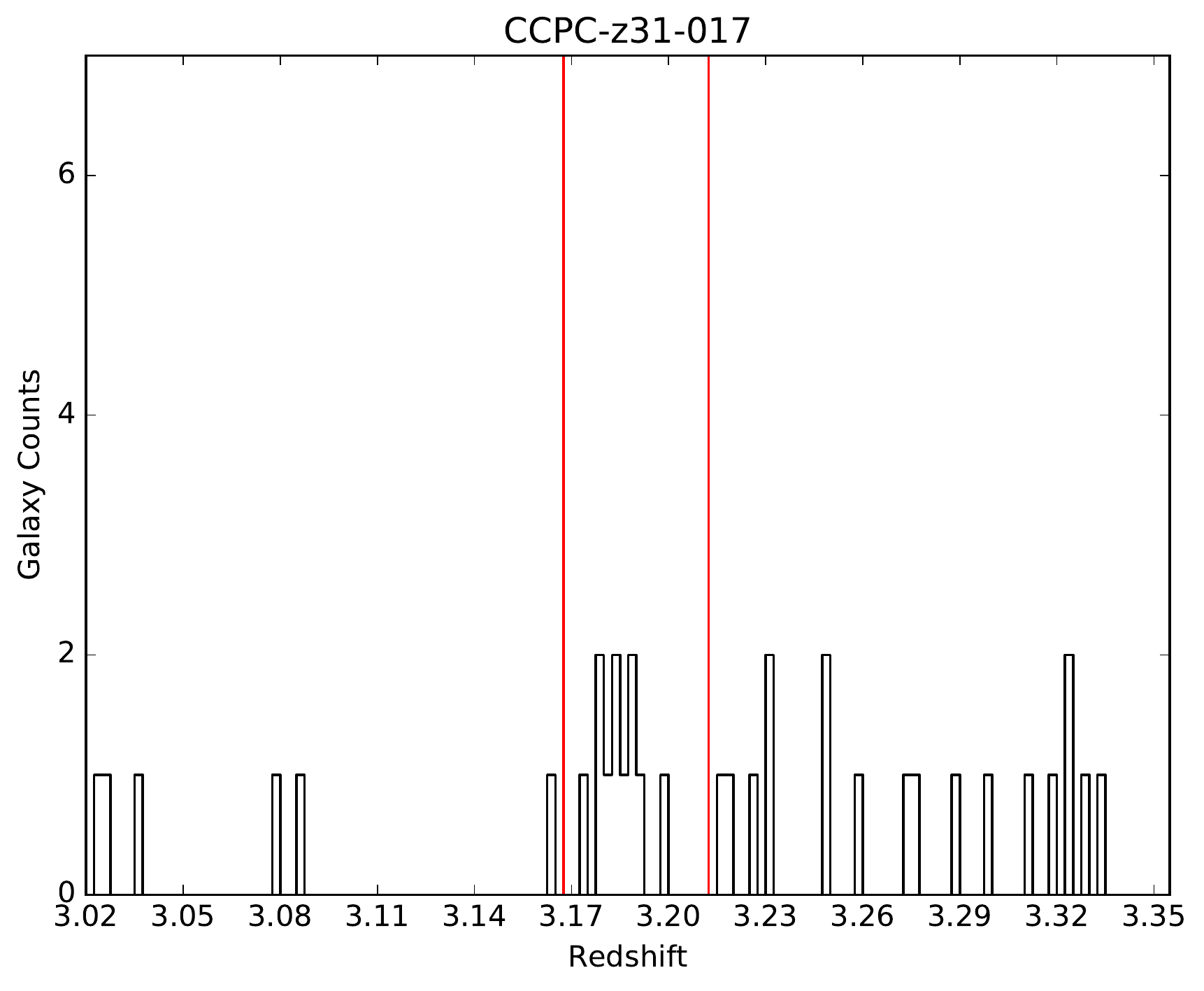}
\label{fig:CCPC-z31-017}
\end{subfigure}
\hfill
\end{figure*}
\clearpage 

\begin{figure*}
\centering
\begin{subfigure}
\centering
\includegraphics[height=7.5cm,width=7.5cm]{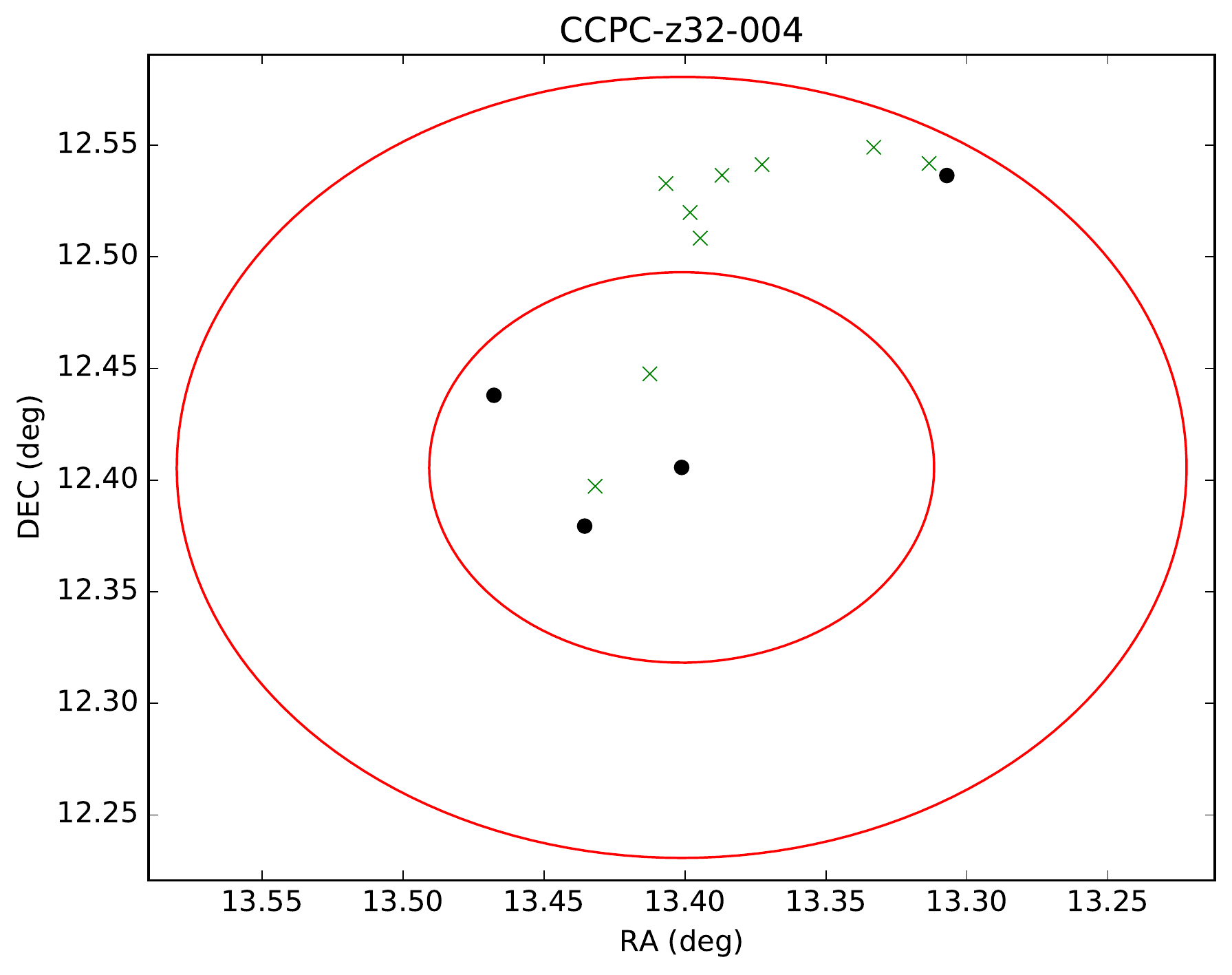}
\label{fig:CCPC-z32-004_sky}
\end{subfigure}
\hfill
\begin{subfigure}
\centering
\includegraphics[scale=0.52]{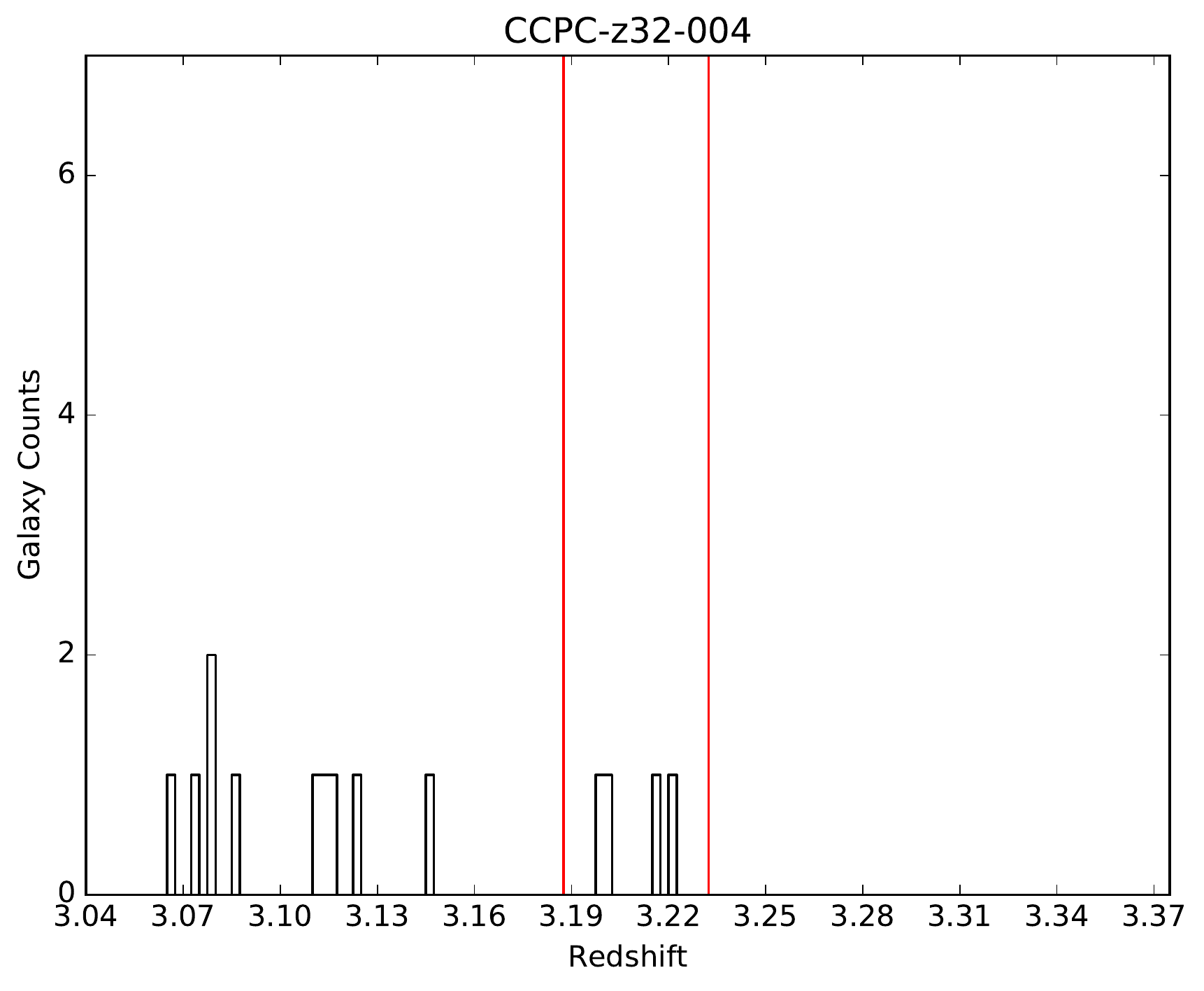}
\label{fig:CCPC-z32-004}
\end{subfigure}
\hfill
\end{figure*}

\begin{figure*}
\centering
\begin{subfigure}
\centering
\includegraphics[height=7.5cm,width=7.5cm]{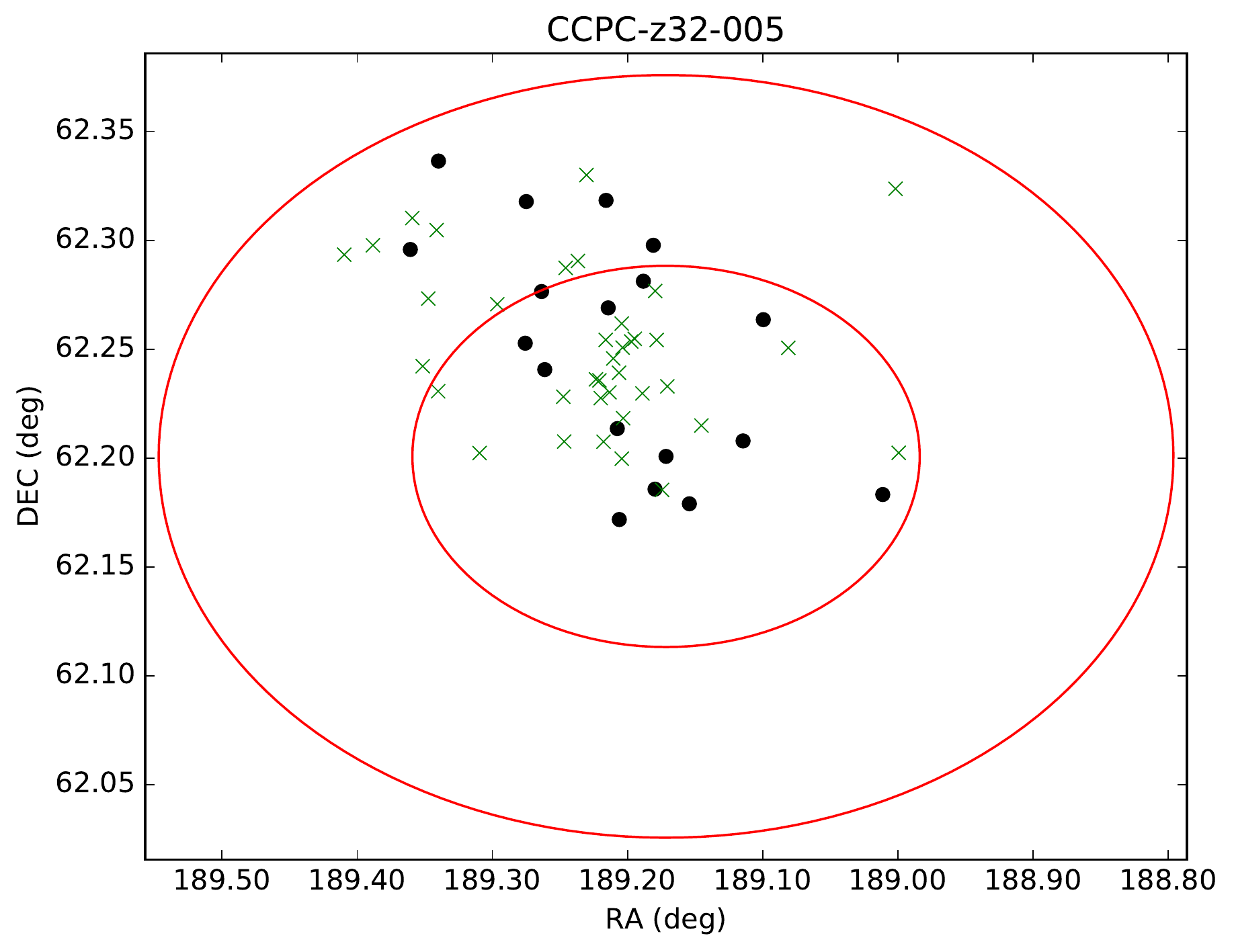}
\label{fig:CCPC-z32-005_sky}
\end{subfigure}
\hfill
\begin{subfigure}
\centering
\includegraphics[scale=0.52]{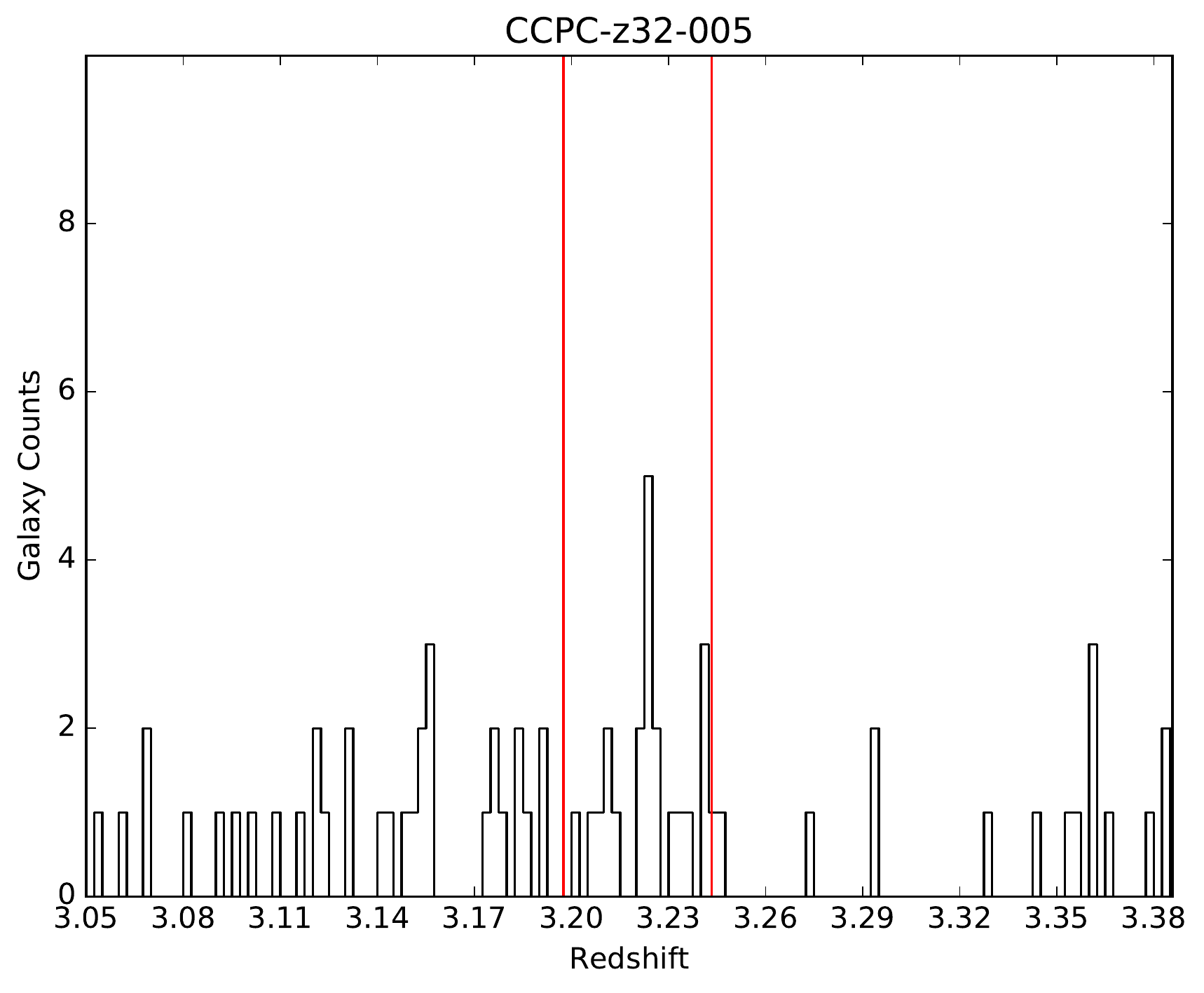}
\label{fig:CCPC-z32-005}
\end{subfigure}
\hfill
\end{figure*}

\begin{figure*}
\centering
\begin{subfigure}
\centering
\includegraphics[height=7.5cm,width=7.5cm]{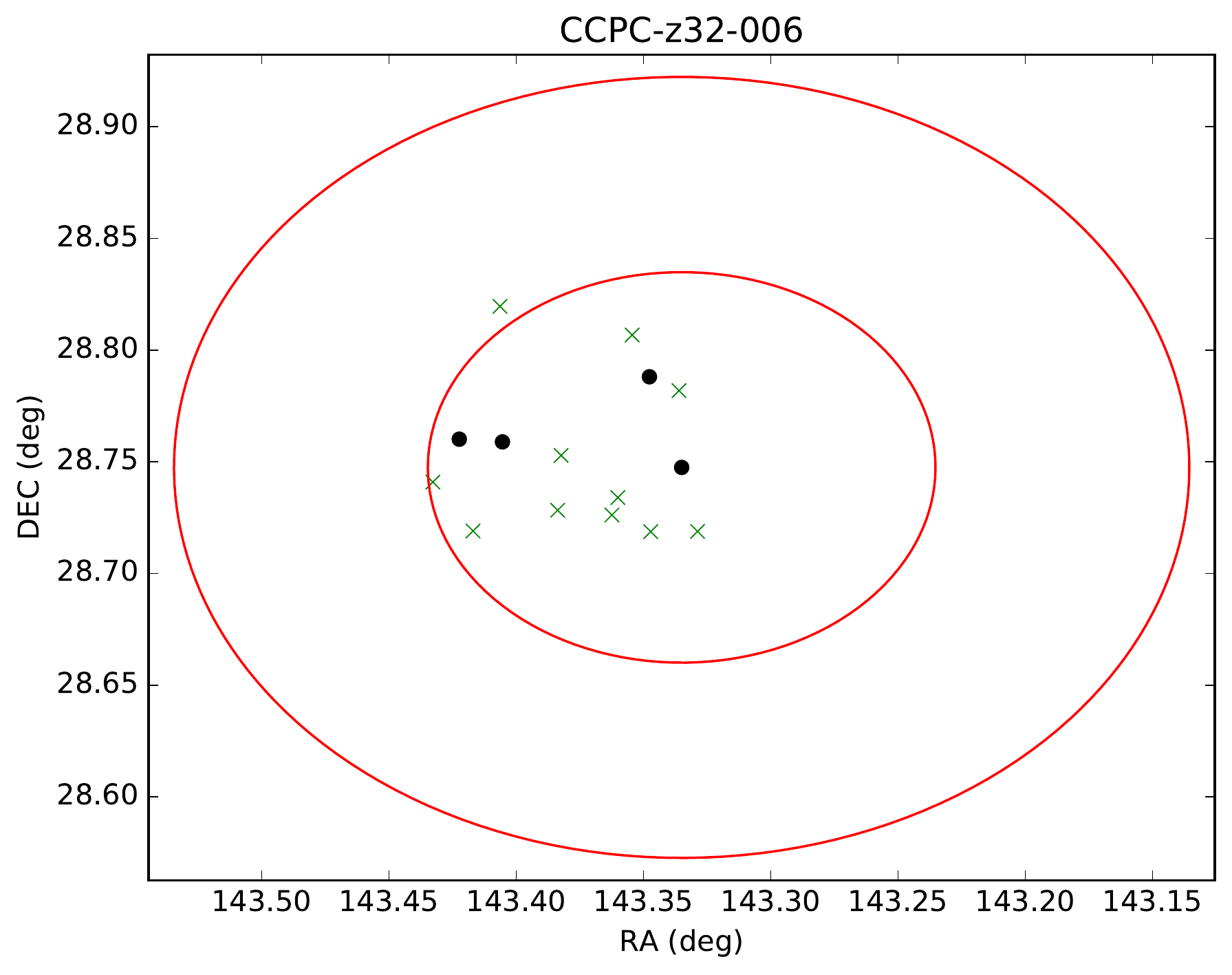}
\label{fig:CCPC-z32-006_sky}
\end{subfigure}
\hfill
\begin{subfigure}
\centering
\includegraphics[scale=0.52]{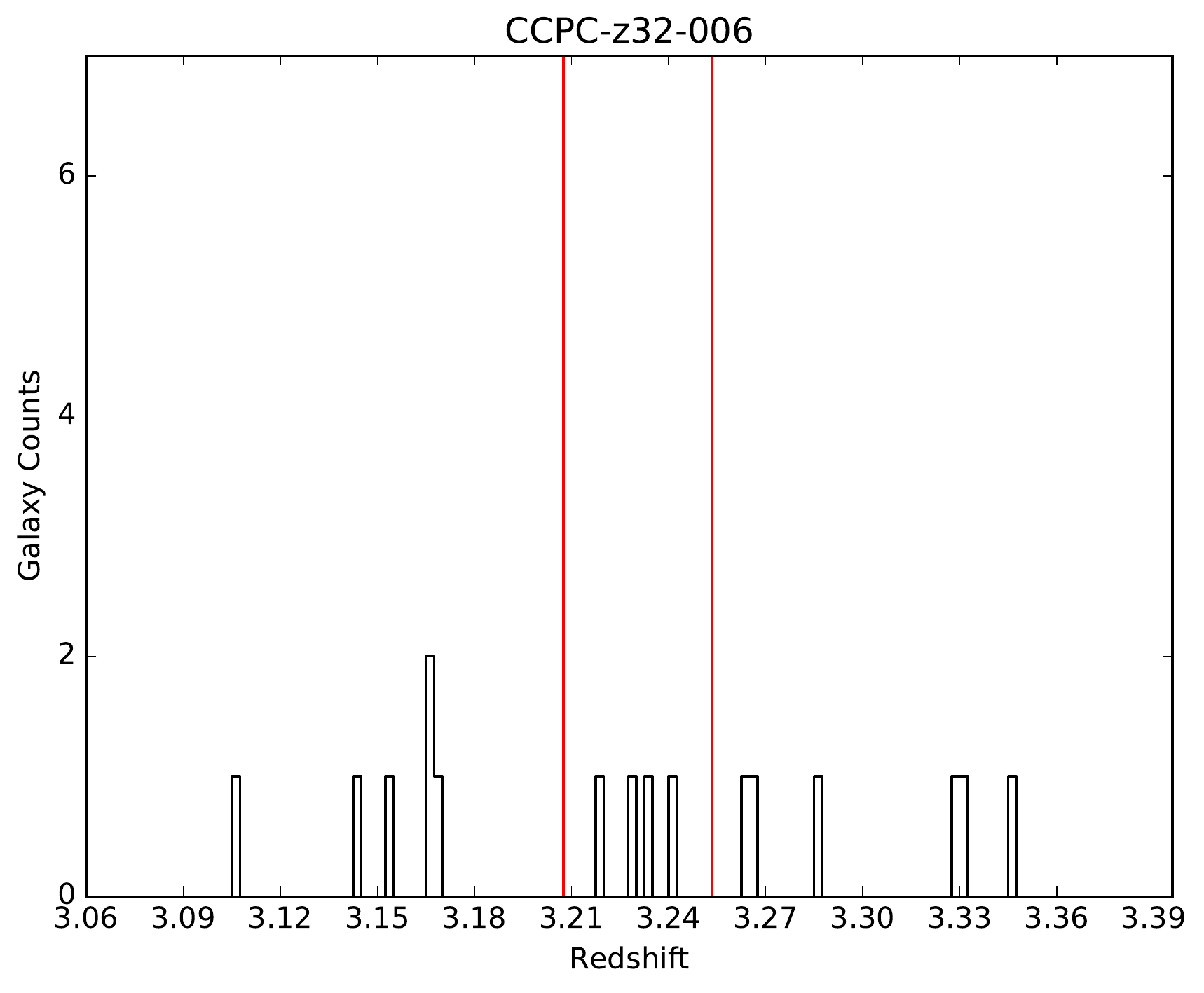}
\label{fig:CCPC-z32-006}
\end{subfigure}
\hfill
\end{figure*}
\clearpage 

\begin{figure*}
\centering
\begin{subfigure}
\centering
\includegraphics[height=7.5cm,width=7.5cm]{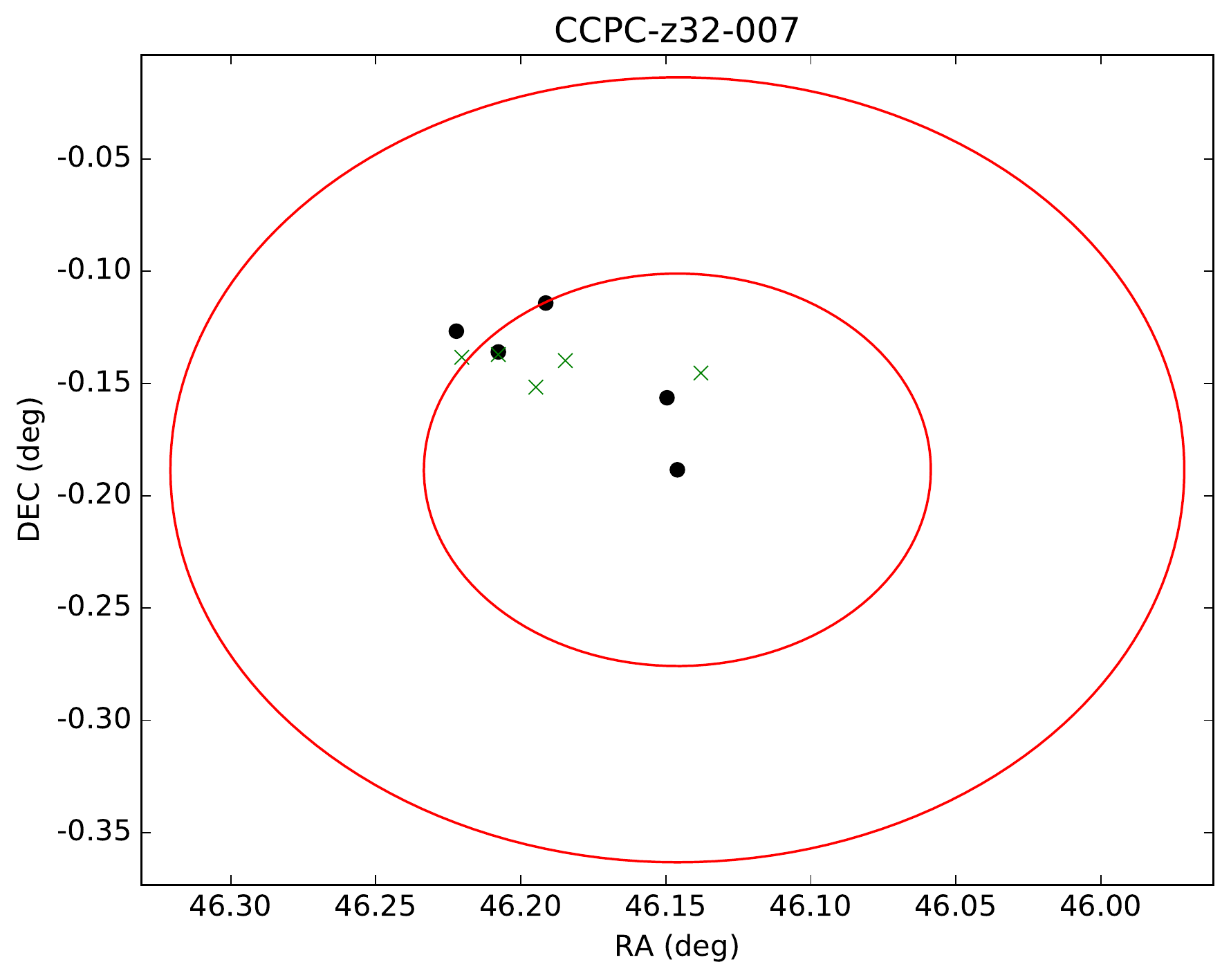}
\label{fig:CCPC-z32-007_sky}
\end{subfigure}
\hfill
\begin{subfigure}
\centering
\includegraphics[scale=0.52]{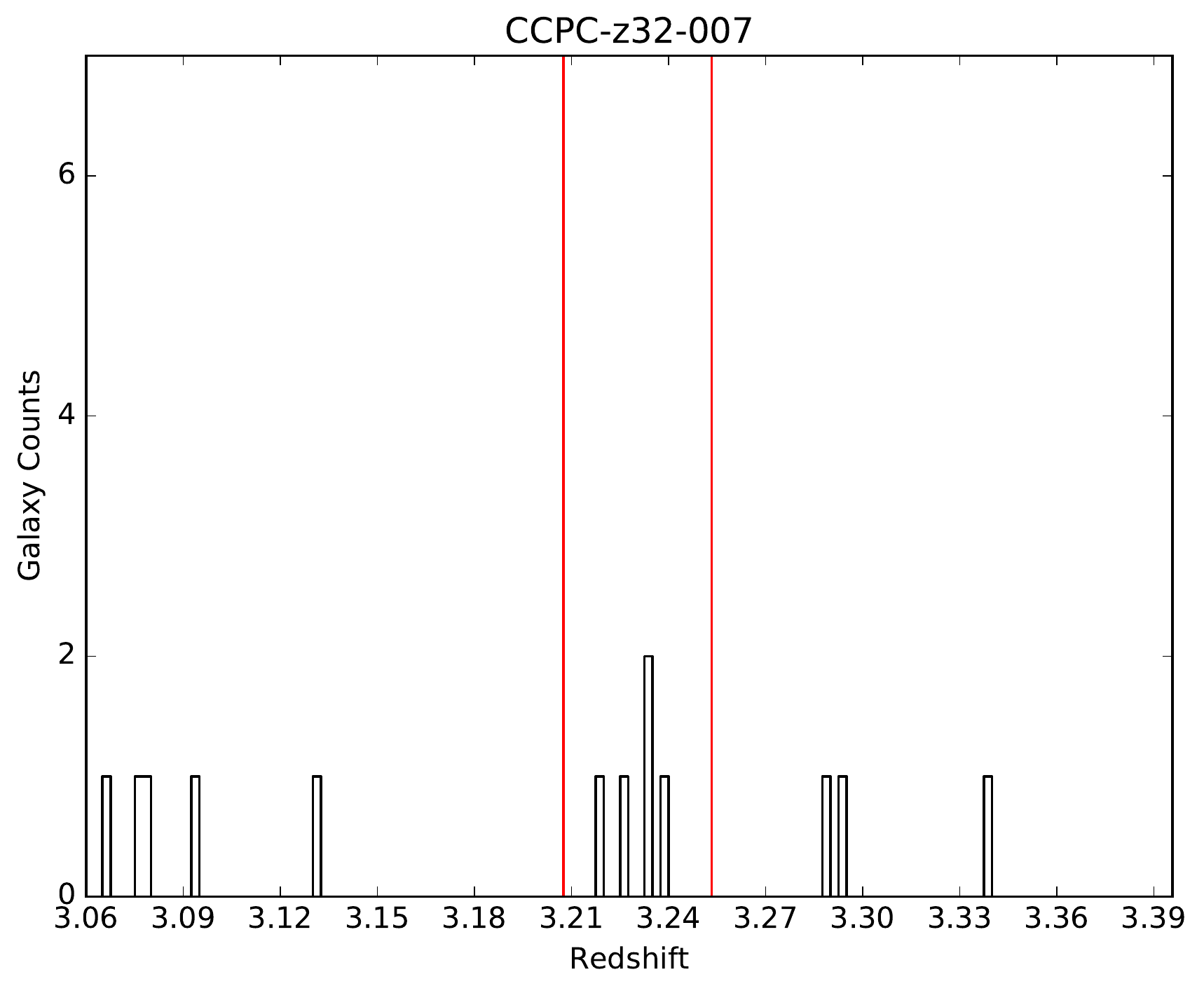}
\label{fig:CCPC-z32-007}
\end{subfigure}
\hfill
\end{figure*}

\begin{figure*}
\centering
\begin{subfigure}
\centering
\includegraphics[height=7.5cm,width=7.5cm]{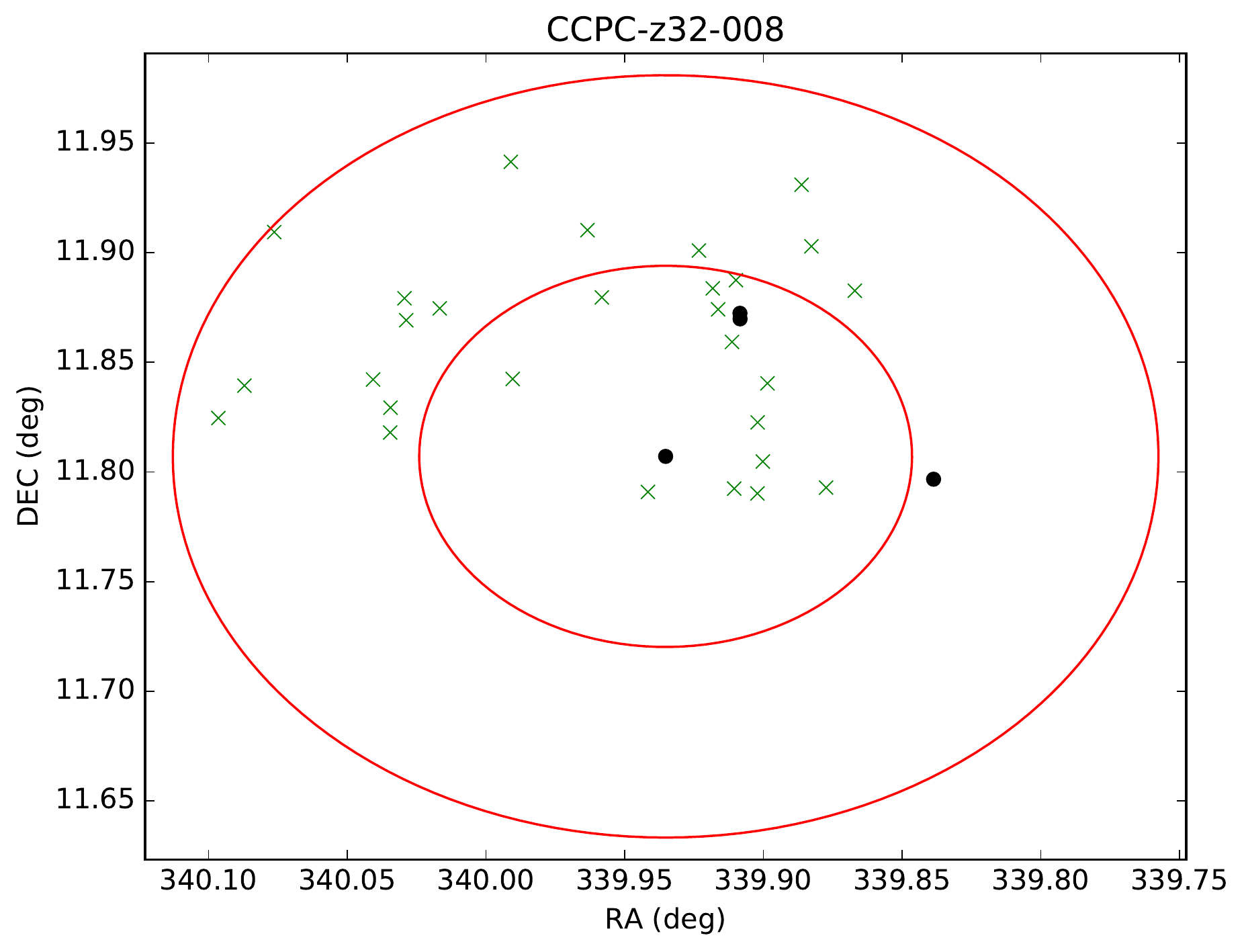}
\label{fig:CCPC-z32-008_sky}
\end{subfigure}
\hfill
\begin{subfigure}
\centering
\includegraphics[scale=0.52]{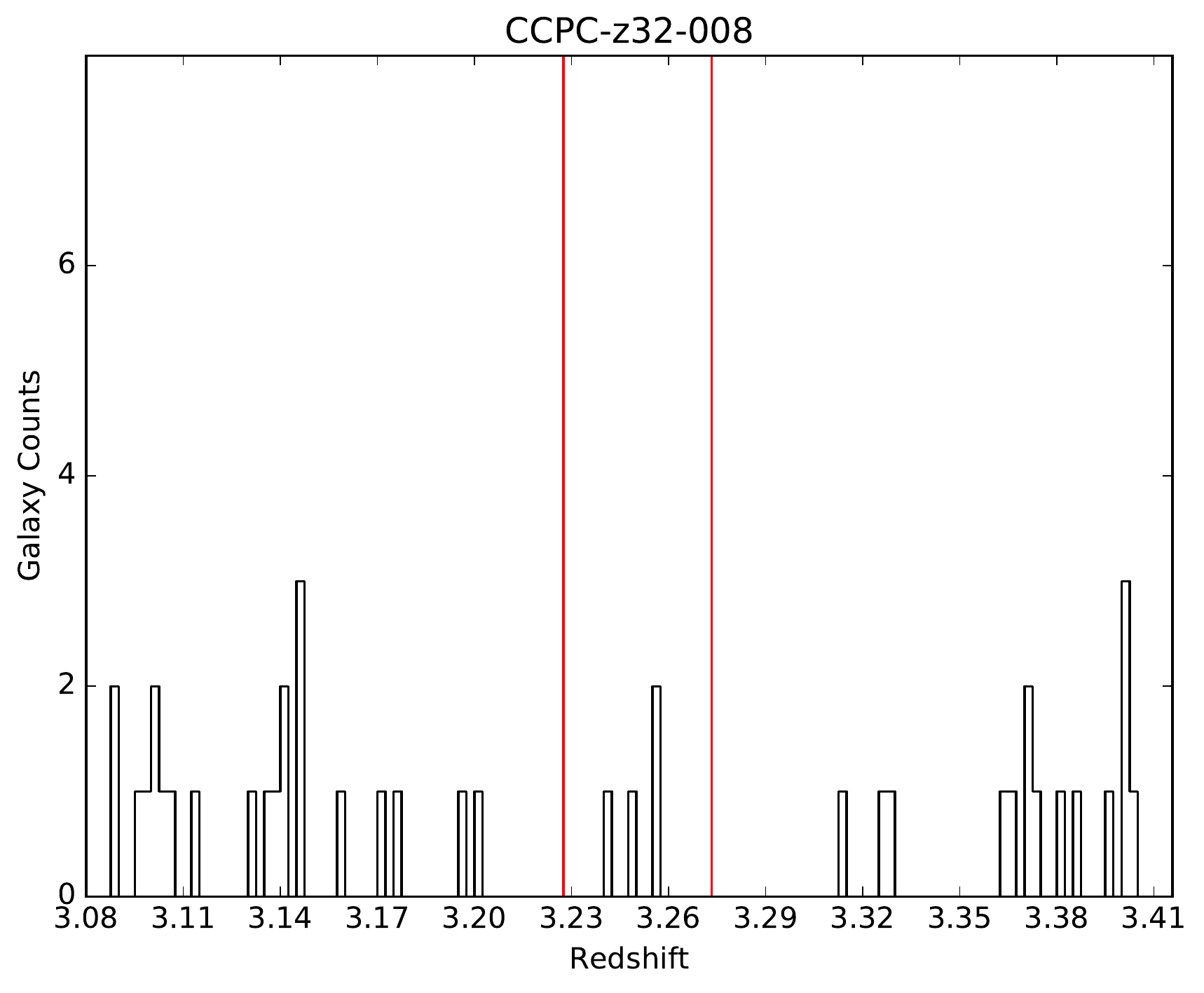}
\label{fig:CCPC-z32-008}
\end{subfigure}
\hfill
\end{figure*}

\begin{figure*}
\centering
\begin{subfigure}
\centering
\includegraphics[height=7.5cm,width=7.5cm]{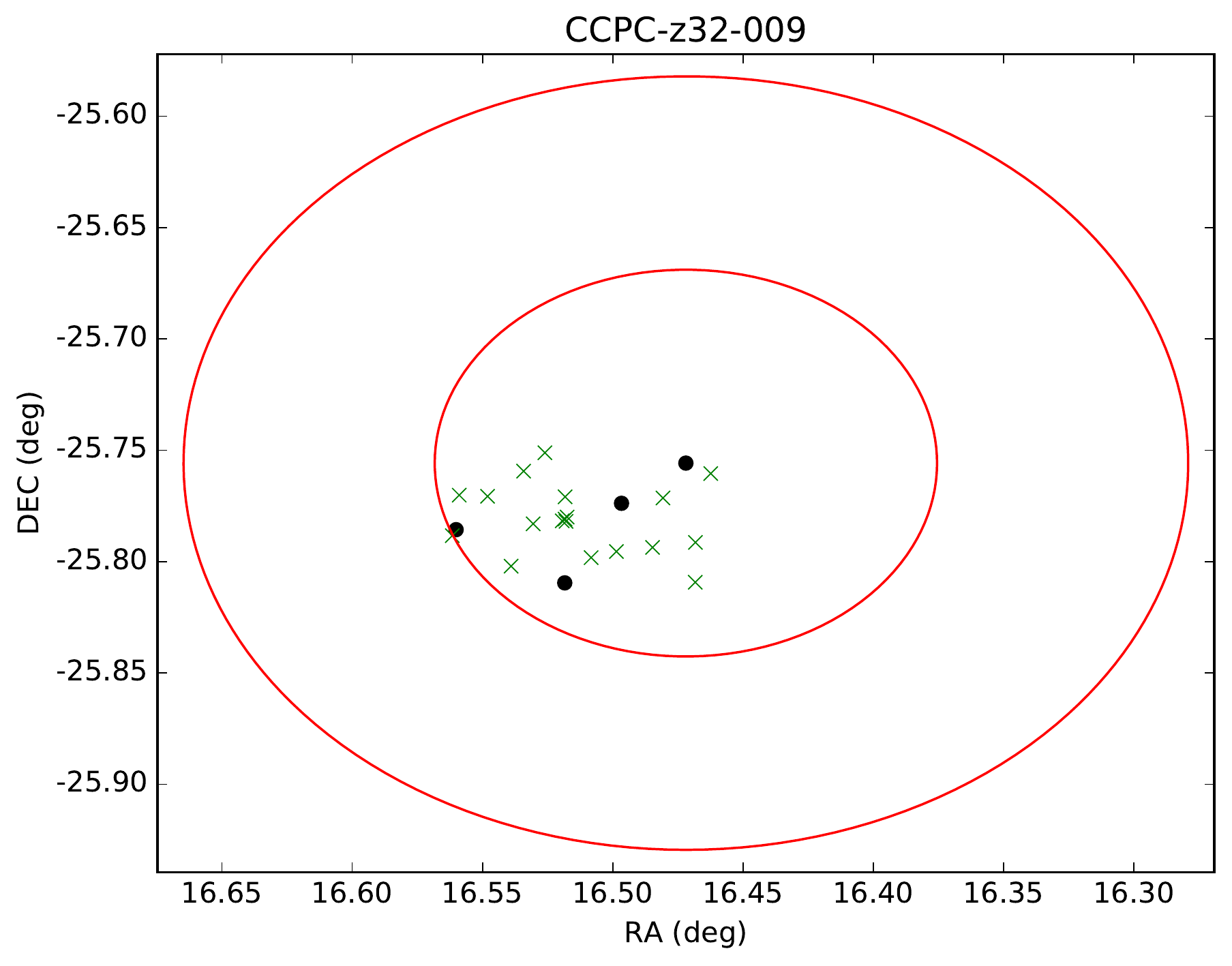}
\label{fig:CCPC-z32-009_sky}
\end{subfigure}
\hfill
\begin{subfigure}
\centering
\includegraphics[scale=0.52]{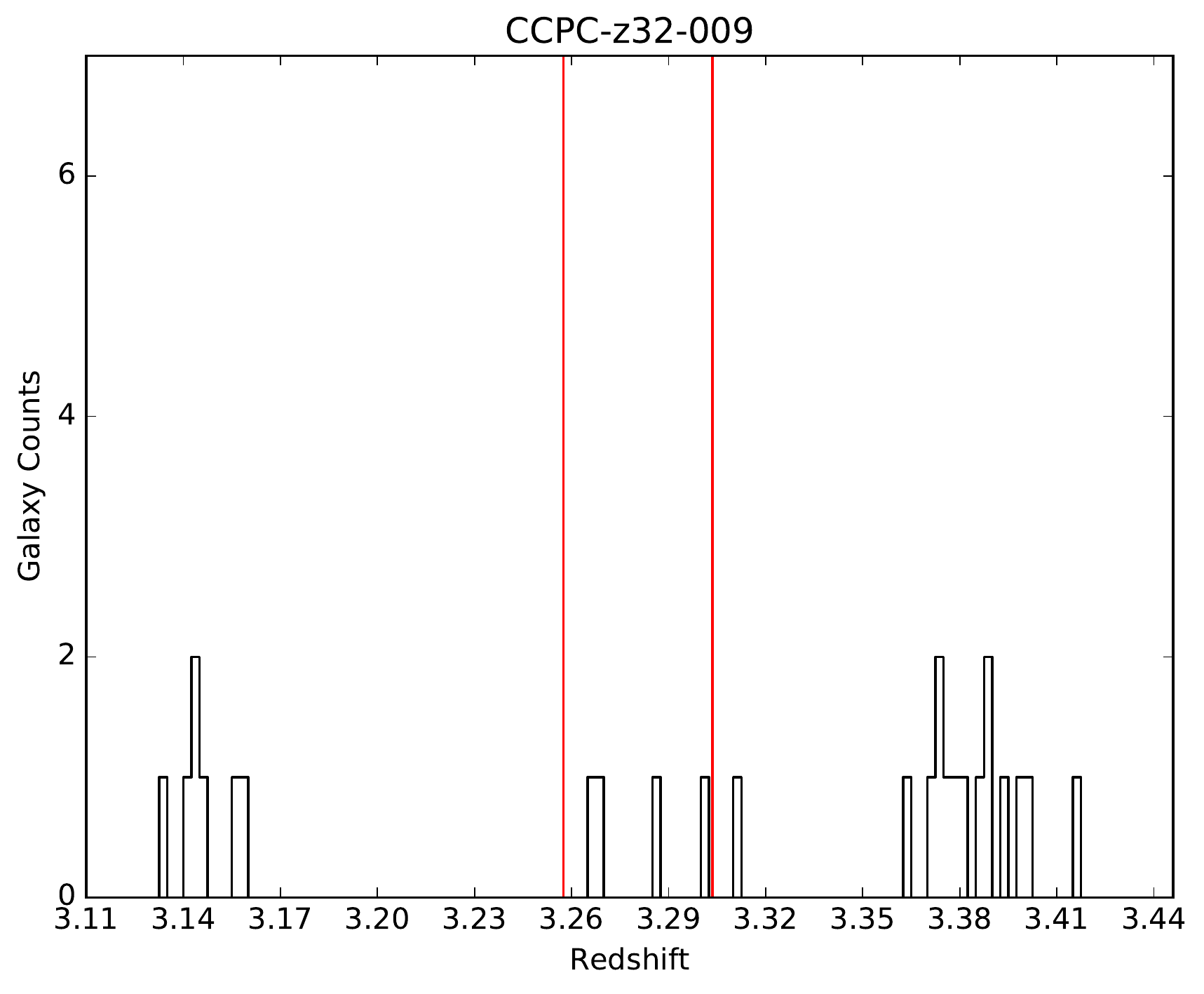}
\label{fig:CCPC-z32-009}
\end{subfigure}
\hfill
\end{figure*}
\clearpage 

\begin{figure*}
\centering
\begin{subfigure}
\centering
\includegraphics[height=7.5cm,width=7.5cm]{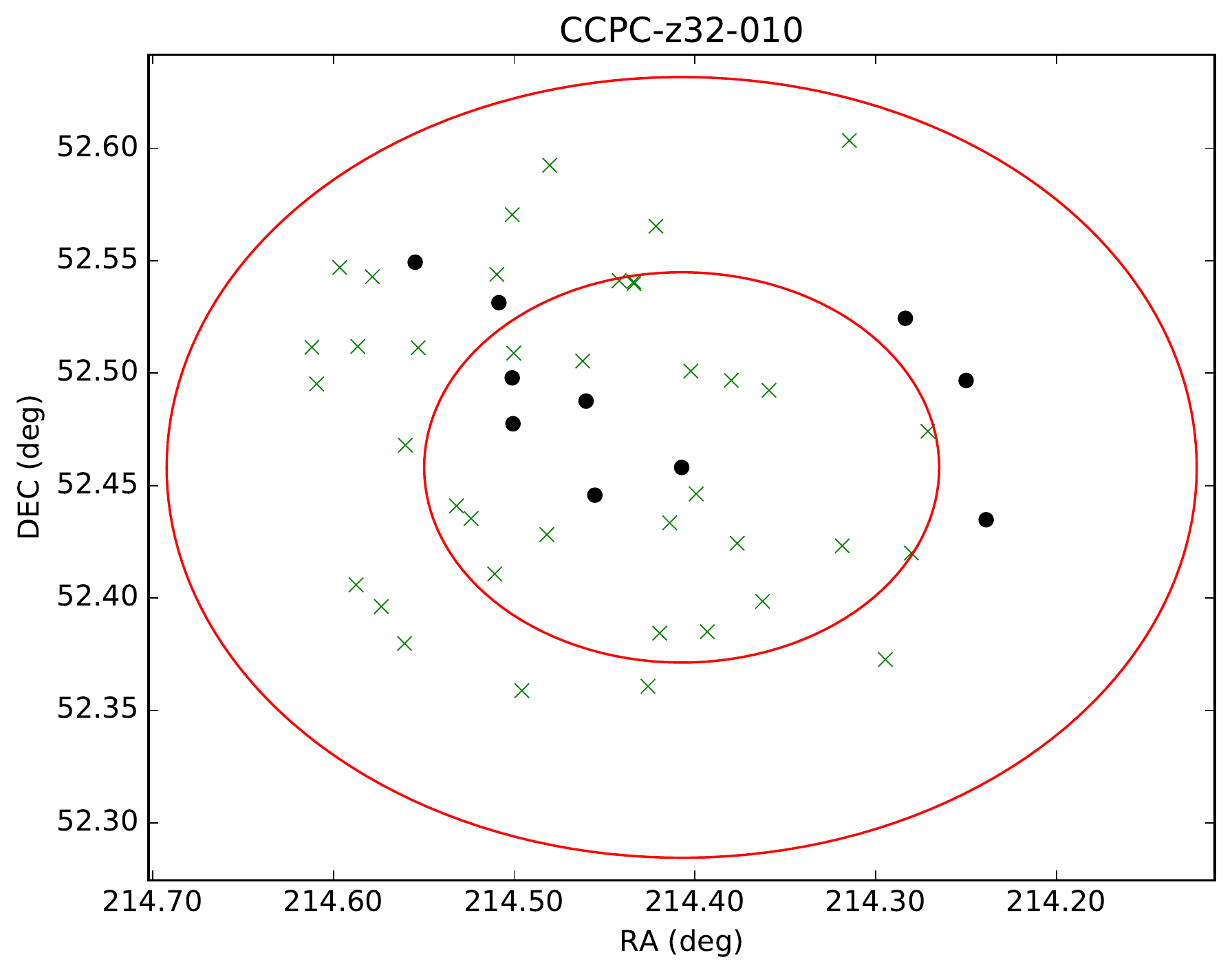}
\label{fig:CCPC-z32-010_sky}
\end{subfigure}
\hfill
\begin{subfigure}
\centering
\includegraphics[scale=0.52]{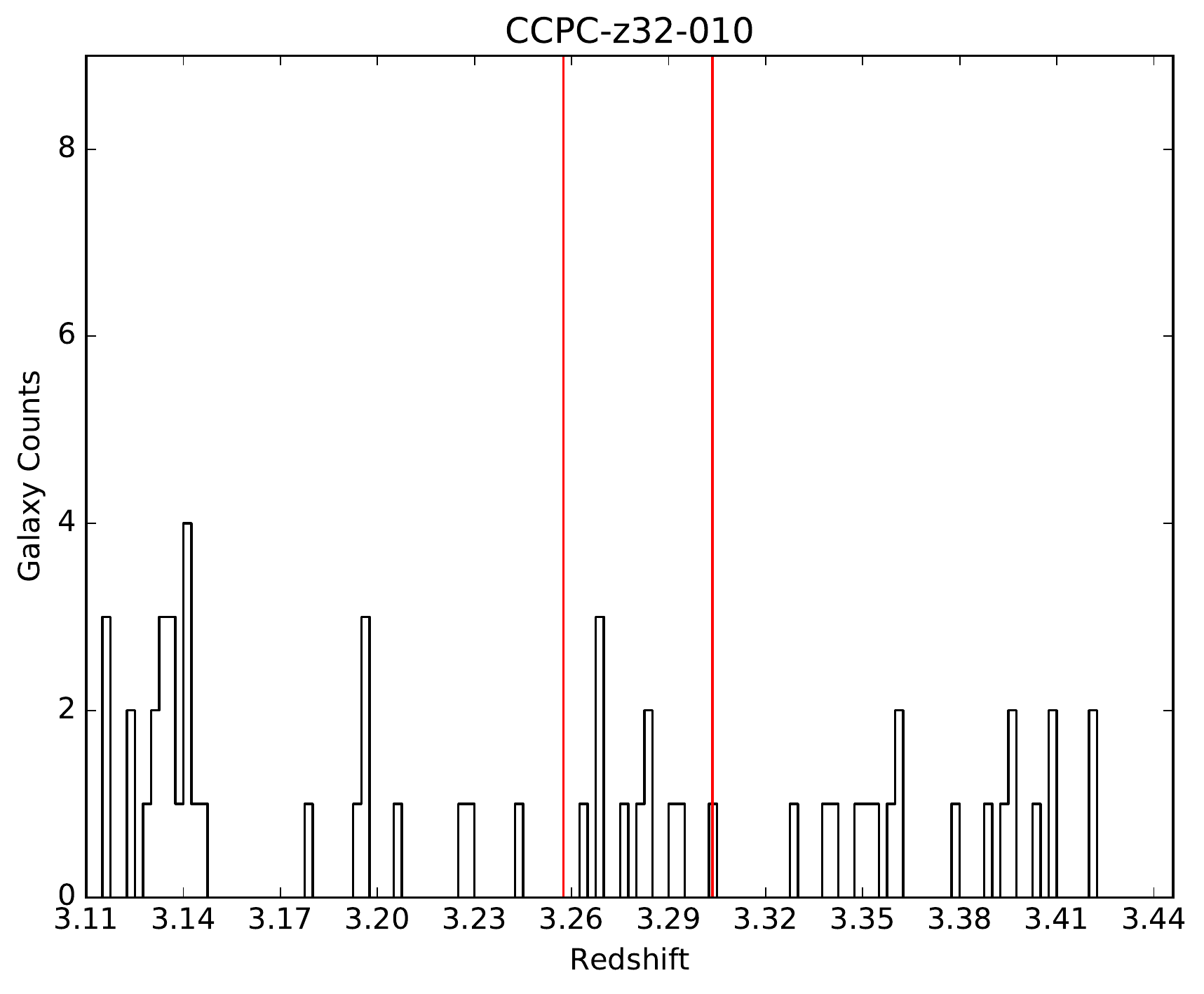}
\label{fig:CCPC-z32-010}
\end{subfigure}
\hfill
\end{figure*}

\begin{figure*}
\centering
\begin{subfigure}
\centering
\includegraphics[height=7.5cm,width=7.5cm]{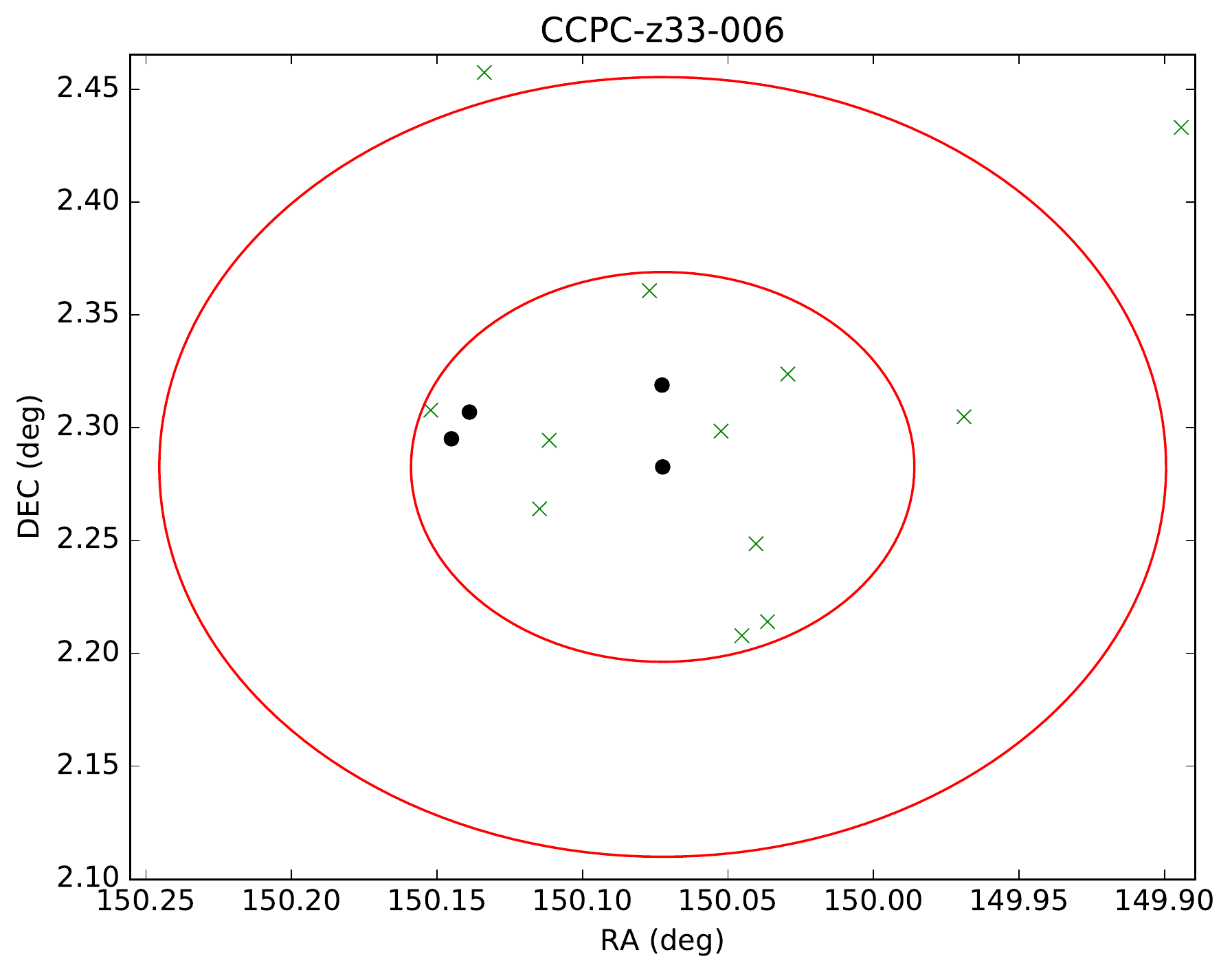}
\label{fig:CCPC-z33-006_sky}
\end{subfigure}
\hfill
\begin{subfigure}
\centering
\includegraphics[scale=0.52]{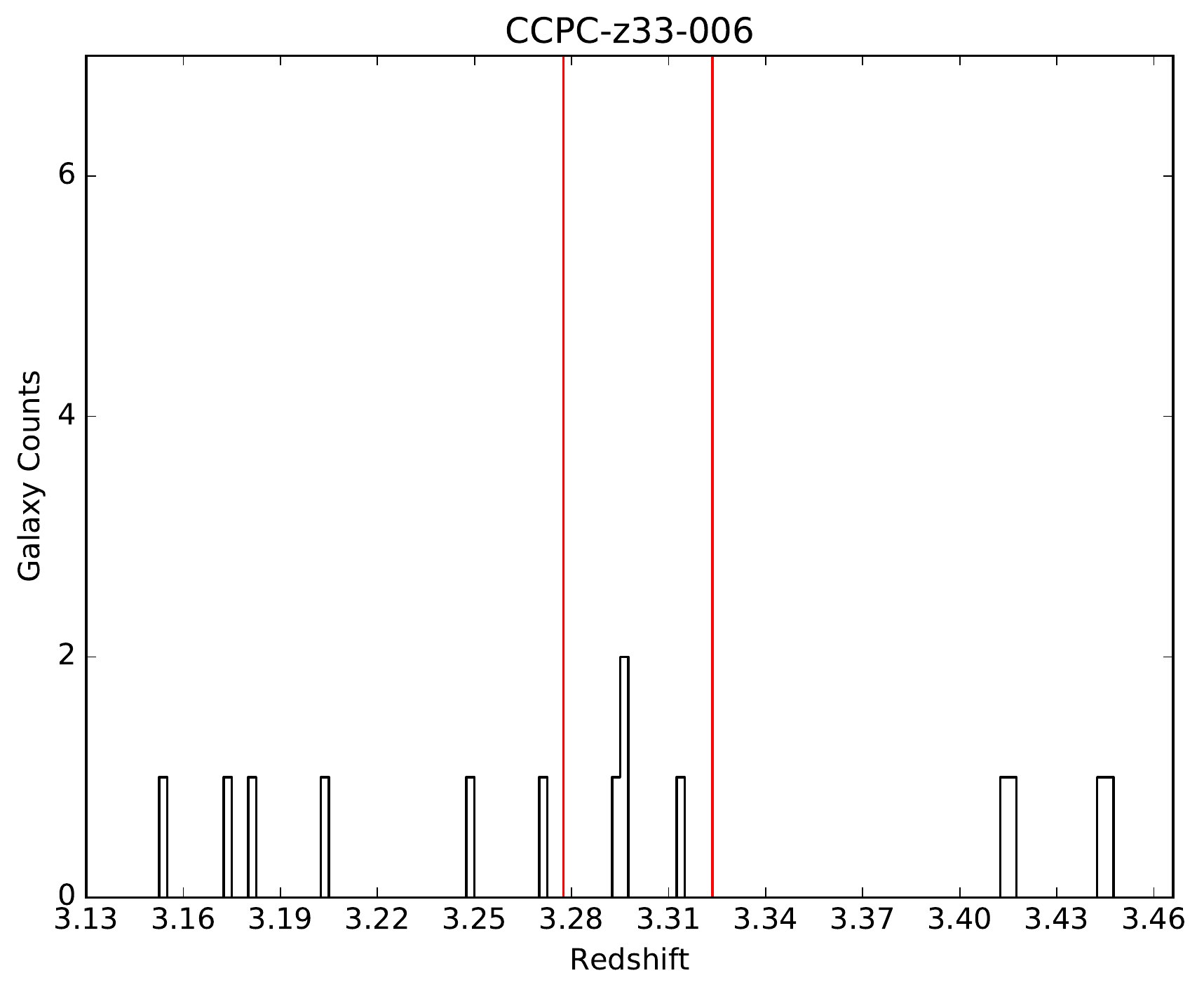}
\label{fig:CCPC-z33-006}
\end{subfigure}
\hfill
\end{figure*}

\begin{figure*}
\centering
\begin{subfigure}
\centering
\includegraphics[height=7.5cm,width=7.5cm]{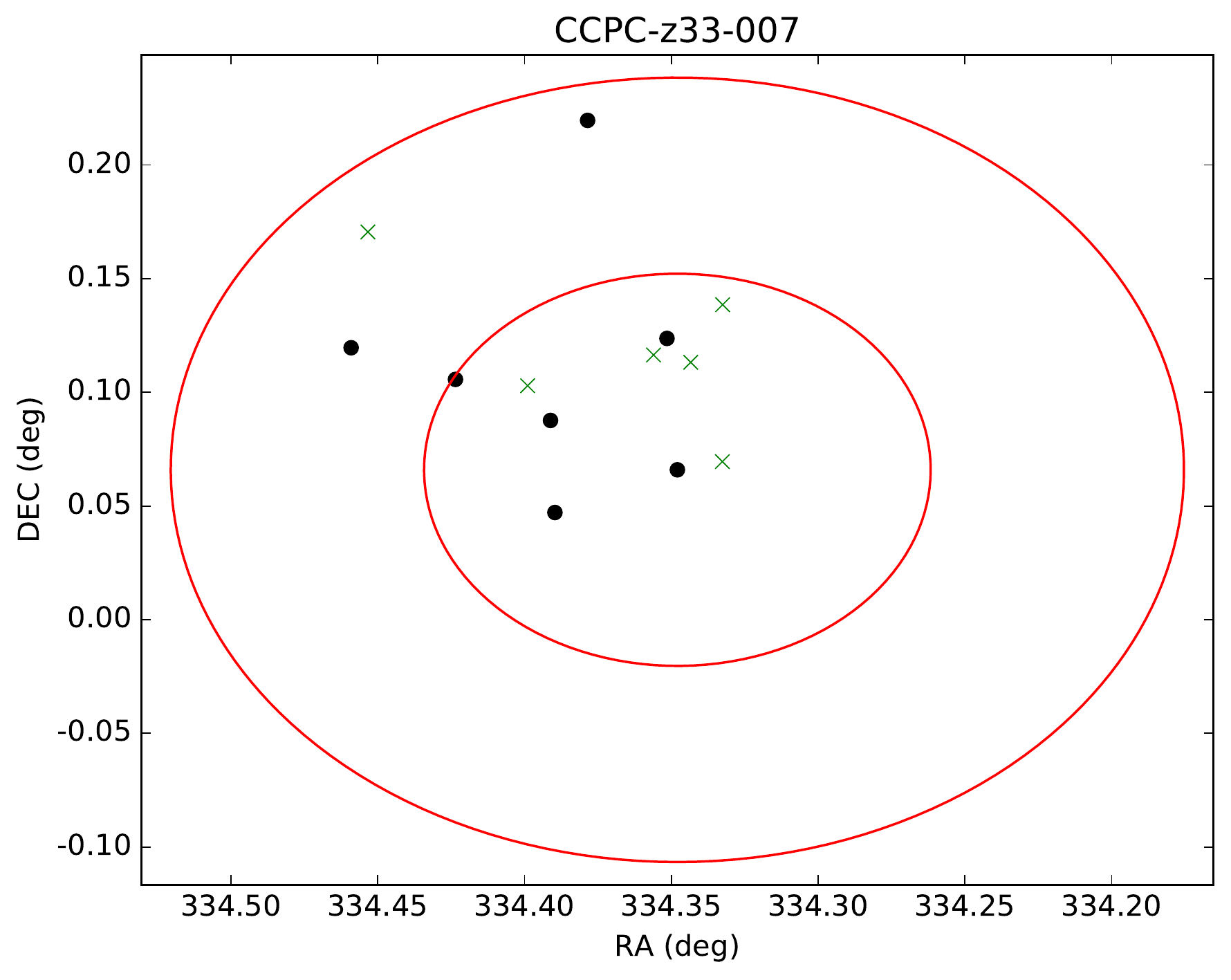}
\label{fig:CCPC-z33-007_sky}
\end{subfigure}
\hfill
\begin{subfigure}
\centering
\includegraphics[scale=0.52]{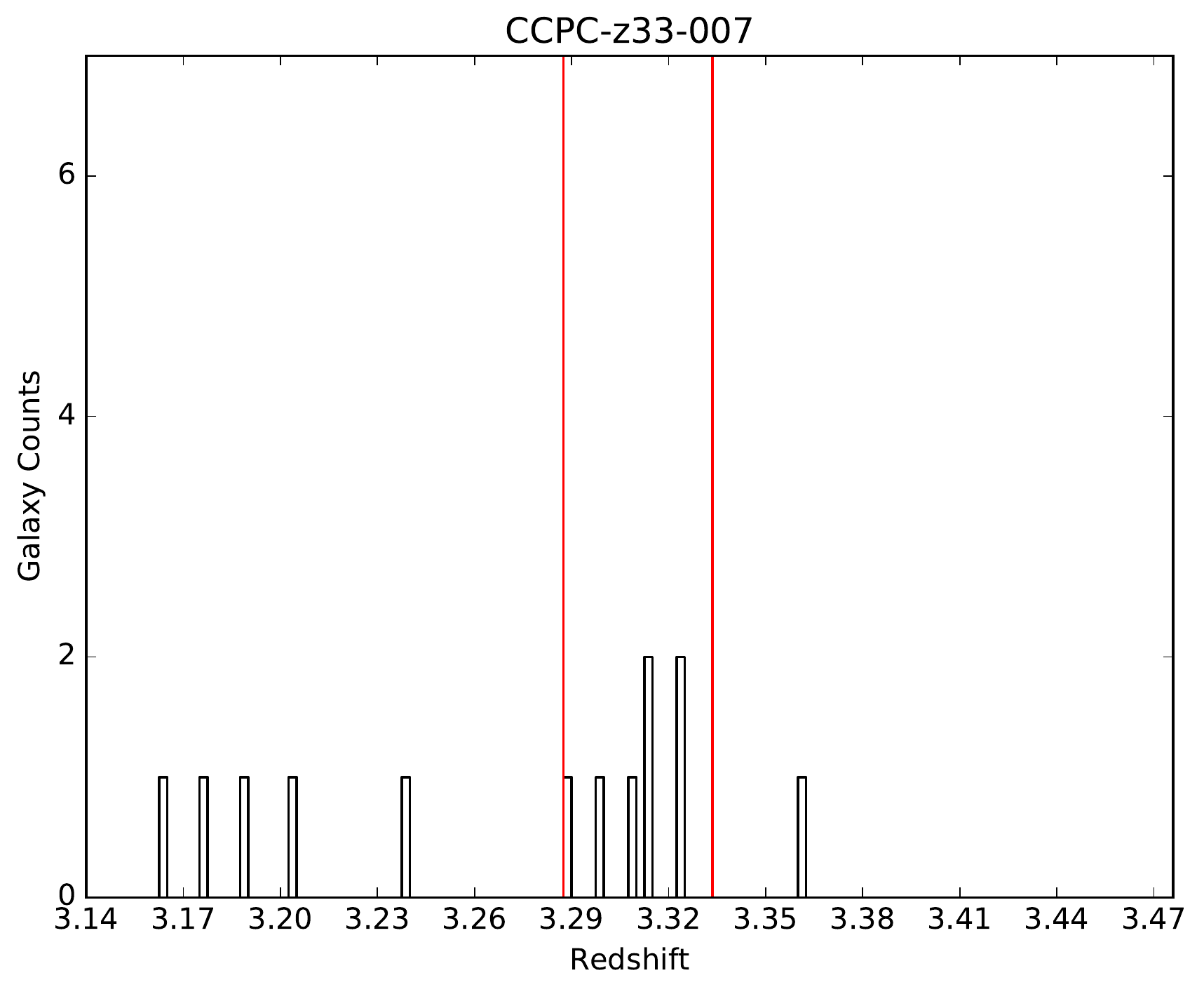}
\label{fig:CCPC-z33-007}
\end{subfigure}
\hfill
\end{figure*}
\clearpage 

\begin{figure*}
\centering
\begin{subfigure}
\centering
\includegraphics[height=7.5cm,width=7.5cm]{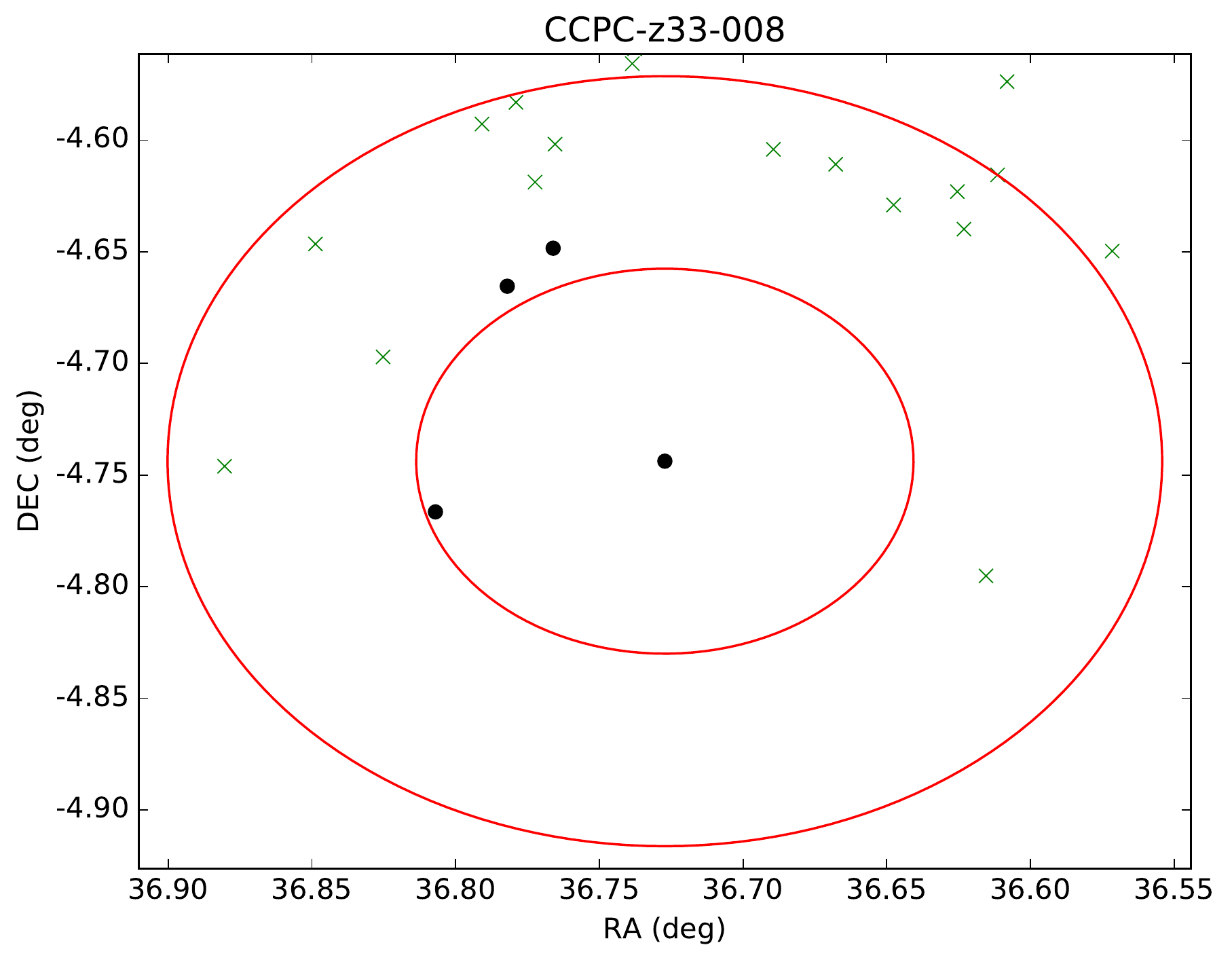}
\label{fig:CCPC-z33-008_sky}
\end{subfigure}
\hfill
\begin{subfigure}
\centering
\includegraphics[scale=0.52]{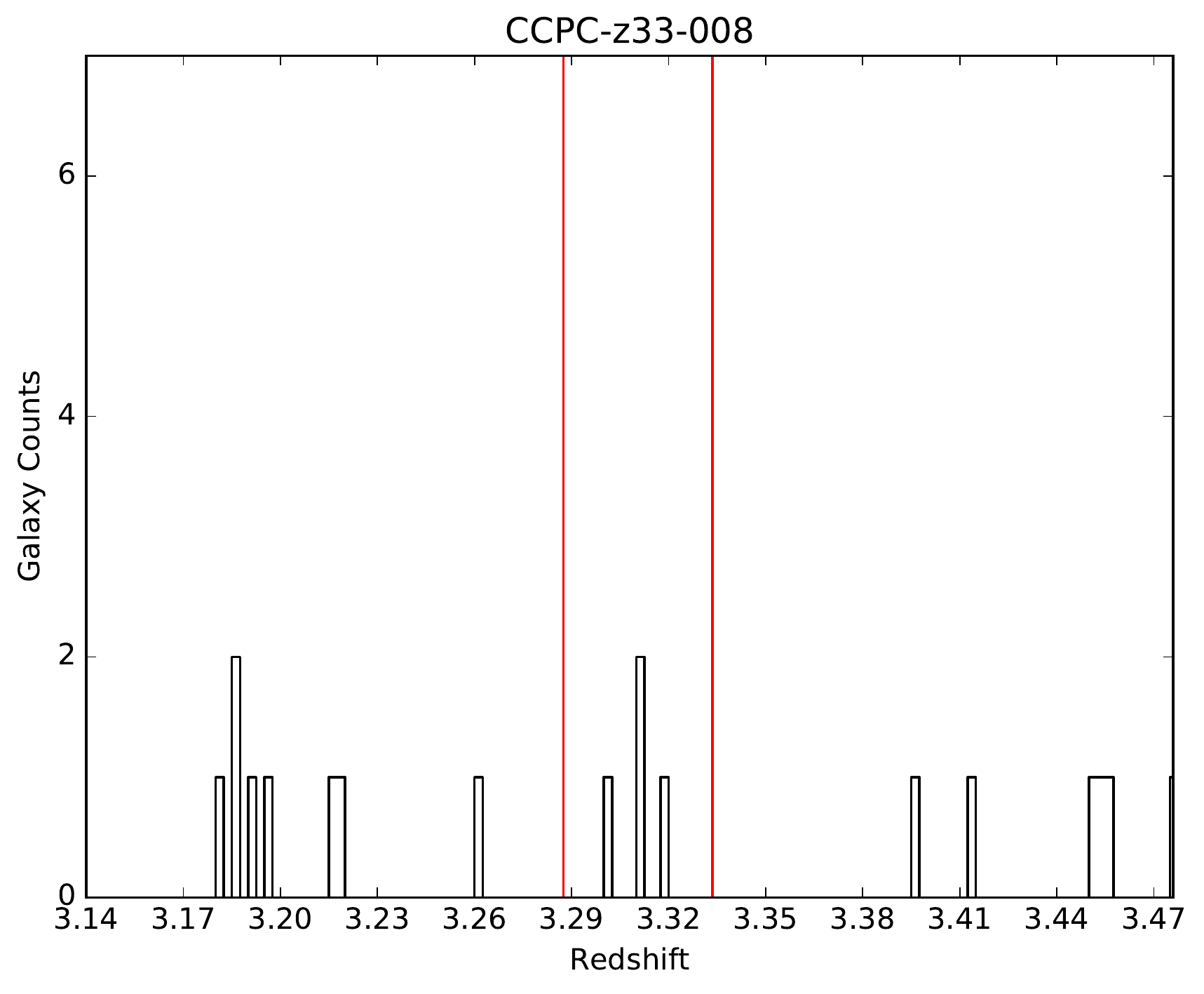}
\label{fig:CCPC-z33-008}
\end{subfigure}
\hfill
\end{figure*}

\begin{figure*}
\centering
\begin{subfigure}
\centering
\includegraphics[height=7.5cm,width=7.5cm]{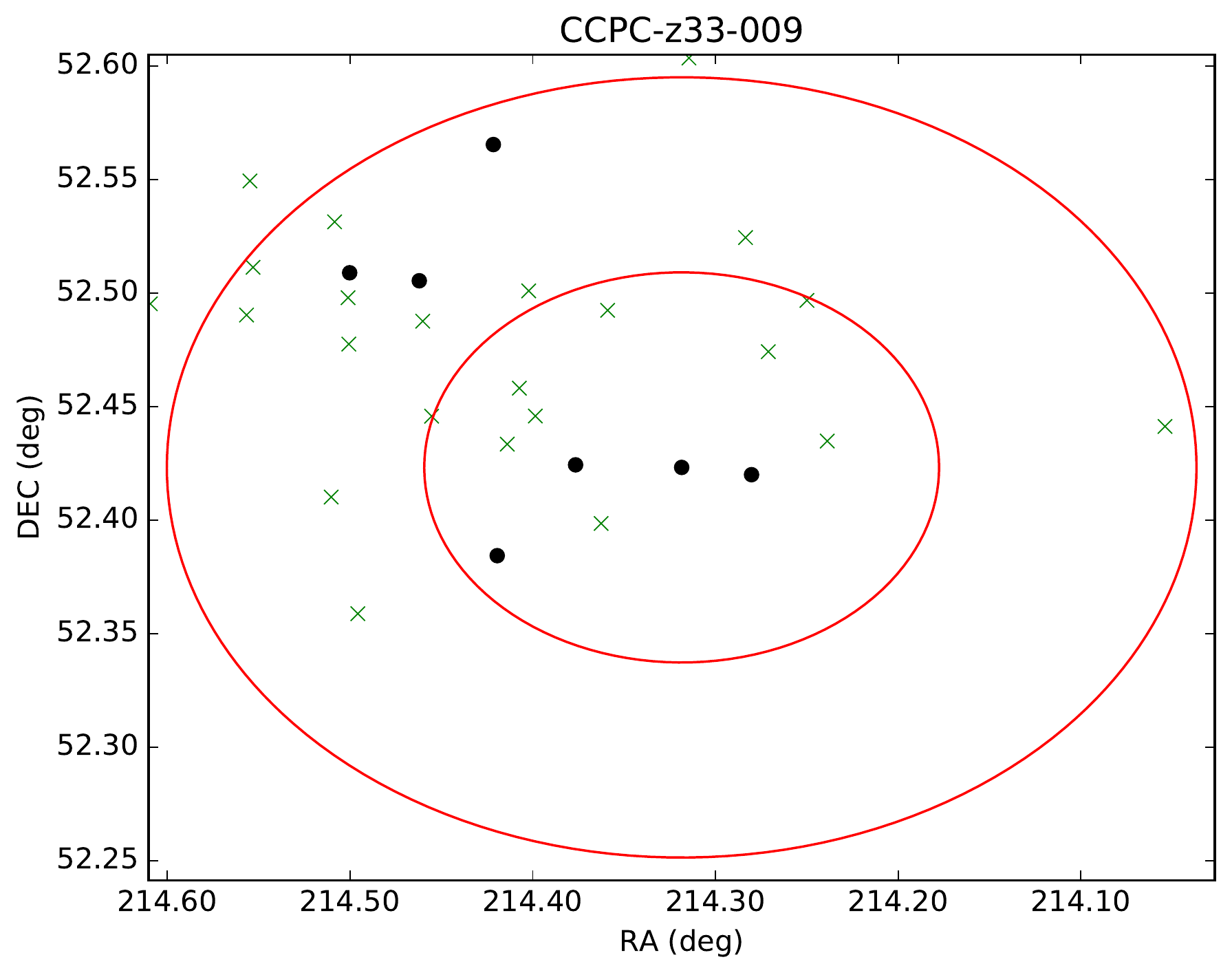}
\label{fig:CCPC-z33-009_sky}
\end{subfigure}
\hfill
\begin{subfigure}
\centering
\includegraphics[scale=0.52]{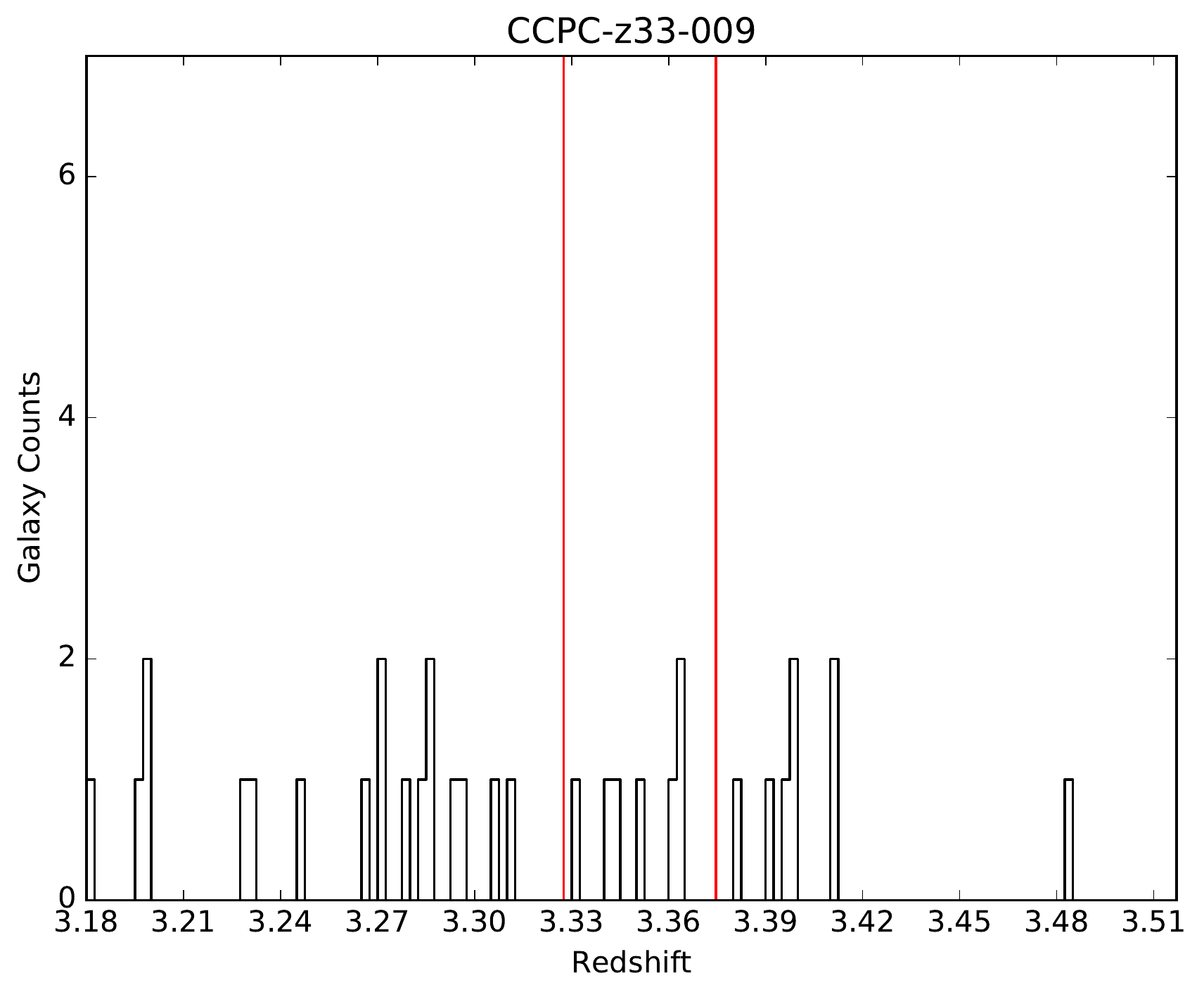}
\label{fig:CCPC-z33-009}
\end{subfigure}
\hfill
\end{figure*}

\begin{figure*}
\centering
\begin{subfigure}
\centering
\includegraphics[height=7.5cm,width=7.5cm]{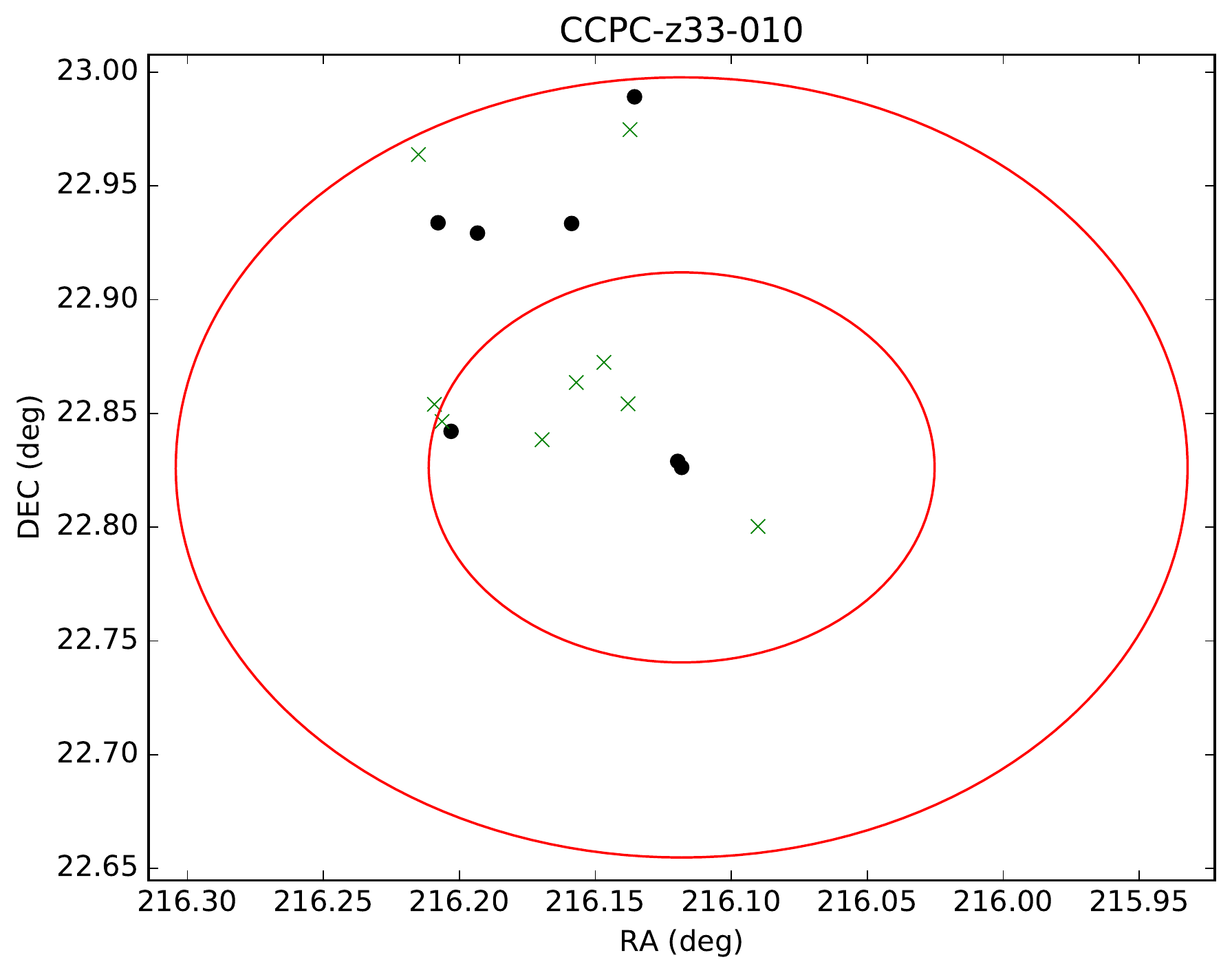}
\label{fig:CCPC-z33-010_sky}
\end{subfigure}
\hfill
\begin{subfigure}
\centering
\includegraphics[scale=0.52]{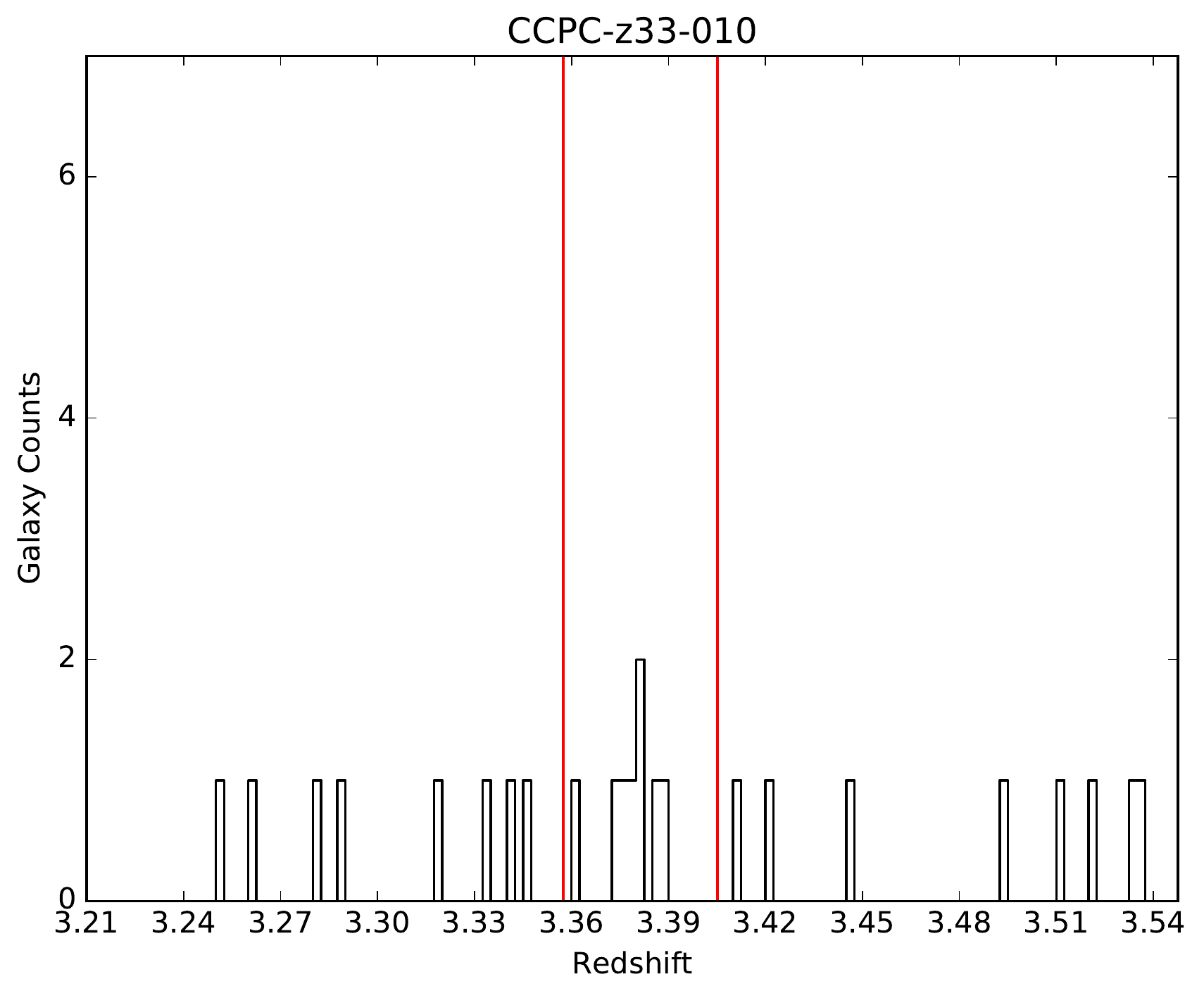}
\label{fig:CCPC-z33-010}
\end{subfigure}
\hfill
\end{figure*}
\clearpage 

\begin{figure*}
\centering
\begin{subfigure}
\centering
\includegraphics[height=7.5cm,width=7.5cm]{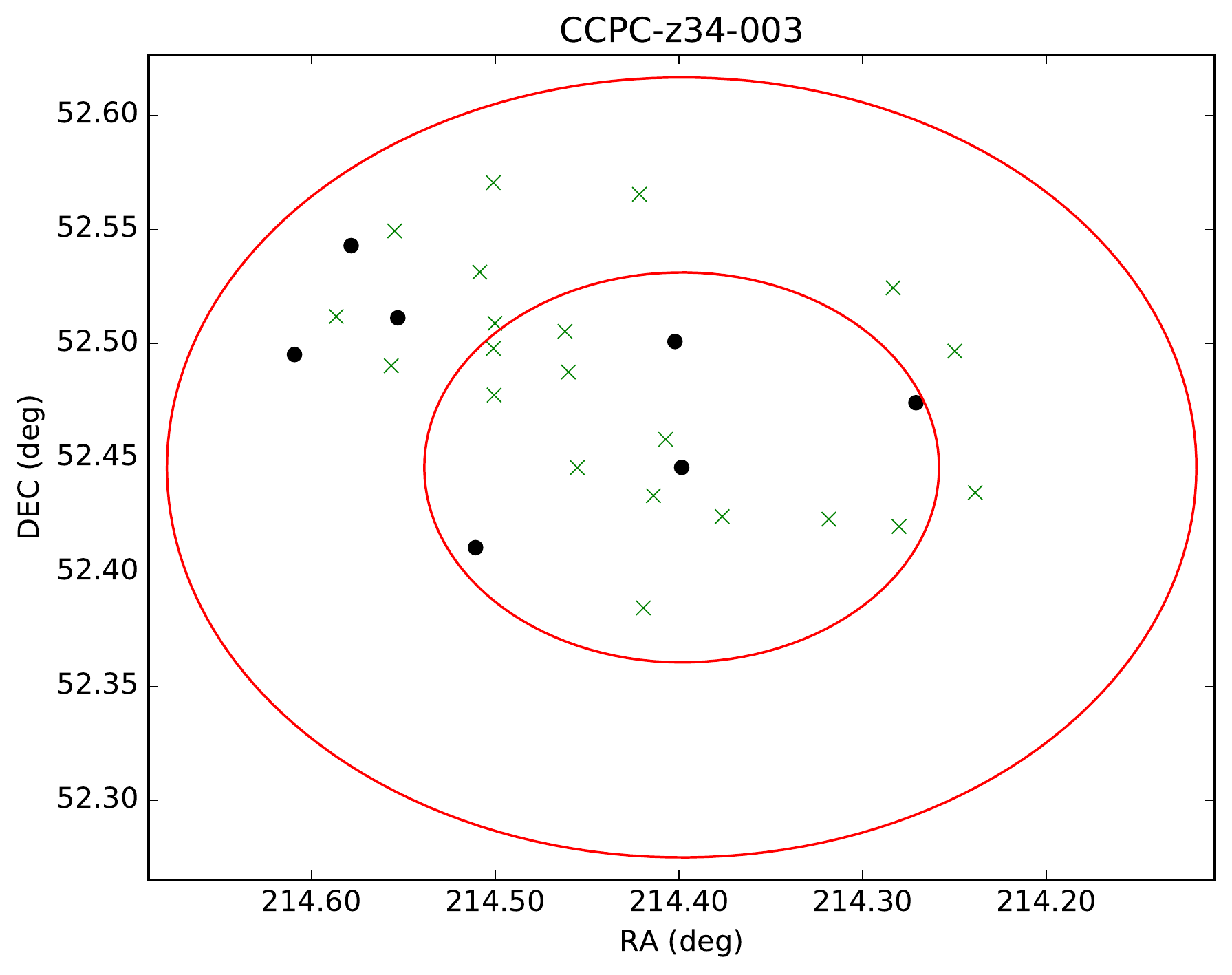}
\label{fig:CCPC-z34-003_sky}
\end{subfigure}
\hfill
\begin{subfigure}
\centering
\includegraphics[scale=0.52]{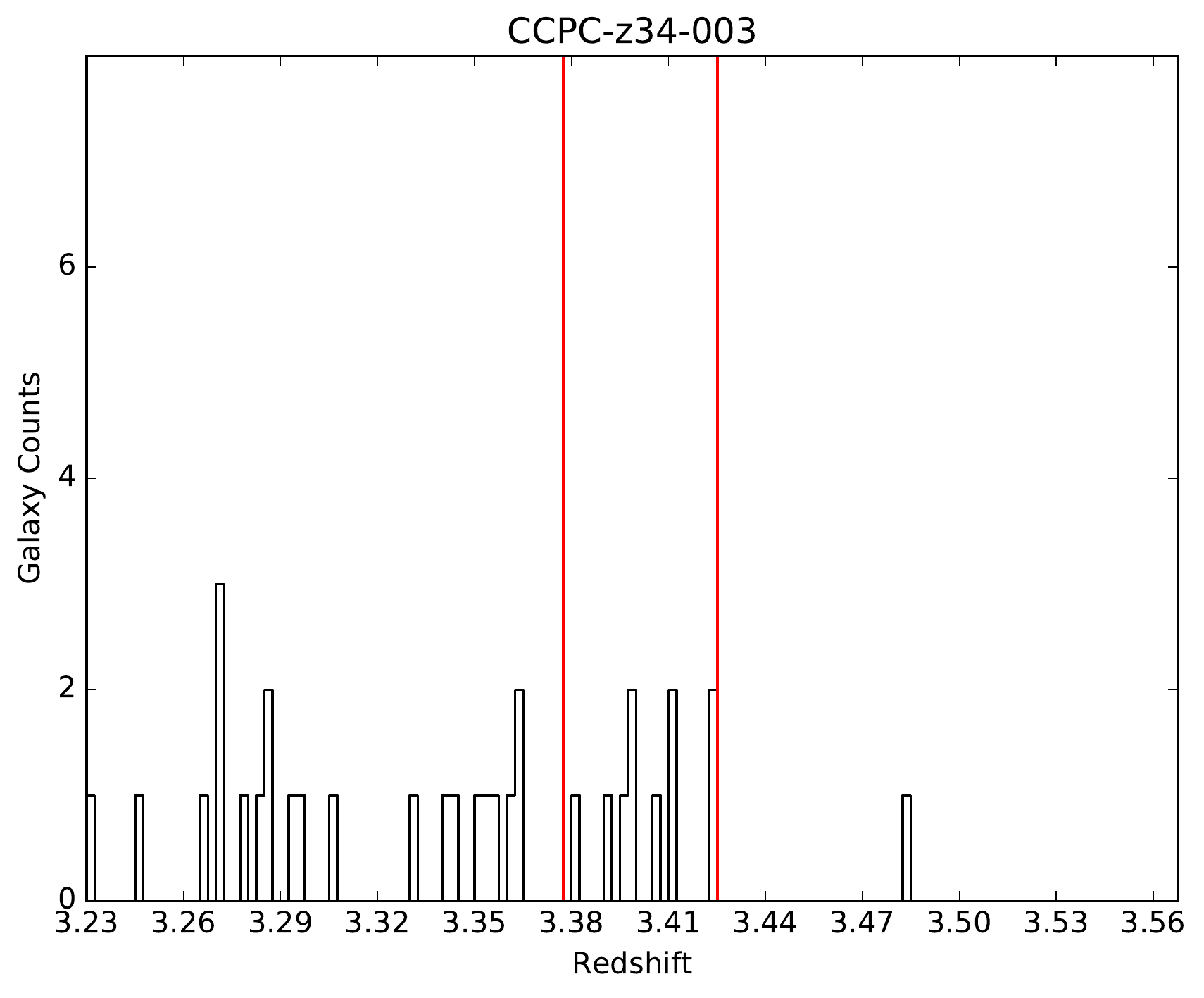}
\label{fig:CCPC-z34-003}
\end{subfigure}
\hfill
\end{figure*}

\begin{figure*}
\centering
\begin{subfigure}
\centering
\includegraphics[height=7.5cm,width=7.5cm]{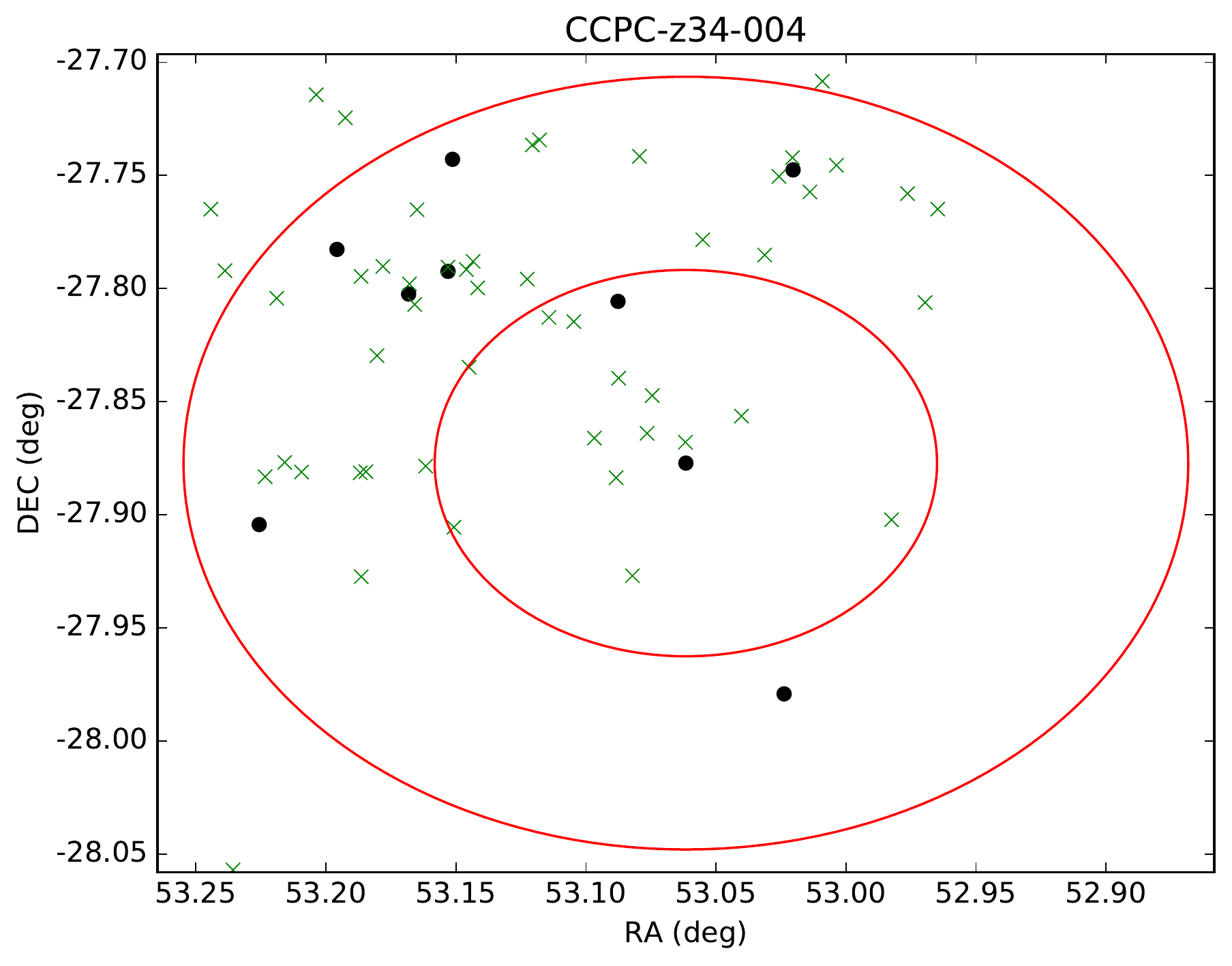}
\label{fig:CCPC-z34-004_sky}
\end{subfigure}
\hfill
\begin{subfigure}
\centering
\includegraphics[scale=0.52]{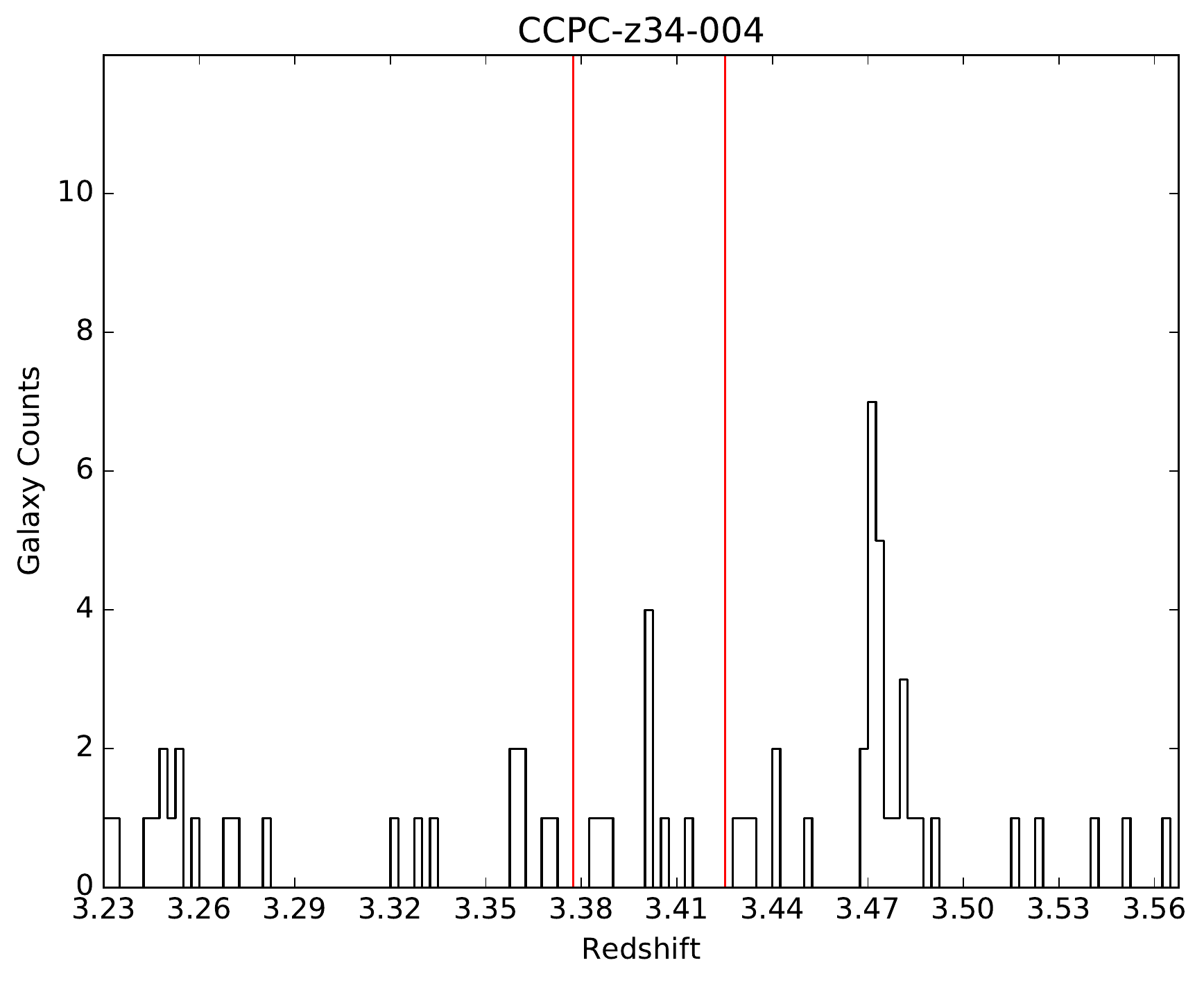}
\label{fig:CCPC-z34-004}
\end{subfigure}
\hfill
\end{figure*}

\begin{figure*}
\centering
\begin{subfigure}
\centering
\includegraphics[height=7.5cm,width=7.5cm]{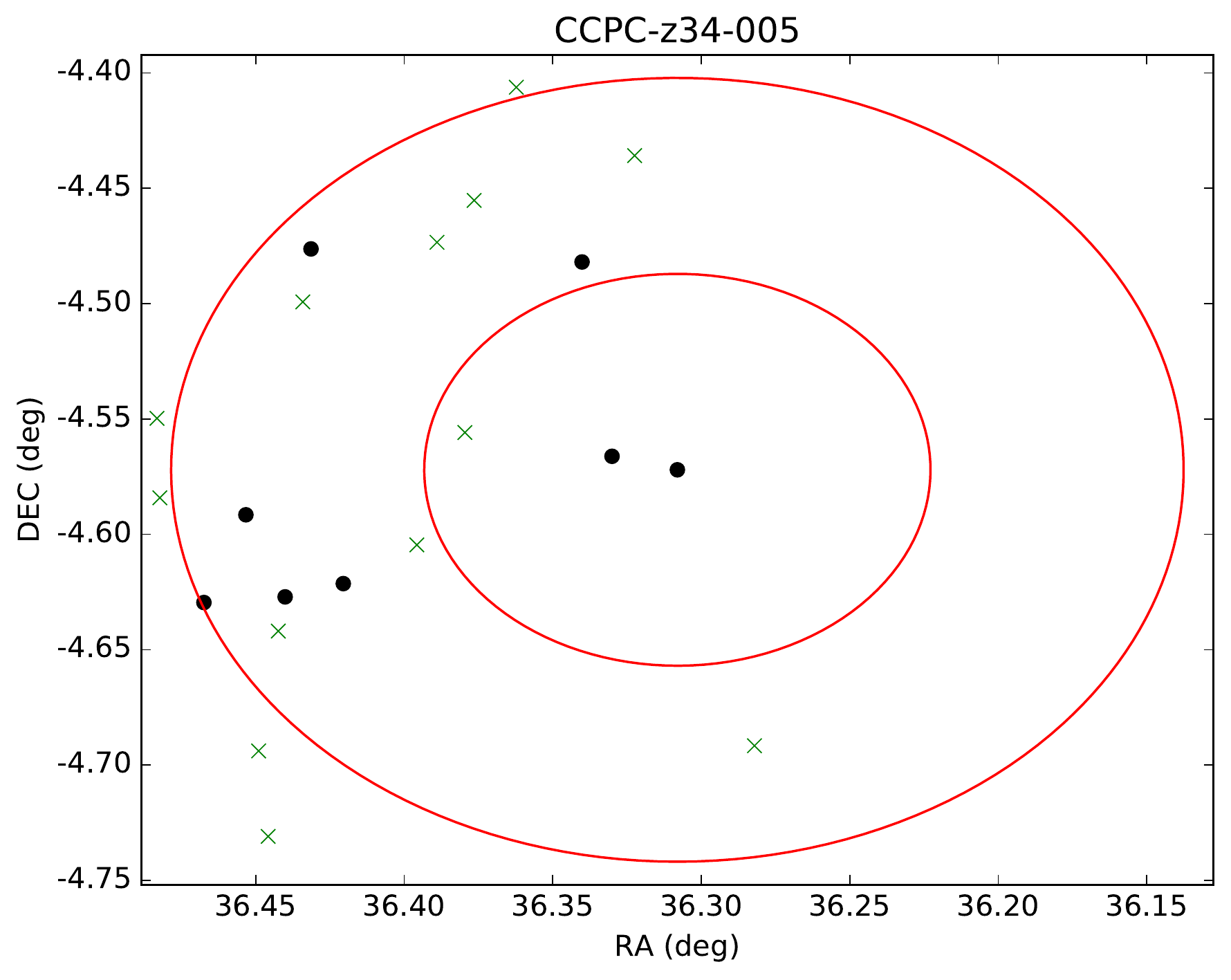}
\label{fig:CCPC-z34-005_sky}
\end{subfigure}
\hfill
\begin{subfigure}
\centering
\includegraphics[scale=0.52]{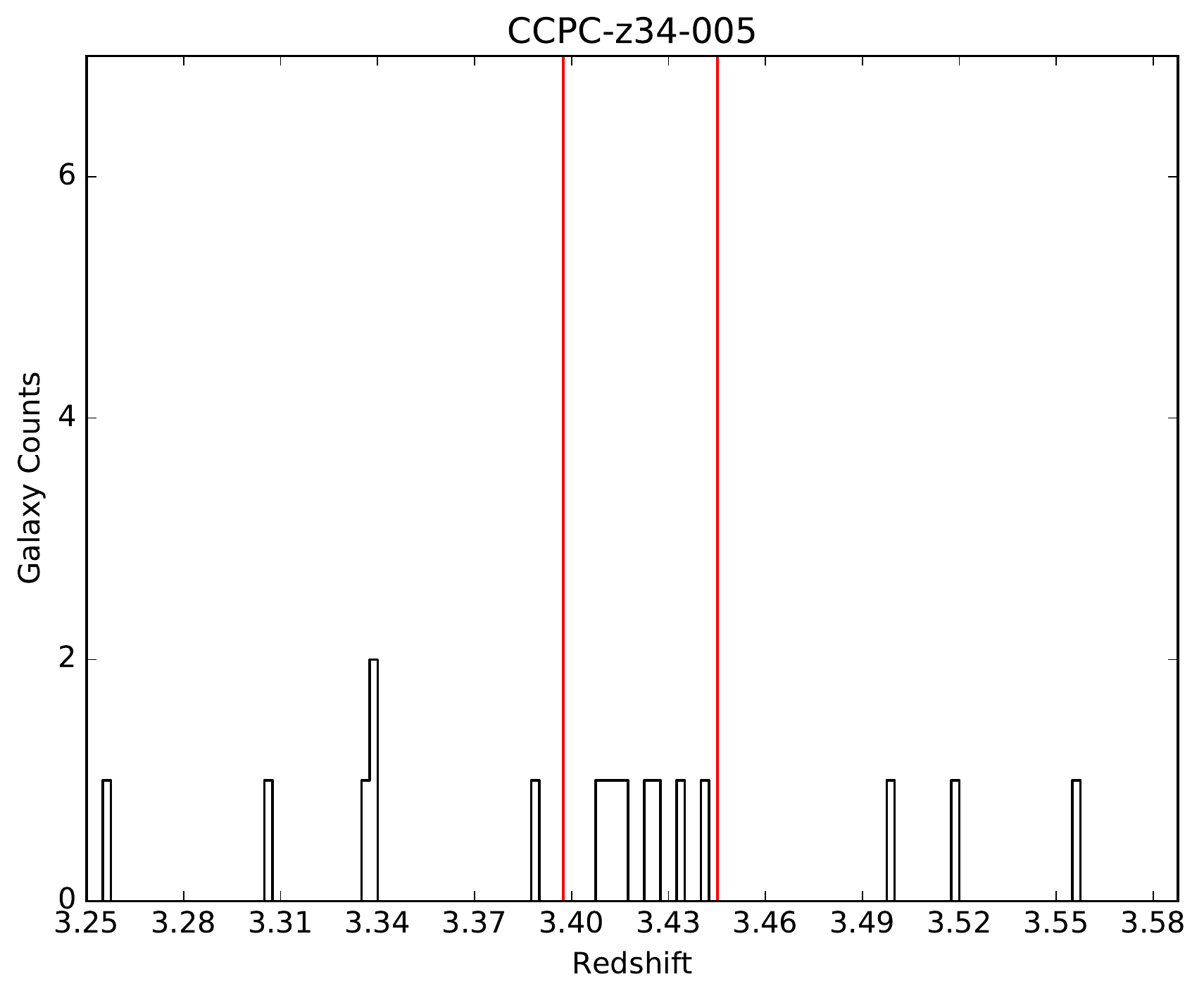}
\label{fig:CCPC-z34-005}
\end{subfigure}
\hfill
\end{figure*}
\clearpage 

\begin{figure*}
\centering
\begin{subfigure}
\centering
\includegraphics[height=7.5cm,width=7.5cm]{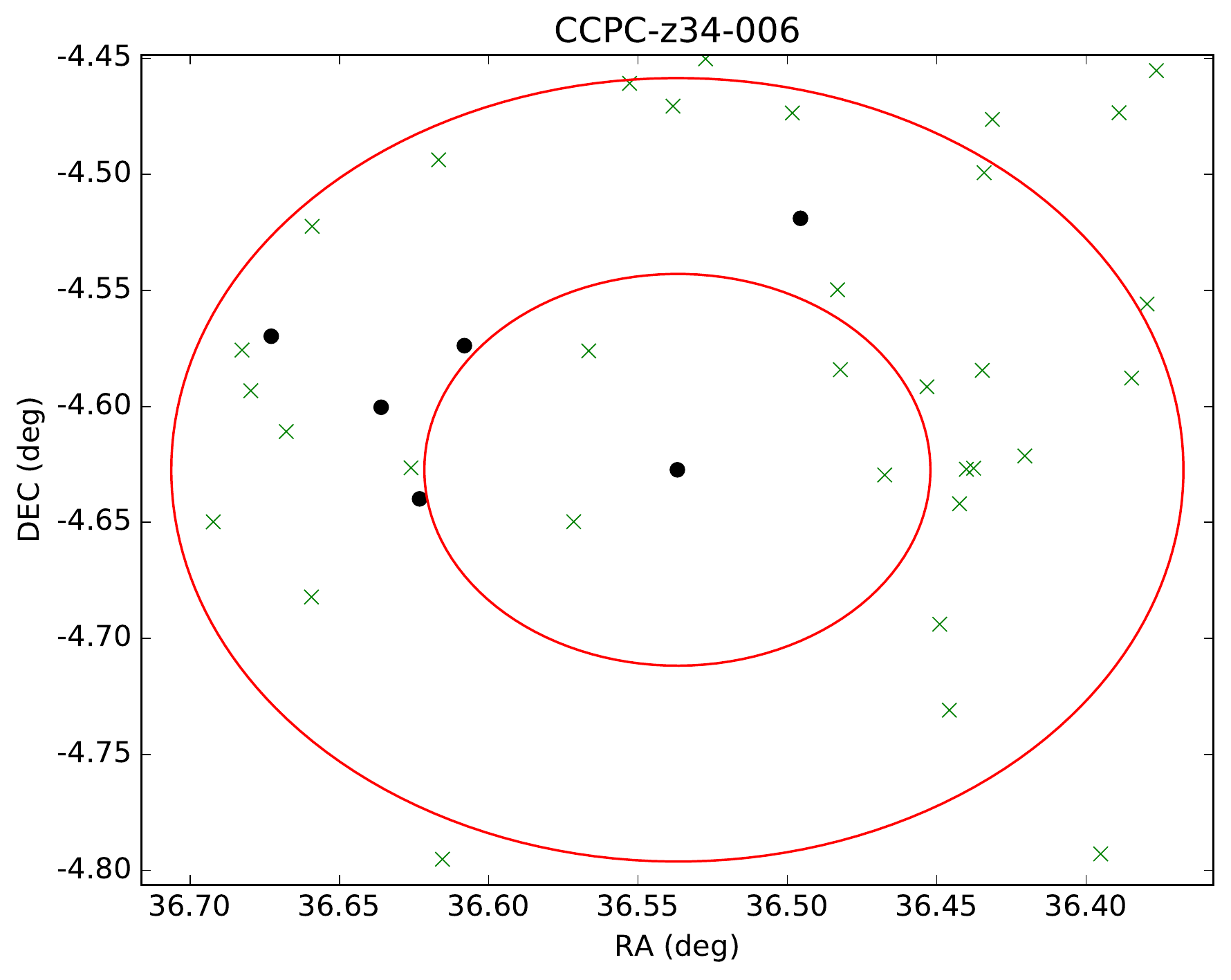}
\label{fig:CCPC-z34-006_sky}
\end{subfigure}
\hfill
\begin{subfigure}
\centering
\includegraphics[scale=0.52]{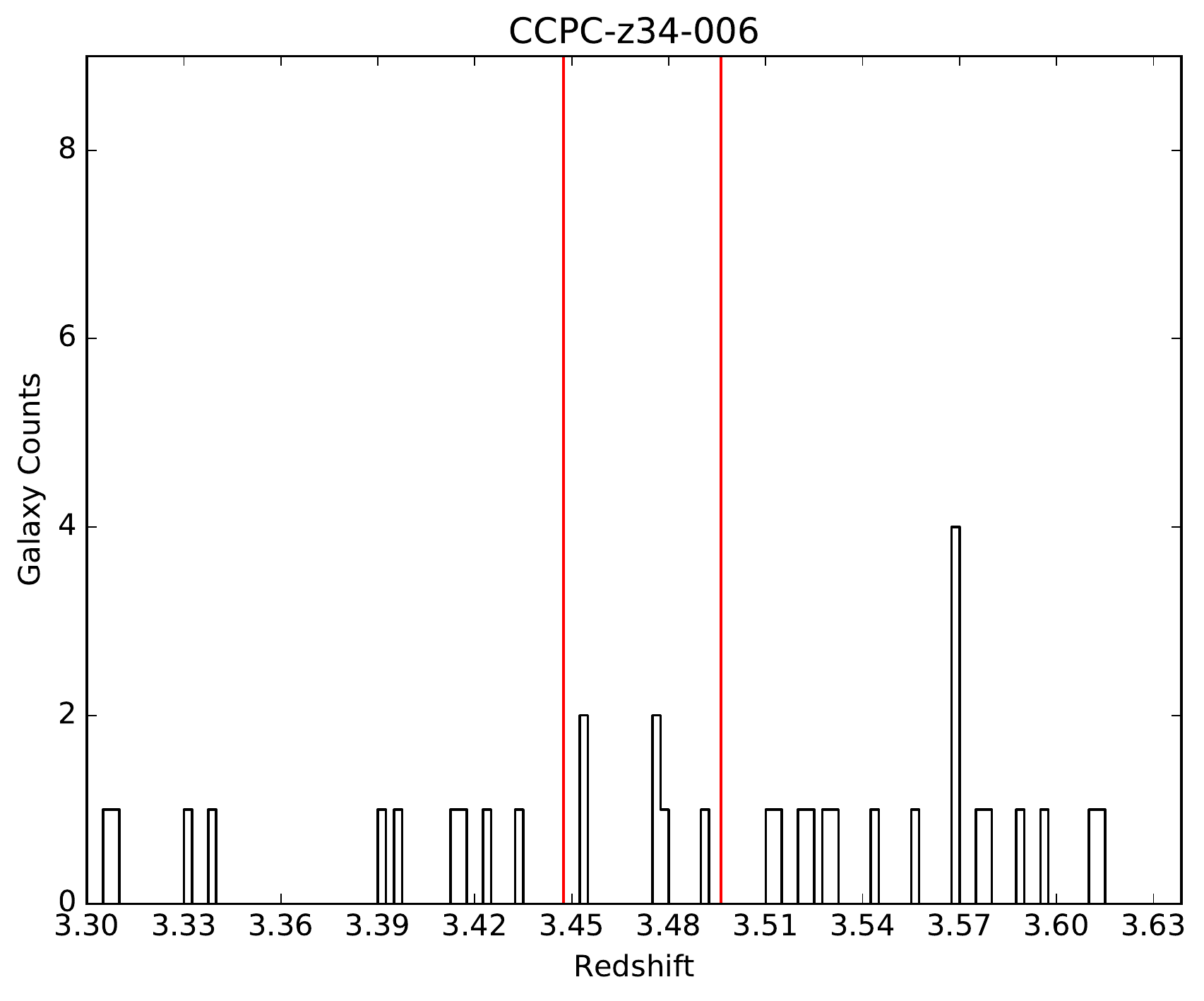}
\label{fig:CCPC-z34-006}
\end{subfigure}
\hfill
\end{figure*}

\begin{figure*}
\centering
\begin{subfigure}
\centering
\includegraphics[height=7.5cm,width=7.5cm]{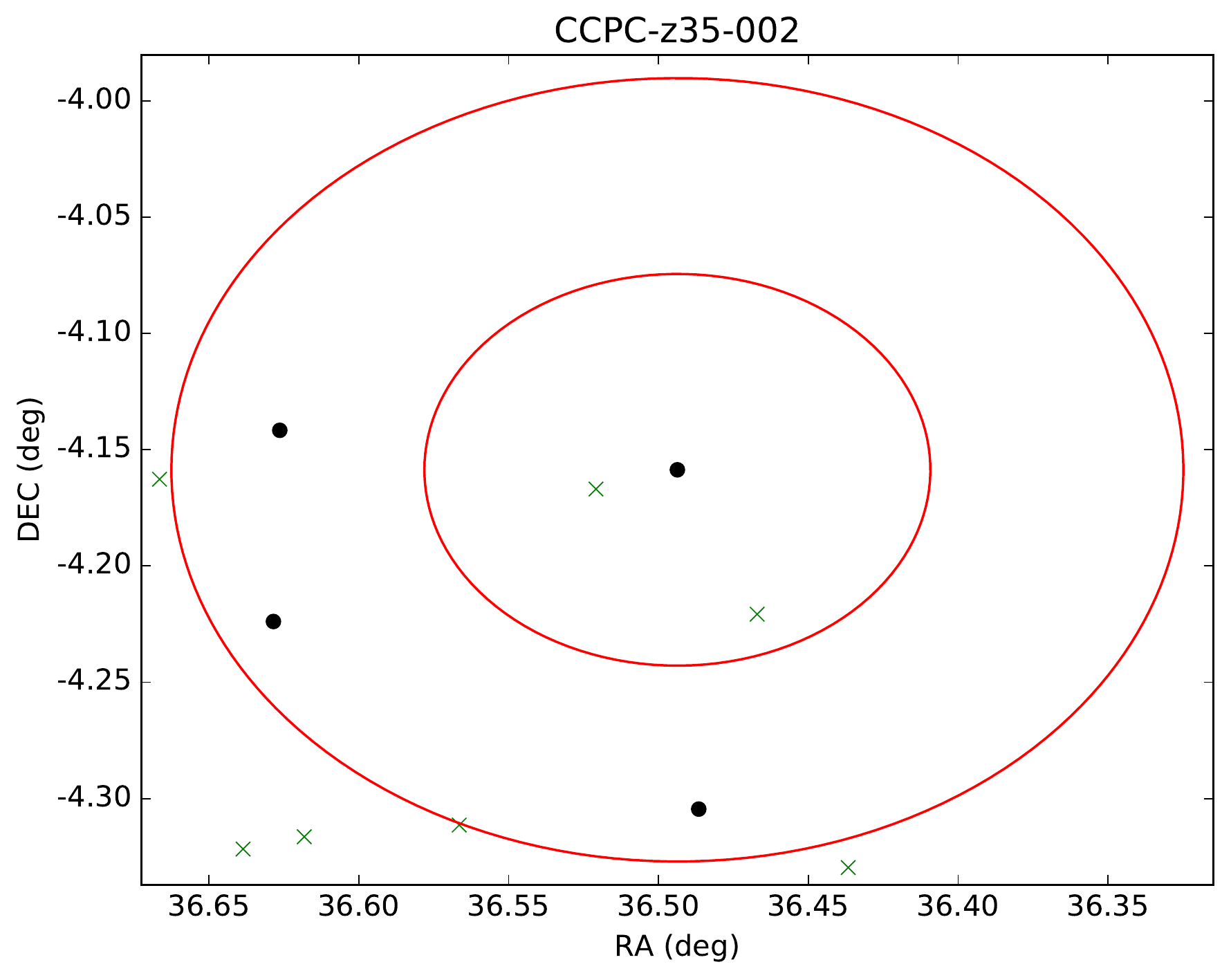}
\label{fig:CCPC-z35-002_sky}
\end{subfigure}
\hfill
\begin{subfigure}
\centering
\includegraphics[scale=0.52]{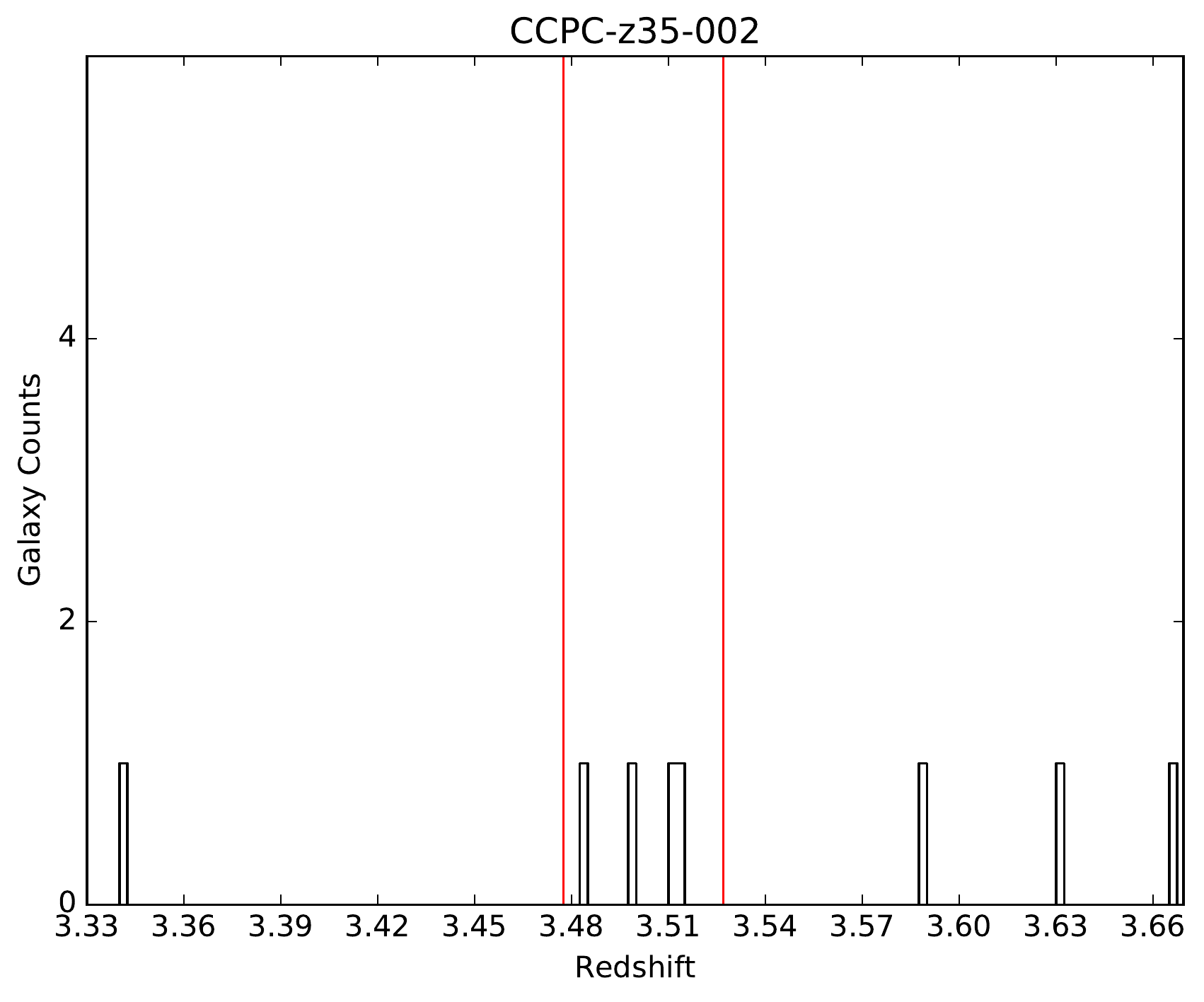}
\label{fig:CCPC-z35-002}
\end{subfigure}
\hfill
\end{figure*}

\begin{figure*}
\centering
\begin{subfigure}
\centering
\includegraphics[height=7.5cm,width=7.5cm]{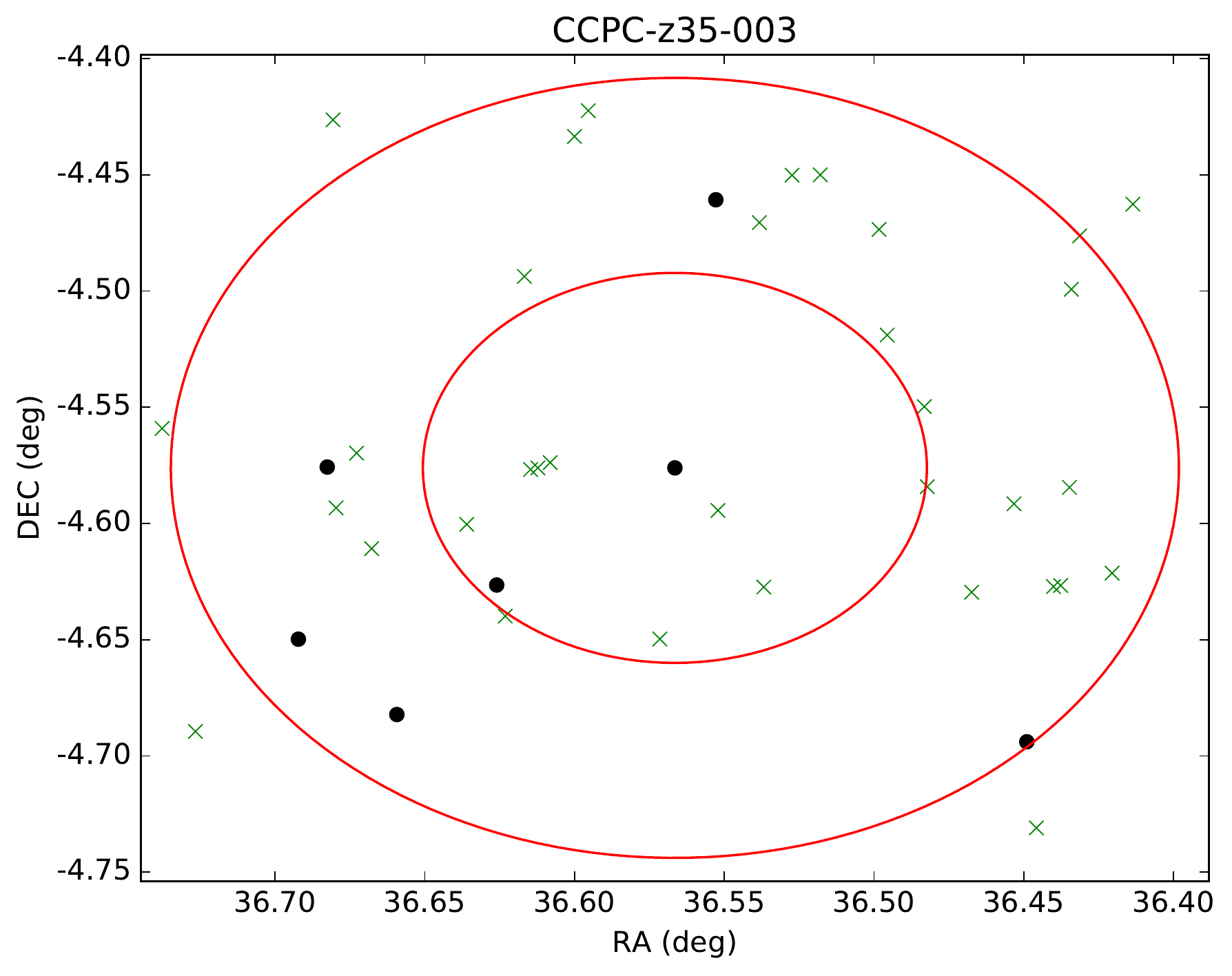}
\label{fig:CCPC-z35-003_sky}
\end{subfigure}
\hfill
\begin{subfigure}
\centering
\includegraphics[scale=0.52]{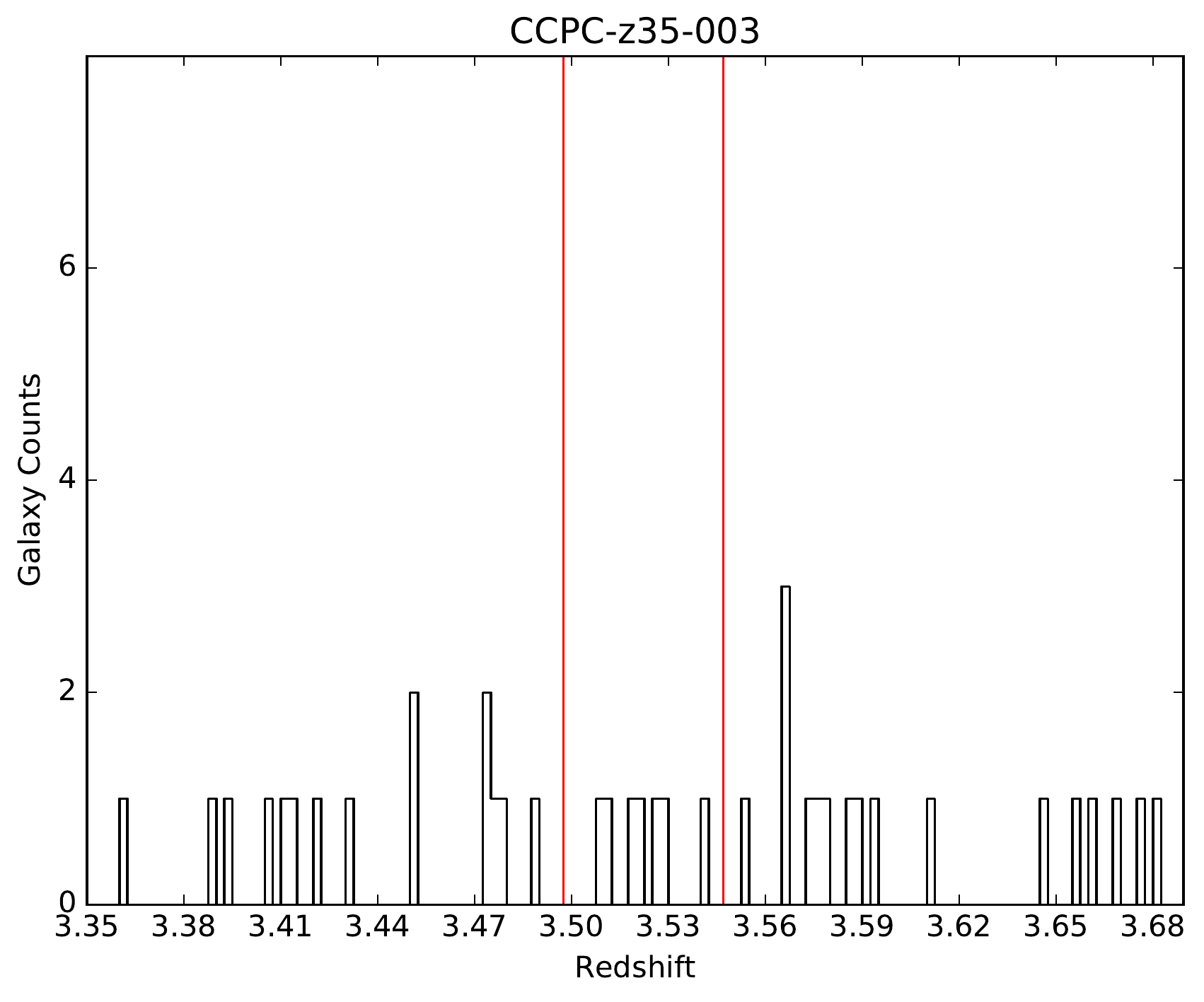}
\label{fig:CCPC-z35-003}
\end{subfigure}
\hfill
\end{figure*}
\clearpage 

\begin{figure*}
\centering
\begin{subfigure}
\centering
\includegraphics[height=7.5cm,width=7.5cm]{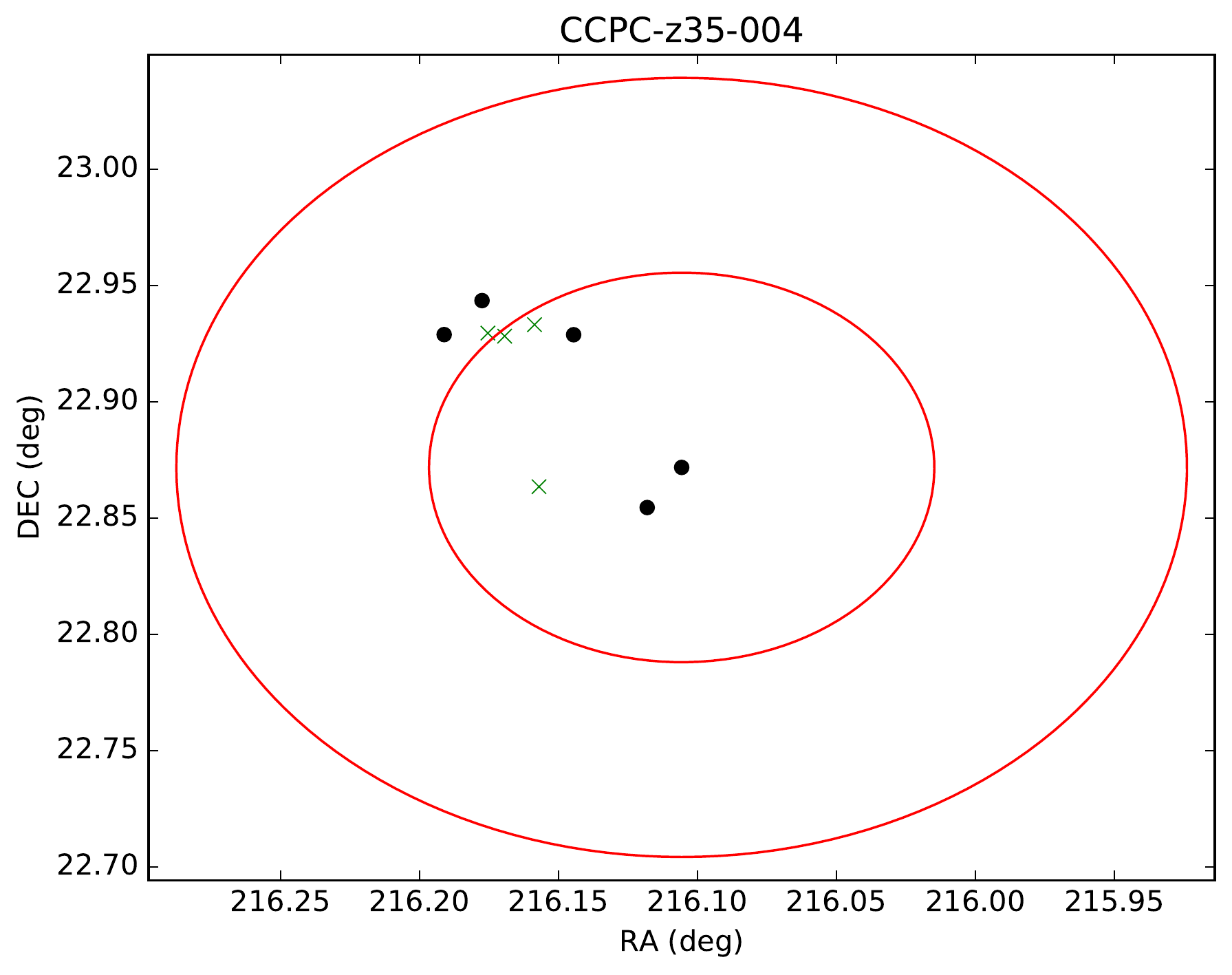}
\label{fig:CCPC-z35-004_sky}
\end{subfigure}
\hfill
\begin{subfigure}
\centering
\includegraphics[scale=0.52]{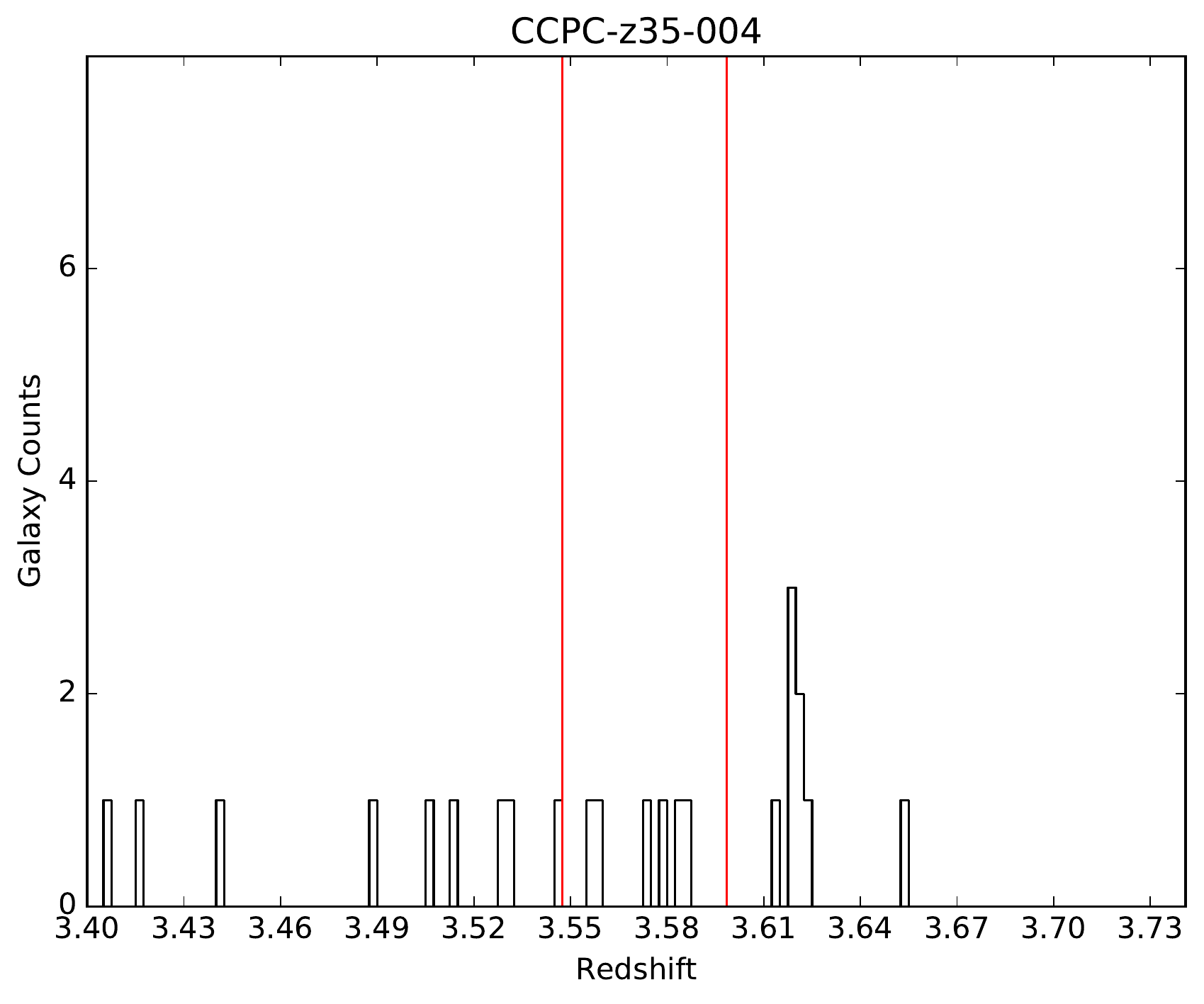}
\label{fig:CCPC-z35-004}
\end{subfigure}
\hfill
\end{figure*}

\begin{figure*}
\centering
\begin{subfigure}
\centering
\includegraphics[height=7.5cm,width=7.5cm]{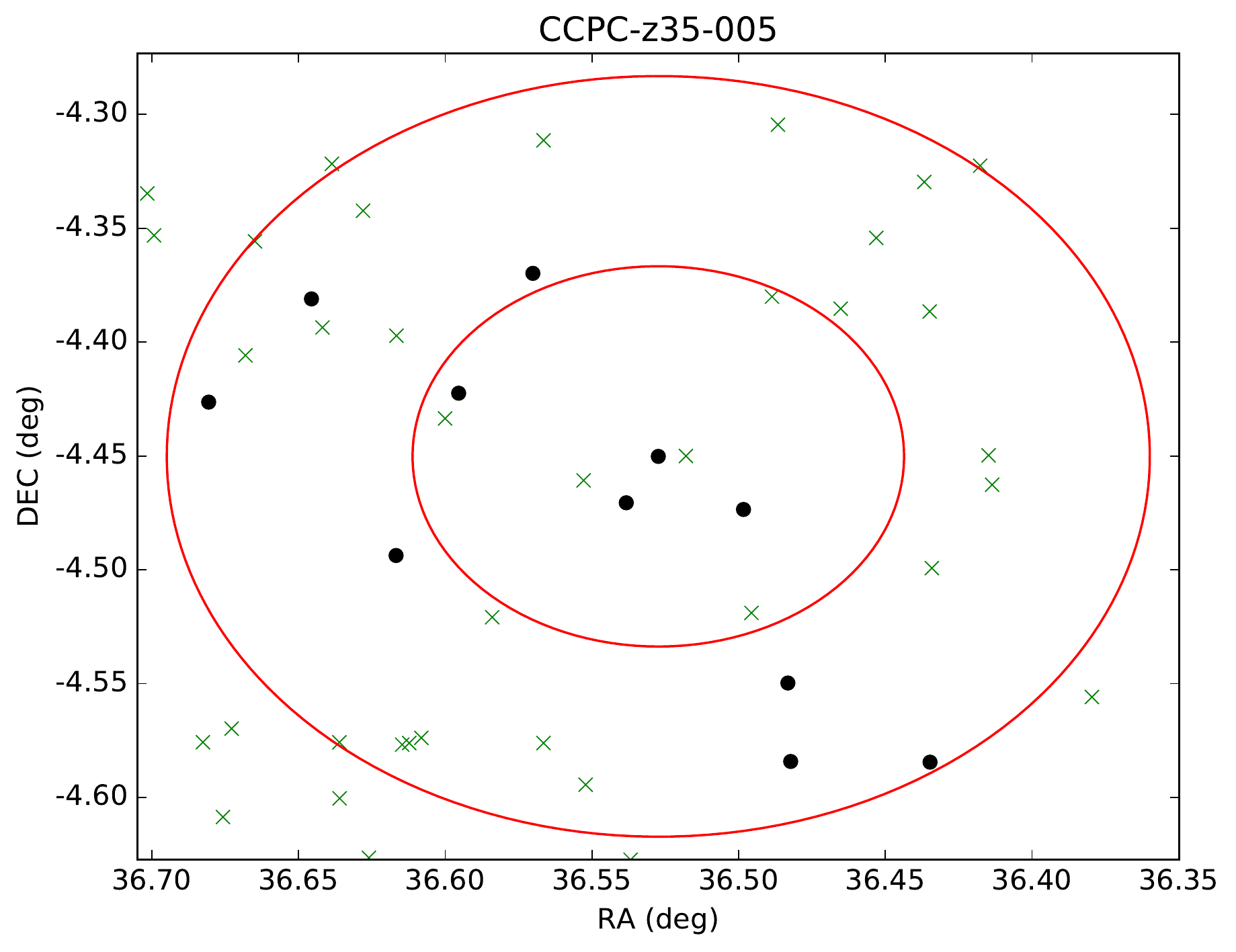}
\label{fig:CCPC-z35-005_sky}
\end{subfigure}
\hfill
\begin{subfigure}
\centering
\includegraphics[scale=0.52]{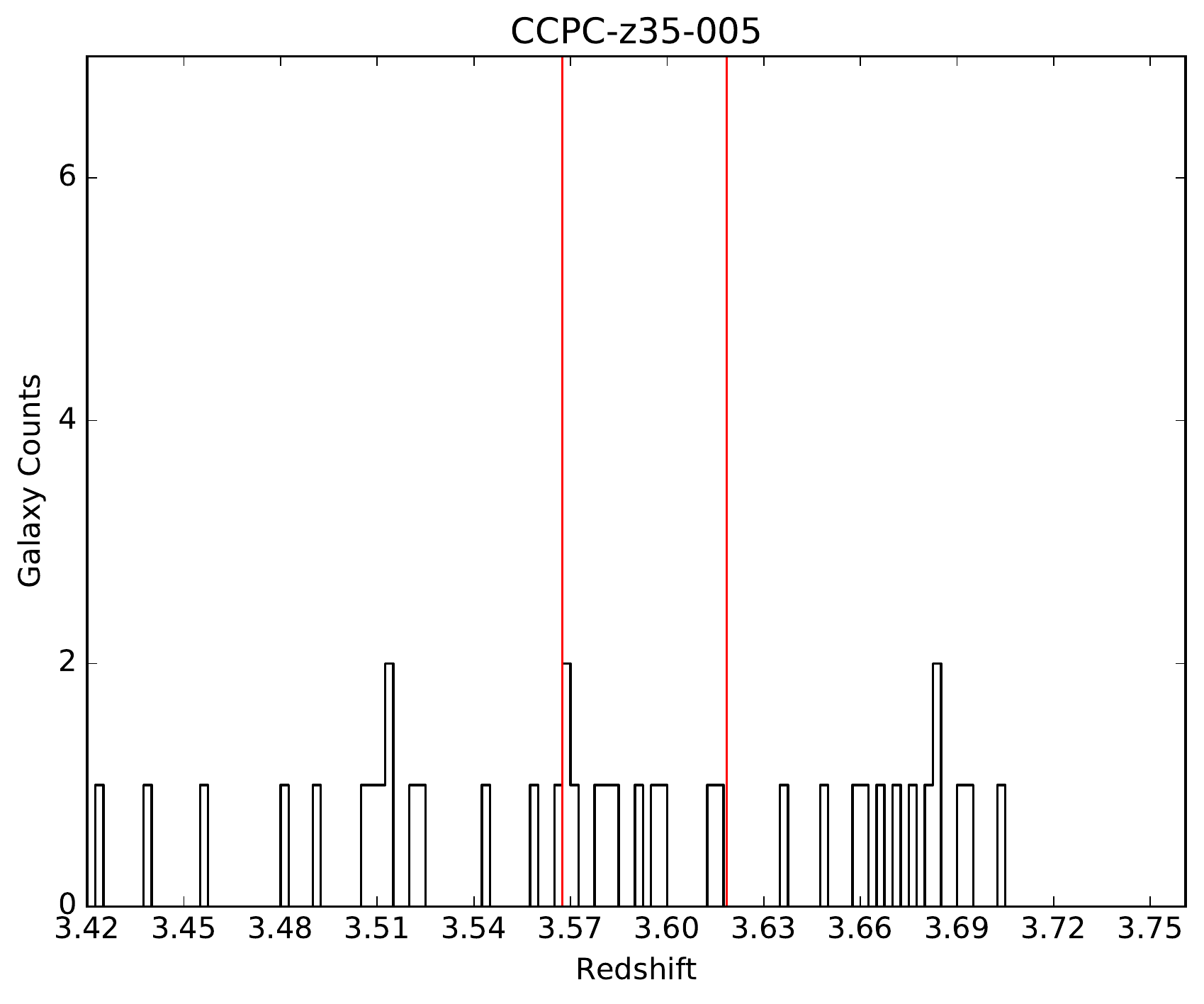}
\label{fig:CCPC-z35-005}
\end{subfigure}
\hfill
\end{figure*}

\begin{figure*}
\centering
\begin{subfigure}
\centering
\includegraphics[height=7.5cm,width=7.5cm]{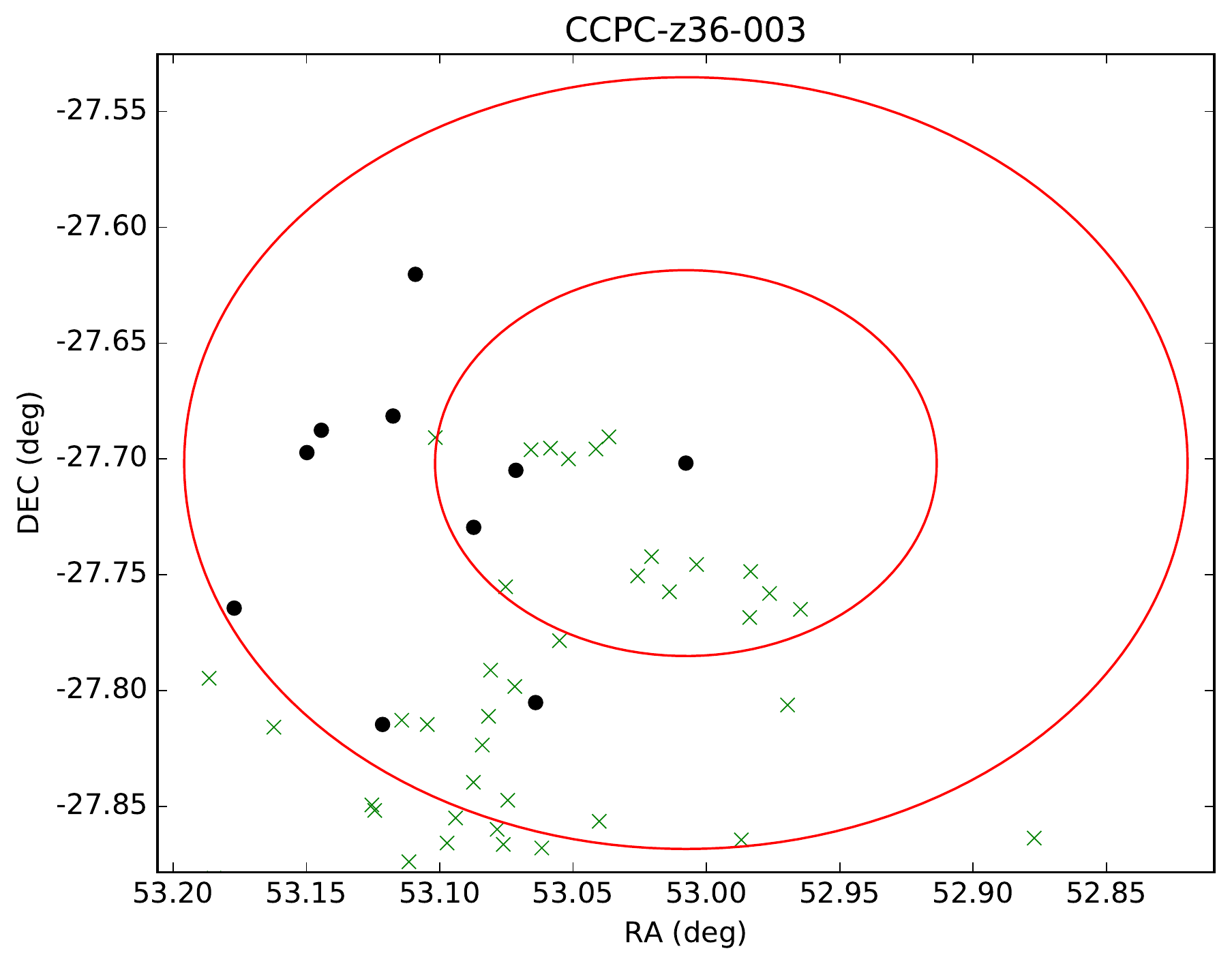}
\label{fig:CCPC-z36-003_sky}
\end{subfigure}
\hfill
\begin{subfigure}
\centering
\includegraphics[scale=0.52]{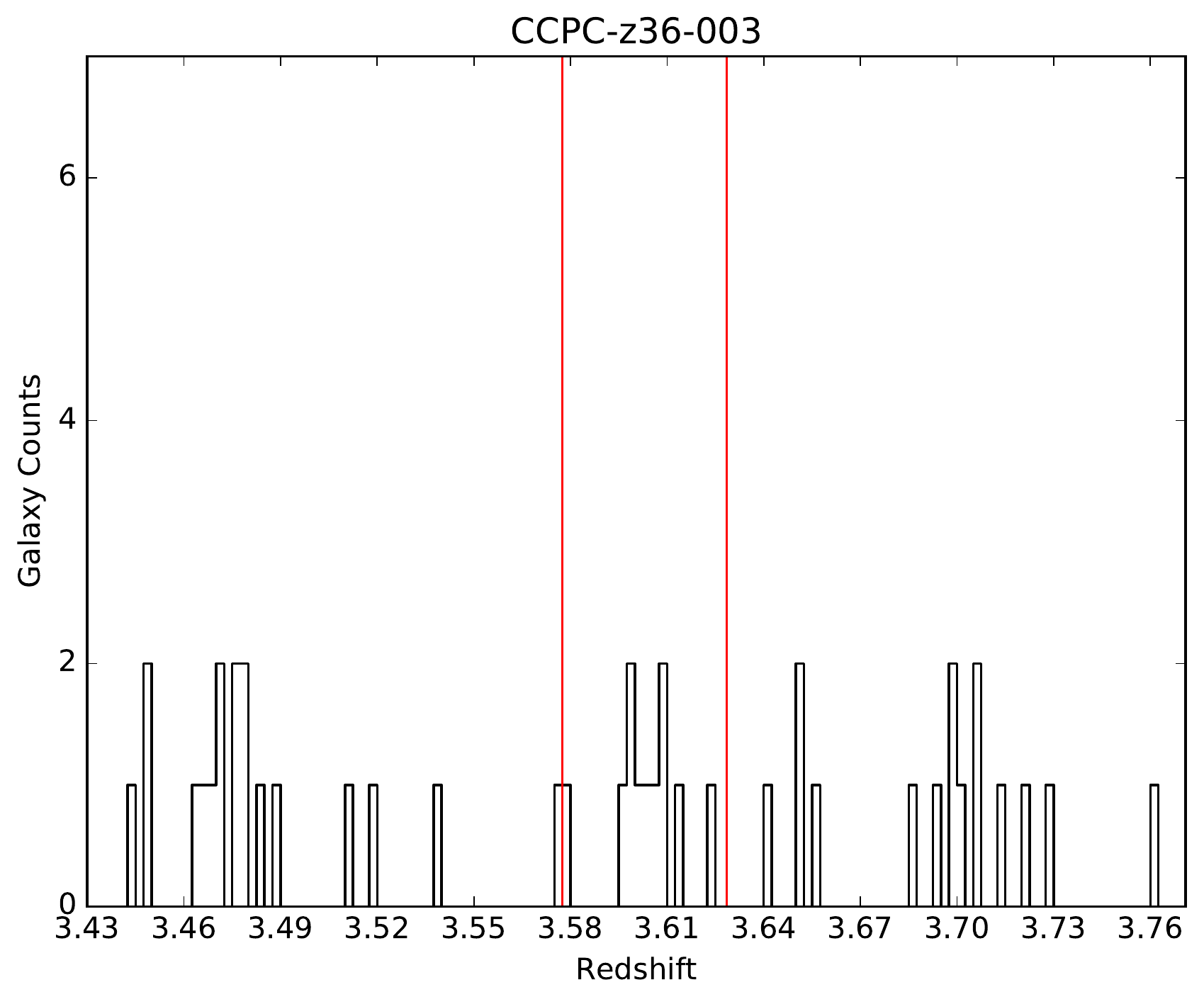}
\label{fig:CCPC-z36-003}
\end{subfigure}
\hfill
\end{figure*}
\clearpage 

\begin{figure*}
\centering
\begin{subfigure}
\centering
\includegraphics[height=7.5cm,width=7.5cm]{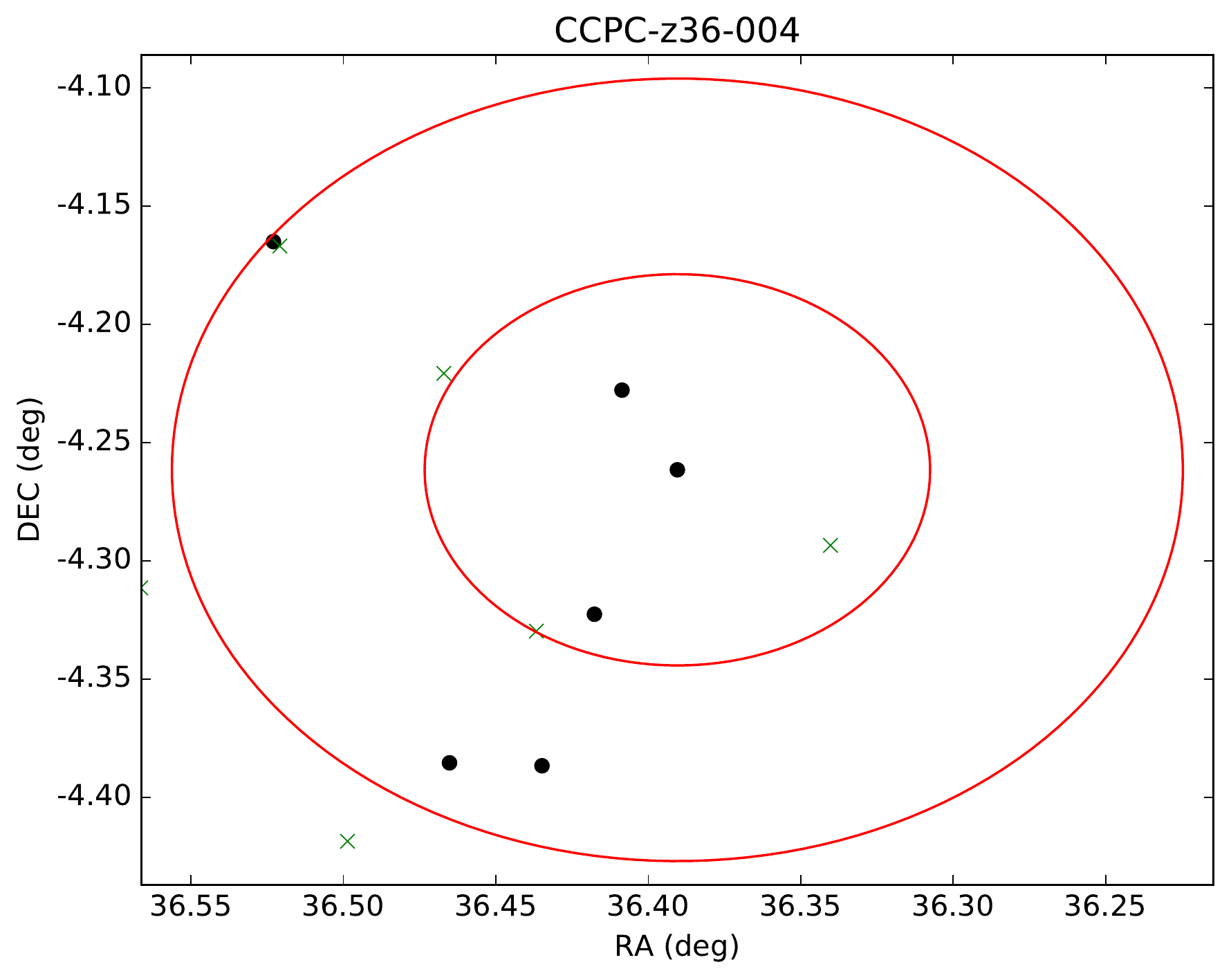}
\label{fig:CCPC-z36-004_sky}
\end{subfigure}
\hfill
\begin{subfigure}
\centering
\includegraphics[scale=0.52]{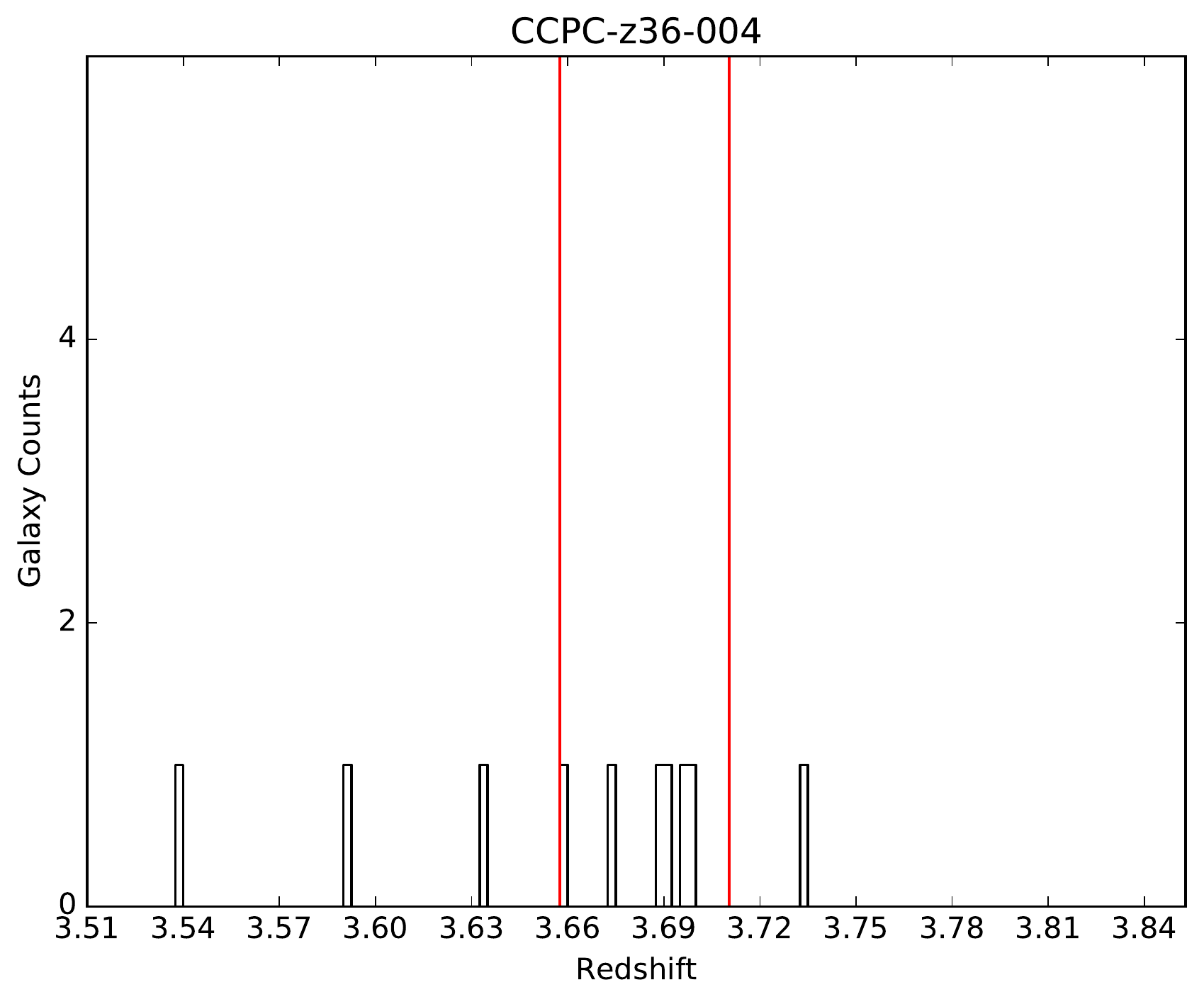}
\label{fig:CCPC-z36-004}
\end{subfigure}
\hfill
\end{figure*}

\begin{figure*}
\centering
\begin{subfigure}
\centering
\includegraphics[height=7.5cm,width=7.5cm]{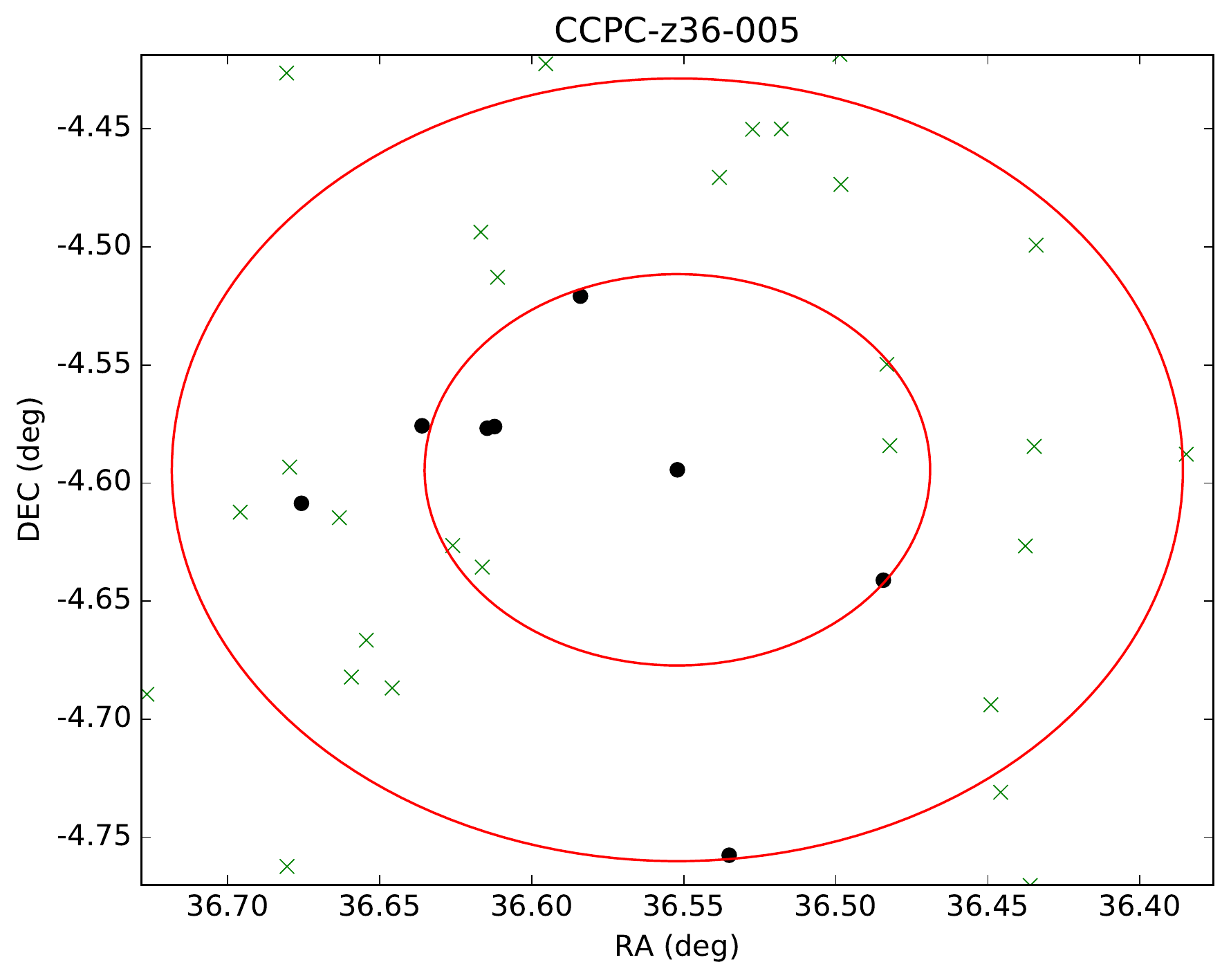}
\label{fig:CCPC-z36-005_sky}
\end{subfigure}
\hfill
\begin{subfigure}
\centering
\includegraphics[scale=0.52]{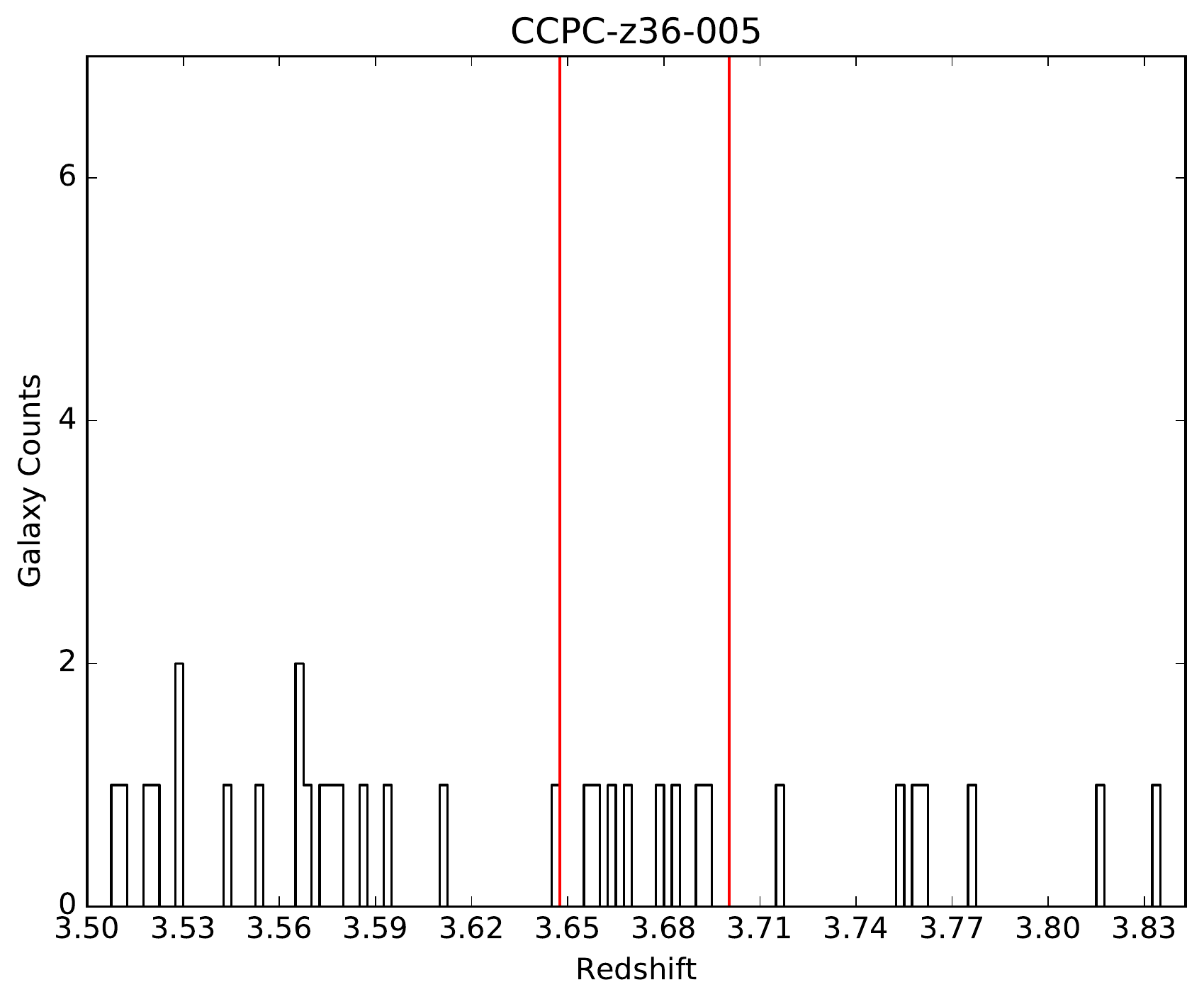}
\label{fig:CCPC-z36-005}
\end{subfigure}
\hfill
\end{figure*}

\begin{figure*}
\centering
\begin{subfigure}
\centering
\includegraphics[height=7.5cm,width=7.5cm]{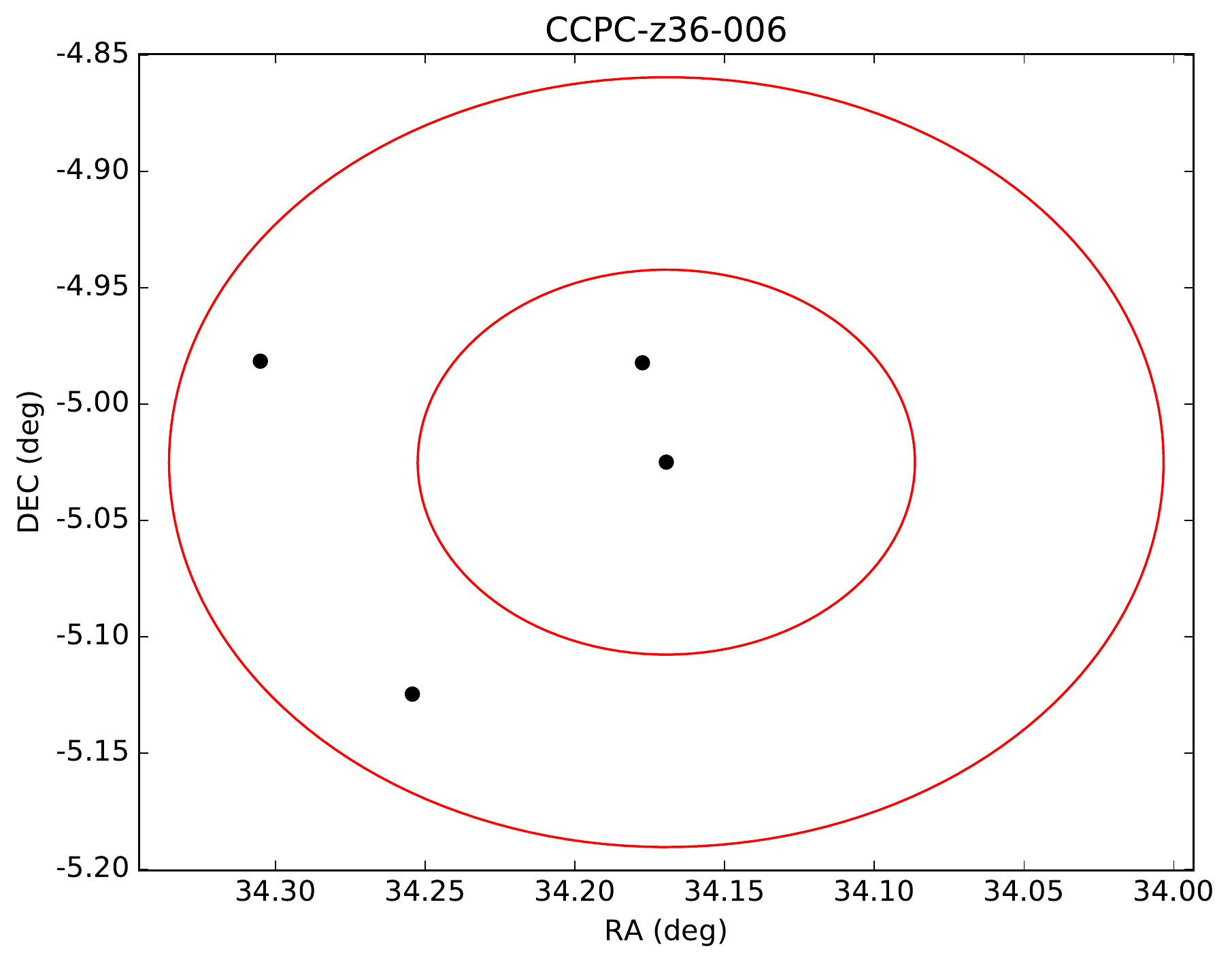}
\label{fig:CCPC-z36-006_sky}
\end{subfigure}
\hfill
\begin{subfigure}
\centering
\includegraphics[scale=0.52]{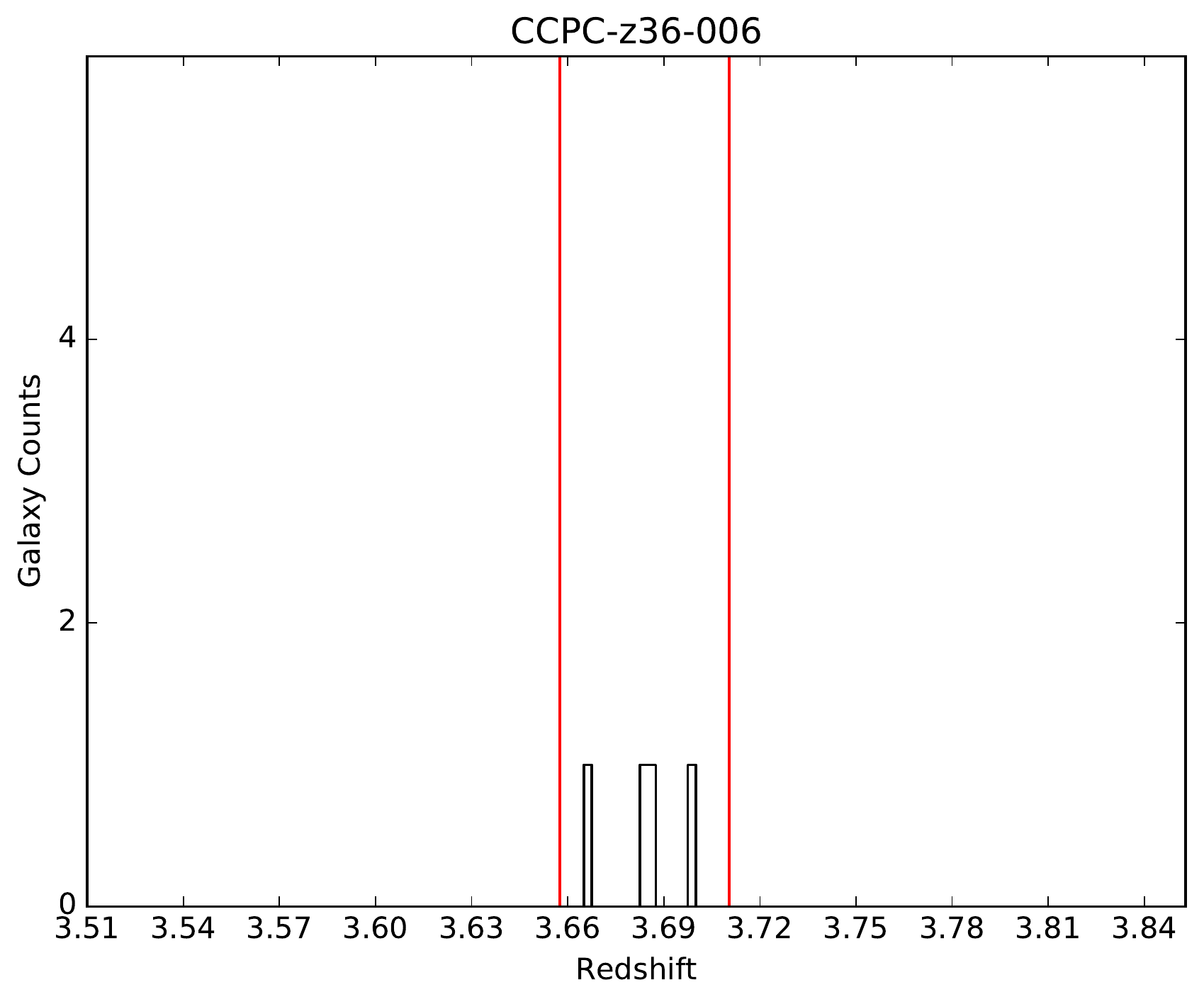}
\label{fig:CCPC-z36-006}
\end{subfigure}
\hfill
\end{figure*}
\clearpage 

\begin{figure*}
\centering
\begin{subfigure}
\centering
\includegraphics[height=7.5cm,width=7.5cm]{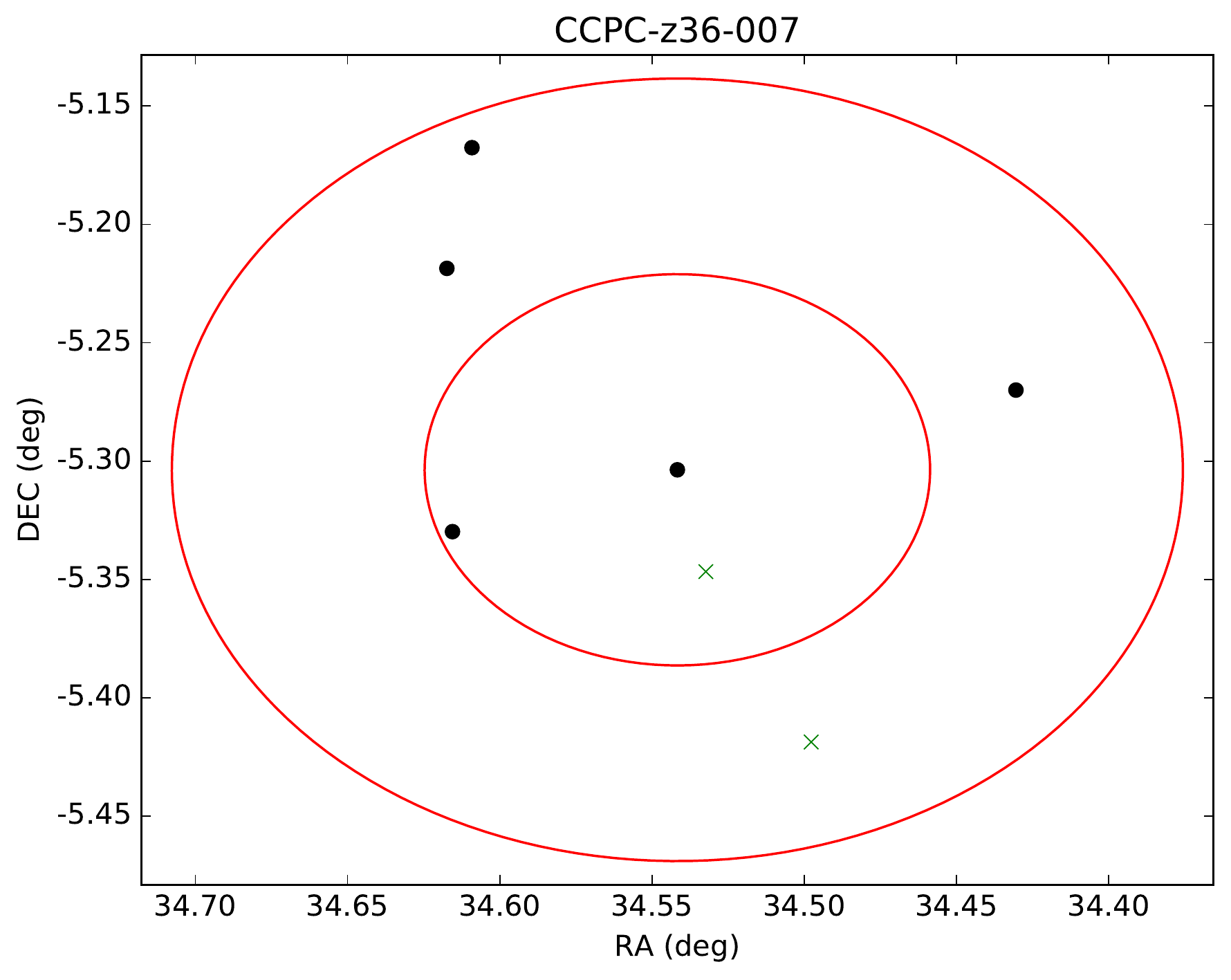}
\label{fig:CCPC-z36-007_sky}
\end{subfigure}
\hfill
\begin{subfigure}
\centering
\includegraphics[scale=0.52]{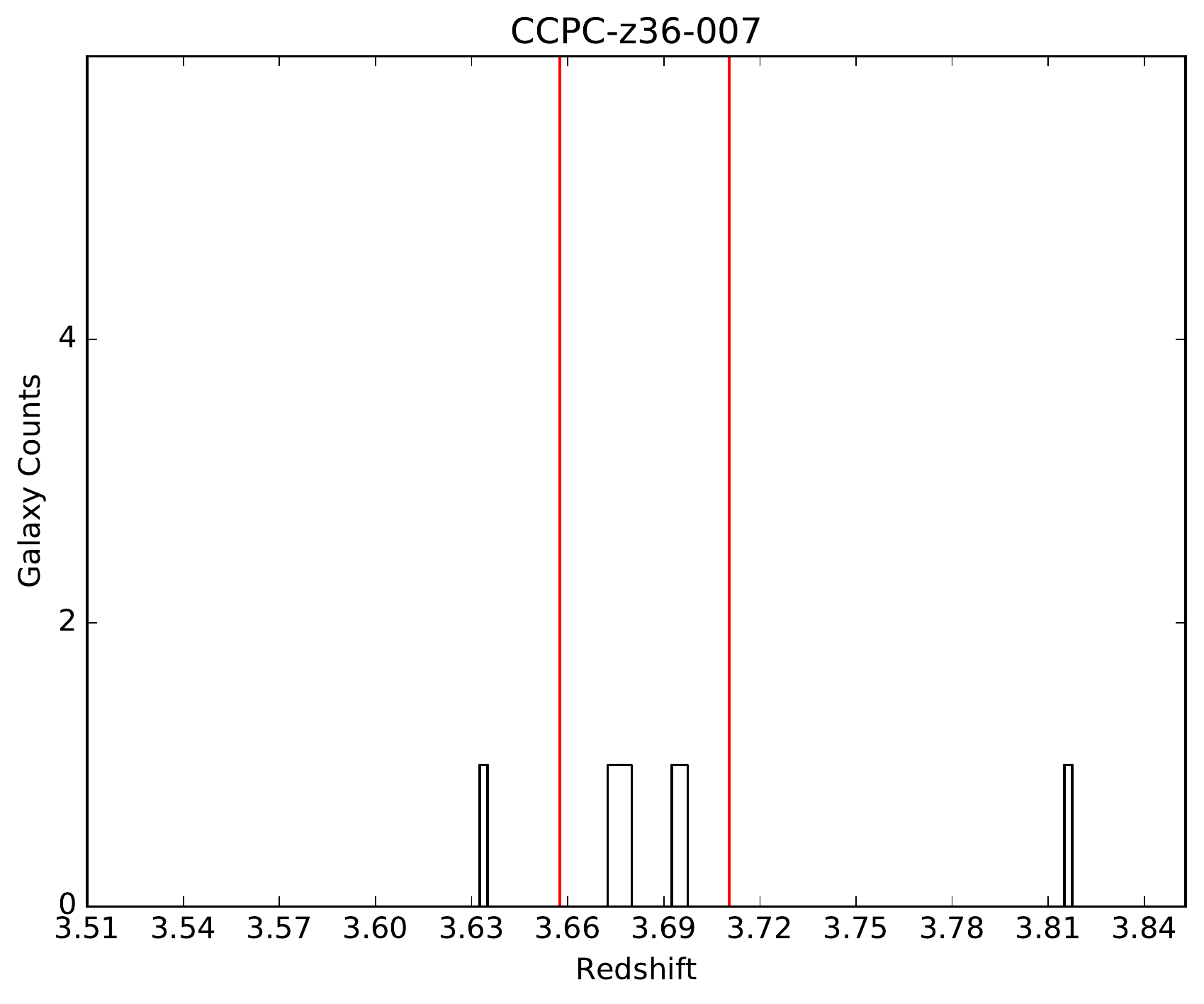}
\label{fig:CCPC-z36-007}
\end{subfigure}
\hfill
\end{figure*}

\begin{figure*}
\centering
\begin{subfigure}
\centering
\includegraphics[height=7.5cm,width=7.5cm]{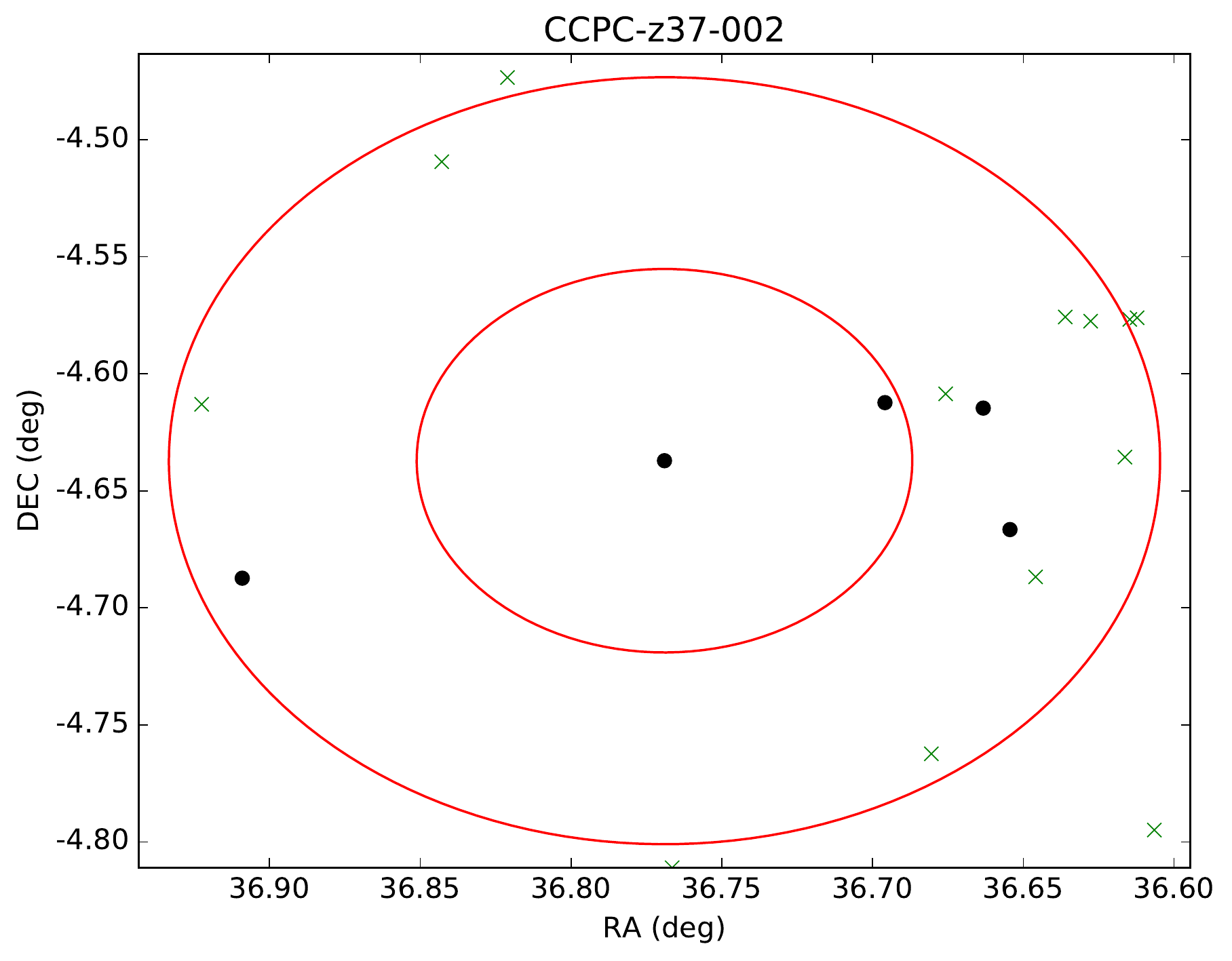}
\label{fig:CCPC-z37-002_sky}
\end{subfigure}
\hfill
\begin{subfigure}
\centering
\includegraphics[scale=0.52]{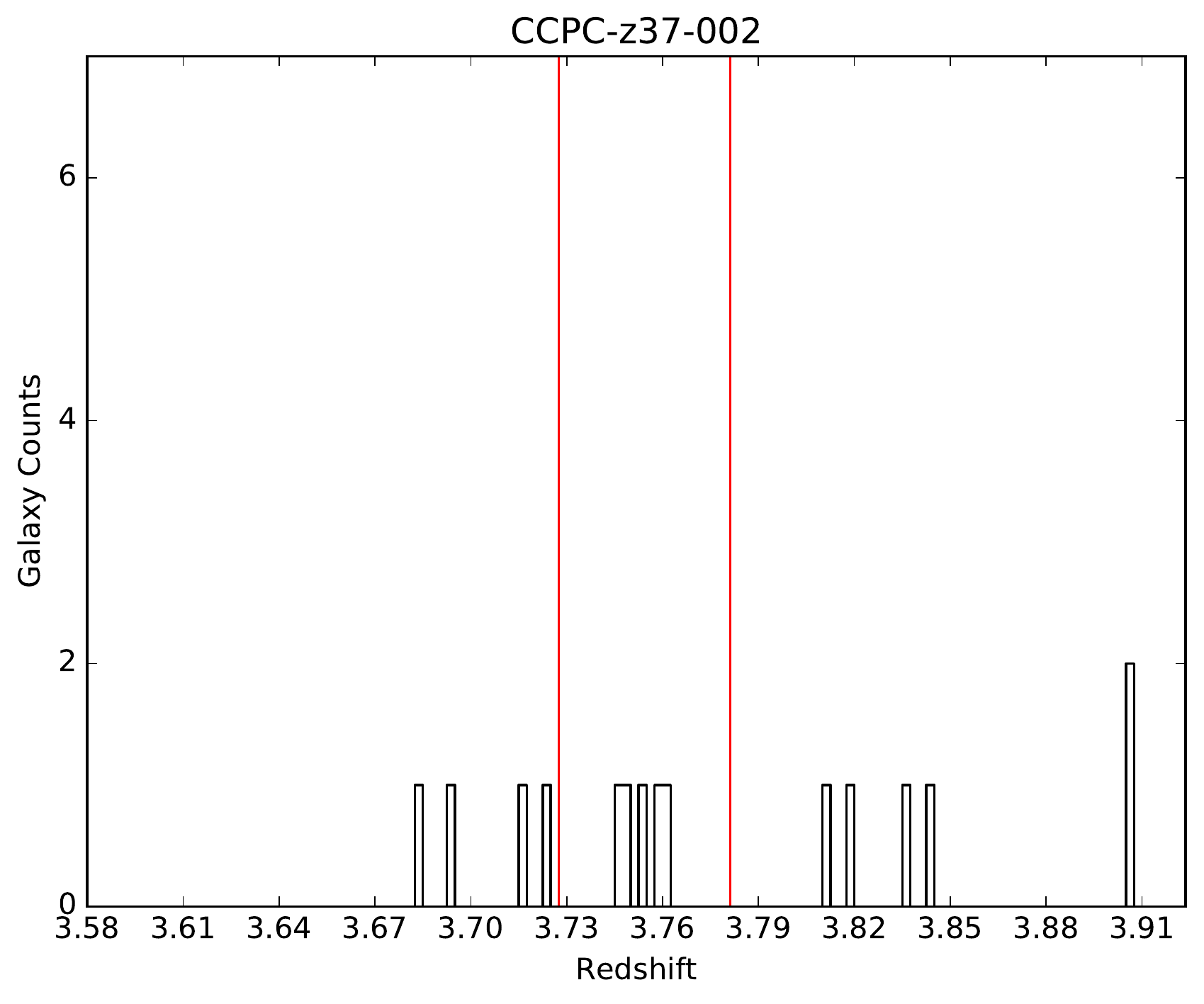}
\label{fig:CCPC-z37-002}
\end{subfigure}
\hfill
\end{figure*}

\begin{figure*}
\centering
\begin{subfigure}
\centering
\includegraphics[height=7.5cm,width=7.5cm]{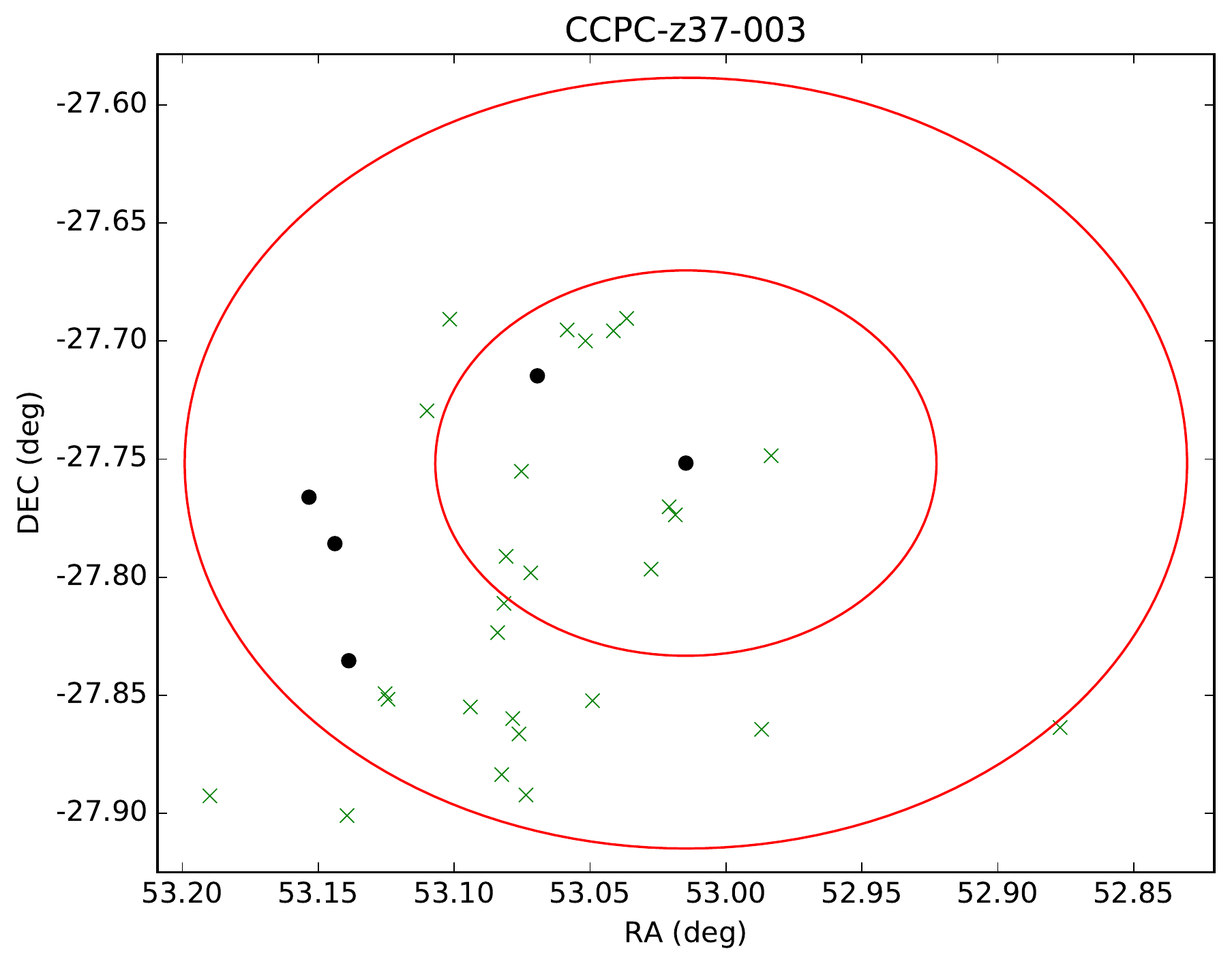}
\label{fig:CCPC-z37-003_sky}
\end{subfigure}
\hfill
\begin{subfigure}
\centering
\includegraphics[scale=0.52]{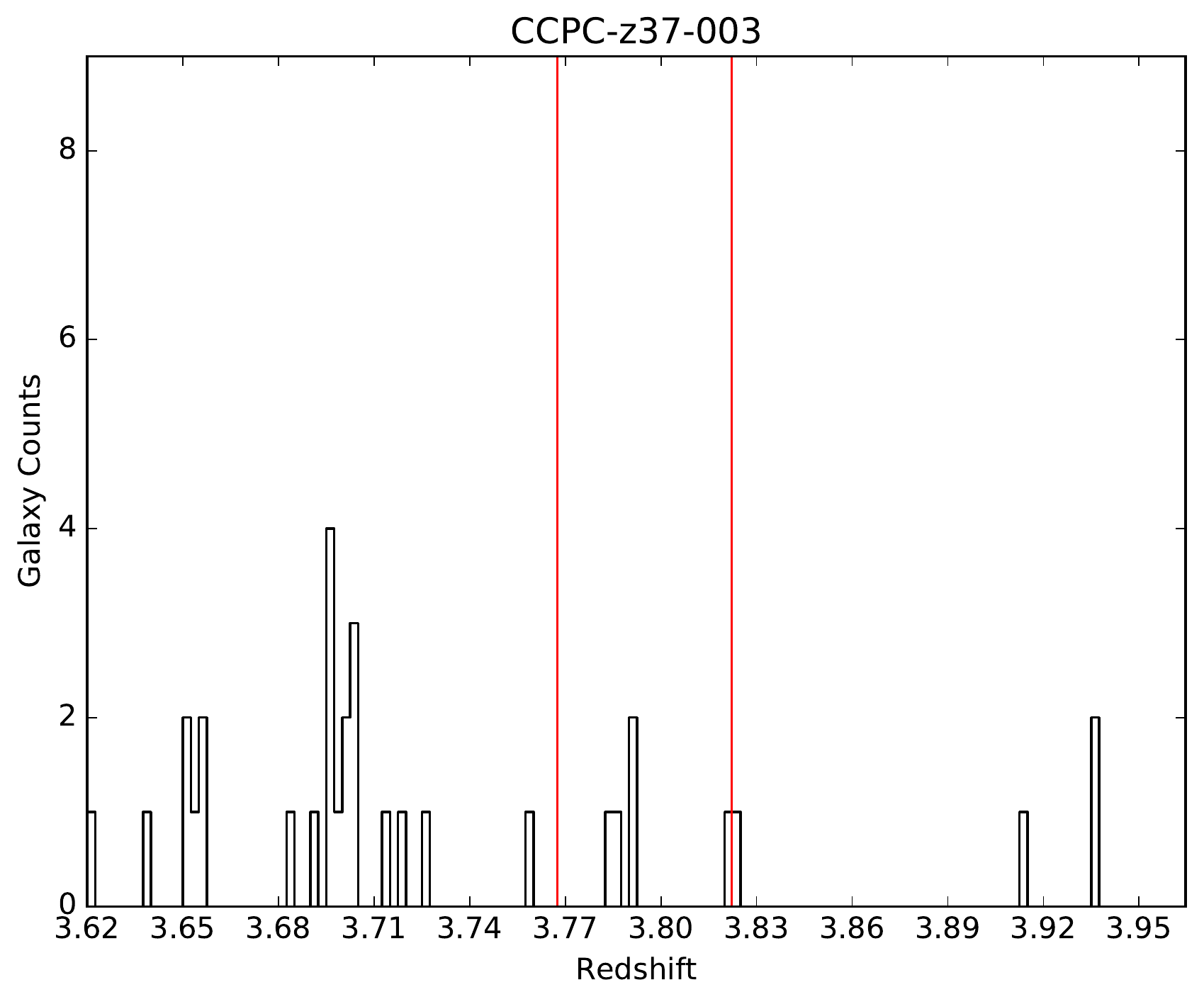}
\label{fig:CCPC-z37-003}
\end{subfigure}
\hfill
\end{figure*}
\clearpage 

\begin{figure*}
\centering
\begin{subfigure}
\centering
\includegraphics[height=7.5cm,width=7.5cm]{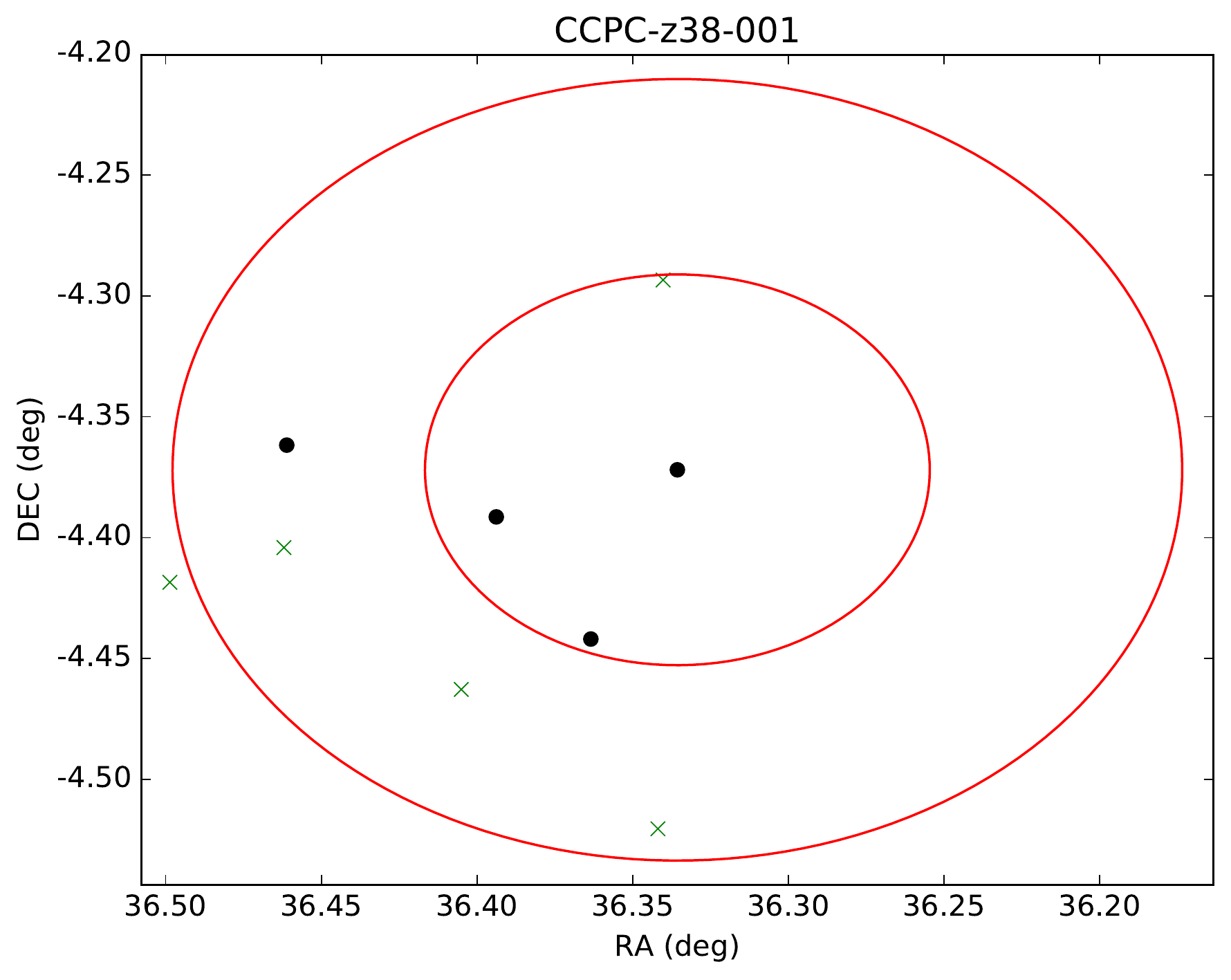}
\label{fig:CCPC-z38-001_sky}
\end{subfigure}
\hfill
\begin{subfigure}
\centering
\includegraphics[scale=0.52]{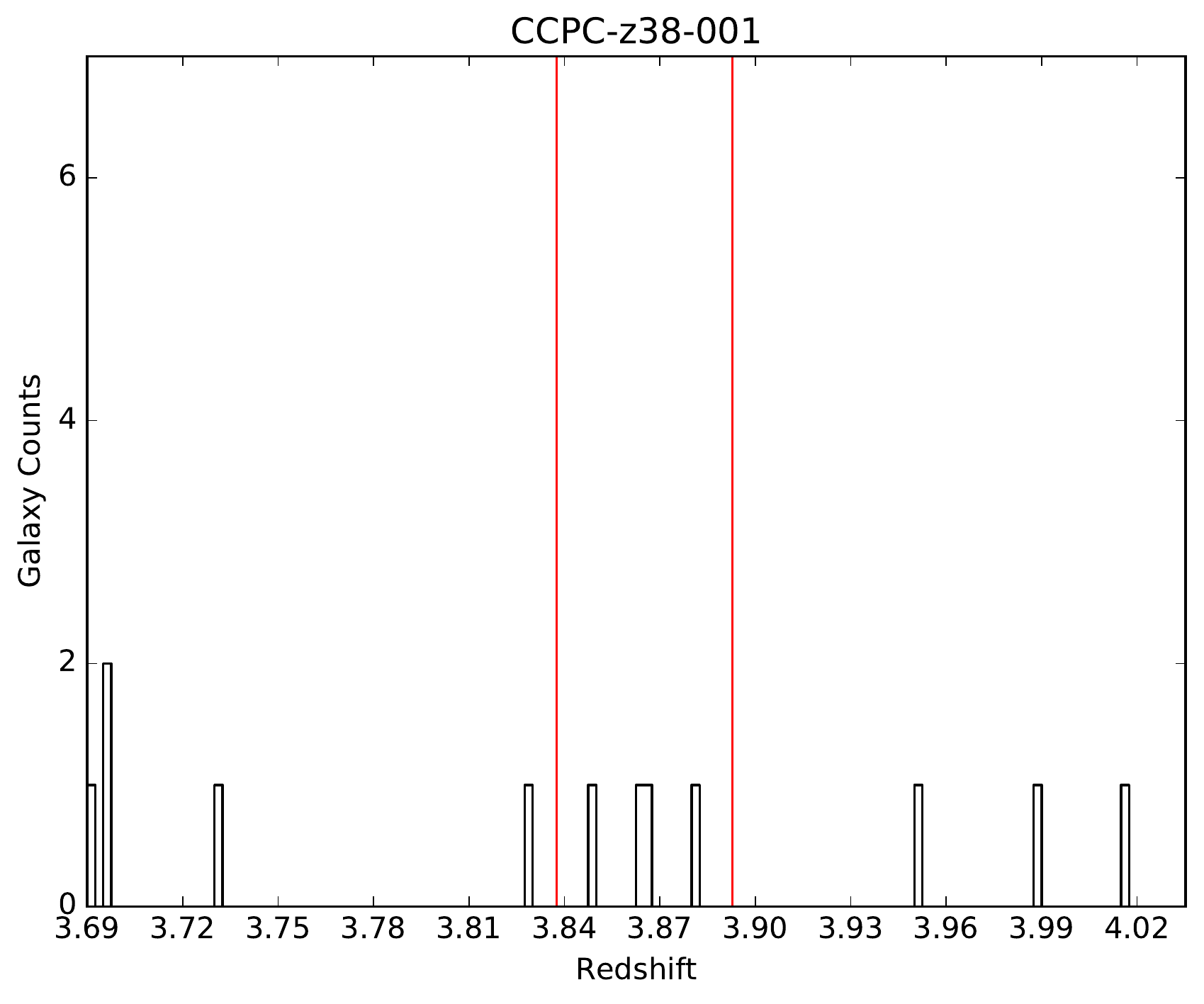}
\label{fig:CCPC-z38-001}
\end{subfigure}
\hfill
\end{figure*}

\begin{figure*}
\centering
\begin{subfigure}
\centering
\includegraphics[height=7.5cm,width=7.5cm]{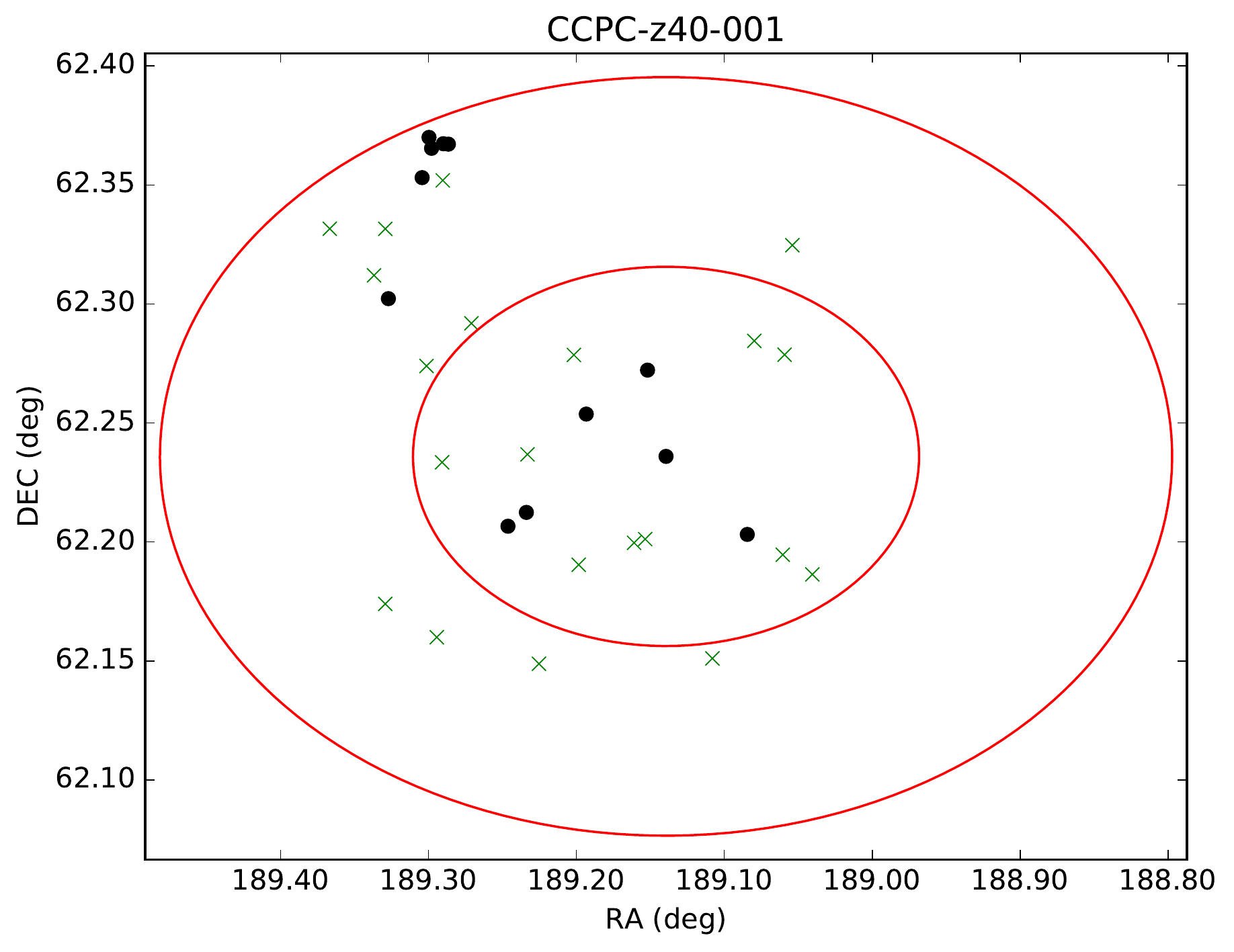}
\label{fig:CCPC-z40-001_sky}
\end{subfigure}
\hfill
\begin{subfigure}
\centering
\includegraphics[scale=0.52]{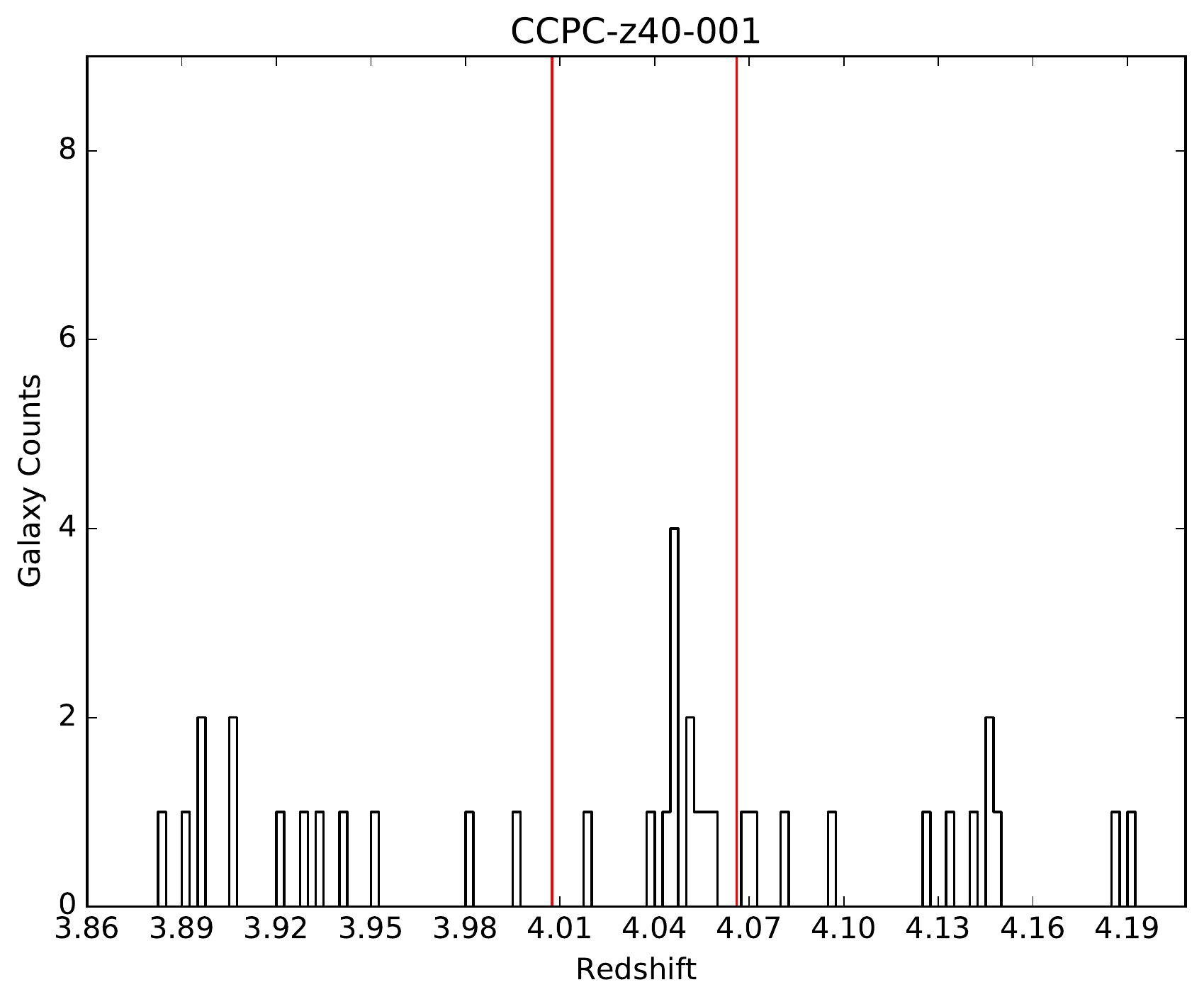}
\label{fig:CCPC-z40-001}
\end{subfigure}
\hfill
\end{figure*}

\begin{figure*}
\centering
\begin{subfigure}
\centering
\includegraphics[height=7.5cm,width=7.5cm]{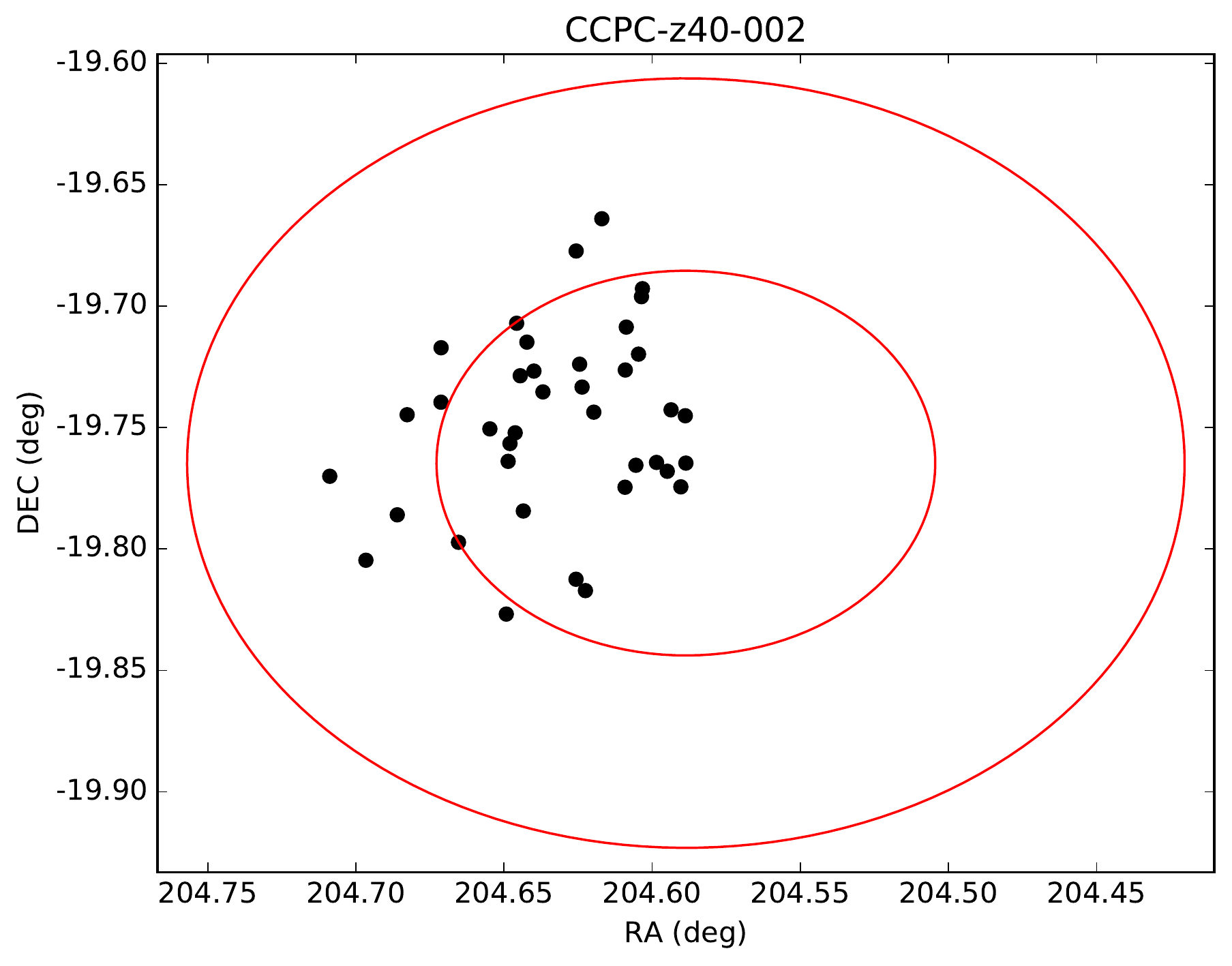}
\label{fig:CCPC-z40-002_sky}
\end{subfigure}
\hfill
\begin{subfigure}
\centering
\includegraphics[scale=0.52]{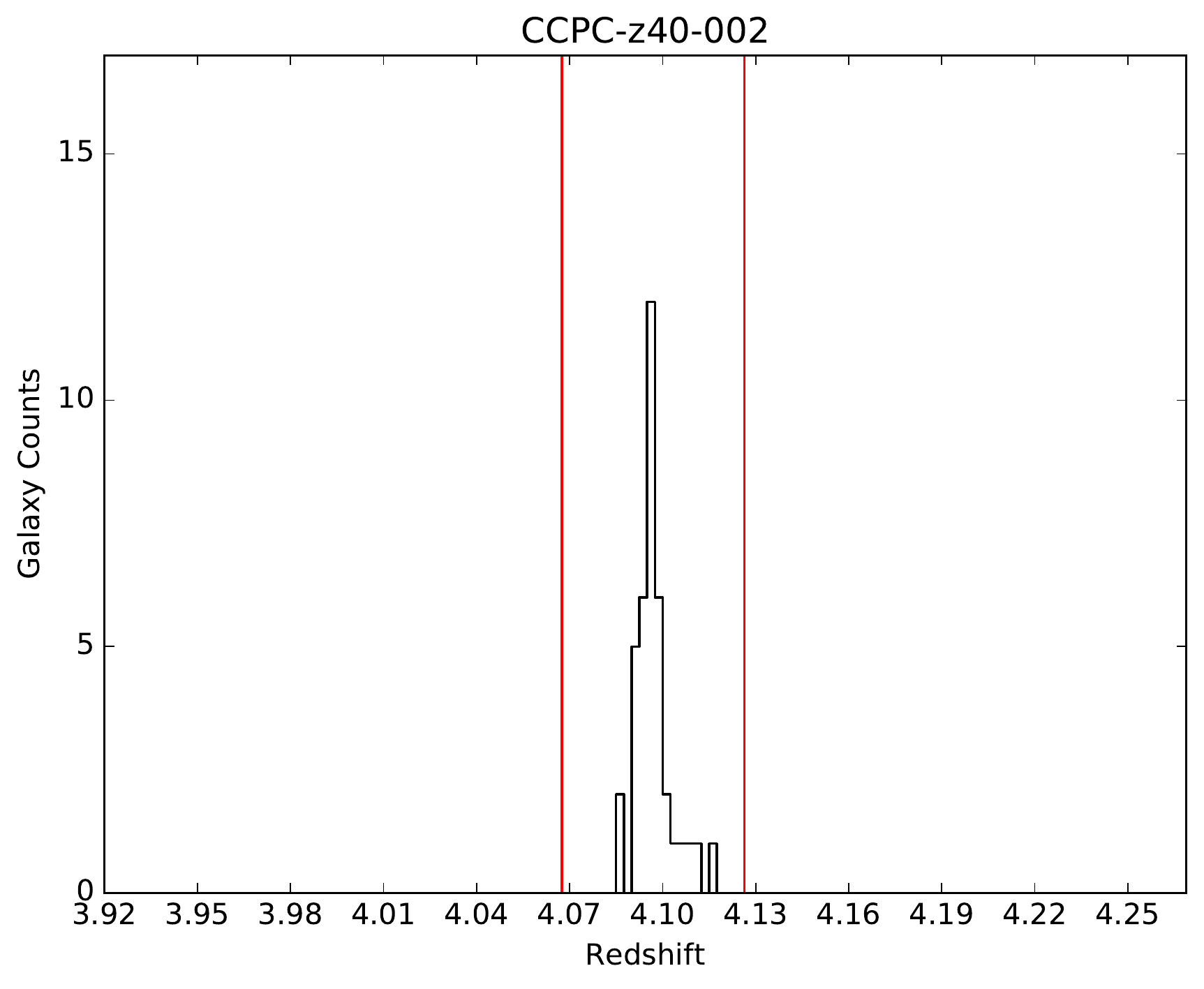}
\label{fig:CCPC-z40-002}
\end{subfigure}
\hfill
\end{figure*}
\clearpage 

\begin{figure*}
\centering
\begin{subfigure}
\centering
\includegraphics[height=7.5cm,width=7.5cm]{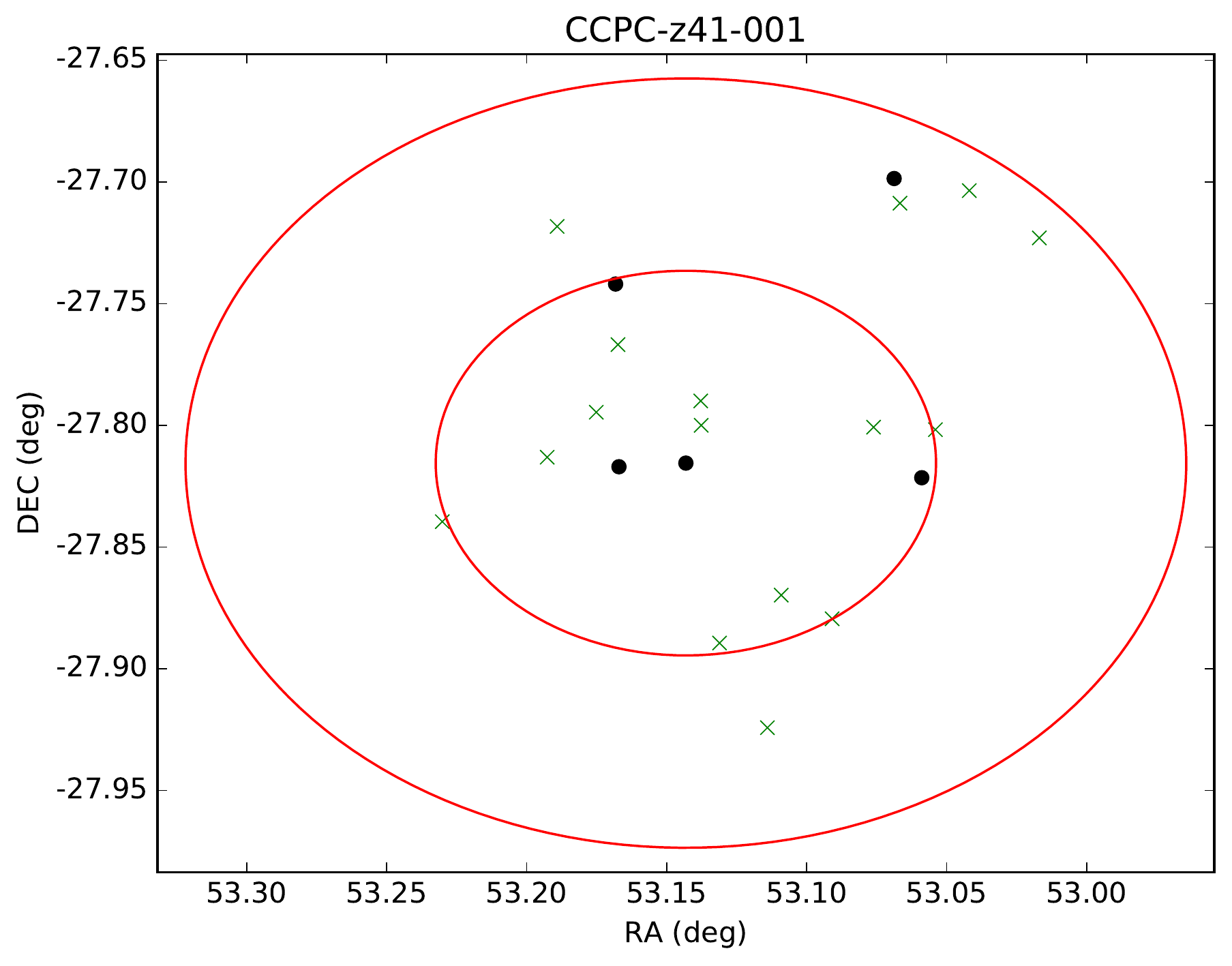}
\label{fig:CCPC-z41-001_sky}
\end{subfigure}
\hfill
\begin{subfigure}
\centering
\includegraphics[scale=0.52]{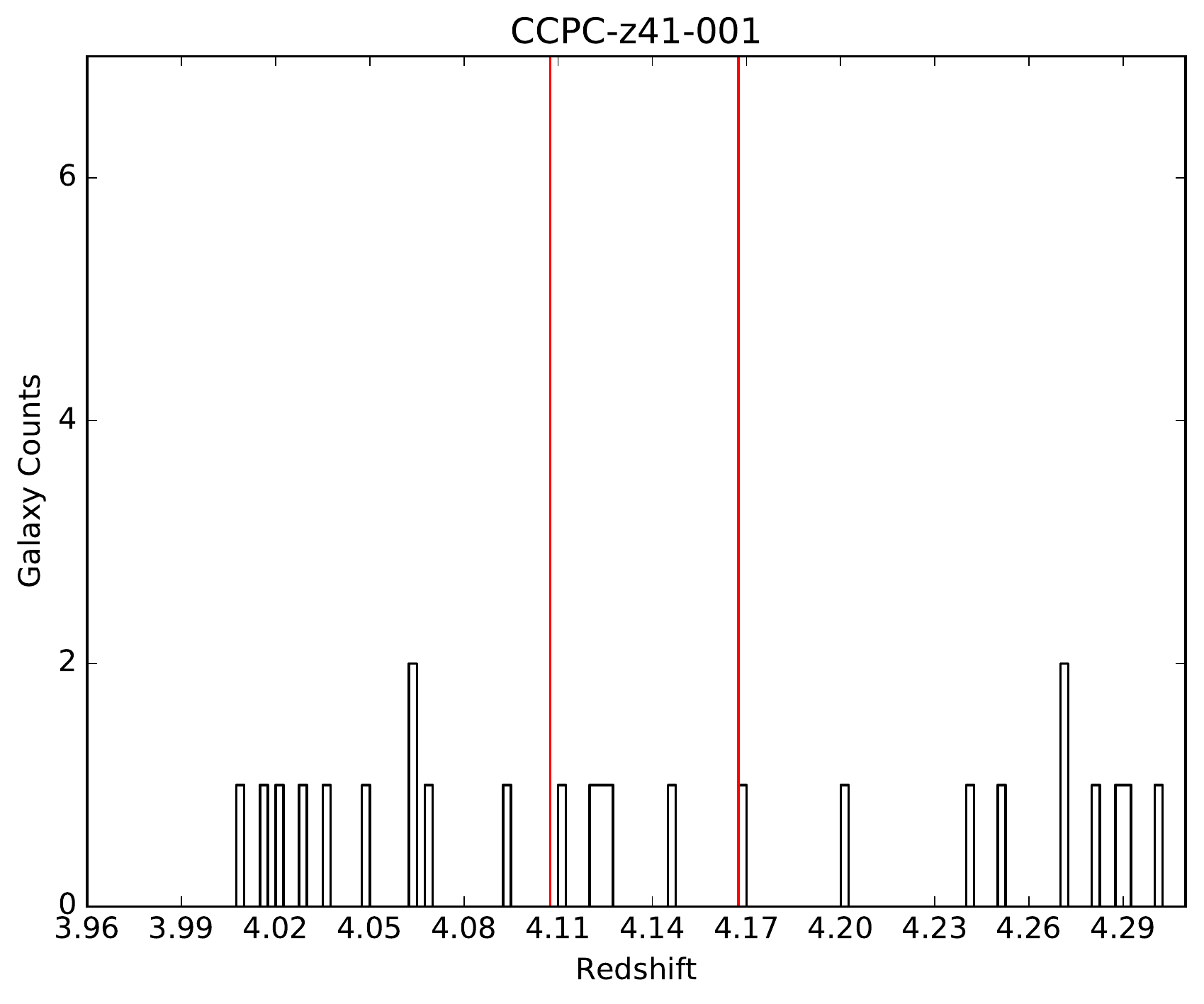}
\label{fig:CCPC-z41-001}
\end{subfigure}
\hfill
\end{figure*}

\begin{figure*}
\centering
\begin{subfigure}
\centering
\includegraphics[height=7.5cm,width=7.5cm]{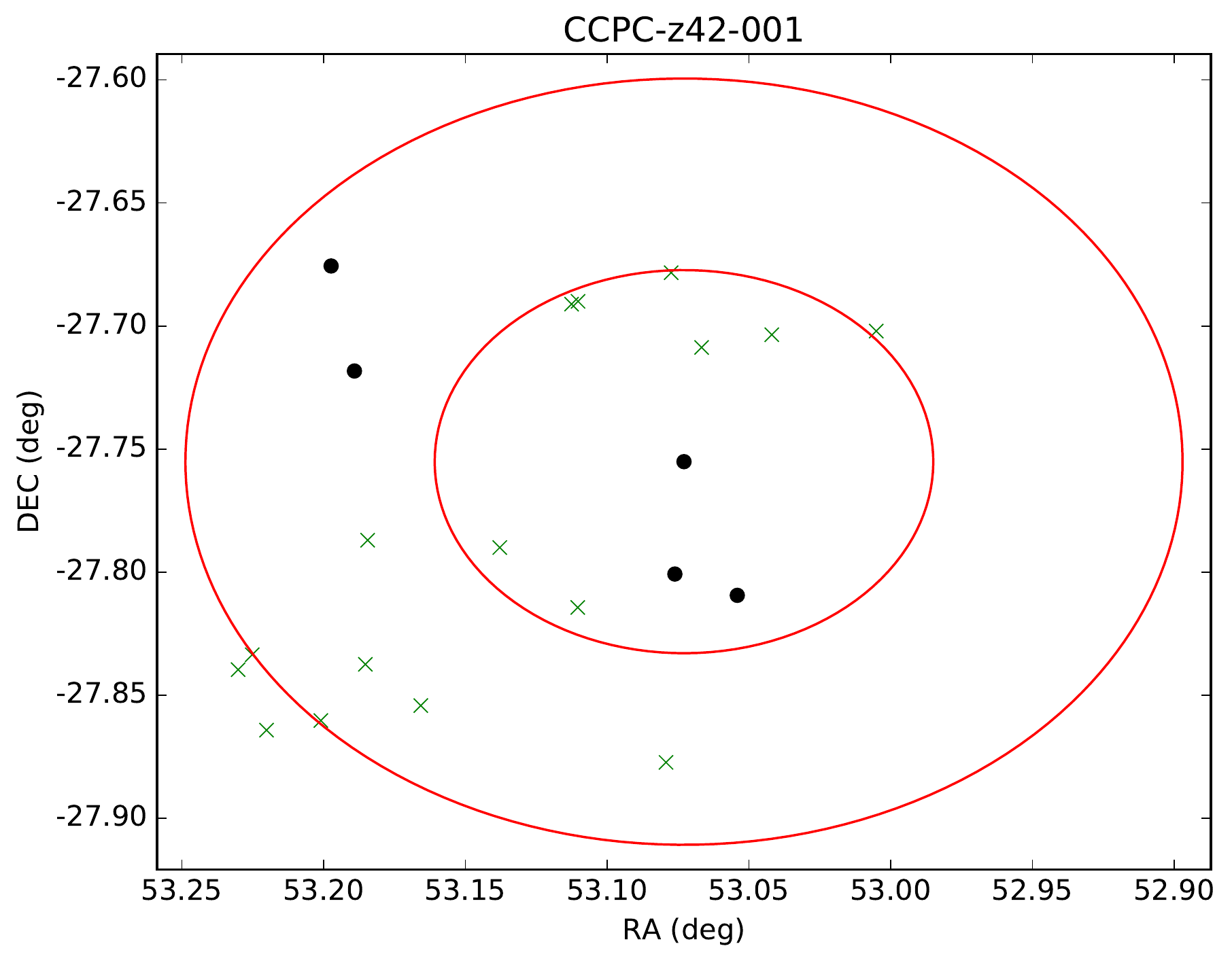}
\label{fig:CCPC-z42-001_sky}
\end{subfigure}
\hfill
\begin{subfigure}
\centering
\includegraphics[scale=0.52]{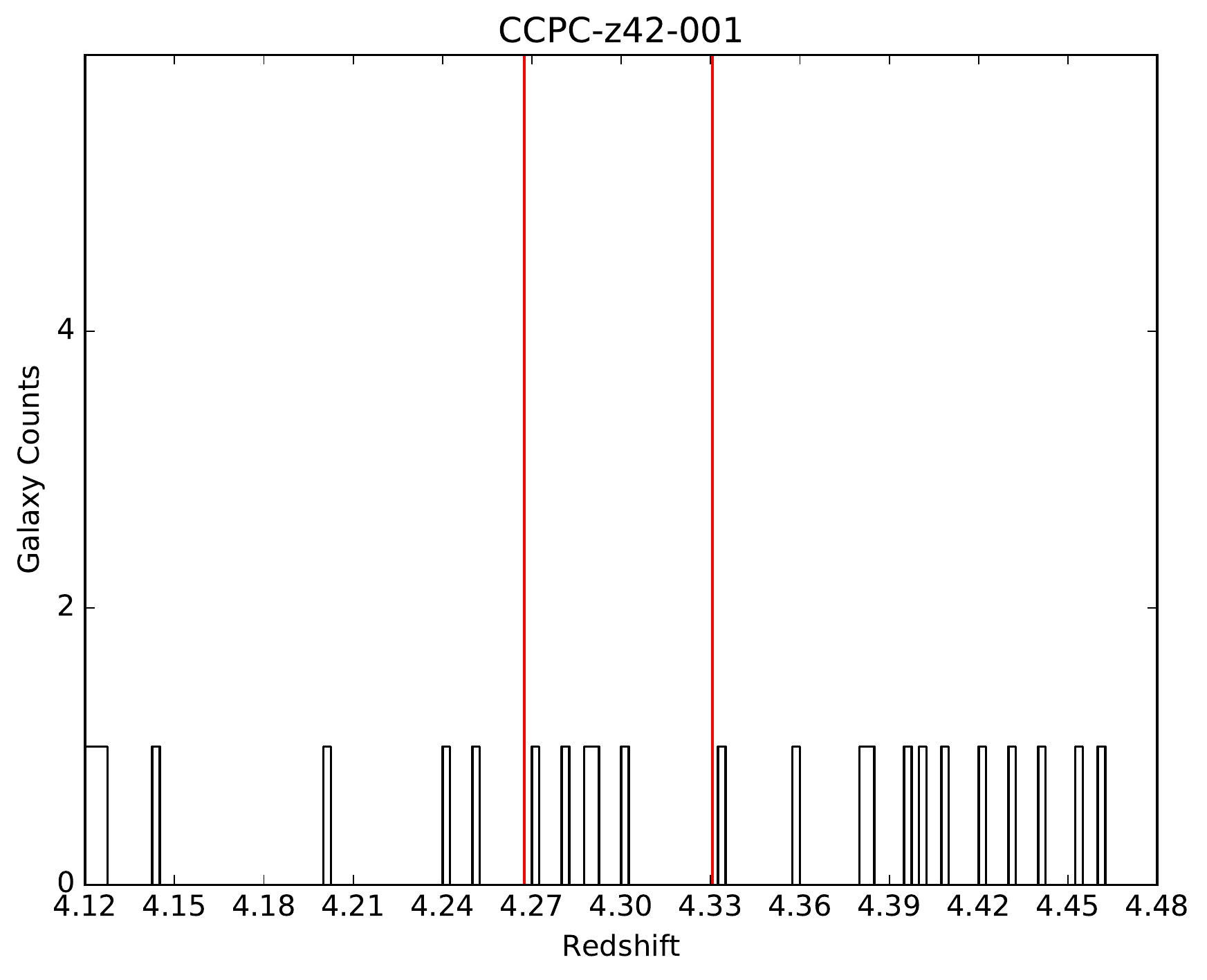}
\label{fig:CCPC-z42-001}
\end{subfigure}
\hfill
\end{figure*}

\begin{figure*}
\centering
\begin{subfigure}
\centering
\includegraphics[height=7.5cm,width=7.5cm]{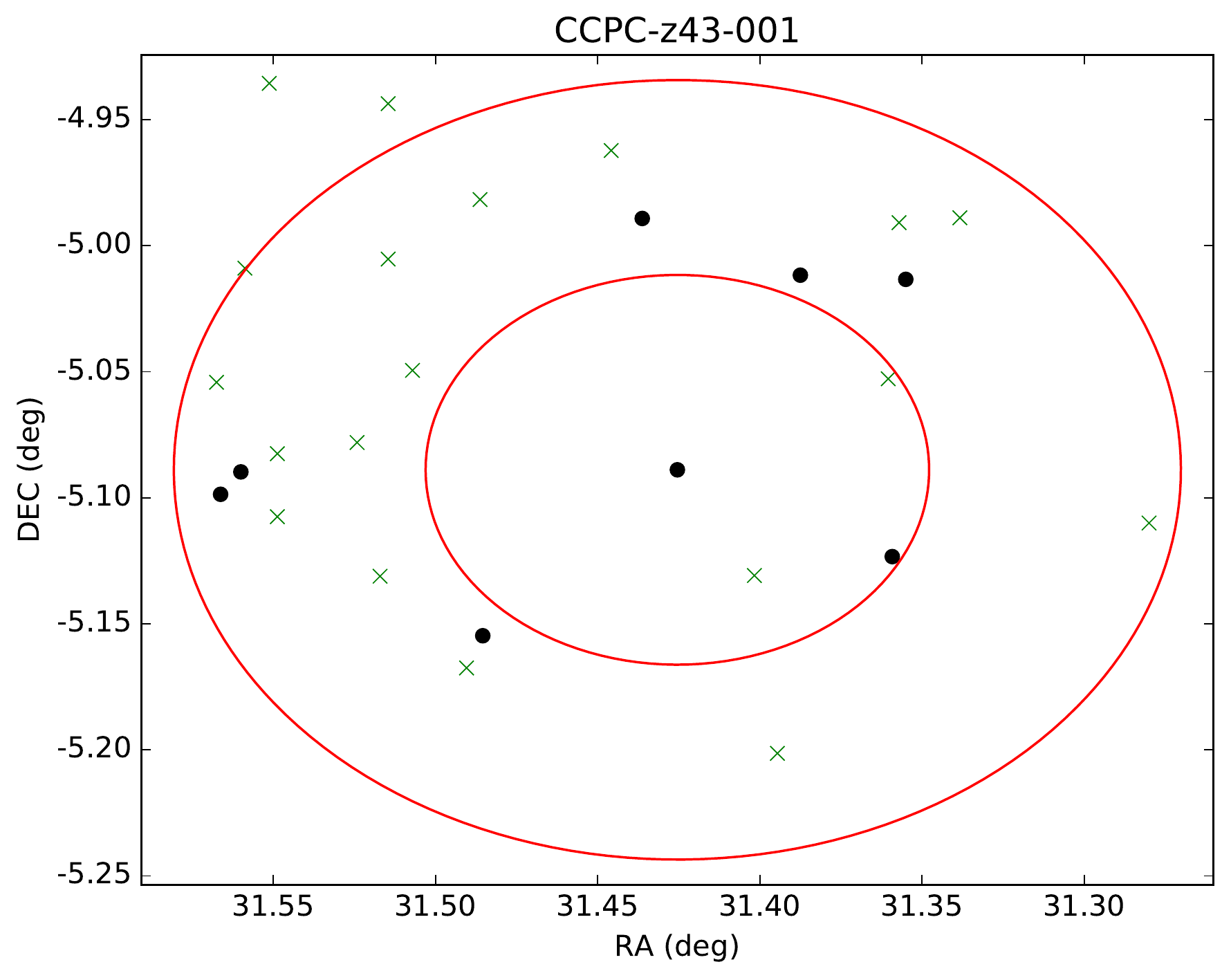}
\label{fig:CCPC-z43-001_sky}
\end{subfigure}
\hfill
\begin{subfigure}
\centering
\includegraphics[scale=0.52]{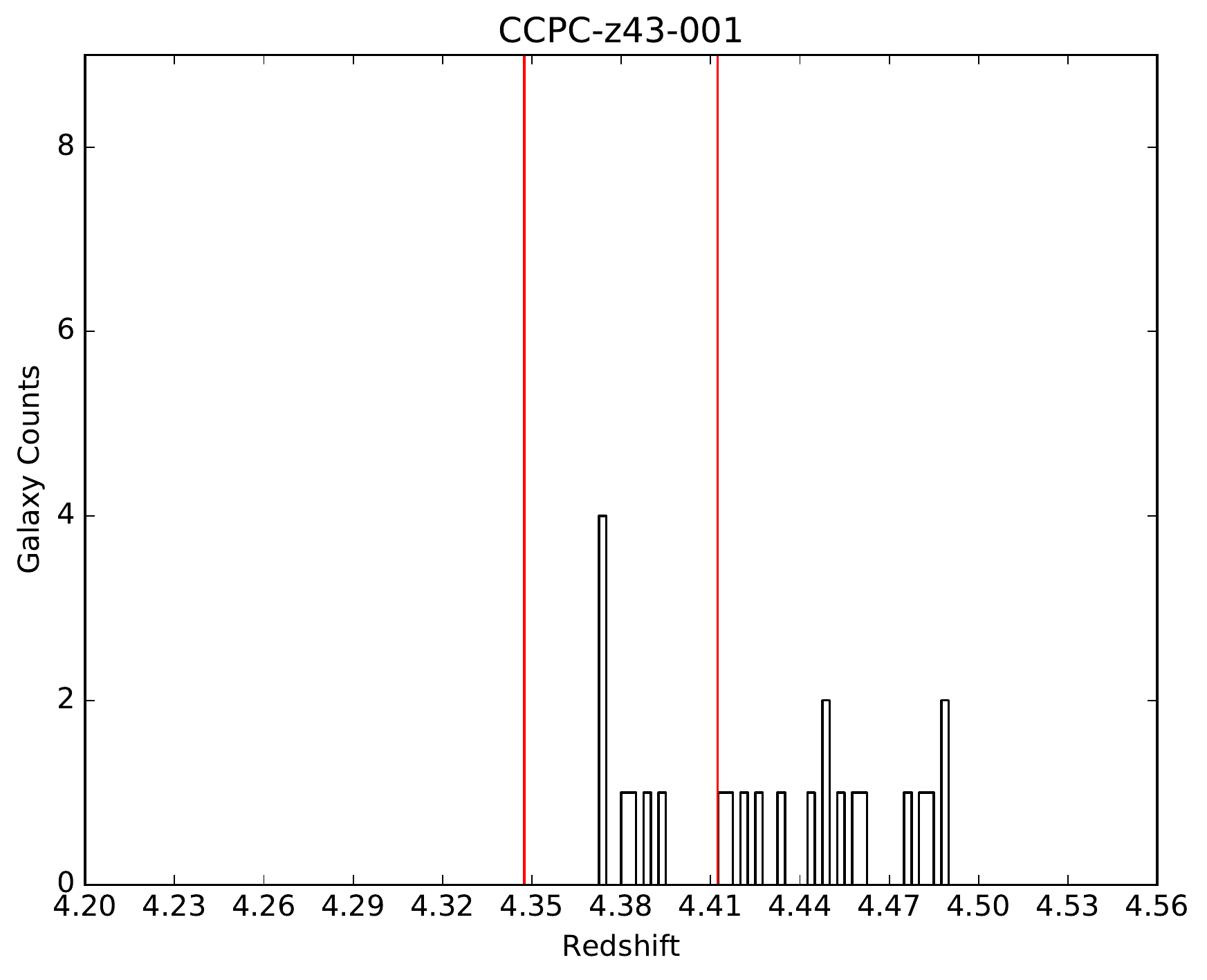}
\label{fig:CCPC-z43-001}
\end{subfigure}
\hfill
\end{figure*}
\clearpage 

\begin{figure*}
\centering
\begin{subfigure}
\centering
\includegraphics[height=7.5cm,width=7.5cm]{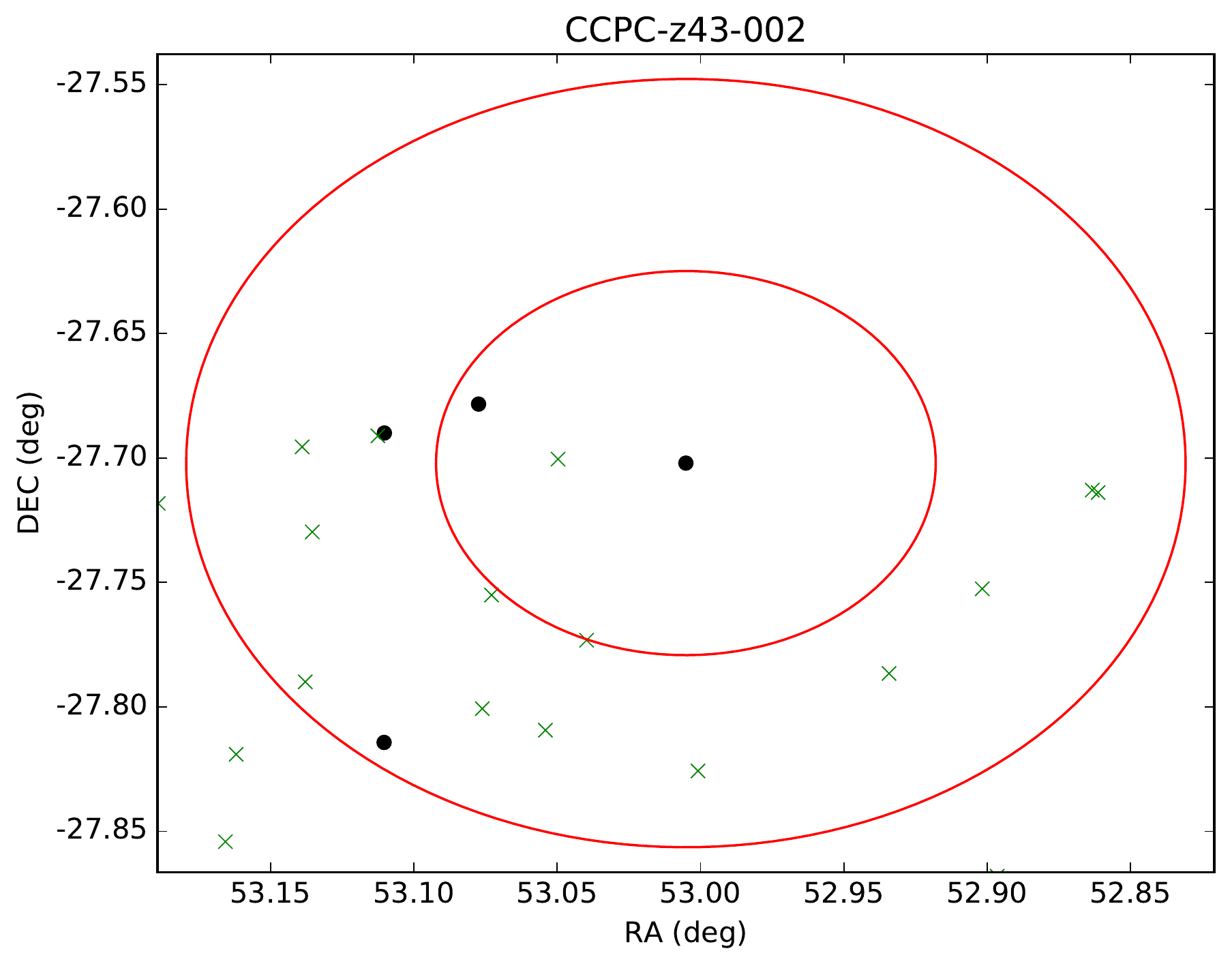}
\label{fig:CCPC-z43-002_sky}
\end{subfigure}
\hfill
\begin{subfigure}
\centering
\includegraphics[scale=0.52]{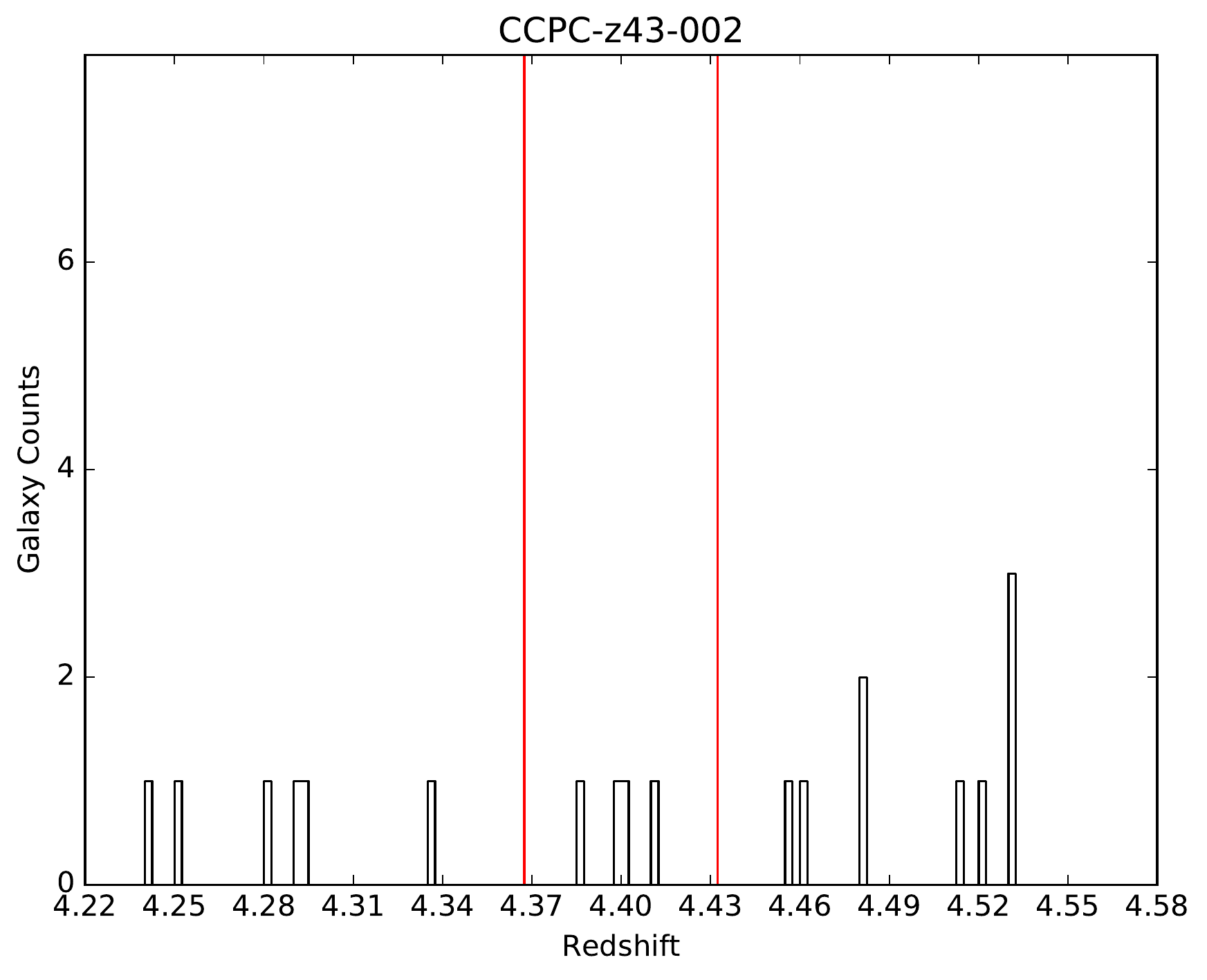}
\label{fig:CCPC-z43-002}
\end{subfigure}
\hfill
\end{figure*}

\begin{figure*}
\centering
\begin{subfigure}
\centering
\includegraphics[height=7.5cm,width=7.5cm]{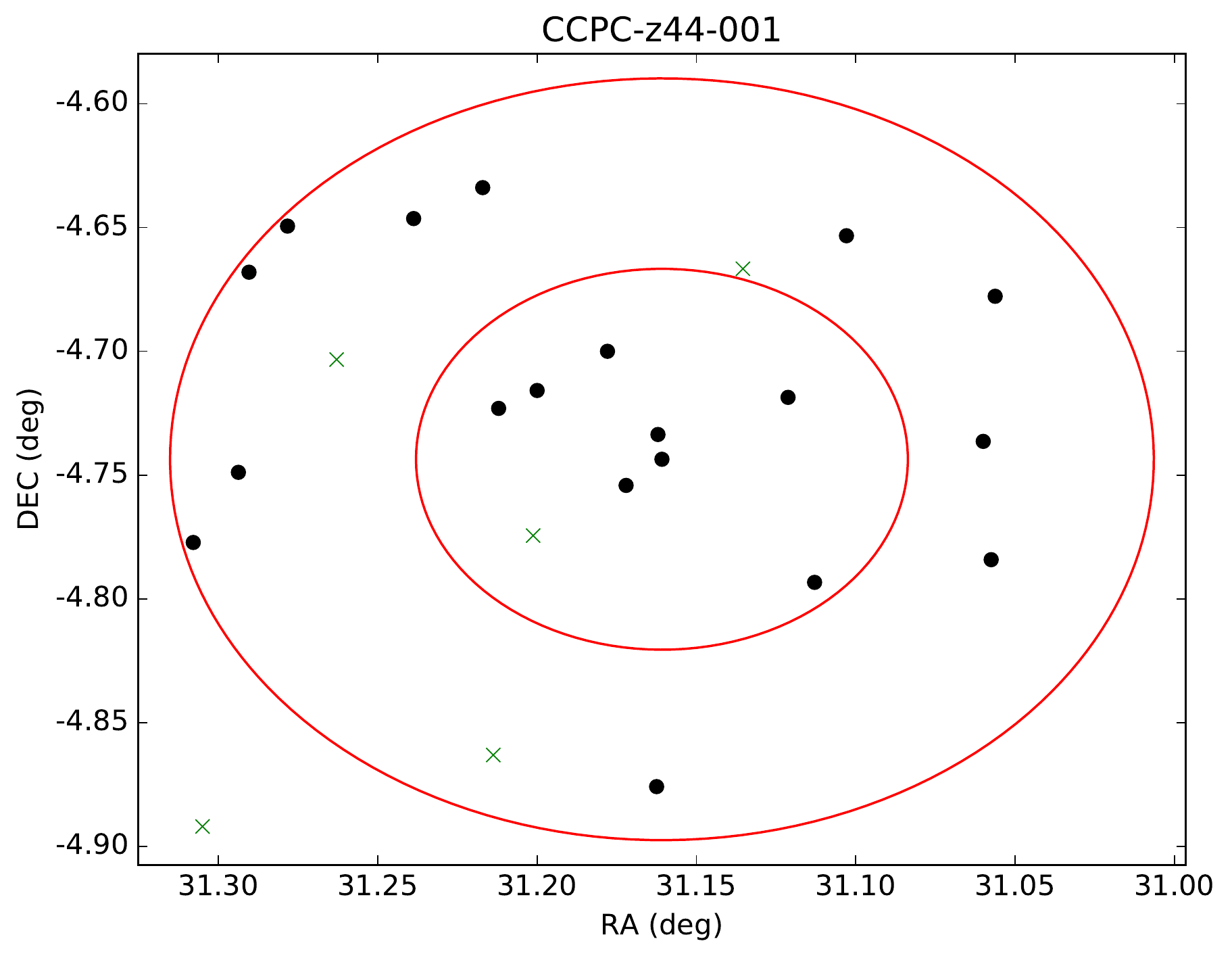}
\label{fig:CCPC-z44-001_sky}
\end{subfigure}
\hfill
\begin{subfigure}
\centering
\includegraphics[scale=0.52]{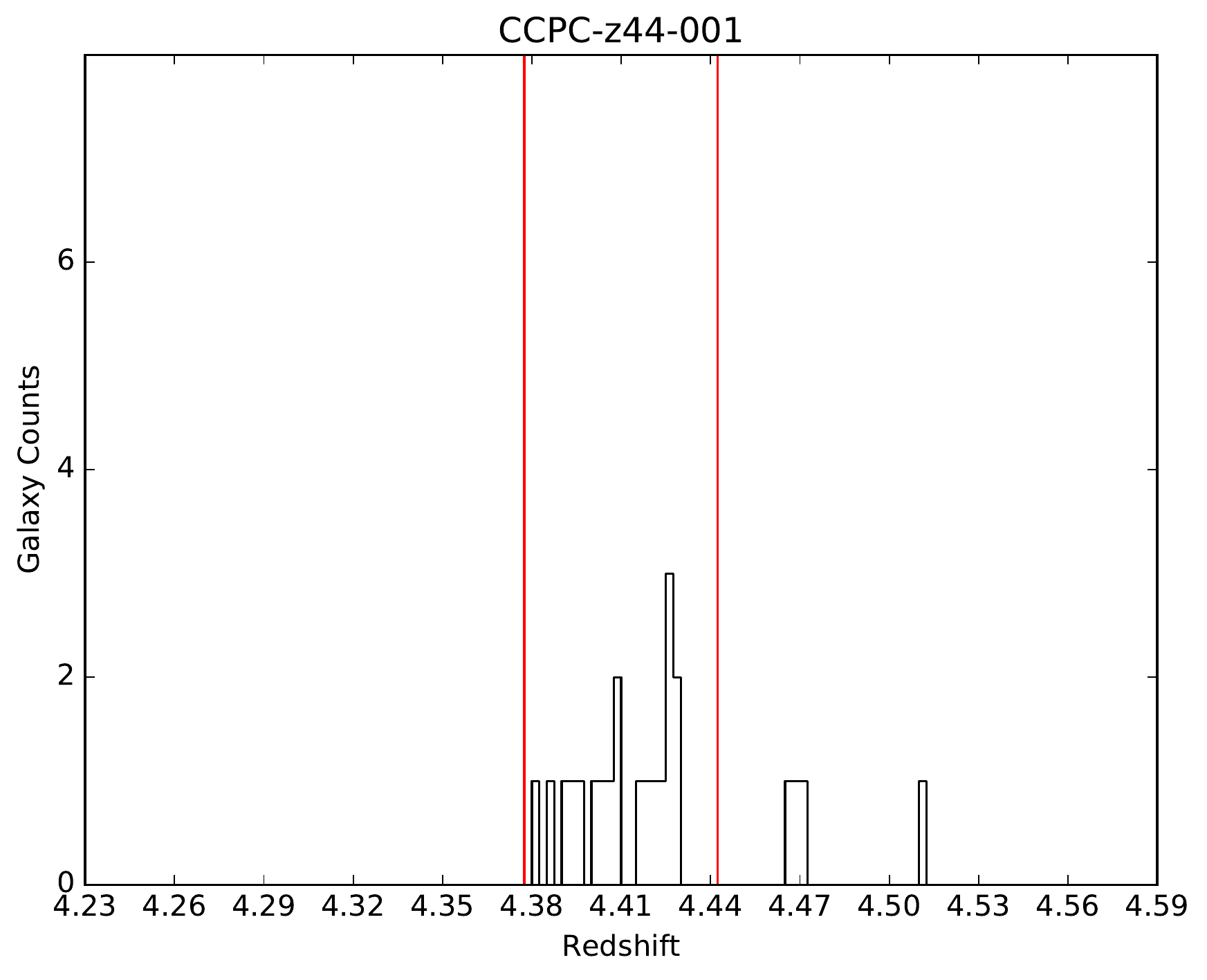}
\label{fig:CCPC-z44-001}
\end{subfigure}
\hfill
\end{figure*}

\begin{figure*}
\centering
\begin{subfigure}
\centering
\includegraphics[height=7.5cm,width=7.5cm]{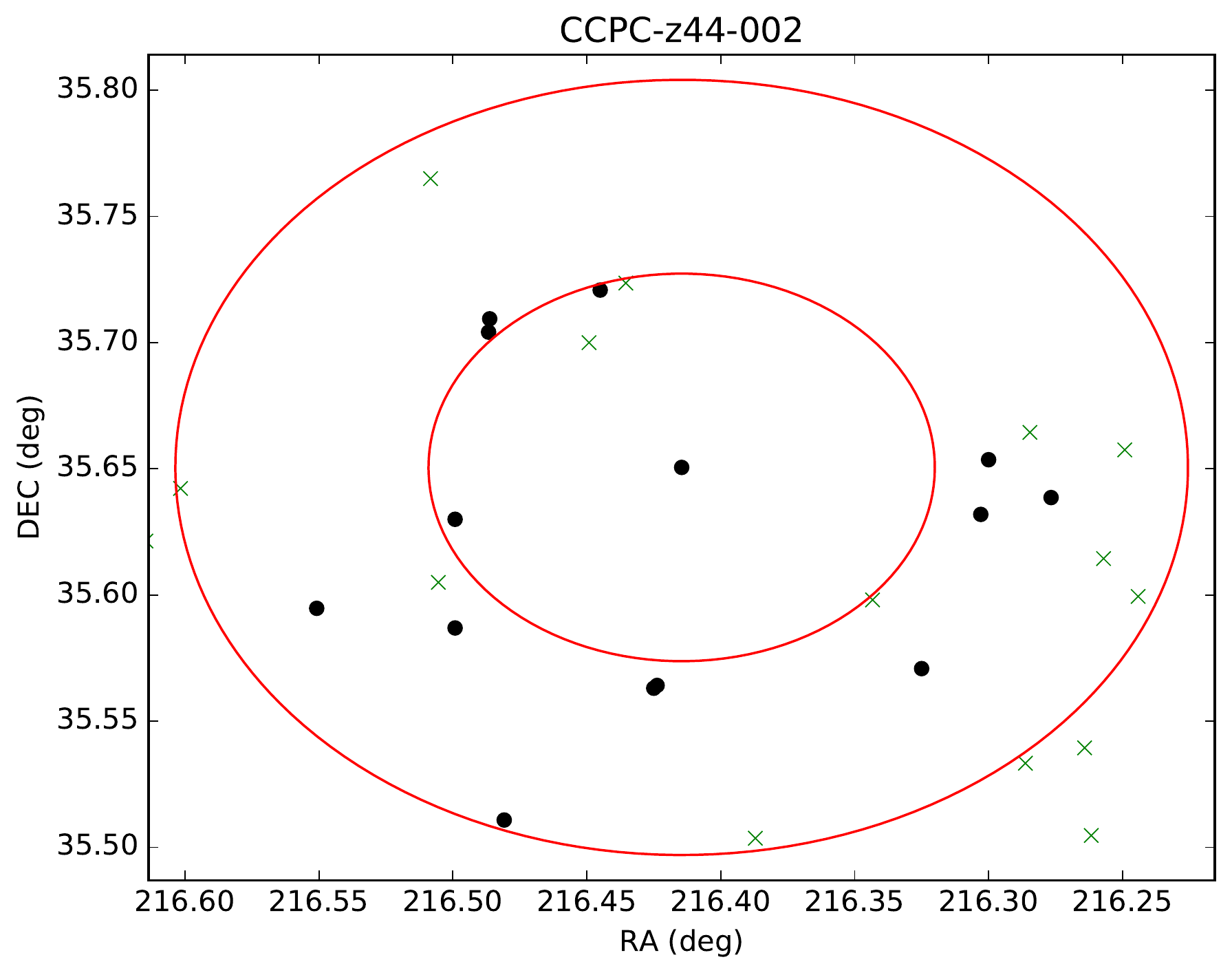}
\label{fig:CCPC-z44-002_sky}
\end{subfigure}
\hfill
\begin{subfigure}
\centering
\includegraphics[scale=0.52]{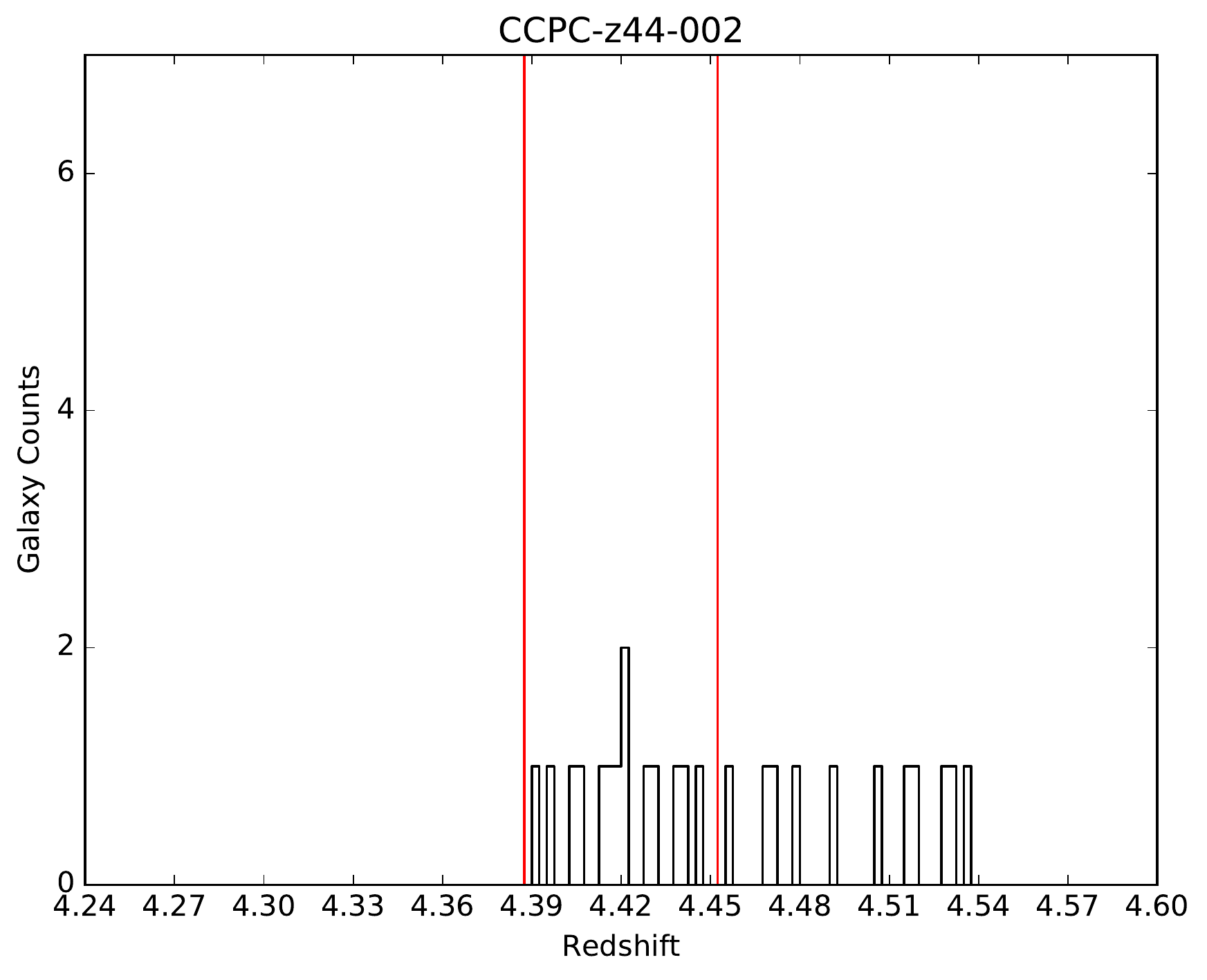}
\label{fig:CCPC-z44-002}
\end{subfigure}
\hfill
\end{figure*}
\clearpage 

\begin{figure*}
\centering
\begin{subfigure}
\centering
\includegraphics[height=7.5cm,width=7.5cm]{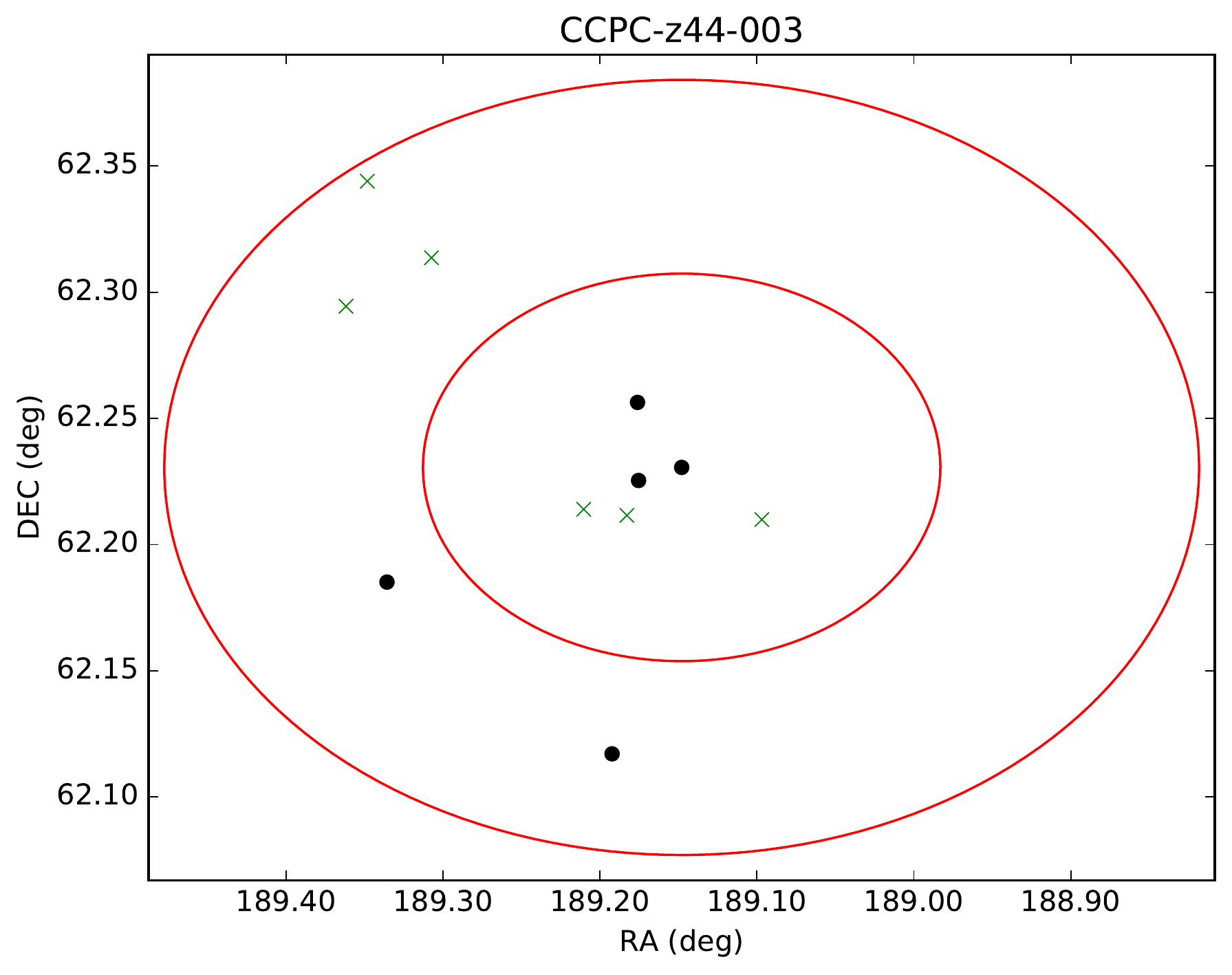}
\label{fig:CCPC-z44-003_sky}
\end{subfigure}
\hfill
\begin{subfigure}
\centering
\includegraphics[scale=0.52]{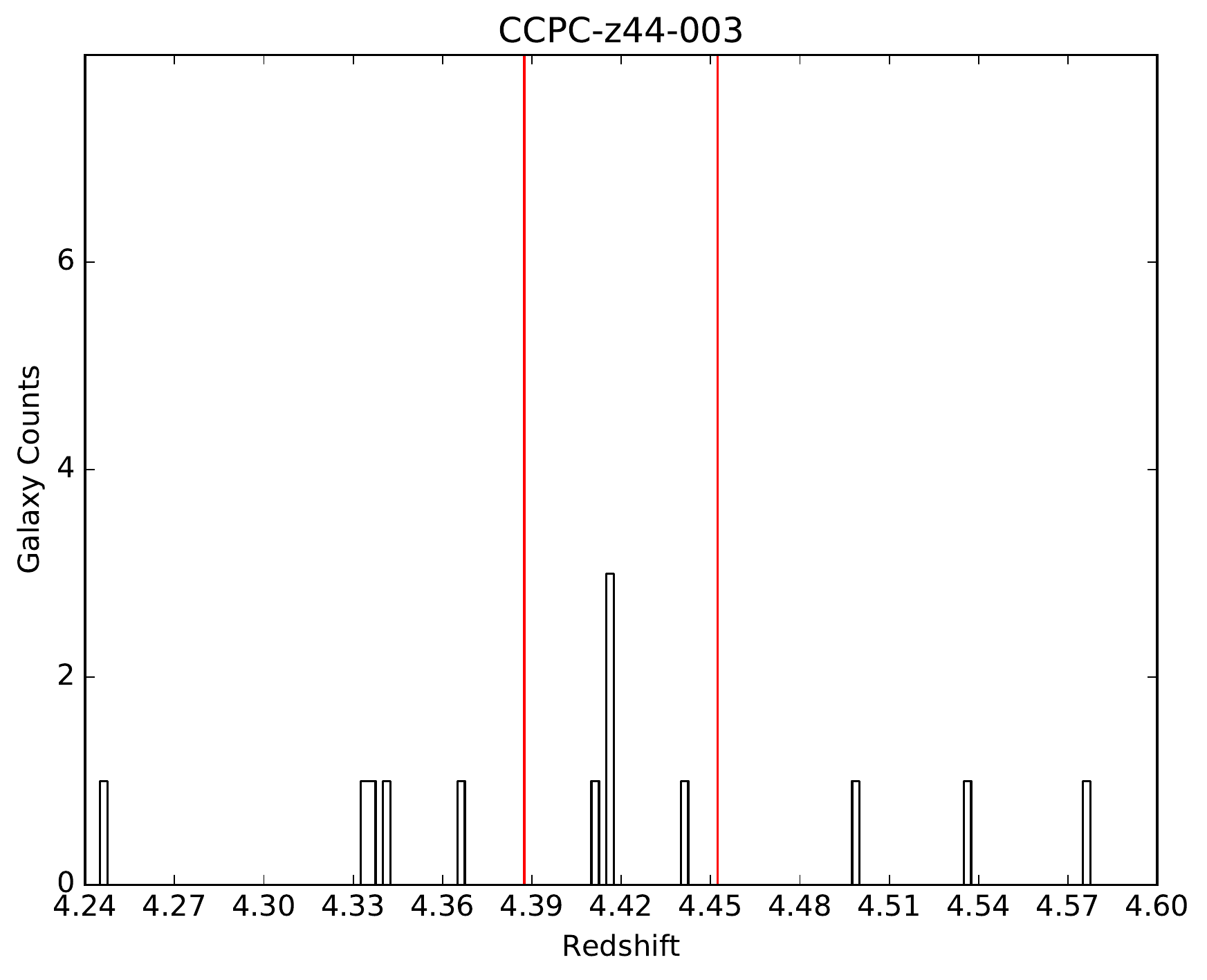}
\label{fig:CCPC-z44-003}
\end{subfigure}
\hfill
\end{figure*}

\begin{figure*}
\centering
\begin{subfigure}
\centering
\includegraphics[height=7.5cm,width=7.5cm]{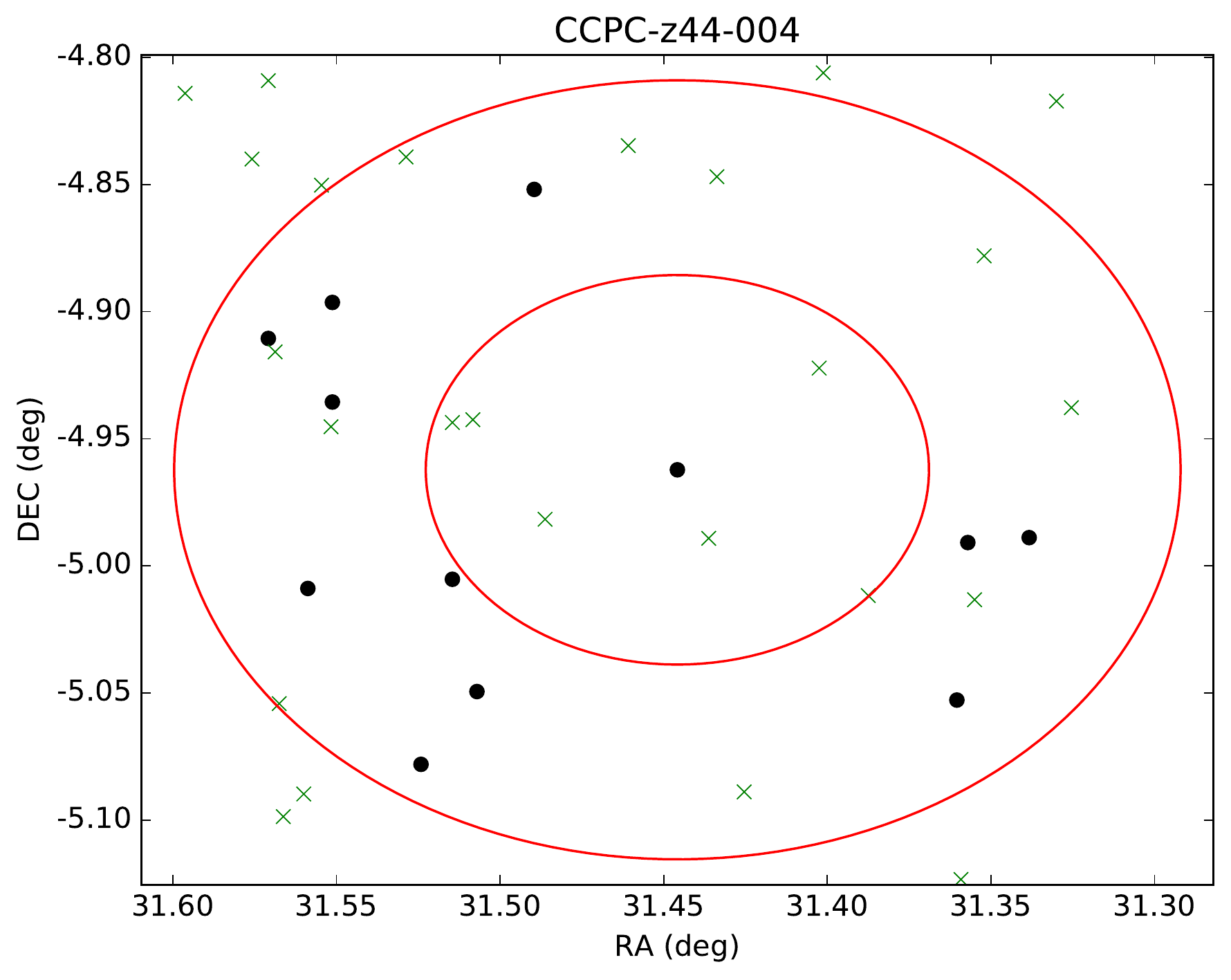}
\label{fig:CCPC-z44-004_sky}
\end{subfigure}
\hfill
\begin{subfigure}
\centering
\includegraphics[scale=0.52]{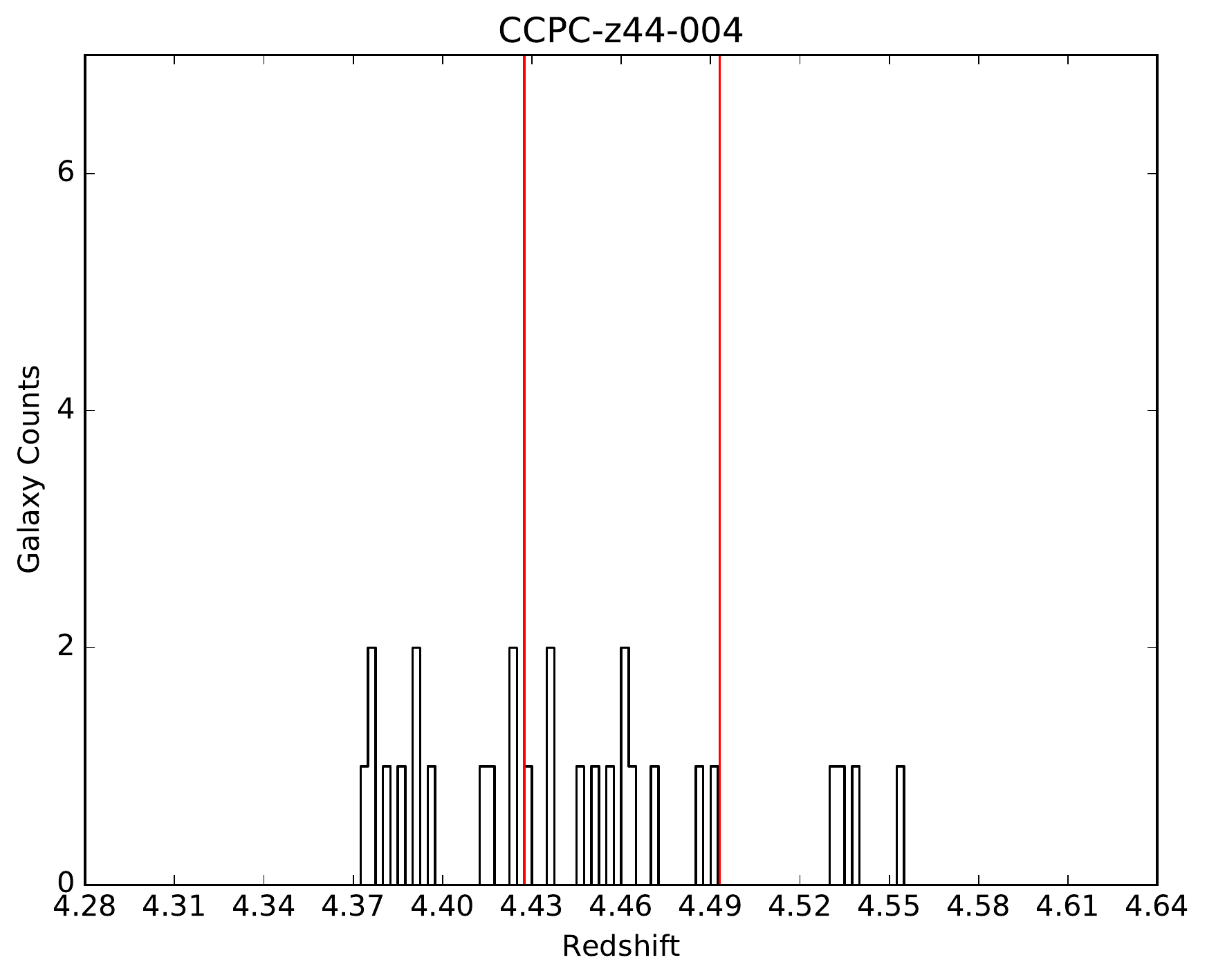}
\label{fig:CCPC-z44-004}
\end{subfigure}
\hfill
\end{figure*}

\begin{figure*}
\centering
\begin{subfigure}
\centering
\includegraphics[height=7.5cm,width=7.5cm]{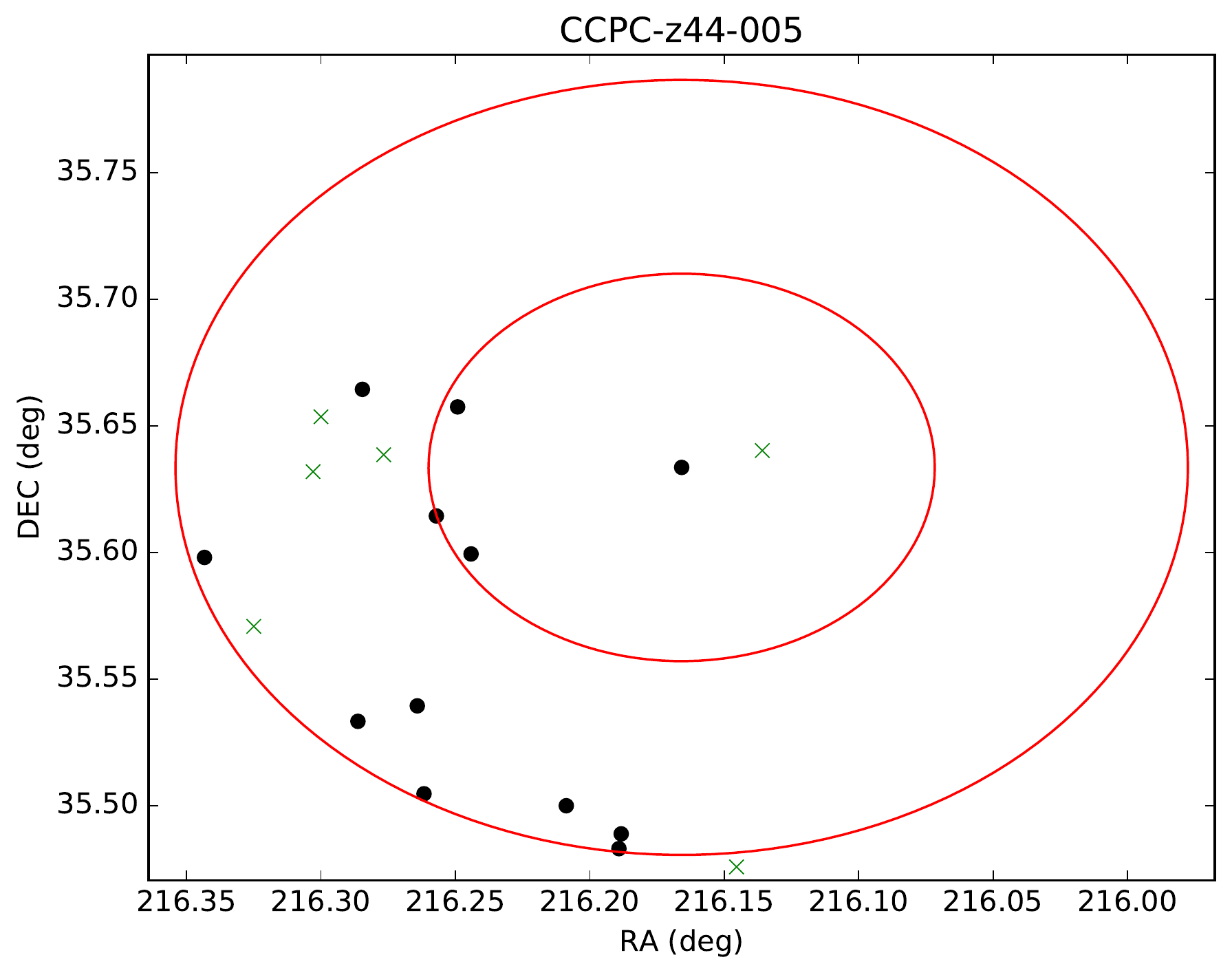}
\label{fig:CCPC-z44-005_sky}
\end{subfigure}
\hfill
\begin{subfigure}
\centering
\includegraphics[scale=0.52]{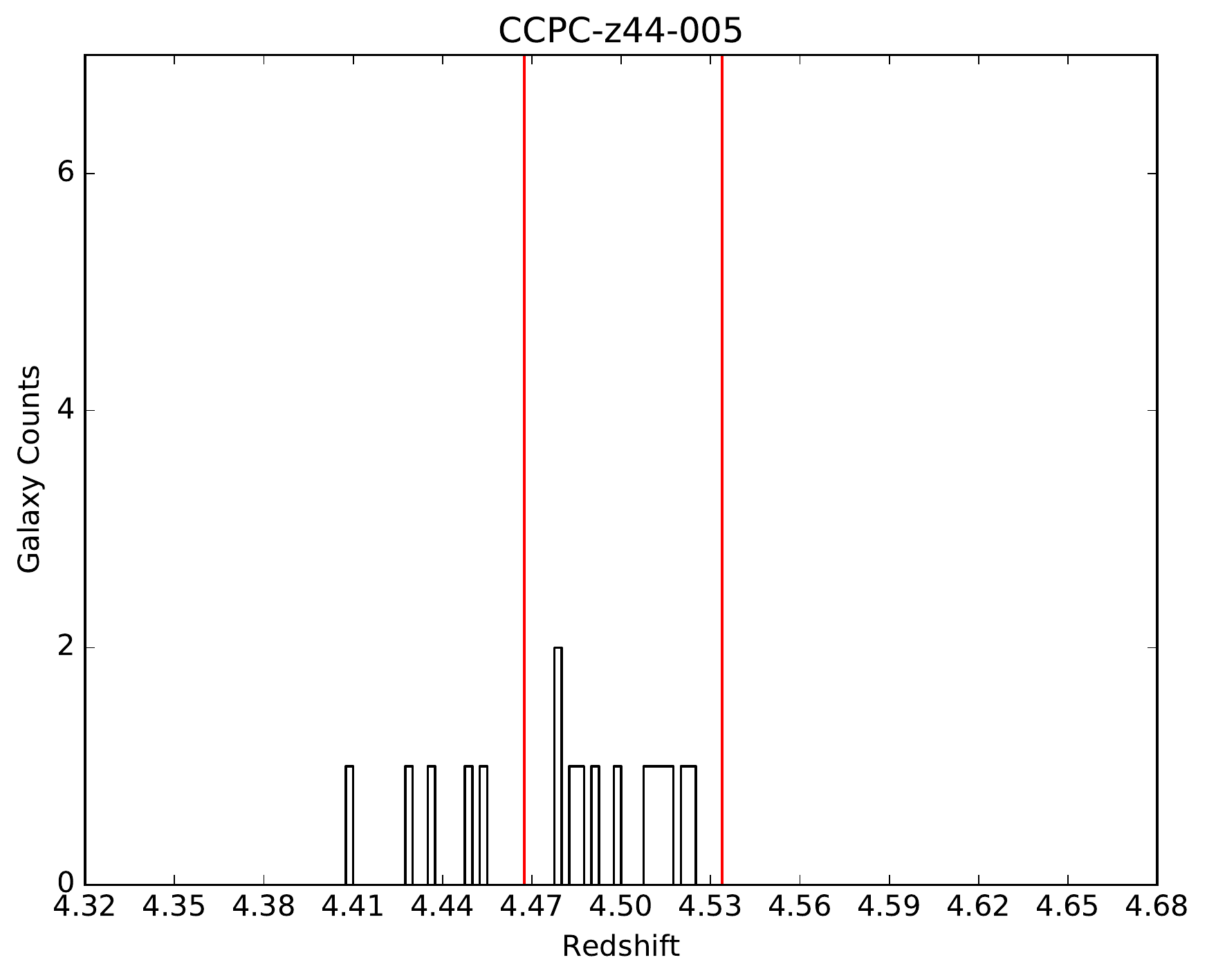}
\label{fig:CCPC-z44-005}
\end{subfigure}
\hfill
\end{figure*}
\clearpage 

\begin{figure*}
\centering
\begin{subfigure}
\centering
\includegraphics[height=7.5cm,width=7.5cm]{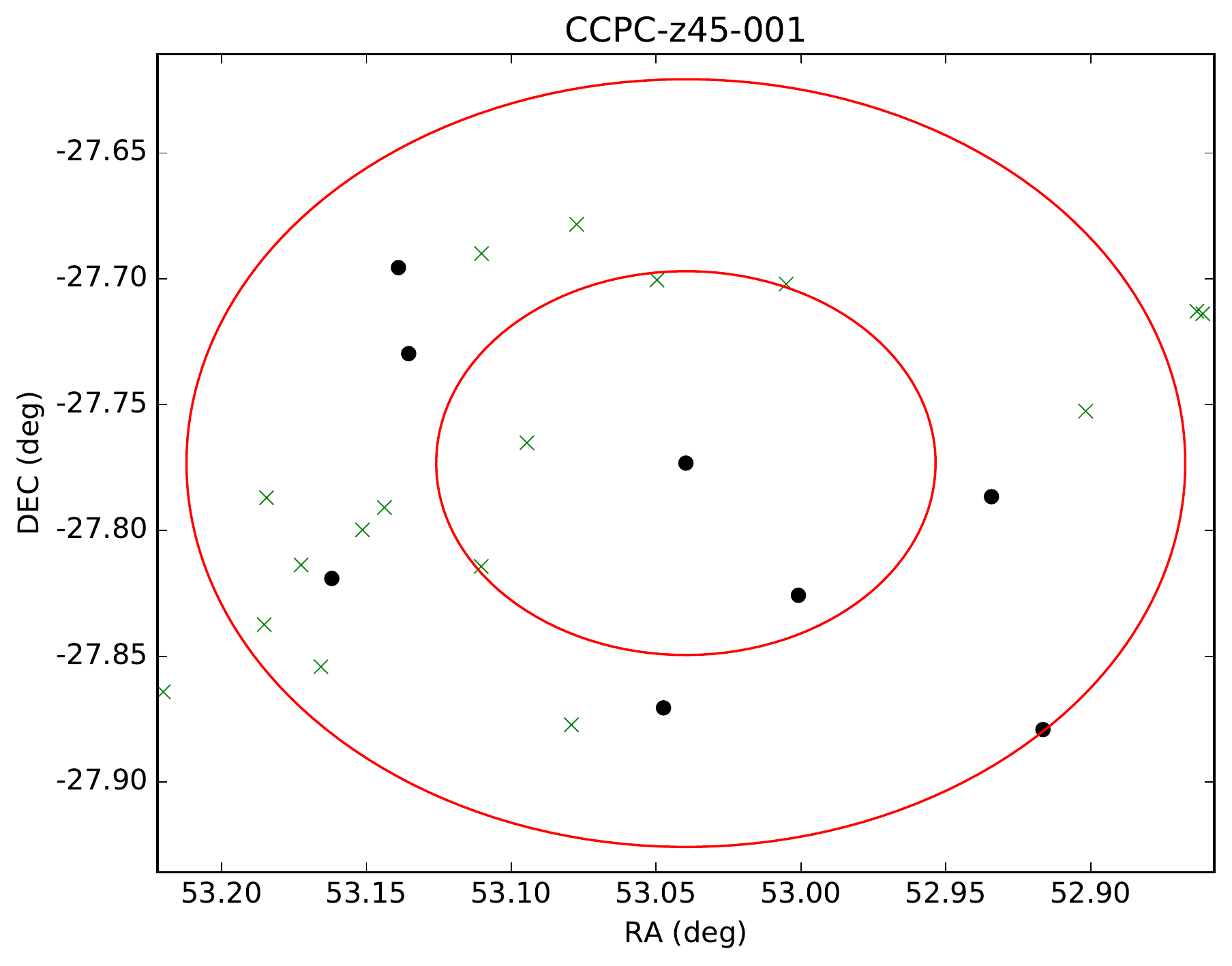}
\label{fig:CCPC-z45-001_sky}
\end{subfigure}
\hfill
\begin{subfigure}
\centering
\includegraphics[scale=0.52]{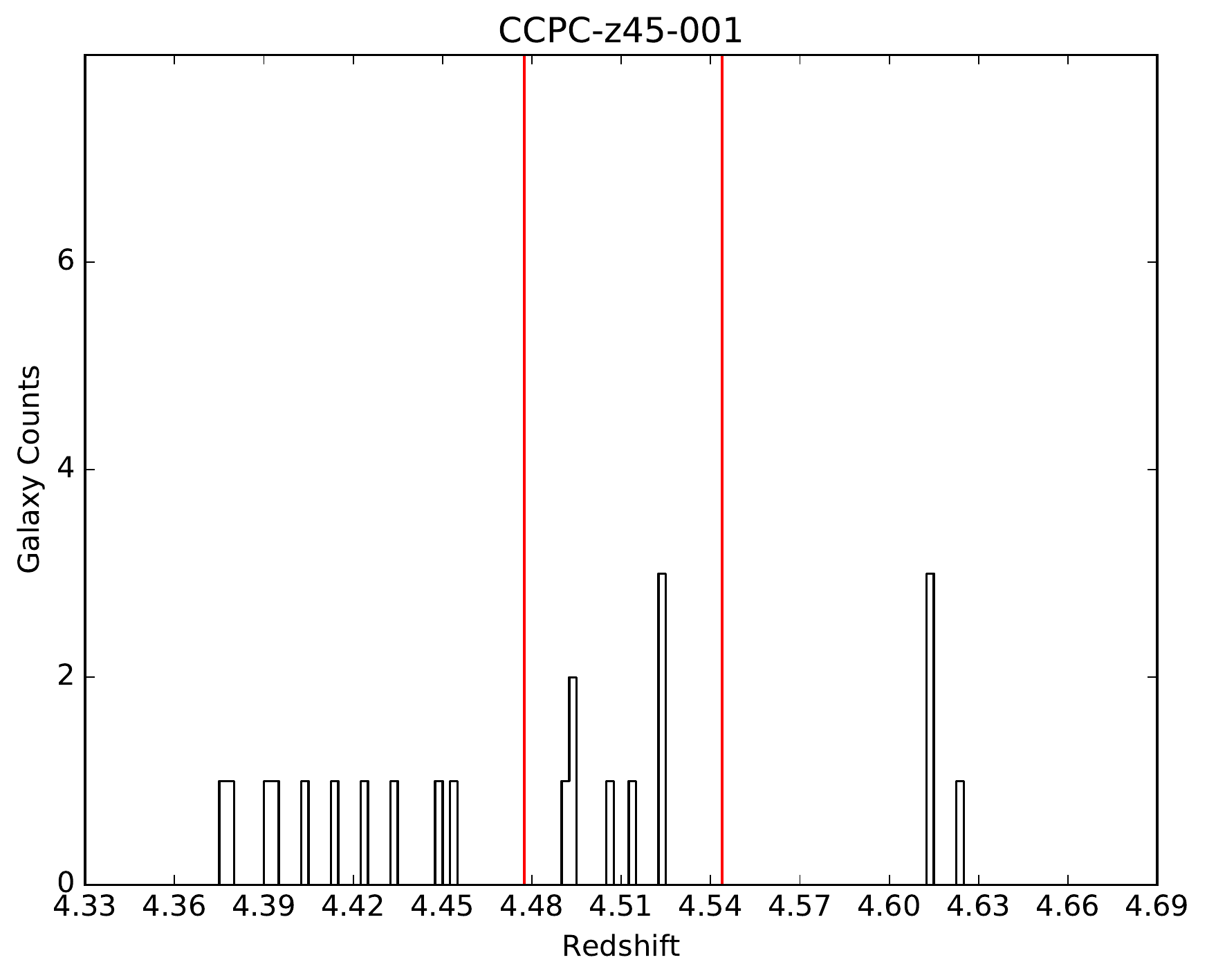}
\label{fig:CCPC-z45-001}
\end{subfigure}
\hfill
\end{figure*}

\begin{figure*}
\centering
\begin{subfigure}
\centering
\includegraphics[height=7.5cm,width=7.5cm]{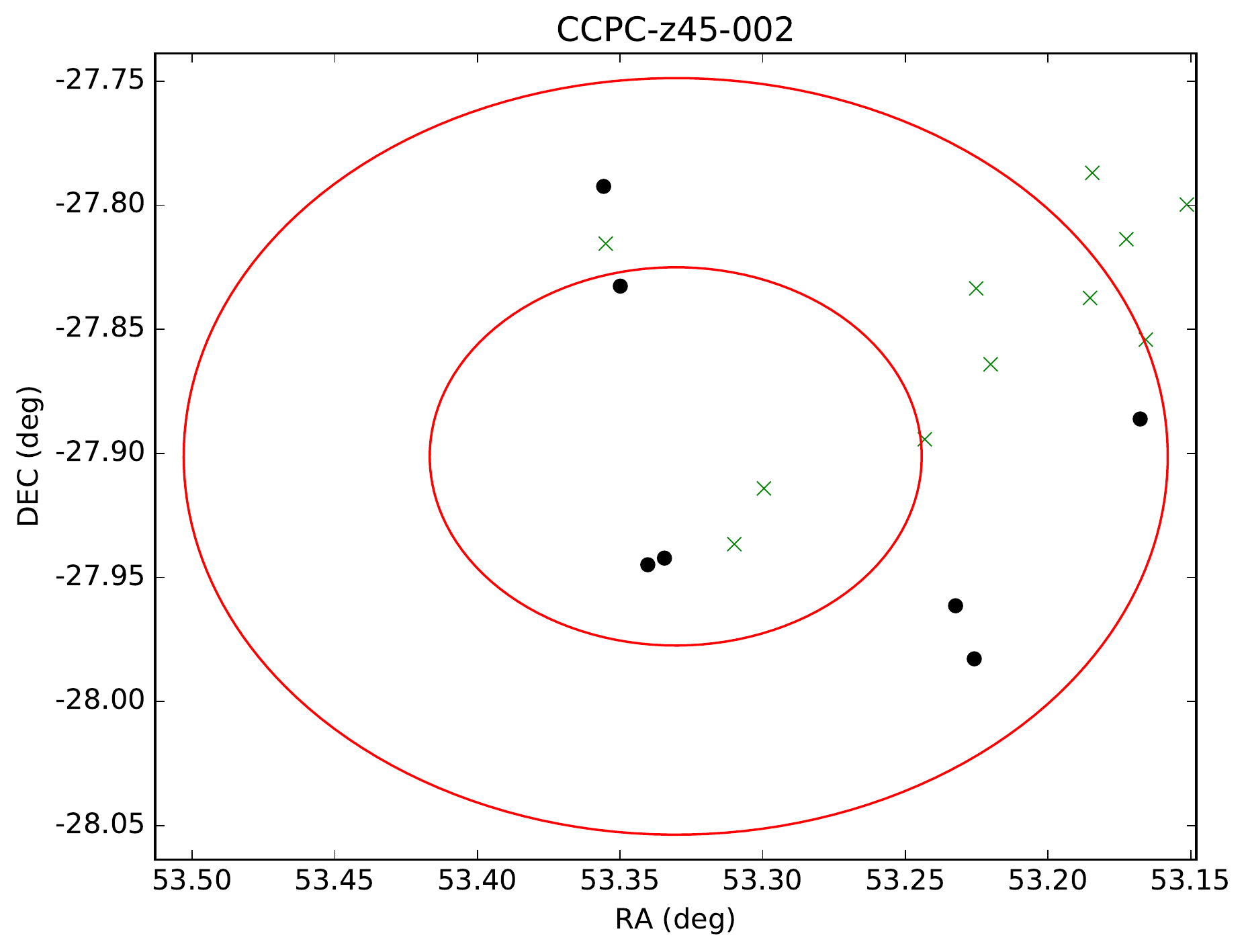}
\label{fig:CCPC-z45-002_sky}
\end{subfigure}
\hfill
\begin{subfigure}
\centering
\includegraphics[scale=0.52]{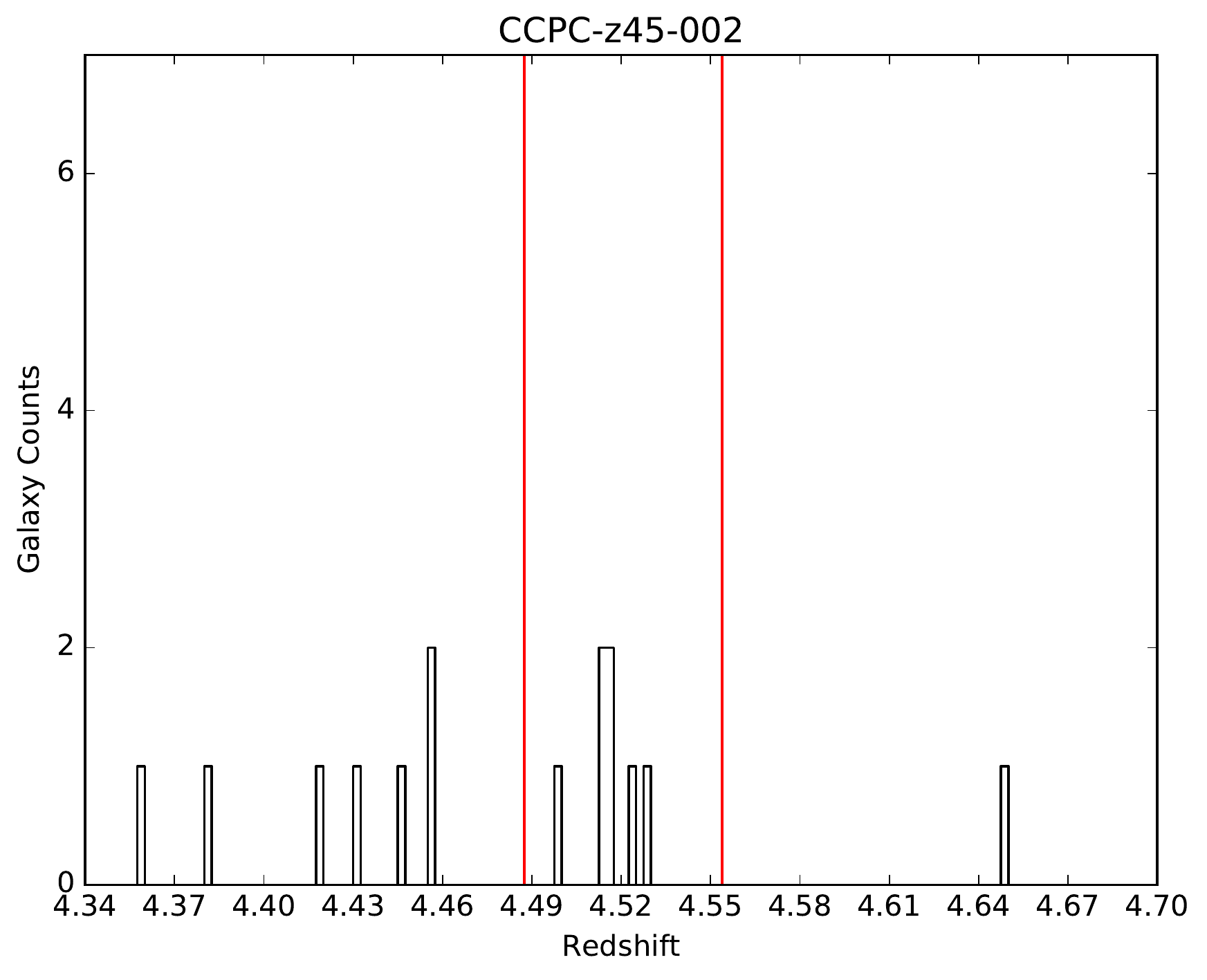}
\label{fig:CCPC-z45-002}
\end{subfigure}
\hfill
\end{figure*}

\begin{figure*}
\centering
\begin{subfigure}
\centering
\includegraphics[height=7.5cm,width=7.5cm]{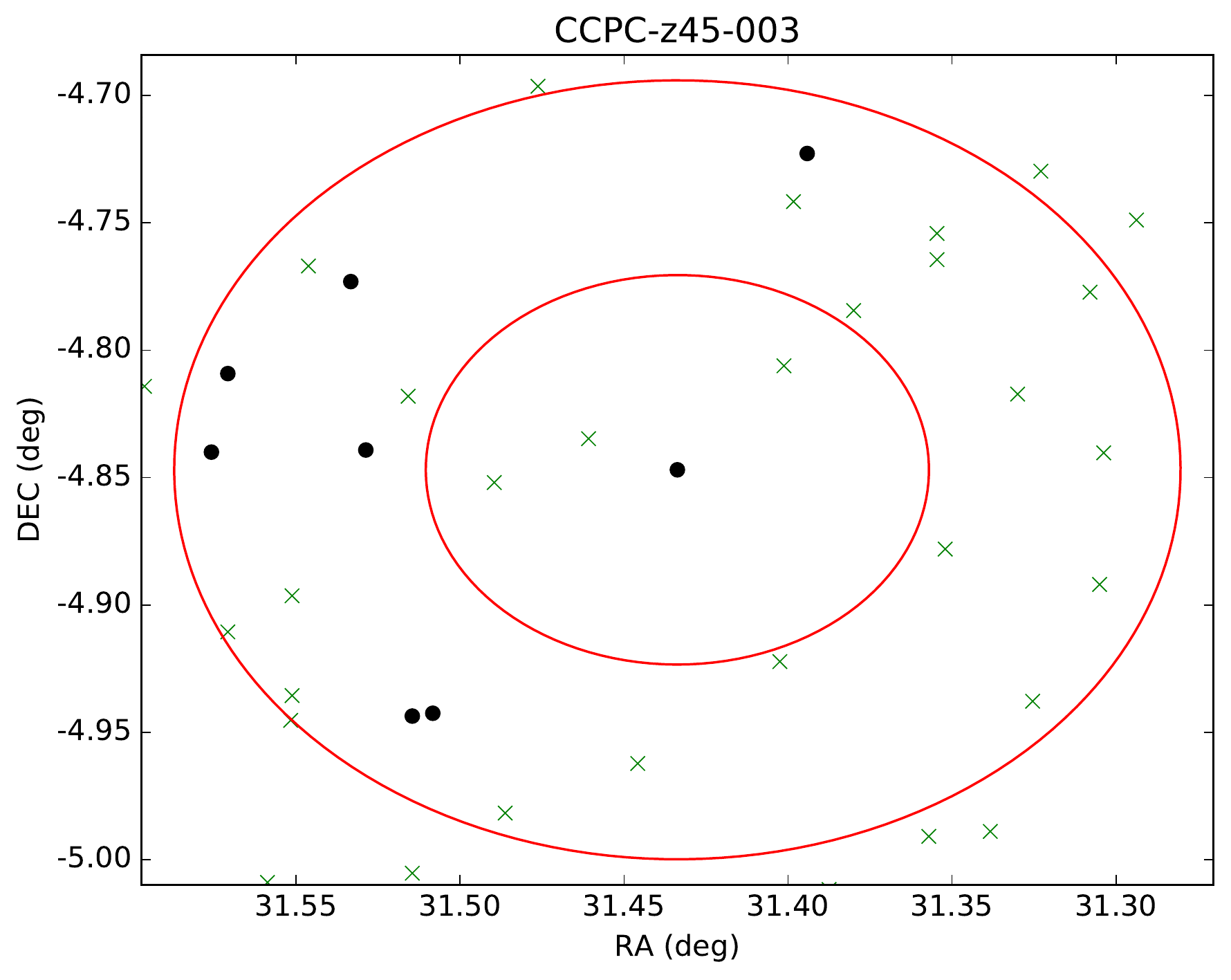}
\label{fig:CCPC-z45-003_sky}
\end{subfigure}
\hfill
\begin{subfigure}
\centering
\includegraphics[scale=0.52]{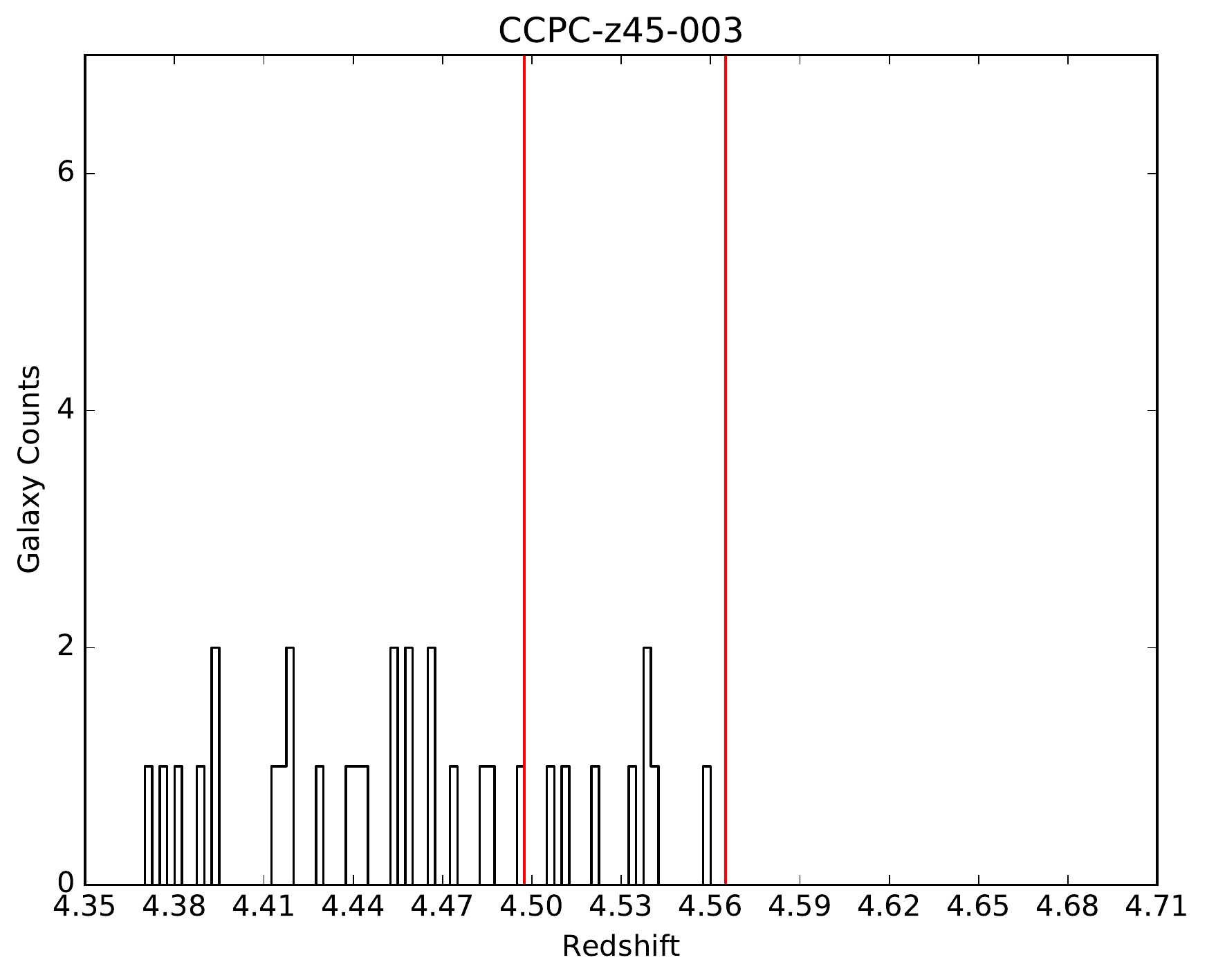}
\label{fig:CCPC-z45-003}
\end{subfigure}
\hfill
\end{figure*}
\clearpage 

\begin{figure*}
\centering
\begin{subfigure}
\centering
\includegraphics[height=7.5cm,width=7.5cm]{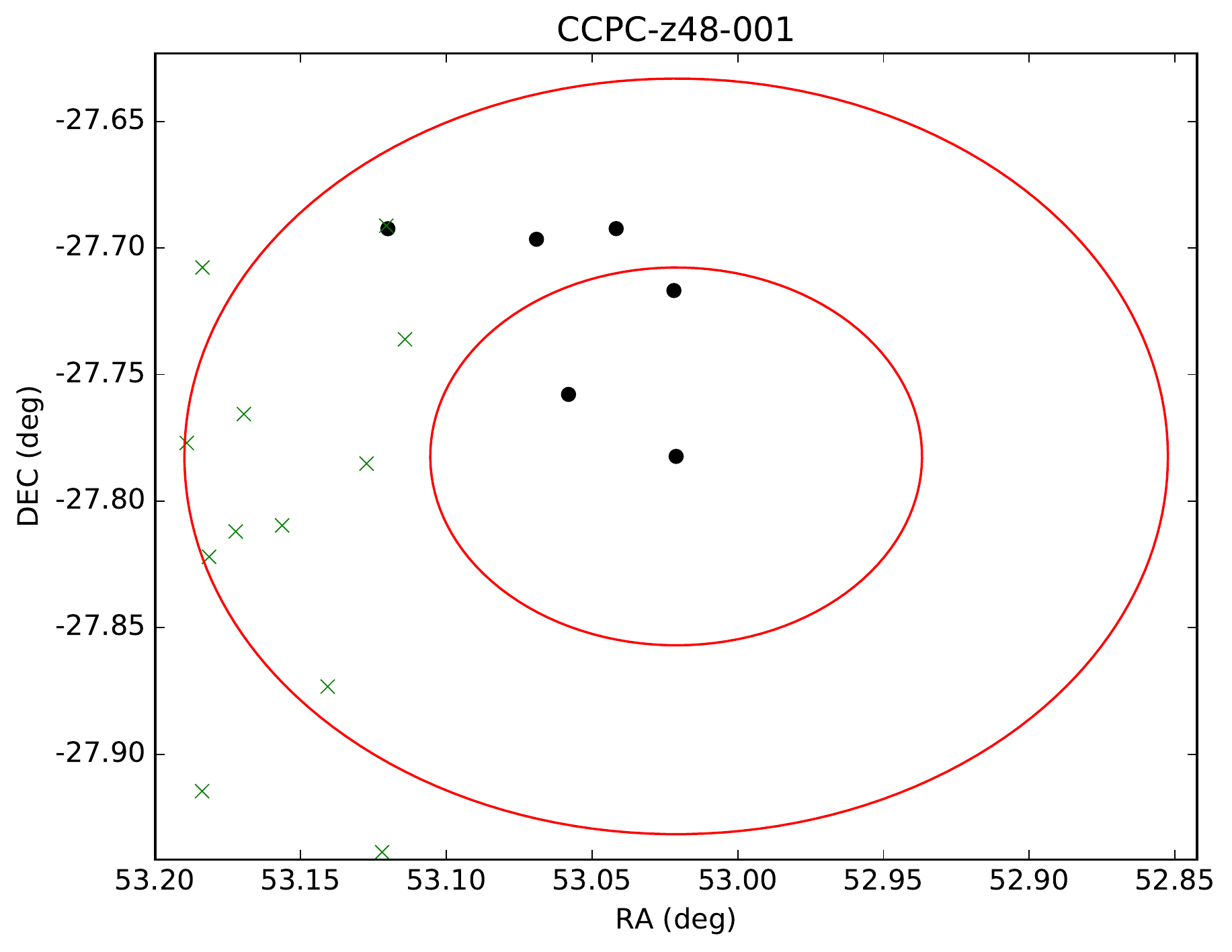}
\label{fig:CCPC-z48-001_sky}
\end{subfigure}
\hfill
\begin{subfigure}
\centering
\includegraphics[scale=0.52]{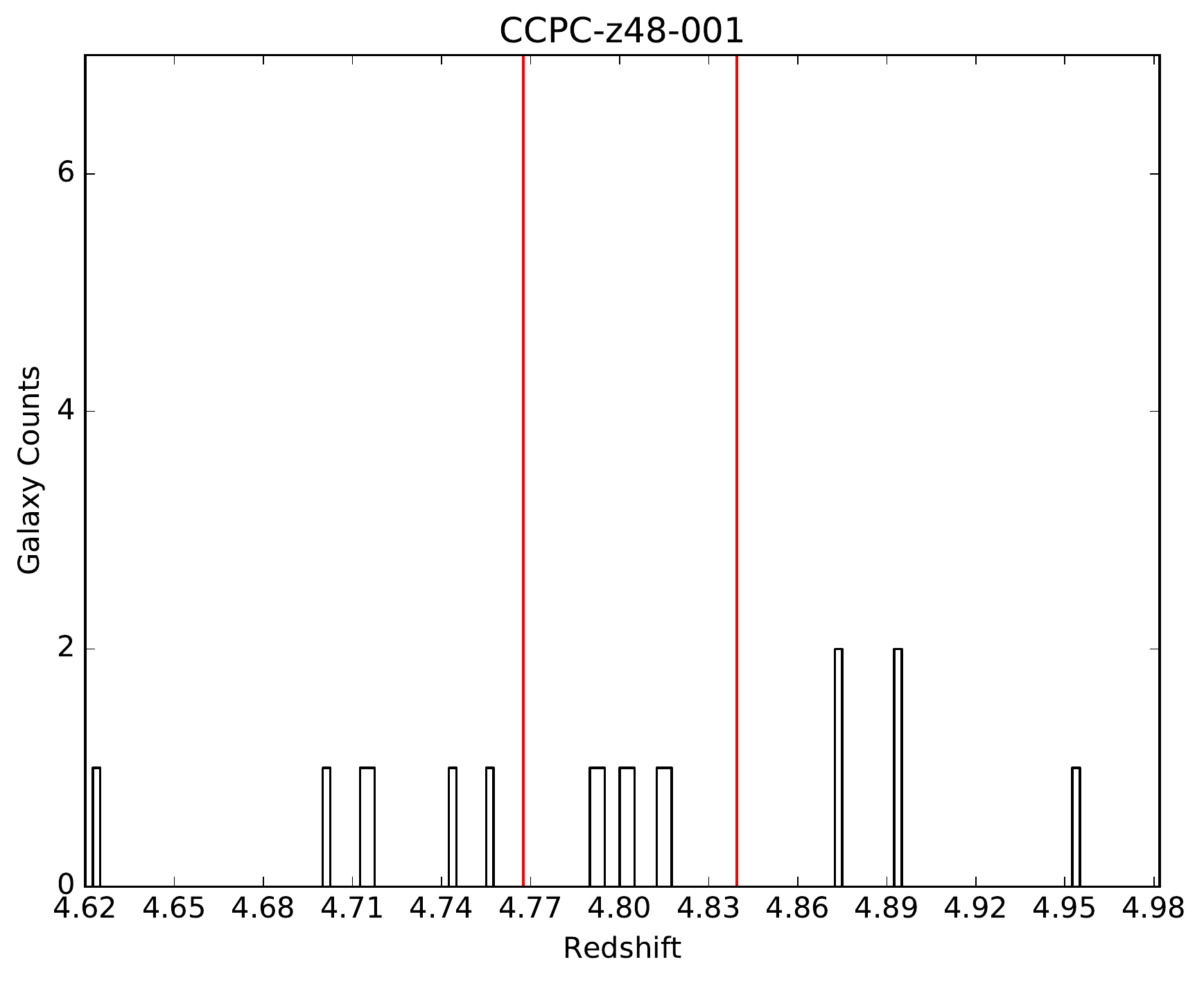}
\label{fig:CCPC-z48-001}
\end{subfigure}
\hfill
\end{figure*}

\begin{figure*}
\centering
\begin{subfigure}
\centering
\includegraphics[height=7.5cm,width=7.5cm]{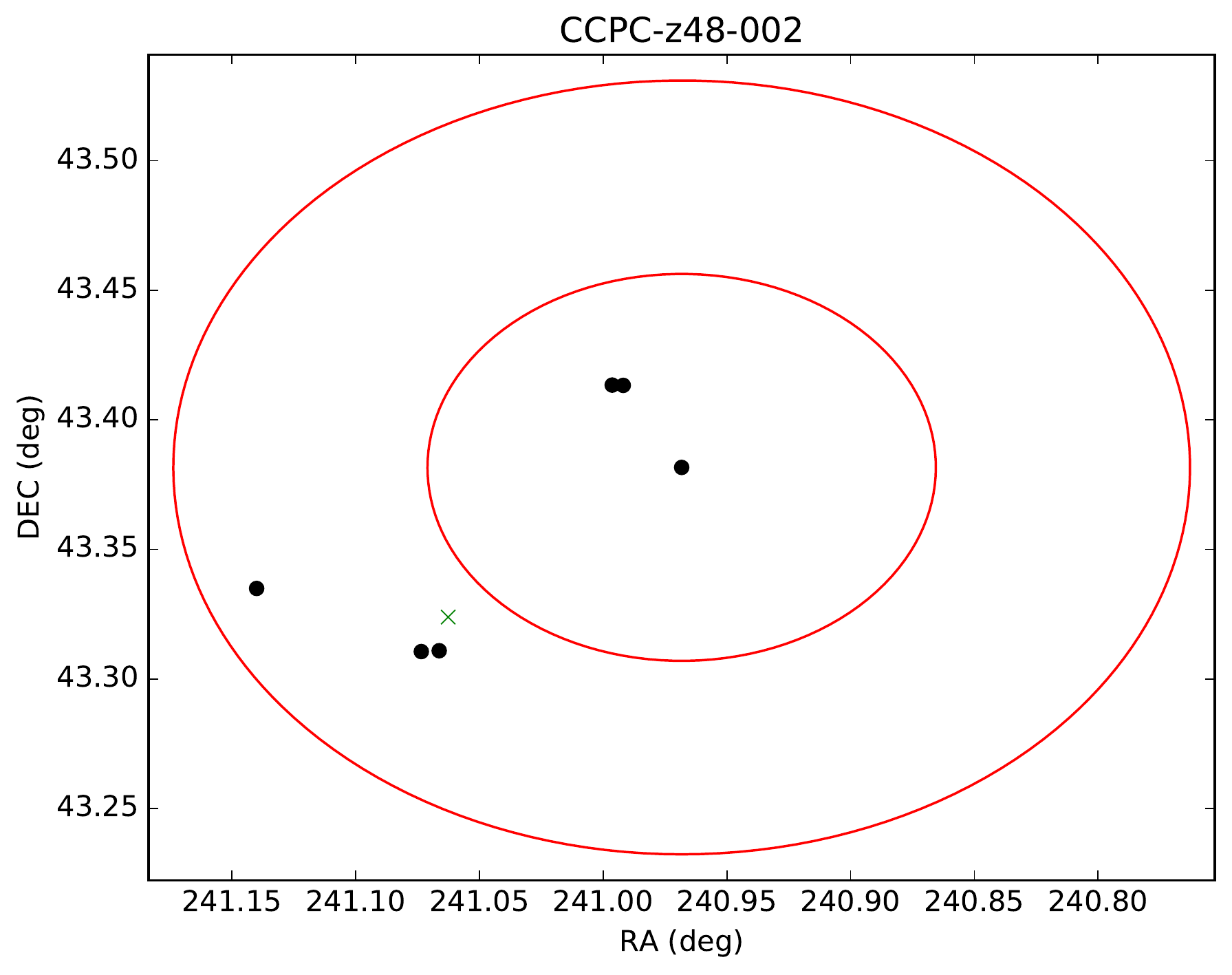}
\label{fig:CCPC-z48-002_sky}
\end{subfigure}
\hfill
\begin{subfigure}
\centering
\includegraphics[scale=0.52]{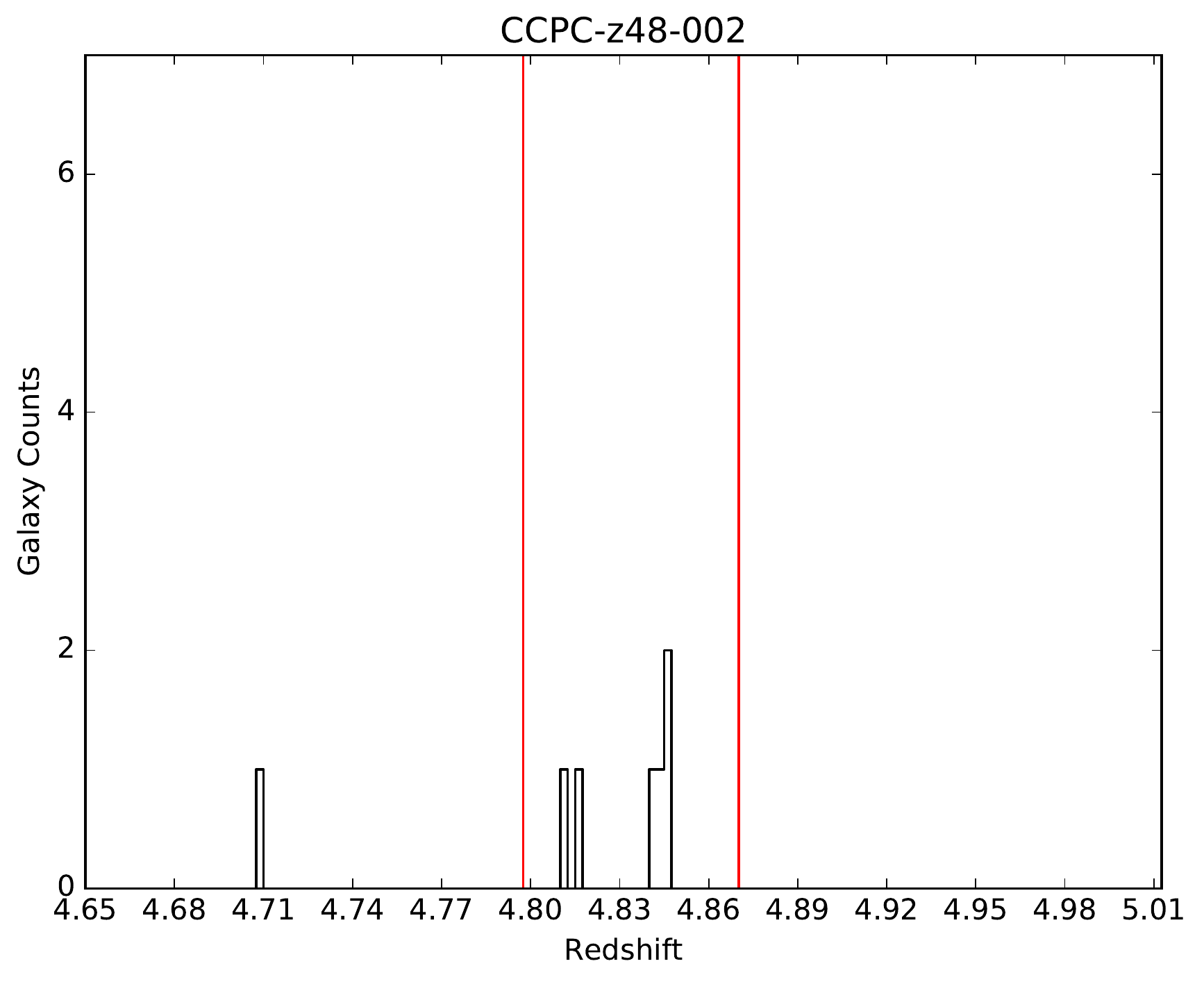}
\label{fig:CCPC-z48-002}
\end{subfigure}
\hfill
\end{figure*}

\begin{figure*}
\centering
\begin{subfigure}
\centering
\includegraphics[height=7.5cm,width=7.5cm]{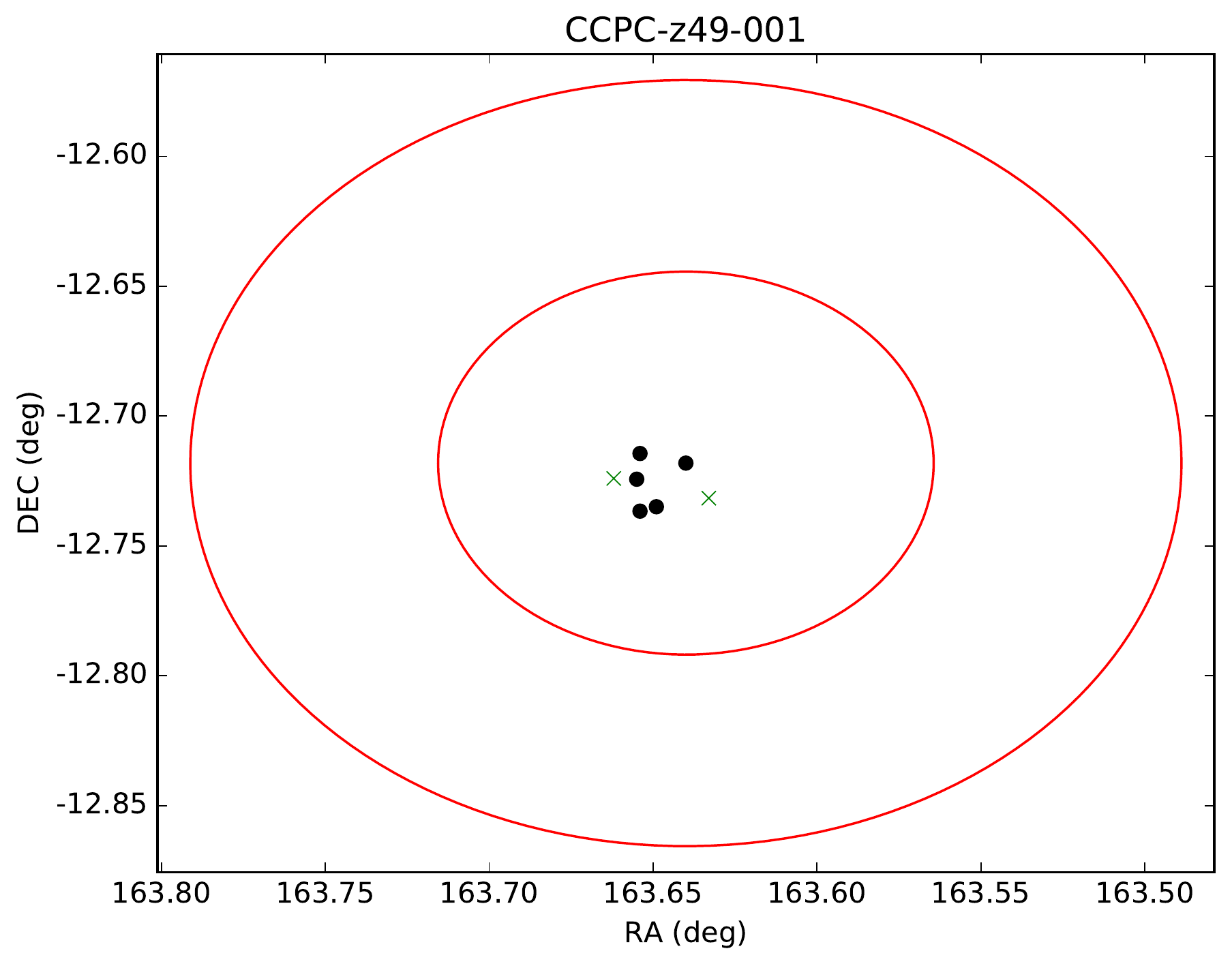}
\label{fig:CCPC-z49-001_sky}
\end{subfigure}
\hfill
\begin{subfigure}
\centering
\includegraphics[scale=0.52]{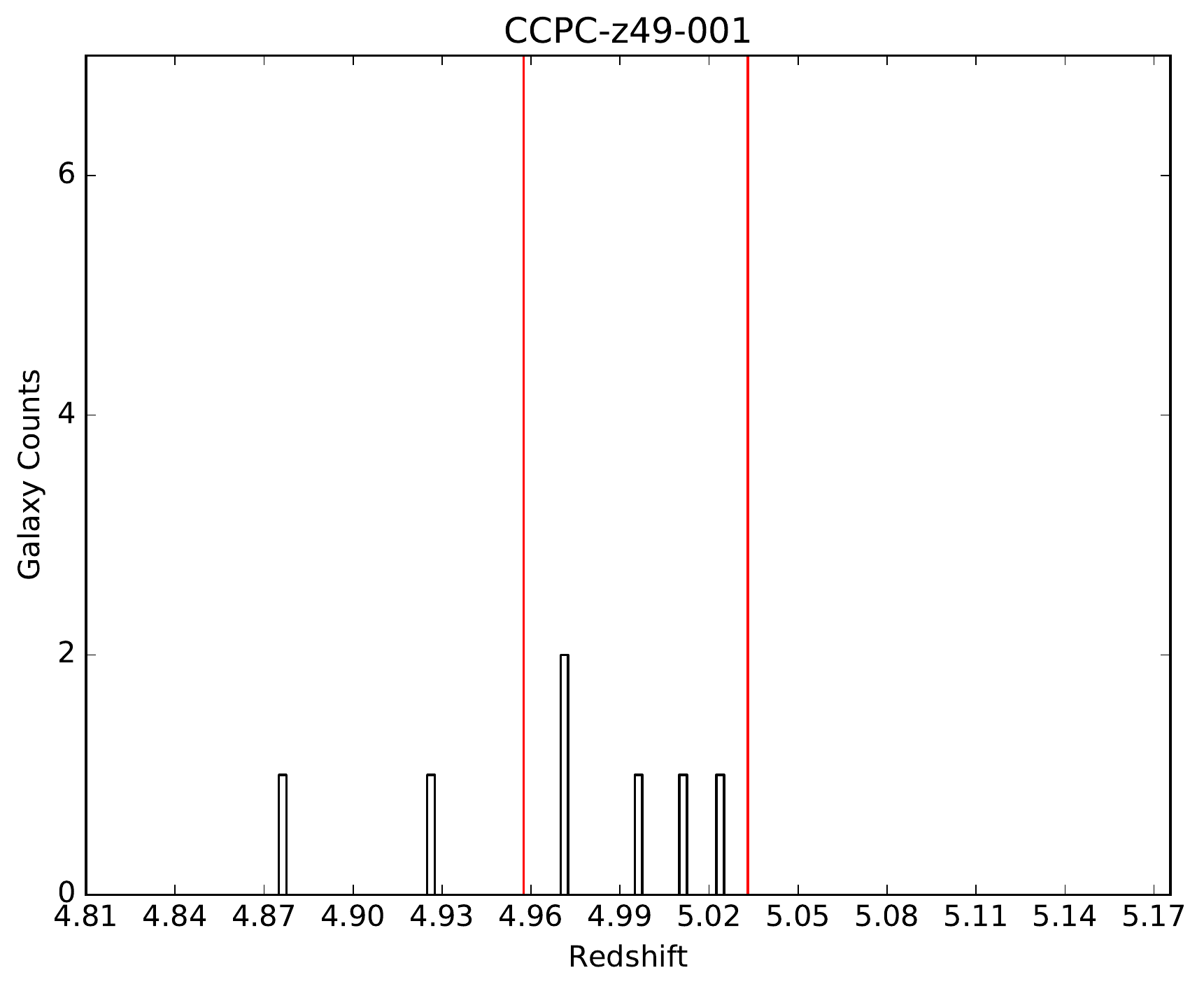}
\label{fig:CCPC-z49-001}
\end{subfigure}
\hfill
\end{figure*}
\clearpage 

\begin{figure*}
\centering
\begin{subfigure}
\centering
\includegraphics[height=7.5cm,width=7.5cm]{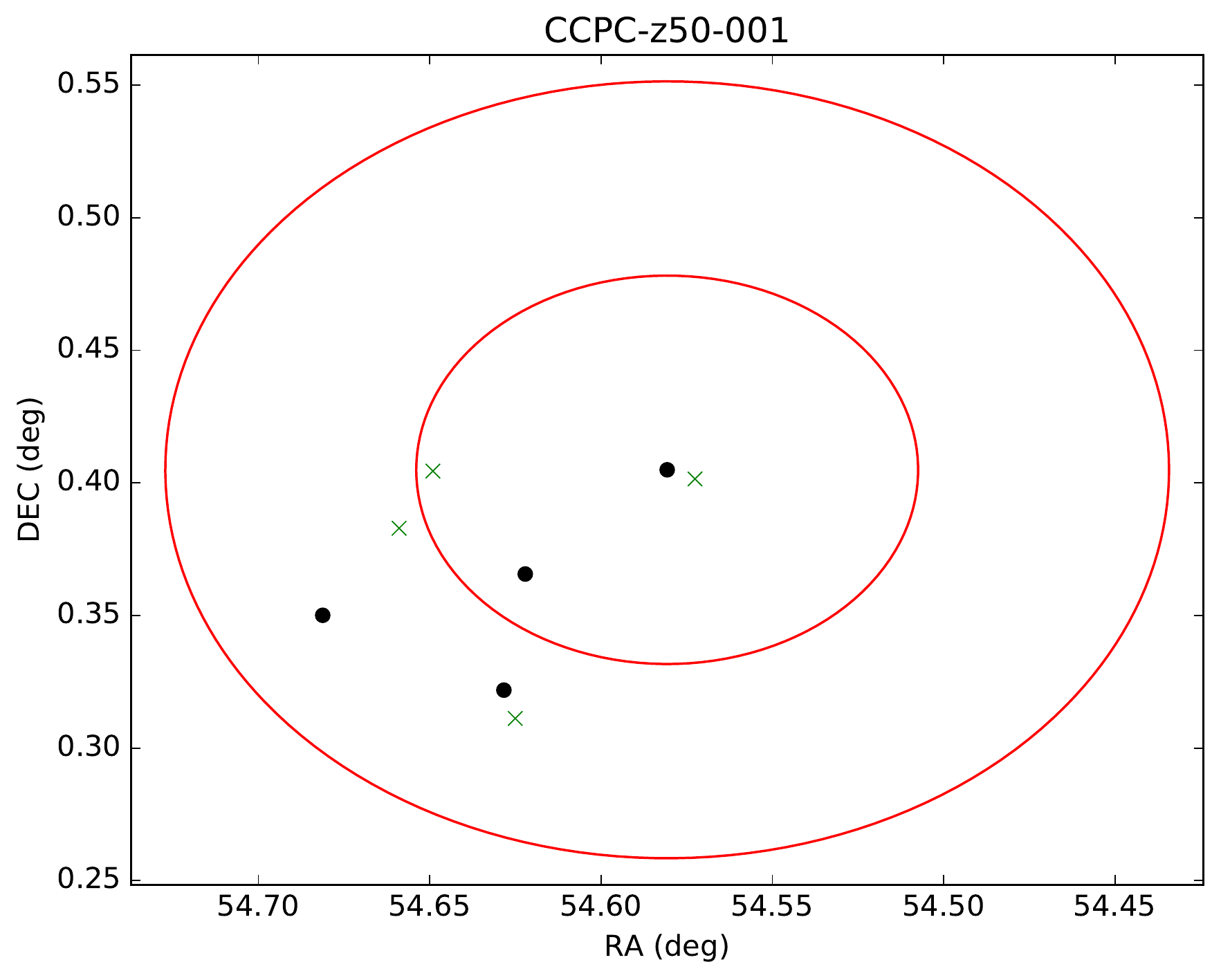}
\label{fig:CCPC-z50-001_sky}
\end{subfigure}
\hfill
\begin{subfigure}
\centering
\includegraphics[scale=0.52]{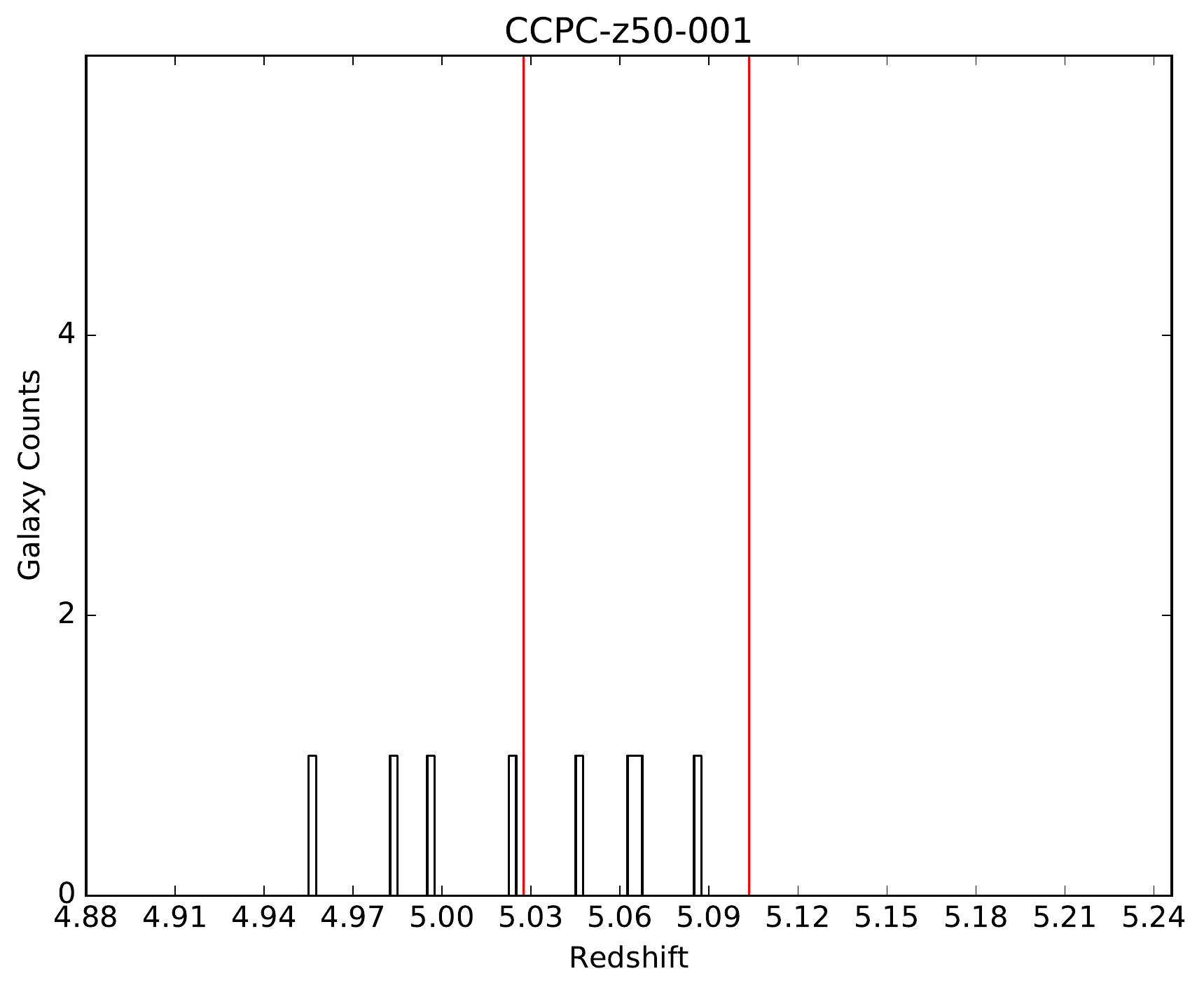}
\label{fig:CCPC-z50-001}
\end{subfigure}
\hfill
\end{figure*}

\begin{figure*}
\centering
\begin{subfigure}
\centering
\includegraphics[height=7.5cm,width=7.5cm]{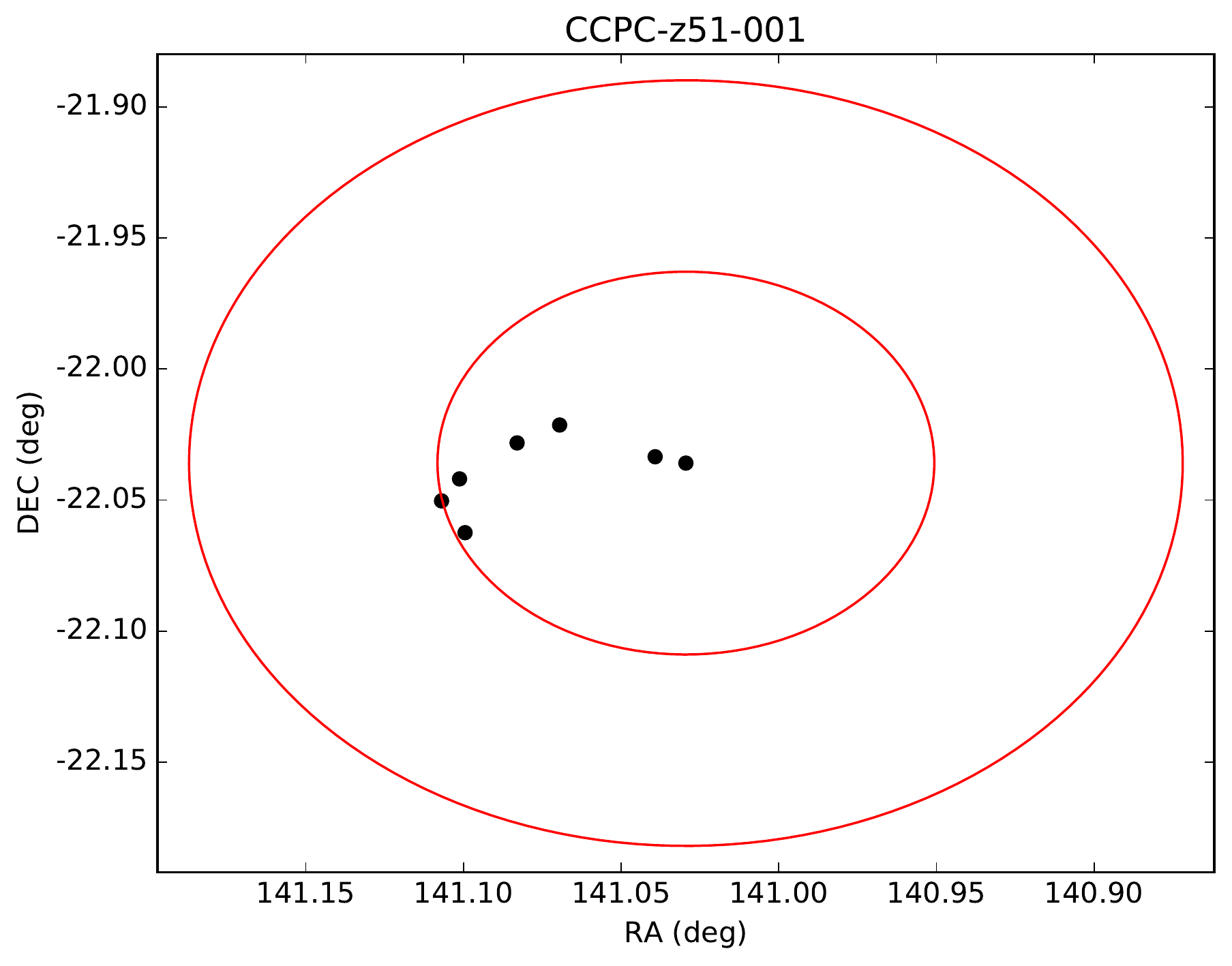}
\label{fig:CCPC-z51-001_sky}
\end{subfigure}
\hfill
\begin{subfigure}
\centering
\includegraphics[scale=0.52]{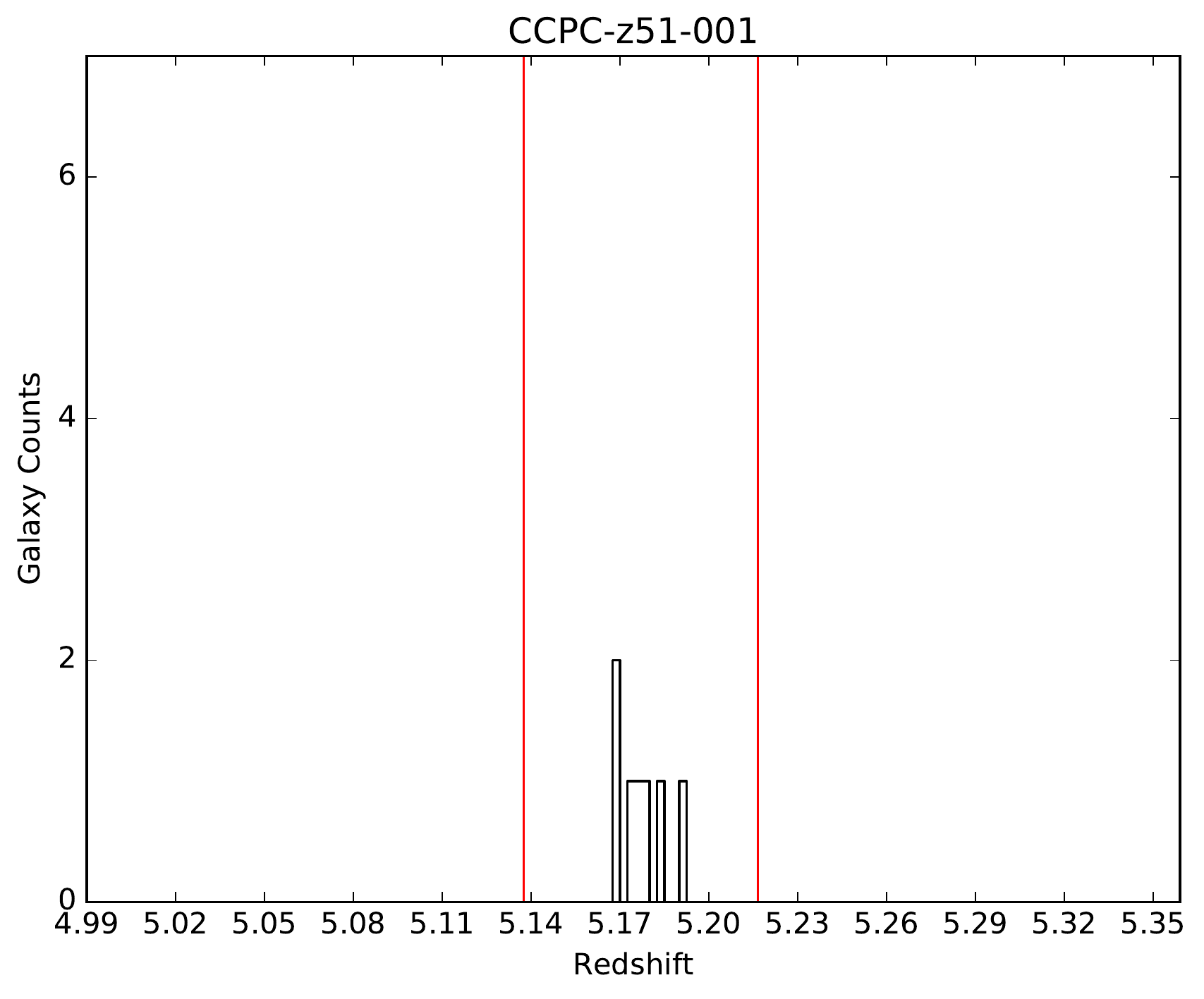}
\label{fig:CCPC-z51-001}
\end{subfigure}
\hfill
\end{figure*}

\begin{figure*}
\centering
\begin{subfigure}
\centering
\includegraphics[height=7.5cm,width=7.5cm]{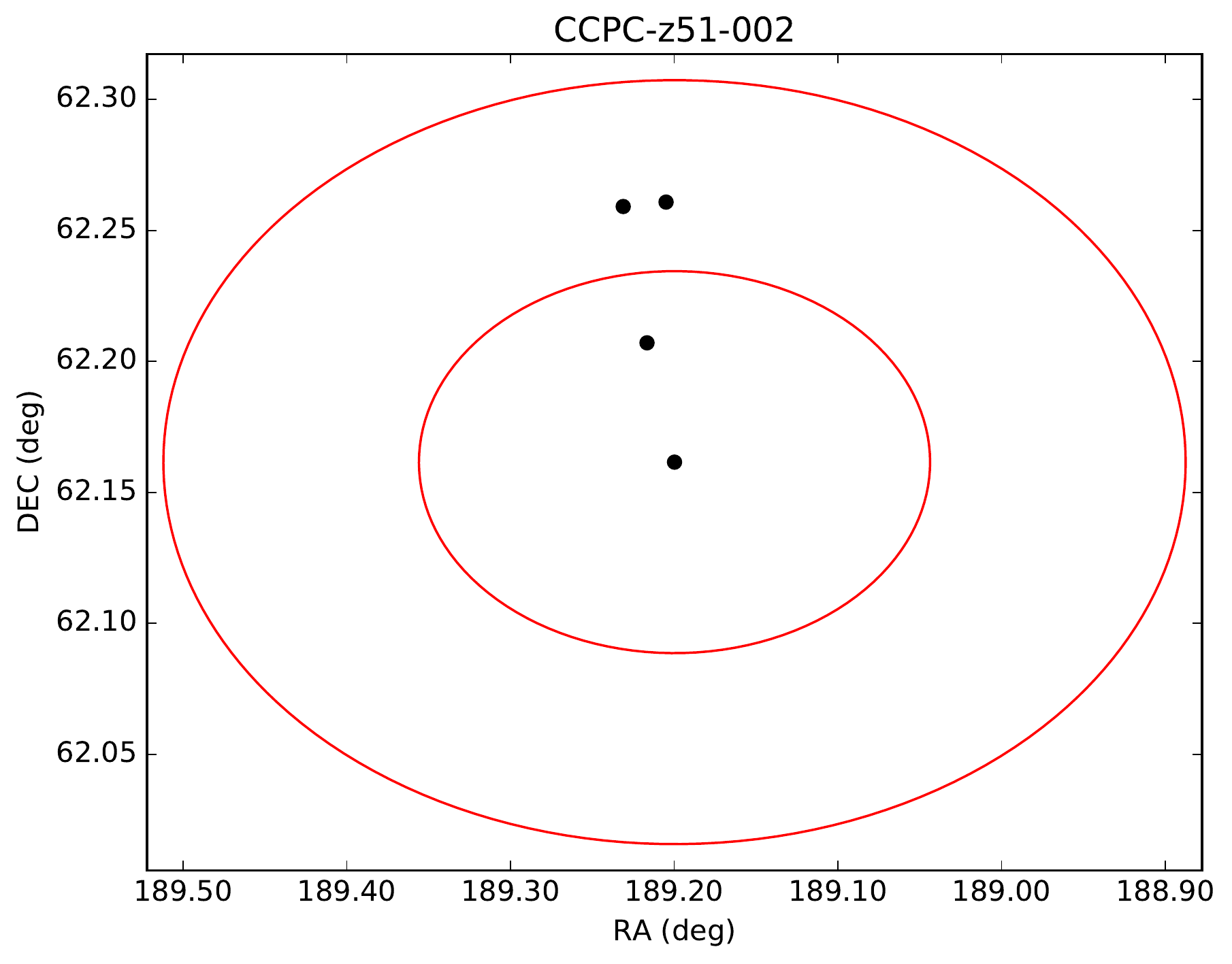}
\label{fig:CCPC-z51-002_sky}
\end{subfigure}
\hfill
\begin{subfigure}
\centering
\includegraphics[scale=0.52]{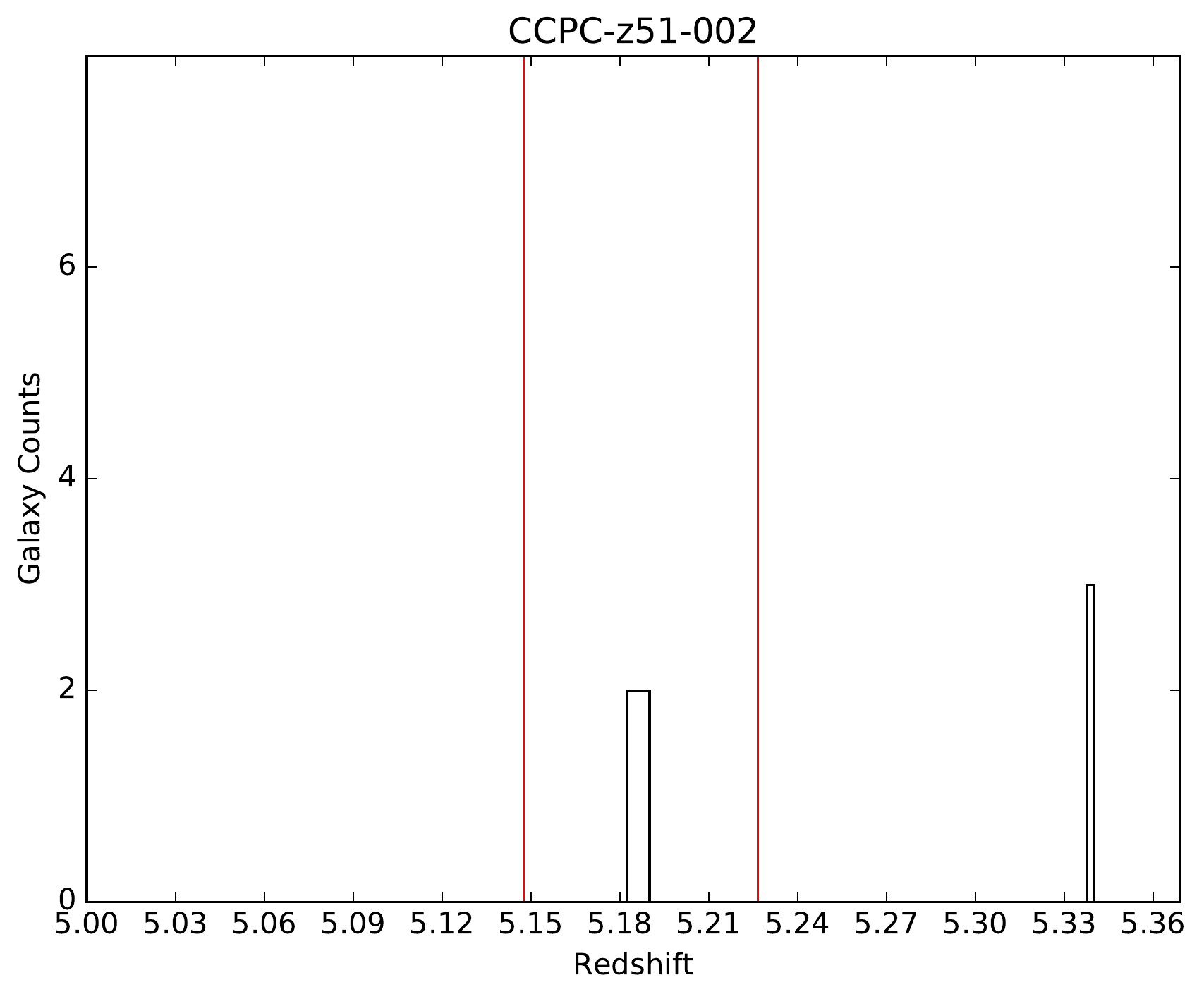}
\label{fig:CCPC-z51-002}
\end{subfigure}
\hfill
\end{figure*}
\clearpage 

\begin{figure*}
\centering
\begin{subfigure}
\centering
\includegraphics[height=7.5cm,width=7.5cm]{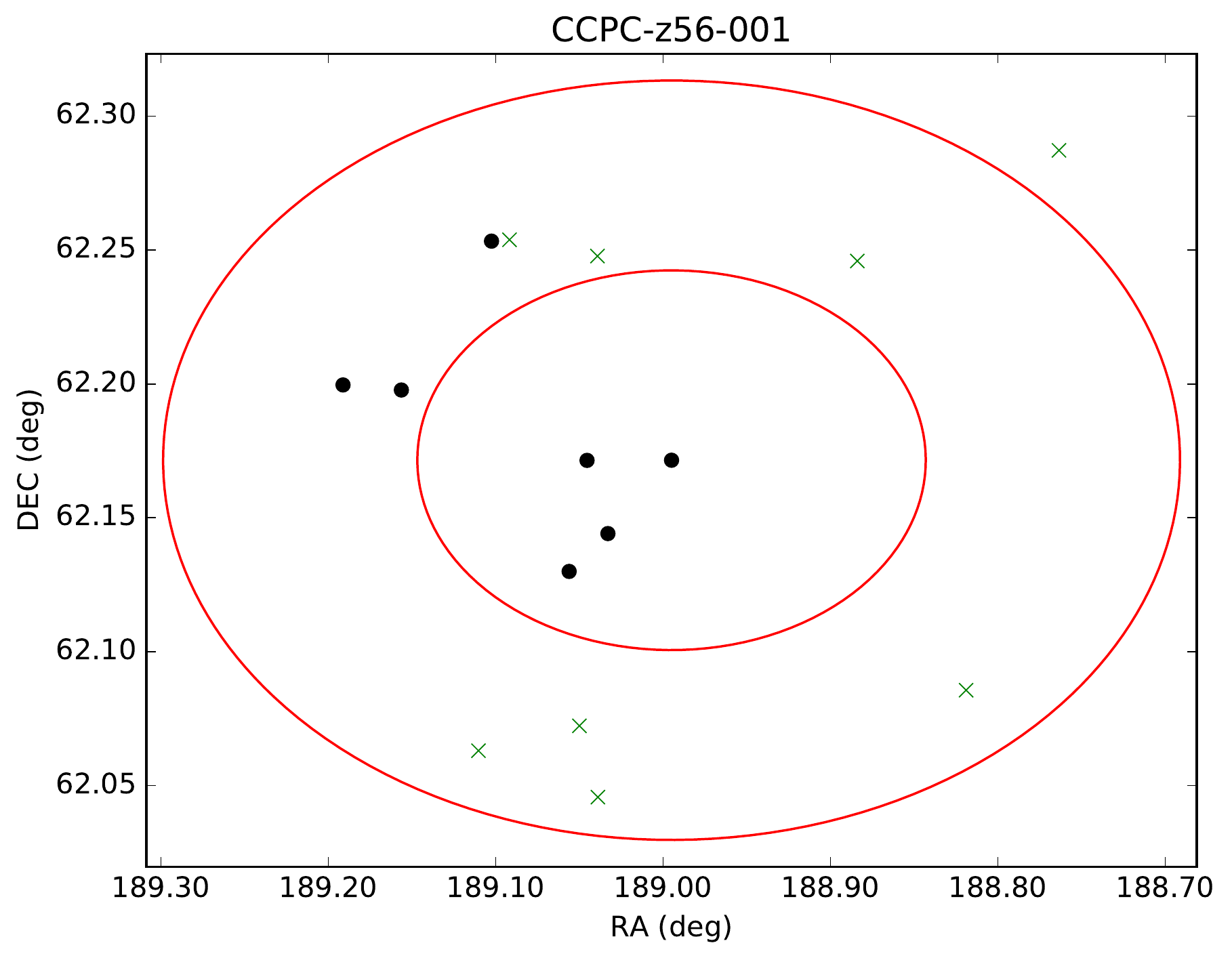}
\label{fig:CCPC-z56-001_sky}
\end{subfigure}
\hfill
\begin{subfigure}
\centering
\includegraphics[scale=0.52]{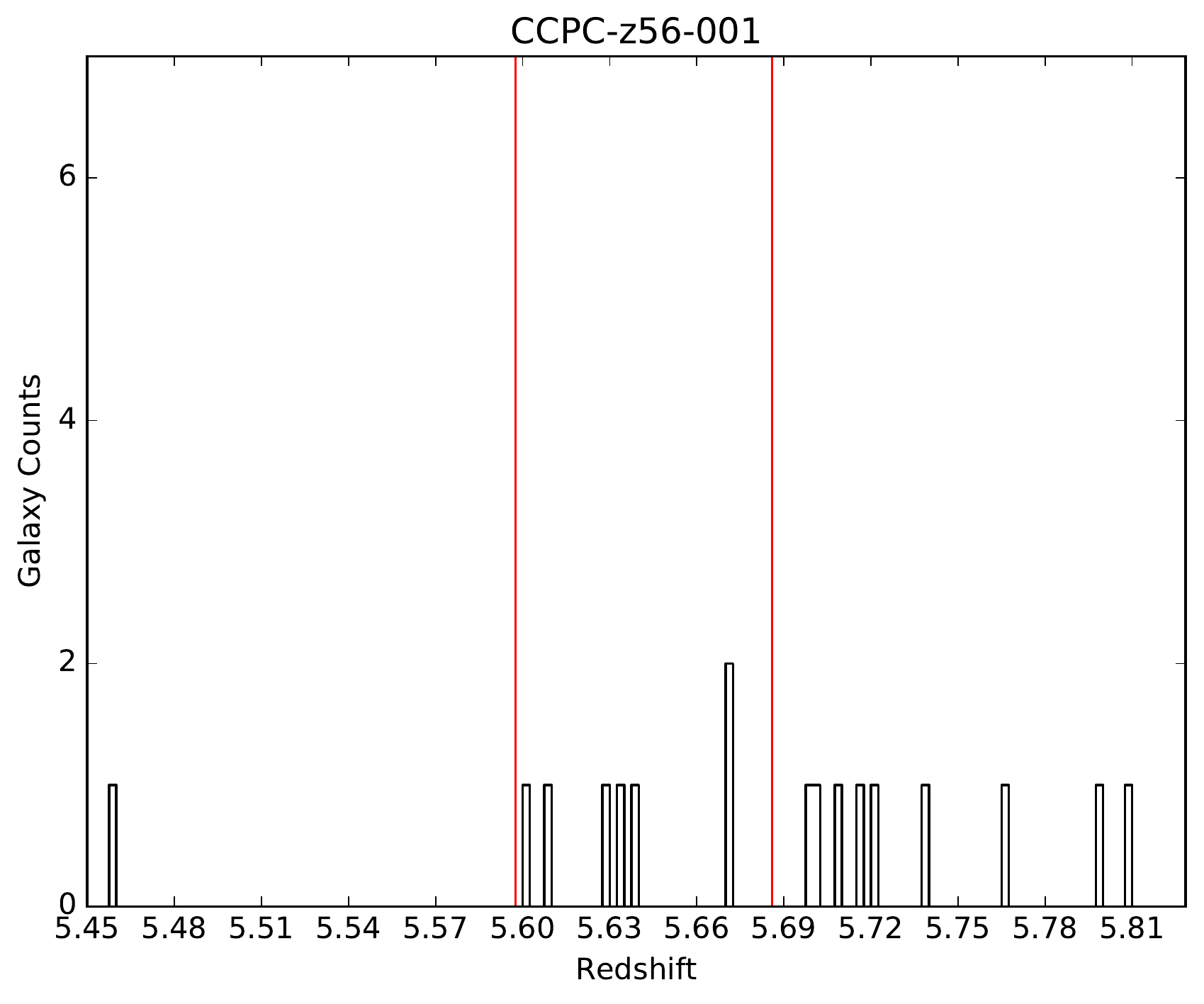}
\label{fig:CCPC-z56-001}
\end{subfigure}
\hfill
\end{figure*}

\begin{figure*}
\centering
\begin{subfigure}
\centering
\includegraphics[height=7.5cm,width=7.5cm]{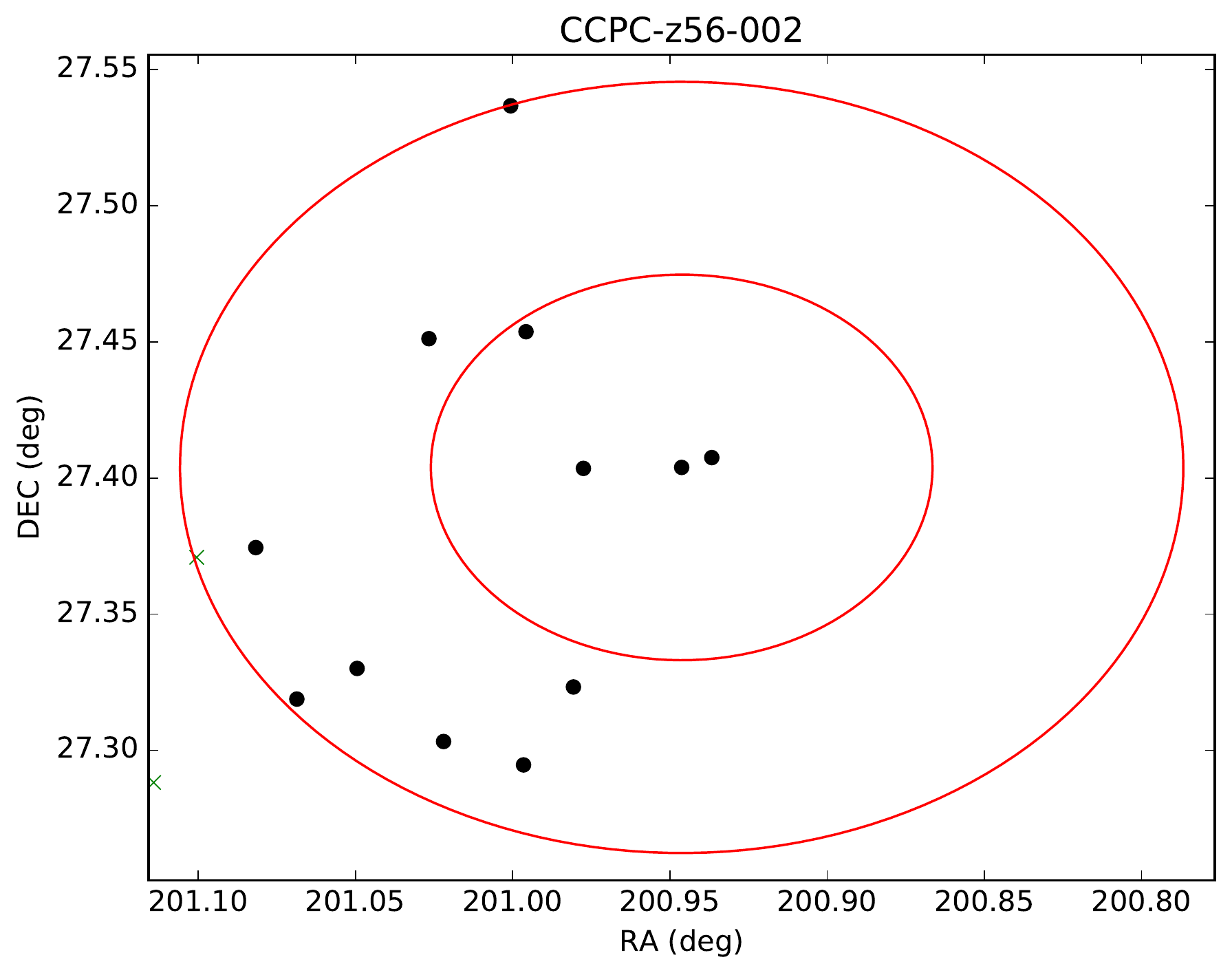}
\label{fig:CCPC-z56-002_sky}
\end{subfigure}
\hfill
\begin{subfigure}
\centering
\includegraphics[scale=0.52]{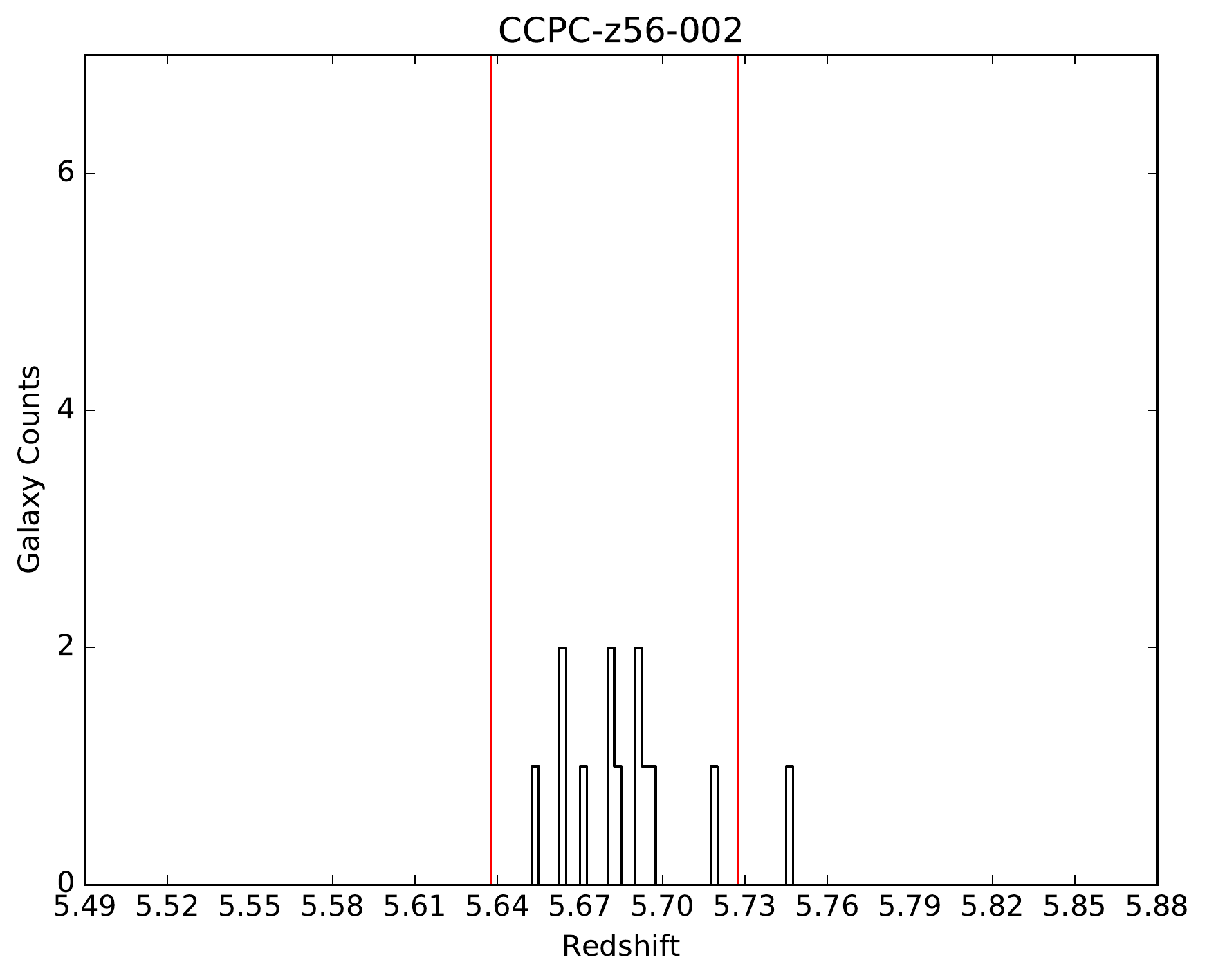}
\label{fig:CCPC-z56-002}
\end{subfigure}
\hfill
\end{figure*}

\begin{figure*}
\centering
\begin{subfigure}
\centering
\includegraphics[height=7.5cm,width=7.5cm]{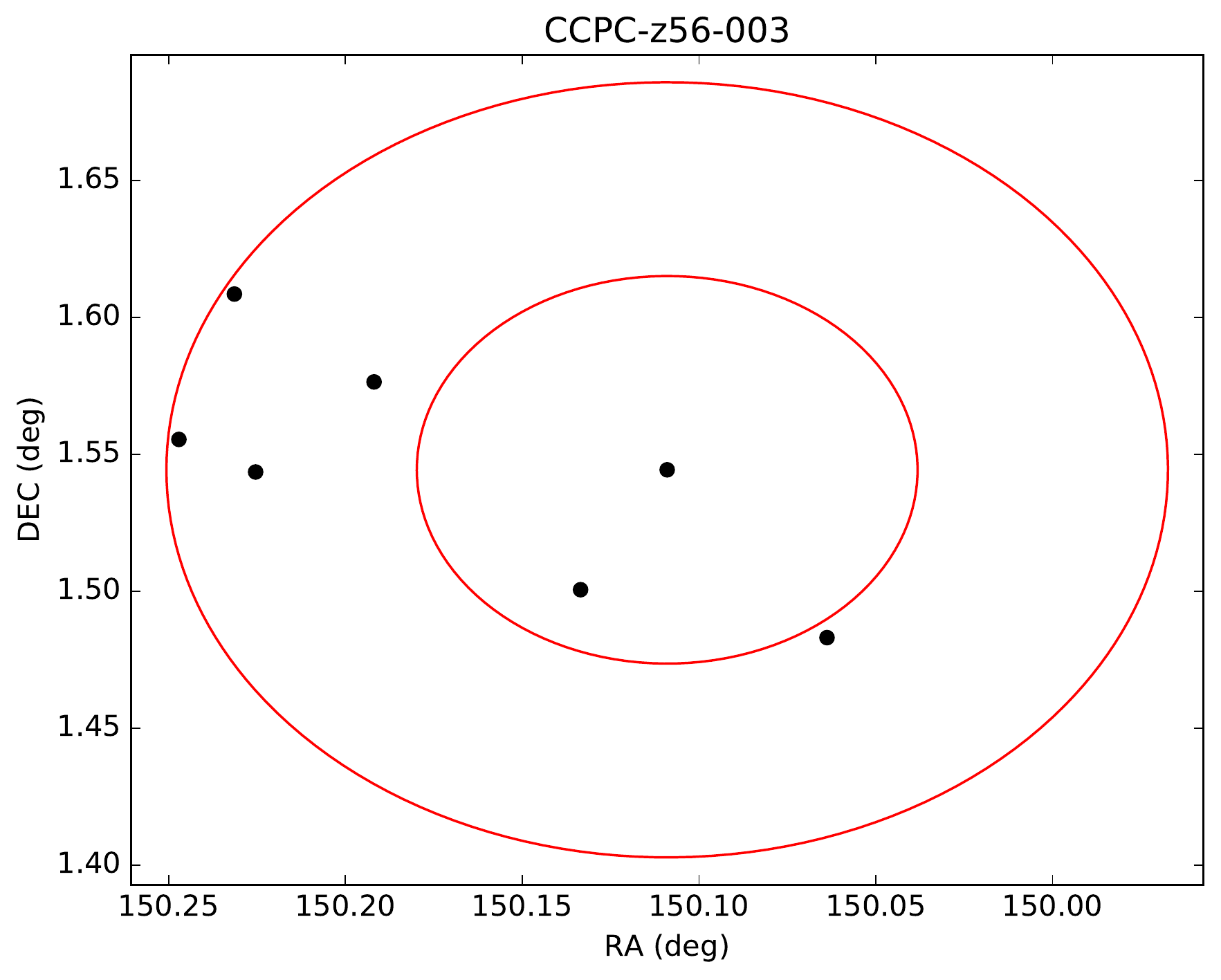}
\label{fig:CCPC-z56-003_sky}
\end{subfigure}
\hfill
\begin{subfigure}
\centering
\includegraphics[scale=0.52]{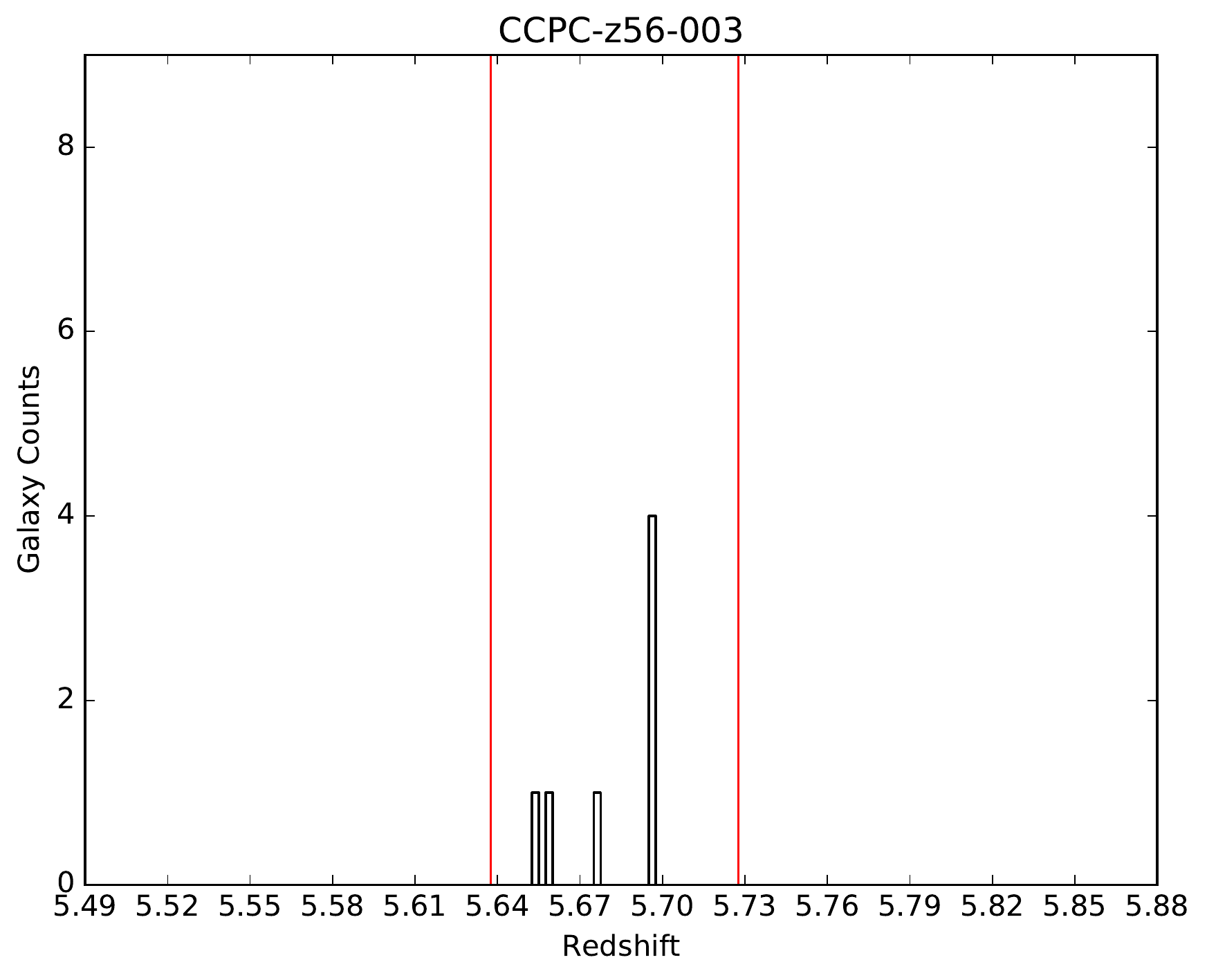}
\label{fig:CCPC-z56-003}
\end{subfigure}
\hfill
\end{figure*}
\clearpage 

\begin{figure*}
\centering
\begin{subfigure}
\centering
\includegraphics[height=7.5cm,width=7.5cm]{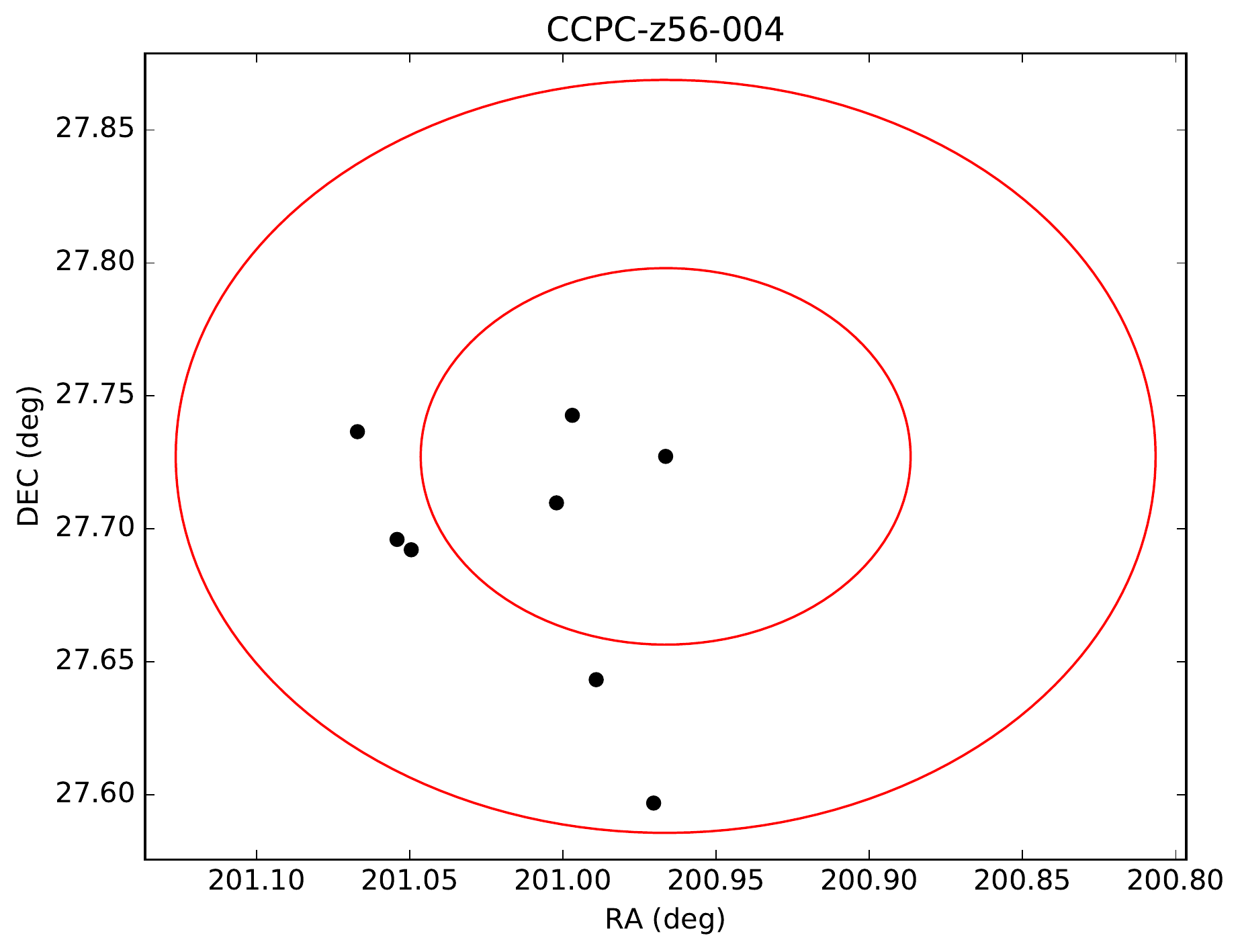}
\label{fig:CCPC-z56-004_sky}
\end{subfigure}
\hfill
\begin{subfigure}
\centering
\includegraphics[scale=0.52]{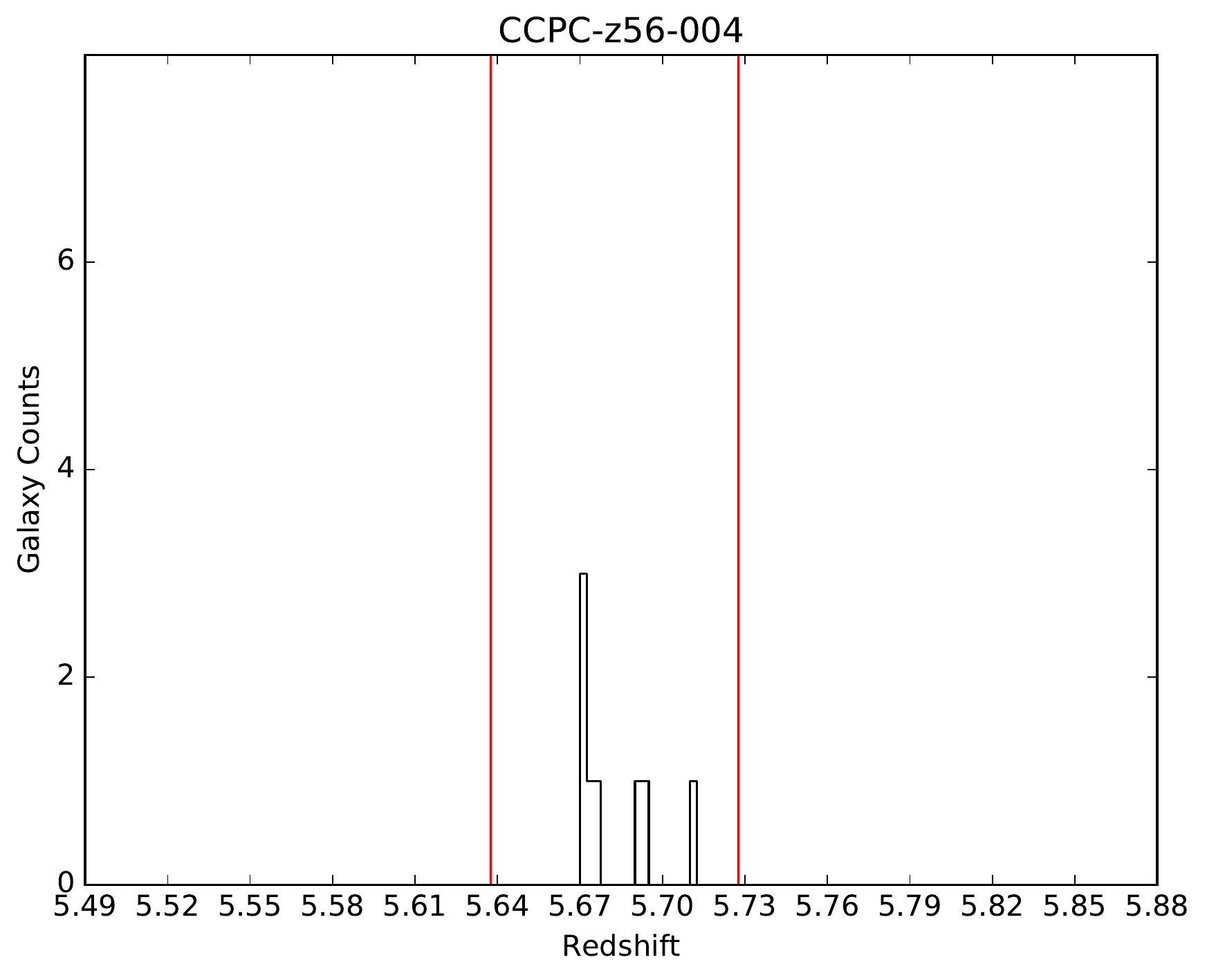}
\label{fig:CCPC-z56-004}
\end{subfigure}
\hfill
\end{figure*}

\begin{figure*}
\centering
\begin{subfigure}
\centering
\includegraphics[height=7.5cm,width=7.5cm]{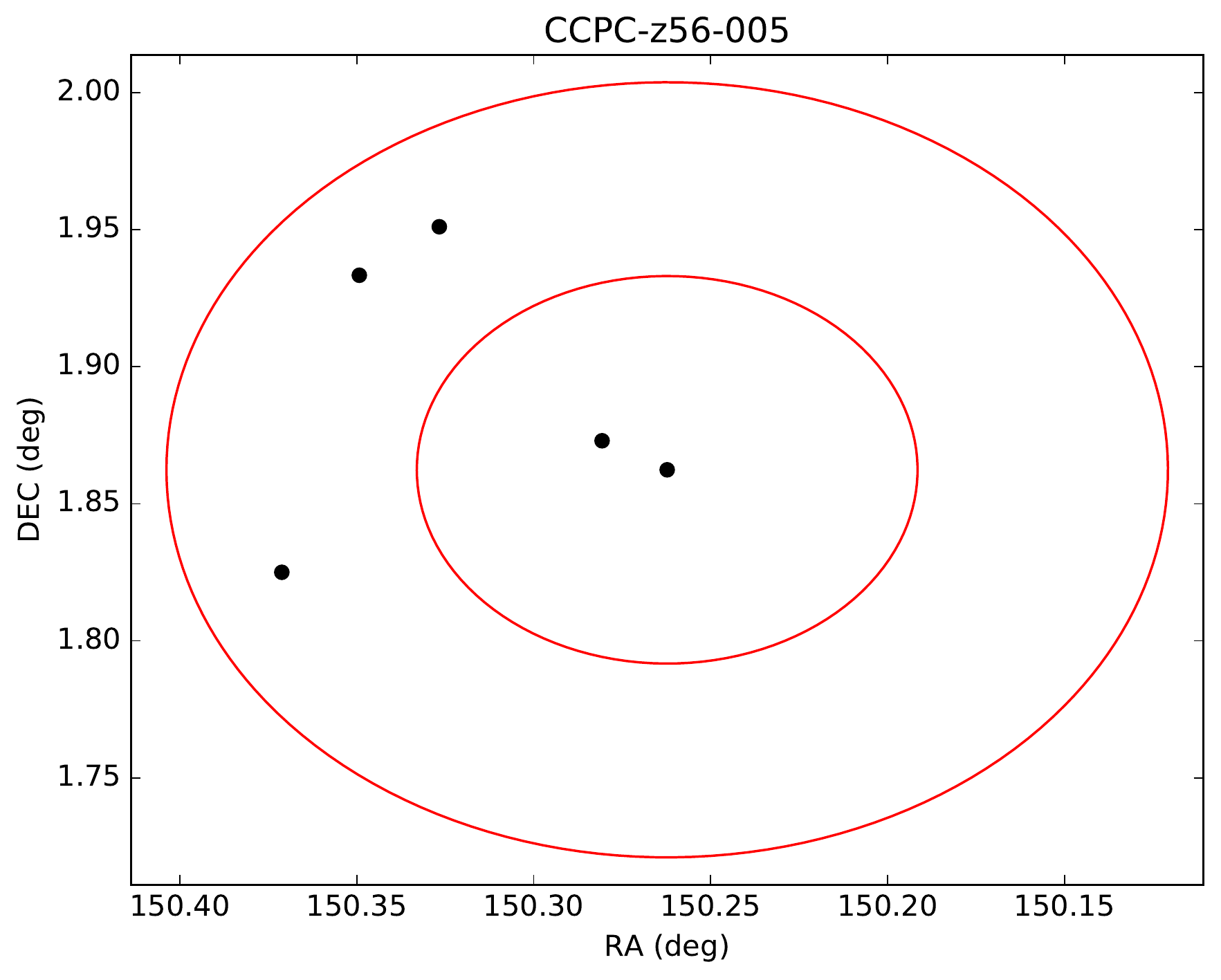}
\label{fig:CCPC-z56-005_sky}
\end{subfigure}
\hfill
\begin{subfigure}
\centering
\includegraphics[scale=0.52]{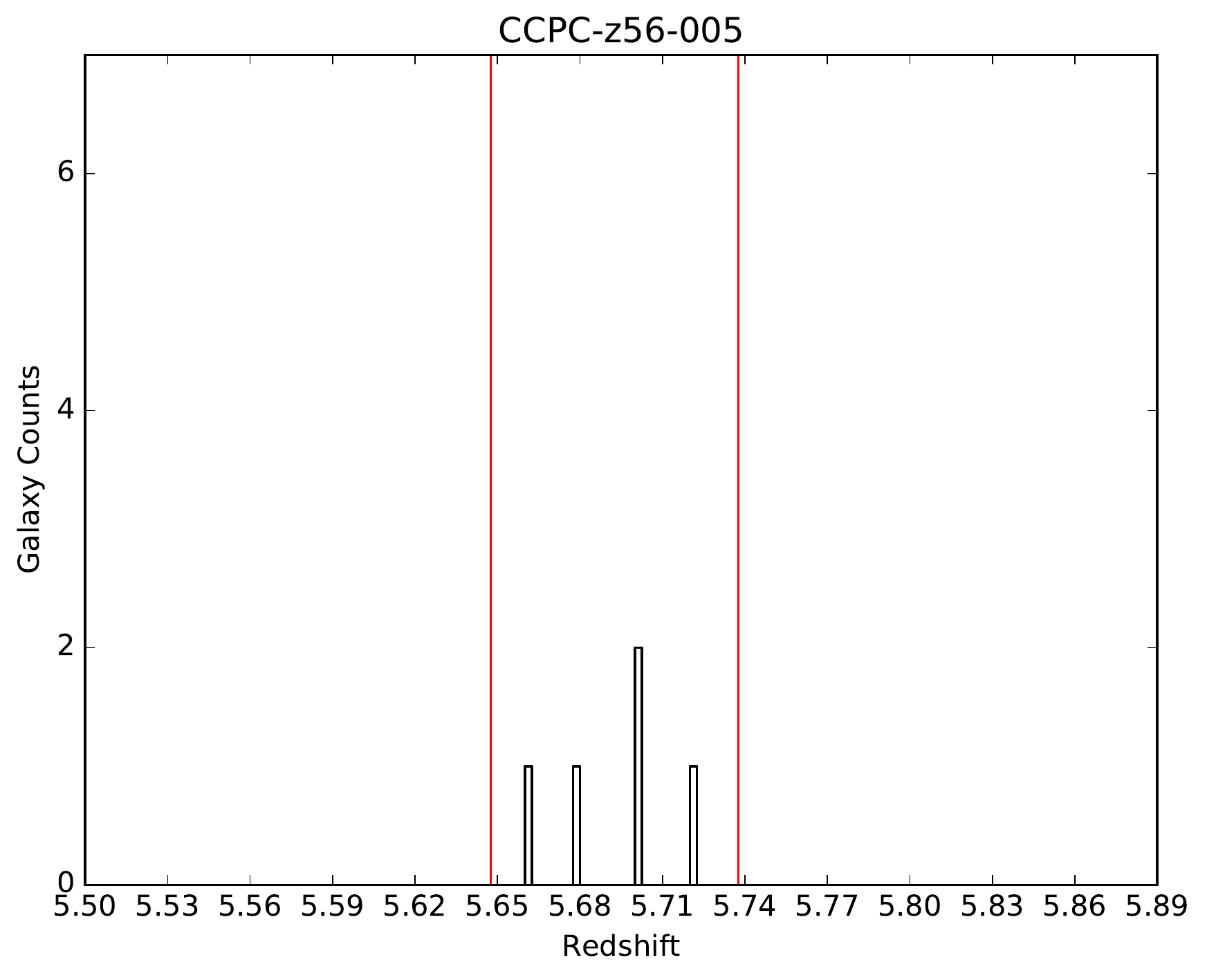}
\label{fig:CCPC-z56-005}
\end{subfigure}
\hfill
\end{figure*}

\begin{figure*}
\centering
\begin{subfigure}
\centering
\includegraphics[height=7.5cm,width=7.5cm]{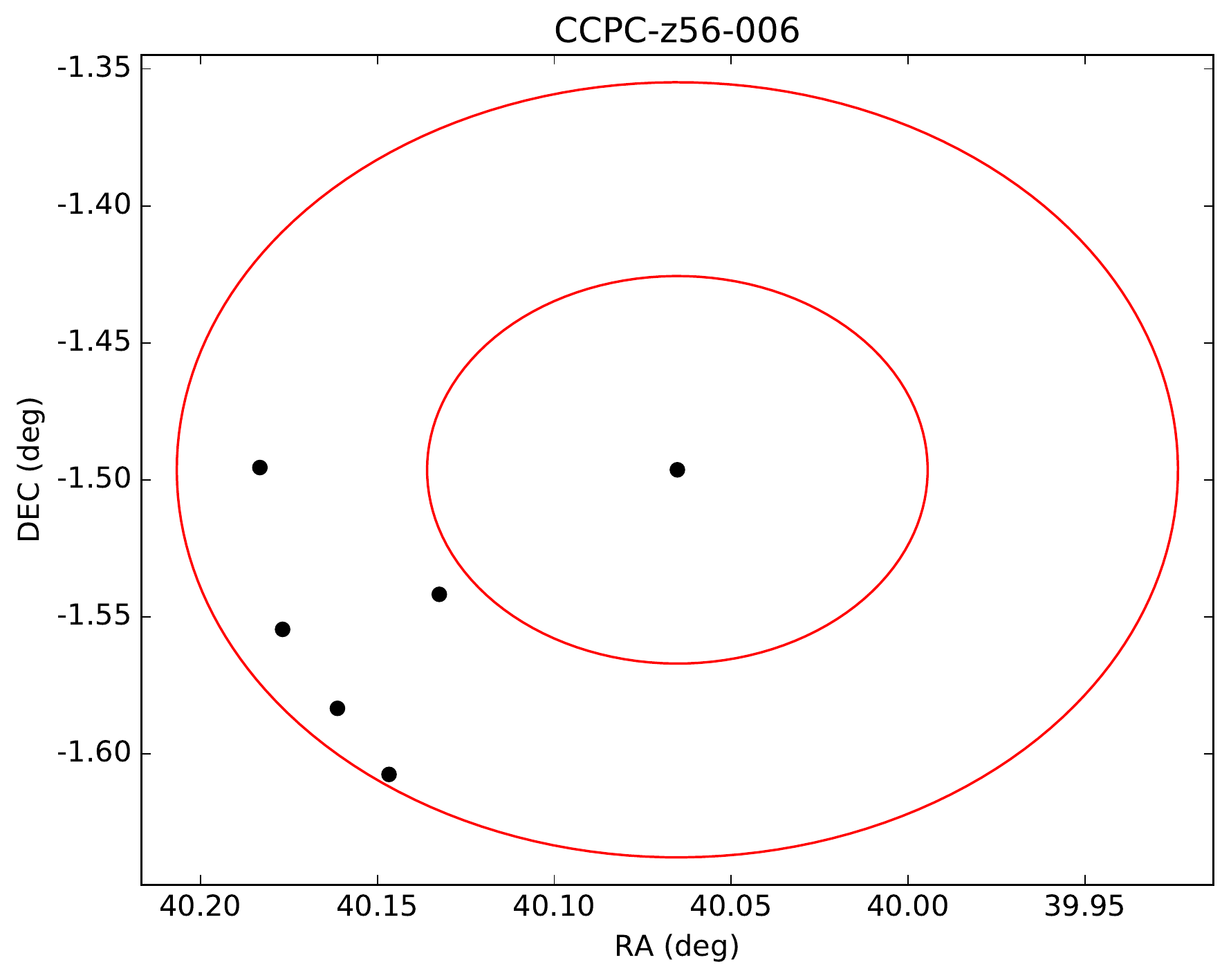}
\label{fig:CCPC-z56-006_sky}
\end{subfigure}
\hfill
\begin{subfigure}
\centering
\includegraphics[scale=0.52]{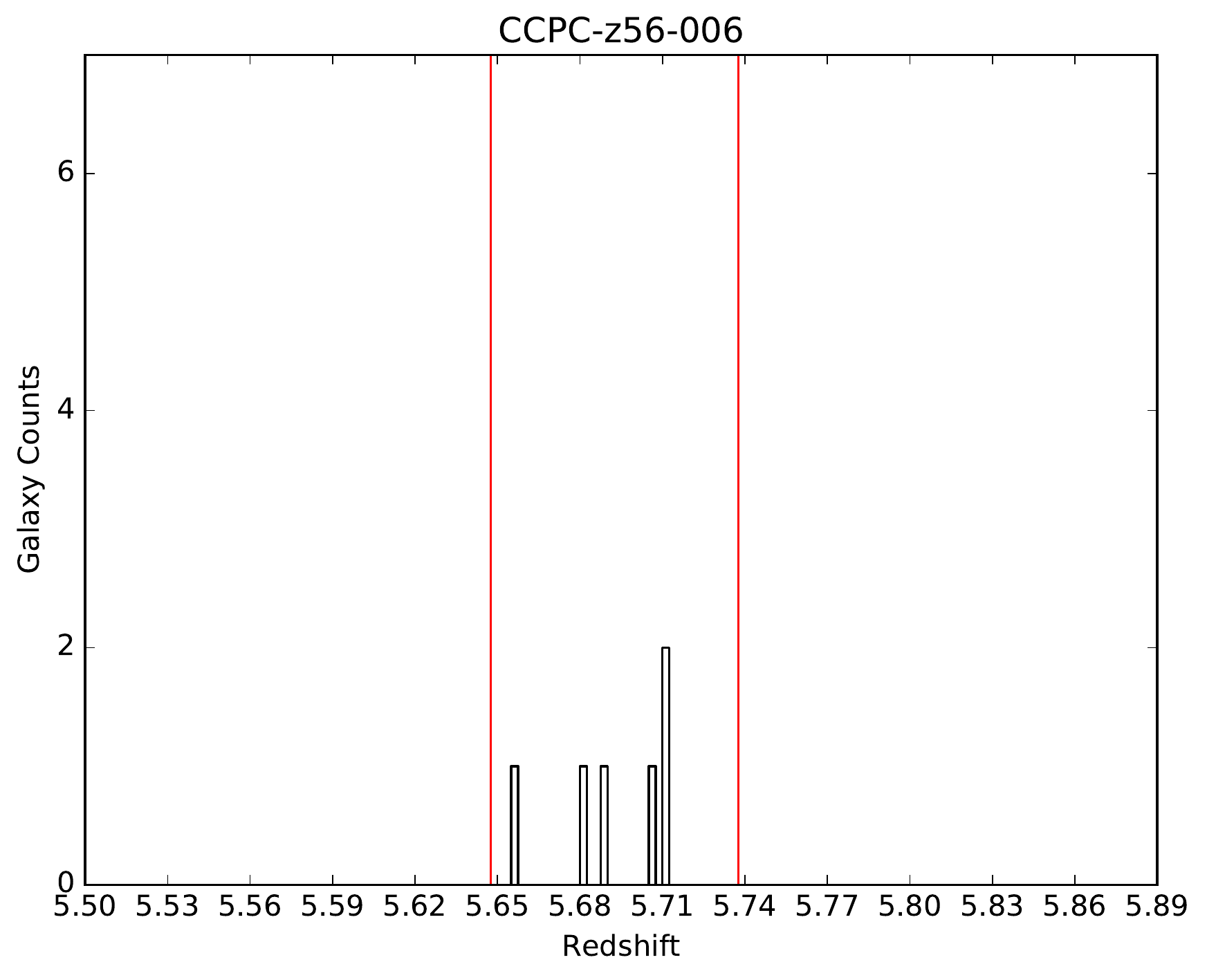}
\label{fig:CCPC-z56-006}
\end{subfigure}
\hfill
\end{figure*}
\clearpage 

\begin{figure*}
\centering
\begin{subfigure}
\centering
\includegraphics[height=7.5cm,width=7.5cm]{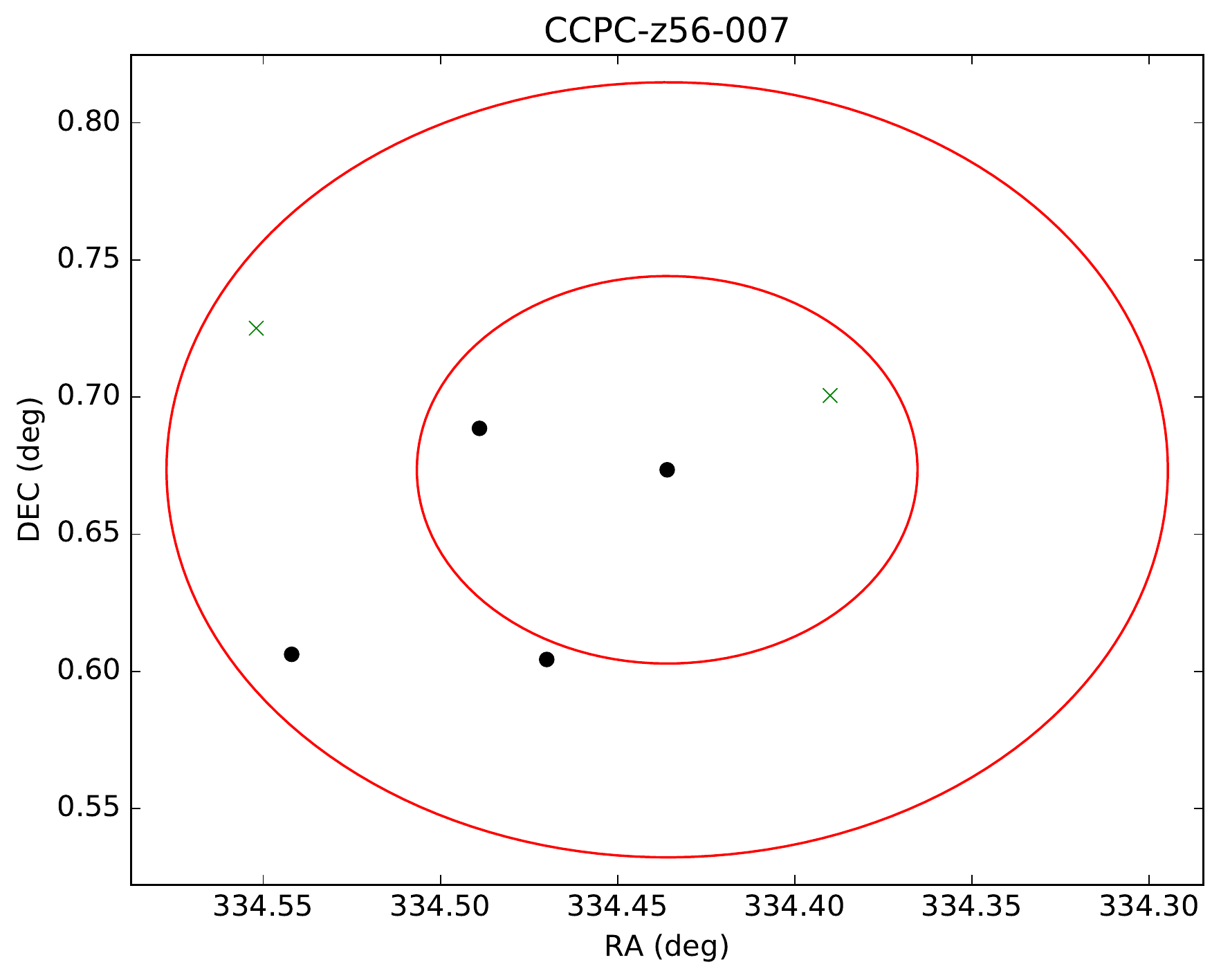}
\label{fig:CCPC-z56-007_sky}
\end{subfigure}
\hfill
\begin{subfigure}
\centering
\includegraphics[scale=0.52]{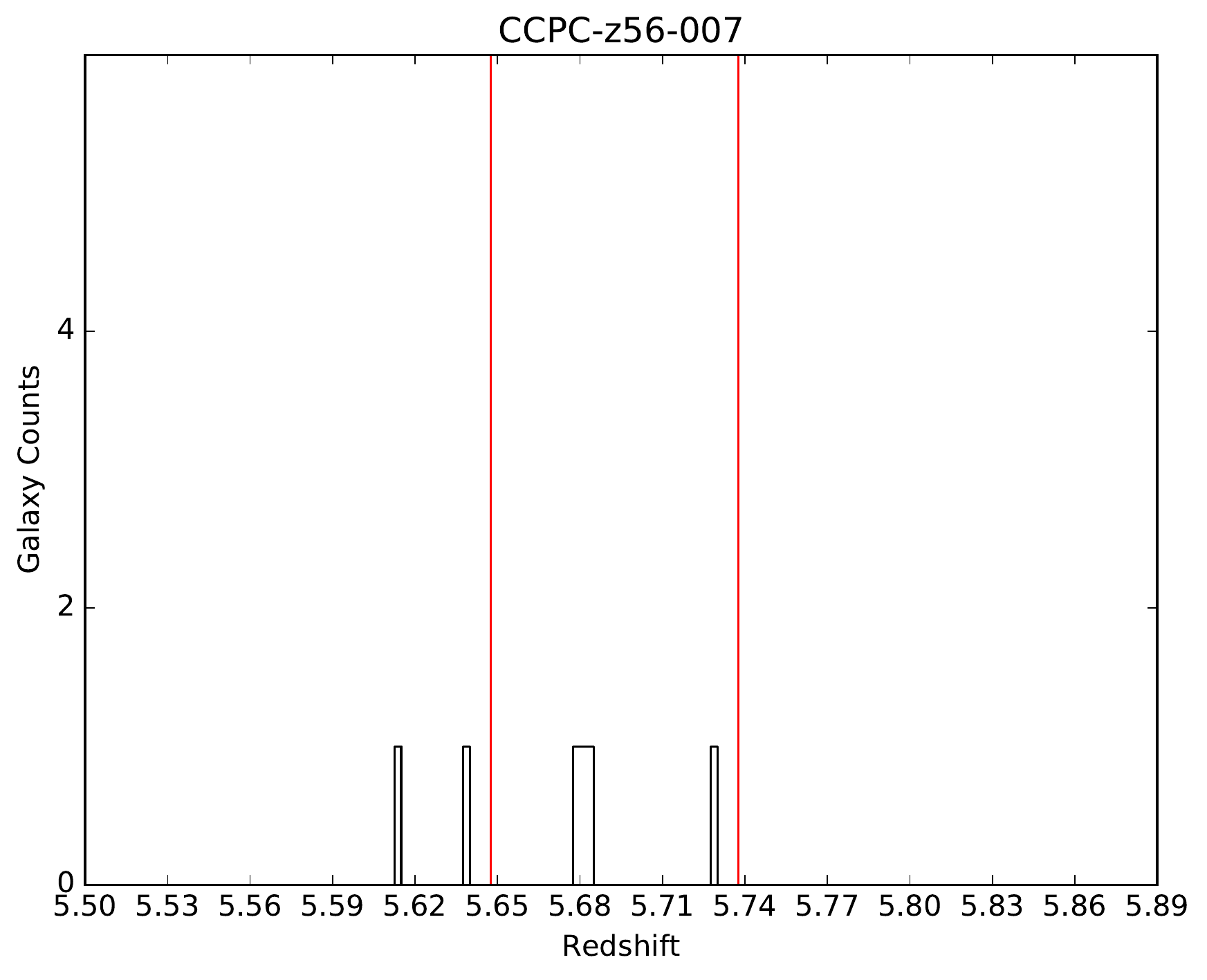}
\label{fig:CCPC-z56-007}
\end{subfigure}
\hfill
\end{figure*}

\begin{figure*}
\centering
\begin{subfigure}
\centering
\includegraphics[height=7.5cm,width=7.5cm]{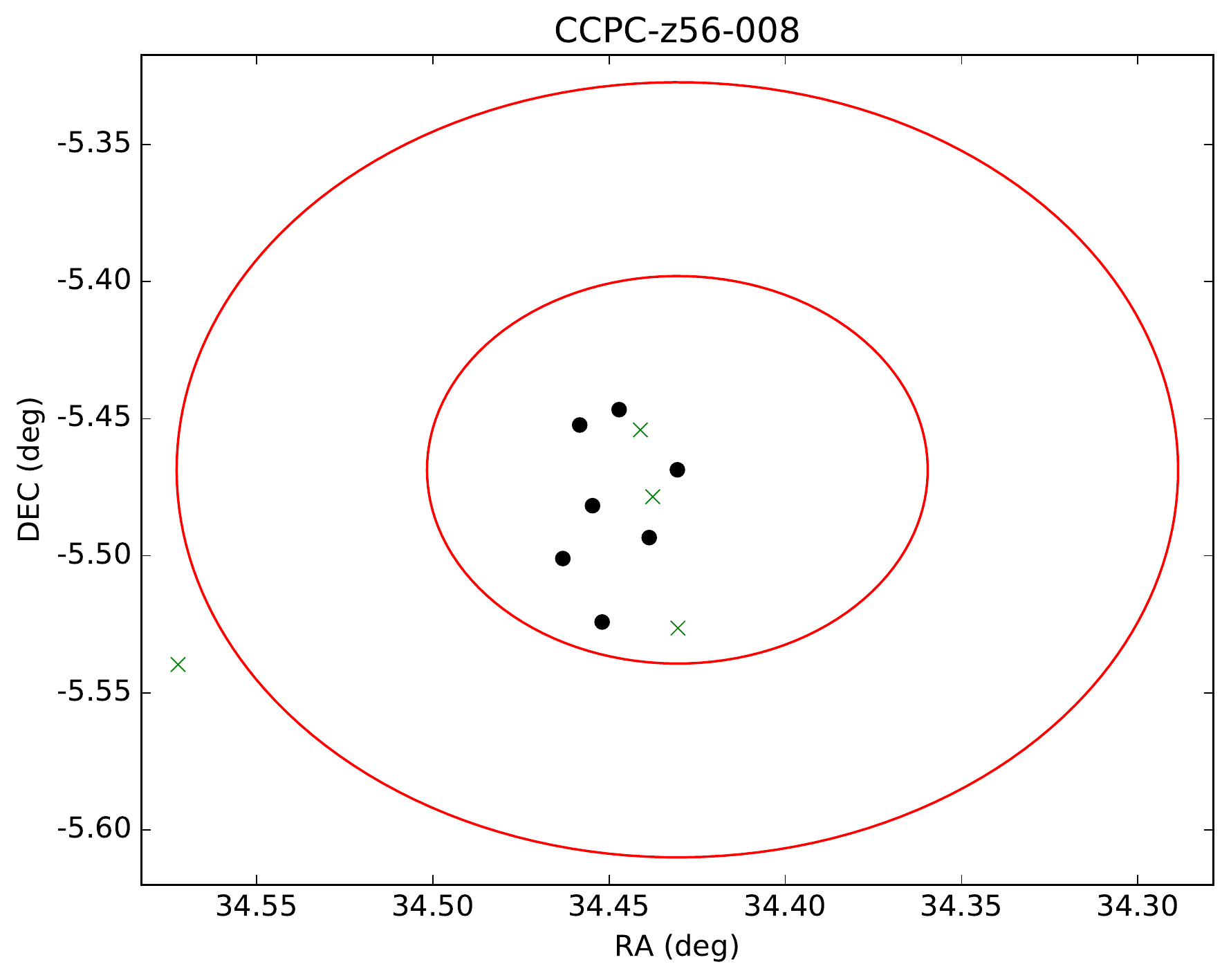}
\label{fig:CCPC-z56-008_sky}
\end{subfigure}
\hfill
\begin{subfigure}
\centering
\includegraphics[scale=0.52]{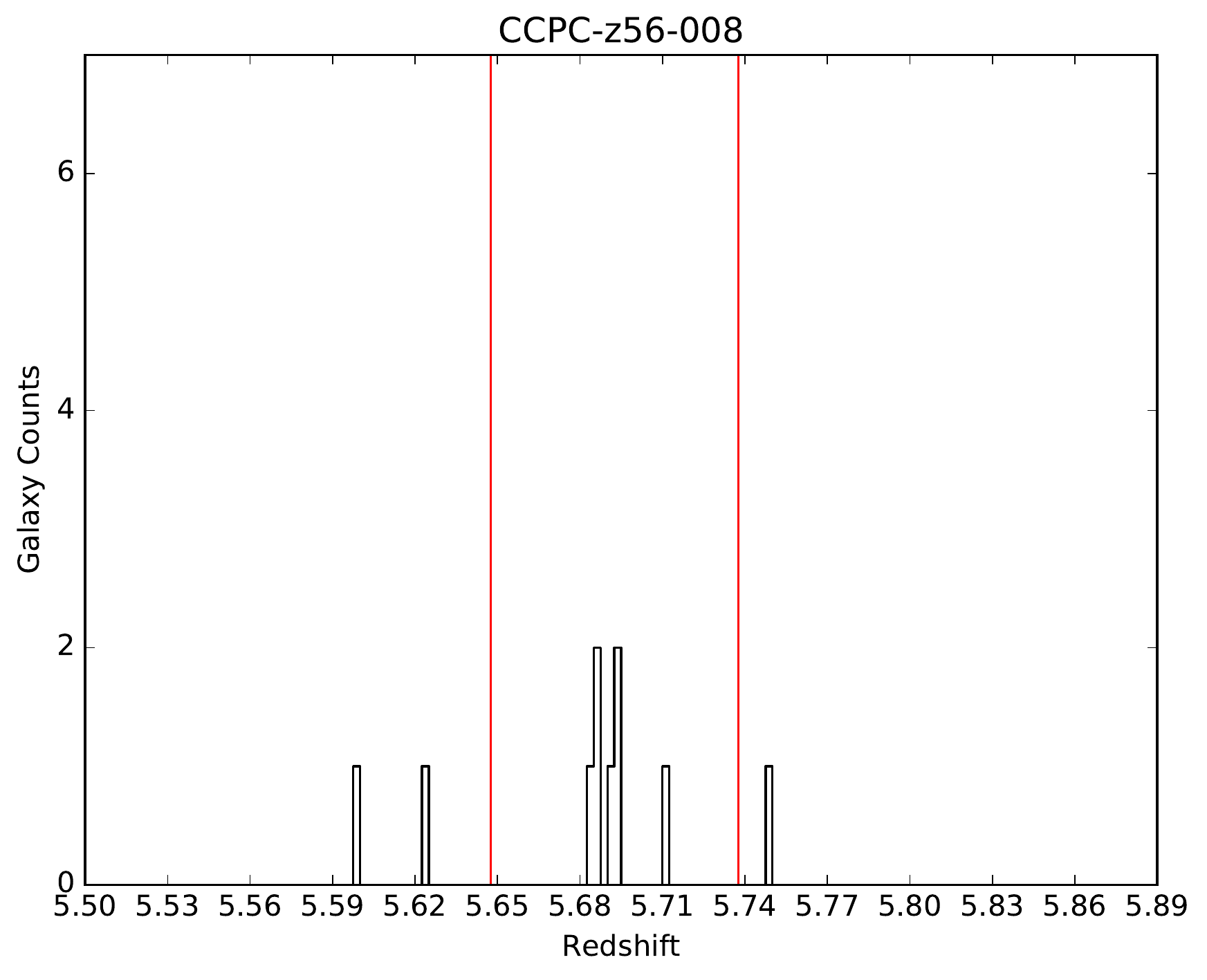}
\label{fig:CCPC-z56-008}
\end{subfigure}
\hfill
\end{figure*}

\begin{figure*}
\centering
\begin{subfigure}
\centering
\includegraphics[height=7.5cm,width=7.5cm]{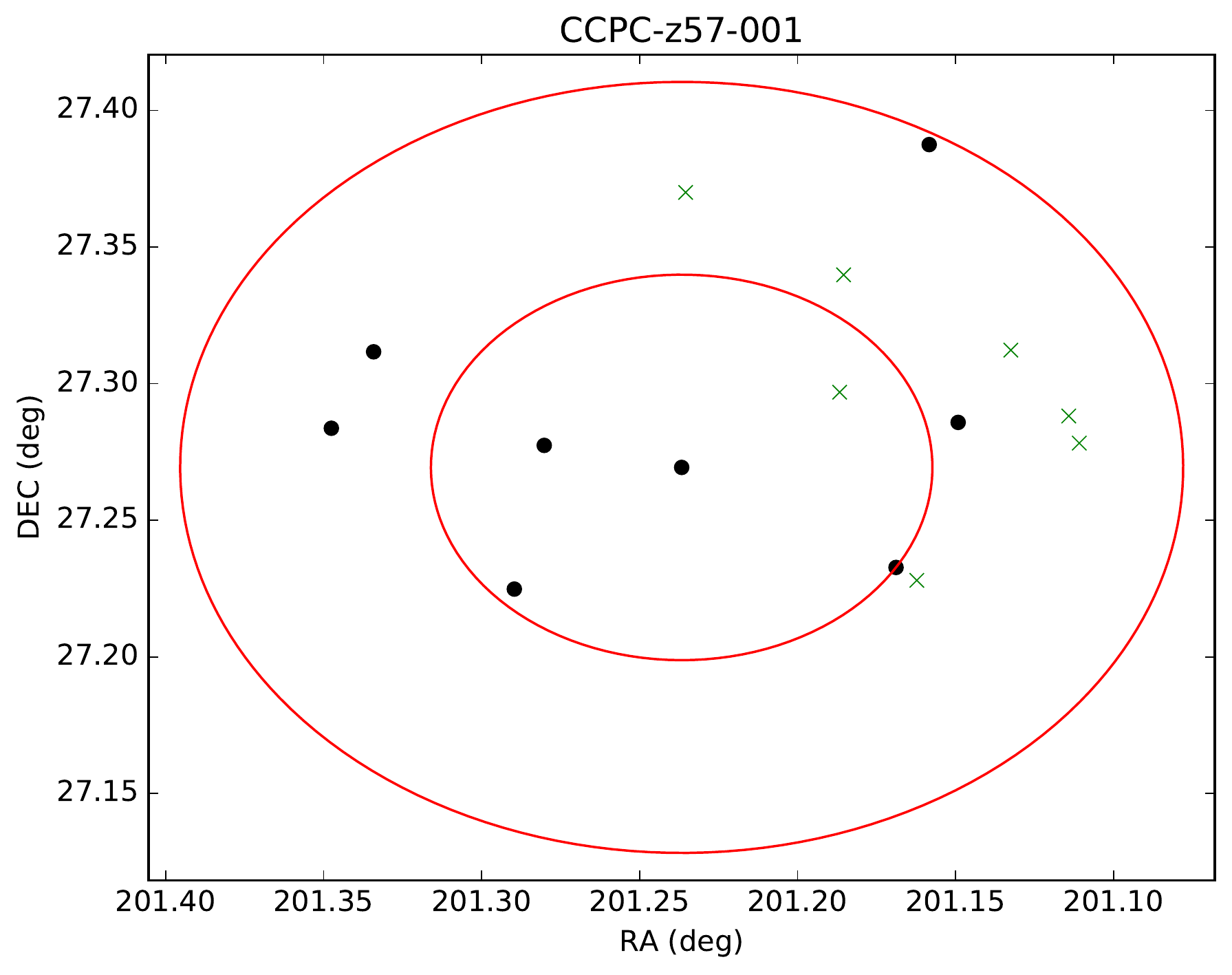}
\label{fig:CCPC-z57-001_sky}
\end{subfigure}
\hfill
\begin{subfigure}
\centering
\includegraphics[scale=0.52]{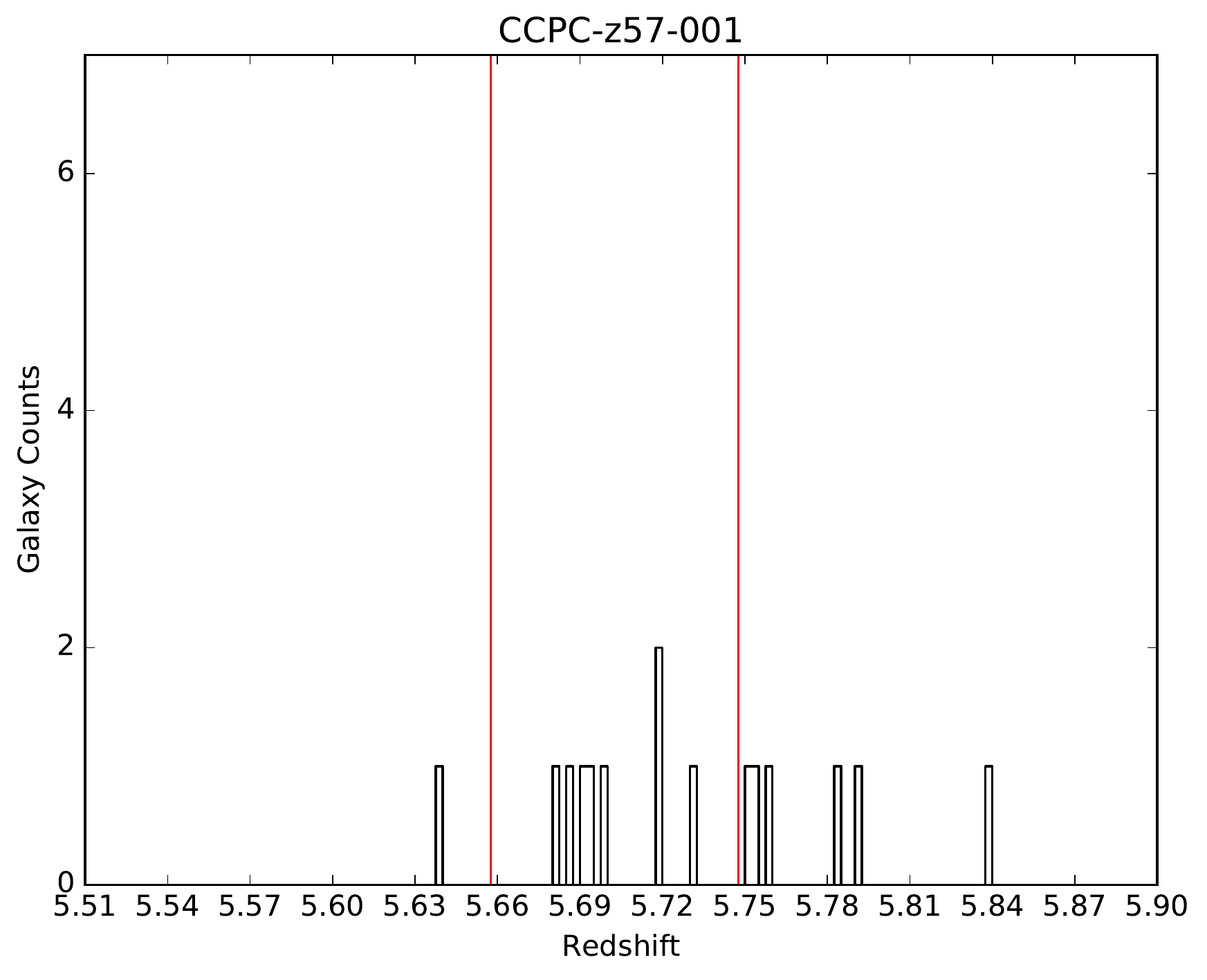}
\label{fig:CCPC-z57-001}
\end{subfigure}
\hfill
\end{figure*}
\clearpage 

\begin{figure*}
\centering
\begin{subfigure}
\centering
\includegraphics[height=7.5cm,width=7.5cm]{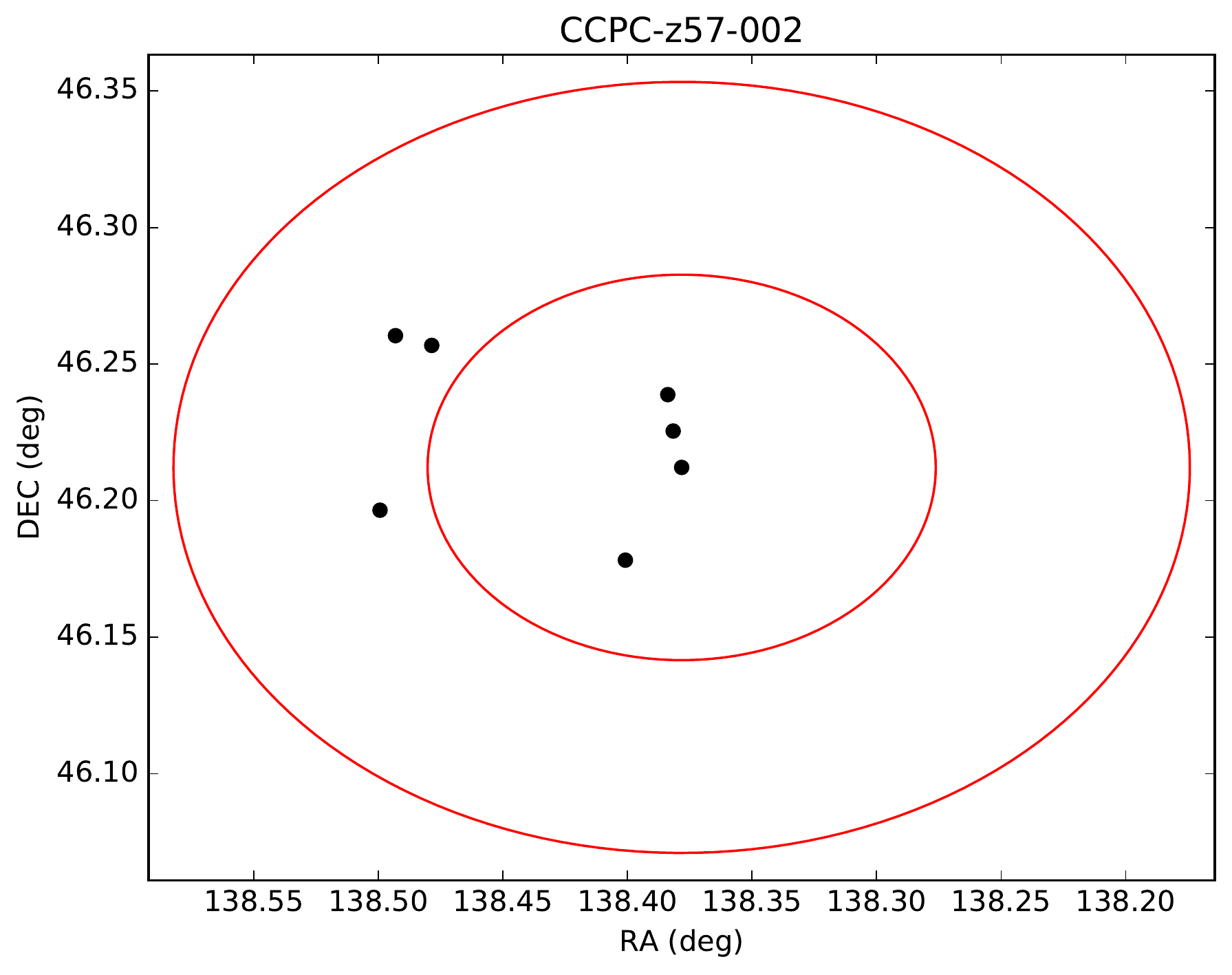}
\label{fig:CCPC-z57-002_sky}
\end{subfigure}
\hfill
\begin{subfigure}
\centering
\includegraphics[scale=0.52]{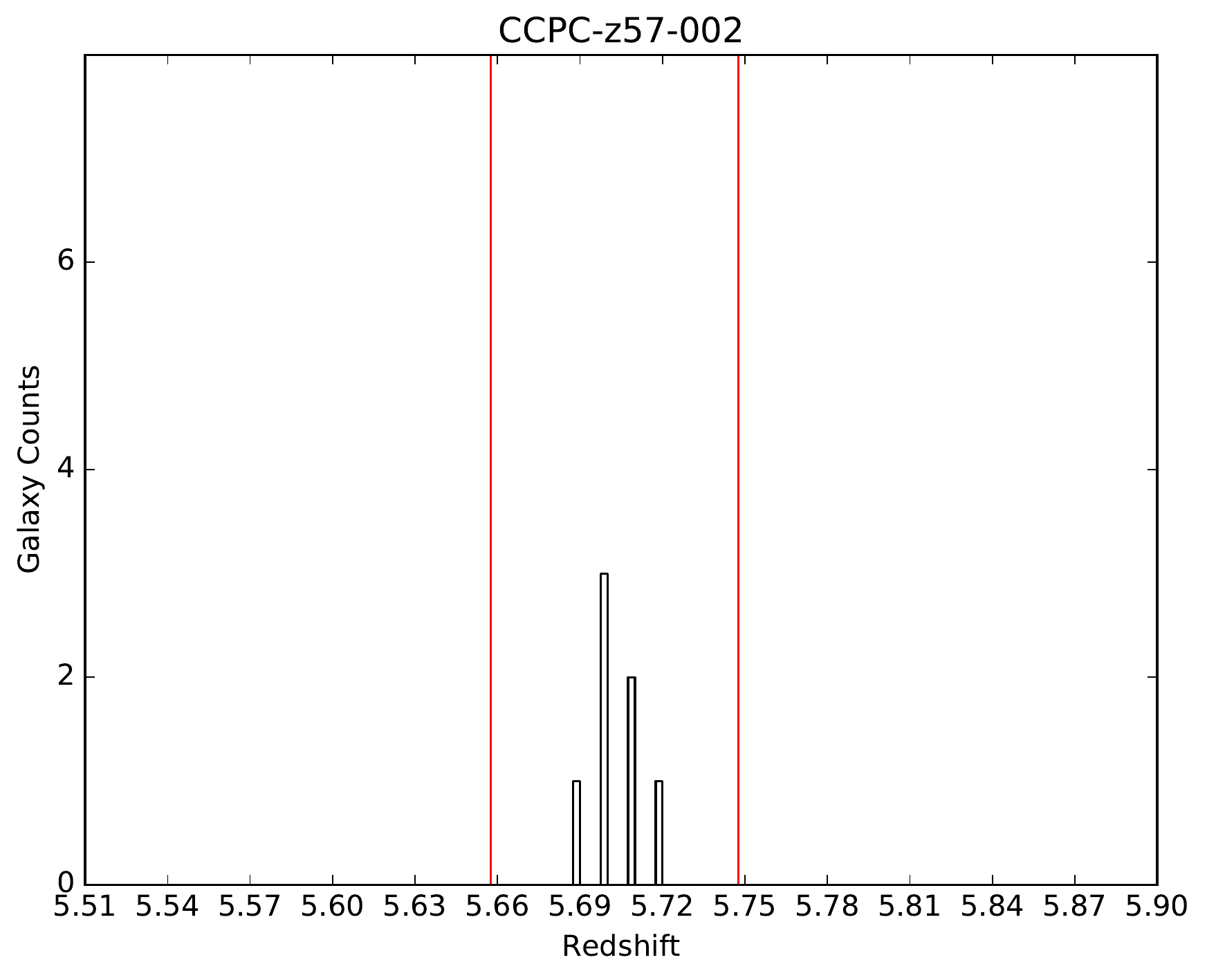}
\label{fig:CCPC-z57-002}
\end{subfigure}
\hfill
\end{figure*}

\begin{figure*}
\centering
\begin{subfigure}
\centering
\includegraphics[height=7.5cm,width=7.5cm]{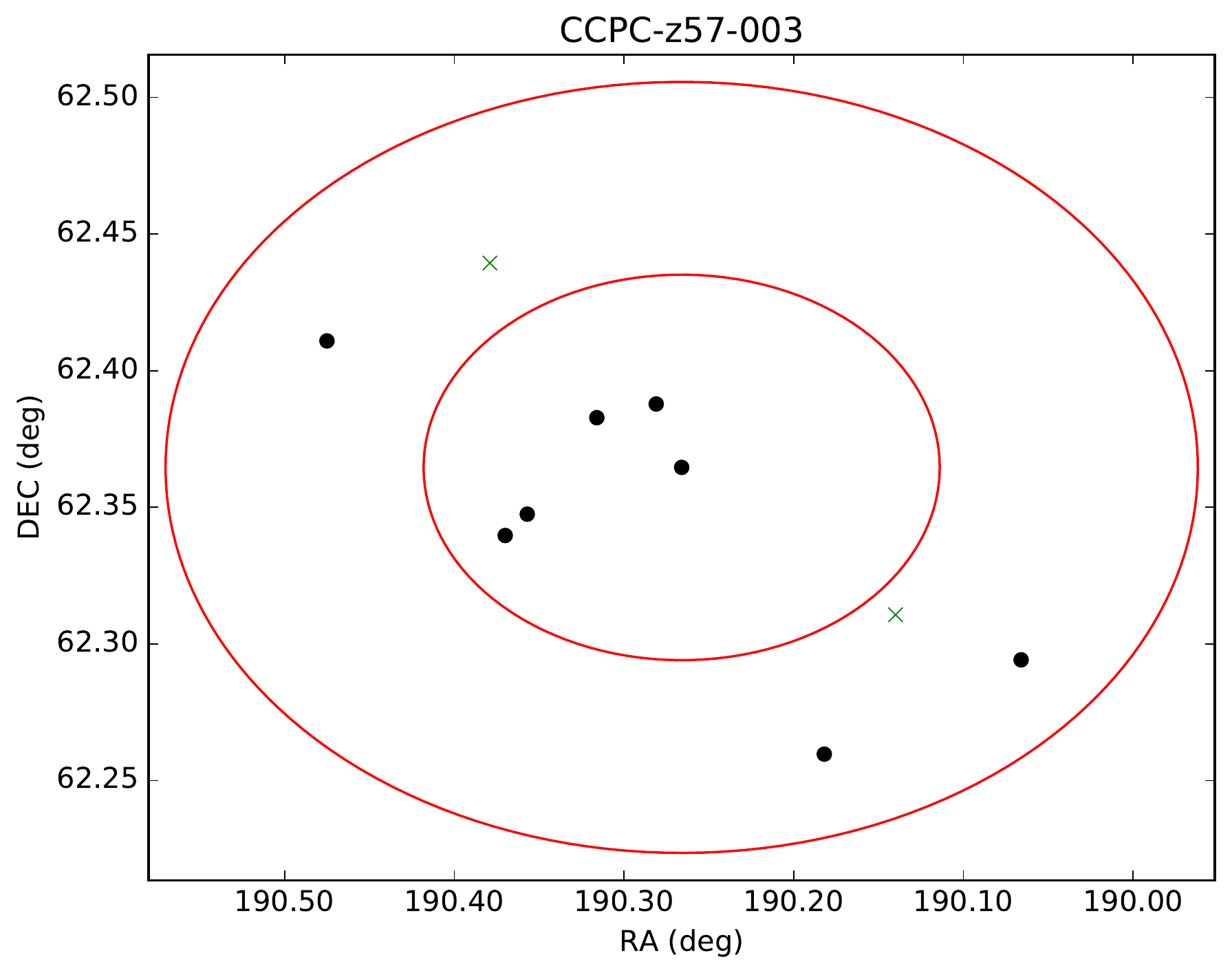}
\label{fig:CCPC-z57-003_sky}
\end{subfigure}
\hfill
\begin{subfigure}
\centering
\includegraphics[scale=0.52]{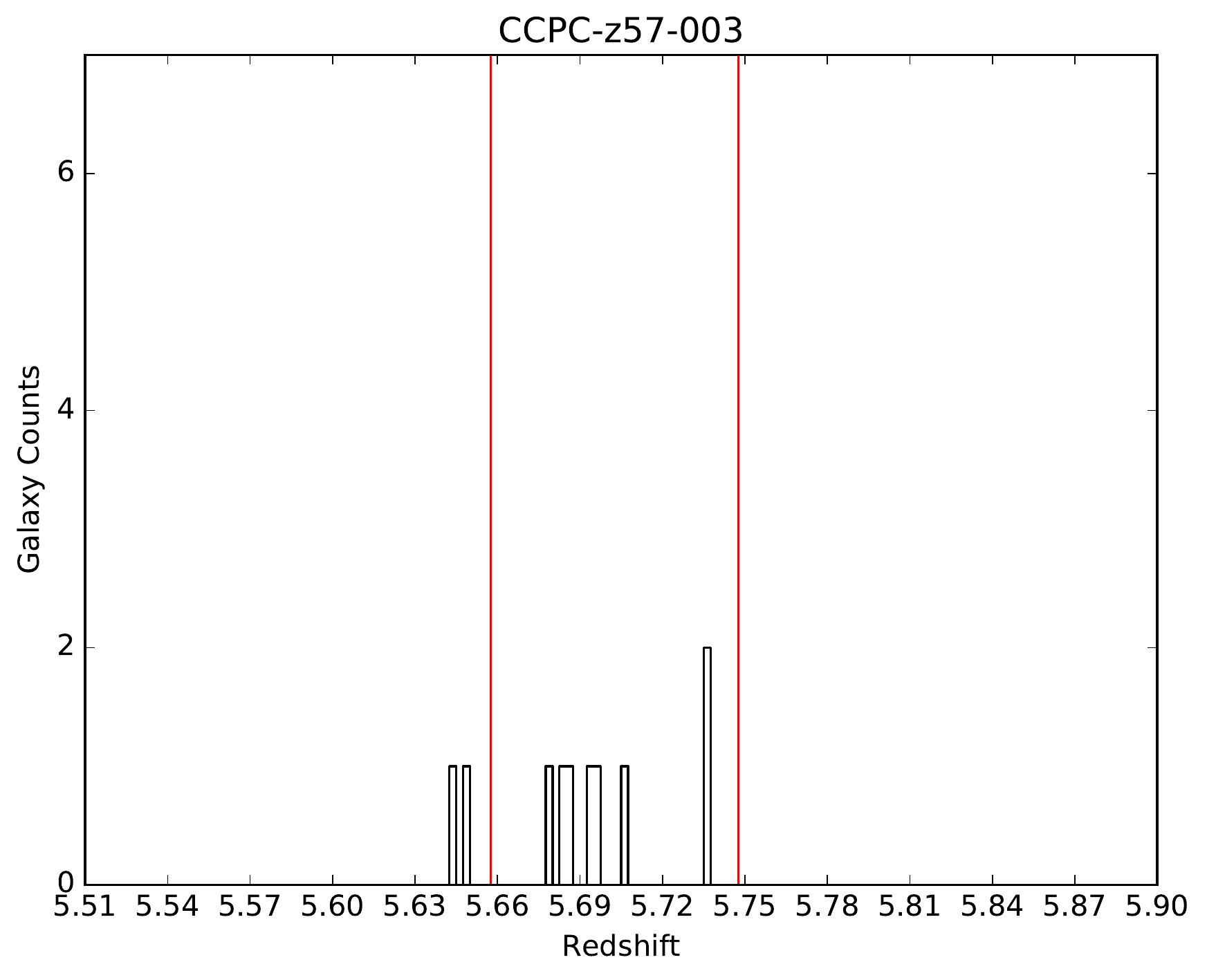}
\label{fig:CCPC-z57-003}
\end{subfigure}
\hfill
\end{figure*}

\begin{figure*}
\centering
\begin{subfigure}
\centering
\includegraphics[height=7.5cm,width=7.5cm]{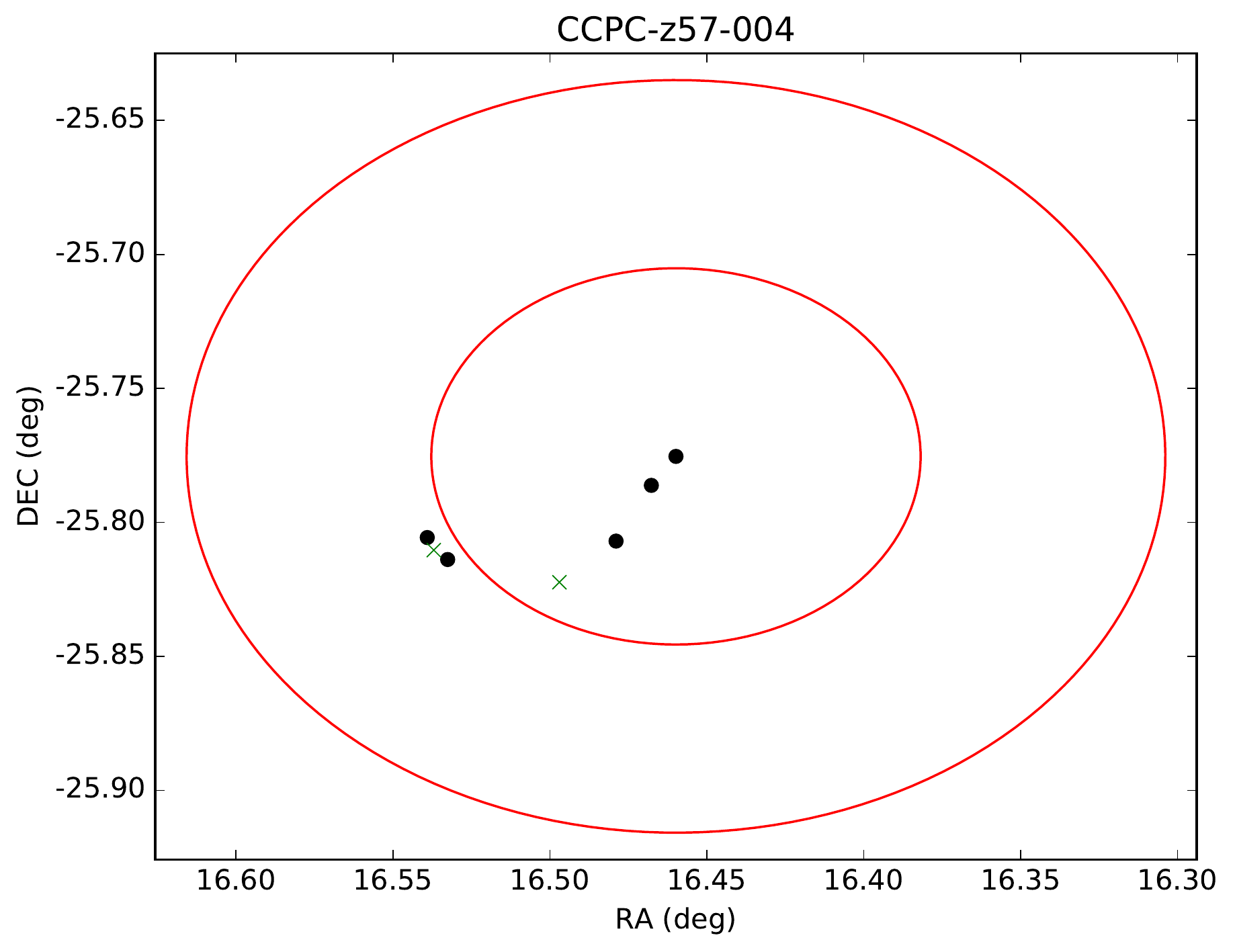}
\label{fig:CCPC-z57-004_sky}
\end{subfigure}
\hfill
\begin{subfigure}
\centering
\includegraphics[scale=0.52]{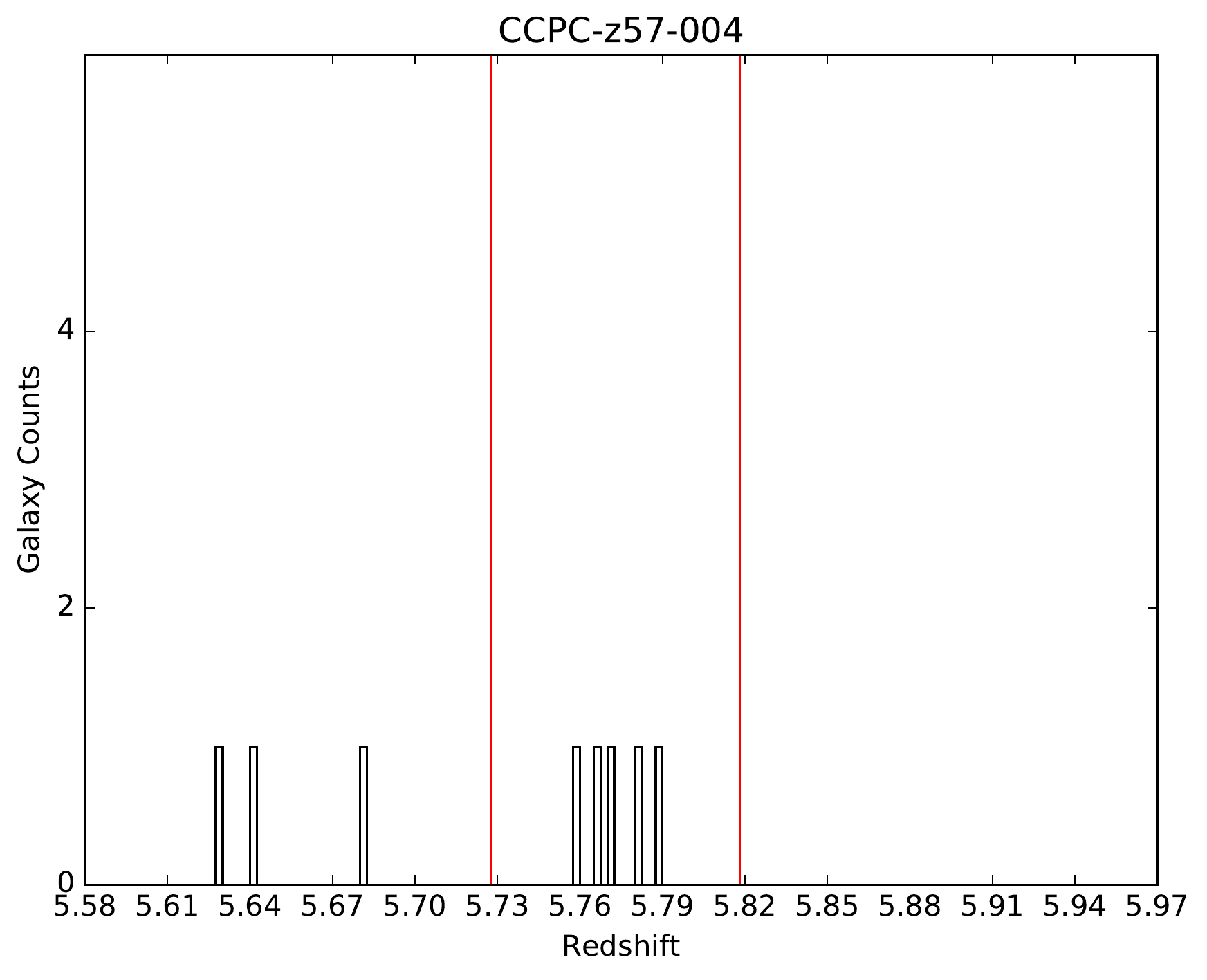}
\label{fig:CCPC-z57-004}
\end{subfigure}
\hfill
\end{figure*}
\clearpage 

\begin{figure*}
\centering
\begin{subfigure}
\centering
\includegraphics[height=7.5cm,width=7.5cm]{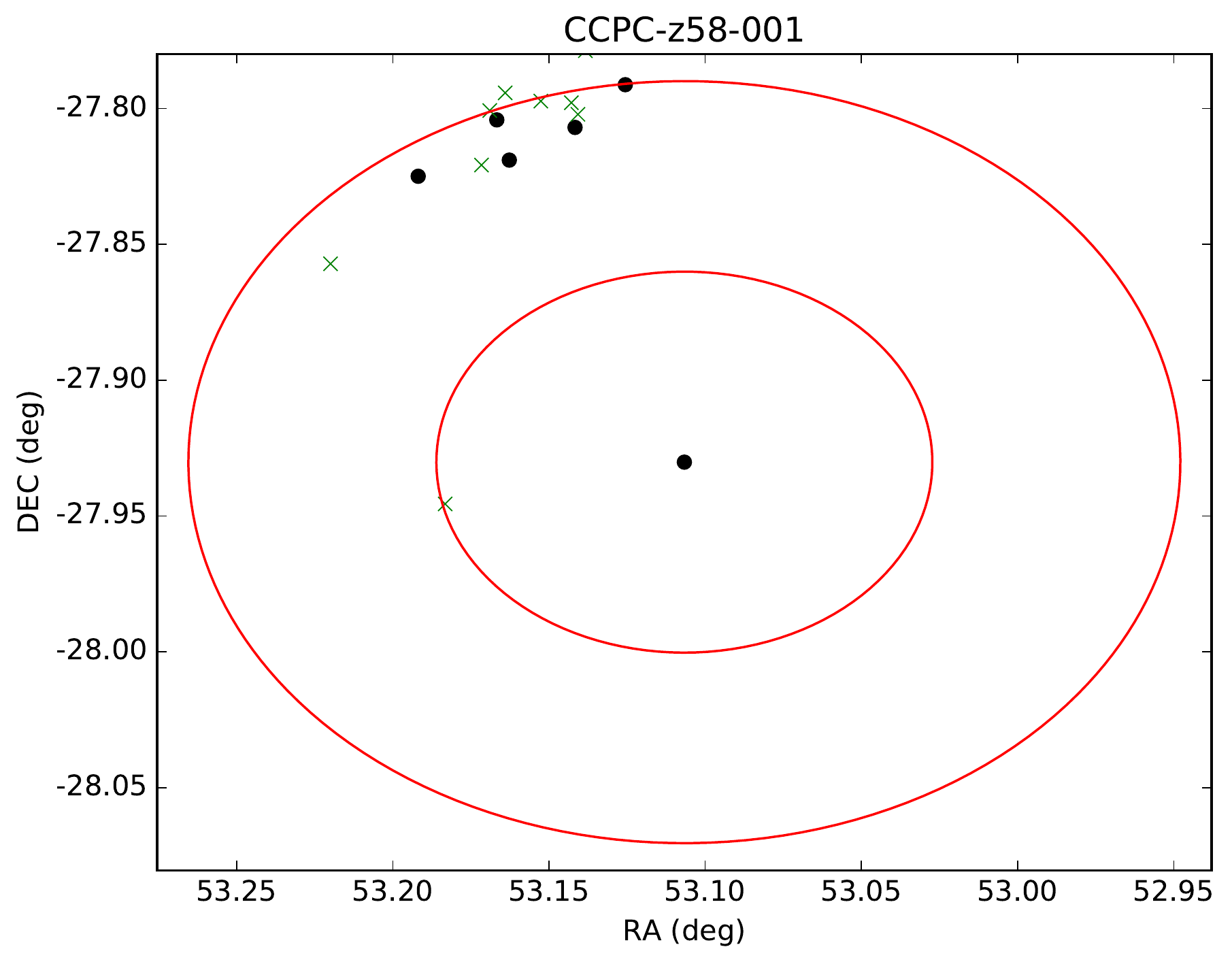}
\label{fig:CCPC-z58-001_sky}
\end{subfigure}
\hfill
\begin{subfigure}
\centering
\includegraphics[scale=0.52]{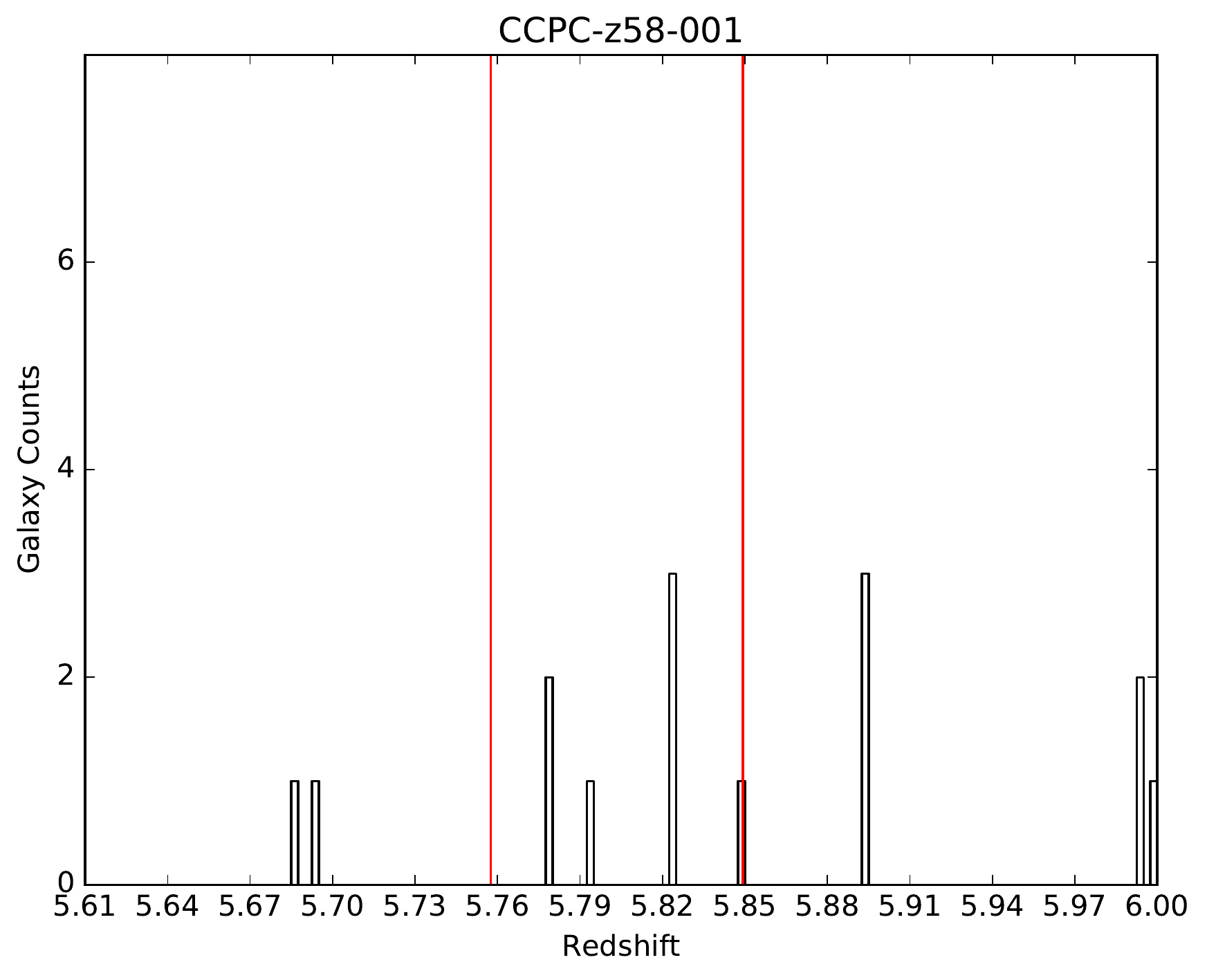}
\label{fig:CCPC-z58-001}
\end{subfigure}
\hfill
\end{figure*}

\begin{figure*}
\centering
\begin{subfigure}
\centering
\includegraphics[height=7.5cm,width=7.5cm]{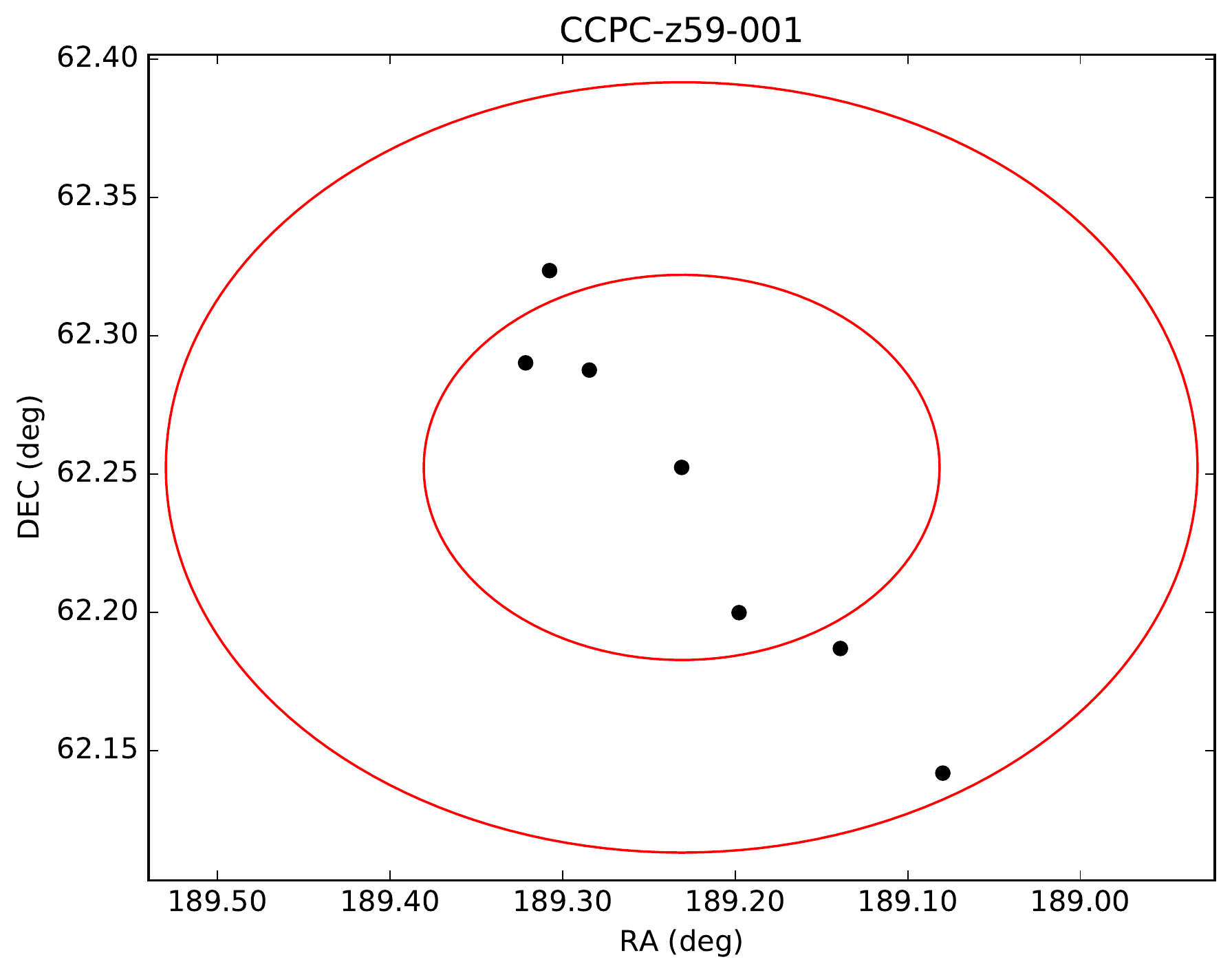}
\label{fig:CCPC-z59-001_sky}
\end{subfigure}
\hfill
\begin{subfigure}
\centering
\includegraphics[scale=0.52]{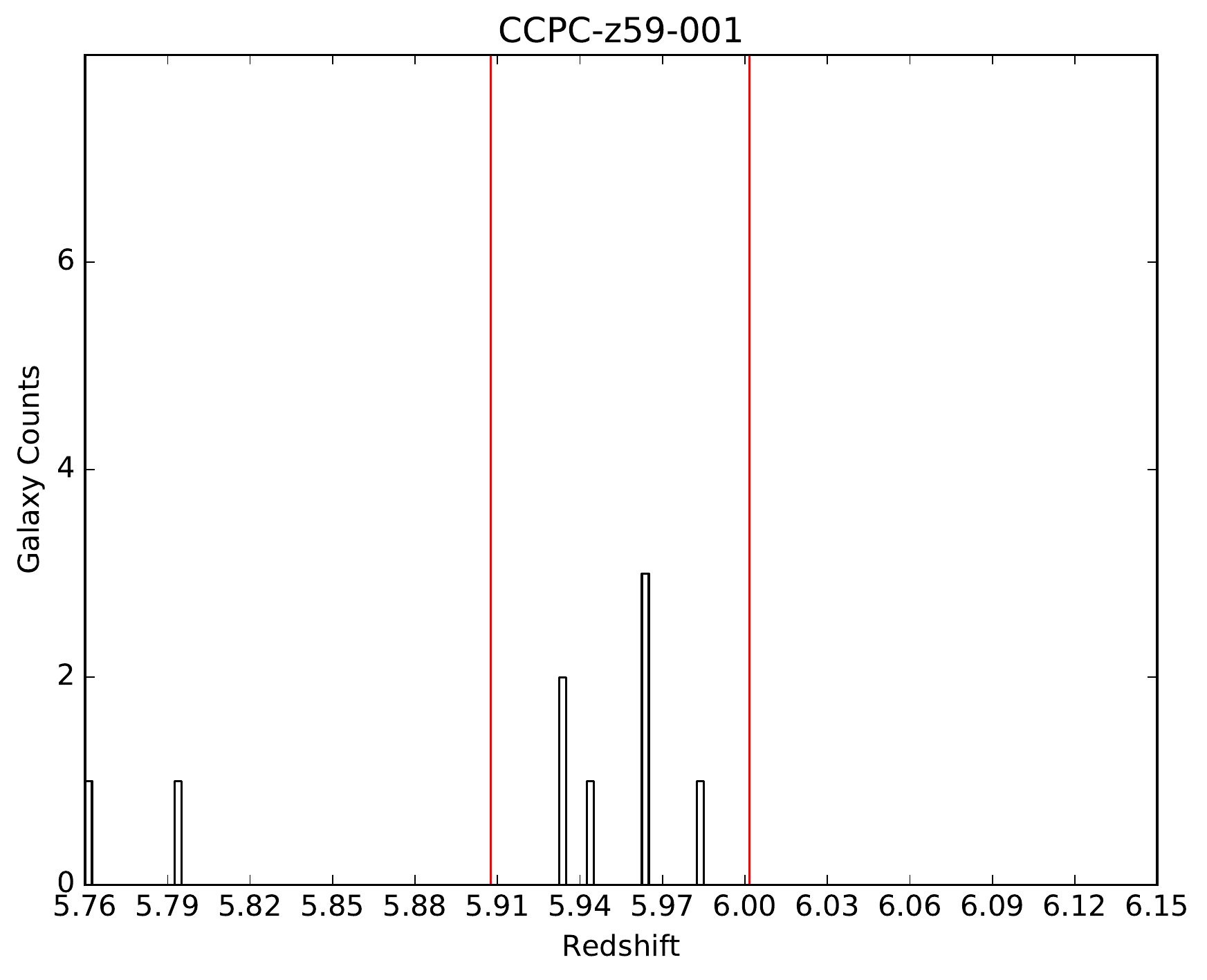}
\label{fig:CCPC-z59-001}
\end{subfigure}
\hfill
\end{figure*}

\begin{figure*}
\centering
\begin{subfigure}
\centering
\includegraphics[height=7.5cm,width=7.5cm]{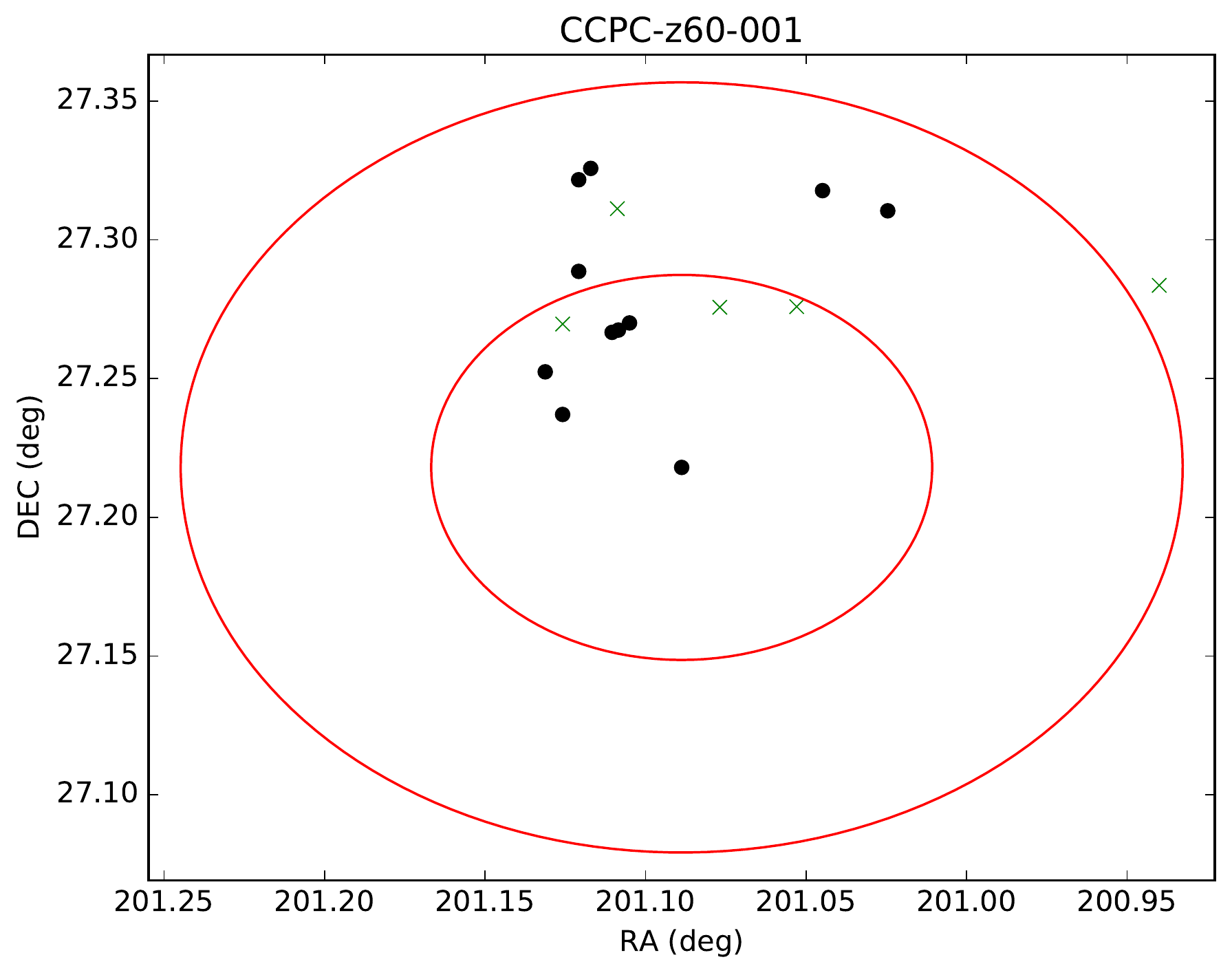}
\label{fig:CCPC-z60-001_sky}
\end{subfigure}
\hfill
\begin{subfigure}
\centering
\includegraphics[scale=0.52]{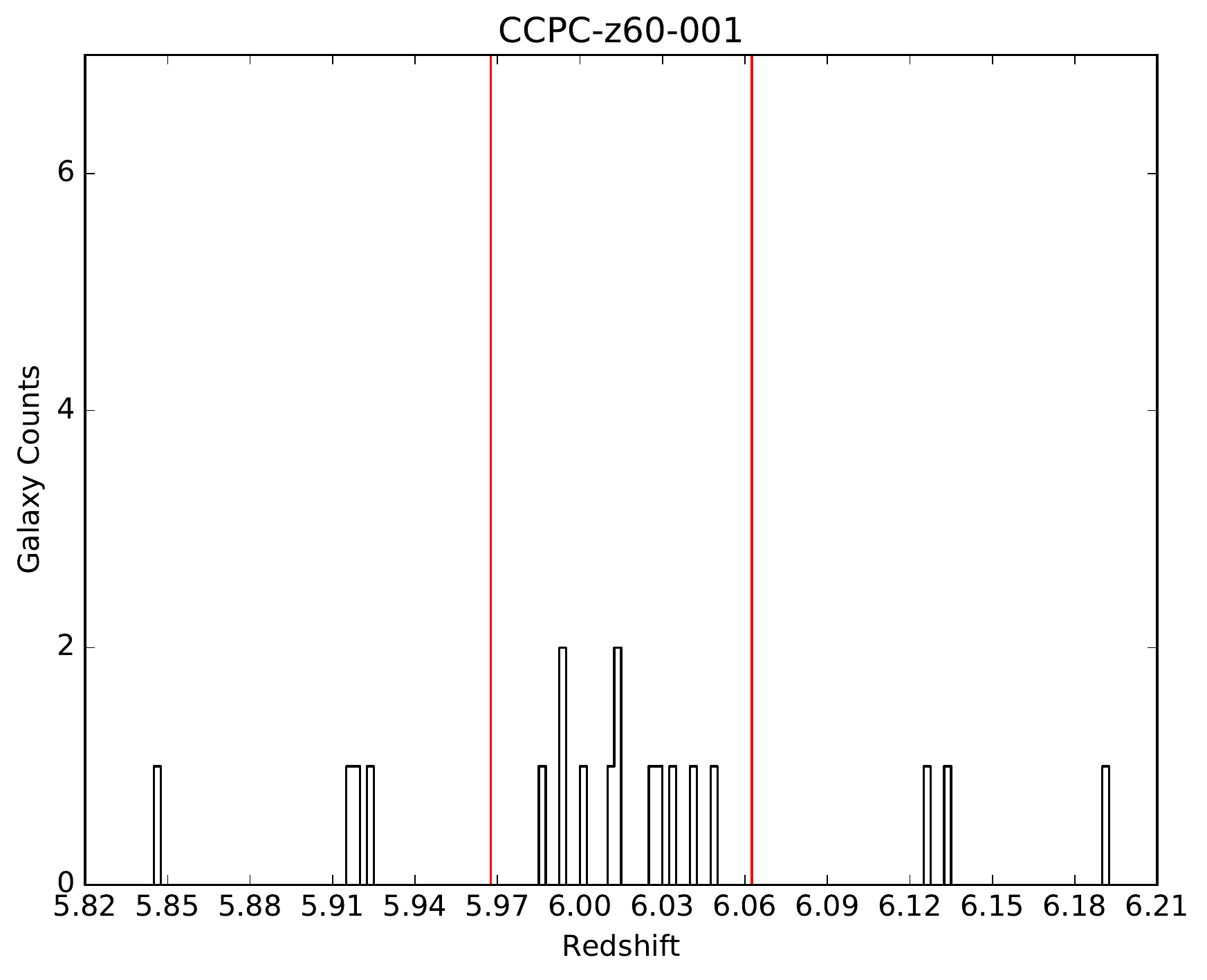}
\label{fig:CCPC-z60-001}
\end{subfigure}
\hfill
\end{figure*}
\clearpage 

\begin{figure*}
\centering
\begin{subfigure}
\centering
\includegraphics[height=7.5cm,width=7.5cm]{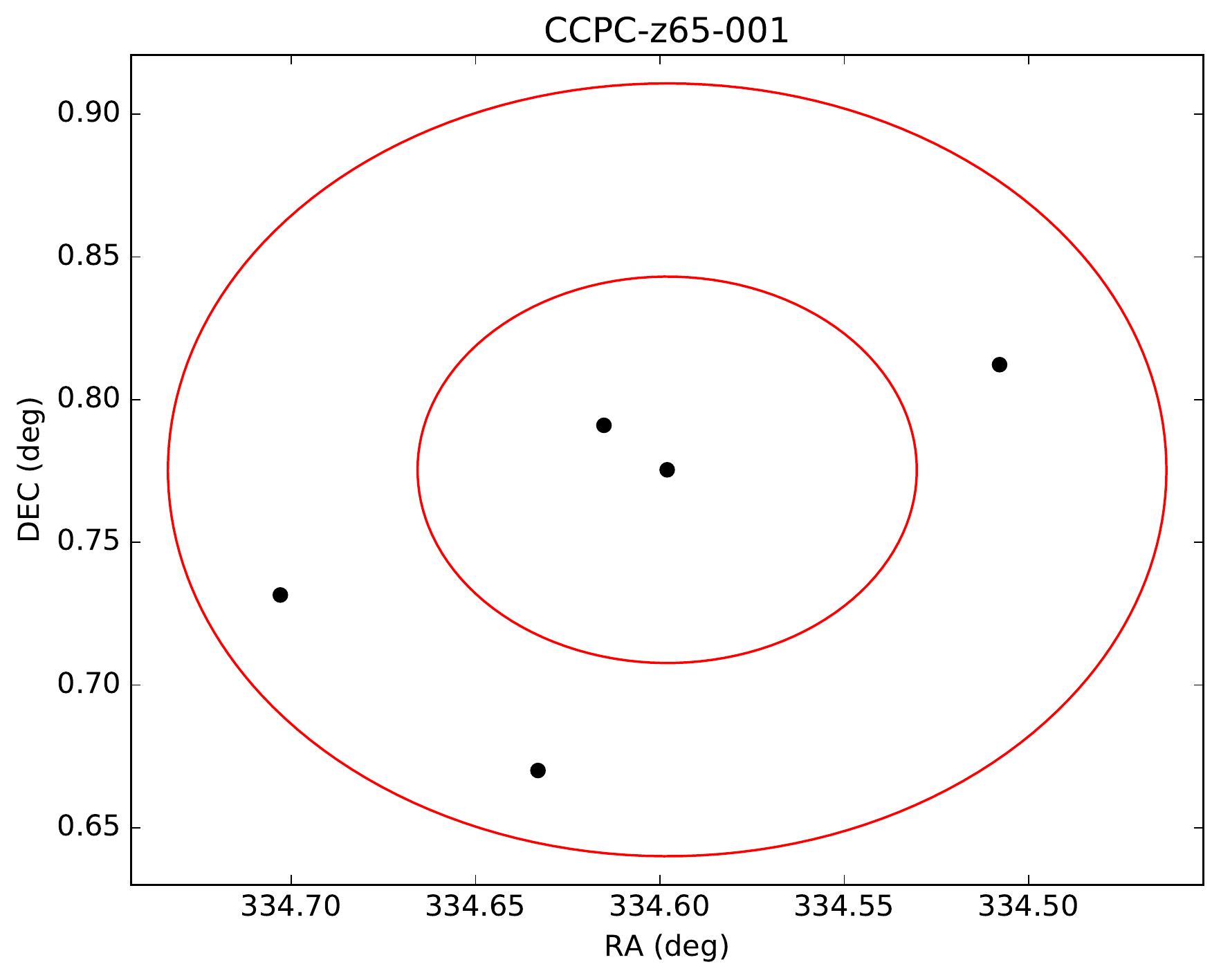}
\label{fig:CCPC-z65-001_sky}
\end{subfigure}
\hfill
\begin{subfigure}
\centering
\includegraphics[scale=0.52]{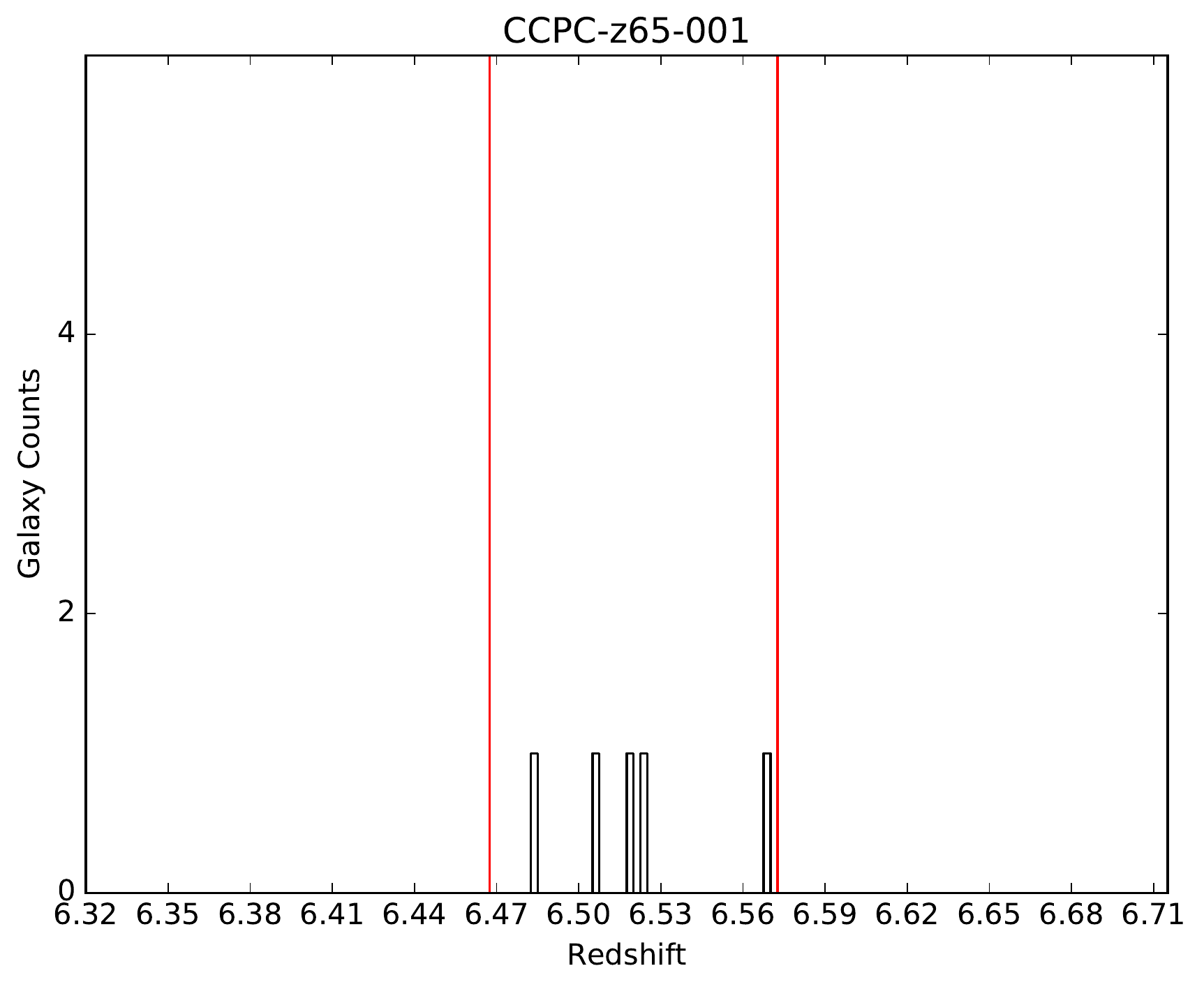}
\label{fig:CCPC-z65-001}
\end{subfigure}
\hfill
\end{figure*}

\begin{figure*}
\centering
\begin{subfigure}
\centering
\includegraphics[height=7.5cm,width=7.5cm]{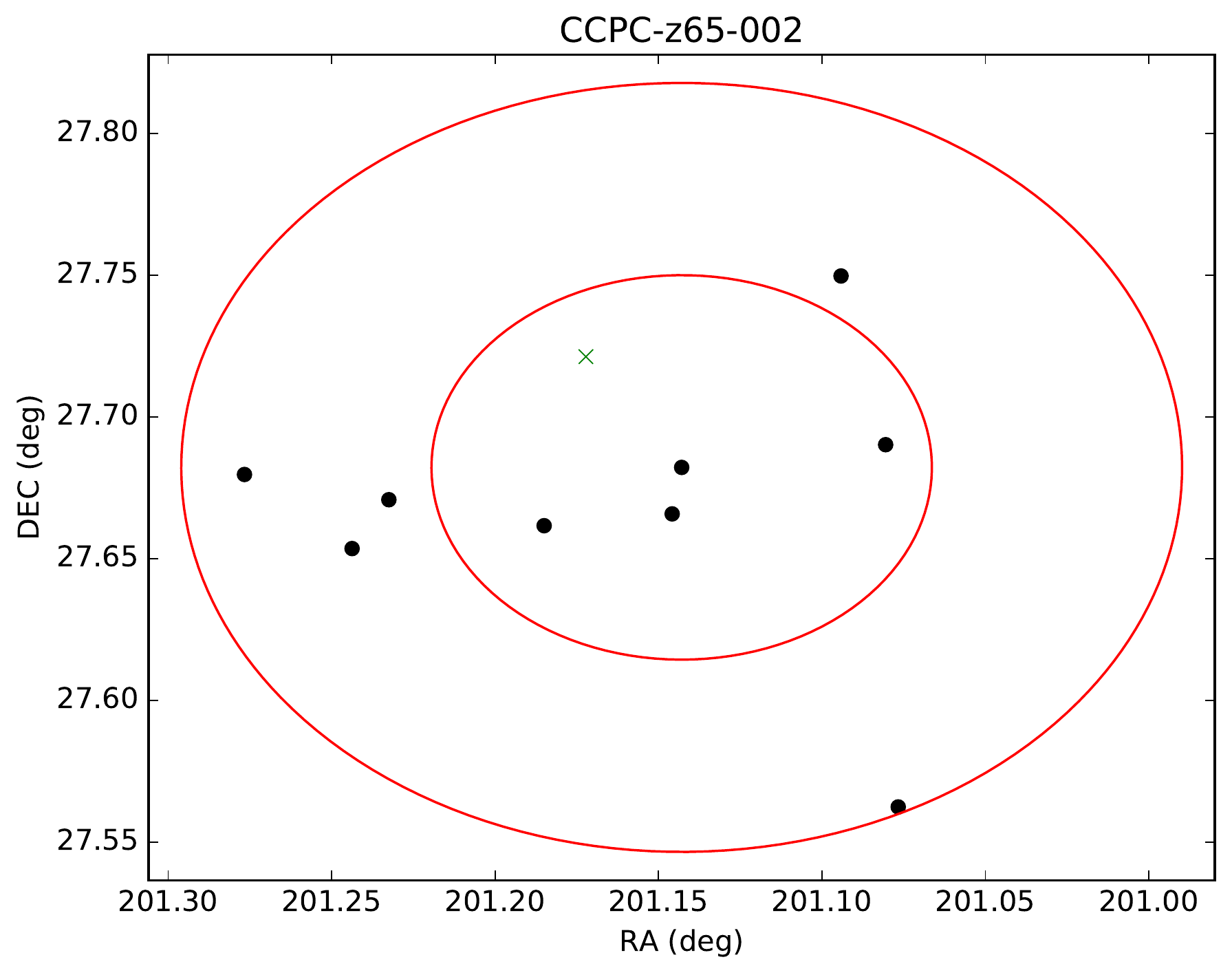}
\label{fig:CCPC-z65-002_sky}
\end{subfigure}
\hfill
\begin{subfigure}
\centering
\includegraphics[scale=0.52]{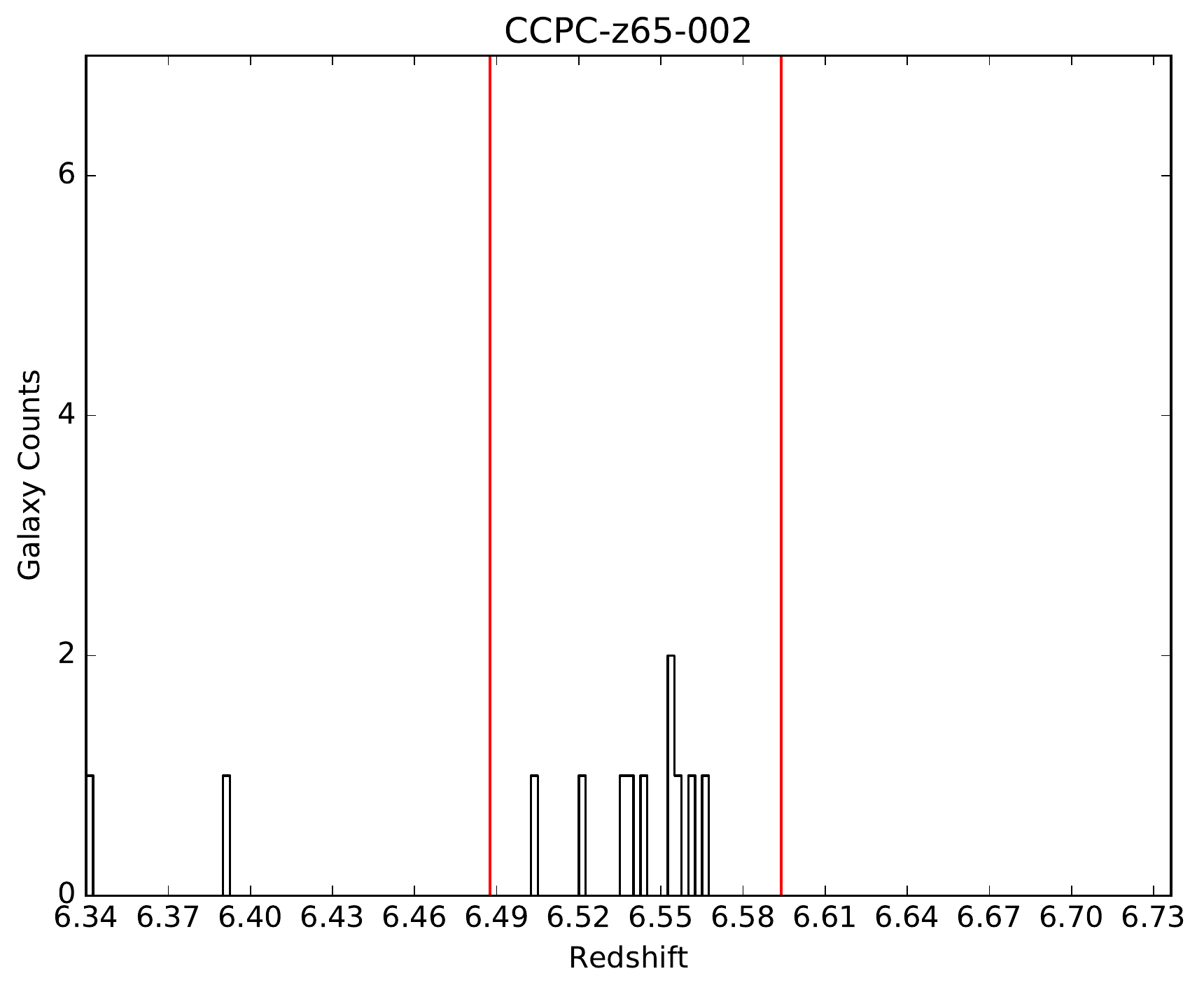}
\label{fig:CCPC-z65-002}
\end{subfigure}
\hfill
\end{figure*}

\begin{figure*}
\centering
\begin{subfigure}
\centering
\includegraphics[height=7.5cm,width=7.5cm]{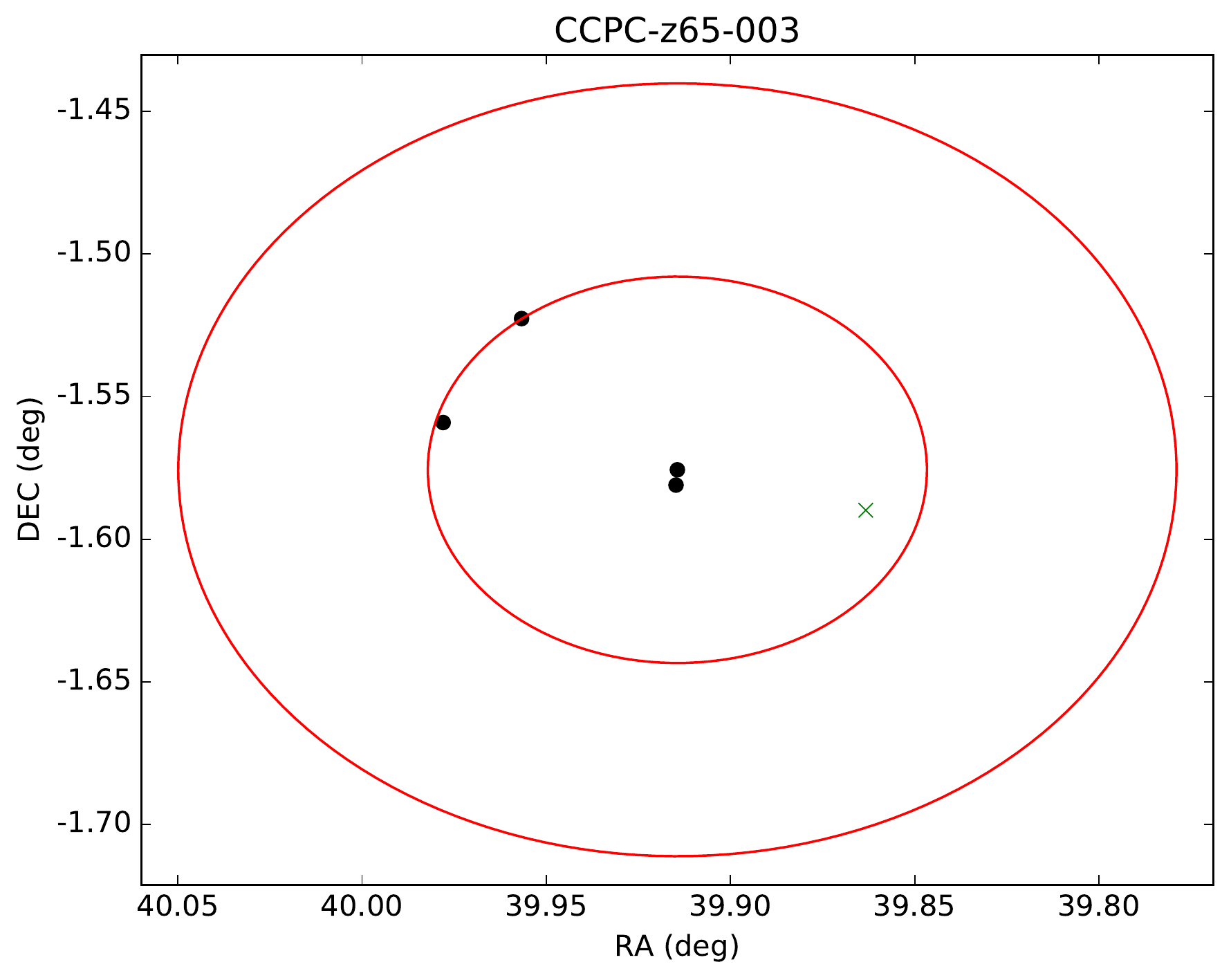}
\label{fig:CCPC-z65-003_sky}
\end{subfigure}
\hfill
\begin{subfigure}
\centering
\includegraphics[scale=0.52]{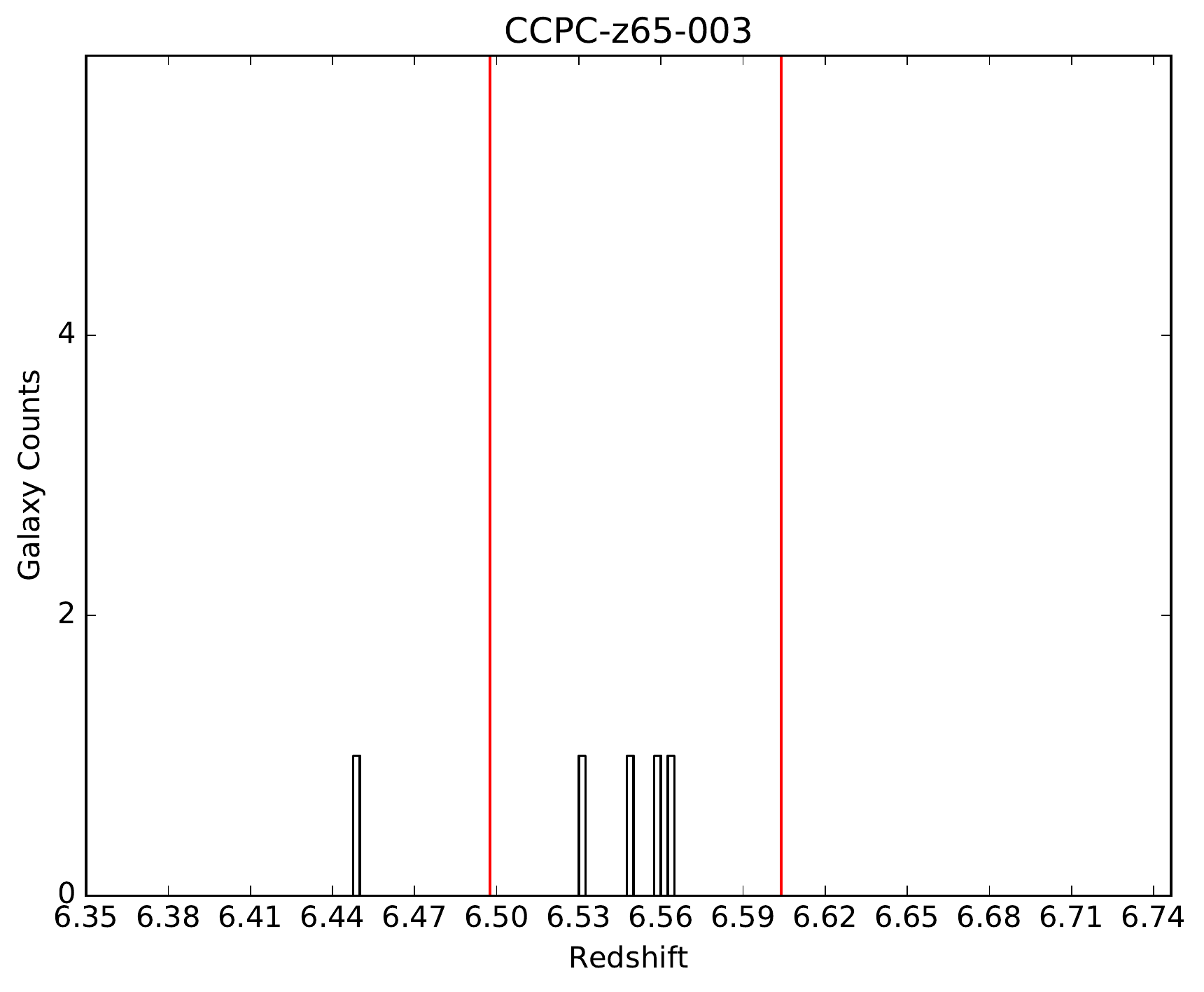}
\label{fig:CCPC-z65-003}
\end{subfigure}
\hfill
\end{figure*}
\clearpage 

\begin{figure*}
\centering
\begin{subfigure}
\centering
\includegraphics[height=7.5cm,width=7.5cm]{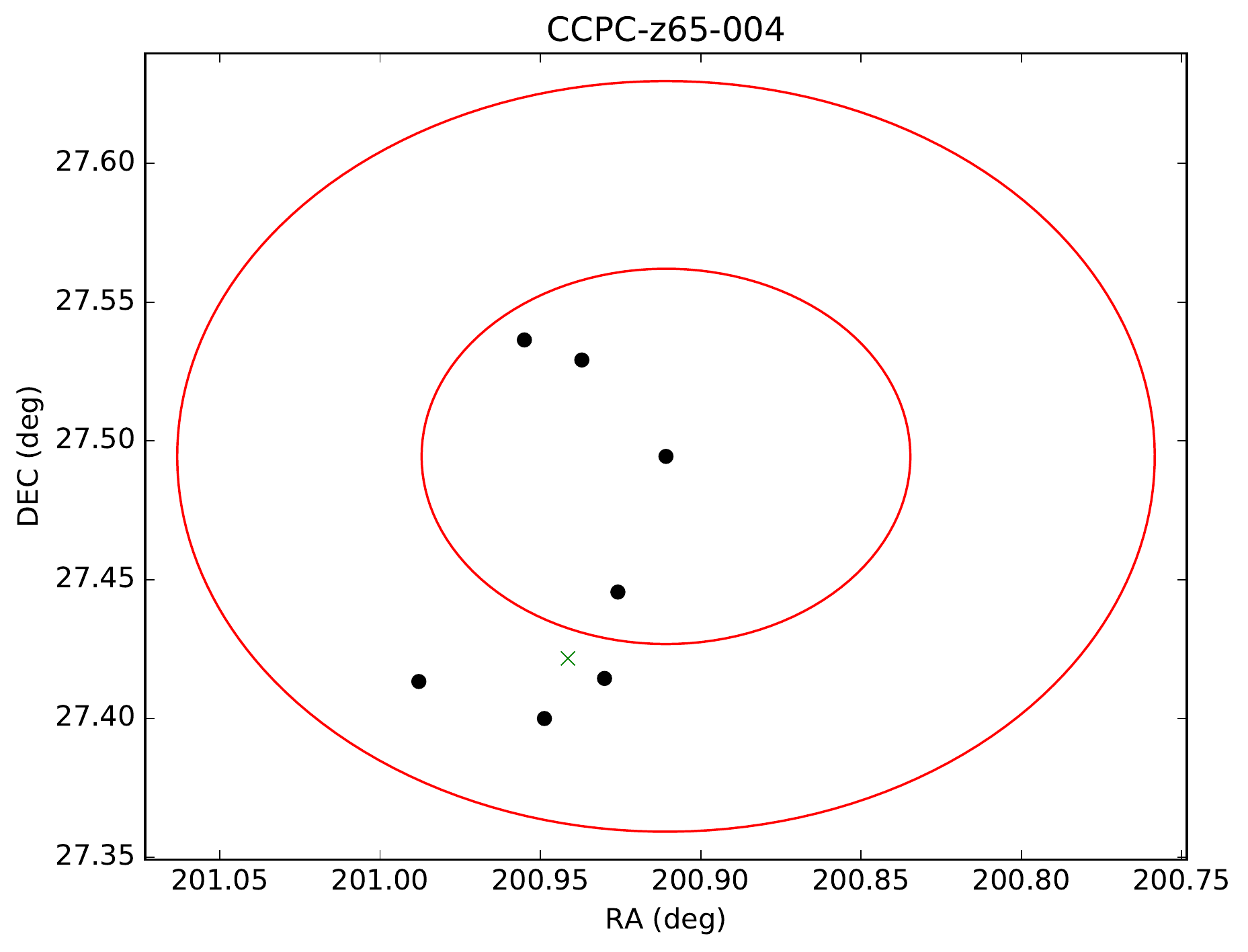}
\label{fig:CCPC-z65-004_sky}
\end{subfigure}
\hfill
\begin{subfigure}
\centering
\includegraphics[scale=0.52]{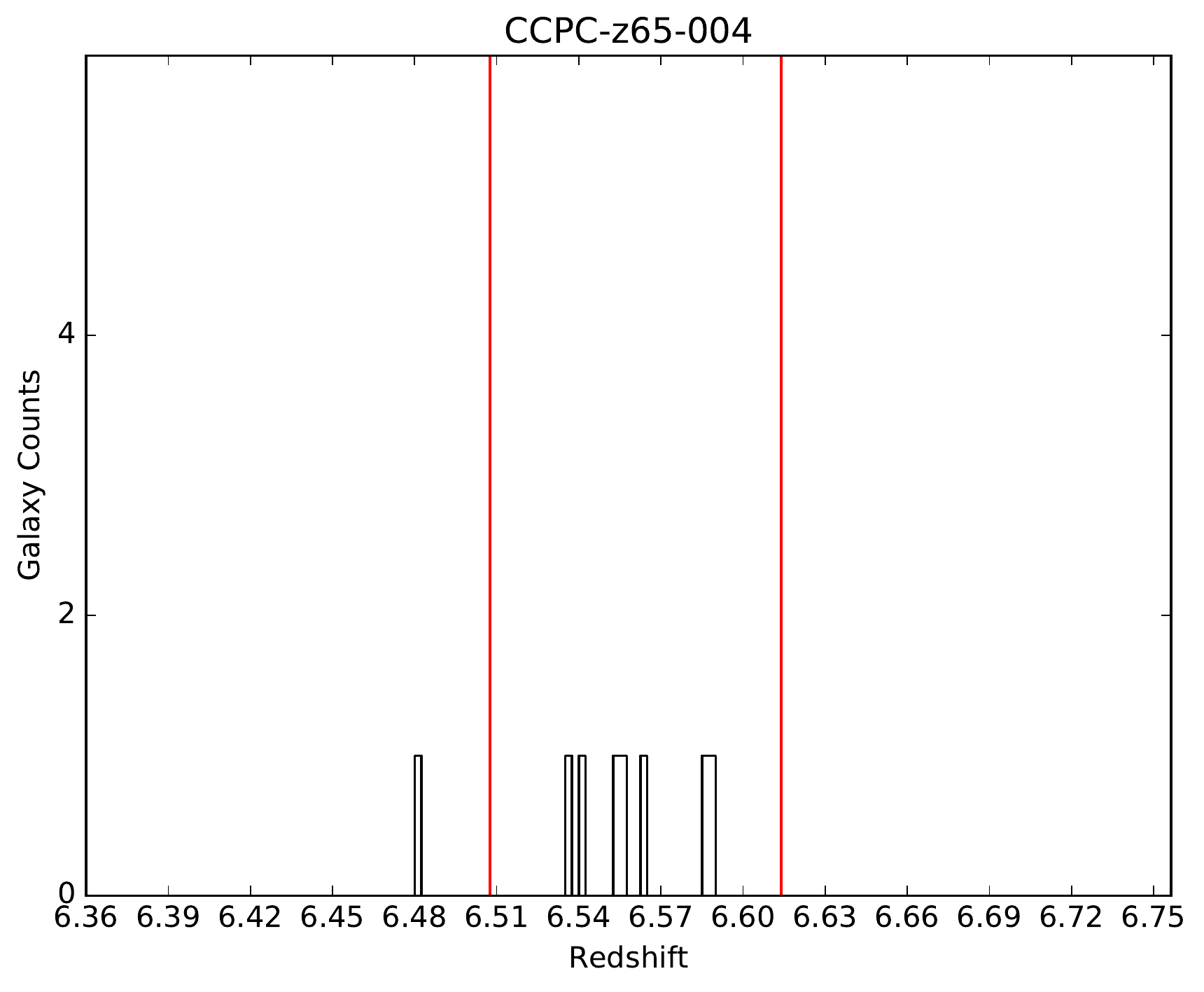}
\label{fig:CCPC-z65-004}
\end{subfigure}
\hfill
\end{figure*}

\begin{figure*}
\centering
\begin{subfigure}
\centering
\includegraphics[height=7.5cm,width=7.5cm]{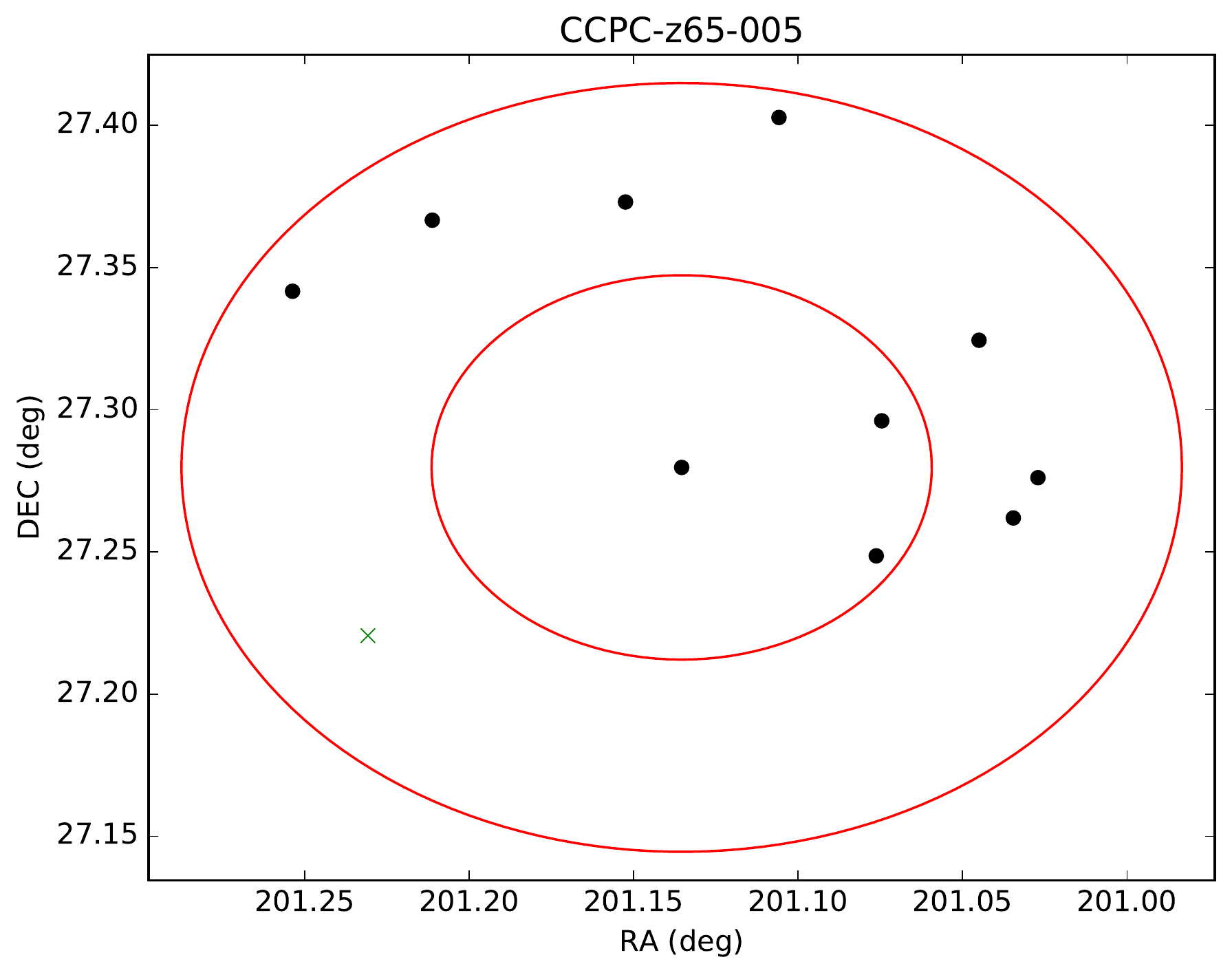}
\label{fig:CCPC-z65-005_sky}
\end{subfigure}
\hfill
\begin{subfigure}
\centering
\includegraphics[scale=0.52]{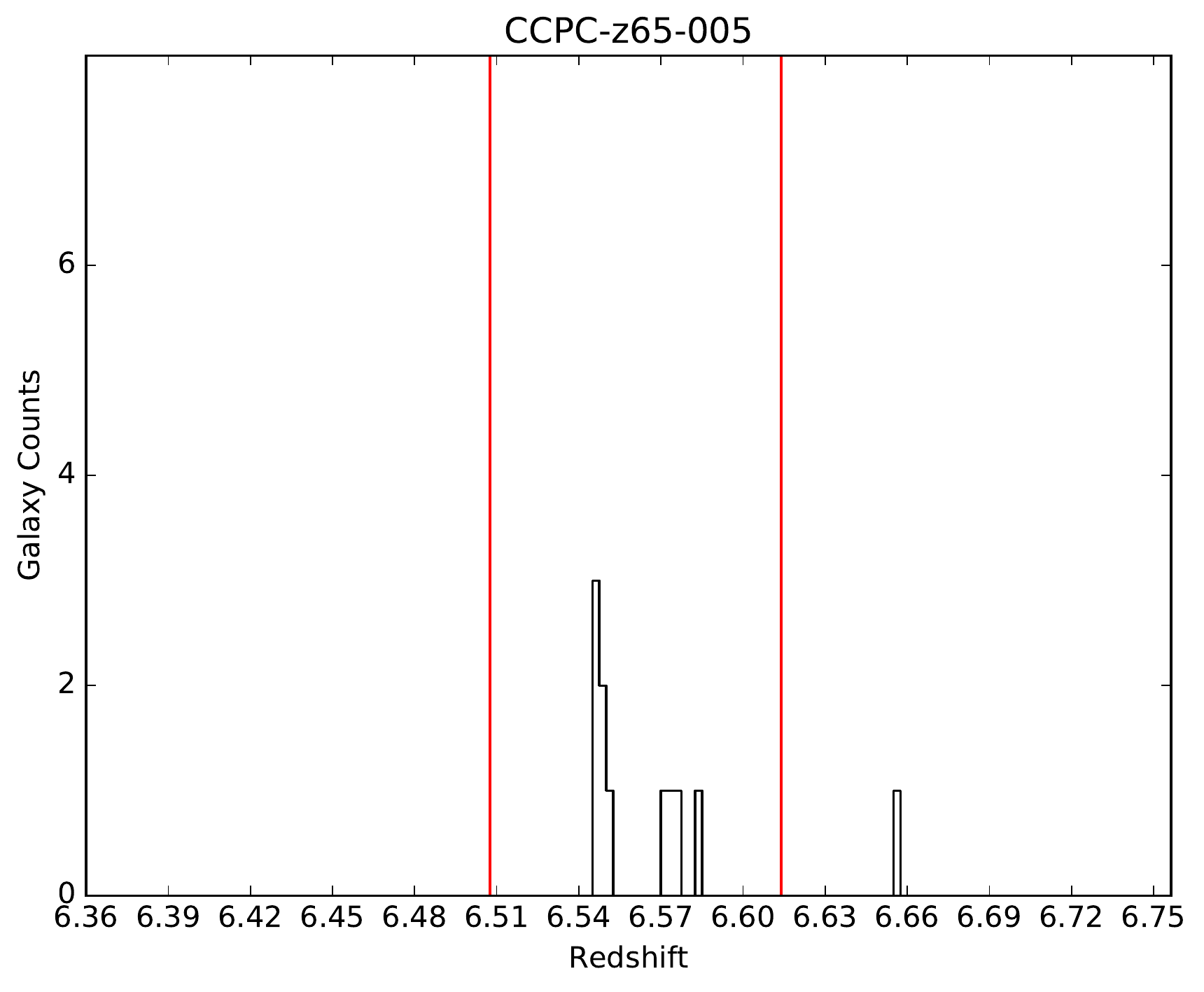}
\label{fig:CCPC-z65-005}
\end{subfigure}
\hfill
\end{figure*}

\end{document}